\documentclass[pdftex]{panda_book}
\usepackage[numbers]{natbib}
\usepackage{chapterbib}
\usepackage{float}
\pdfoutput=1
\markpanda{Strong interaction studies with antiprotons}{Technical Design Report
- TRK}
\begin{document}
%
% TRK TDR - Main Sections
%
\pagenumbering{roman}
\onecolumn
%
% PAGE I - Title and Abstract and Figure
%
%\vspace*{0.5cm}
\begin{center}
{\bfseries \sffamily \huge Technical Design Report for the:\\ \ \\ 
\Panda{}\\  
Straw Tube Tracker \\ \ \\
{\sffamily \small (Anti\underline{P}roton \underline{An}nihilations at \underline{Da}rmstadt)}\\
\ \\ Strong Interaction Studies with Antiprotons}
\vskip 1cm
{\large \Panda{} Collaboration}
%
%removed for final version
%\vskip 0.5cm
%\fbox{\today}
%
\end{center}
\vskip 1cm
\begin{center}
% put a typical picture here
\includegraphics[width=1.7\swidth]{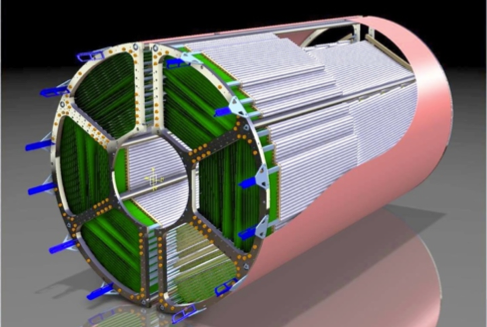}
\end{center}
\begin{center}
\vskip -0.5cm 
{\bfseries \sffamily \Huge STT\\ 
{\large \panda}}
\end{center}
\vfill
%
% PAGE II-IV - Collaboration
%
\newpage
\begin{center}
\vspace*{3mm }
{\LARGE \bfseries \sffamily The \Panda{} Collaboration}
\vskip 7mm
%
% list of institutions
%
\institem{Universit\"at {\bf Basel}, Switzerland}
\authitem{W.~Erni},
\authitem{I.~Keshelashvili},
\authitem{B.~Krusche},
\authitem{M.~Steinacher}\lastitem
\institem{Institute of High Energy Physics, Chinese Academy of Sciences, {\bf Beijing}, China}
\authitem{Y.~Heng},
\authitem{Z.~Liu},
\authitem{H.~Liu},
\authitem{X.~Shen},
\authitem{Q.~Wang},
\authitem{H.~Xu}\lastitem
\institem{Universit\"at {\bf Bochum}, I. Institut f\"ur Experimentalphysik, Germany}
\authitem{A.~Aab},
\authitem{M.~Albrecht},
\authitem{J.~Becker},
\authitem{A.~Csap\'o},
\authitem{F.~Feldbauer},
\authitem{M.~Fink},
\authitem{P.~Friedel},
\authitem{F.H.~Heinsius},
\authitem{T.~Held},
\authitem{L.~Klask},
\authitem{H.~Koch},
\authitem{B.~Kopf},
\authitem{S.~Leiber}
\authitem{M.~Leyhe},
\authitem{C.~Motzko},
\authitem{M.~Peliz\"aus},
\authitem{J.~Pychy},
\authitem{B.~Roth},
\authitem{T.~Schr\"oder},
\authitem{J.~Schulze},
\authitem{C.~Sowa},
\authitem{M.~Steinke},
\authitem{T.~Trifterer},
\authitem{U.~Wiedner},
\authitem{J.~Zhong}\lastitem
% ...
%\centerline{$^\dagger$\ Membership to be approved by the Coordination Board}
%
% EOF
%

\end{center}
%
% Spokespersons
%
\vfill
\hrulefill\\
\begin{tabbing}
Editors:  \hspace{3cm} \= Paola Gianotti  \hspace{1cm}  \= Email: \verb$gianotti@lnf.infn.it$ \\
                       \> Peter Wintz \>  Email: \verb$p.wintz@fz-juelich.de$ \\ \ \\
%                       \> Jerzy Smyrski \> Email: \verb$jerzy.smyrski@uj.edu.pl$ \\ \ \\
Technical Coordinator: \> Lars Schmitt \> Email: \verb$l.schmitt@gsi.de$ \\ 
Deputy:  \> Bernd Lewandowski \> Email: \verb$b.lewandowski@gsi.de$ \\ \ \\
Physics Coordinator: \> Diego Bettoni \> Email: \verb$diego.bettoni@fe.infn.it$\\
Deputy: \> Albrecht Gillitzer \> Email: \verb$a.gillitzer@fz-juelich.de$\\ \ \\
Computing coordinator: \> Johan Messchendorp \> Email: \verb$messchendorp@kvi.nl$\\
Deputy: \> Stefano Spataro \> Email: \verb$stefano.spataro@to.infn.it$\\ \ \\

Spokesperson: \>  Ulrich Wiedner \> Email: \verb$ulrich.wiedner@ruhr-uni-bochum.de$ \\
Deputy:  \> Paola Gianotti  \> Email: \verb$paola.gianotti@lnf.infn.it$ \\
\end{tabbing}
\hrulefill\\
%STT Group:
%\begin{tabbing}
%\hspace{0.1\textwidth}\=\hspace{0.3\textwidth}\= \hspace{0.3\textwidth} \=\hspace{0.3\textwidth}\\
%\>W.~Bardan$^1$ \>             $^2$ \>             M.~Ramm$^3$ \\
%
%
%\end{tabbing}
%$^1$ Jagiellonian University, {\bf Cracow}, Poland\\
%\hrulefill\\
%
%
% draft: statistics
%
%\COM{...Statistics for this draft version: \arabic{auth} authors in \arabic{inst} institutes...}
\vfill
%
% preamble
%
\cleardoublepage
% preamble.tex
%
\begin{center}
\vspace*{2cm}
{\Large \bfseries \sffamily Preface}\addcontentsline{toc}{chapter}{Preface}
\vskip 2cm
\begin{minipage}[t]{8cm}
\sloppy\large
This document describes the technical layout and the expected performance of the Straw Tube Tracker (STT), the 
main tracking detector of the \Panda target spectrometer. The STT encloses a Micro-Vertex-Detector (MVD) for the inner 
tracking and is followed in beam direction by a set of GEM-stations. The tasks of the STT are the measurement of the 
particle momentum from the reconstructed trajectory and the measurement of the specific energy-loss for a particle identification. 
Dedicated simulations with full ana\-ly\-sis studies of certain proton-antiproton reactions, identified as being benchmark tests for the 
whole \Panda scientific program, have been performed to test the STT layout and performance. The results are presented, and the time 
lines to construct the STT are described.  
\end{minipage}
\end{center}
\vspace*{2cm}
\centerline{
% add a detector picture here
%\includegraphics[width=0.8\dwidth]{./main/STT.pdf}
}
\clearpage
\vspace*{18cm}
\hrulefill\\
\vspace*{2cm}\\
\begin{minipage}[t]{10cm}
\sloppy
The use of registered names, trademarks, \etc in this publication does not
imply, even in the absence of specific statement, that such names are exempt
from the relevant laws and regulations and therefore free for general use.
\end{minipage}
\vfill
% EOF
%

%
% Table of contents
%
\cleardoublepage
\tableofcontents
%
% EOF
%

%
% Introduction
%
% FILE: stt_tdr_int.tex
%
% PANDA STT TDR
% Main File for Chapter
% ``Introduction''
%
\cleardoublepage
\pagenumbering{arabic}
\setcounter{page}{1}
%\chapter{Introduction}
%\label{sec:int}

%\begin{bibunit}
\chapter{The \Panda Experiment and its Tracking Concept}
%\addcontentsline{toc}{chapter}{Overview of the \Panda experiment and the overall tracking concept}

%Author: T. Würschig , contact: t.wuerschig-at-hiskp.uni-bonn.de

The following sections contain a general introduction to the \Panda experiment
and, in particular, a short description of the implemented overall tracking concept.
They belong to a common introductory part for the volumes 
of all individual tracking systems.

\section{The \Panda Experiment}
%\section*{I.\hspace{6mm} The \Panda experiment}
%\addcontentsline{toc}{section}{I.\hspace{6mm} The \Panda experiment}

\begin{figure}[!b]
\begin{center}
\vspace{-4mm}
 \includegraphics[width=7.5 cm]{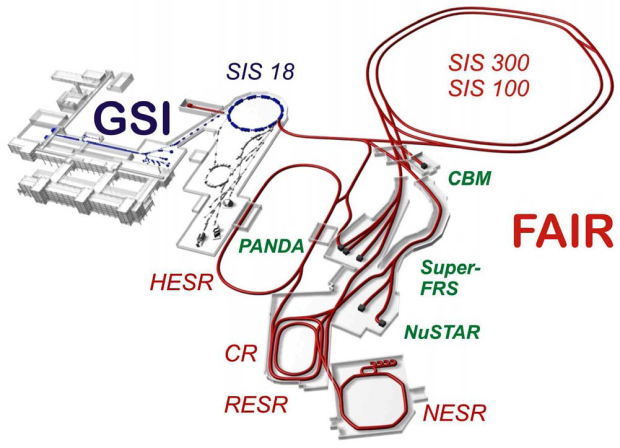}
% Pic3-01_MVD-BTS-Geometry.png: 1239x931 pixel, 125dpi, 25.18x18.92 cm, bb=0 0 714 536 
\caption[Overview of the future \Fair facility]
{Overview of the future \Fair facility. 
The upgraded accelerators of the existing GSI facility will act as injectors. 
New accelerator and storage rings are highlighted in red, 
experimental sites are indicated with green letters.
}
\label{Pic-FAIR}
\end{center}
\end{figure}

The \Panda (Anti\textbf{P}roton \textbf{AN}nihilation 
at \textbf{DA}rmstadt) experiment~\cite{PANDA-LOI}
is one of the key projects at the future 
\textbf{F}acility for \textbf{A}ntiproton and \textbf{I}on \textbf{R}esearch (\Fair)~\cite{FAIR1}\,\cite{Greenpaper}, 
which is currently under construction at GSI, Darmstadt. 
For this new facility the present GSI accelerators 
will be upgraded and further used as injectors. 
The completed accelerator facility will feature 
a complex structure of new accelerators and storage rings. 
An overview of the \Fair facility is given in \Reffig{Pic-FAIR}. 
Further details of the accelerator complex are described in~\cite{FAIR-Accelerator}. 
The \Fair accelerators will deliver primary proton and ion beams 
as well as secondary beams of antiprotons 
or radioactive ions, all with high energy, high intensity and high quality. 
Experiments to be installed at the facility will address 
a wide range of physics topics in the fields of 
nuclear and hadron physics as well as in atomic and plasma physics. 
An executive summary of the main \Fair projects can be found in~\cite{FAIR1} and~\cite{FAIR2}.

\begin{figure*}[tp]
\begin{center}
\includegraphics[width=\dwidth]{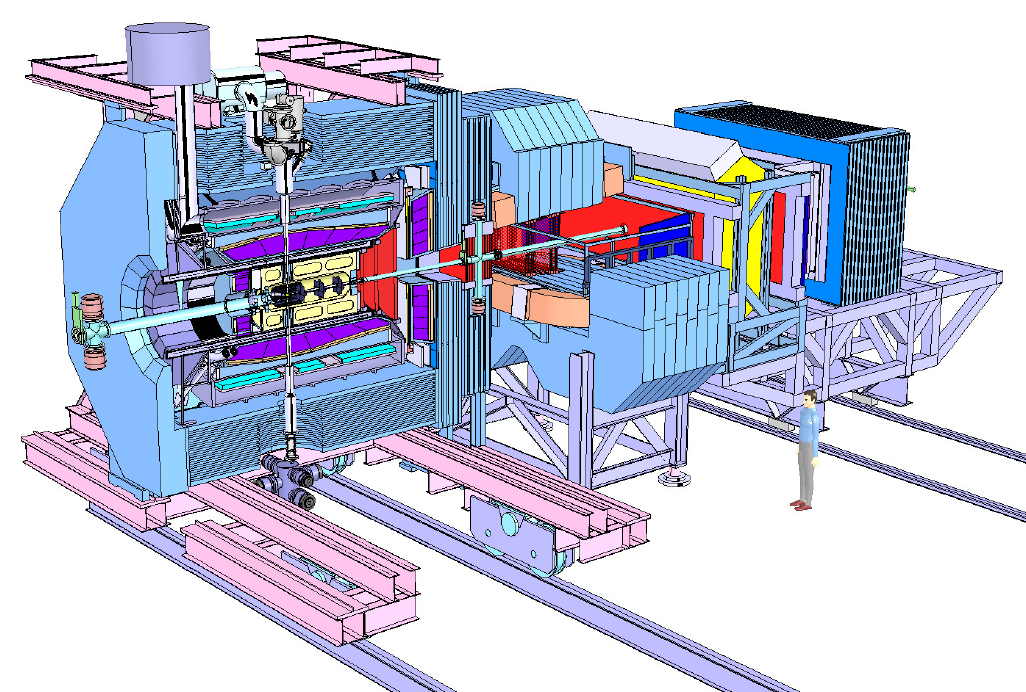}
\caption[Layout of the \Panda detector]
{
Layout of the \Panda detector consisting of a Target Spectrometer, 
surrounding the interaction region, and a 
Forward Spectrometer to detect particles emitted in the forward region. 
The \HESR antiproton beam enters the apparatus from the left side.
}
\label{fig:int:det}
\end{center}
\end{figure*}   

The \Panda experiment will perform precise studies of antiproton-proton annihilations 
and reactions of antiprotons with nucleons of heavier nuclear targets. 
It will benefit from antiproton beams with unprecedented intensity and quality. 
The covered centre-of-mass energy between 2.3~\gev and 5.5~\gev
allows for very accurate measurements, especially in the charm quark sector. 
Based on a broad physics program, studying the non-pertubative regime,
it will be possible to explore the nature of the strong interaction
and to obtain a significant progress in our understanding 
of the QCD spectrum and hadron structure.

Nowadays these studies are carried out mainly at electron-positron machines that offer 
the advantage of kinematically clean reactions 
but at the price of a reduced set of final states and reduced cross-sections.
Also the future experiments currently planned as upgrade 
at existing high-energy physics facilities will not deliver 
high-precision data over the full charm spectrum.
In this context, the \Panda experiment will be a unique tool 
to improve both statistics and precision of existing data 
and to further explore the physics in the charm quark sector.
Moreover, the \Panda collaboration is in the ideal situation 
to be able to benefit from the expertise gained during
the construction of the LHC detectors and of the B-factory experiments,
which have determined a significant progress in the detector technology
due to the performed optimisation or the introduction of
completely new concepts and instruments.

In the first section of this chapter 
the scientific program of \Panda will be summarised. 
It ranges from charmonium spectroscopy 
to the search for exotic hadrons and the study of nucleon structure, 
from the study of in-medium modifications of hadron masses to the physics of 
hypernuclei.
Therefore, antiproton beams in the momentum range from 
1.5~\gevc to 15~\gevc
will be provided by the high-energy storage ring (\HESR) to the experiment.
An overview of this accelerator and storage ring will be given in the second section.
To explore the broad physics program, the \Panda collaboration wants 
to build a state-of-the-art general purpose detector 
studying annihilation reactions of antiprotons with protons (\pbarp) 
and in nuclear matter (\pbarA). 
The different target systems will be discussed in section~\ref{intro:target}.
The \Panda apparatus consists of a set of systems surrounding 
an internal target placed in one of the two straight sections 
of the \HESR.
\Reffig{fig:int:det} shows the layout of the \Panda detector. 
It consists of a 4~m long and 2~T strong superconducting solenoid 
instrumented to detect both charged and neutral particles 
emitted at large and backward angles (Target Spectrometer, TS) 
and of a 2~Tm resistive dipole magnetic spectrometer 
to detect charged and neutral particles emitted at angles between 
zero and twenty degrees (Forward Spectrometer, FS) with respect to the beam axis.
A complex detector arrangement is necessary in order to reconstruct 
the complete set of final states, relevant to achieve the proposed physics goals.
With the installed setup, a good particle identification with an
almost complete solid angle will be combined 
with excellent mass, momentum and spatial resolution.  
More details of the \Panda detector will be described in section~\ref{s:over:panda}.

% FILE: physicsintro.tex
%
\subsection{The Scientific Program}
%\subsection*{I.1\hspace{4mm} The scientific program}
%\addcontentsline{toc}{subsection}{I.1\hspace{4mm} The scientific program}
%\COM{Author(s): D. Bettoni}
\label{sec:sciprog}
One of the most challenging and fascinating 
goals of modern physics is the achievement of a fully quantitative
understanding of the strong interaction, which is the subject of hadron physics.
Significant progress has been achieved over the past few years thanks to
considerable advances in experiment and theory. New experimental results
have stimulated a very intense theoretical activity and a refinement of
the theoretical tools. 

Still there are many fundamental questions which remain basically
unanswered.
Phenomena such as the confinement of quarks, the existence of glueballs 
and hybrids, the origin of the masses of hadrons in the context of the
breaking of chiral symmetry are long-standing puzzles and represent
the intellectual challenge in our attempt to understand the nature of the
strong interaction and of hadronic matter.

Experimentally, studies of hadron structure can be performed with different
probes such as electrons, pions, kaons, protons or antiprotons.
In antiproton-proton annihilation, particles with gluonic degrees of freedom
as well as particle-antiparticle pairs are copiously produced,
allowing spectroscopic studies with very high statistics and precision.
Therefore, antiprotons are an excellent tool to address the open problems.

%The \FAIR facility will provide antiproton beams
%of the highest quality in terms of intensity and resolution, which will
%provide an excellent tool to answer these fundamental questions.

%The \PANDA experiment will use the
%antiproton beam from the \HESR colliding with an
%internal proton
%target and a general purpose spectrometer to carry out a rich and
%diversified hadron physics program.

The \PANDA experiment is being designed to fully exploit the extraordinary 
physics potential arising from the availability of high-intensity, cooled
antiproton beams.
%The aim of the rich experimental program is to improve our knowledge of the 
%strong interaction and of hadron structure.
%Significant progress beyond the present understanding of the field is expected
%thanks to improvements in statistics and precision of the data.
Main experiments of the rich and diversified hadron physics program
are briefly itemised in the following.
More details can be found in the \panda physics booklet~\cite{PANDA:PhysBooklet}.
%Many experiments are foreseen in \PANDA.
\begin{itemize}
\item \textbf{Charmonium Spectroscopy}\\
%While the \ccbar spectrum below the \DDbar threshold can be computed within 
%the framework of non-relativistic potential models and, more recently, in Lattice QCD,
%the same tools fail above that threshold.
A precise measurement of all states below and above the
open charm threshold is of fundamental importance for a better understanding of QCD.
All charmonium states can be formed directly in \pbarp annihilation.
At full luminosity \PANDA will be able 
to collect several thousand \ccbar states per day.
By means of fine scans it will be possible to measure masses with accuracies 
of the order of 100~\kev and widths to 10\% or better.
The entire energy region below and above the open charm threshold will be explored.

\item \textbf{Search for Gluonic Excitations}\\
One of the main challenges of hadron physics
is the search for gluonic excitations, 
i.e.~hadrons in which the gluons can act as principal components.
These gluonic hadrons fall into two main categories: glueballs, 
i.e.~states of pure glue, and hybrids, which consist of a \qqbar pair and excited glue.
The additional degrees of freedom carried by gluons allow these hybrids and glueballs
to have \JPC exotic quantum numbers: in this case mixing effects with nearby
\qqbar states are excluded and this makes their experimental identification easier.
The properties of glueballs and hybrids are determined by the long-distance features
of QCD and their study will yield fundamental insight into the structure of the QCD
vacuum.
Antiproton-proton annihilations provide 
a very favourable environment in which to look for gluonic hadrons. 

\item \textbf{Study of Hadrons in Nuclear Matter}\\
The study of medium modifications of hadrons embedded in hadronic matter
is aiming at understanding the origin of hadron masses in the context
of spontaneous chiral symmetry breaking in QCD and its partial restoration
in a hadronic environment. 
So far experiments have been focussed on the light quark sector.
The high-intensity \pbar beam of up to 15~\gevc will allow an extension of
this program to the charm sector both for hadrons with hidden and open charm.
The in-medium masses of these states are expected to be affected primarily
by the gluon condensate.

Another study which can be carried out in \PANDA is the measurement of \jpsi
and \D meson production cross sections in \pbar annihilation
on a series of nuclear targets. The
comparison of the resonant \jpsi yield obtained from \pbar annihilation
on protons and different nuclear targets allows to deduce the
\jpsi-nucleus dissociation cross section, a fundamental parameter to 
understand \jpsi suppression in relativistic heavy ion collisions interpreted
as a signal for quark-gluon plasma formation. 

\item \textbf{Open Charm Spectroscopy}\\ 
The \HESR, running at full luminosity and at \pbar
momenta larger than 6.4~\gevc, would produce a large number
of \D meson pairs.
The high yield and the well defined production kinematics of \D meson pairs 
would allow to carry out a significant charmed meson spectroscopy program 
which would include, for example, the rich \D and \Ds meson spectra.

\item \textbf{Hypernuclear Physics}\\
Hypernuclei are systems in which 
neutrons or protons are replaced by hyperons. 
In this way a new quantum number, strangeness, 
is introduced into the nucleus.
Although single and double $\Lambda$-hypernuclei were discovered many decades ago, only 6 double 
$\Lambda$-hypernuclei are presently known. 
The availability of \pbar beams at \FAIR will allow efficient production
of hypernuclei with more than one strange hadron, making \PANDA competitive
with planned dedicated facilities. This will open new perspectives for
nuclear structure spectroscopy and for studying the forces between hyperons
and nucleons. 

\item \textbf{Electromagnetic Processes}\\
In addition to the spectroscopic studies described above, \PANDA will be able to
investigate the structure of the nucleon using electromagnetic processes, such as
Deeply Virtual Compton Scattering (DVCS) and the process \pbarp~$\to$~\ee, which 
will allow the determination of the electromagnetic form factors of the proton 
in the timelike region over an extended $q^2$ region.
Furthermore, measuring the Drell Yan production of muons 
would give access to the transverse nucelon structure.
\end{itemize}

\begin{figure*}[thb]
\begin{center}
\vspace{-2mm}
\includegraphics[width=14.5cm]{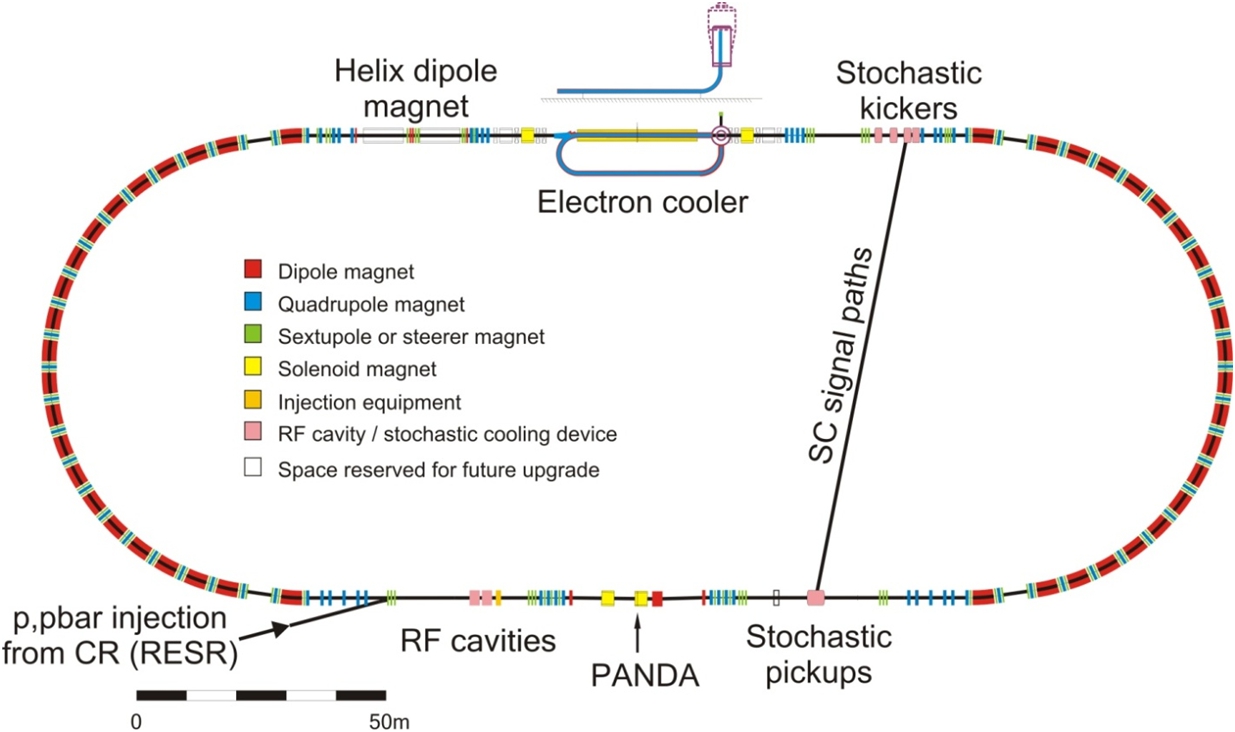}
\vspace{-1mm}
\caption[Layout of the High Energy Storage Ring \HESR]
{Layout of the High Energy Storage Ring \HESR. 
The beam is injected from the left into the lower straight section. %Stochastic cooling and electron cooling is foreseen.  
The location of the \PANDA target is indicated with an arrow.}
\label{f:over:hesr}
\end{center}
\end{figure*}

\subsection{High Energy Storage Ring -- \HESR}
%
%\subsection*{I.2 \hspace{4mm} High Energy Storage Ring -- \HESR}
%\addcontentsline{toc}{subsection}{I.2 \hspace{4mm} High Energy Storage Ring -- \HESR}
\label{s:over:hesr}

\begin{table*}
\caption[Experimental requirements and operation modes of \HESR]{Experimental requirements and operation modes of \HESR for the full \Fair version.}
\smallskip
\begin{center}
\scalebox{0.85}{
\begin{tabular}{l l} 
\hline
  \multicolumn{2}{c}{\textbf{Experimental Requirements}}\\ 
\hline
Ion species & Antiprotons \\
$\bar{p}$ production rate &
$2\cdot 10^7 ${/s} ($1.2\cdot 10^{10}$ per 10~min) \\
Momentum / Kinetic energy range &
1.5 to 15~\gevc / 0.83 to 14.1~\gev \\
Number of particles &
$10^{10}$ to $10^{11}$ \\
%Target thickness &
%$4\times 10^{15} ${atoms/cm$^2$} (H$_2$ pellets) \\
%Transverse emittance &
%$<$ 1 {mm$\times$mrad} \\
%Beam size (radius) at IP & $\sim$ 1\,mm (RMS) \\
Betatron amplitude at IP &
1~m to 15~m \\
Betatron amplitude E-Cooler &
25~m to 200~m \\
\vspace{-2mm} & \\ 
%\hline
\hline
  \multicolumn{2}{c}{\textbf{Operation Modes}}\\ 
\hline
High resolution (HR) & Peak Luminosity of 2$\cdot
10^{31}${cm$^{-2}$s$^{-1}$} for $10^{10}\
\bar{p}$ \\
 & assuming $\rho_{target}$ = $4\cdot 10^{15}$ atoms/cm$^2$\\
 & RMS momentum spread $\sigma_p / p \leq 4\cdot 10^{-5}$, \\
 & 1.5 to 8.9 \gevc \\
High luminosity (HL) & Peak Luminosity up to 2$\cdot
10^{32}${cm$^{-2}$s$^{-1}$} for $10^{11}\
\bar{p}$ \\
 & assuming $\rho_{target}$ = $4\cdot 10^{15}$ atoms/cm$^2$\\
 & RMS momentum spread  $\sigma_p / p \sim 10^{-4}$,\\
 & 1.5 to {15} \gevc  \\
 \hline
\end{tabular}
}%scalebox
\label{t:hesr1}
\end{center}
\end{table*}

The \HESR is dedicated to supply \PANDA with 
high intensity and high quality antiproton
beams over a broad momentum range from 1.5~\gevc to 15~\gevc~\cite{HESR:FAIR:2008}.
\Reftbl{t:hesr1} summarises the experimental requirements 
and main parameters of the two operation modes for the full \Fair version.
The High Luminosity (HL) and the High Resolution (HR) mode 
are established to fulfil all challenging specifications
for the experimental program of \PANDA~\cite{hesr:lehrach:2009}.
The HR mode is defined in the momentum range from 1.5~\gevc to 9~\gevc. 
To reach a relative momentum spread down to the order of
$10^{-5}$, only $10^{10}$ circulating particles in the ring are
anticipated. 
The HL mode requires an order of magnitude higher beam
intensity with reduced momentum resolution to reach a peak luminosity
of 2$\cdot 10^{32}${cm$^{-2}$s$^{-1}$} 
in the full momentum range up to 15~\gevc.
To reach these beam parameters a very powerful phase-space cooling is needed.
Therefore, high-energy electron cooling~\cite{Ecool08} 
and high-bandwidth stochastic cooling~\cite{HESR-CoolingScenario} will be utilised.

The \HESR lattice is designed as a racetrack shaped ring with a
maximum beam rigidity of 50~Tm (see \Reffig{f:over:hesr}). 
It consists of two 180$^{\circ}$ arcs and two 155~m long straight sections 
with a total circumference of 575~m~\cite{HESR-Lattice}. 
The arc quadrupole magnets will allow for a flexible adjustment of transition energy, 
horizontal and vertical betatron tune as well as horizontal dispersion.
In the straight section opposite to the injection point, an electron cooler will be installed. 
The \panda detector with the internal target is placed at the other side. 
Further components in the straight \panda section are 
beam injection kickers, septa and multi-harmonic RF cavities. 
The latter allow for a compensation of energy losses due to the beam-target interaction, 
a bunch rotation and the decelerating 
or accelerating of the beam. 
Stochastic cooling is implemented via several kickers and opposing high-sensitivity pick-ups 
on either side of the straight sections. 

\begin{figure}[!b]
\vspace{-2mm}
\begin{center}
\includegraphics[width=\swidth]{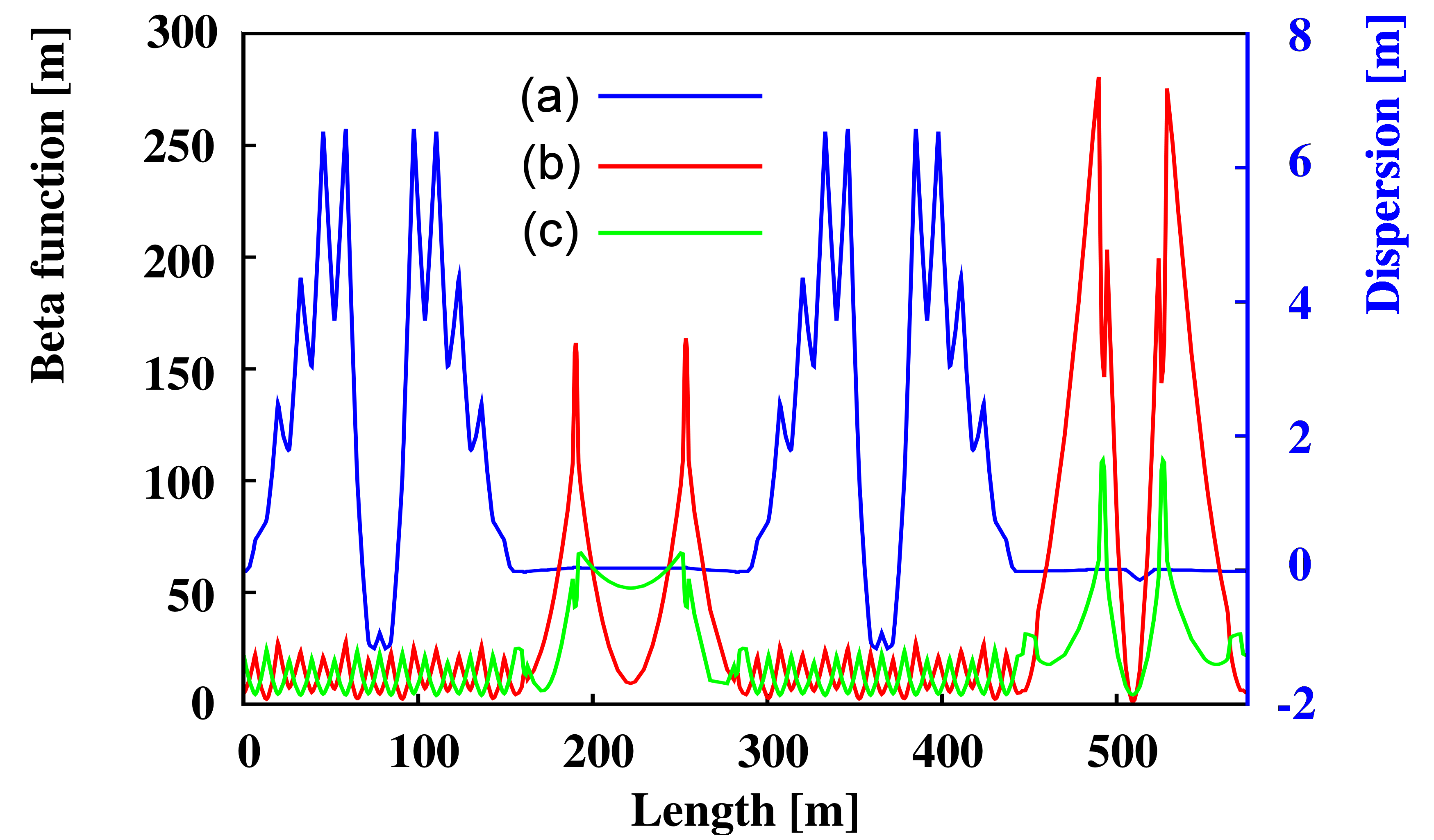}
\vspace{-2mm}
\caption[Optical functions of \HESR lattice for $\gamma_{tr}$ = 6.2]
{Optical functions of the $\gamma_{tr}$ = 6.2 lattice:
Horizontal dispersion (a), horizontal (b) and vertical (c) betatron function. 
Electron cooler and target are located at
a length of 222~m and 509~m, respectively.
}
\label{f:hesr:optics}
\end{center}
\end{figure}

Special requirements for the lattice are low dispersion in the
straight sections and small betatron amplitudes in the range between 
1~m and 15~m at the internal interaction point (IP) of the \PANDA detector. 
In addition, the betatron amplitude at the electron cooler 
must be adjustable within a large range between 25~m and 200~m.
%Both betatron tunes will roughly be 7.62 for different optical
%settings and natural chromaticities will be ranging 
%$-$12 to $-$17 in X and from $-$10 to $-$13 in Y.  
%There are by now four defined optical settings: 
%Injection, $\gamma_{tr}$~$=$~6.2,\, $\gamma_{tr}$~$=$~13.4,\, $\gamma_{tr}$~=~33.2.
Examples of the optical functions for one of the defined optical settings are shown in \Reffig{f:hesr:optics}. 
The deflection of the %large aperture 
spectrometer dipole magnet of the \Panda detector 
will be compensated by two dipole magnets that create a beam chicane.
These will be placed 4.6~m upstream and 13~m downstream the \Panda IP
thus defining a boundary condition for the quadrupole elements closest to the experiment.
For symmetry reasons, they have to be placed at $\pm$14~m with respect to the IP. 
The asymmetric placement of the chicane dipoles will result 
in the experiment axis occurring at a small angle with respect
to the axis of the straight section.
The \PANDA solenoid will be compensated by one solenoid magnet. 
%Another four 
Additional correction dipoles 
have to be included around the electron cooler 
%to correct the beam deflection 
due to the toroids  that will be used to overlap the electron beam with the antiproton beam.
Phase-space coupling induced by the electron cooler solenoid will be 
compensated by two additional solenoid magnets.

Closed orbit correction and local orbit bumps at dedicated locations in the ring 
are crucial to meet requirements for the beam-target interaction 
in terms of maximised ring acceptance and optimum beam-target overlap \cite{Wel08}.
The envisaged scheme aims on a reduction of
maximum closed orbit deviations to below 5~mm 
while not exceeding 1~mrad of corrector strength.
Therefore, 64 beam position monitors and 
48 orbit correction dipoles are intended to be used.
Because a few orbit bumps will have to be used %at a few positions 
in the straight parts of the \HESR,
all correction dipoles therein %for orbit corrections in these sections 
are designed to provide an additional deflection strength of 1~mrad. 

Transverse and longitudinal cooling will be used to compensate a transverse beam blow up 
and to achieve a low momentum spread, respectively. %~\cite{HESR-CoolingScenario}.
While stochastic cooling will be applicable in the whole momentum range, 
electron cooling is foreseen in a range from 1.5~\gevc to 8.9~\gevc with a possible upgrade to 15~\gevc.
The relative momentum spread can be further improved by combining both cooling systems.
Beam losses are dominated by hadronic interactions between antiprotons and target protons, 
single large-angle Coulomb scattering in the target and energy straggling induced 
by Coulomb interactions of the antiprotons with target electrons. 
%The relative momentum acceptance of the \HESR ring is restricted to about 1$\cdot$10$^{-3}$. 
Mean beam lifetimes for the \HESR range between 1540~s and 7100~s. 
The given numbers correspond to the time, after which the initial beam intensity is reduced by a factor of $1/e$.
A detailed discussion of the beam dynamics and beam equliibria for the \HESR can be found 
in~\cite{hesr:lehrach:2009,lehrach:2006,hinterberger:2006,BoineFrankenheim:2006ci}.
Advanced simulations have been performed for both cooling scenarios.
In case of electron cooled beams the RMS relative momentum spread obtained for the HR mode 
ranges from $7.9\cdot 10^{-6}$ (1.5~\gevc) to $2.7\cdot 10^{-5}$ (8.9~\gevc), 
and $1.2\cdot 10^{-4}$ (15~\gevc) \cite{Rei07}.
%Results were obtained with an empirical magnetised cooling force formula~\cite{parkhomchuk:2000} 
%and an analytical description for intra-beam scattering~\cite{sorensen:1987}.
%Beam heating by beam-target interaction is described by transverse and
%longitudinal emittance growth due to Coulomb scattering and energy
%straggling~\cite{hinterberger:1989a,hinterberger:1989b}.
With stochastic cooling in a bandwidth of 2~GHz to 6~GHz, 
%are based on a Fokker-Planck approach.
%In this case, 
the RMS relative momentum spread for the HR mode results in 
$5.1\cdot 10^{-5}$ (3.8~\gevc), $5.4\cdot 10^{-5}$ (8.9~\gevc)
and $3.9\cdot 10^{-5}$ (15~\gevc)~\cite{Sto08}.
In the HL mode a RMS relative momentum spread of roughly $10^{-4}$ can be expected.
Transverse stochastic cooling can be adjusted independently to
ensure sufficient beam-target overlap.

\begin{figure*}[thb]
\begin{center}
\includegraphics[width=15 cm]{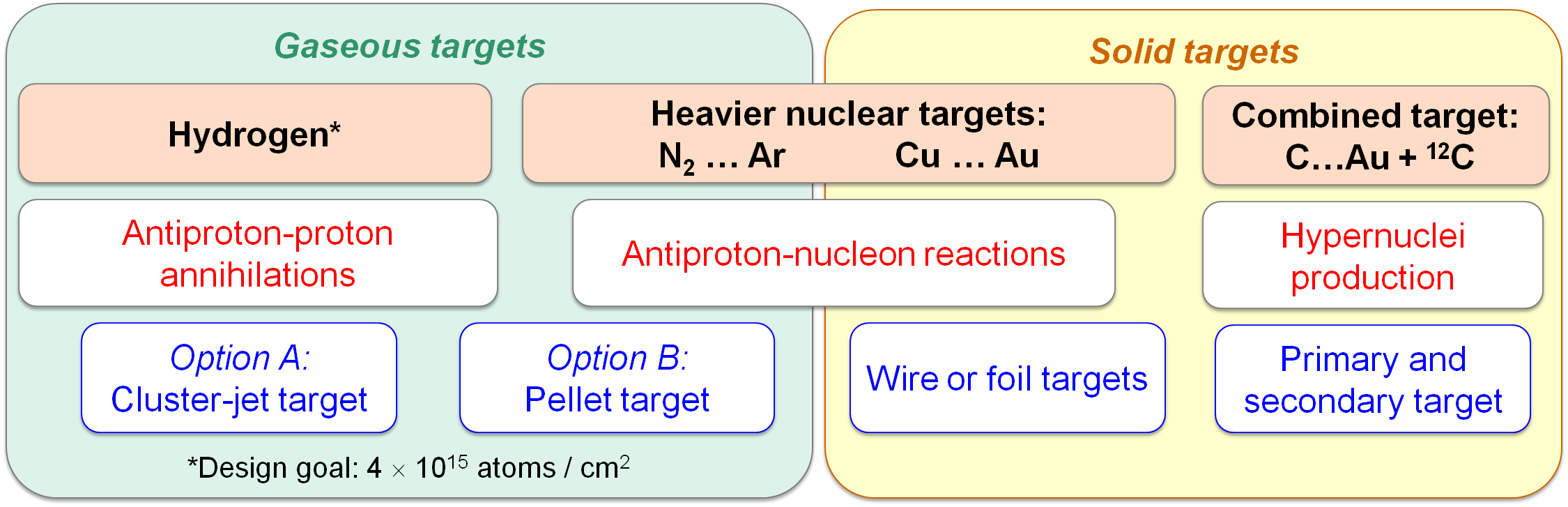}
\caption
[Summary of the different target options foreseen at \panda]
{Summary of the different target options foreseen at \panda.}
\label{pic-target:options}
\end{center}
\end{figure*}

\subsection{Targets}
\label{intro:target}
%\subsection*{I.3 \hspace{4mm} Targets}
%\addcontentsline{toc}{subsection}{I.3 \hspace{4mm} Target systems}
The design of the solenoid magnet allows for an implementation of different target systems. 
\panda will use both gaseous and non-gaseous targets. 
A very precise positioning of the target is crucial for the exact definition of 
the primary interaction vertex. 
In this context, big challenges for either system result from the long distance of 
roughly 2~m between the target injection point and the dumping system. 
Hydrogen target systems will be used for the study of antiproton-proton reactions. 
A high effective target density of about $4\cdot10^{15}$ hydrogen atoms per square centimetre 
must be achieved to fulfill the design goals of the high luminosity mode. 
Besides the application of hydrogen as target material, an extension to heavier gases 
such as deuterium, nitrogen or argon is planned 
for complementary studies with nuclear targets. 

At present, two different solutions are under development: 
a cluster-jet and a pellet target. 
Both will potentially provide sufficient target thickness but exhibit different properties 
concerning their effect on the beam quality and the definition of the IP. 
Solid targets are foreseen for hyper-nuclear studies and the study of antiproton-nucleus 
interaction using heavier nuclear targets. 
The different target options are shortly described in the following.
\Reffig{pic-target:options} gives an overview to all target option foreseen at \panda.

\subsubsection*{Cluster Jet Target}
%\addcontentsline{toc}{subsection}{\hspace{4mm}...  Cluster Jet Target}

Cluster jet targets provide a homogeneous and adjustable target density without any time structure. 
Optimum beam conditions can be applied in order to achieve highest luminosity. 
The uncertainty of the IP in a plane perpendicular to the beam axis is defined by 
the optimised focus of the beam only. 
An inherent disadvantage of cluster-jet targets is the lateral spread of the cluster jet leading 
to an uncertainty in the definition of the IP along the beam axis of several millimetres. 

For the target production a pressurised cooled gas is injected into vacuum through a nozzle. 
The ejected gas immediately condensates and forms a narrow supersonic jet of molecule clusters.  
The cluster beam typically exposes a broad mass distribution which strongly depends on the gas 
input pressure and temperature.
In case of hydrogen, the average number of molecules per cluster varies from 10$^3$ to 10$^6$. 
The cluster-jets represent a highly diluted target and offer a very homogenous density profile. 
Therefore, they may be seen as a localised and homogeneous monolayer of hydrogen atoms
being passed by the antiprotons once per revolution, i.e.~the antiproton beam can be focused at highest phase space density.
The interaction point is thus defined transversely 
but has to be reconstructed longitudinally in beam direction. 
%The possibility of adjusting the target density along
%with the gradual consumption of antiprotons for running at constant
%luminosity will be an important feature.
At a dedicated prototype cluster target station an effective target density 
of $1.5\cdot 10^{15}$ hydrogen atoms per square centimetre 
has been achieved using the exact \panda geometry \cite{taeschner:2011}.
This value is close to the maximum number required by \panda. 
Even higher target densities seem to be feasible and are topic of ongoing R\&D work.

%\paragraph*{Cluster-Jet Target}
%The expansion of pressurised cold hydrogen gas into vacuum through a
%Laval-type nozzle leads to a condensation of hydrogen molecules
%forming a narrow supersonic jet of hydrogen clusters. The cluster size varies
%from $10^{3}$ to $10^{6}$ hydrogen molecules tending to become larger at higher
%inlet pressure and lower nozzle temperatures.  Such a cluster-jet
%with density of $10^{15}$ atoms/cm$^2$ acts as a very diluted target since it
%
%Fulfilling the luminosity demand for \PANDA still requires a density
%increase compared to current applications. Additionally, due to
%detector constraints, the distance between the cluster-jet nozzle and
%the target will be larger than usual. The great advantage of cluster 
%targets is the
%homogeneous density profile and 

\subsubsection*{Hydrogen Pellet Target}
%\addcontentsline{toc}{subsection}{\hspace{4mm}...  Hydrogen Pellet Target}

Pellet targets provide a stream of frozen molecule droplets, called pellets, 
which drip with a fixed frequency off from a fine nozzle into vacuum.
The use of pellet targets gives access to high effective target densities. 
The spatial resolution of the interaction zone can be reduced by skimmers to a few millimetres. 
A further improvement of this resolution can be achieved  by tracking the individual pellets. 
However, pellet targets suffer from a non-uniform time distribution, 
which results in larger variations of the instantaneous luminosity as compared to a cluster-jet target. 
The maximum achievable average luminosity is very sensitive to deviations 
of individual pellets from the target axis. 
The beam must be widened in order to warrant a beam crossing of all pellets. 
Therefore, an optimisation between the maximum pellet-beam crossing time on the one hand 
and the beam focusing on the other is necessary.   

The design of the planned pellet target is based on the one currently used at the WASA-at-COSY 
experiment~\cite{WASA-Pellet}.
The specified design goals for the pellet size and the mean lateral spread of the pellet train 
are given by a radius of 25~$\mu$m to 40~$\mu$m and 
a lateral RMS deviation in the pellet train
of approximately 1~mm, respectively. 
At present, typical variations of the interspacing of individual pellets range 
between 0.5~mm and 5~mm. 
A new test setup with an improved performance has been constructed~\cite{PANDA-PelletTarget}. 
First results have demonstrated the mono-disperse and satellite-free droplet production 
for cryogenic liquids of H$_2$, N$_2$ and Ar~\cite{PANDA-PelletTarget2009}. 
However, the prototype does not fully include the \panda geometry. 
The handling of the pellet train over a long distance still has to be investigated in detail. 
The final resolution on the interaction point is envisaged to be in the order of 50~$\mu$m. 
Therefore, an additional pellet tracking system is planned.

%\paragraph*{Pellet Target} 
%The pellet target features a stream of frozen hydrogen micro-spheres,
%called pellets, traversing the antiproton beam perpendicularly. 
%Typical parameters for pellets at the interaction point are the rate
%of $15 -150\times 10^{3} s^{-1}$, the pellet size of 
%$10 - 30 \mu m$, and
%the velocity of about 60 m/s. At the interaction point the pellet train
%has a lateral spread of $\approx 1.5$ mm and an interspacing of
%pellets that varies between $0.5$ to 4 mm. With proper adjustment of
%the $\beta$-function of the coasting antiproton beam at the target
%position, the design luminosity for \PANDA can be reached in time
%average. The present R\&D is concentrating on minimising the
%luminosity variations such that the instantaneous interaction rate
%does not exceed the rate capability of the detector systems. Due to the large 
%number of interactions expected in every pellet, and thanks to the
%foreseen pellet tracking system, a resolution in the vertex position 
%of 50 $\mu m$ will be possible with this target.

\subsubsection*{Other Target Options}
%\addcontentsline{toc}{subsection}{\hspace{4mm}...  Other Target Options}

In case of solid target materials the use of wire targets is planned. 
The hyper-nuclear program requires a separate target station in upstream position. 
It will comprise a primary and secondary target. 
The latter must be instrumented with appropriate detectors. 
Therefore, a re-design of the innermost part of the \panda spectrometer becomes necessary.
This also includes the replacement of the MVD. 
%However, measurements with this modified setup are only foreseen at 
%a later stage of the experiment.

%\paragraph*{Other Targets} are under consideration for the
%hypernuclear studies where a separate target station upstream will
%comprise primary and secondary target and detectors. Moreover, current
%R\&D is being undertaken for the development of a liquid helium target. 
%A wire target may be employed to study antiproton-nucleus interactions.

\subsection{Luminosity Considerations}
%\subsection*{I.4 \hspace{4mm} Luminosity Considerations}
%\addcontentsline{toc}{subsection}{I.4 \hspace{4mm} Luminosity considerations}
\label{lumi-considerations}

%All luminosity considerations in this section are discussed for the full \FAIR version. 
The luminosity $L$ describes the flux of beam particles convolved with the target opacity. 
Hence, an intense beam, a highly effective target thickness 
and an optimised beam-target overlap are essential to yield 
a high luminosity in the experiment.
The product of $L$ and the total hadronic cross section $\sigma_H$ 
delivers the interaction rate $R$, 
i.e.~the number of antiproton-proton interactions in a specified time interval,
which determines the achievable number of events for 
all physics channels and allows the extraction 
of occupancies in different detector regions. 
These are needed as input for the associated hardware development. 

Obviously, the achievable luminosity is directly linked 
with the number of antiprotons in the \HESR.
The particles are injected at discrete time intervals. 
The maximum luminosity thus depends on the antiproton production rate $R_{\bar p}$~=~d$N_{\bar p}$/d$t$.
Moreover, a beam preparation must be performed before the target can be switched on.
It includes pre-cooling to equilibrium, 
the ramping to the desired beam momentum and a fine-tuned focusing 
in the target region as well as in the section for the electron cooler.
Therefore, the operation cycle of the \HESR can be separated into two sequences
related to the beam preparation time  $t_{\textsf{\tiny{prep}}}$ (target off) 
and the time for data taking $t_{\textsf{\tiny{exp}}}$ (target on), respectively.
The  beam preparation time  $t_{\textsf{\tiny{prep}}}$ also contains 
the period between the target switch-off and the injection,
at which the residual antiprotons are either dumped 
or transferred back to the injection momentum.

\subsubsection*{Macroscopic Luminosity Profile}
\begin{figure}[hbtp]
  \centering
  \includegraphics[width=\swidth]{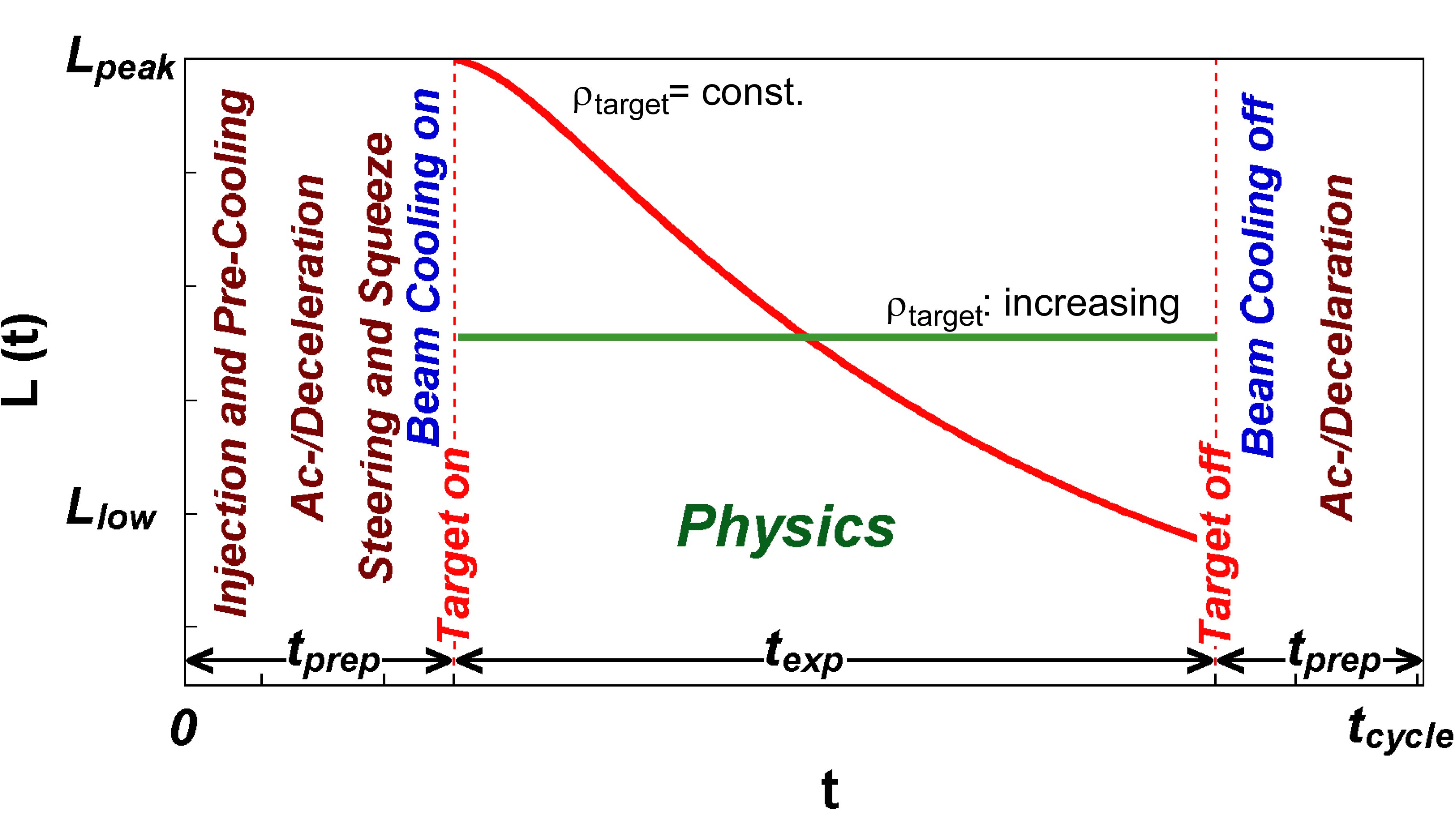}
  \caption[Time dependent macroscopic luminosity profile~$L(t)$ in one operation cycle]
{
Time dependent macroscopic luminosity profile~$L(t)$ in one operation cycle 
for constant (solid red) and increasing (green dotted) target density $\rho_{\textsf{\tiny{target}}}$.
Different measures for beam preparation are indicated.
Pre-cooling is performed at 3.8 \gevc. 
A maximum ramp of 25~mT/s is specified for beam ac-/deceleration.} 
%To calculate the cycle average luminosity, machine cycles and beam
%preparation times have to be specified. After injection, the beam
%is pre-cooled to equilibrium (with target off) at 3.8 \gevc.
%The beam is then ac-/decelerated to the desired beam momentum. A
%maximum ramp rate of 25~mT/s is specified. After reaching the
%final momentum beam steering and focusing in the target and beam
%cooler region takes place. Total beam preparation time $t_{prep}$
%ranges from 120~s for 1.5 \gevc to 290~s for 15 \gevc. 
  \label{fig:lumcycle}
\end{figure}

A schematic illustration of the luminosity profile during one operation cycle
is  given in \Reffig{fig:lumcycle}.
The maximum luminosity is obtained directly after the target is switched on.  
During data taking the luminosity decreases due to hadronic interactions,
single Coulomb scattering and energy straggling of the circulating beam in the target.
Compared to beam-target interaction, minor contributions are related to single intra-beam scattering (Touschek effect).
Beam losses caused by residual gas scattering can be neglected, if the vacuum is better than 10$^{-9}$~mbar. 
A detailed analysis of all beam loss processes can be found in~\cite{lehrach:2006,hinterberger:2006}.
The relative beam loss rate $R_{\textsf{\tiny{loss}}}$ for the total cross section $\sigma_{\textsf{\tiny{tot}}}$ 
is given by the expression
\begin{equation}
R_{\textsf{\tiny{loss}}} = \tau^{-1} = f_0 \cdot n_t \cdot \sigma_{\textsf{\tiny{tot}}}
\end{equation}
where $\tau$ corresponds to the mean (1/$e$) beam lifetime, 
$f_0$ is the revolution frequency of the antiprotons in the ring
and $n_t$ is the effective target thickness 
defined as an area density given in atoms per square centimetre. 
For beam-target interactions, the beam lifetime is independent of the beam intensity.
The Touschek effect depends on the beam equilibria and beam intensity. 
At low momenta the beam cooling scenario and the ring acceptance have large impact 
on the achievable beam lifetime.
%Beam lifetimes are ranging from 1540~s to 7100~s. 

\begin{table*}
\caption[Maximum achievable cycle averaged luminosity for different H$_2$ target setups]
{
Calculation of the maximum achievable cycle averaged luminosity
for three different beam momenta: 
Input parameters and final results for different H$_2$ target setups.
}
\smallskip
\centering {\small
\begin{tabular}{lccc}
 & 1.5 \gevc & 9 \gevc & 15 \gevc \\ \hline
\vspace{-3mm}&&&\\
Total hadronic cross section/ mbarn &
$100$&
$57$&
$51$\\ 
\hline
\multicolumn{4}{c}{\parbox[0pt][4mm][c]{10cm}{\textbf{Cluster jet target}}}\\ 
\hline
\vspace{-3mm}&&&\\
Target density: /cm$^{-2}$ &
$8 \cdot 10^{14}$ &
$8 \cdot  10^{14}$&
$8 \cdot  10^{14}$\\ 
Antiproton production rate: /s$^{-1}$ &
$2 \cdot  10^{7}$ &
$2 \cdot  10^{7}$ &
$2 \cdot  10^{7}$ \\
Beam preparation time: /s &
$120$ &
$140$ &
$290$ \\
Optimum cycle duration: /s &
$1280$ &
$2980$ &
$4750$ \\ 
Mean beam lifetime: /s 
&
$\sim 5920$ &
$\sim 29560$ &
$\sim 35550$ \\ 
Max Cycle Averaged Luminosity: /cm$^{-2}$s$^{-1}$ &
$0.29 \cdot  10^{32}$ &
$0.38 \cdot  10^{32}$ &
$0.37 \cdot  10^{32}$ \\
%\end{tabular}
%}
%\end{table*}

%\begin{table*}
%\caption{Max. cycle averaged luminosity for a H$_2$ pellet target.}
%\label{tab:lifetimes} \centering {\small
%\begin{tabular}{lccc}
 % & \multicolumn{3}{c}{$(\tauloss^{-1})$ / s$^{-1}$} \\
\hline
\multicolumn{4}{c}{\parbox[0pt][4mm][c]{10cm}{\textbf{Pellet target}}}\\ 
\hline
\vspace{-3mm}&&&\\
Target density: / cm$^{-2}$&
$4 \cdot 10^{15}$ &
$4 \cdot 10^{15}$&
$4 \cdot 10^{15}$ \\
Antiproton production rate: /s$^{-1}$&
$2 \cdot  10^{7}$ &
$2 \cdot  10^{7}$ &
$2 \cdot  10^{7}$ \\
Beam preparation time: /s &
$120$ &
$140$ &
$290$ \\
Optimum cycle duration: /s &
$4820$ &
$1400$ &
$2230$ \\ 
Mean beam lifetime: /s 
&
$\sim 1540$ &
$\sim 6000$ &
$\sim 7100$ \\ 
Max Cycle Averaged Luminosity: /cm$^{-2}$s$^{-1}$ &
$0.53 \cdot  10^{32}$ &
$1.69 \cdot  10^{32}$ &
$1.59 \cdot  10^{32}$ \\ \hline
\end{tabular}
}
\label{table_HydroAvLumi}
\end{table*}

\subsubsection*{Cycle Average Luminosity}

\begin{table*}[t]
\caption[Expected luminosities for heavier nuclear targets at \Panda]
{Expected maximum average luminosities, $\bar L$, 
and required effective target thickness, $n_t$, 
for heavier nuclear targets at \panda
at minimum and maximum beam momentum $p_{\textsf{\tiny{beam}}}$.
Given numbers refer to an assumed number of $10^{11}$ antiprotons in the \HESR.}
\smallskip
\begin{center}
\small
\begin{tabular}{|c|c|c|c|}
\hline
\vspace{-3mm}
  & \hspace{1.5cm}
    & \hspace{1.5cm}
      & \hspace{1.5cm} \\
\vspace{-3mm}&&&\\
Target material 
  & $\bar L$ ($p_{\textsf{\tiny{beam}}}$=1.5~\gevc)
    & $\bar L$ ($p_{\textsf{\tiny{beam}}}$=15~\gevc) 
      & $n_t$\\
 & [cm$^{-2}$s$^{-1}$] 
    & [cm$^{-2}$s$^{-1}$] 
      & [atoms/cm$^2$]
\parbox[0pt][6mm][t]{0cm}{}\\
\hline
\vspace{-3mm}&&&\\
deuterium 
  & $5\cdot 10^{31}$ 
    & $1.9\cdot 10^{32}$
      & $3.6\cdot 10^{15}$\\
\vspace{-3mm}&&&\\
argon 
  & $4\cdot 10^{29}$
    & $2.4\cdot 10^{31}$
      & $4.6\cdot 10^{14}$\\
\vspace{-3mm}&&&\\
gold 
  & $4\cdot 10^{28}$
    & $2.2\cdot 10^{30}$
      & $4.1\cdot 10^{13}$
\parbox[0pt][6mm][t]{0cm}{}\\
\hline
\end{tabular}
%\vspace{-6mm}
\label{table_NuclearAvLumi}
\end{center}
\end{table*} 

In physics terms, the time-averaged cycle luminosity is most relevant.
The maxi\-mum average luminosity depends on the ratio of the antiproton production rate 
to the loss rate and is thus inversely proportional to the total cross section.
It can be increased if the residual antiprotons after each cycle 
are transferred back to the injection momentum and then merged with the newly injected particles.
Therefore, a bucket scheme utilising broad-band cavities is foreseen 
for beam injection and the refill procedure. 
Basically, the cycle average luminosity $\bar L$ reads as:
\begin{equation}
\label{eq-AvLumiGeneralDefinition}
\bar L = N_{\bar p,\textsf{\tiny{0}}} \cdot f_0 \cdot n_t \cdot 
{{\tau \left[1-e^{-{t_{\textsf{\tiny{exp}}}\over{\tau}}}\right]}\over t_{\textsf{\tiny{exp}}} + t_{\textsf{\tiny{prep}}}}
\end{equation}
where $N_{\bar p,\textsf{\tiny{0}}}$ corresponds to the number of available particles at the start of the target insertion.

For the calculations, machine cycles and beam preparation times have to be specified.
The maximum cycle average luminosity is achieved by an optimisation of 
the cycle time $t_{\textsf{\tiny{cycle}}}=t_{\textsf{\tiny{exp}}} + t_{\textsf{\tiny{prep}}}$.
Constraints are given by the restricted number antiprotons in the \HESR,
the achievable effective target thickness and the specified antiproton production rate 
of $R_{\bar p} = 2\cdot10^7$~s$^{-1}$ at \FAIR.
%A further limitation occurs if the antiproton production rate is not high enough 
%to compensate beam losses within one mean beam lifetime.

Main results of calculations performed for different hydrogen targets are summarised in \Reftbl{table_HydroAvLumi}.
The total hadronic cross section, $\sigma_H^{\bar p p}$, 
decreases with higher beam momentum from approximately 100~mbarn at 1.5~\gevc 
to 50~mbarn at 15~\gevc.
With the limited number of 10$^{11}$ antiprotons, as specified for the high-luminosity mode, 
cycle averaged luminosities of up to $1.6 \cdot 10^{32}$~cm$^{-2}$s$^{-1}$ can be achieved at
15~\gevc for cycle times of less than one beam lifetime.
Due to the very short beam lifetimes at lowest beam momenta more than $10^{11}$ 
particles can not be provided in average.
As a consequence, the average luminosity drops below the envisaged design value at around 2.4~\gevc 
to finally roughly 5\,$\,\cdot\,$\,$10^{31}$ s$^{-1}$cm$^{-2}$ at 1.5~\gevc.
%As a consequence, cycle average luminosities are below 10$^{32}$ cm$^{-2}$ s$^{-1}$.
Due to the lower assumed target density the achievable luminosity of the cluster-jet target 
is smaller compared to the pellet operation.

\begin{figure}[!b]
\begin{center}
\includegraphics[width=7.5 cm]{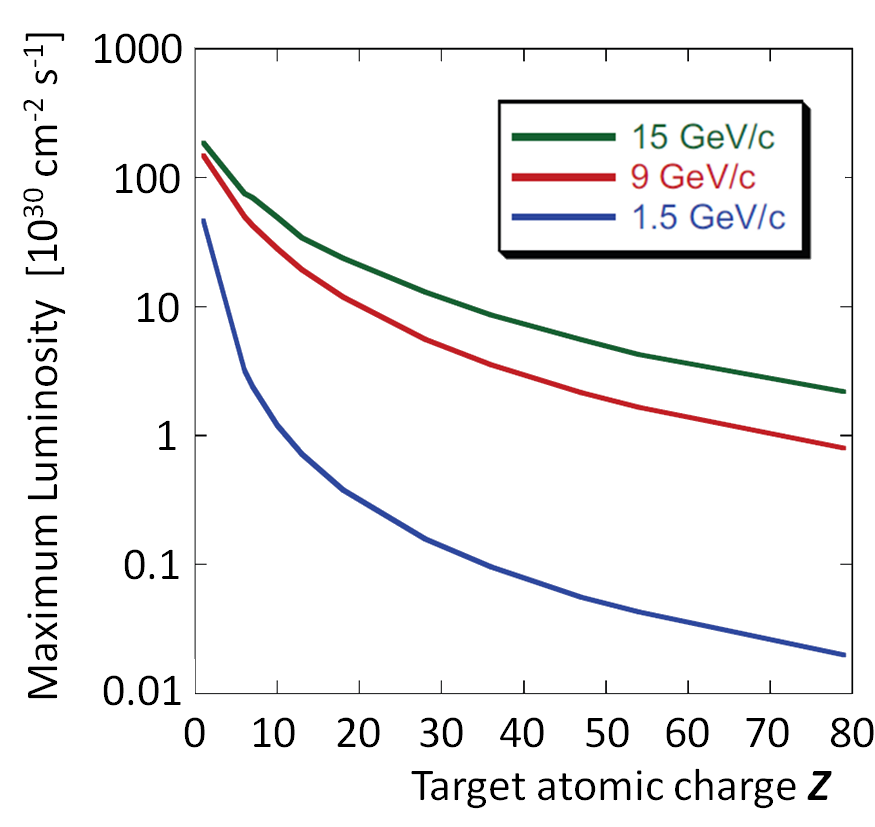}
% Pic3-01_MVD-BTS-Geometry.png: 1239x931 pixel, 125dpi, 25.18x18.92 cm, bb=0 0 714 536 
\caption[Maximum average luminosity vs.~atomic charge $Z$ of the target 
for three different beam momenta]
{Maximum average luminosity vs. atomic charge, $Z$, of the target
for three different beam momenta.}
\label{Pic-HESR_NuclearAvLumi}
\end{center}
\end{figure}

In case of nuclear targets the total hadronic cross section 
for the interaction of antiprotons with target nucleons 
can be estimated from geometric considerations taking into account
the proton radius of $r_p =0.9$ fm and the radius of a spherical nucleus $R_A$,
which can be roughly approximated as  $R_A =\,$ $r_0 A^{1/3}$, 
where $r_0  = 1.2$ fm and $A$ is the mass number.
With the assumption that $\sigma_H^{\bar p p}= \pi r_p^2$, 
the required total hadronic cross section, $\sigma_H^{\bar p A}$, 
for a nucleus of mass number $A$ can be extracted from the given values of 
$\sigma_H^{\bar p p}$ for antiproton-proton collisions as follows:

\begin{equation}
\label{eq-pbarNucleonXsection}
\sigma_H^{\bar p A} = \pi (R_A + r_p)^2 = \sigma_H^{\bar p p} \cdot \left({R_A\over r_p}+1\right)^2 
\end{equation}

Simulation results on maximum average luminosities based 
on equation~\ref{eq-pbarNucleonXsection} are shown 
in \Reffig{Pic-HESR_NuclearAvLumi}. 
They include adapted beam losses in the target due to 
single Coulomb scattering and energy straggling. 
Compared to antiproton-proton experiments, the maximum average luminosity for nuclear targets decreases rapidly 
with both, higher atomic charge $Z$ and lower beam momenta, by up to three orders of magnitude.  
Specific values for selected nuclear targets are given in \Reftbl{table_NuclearAvLumi} 
with the effective target thickness required to reach these numbers.

\begin{table*}[t]
\caption[Estimate on the expected event rates at \panda]
{Summary of expected event rates at \panda. 
Numbers for the hydrogen target correspond to the pellet system (see \Reftbl{table_HydroAvLumi}).
The given ratio $\bar L_{\textsf{\tiny{peak}}}/\bar L_{\textsf{\tiny{exp}}}$ 
corresponds to the maximum value to achieve the nominal interaction rate of $R_{\textsf{\tiny{nom}}} = 2 \cdot10^7$~s$^{-1}$.
Rough estimates for nuclear targets are based on the numbers given in \Reftbl{table_NuclearAvLumi},
with $\bar L = \bar L_{\textsf{\tiny{exp}}}$, and  $\sigma_H$ calculated according to equation~\ref{eq-pbarNucleonXsection}.
}
\smallskip
\begin{center}
\small
\begin{tabular}{|c|c|c|c|c|c|c|}
\hline
&&&&&&\\
Target  & $p_{\textsf{\tiny{beam}}}$  & $\bar L_{\textsf{\tiny{exp}}}$  & $L_{\textsf{\tiny{inst}}}$ 
 & $\sigma_H$  & $\bar R_{\textsf{\tiny{exp}}}$  & $\bar L_{\textsf{\tiny{peak}}}/\bar L_{\textsf{\tiny{exp}}}$ \\ 
material & [\gevc] & [cm$^{-2}$s$^{-1}$] & [cm$^{-2}$s$^{-1}$] & [mbarn] & [s$^{-1}$] & \newline \footnotesize $(R_{\textsf{\tiny{nom}}})$ \\
&&&&&&\\
\hline
&&&&&&\\
\multirow{2}{*}{hydrogen}
  & 1.5
    & $5.4\cdot 10^{31}$
      & ($5.9 \pm 0.6$)$\,\cdot \,10^{31}$
	& 100
	  & $5.4\cdot 10^{6}$
	    & 3.7 \\
  & 15
    & $1.8\cdot 10^{32}$
      & ($2.0 \pm 0.2$)$\,\cdot \, 10^{32}$
	& 51
	  & $9.7\cdot 10^{6}$
	    & 2.1 \\
&&&&&&\\
\hline
&&&&&&\\
\multirow{2}{*}{argon}
  & 1.5
    & $4.0\cdot 10^{29}$
      & ($4.4 \pm 0.4$)$\,\cdot \, 10^{29}$
	& 2020
	  & $8.1\cdot 10^{5}$
	    & \multirow{2}{*}{--} \\
  & 15
    & $2.4\cdot 10^{31}$
      & ($2.6 \pm 0.3$)$\,\cdot \, 10^{31}$
	& 1030
	  & $2.5\cdot 10^{7}$
	    &  \\
\multirow{2}{*}{gold}
  & 1.5
    & $4.0\cdot 10^{28}$
      & ($4.4 \pm 0.4$)$\,\cdot \, 10^{28}$
	& 7670
	  & $3.1\cdot 10^{6}$
	    & \multirow{2}{*}{--} \\
  & 15
    & $2.2\cdot 10^{30}$
      & ($2.6 \pm 0.3$)$\,\cdot \, 10^{30}$
	& 3911
	  & $8.6\cdot 10^{6}$
	    & \\
&&&&&&\\
\hline
\end{tabular}
%\vspace{-6mm}
\label{table_EventRates}
\end{center}
\end{table*} 

\subsubsection*{Event Rates}
\label{event-rates}

Besides the cycle-averaged luminosity an evaluation 
of the instantaneous luminosity during the data taking 
is indispensable for performance studies of the \panda detector.
Associated event rates define the maximum data load
to be handled at different timescales by the individual subsystems.
The discussions in this section are based on the following assumptions:
\begin{itemize}
 \item Nominal antiproton production rate at \FAIR:  \\ $R_{\bar p}=2\cdot10^7$~s$^{-1}$
 \item Effective target density: \\ $n_t=4\cdot10^{15}$~atoms/cm$^2$
 \item Maximum number of antiprotons in the \HESR: \\ $N_{\bar p,\textsf{\tiny{max}}}$~=~10$^{11}$
 \item Recycling of residual antiprotons at the end of each cycle
\end{itemize}

As indicated in \Reffig{fig:lumcycle}
the instantaneous luminosity during the cycle changes on a macroscopic timescale.
One elegant way to provide constant event rates in case of a cluster-jet target 
is given by the possibility to compensate the antiproton consumption during an accelerator cycle 
by the increase of the effective target density. 
Alternatively, using a constant target beam density the beam-target overlap might be increased
adequately to the beam consumption.  
With these modifications the instantaneous luminosity during the cycle 
is expected to be kept constant to a level of 10\%.

The values for the luminosity as given in \Reftbl{table_HydroAvLumi} 
are averaged over the full cycle time. 
However, to extract the luminosity during data taking, $\bar L_{\textsf{\tiny{exp}}}$,
these numbers must be rescaled to consider the time average over the experimental time:
\begin{equation}
\label{eq-LumiExp}
\bar L_{\textsf{\tiny{exp}}} = (t_{\textsf{\tiny{cycle}}}/t_{\textsf{\tiny{exp}}}) \cdot \bar L
\end{equation}
In addition to the fluctuation of the instantaneous luminosity during the operation cycle as dicussed above
($\Delta L_{\textsf{\tiny{inst}}}/L_{\textsf{\tiny{inst}}}\leq 10\%$),
it must be considered that the \HESR will be only filled by 90\% in case of using a barrier-bucket system.
As a consequence, values for $L_{\textsf{\tiny{inst}}}$ during data taking 
are 10\% higher than the ones for $\bar L_{\textsf{\tiny{exp}}}$.

An estimate of peak luminosities, $L_{\textsf{\tiny{peak}}}> L_{\textsf{\tiny{inst}}}$, 
must further include possible effects on a short timescale.
Contrary to homogeneous cluster beams, 
a distinct time structure is expected 
for the granular volume density distribution of a pellet beam.
Such time structure depends on the transverse and longitudinal overlap between 
single pellets and the circulating antiproton beam in the interaction region. 
Deviations of the instantaneous luminosity on a microsecond timescale 
are caused by variations of the pellet size, the pellet trajectory 
and the interspacing between consecutive pellets. 
The latter must be well controlled to avoid the possible presence of 
more than one pellet in the beam at the same instant. 
The resulting ratio $L_{\textsf{\tiny{peak}}}/L_{\textsf{\tiny{exp}}}$ depends on the pellet size.
First studies on the expected peak values for the \panda pellet target have been performed~\cite{Smirnov:2009}.
Results indicate that the peak luminosity stays below $10^{33}$~cm$^{-2}$s$^{-1}$
if the pellet size is not bigger than 20~$\mu$m.

Finally, for the extraction of event rates the obtained luminosities are multiplied with the hadronic cross section.
\Reftbl{table_EventRates} summarises the main results for a hydrogen target based on a pellet system,
which is expected to deliver upper limits for the occuring event rates.
In addition, a rough estimate for nuclear targets based on the input of \Reftbl{table_NuclearAvLumi} 
and equation~\ref{eq-pbarNucleonXsection} is given.
Even though these values still must be verified by detailed studies, 
it can be seen that the reduced average luminosity for heavier nuclear targets 
is counter-balanced by an increased cross-section that results in comparable event rates.

Based on the given assumptions and caveats, as discussed in this section, 
a nominal interaction rate of $R_{\textsf{\tiny{nom}}} = 2 \cdot10^7$~s$^{-1}$
can be defined that all detector systems have to be able to handle.
This specification includes the requirement that density fluctuations of the beam-target overlap
have to be smaller than a factor of two ($\bar L_{\textsf{\tiny{peak}}}/\bar L_{\textsf{\tiny{exp}}}$).
However, in order to avoid data loss it might be important to introduce a generic safety factor 
that depends on special features of the individual detector subsystems and their position 
with respect to the interaction region.

\section{The \PANDA Detector}
%\svnInfo $Id: panda.tex 709 2009-02-10 09:03:42Z IntiL $
%\section*{I.5 \hspace{4mm} The \PANDA Detector}
%\addcontentsline{toc}{subsection}{I.5 \hspace{4mm} The \PANDA Detector}
\label{s:over:panda}

\begin{figure*}
%[!b]
\centering
\includegraphics[width=0.85\dwidth]{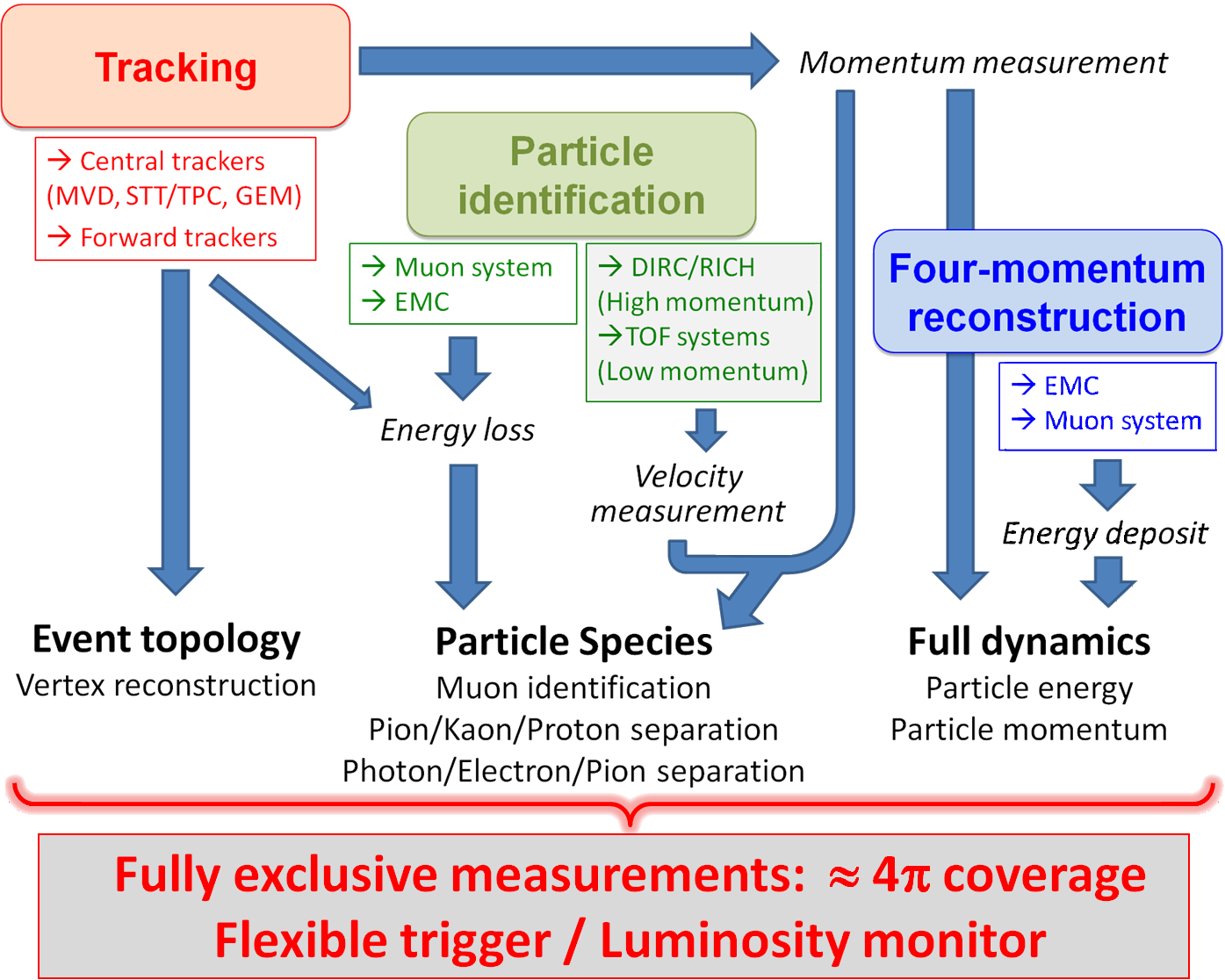}
\caption[Basic detection concept]
{Basic detection concept. The main components are described in 
sections~\ref{sec:det:ts} and \ref{sec:det:fs}.
}
\label{pic-DetConcept}       
\end{figure*}

The main objectives of the design of the \PANDA experiment 
are to achieve $4\pi$ acceptance, high resolution
for tracking, particle identification and calorimetry, high rate
capabilities and a versatile readout and event selection. 
To obtain a good momentum resolution the detector will be composed of 
two magnetic spectrometers: 
the {\em Target Spectrometer (TS)}, based on a superconducting solenoid magnet surrounding
the interaction point, which will be used to measure at large polar angles 
and the {\em Forward Spectrometer (FS)}, based on a dipole magnet, for small angle tracks. 
An overview of the detection concept is shown in \Reffig{pic-DetConcept}.

It is based on a complex setup of modular subsystems including tracking detectors
(MVD, STT, GEM), electromagnetic calorimeters (EMC), a muon system,
Cherenkov detectors (DIRC and RICH) and a time-of-flight (TOF) system.
A sophisticated concept for the data acquisition with a flexible trigger is planned 
in order to exploit at best the set of final states relevant for the \PANDA physics objectives. 

% OLD origunal: A silicon vertex detector will surround the interaction point. 
%In both spectrometer parts, tracking, charged particle identification,
%electromagnetic calorimetry and muon identification will be available to
%allow to detect the complete spectrum of final states relevant for the
%\PANDA physics objectives.

%EXA proceeding: The proposed installation facilitates exclusive measurements. In addition, the
%high modularity of this design allows for a variety of complementary measurements
%from nuclear to particle physics. 
%Moreover, an absolute measurement of total cross sections is planned. 
%Therefore, a luminosity monitor measuring elastically scattered
%antiprotons will be installed in very forward direction. 
%\addcontentsline{toc}{subsection}{\hspace{4mm}...  Target Spectrometer}

\begin{figure*}
\begin{center}
\includegraphics[width=0.85\dwidth]{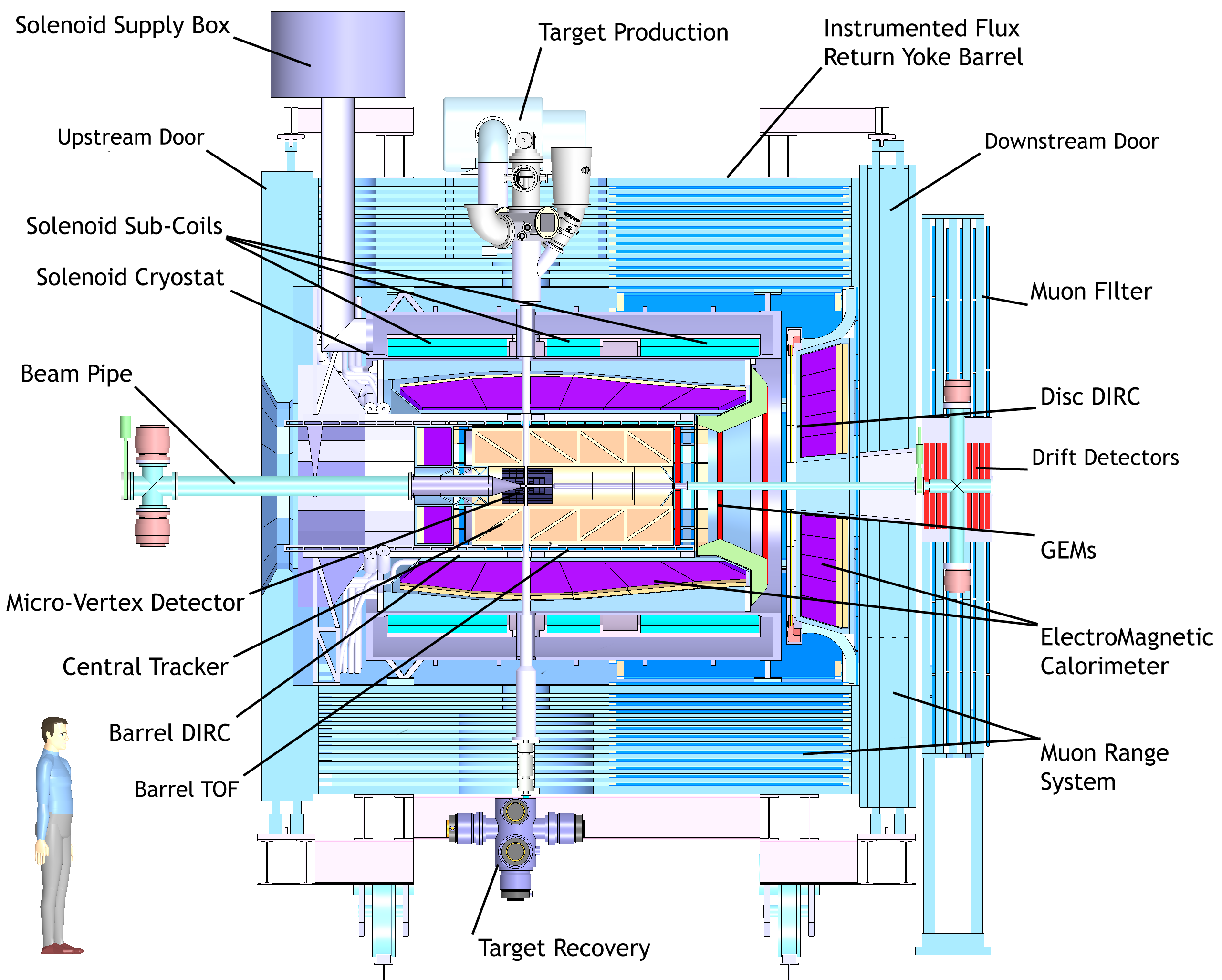}
\caption[Artistic side view of the Target Spectrometer (TS) of \PANDA]
{Artistic side view of the Target Spectrometer (TS) of \PANDA. 
  To the right of this the Forward Spectrometer (FS) follows, which is
illustrated in \Reffig{f:over:fs_view}.}
\label{f:over:ts_view}
\end{center}
\end{figure*}

The Target Spectrometer will surround the interaction point and measure
charged tracks in a highly homogeneous solenoidal field. 
In the manner of a collider detector it will contain
detectors in an onion shell like configuration. Pipes for the
injection of target material will have to cross the spectrometer
perpendicular to the beam pipe.

The Target Spectrometer will be arranged in three parts: the barrel
covering angles between 22$\degrees$ and 140$\degrees$, the forward
end cap extending the angles down to 5$\degrees$ and 10$\degrees$ in
the vertical and horizontal planes, respectively, and the backward end
cap covering the region between about 145$\degrees$ and 170$\degrees$.
Please refer to \Reffig{f:over:ts_view} for a complete overview.

\subsection{Target Spectrometer}
\label{sec:det:ts}
\subsubsection*{Beam-Target System}
\label{sec:det:ts:tgt}

The beam-target system consists of the apparatus for the target production 
and the corresponding vacuum system for the interaction region.  
The beam and target pipe cross sections inside the target spectrometer are decreased 
to an inner diameter of 20~mm close to the interaction region.  
%A beam-target cross is currently foreseen for intersection of the target cross section 
%with the beam pipe. 
The innermost parts are planned to be made of beryllium, titanium or a suited alloy 
which can be thinned to wall thicknesses of 200~$\mu$m. 
Due to the limited space and the constraints on the material budget close to the IP, 
vacuum pumps along the beam pipe can only be placed outside the target spectrometer. 
Insections are foreseen in the iron yoke of the magnet which allow the integration 
of either a pellet or a cluster-jet target. 
The target material will be injected from the top. 
Dumping of the target residuals after beam crossing 
is mandatory to prevent backscattering into 
the interaction region. 
%which would increase the residual gas concentration 
%and dramatically worsen the overall performance of the experiment.
The entire vacuum system is kept variable and allows an operation of both target types. 
Moreover, an adaptation to non-gaseous nuclear wire targets is possible. 
For the targets of the planned hypernuclear experiment the whole upstream end cap 
and parts of the inner detector geometry will be modified.
%The compact design of the detector layers nested inside the
%solenoidal magnetic field, combined with the request of minimal
%distance from the interaction point to the vertex tracker, leaves only a very
%restricted space for the target installations.  
A detailed discussion of the different target options can be found in chapter~\ref{intro:target}.
%In order to reach the
%design luminosity of $2\times 10^{32}$ s$^{-1}$cm$^{-2}$ a target thickness
%of about $4\times 10^{15}$ hydrogen atoms per cm$^2$ is required assuming
%$10^{11}$ stored anti-protons in the \HESR ring.
%These conditions pose a real challenge for an internal target
%inside a storage ring. 
%For the target 
%The design of the primary target setup is compatible with different options 
%that are foreseen to be used at \PANDA

%At present, two different, complementary
%techniques for the internal target are being developed further: the
%cluster-jet target and the pellet target. Both techniques are capable
%of providing sufficient densities for hydrogen at the interaction
%point, but exhibit different properties concerning their effect on the
%beam quality and the definition of the interaction point. In addition,
%internal targets also of heavier gases, deuterium,
%nitrogen or argon can be made available.

%For non-gaseous nuclear targets the situation is different, in
%particular in the case of the planned hypernuclear experiment. 
%In these studies, the whole upstream end cap and part of the inner detector
%geometry will be modified.

\subsubsection*{Solenoid Magnet}

The solenoid magnet of the TS will deliver a very homogeneous solenoid field 
of 2~T with fluctuations of less than $\pm$2\%.
In addition, a limit of $\mathrm{\int} B_r/B_z \mathrm{d}z<2$~mm is specified for
the normalised integral of the radial field component.
The superconducting coil of the magnet has a length of 2.8~m and an inner radius of 90~cm,
using a laminated iron yoke for the flux return.
The cryostat for the solenoid coils is required to have two warm bores of 
100 mm diameter, one above and one below the target position, to
allow for insertion of internal targets.
The load of the integrated inner subsystems can be picked up at defined fixation points.
A precise description of the magnet system and detailed field strength calculations 
can be found in~\cite{PANDA-MagnetTDR}. 

\subsubsection*{Micro Vertex Detector}

\begin{figure}[!b]
\begin{center}
\includegraphics[trim=0 0.1cm 0 0.3cm, clip, width=7. cm]{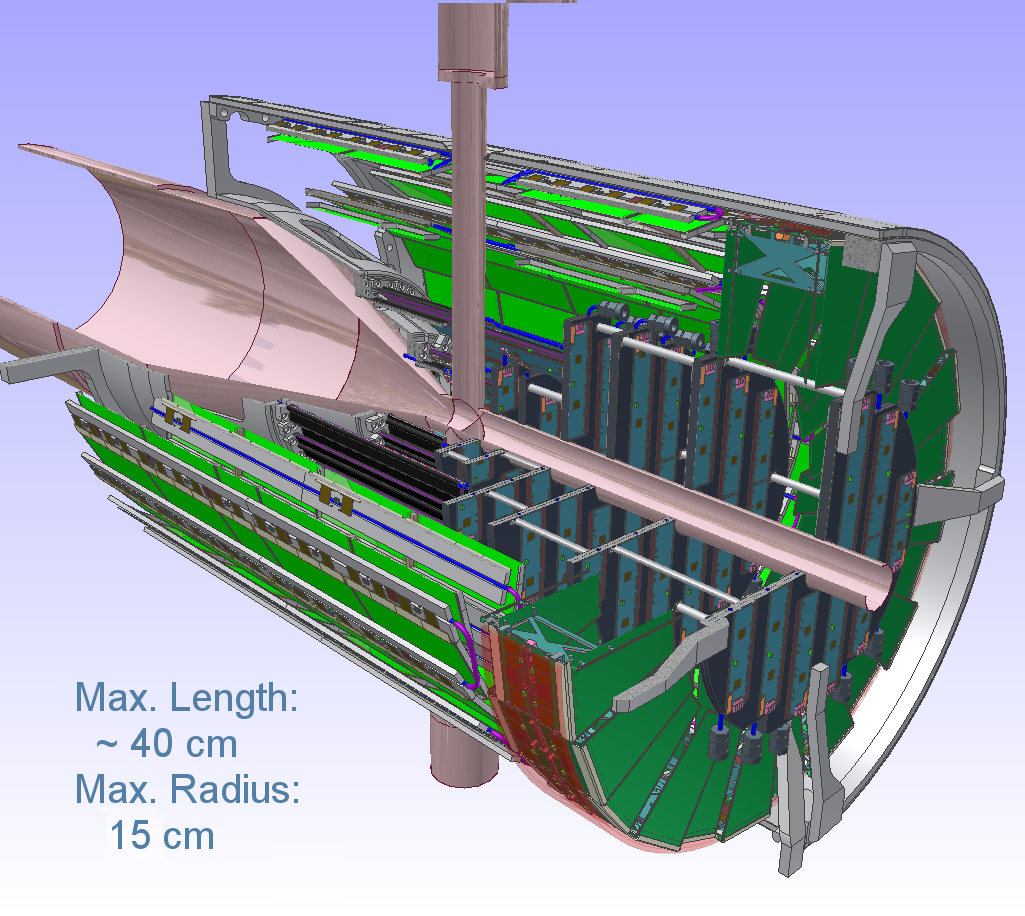}
\caption[The Micro Vertex Detector (MVD) of the Target Spectrometer]
  {The Micro Vertex Detector (MVD) of the Target Spectrometer
  surrounding the beam and target pipes seen from downstream.  
  To allow a look inside the detector a three-quarters portraits is chosen.}
\label{fig:det:mvd}
\end{center}
\end{figure}

The design of the Micro Vertex Detector (\Mvd) for the Target
Spectrometer is optimised for the detection of secondary decay vertices from
charmed and strange hadrons and for a maximum acceptance close to the interaction
point. It will also strongly improve the transverse momentum
resolution. The setup is depicted in \Reffig{fig:det:mvd}.

The concept of the \Mvd is based on radiation hard silicon pixel
detectors with fast individual pixel readout circuits and silicon
strip detectors. The layout foresees a four layer barrel detector with
an inner radius of 2.5~cm and an outer radius of 13~cm. The two
innermost layers will consist of pixel detectors and the outer two
layers will be equipped with double-sided silicon strip
detectors.

Six detector wheels arranged perpendicular to the beam will achieve
the best acceptance for the forward part of the particle spectrum.
While the inner four layers will be made entirely of pixel detectors,
the following two will be a combination of strip detectors on the outer
radius and pixel detectors closer to the beam pipe. 

\subsubsection*{Additional Forward Disks}

Two additional silicon disk layers are considered further downstream
at around 40~cm and 60~cm 
to achieve a better acceptance of hyperon cascades.
They are intended to be made entirely of silicon strip detectors.
Even though they are not part of the central MVD 
it is planned, as a first approach, to follow 
the basic design as defined for the strip disks of the MVD.
However, an explicit design optimisation still has to be performed.
Two of the critical points to be checked are related 
to the increased material budget caused by these layers 
and the needed routing of cables and supplies for these additional disks
inside the very restricted space left by the adjacent detector systems.

\subsubsection*{Straw Tube Tracker (\Stt)}
\label{sec:det:ts:stt}

\begin{figure}[htb]
\begin{center}
\includegraphics[width=\swidth]{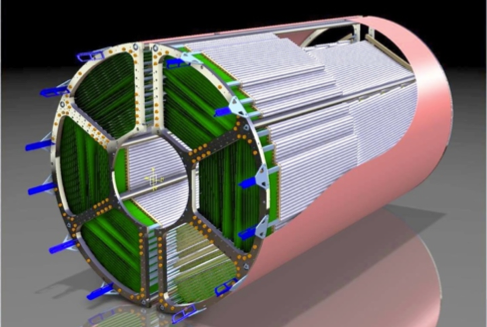} 
\caption[Straw Tube Tracker (STT) of the Target Spectrometer]
  {The Straw Tube Tracker (STT) of the Target Spectrometer seen from upstreams.}
\label{fig:exp:ts:stt}
\end{center}
\end{figure}

This detector will consist of aluminised Mylar tubes called {\em straws}. 
These will be stiffened by operating them at an overpressure of 1~bar 
which makes them self-supporting.
The straws are to be arranged in planar
layers which are mounted in a hexagonal shape around the \Mvd as shown
in \Reffig{fig:exp:ts:stt}. In total there are 27 layers of which
the 8 central ones are skewed, to achieve an acceptable resolution of
3 mm also in $z$ (parallel to the beam). The gap to the
surrounding detectors will be filled with further individual
straws. In total there will be 4636 straws around the beam pipe at
radial distances between 15~cm and 41.8~cm with an overall length
of 150~cm. All straws have a diameter of 10~mm 
and are made of a 27~$\mu$m thick Mylar foil.
Each straw tube is constructed with a single anode wire in the centre
that is made of 20~$\mu$m thick gold plated tungsten 
The gas mixture used will be Argon based with CO$_2$ as quencher. 
It is foreseen to have a gas gain not greater than 10$^5$ in order to
warrant long term operation. With these parameters, a resolution in
$x$ and $y$ coordinates of less than 150~$\mu$m is expected.
A thin and light space frame will hold the straws in place, 
the force of the wire however is kept solely by the straw itself. 
This overall design results in a material budget of 1.2\% 
of one radiation length.

\subsubsection*{Forward GEM Detectors}
\begin{figure}[h]
\includegraphics[width=\swidth]{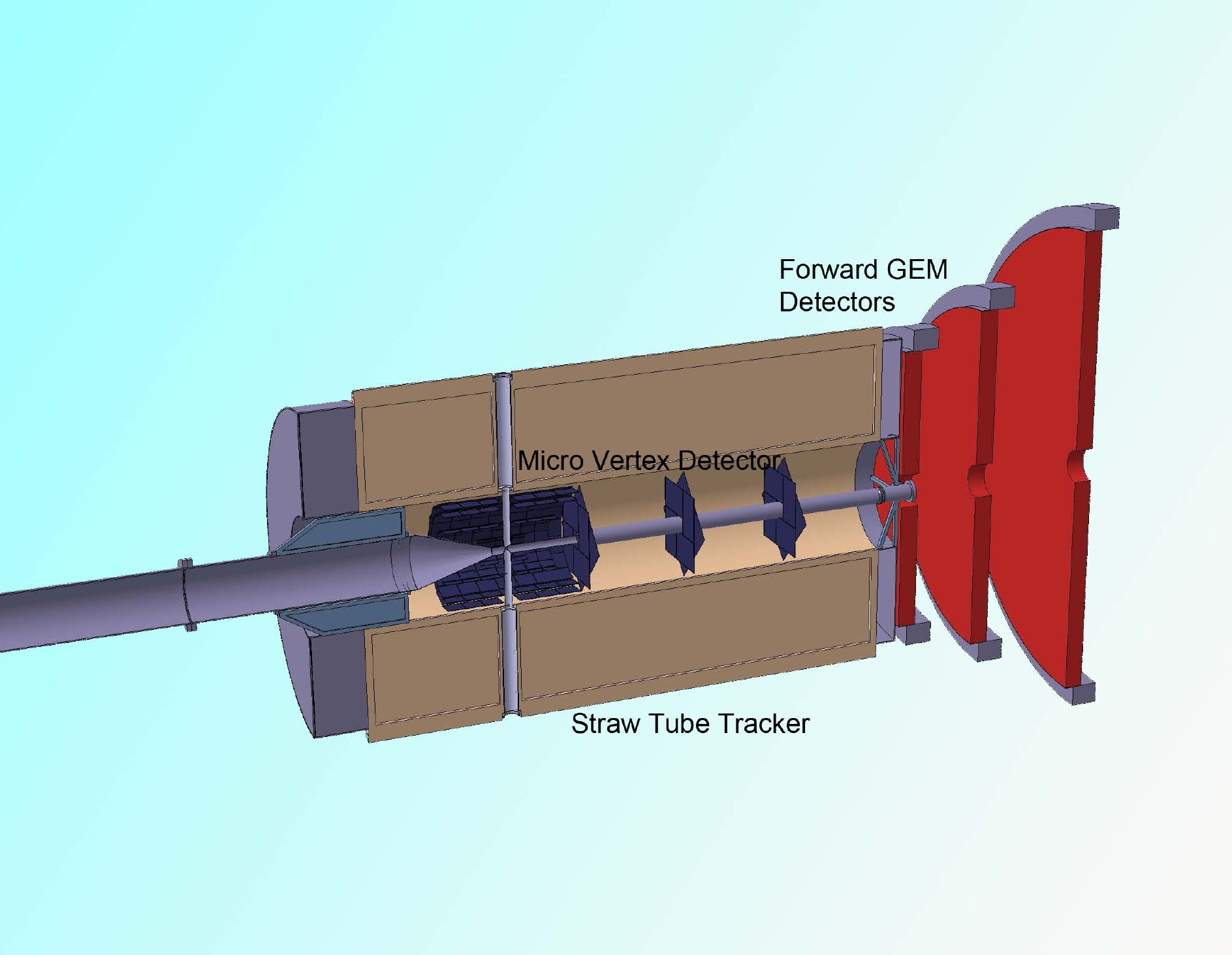}
\caption[Tracker assembly]
  {Schematic drawing of the tracking detectors of the Target Spectrometer.}
\label{fig:exp:ts:tracking}
\end{figure}

\Reffig{fig:exp:ts:tracking} shows the components of the tracking system of the Target Spectrometer.

Particles emitted at angles below 22\degrees{} which are not covered
fully by the STT will be tracked by three planar stations placed approximately 
1.1~m, 1.4~m and 1.9~m downstream of the target.  
Each of the station consists of double planes with two
projections per plane.
The stations will be equipped with Gaseous micro-pattern detectors 
based on Gas Electron Multiplier (GEM) foils as amplification stages. 
The chambers have to sustain a high counting rate of particles peaked 
at the most forward angles due to the relativistic boost of the reaction products 
as well as due to the small angle \pbarp elastic scattering. 
The maximum expected particle flux in the first chamber in the vicinity 
of the 5~cm diameter beam pipe will be about 
$3 \cdot 10^{4}$~cm$^{-2}$s$^{-1}$.

%\subsubsection*{Cherenkov Detectors and Time-of-Flight}

%Charged particle identification of hadrons and leptons over a large
%range of angles and momenta is an essential requirement for
%%meeting the physics objectives of \Panda. There will be several
%dedicated systems which, complementary to the other detectors, will
%provide means to identify particles. The main part of the momentum
%spectrum above 1 GeV/$c$ will be covered by Cherenkov detectors. 
%Below the Cherenkov threshold of kaons several other processes have 
%to be employed for particle identification.
%In addition a time-of-flight barrel will identify slow particles.

\subsubsection*{Barrel DIRC}

At polar angles between 22\degrees{} and 140\degrees{}, particle
identification will be performed by the Detection of Internally
Reflected Cherenkov (\Dirc) light as realised in the {\INST{BaBar}}
detector~\cite{Staengle:1997xp}.  It will consist of 1.7~cm thick
fused silica (artificial quartz) slabs surrounding the beam line
at a radial distance of 45~cm to 54~cm. At {\INST{BaBar}} the light was imaged across a large
stand-off volume filled with water onto 11,000 photomultiplier
tubes. At \Panda, it is intended to focus the images by lenses onto
Micro-Channel Plate PhotoMultiplier Tubes (MCP PMTs) which are
insensitive to magnet fields. This fast light detector type allows a
more compact design and the readout of two spatial coordinates. 
%The DIRC design with its compact radiator mounted close to the
%\Emc will minimise the conversions. Part of these conversions will be
%recovered with information from the DIRC detector, as was shown by 
%{\INST{BaBar}}~\cite{Adametz05}.

\subsubsection*{Forward End-Cap DIRC} 

A similar concept is considered to be employed in the forward direction for
particles at polar angles between 5\degrees{} and 22\degrees{}. The same radiator,
fused silica, is to be employed, however in shape of a disk. The
radiator disk will be 2~cm thick and will have a radius of
110~cm. It will be placed directly upstream of the forward end cap
calorimeter. At the rim around the disk the Cherenkov light will be
measured by focusing elements. In addition measuring the time of
propagation the expected light pattern can be distinguished in a
3-dimensional parameter space. Dispersion correction is achieved by
the use of alternating dichroic mirrors transmitting and reflecting
different parts of the light spectrum. As photon detectors either
silicon photomultipliers or microchannel plate PMTs are considered.

\subsubsection*{Scintillator Tile Barrel (Time-of-Flight)}

For slow particles at large polar angles, particle identification will
be provided by a time-of-flight (TOF) detector positioned 
just outside the Barrel DIRC,
where it can be also used to detect photon conversions in the DIRC radiator.
The detector is based on scintillator tiles of $\mathrm{28.5 \times 28.5~mm^2}$ size, 
individually read out by two Silicon PhotoMultipliers per tile.
The full system consists of 5,760 tiles in the barrel part 
and can be augmented also by approximately 1,000 tiles 
in forward direction just in front of the endcap disc DIRC.
Material budget and the dimension of this system are optimised such  
that a value of less than 2\% of one radiation length, including readout
and mechanics and less than 2~cm radial thickness will be reached, respectively.
The expected time resolution of 100~ps will allow precision timing of tracks 
for event building and fast software triggers.
The detector also provides well timed input with a good spatial resolution
for online pattern recognition.

\subsubsection*{Electromagnetic Calorimeters}

%\begin{figure*}[htb]
%\begin{center}
%\includegraphics[width=0.75\dwidth]{./int/emc}
%\caption[Barrel and forward end cap of the Electro-Magnetic Calorimeter.]
%{Barrel and forward end cap of the Electro-Magnetic Calorimeter 
%  (EMC) with its mounting structures and cooling pipes.  These
%  structures will be mounted directly inside the cryostat and forward
%  end of the flux return yoke.}
%\label{fig:emc}
%\end{center}
%\end{figure*}

Expected high count rates and a geometrically compact design of the
Target Spectrometer require a fast scintillator material with a short
radiation length and Moli\`ere radius for the construction of the
electromagnetic calorimeter (\Emc). Lead tungsten (PbWO$_4$) is a
high density inorganic scintillator with sufficient energy and time
resolution for photon, electron, and hadron detection even at
intermediate energies~\cite{Mengel:1998si,Novotny:2000zg,Hoek:2002ss}.

The crystals will be 20~cm long, i.e.~approximately 22~$X_0$, in
order to achieve an energy resolution below 2\percent{} at
1~$\gev$~\cite{Mengel:1998si,Novotny:2000zg,Hoek:2002ss} at a
tolerable energy loss due to longitudinal leakage of the shower.
Tapered crystals with a front size of  $\mathrm{2.1 \times 2.1~cm^2}$ will be
mounted in the barrel EMC with an inner radius of 57~cm. This
implies 11,360 crystals for the barrel part of the calorimeter.  The
forward end cap EMC will be a planar arrangement of 3,600 tapered
crystals with roughly the same dimensions as in the barrel part, and
the backward end cap EMC comprises of 592 crystals. The readout of
the crystals will be accomplished by large area avalanche photo diodes
in the barrel and in the backward end cap, vacuum photo-triodes will be used 
in the forward end cap. The light yield can be increased by a factor of 
about 4 compared to room temperature by cooling the crystals down to 
$-25$~$\degC$. 
%The arrangement of the barrel and forward end cap calorimeters 
%is shown in \Reffig{fig:emc}.

The EMC will allow to achieve an $e$/$\pi$ ratio of 10$^3$ for momenta above 0.5~\gevc.  
Therefore, $e$/$\pi$-separation will not require an additional gas Cherenkov detector 
in favour of a very compact geometry of the EMC. 
A detailed description of the detector system can be found in~\cite{PANDA:TDR:EMC}.

\subsubsection*{Muon Detectors}

The laminated yoke of the solenoid magnet acts as a range system for the detection of muons. 
There are 13 sensitive layers, each 3~cm thick (layer ``zero" is a double-layer). They alternate with 3~cm thick iron absorber layers (first and last iron layers are 6~cm thick), introducing enough material for the absorption of pions in the \PANDA momentum range and angles.
In the forward end cap more material is needed due to the higher momenta of the occurring particles. 
Therefore, six detection layers will be placed around five iron layers of 6~cm each 
within the downstream door of the return yoke, 
and a removable muon filter with additional five layers of 6~cm iron and corresponding detection layers
will be moved in the space between the solenoid and the dipole. 

As sensors between the absorber layers, rectangular aluminum Mini Drift Tubes (MDT) are foreseen.
%which are similar to the ones applied at the COMPASS experiment~\cite{COMPASS}.
Basically, these are drift tubes with additional capacitive
coupled strips, read out on both ends to obtain the longitudinal coordinate.
All together, the laminated yoke of the solenoid magnet and the additional muon filters 
will be instrumented with 2,600 MDTs and 700 MDTs, respectively.

\begin{figure*}[htb]
\begin{center}
\includegraphics[width=0.9\dwidth]{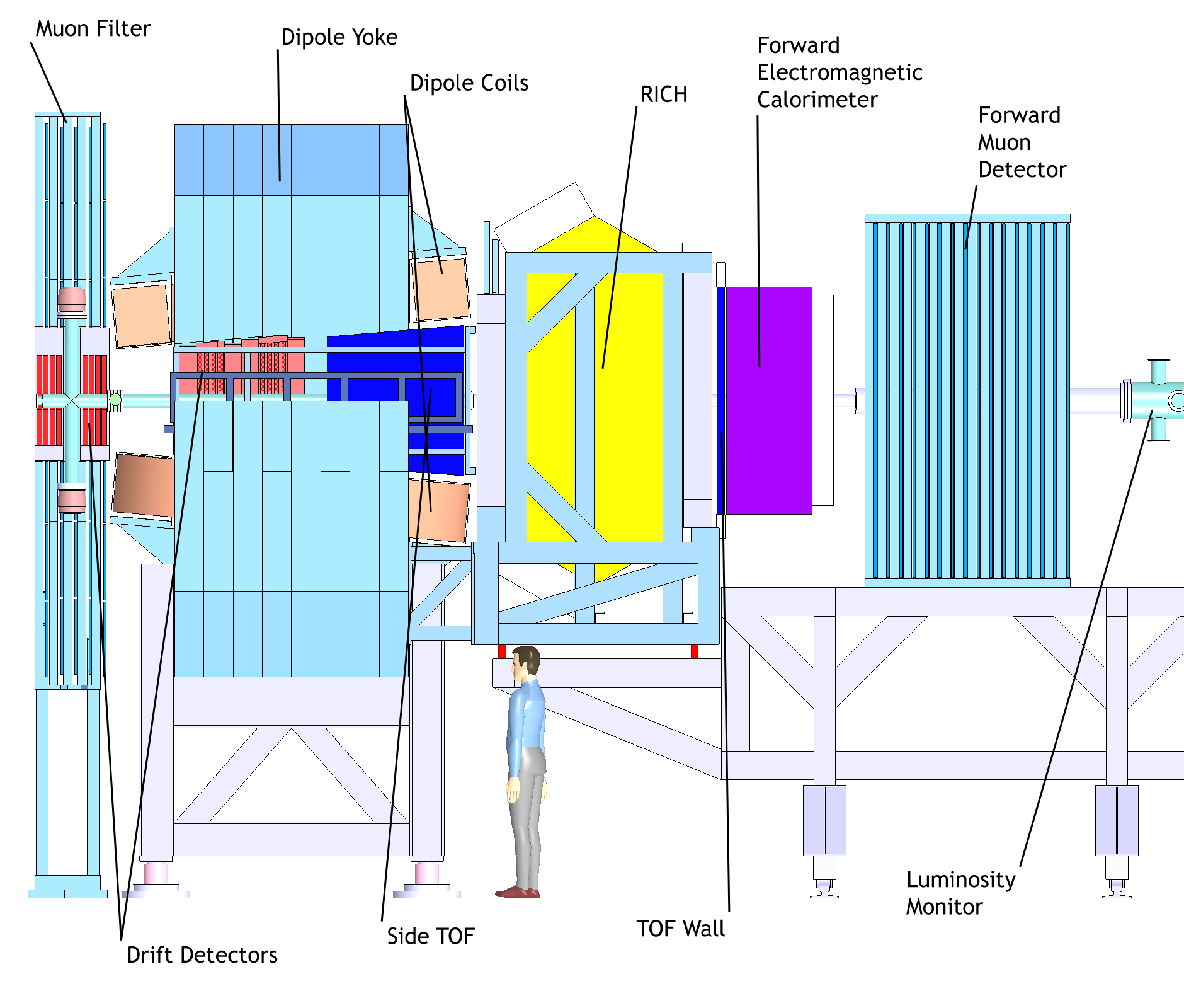}
\caption[Artistic side view of the Forward Spectrometer (FS) of \PANDA]
{Artistic side view of the Forward Spectrometer (FS) of \PANDA. 
  It is preceded on the left by the Target Spectrometer (TS),
which is illustrated in \Reffig{f:over:ts_view}.}
\label{f:over:fs_view}
\end{center}
\end{figure*}

\subsubsection*{Hypernuclear Detector}

The hypernuclei study will make use of the modular structure of
\Panda. Removing the backward end cap calorimeter and the \Mvd will allow to
add a dedicated nuclear target station and the required additional
detectors for $\gamma$ spectroscopy close to the entrance of
\Panda. While the detection of hyperons and low momentum $K^{\pm}$
can be ensured by the universal detector and its PID system, a
specific target system and a $\gamma$-detector are additional
components required for the hypernuclear studies.
%\paragraph*{Active Secondary Target}
%compact high resolution solid state tracker} 

The production of hypernuclei proceeds as a two-stage process. 
First hyperons, in particular $\Cascade \bar \Cascade$, are produced on a nuclear target. 
%In some cases the $\Xi$ will be slow enough to be captured in
In addition, a secondary target is needed for the formation of a double hypernucleus.
%a secondary target, where it reacts in a nucleus to form a .
The geometry of this secondary target is determined 
by the short mean life of the $\Cascade^-$ of only 0.164~ns. %$\Xi^-$
This limits the required thickness of the active secondary target to about 25~mm to 30~mm. 
It will consist of a compact sandwich structure of silicon micro-strip detectors and absorbing material. 
In this way the weak decay cascade of the hypernucleus can be detected in the sandwich structure.
%\paragraph*{Germanium Array} 

An existing germanium-array with refurbished readout will be used for
the $\gamma$-spectroscopy of the nuclear decay cascades of
hypernuclei. The main limitation will be the load due to neutral or
charged particles traversing the germanium detectors. Therefore,
readout schemes and tracking algorithms are presently being developed
which will enable high resolution $\gamma$-spectroscopy in an
environment of high particle flux.

%%%%%%%%%%%%%  

\subsection{Forward Spectrometer}
%\addcontentsline{toc}{subsection}{\hspace{4mm}...  Forward Spectrometer}
\label{sec:det:fs}

The Forward Spectrometer (FS) will cover all particles emitted in
vertical and horizontal angles below $\pm \mathrm{5}^\circ$ and $\pm \mathrm{10}^\circ$,
respectively.  Charged particles will be deflected by an integral
dipole field.  Cherenkov detectors, calorimeters and
muon counters ensure the detection of all particle types.  
\Reffig{f:over:fs_view} gives an overview to the instrumentation 
of the FS.

\subsubsection*{Dipole Magnet}
A 2~Tm dipole magnet with a window frame, a 1~m gap, and more
than 2~m aperture will be used for the momentum analysis of charged
particles in the FS.  In the current planning, the
magnet yoke will occupy about 1.6~m in beam direction starting from
3.9~m downstream of the target.  Thus, it covers the entire angular
acceptance of the TS of $\pm$10\degrees{} and
$\pm$5\degrees{} in the horizontal and in the vertical direction,
respectively.  The bending power of the dipole on the beam line causes
a deflection of the antiproton beam at the maximum momentum of
15~$\gevc$ of 2.2\degrees{}. 
%/The designed acceptance for charged
%/particles covers a dynamic range of a factor 15 with the detectors
%/downstream of the magnet. 
For particles with lower momenta, detectors
will be placed inside the yoke opening. The beam deflection will be
compensated by two correcting dipole magnets, placed around
the \Panda detection system.
The dipole field will be ramped during acceleration in the \HESR 
and the final ramp maximum scales with the selected beam momentum.

%For more details on the dipole magnet see~\cite{PANDA-MagnetTDR}.

\subsubsection*{Forward Trackers}
\label{sec:det:fs:trk}

%\COM[Inti]{I am not sure if my additions are correct...}

\begin{figure}[htb]
\begin{center}
\includegraphics[width=0.90\swidth]{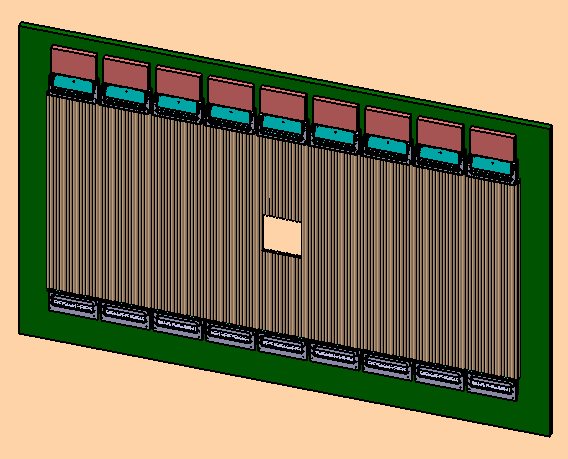} 
\caption[Double layer of straw tubes in the tracker of the Forward Spectrometer]
{Double layer of straw tubes with preamplifier cards and gas manifolds
mounted on rectangular support frame.
The opening in the middle of the detector is foreseen for the beam pipe.}
\label{fig:det:fs:trk:dc1}
\end{center}
\end{figure}

The deflection of particle trajectories in the field of the dipole
magnet will be measured with three pairs of tracking drift detectors.
The first pair will be placed in front, the second within and the
third behind the dipole magnet.  Each pair will contain two autonomous
detectors, thus, in total, 6 independent detectors
will be mounted.  Each tracking detector will consist of four
double-layers of straw tubes (see \Reffig{fig:det:fs:trk:dc1}), two
with vertical wires and two with wires inclined by a few degrees.  The
optimal angle of inclination with respect to vertical direction will
be chosen on the basis of ongoing simulations.  The planned
configuration of double-layers of straws will allow to reconstruct
tracks in each pair of tracking detectors separately, also in case of
multi-track events.
 
\subsubsection*{Forward Particle Identification}
%\paragraph*{RICH Detector}
To enable the $\pi$/$K$ and $K$/$p$ separation also at the highest
momenta a \Rich detector is proposed. The favoured design is a dual
radiator \Rich detector similar to the one used at
\INST{HERMES}~\cite{Akopov:2000qi}. Using two radiators, silica
aerogel and C$_4$F$_{10}$ gas, provides $\pi$/$K$/$p$ separation in a
broad momentum range from 2 to 15~$\gevc$.  The two different indices
of refraction are 1.0304 and 1.00137, respectively.  The total
thickness of the detector is reduced to the freon gas radiator
\mbox{(5\%$X_0$),} the aerogel radiator (2.8\%$X_0$), and the
aluminum window (3\%$X_0$) by using a lightweight mirror focusing
the Cherenkov light on an array of photo-tubes placed outside the
active volume.

%\paragraph*{Time-of-Flight Wall}
A wall of slabs made of plastic scintillator and read out on both ends
by fast photo-tubes will serve as time-of-flight stop counter placed at
about 7~m from the target. 
Similar detectors will be placed inside the dipole magnet opening 
to detect low momentum particles which do not exit the dipole magnet.
The time resolution is expected to be in the order of 50~ps
thus allowing a good $\pi$/$K$ and  $K$/$p$ separation 
up to momenta of 2.8~\gevc and 4.7~\gevc, respectively.

\subsubsection*{Forward Electromagnetic Calorimeter}

For the detection of photons and electrons a {Shashlyk}-type
calorimeter with high resolution and efficiency will be employed. 
The detection is based on lead-scintillator sandwiches read out with
wave-length shifting fibres passing through the block and coupled to
photo-multipliers. 
The lateral size of one module is $\mathrm{110~mm \times 110~mm}$ and a length of 680~mm ($=20X_0$).
A higher spatial resolution will be achieved by sub-dividing each module 
into 4 channels of 55~mm$\times$55~mm size coupled to 4 PMTs.
To cover the forward acceptance, 351
such modules, arranged in 13 rows and 27 columns at a distance of
7.5~m from the target, are required.
%The technique has already been successfully used in
%the \INST{E865} experiment~\cite{bib:emc:E865} and it has been adopted
%for various other experiments like PHENIX and LHCb.  
With similar modules, based on the same technique as proposed for \PANDA,
an energy resolution of $4\%/\sqrt{E}$~\cite{bib:emc:KOP99} has been achieved.
%A view of a 3x3 matrix of Shashlyk modules with 
%is shown in
%\Reffig{fig:det:fs:shashlyk}.  

%\begin{figure}[htb]
%\begin{center}
%\includegraphics[width=\swidth]{./int/Shashlyk} 
%\caption[Shashlyk module for the FS calorimeter]
%{$3 \times 3$ matrix of prototype Shashlyk modules as they should 
%  be employed for the \PANDA Forward Electromagnetic Calorimeter.}
%\label{fig:det:fs:shashlyk}
%\end{center}
%\end{figure}

\subsubsection*{Forward Muon Detectors}

For the very forward part of the muon spectrum, a further range
tracking system consisting of interleaved absorber layers and
rectangular aluminium drift-tubes is being designed, similar to the
muon system of the TS, but laid out for higher
momenta. The system allows discrimination of pions from muons,
detection of pion decays and, with moderate resolution, also the
energy determination of neutrons and anti-neutrons.
The forward muon system will be placed at about 9~m from the target.

\subsubsection*{Luminosity Monitor}
%\addcontentsline{toc}{subsection}{\hspace{4mm}...  Luminosity monitor}
The luminosity at \PANDA will be determined by using elastic
antiproton-proton scattering as a reference channel. 

At very small transferred momentum, corresponding to small polar angles, the elastic
cross section is dominated by the Coulomb component which is exactly
calculable. Taking the beam divergence into account, the angular
distribution of scattered antiprotons will be measured in the range of 3-8 mrad, 
corresponding to the Coulomb-nuclear interference region. The angle of each 
scattered antiproton will be measured by four layers of thin silicon microstrip
detectors placed about 11 m behind the interaction point, behind the Forward 
Spectrometer. The planes are positioned as close to the beam axis as possible 
and separated by 10-20 cm along the beam direction. The current design foresees 
that every plane consists of 8 sensors in trapezoidal shape, covering the whole
azimuthal angle, in order to suppress systematic effects from {\it e.g.} the
forward dipole magnet and potential misalignment of the beam. The silicon
sensors will be located inside the vacuum to minimize scattering of the 
antiprotons before traversing the tracking planes. With the proposed detector 
setup an absolute precision of 3\% for the time integrated luminosity is expected.

\subsection{Data Acquisition}
%\addcontentsline{toc}{subsection}{\hspace{4mm}...  Data Acquisition}

In \PANDA, a data acquisition concept is being developed to be
as much as possible matched to the complexity of the
experiment and the diversity of physics objectives and the rate
capability of at least $2\cdot10^{7}$ events/s.
Therefore, every sub-detector system is a self-triggering entity.
Signals are detected autonomously by the sub-systems and are preprocessed.
Only the physically relevant information is extracted and transmitted.
This requires hit-detection, noise-suppression and clusterisation at
the readout level. 
The data related to a particle hit, with a substantially reduced rate
in the preprocessing step, is marked by a precise time stamp and
buffered for further processing.
The trigger selection finally occurs in computing nodes which
access the buffers via a high-bandwidth network fabric. The new
concept provides a high degree of flexibility in the choice of trigger
algorithms. It makes trigger conditions available which are
outside the capabilities of the standard approach.

\subsection{Infrastructure}
%\addcontentsline{toc}{subsection}{\hspace{4mm}...  Infrastructure}

\label{s:over:infra}
The \PANDA experimental hall will be located in the east straight section of \HESR.
The planned floor space in the hall will be of $\mathrm{43~m \times 29~m}$. 
Within the cave, the \Panda detector, the auxiliary equipment, the beam steering magnets 
and the focusing elements will be housed. To allow for access during \HESR operation, 
the area of the beam line and the detector will be shielded with movable concrete blocks. 
Controlled access will be provided via a properly designed chicane in the concrete wall.
In addition, the experimental hall will provide additional space for components storage 
and detector parts assembly. The \PANDA hall will feature an overhead crane, spanning the 
whole area and with a maximum load capacity of $25~t$.
The shielded space for the \PANDA detector and the beam line will have an area of 
$\mathrm{37~m \times 9.4~m}$ and a height of 8.5~m. The beam line at a height of 3.5~m.  
The floor level in the \HESR tunnel will be 2~m higher. The TS with its front-end electronics 
will be mounted on rails and movable from the on-beam position to outside the shielded area, 
to allow simultaneous detector and accelerator maintenance.

In the south-west corner of the \PANDA hall, the experiment counting house complex is foreseen.
It will be a complex made of five floors. At the first floor, the supplies for power, high voltage, 
cooling water, gases and other services will be housed. The second floor will provide space for the 
readout electronics and data processing and the online processing farm will be housed at the third floor. 
The hall electricity supply and ventilation will be hosted at the fifth floor, whereas at the fourth floor 
there will be the space for the shift crew: the control room and a meeting room, with some service rooms, 
will be at the same level of the surrounding ground.
The \PANDA experiment will need liquid helium for the TS solenoid and for the compensation solenoid. 
The refrigeration scheme will be similar to the one used for the BaBar magnet \cite{babar:magnet}.
The cryogenic plant will be built and characterised at FZ J\"ulich and moved to FAIR with the magnet. 
In the natural convection refrigeration scheme that has been proposed, the storage-liquefaction Dewar close to the 
liquefier acts as buffer for the system. With the projected LHe consumption (safety factor on cryogenic losses included), 
a $2000~l$ storage will allow $\sim 10~h$ of operation in case of liquefier failure, giving ample time margin for the 
magnet discharge. The supply point will be at the north-east area of the building. From that point, the LHe will be 
delivered to the control Dewar, which can be chosen sufficiently small ($\sim 30~l$) to minimise the LHe inventory 
in the \PANDA hall. The helium gas coming out from the thermal shields would be recovered at room temperature and 
pressure in the low pressure recovery system.

All other cabling, which will be routed starting at the counting house, will join the LHe supply lines at the end 
of the rails system of the TS at the eastern wall. The temperature of the building will be moderately controlled. 
More stringent requirements with respect to temperature and humidity for the detectors have to be maintained locally. 
To facilitate cooling and avoid condensation, the Target Spectrometer will be kept in a tent with dry air at a controlled temperature.

%EOF detector.tex

\begin{figure*}[tp]
\begin{center}
\includegraphics[width=\dwidth]{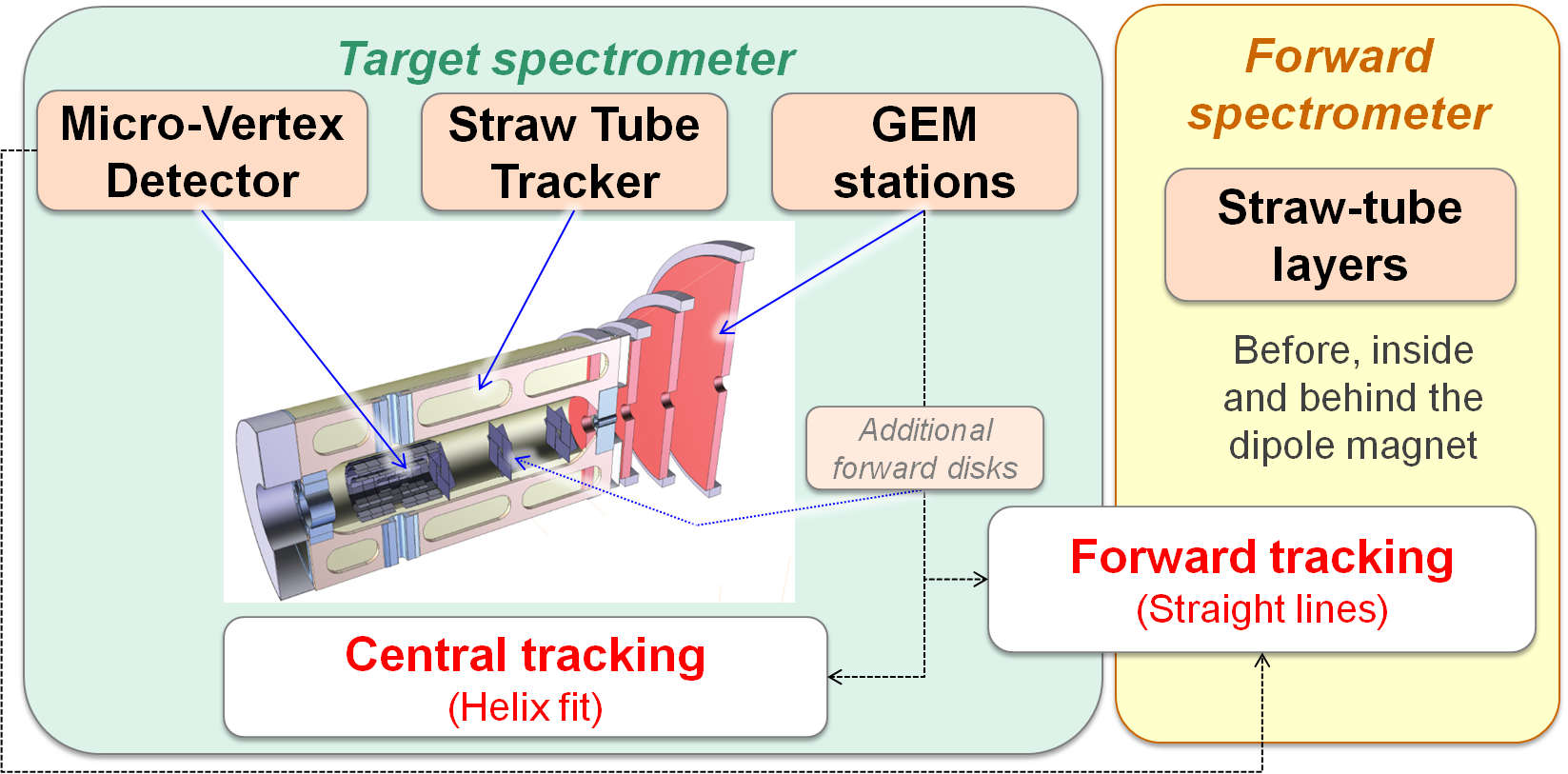}
\caption[Overview of the \PANDA tracking system.]
{
Overview of the \PANDA tracking system, including the option of the additional forward disks.
}
\label{fig:int:tracking}
\end{center}
\end{figure*}   

\section{The Charged Particle Tracking System}
%\section*{II.\hspace{4mm} The charged particle tracking system}
%\addcontentsline{toc}{section}{II.\hspace{4mm} The charged particle tracking system}

There are different tracking systems for charged particles at \PANDA,
positioned inside the target spectrometer 
and in the forward region around the dipole magnet. 
Main tasks of the global tracking system are 
the accurate determination of the particle momenta, 
a high spatial resolution of the primary interaction vertex 
and the detection of displaced secondary vertices. 
Therefore, measurements of different subdetectors 
have to be merged in order to access the full tracking information.

\subsection{Basic Approach}

The magnetic solenoid field in the target spectrometer results in 
a circular transverse motion of charged particles with non-zero transverse momentum. 
The particle momentum then can be extracted via the determination of the bending radius. 
However, tracks with a small polar angle will exit the solenoid field too soon to be measured properly.
For this case, the particle deflection induced by the subsequent dipole magnet 
is used to measure the particle momentum. 
Basically it can be deduced from a combined straight line fit before and after the dipole.

Due to the different analysing magnets, 
different track fitting algorithms 
have to be applied for central and forward tracks.  
Central tracks are reconstructed by combining hit points in the MVD layers 
with the hit information of the STT or the GEM stations. 
For the reconstruction of small angle tracks the straw tube layers 
in the forward spectrometer have to be used.
In overlap regions the MVD, the additional forward disks or the GEM stations 
can contribute to the forward tracking
because the delivery of an additional track point closer to the IP 
significantly improves the precision of the fitting results. 
After the global identification of individual tracks 
an event mapping have to be performed to match  
different tracks of the same event to a common vertex 
which either corresponds to the primary interaction vertex 
or a delayed decay of short-lived particles. 
%The central tracking system is of particular importance 
%for the reconstruction of charmed hadrons
%because associated charged decay modes at \PANDA often release 
%a high transverse momentum of up to \unit[1.5]{GeV/c}, 
%which leads to a large opening angle of the daughter particles 
%in the laboratory reference frame. 
%In this context, the MVD is essential 
%for the reconstruction of displaced vertices 
%close to the interaction region which 
%is needed e.g.~for the tagging of $D$ mesons

%In this way it is possible to obtain information about the event topology. 

%In many instances associated charged decay modes release 
%a high transverse momentum of up to \unit[1.5]{GeV/c}, 
%which leads to a large opening angle of the daughter particles 
%in the laboratory reference frame. 
%A particular task of the tracking system is the tagging of $D$ mesons via their displaced vertex. 
%It is crucial for many applications of the physics program. 

The luminosity monitor at the downstream end of the experiment 
is a tracking device of its own right.
It was introduced to measure the time integrated luminosity, 
which is essential for the determination of cross sections for different physics processes. 
Therefore, elastically scattered antiprotons are measured 
under small angles corresponding to small momentum transfers. 
The associated differential cross sections are well known 
and thus provide an ideal reference channel. 
Additional information from the MVD will eventually improve 
the measurement by taking advantage of the reconstructed  
slow recoil proton at polar angles of around 90$^{\circ}$, 
which is correlated with the highly energetic antiproton 
detected in the luminosity monitor.

\subsection{Optimisation Criteria}
The different topics of the \PANDA physics program
%each of which is scientifically well justified, 
will impose specific optimisation criteria and
requirements to design and performance of the tracking system.
The optimum design thus depends on the relative weight 
which is given to the different physics aspects. 
Main criteria for the optimisation will be discussed in the following.
\subsubsection*{Acceptance}

Full $2\pi$ azimuthal coverage is mandatory in order to allow
identification of multi-particle final states and studies of
correlations within the produced particles. 
In particular, the spectroscopy program of charmed and strange hadrons 
relies on the measurement of Dalitz plot distributions of three-body final states, 
which requires a smooth acceptance function across the full phase space. 
Particular care has to be taken to avoid gaps in
the acceptance function and to minimise the effect of
discontinuities induced by the transition between adjacent
sub-detector components, by detector frames or by mechanical
support structures.

The fixed-target setup at \PANDA implies a Lorentz boost $\gamma_{CM}$ 
of the centre of mass ranging from 1.20 to 2.92. 
This large dynamic range in the Lorentz boost corresponds 
to a large difference in the typical event topologies
at low and at high antiproton momenta. 
At higher antiproton beam momenta 
the vast majority of the produced particles in the final state 
will be emitted into the forward hemisphere. 
However, light particles like $e^{\pm}$, ${\mu^{\pm}}$ or ${\pi^{\pm}}$ 
may well be emitted into the backward hemisphere 
even at highest beam momentum.
%The critical value of the center of mass momentum 
%that can result in backward emission in the laboratory frame
%is given by 
%$p_{\rm{}crit}=\beta_{CM}\cdot\gamma_{CM}\cdot{}m_x=
%\left(p_{\bar{p}}/\sqrt{s}\right)\cdot{}m_x$. 
As an example, pion backward emission is possible for 
a centre of mass momentum $p_{cm}>93$~\mevc at $p_{\bar{p}}=1.5$~\gevc, 
and for $p_{cm}>380$~\mevc at $p_{\bar{p}}=15$~\gevc. 

%Since muon-pion separation requires
%passage through a large amount of material and thus high momenta
%to efficiently suppress the much more abundant pions, muon
%detection at backward angles doesn't seem to be feasible. 

%However, in the tracking of electrons and pions coverage of a significant
%fraction of the backward hemisphere is required.

Backward charged particle tracking is needed 
for various measurements foreseen at \PANDA.
For instance, for the independent determination of 
the electric and magnetic parts of the time-like proton form factor 
in the reaction \pbarp~$\to$~\ee 
the full angular distribution has to be measured. 
At $q^2=14$~\gevcsq, that is at $p_{\bar{p}}=6.45$~\gevc, 
a polar angle of $160^{\circ}$ in the centre of mass frame 
corresponds to electrons with a momentum of 0.70~\gevc at $\theta_{lab}=113^{\circ}$. 
Detection of pions in the backward hemisphere is important 
in studies of strange, multi-strange and charmed baryon resonances in
\pbarp~$\to$~$Y^{\star}\bar{Y}'$ (+c.c.) reactions where
the excited hyperon $Y^{\star}$ decays by single or double pion
emission. 
Also higher charmonium states may emit pions with decay
energies above the critical value for backward emission in the
laboratory. 
The \PANDA tracking detectors therefore have to cover
the full range of polar angles between $0^{\circ}$ 
%(for particles with $p/q$ different from that of the antiproton beam) 
and about $150^{\circ}$.

Besides the solid angle of the detector 
also the acceptance in momentum space has to be considered. 
Often the final state contains charged particles 
with very large and with very small transverse momentum components 
which need to be reconstructed at the same time. 
Given the strength of the solenoid field of 2~T 
%(at $p_{\bar{p}}\ge{}3.8\,{\rm{}GeV}/c$) 
required to determine the momentum vector of the high transverse momentum particle, 
the radius of the transverse motion of the low transverse momentum particle may be small. 
Sufficient tracking capability already at small distance from the beam axis 
is therefore mandatory. 
As an example one may consider the reaction 
$\bar{p}p\rightarrow{}D^{*+}D^{*-}$ 
close to threshold with $D^{*+}\rightarrow{}D^0\pi^+$ (\& c.c.).
Assuming 39~\mevc momentum of the decay particles in the $D^{*\pm}$ rest frame,
%%and 39~MeV/c momentum of the decay particles in the $D^{*\pm}$ rest frame. 
%The $D^0\rightarrow{}K^-\pi^+$ (\& c.c.) decay particles have 61~MeV/c momentum 
%in the $D^0$/$\bar{D}^0$ rest frame. 
particles of the subsequent decay $D^0\rightarrow{}K^-\pi^+$ (\& c.c.) 
have 61~\mevc momentum in the $D^0$/$\bar{D}^0$ rest frame.
In the solenoid field of the TS,
the charged pions and kaons from the $D^0$/$\bar{D}^0$ decay 
may have helix diameters up to almost 1.5~m.
The transverse motion of the charged pion from the  $D^{*\pm}$ decay
stays within a distance of almost 7~cm from the beam axis
and therefore need to be reconstructed based on the
track information from the MVD only.

\subsubsection*{Delayed Decay Vertex Detection}

An important part of the \PANDA physics program involves final
states consisting of hadrons with open charm or strangeness which
decay by weak interaction and thus have macroscopic decay lengths.
The decay length of charmed hadrons is of the order of 100~$\mu$m
($\approx$~310~$\mu$m for $D^{\pm}$, $\approx$~150~$\mu$m for
$D_s^{\pm}$, $\approx$~120~$\mu$m for $D^0$, $\approx$~130~$\mu$m for
$\Cascade_c^+$, $\approx$~60~$\mu$m for $\Lambda_c^+$, and
$\approx$~30~$\mu$m for $\Cascade_c^0$). 
Therefore, the design of the tracking system aims on a detection
%\PANDA detector is designed
%according to the goal to detect 
of decay vertices of particles with
decay lengths above 100~$\mu$m. 
In order to achieve sufficient separation of the reconstructed decay vertex, 
the inner part of the tracking system has to be located very close to the
interaction point, both in longitudinal and in radial direction.
This requirement is fulfilled in the design of the MVD.

The identification of hyperons and $K_S$ mesons requires the
reconstruction of delayed decay vertices at much larger distances.
$\Lambda$ and $\Cascade$ hyperons have comparatively large decay
lengths of about $8$~cm and $5$~cm, respectively. 
Due to the Lorentz boost this may result in vertices 
which are displaced by tens of centimetres 
from the interaction point mostly in the downstream direction. 
The considerations in the previous section concerning
the required acceptance thus apply with respect to the shifted
emission points of charged particles. 
The inner part of the \PANDA tracking system,
%in particular the Micro-Vertex-Detector (MVD) as the inner part,
therefore needs sufficient extension to the downstream direction
in order to deliver sufficient track information for charged
particle tracks originating from these displaced vertices.

\subsubsection*{Momentum and Spatial Resolution}

The spatial resolution of the tracking detectors is important in
two aspects. In the vicinity of the interaction point it directly
determines the precision to which primary and displaced decay
vertices can be reconstructed. Further on, based on the deflection
of charged particles in both solenoid and dipole magnetic fields,
it is an essential contribution to the momentum resolution of
charged particles in all three coordinates.

The detection of displaced vertices of charmed hadrons imposes
particular requirements to the spatial resolution close the interaction point. 
With a typical Lorentz boost $\beta\gamma\simeq{}2$, $D$ meson decay vertices 
have a displacement of the order of a few hundreds micrometres 
from the primary production point. 
Hence, to distinguish charged daughter particles of $D$ mesons from prompt particles 
a vertex resolution of 100~$\mu$m is required. 
The position resolution is less demanding for the reconstruction of strange hadrons 
having decay lengths on the scale of centimeters. 
In this case a vertex resolution of a few millimetres is sufficient.
Due to the significant Lorentz boost and the small opening angle between the decay particles of hyperons 
the resolution in transverse direction is required to be much better 
than the one for the longitudinal component.

The achievable momentum resolution is a complex function of the
spatial resolution of the tracking sub-detectors, the number of
track-points, the material budget of active and passive components
resulting in multiple scattering, the strength and homogeneity of
the magnetic field, and of the particle species, its momentum and
its emission angle. Due to the respective momentum dependence, it
is generally expected that multiple scattering limits the momentum
resolution of low energy particles, whereas for high energy
particles the smaller curvature of the tracks is the dominant
contribution to the resolution. 

The resolution in the determination of the momentum vectors of the
final state particles directly determines the invariant 
or missing mass resolution of the particles that are to be reconstructed.
Typically, the width of hadrons unstable with respect to strong
interaction (except for certain narrow states like e.g.~charmonium
below the $D\bar{D}$ threshold) is of the order of 10~\mevcc to 100~\mevcc.
As an instrumental mass resolution much below the natural width is
without effect, a value of a few 10~\mevcc seems to be acceptable for
the identification of known states or for the mass measurement of new states. 
With a typical scale of \gevcc for the kinematic particle energy
this translates to a relative momentum resolution $\sigma_p/p$ of
the order of 1\% as design parameter for the \PANDA tracking detectors.

%Simulation results presented in
%this report demonstrate that with the chosen design of the
%tracking detectors the achieved mass and momentum resolution
%fulfills the requirements given above.

%
\subsubsection*{Count Rate Capability}
The expected count rates depend on the event rate as discussed 
in chapter~\ref{lumi-considerations}
and the multiplicity of charged particles produced in the events.
While the total rate is of importance for DAQ design and online
event filtering, the relevant quantity for detector design and
performance is the rate per channel, which is a function of the
granularity per detector layer and of the angular distribution of
the emitted particles. 
The latter depends on the beam momentum and the target material.

The nominal event rate at \PANDA is given by $2 \cdot 10^7$ interactions per second.
In case of $\bar{p}p$ annihilations 
typically only a few charged particles are produced. 
Even if secondary particles are taken into account,
the number of charged particles per event will not be much larger
than 10 in most cases. 
Thus the detector must able to cope with a rate of $2 \cdot 10^8$ particles
per second within the full solid angle.
Particular attention has to be paid to elastic $\bar{p}p$
scattering since this process contributes significantly to the
particle load in two regions of the detector: scattering of
antiprotons at small forward angles and the corresponding emission
of recoil protons at large angles close to $90^{\circ}$. This affects
primarily the inner region of the MVD disc layers and the forward
tracking detector as well as the MVD barrel part and the central
tracker. 

The use of nuclear targets will not create significantly 
higher count rates than obtained with a hydrogen or deuterium target. 
This is due to single Coulomb scattering 
which dramatically increases with the nuclear charge
($\propto{}Z^4$) and results in $\bar{p}$ losses 
with no related signals in the detector. 
In contrast to $\bar{p}p$ collisions in $\bar{p}A$ collisions no
high rate of recoil particles close to $90^{\circ}$ is expected. 
The emission angles of recoil protons from quasi-free $\bar{p}p$
scattering are smeared by Fermi momentum and rescattering, while
recoil nuclei, if they at all survive the momentum transfer, are
too low energetic to pass through the beam pipe.

\subsubsection*{Particle Identification}

Charged particle identification over a wide range of momentum and
emission angle is an essential prerequisite for the capability of
\PANDA to accomplish the envisaged physics program. 
Charged particles with higher momenta will be identified via Cherenkov
radiation by the DIRC detector in the Target Spectrometer  and by
the forward RICH detector in the Forward Spectrometer. 
For positive charged kaon-pion separation in the DIRC 
about 800~\mevc momentum is required. 
While almost all particles emitted within the acceptance of the Forward
Spectrometer are above the Cherenkov threshold due to the forward
Lorentz boost, a number of interesting reaction channels have
final states with heavier charged particles ($K^{\pm},p,\bar{p}$)
at larger angles with momenta below the DIRC threshold. 
In order to separate these low energy kaons
from the much more abundant pions, particle identification
capability based on energy loss information has to be supplied by
the central tracking detector. 
%The required separation power and
%thus quality of the $dE/dx$ measurement in the central tracker
%depends on whether or not a TOF barrel, delivering independent
%information on the particle velocity, will be included in the
%Target Spectrometer.

\subsubsection*{Material Budget}

Any active or passive material inside the detector volume
contributes to multiple scattering of charged particles, electron
bremsstrahlung and photon conversion, and thus reduces the
momentum resolution for charged particles in the tracking
detectors, and detection efficiency and energy resolution for
photons in the EMC. Therefore the material budget has to be kept
as low as possible. Following the more demanding requirements to
meet the performance criteria of the EMC, a total material budget
of MVD and Central Tracker below 10\% is still considered to be
acceptable~\cite{PANDA:TDR:EMC}.

% EOF

%

%
%\newpage
\bibliographystyle{panda_tdr_lit}
%\bibliography{./int/lit_int,./main/lit_main}
\bibliography{./int/lit_int}
%
% EOF: stt_tdr_int.tex 

%
% Include Vol. 3 Files
%
%\include{panda_tdr_stt}
%
% STT TDR
% File for chapter 1 
\chapter{The Straw Tube Tracker - STT}
% FILE: panda_tdr_stt_gen.tex
%
\section{General Overview}
%\COM{Author(s): P. Gianotti/P.Wintz}
This chapter describes the technical layout of the central Straw Tube Tracker (STT) of the
\Panda experiment. The STT is the main tracking detector for charged particles
in the \Panda target spectrometer and consists of 4636 single straw tubes, arranged
in a large cylindrical volume around the beam-target interaction point. 
It encloses the Micro-Vertex-Detector (MVD) for the inner tracking and is followed in beam direction by a vertical
setup of GEM-disks for adding track points in the forward polar angle range, as discussed in the previous section.

The tasks of the STT are the precise spatial reconstruction of the helical trajectories of charged particles in a broad 
momentum range from about a few 100\,MeV/c up to 8\,GeV/c, the measurement of the particle momentum by the reconstructed 
trajectory in the solenoidal magnetic field and the measurement of the specific energy-loss (dE/dx) for particle identification (PID).
The PID information from the STT is needed in particular to separate protons, kaons and pions in the momentum region below about 1\,GeV/c.

Since straw tubes are the basic detector elements of the STT, the next section describes first straw tubes and their properties in general. 
Then the specific straw tube design and chosen gas mixture for the \Panda-STT are described.
The technical layout of the STT, presented in the next sections, is based on the construction and development of several prototype systems. 
The main detector and electronic readout properties have been investigated by various test setups and measurements, including tests with 
high-rate proton beams, which will be discussed in a later chapter.

The presented layout and performance of the STT in the \Panda target spectrometer environment has been checked by dedicated simulations, 
reconstruction and full analysis studies of certain $p\bar{p}$-reactions, identified as being benchmark tests for the whole \Panda scientific 
program. These studies used the official \Panda software framework (PandaRoot) with implemented track-finding, -fitting and ana\-ly\-sis routines 
for primary and secondary tracks in the STT. The results are discussed in detail in a particular chapter.   

The last chapter describes the project organization and summarizes the time lines of the STT construction.

%
%EOF: panda_tdr_stt_gen.tex

% FILE: panda_tdr_stt_des.tex
%
\section{Straw Tube Description}
%\COM{Author(s): P. Wintz}
\label{sec:stt:des}
Straws are gas-filled cylindrical tubes with a conductive inner layer as
cathode and an anode wire stretched along the cylinder axis. An
electric field between the wire and the outer conductor separates electrons and
positive ions produced by a charged particle along its trajectory through
the gas volume. Usually the wire is on positive voltage of a few kV and
collects the electrons while the ions drift to the cathode. By choosing thin
wires, with a diameter of few tens of $\mu$m, the electric field strength near the
wire is high enough to start further gas ionizations by electron collisions with
gas molecules.
Depending on the high voltage and the gas characteristics an
amplification of about $10^{4}-10^{5}$ of the primary charge signal is
possible, which is large enough to read out the signal.
\par
By measuring the drift time of the earliest arriving electrons one gets
the information about the minimum particle track distance from the wire.
The isochrone contains all space points belonging to the same electron drift
time and describes a cylinder around the wire axis.
The characteristic relation between drift time and isochrone   
is given by the electron drift velocity, depending on specific gas parameters, electric and
magnetic field. Therefore, this fundamental relation has to be calibrated
using reference tracks with known space and drift time information.
The particle track is reconstructed by a best fit to the isochrones
measured in a series of several straw tubes with the same orientation.   
Additional skewed straw layers provide a full stereo view of the particle
trajectory.
\par
The specific energy-loss (dE/dx) of a charged particle in 
the straw gas volume can be used to identify the particle species and can be derived from the 
number of ionization electrons per track length (dx) for the generated straw signal. Since the 
specific ionization in gas 
with about 100 ion-electron pairs per cm for minimum ionizing particles is quite low and shows in 
addition a strong fluctuation described by an asymmetric Landau distribution, a higher number of 
measurements is needed to get a sufficient precision for the particles' specific energy-loss. The 
truncated mean method, which rejects from many samples those with the largest energy-losses 
due to the fluctuations, can help to improve the resolution.

\par
Straw detectors exhibit the most simple geometry of highly symmetrical, 
cylindrical tubes and have several advantages which are summarized in the following:
\begin{itemize}
\item robust electrostatic configuration. The shielding tube around each
high voltage wire suppresses signal cross-talk and protects neighbor straws
in case of a broken wire;
\item robust mechanical stability if the straws are arranged in close-packed
multi-layers;
\item high detection efficiency per straw for about 99.5\,\% of the inner tube radius and 
minimal dead zones of a few mm at the tube ends;
\item high tracking efficiency for multi-layers if thin-wall straws are close-packed with 
minimal gaps of about 20\,$\mu$m between adjacent tubes;
\item high spatial resolution, $\sigma _{r\varphi }<$150\,$\mu$m 
depending on the tube diameter and gas characteristics;
\item simple calibration of the space-drift time relation due to the cylindrical
isochrone shape;
\item small radiation length, $X/X_{0}\sim$0.05$\,\%$ per tube, if straws with
thinnest ($\sim$30$\,\mu$m) film tubes are used;
\item the high rate capability can be improved by reducing the occupancy using
smaller tube diameter and/or choosing a fast drift gas.
\end{itemize}

\subsection{Straw Materials}
%\COM{Author(s): P. Wintz}
\label{sec:stt:des:mat}
The straw tubes used for the \Panda STT have a length of 1500\,mm, 10\,mm inner diameter, and a 
total wall thickness of 27\,$\mu$m. They are made of two layers of 12\,$\mu$m thin
aluminized Mylar \cite{bib:stt:des:Myl} films by wrapping two
long film strips around a rotating mandrel and gluing the
two half-overlapping strips together. Then the cylindrical film tube is
stripped off. The aluminization at the inner tube wall is used as the cathode
whereas the aluminization of the second, outer strip layer is used to prevent
light incidence.
\par
\begin{figure}%%[htb]
  \begin{center}
  \includegraphics[width=\swidth]{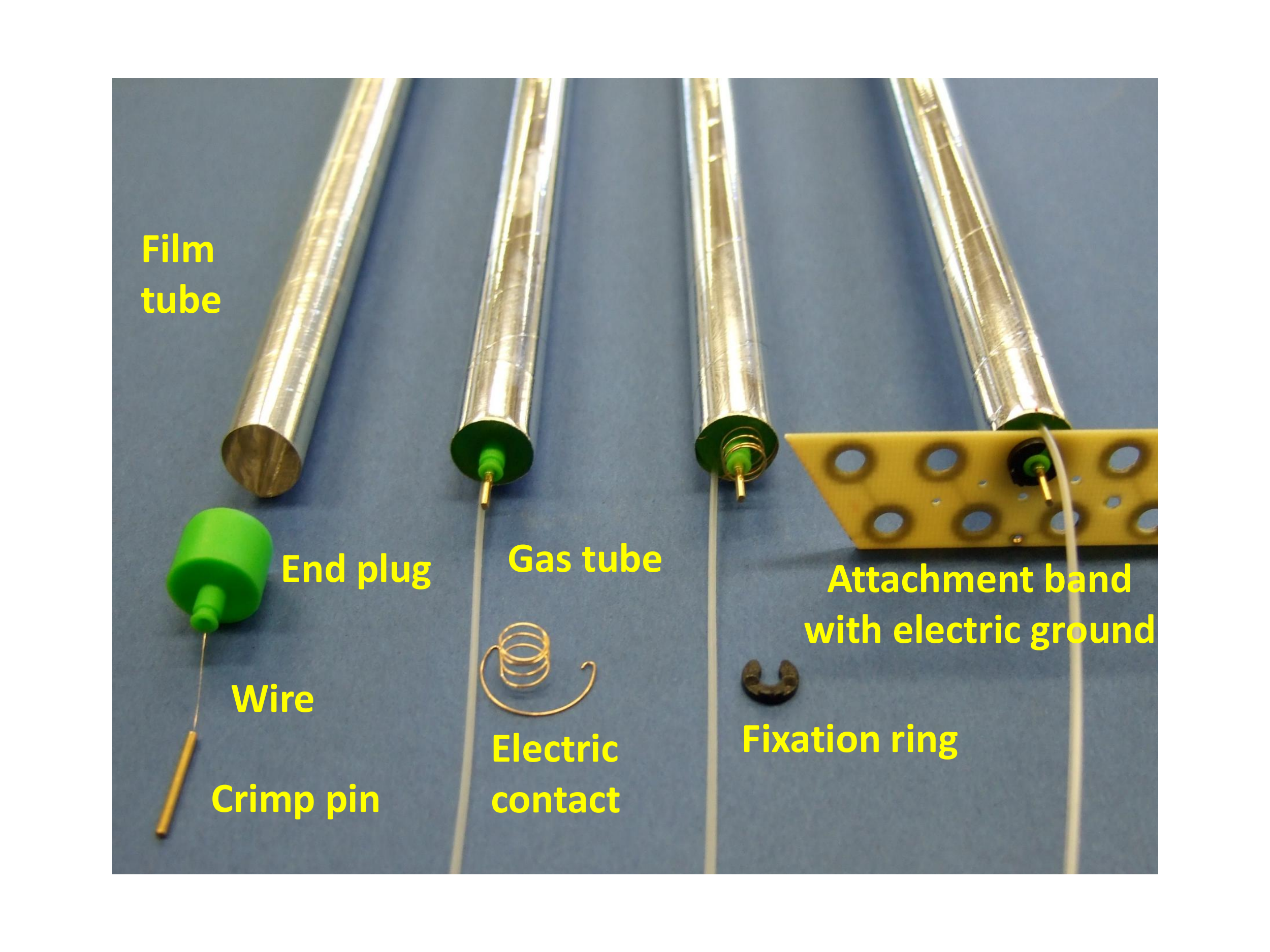}
  \end{center}
 \vspace{-0.5cm}
    \caption[Straw components and assembly steps]{Photograph of all straw components and the straw assembly steps. See the text for a description.}
    \label{fig:stt:des:mat:strawtube}
  \end{figure}

  A gold-plated tungsten-rhenium wire with 20\,$\mu$m diameter is 
  used as anode. Cylindrical precision end plugs made from
  ABS thermoplastic \cite{bib:stt:des:ABS} with a wall thickness of 0.5\,mm close the tube at both 
  ends (see \Reffig{fig:stt:des:mat:strawtube}).
  They are glued to the Mylar film leaving a small 1.5\,mm film overlap on
  both ends. There, a gold-plated copper-beryllium spring wire is inserted to
  provide the electric cathode contacting. 
  The springs allow a 2\,mm tube elongation
  with a typical spring force equivalent to 10\,g. 
  The end plugs have a central hole with a
  3$\,$mm thick cylindrical nose to insert and glue a crimp pin for the wire. 
  A micro PVC (medical quality grade) tube is fed through another hole and glued in the end plugs
  to provide a gas flow through the tube. The total weight of a fully
  assembled straw is 2.5\,g. The anode wire is stretched by a weight of 50$\,$g and crimped in the 
  copper pins at a gas overpressure in the straw tube of 1$\,$bar.   

\Reftbl{tab:stt:des:mat:radlen} lists the different straw components and their thickness in
radiation lengths. The chosen film tubes are
the thinnest used for straw detectors, but still show sufficient mechanical
stability for the assembly to self-supporting multi-layers.
\begin{table*}
\caption[Mean thickness in radiation lengths of the different straw
    tube components]{Mean thickness in radiation lengths of the different straw
    tube components. The number for the gas mixture is evaluated at 20$^{\circ}$\,C and 2$\,$atm.}
\label{tab:stt:des:mat:radlen}
\par\bigskip
\hfill\vbox{{\halign
{\hfil#\quad\hfil&
\hfil#\quad\hfil&
\hfil#\quad\hfil&
\hfil#\quad\hfil&
\hfil#\quad\hfil\cr
\noalign{\smallskip\hrule\smallskip}
\hfil Element \hfil & 
\hfil Material \hfil & 
\hfil X$\left[mm\right]$ \hfil & 
\hfil X$_{0}\left[cm\right]$ \hfil & 
\hfil X/X$_{0}$ \hfil \cr
\noalign{\smallskip\hrule\smallskip}
Film Tube & Mylar, 27$\,\mu$m & 0.085 & 28.7 & 3.0$\times$10$^{-4}$ \cr
Coating & Al, 2$\times$0.03$\,\mu$m & 2$\times$10$^{-4}$ & 8.9 & 2.2$\times 10^{-6}$ \cr
Gas & Ar/CO$_{2}$(10$\,\%$) & 7.85 & 6131 & 1.3$\times$10$^{-4}$ \cr
Wire & W/Re, 20$\,\mu$m & 3$\times$10$^{-5}$ & 0.35 &  8.6$\times$10$^{-6}$ \cr
\noalign{\smallskip\hrule\smallskip}
& & & $\sum_{straw}$ & 4.4$\times 10^{-4}$ \cr
\noalign{\smallskip\hrule\smallskip}
}}}
\hfill\break\par
\end{table*} 
For the proposed \Panda straw tracker the total radiation length of the straw
volume is 1.2$\,\%$ with a maximum number of 27 hit straw layers for a
traversing particle track in radial direction. 

\subsection{Pressurized Straws}
\label{sec:stt:des:pres}
Both, efficiency and resolution of a straw are best for a perfect
cylindrical shape of the film tube and the wire being highly concentrically stretched 
along the cylinder axis. With a wire tension\footnote{Usually given as the mass
weight used to stretch the wire.} of about 50\,g
inside a 1.5\,m long horizontal straw tube the maximum sag due to
gravitation at the middle of the tube is less than 35\,$\mu$m. 
For the 4636 straws of the \Panda central tracker this adds up to a wire tension 
equivalent to about 230\,kg which must be maintained. Usually, this is done
by fixing the straw tubes inside a strong and massive surrounding
frame or by adding reinforcement structures like CF-strips along the
tubes to keep them straight. All methods inevitably increase the
detector thickness given in radiation length by these additional materials.

Therefore a new technique based on self-supporting 
straw double-layers with intrinsic wire tension developed for the COSY-TOF 
straw tracker \cite{bib:stt:des:wintzaip} has been adopted and further developed 
for the \Panda STT. 
Single straw tubes are assembled and the wire is stretched by 50\,g at an 
overpressure of 1\,bar. Then a number of tubes are close-packed and glued
together to planar multi-layers on a reference table which defines a 
precise horizontal tube to tube distance of 10.1\,mm. At the gas overpressure
of 1\,bar the double-layer maintains the nominal wire tension of 50\,g 
for each tube, i.e. becomes self-supporting. 

\begin{figure}%%[htb]
  \centering
  \includegraphics[width=\swidth]{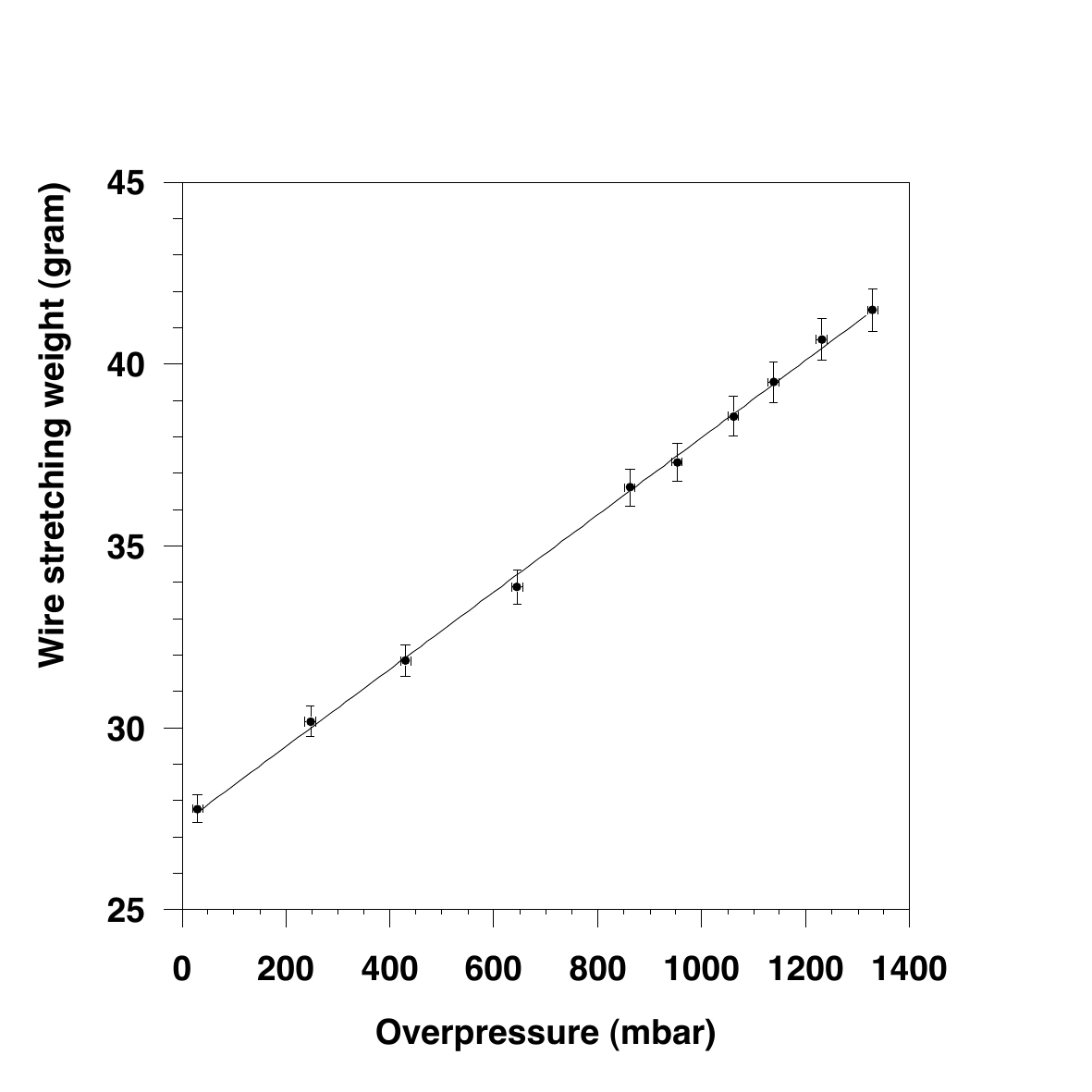}
 \caption[Measured wire tension at different gas overpressures]{Measured wire tension (weight equivalent) at
  different gas overpressures inside a straw. The nominal
  tension is 40\,g at 1.2\,bar overpressure for the COSY-STT straws.}
  \label{fig:stt:des:pres:tension}
\end{figure}

The precision of the tube and wire stretching method by the gas overpressure 
for the used thin film tubes was studied in
detail for the COSY-TOF straw tubes. \Reffig{fig:stt:des:pres:tension} 
shows the measured tension with
decreasing gas overpressure. A well-defined tension is seen, even down to 
vanishing overpressure where only the stiffness of the Mylar film tube 
maintains a wire tension of 28\,g. The nominal
  tension for the COSY-TOF 1\,m long straws was 40\,g at 1.2\,bar overpressure.
For the \Panda 1.5$\,$m long straws the nominal tension is 50\,g at 1.0\,bar overpressure.

\subsection{Gas Mixture}
%\COM{Author(s): P. Wintz, A. Sokolov, P. Gianotti}
\label{sec:stt:des:mix}
The need of high spatial resolution in the STT requires high amplitude anode signals even for the single electron clusters, thus 
requiring high gas gain. On the other side, a high gas gain significantly reduces the chamber lifetime. For the optimum gas  
amplification choice both these factors should be taken into account properly.
\Reftbl{tab:stt:des:gas} shows the main parameters of some of the most used gases and gas mixtures.
\begin{table*}
\begin{center}
\caption[Properties of different gases and gas mixtures]{Properties of different gases and gas mixtures. Z and A are charge and atomic weight, for molecules the total number 
has to be taken, N$_p$ and N$_t$ are the number of primary and total electrons per cm, respectively, E$_x$ and E$_i$ are the 
excitation and ionization energy, respectively, W$_i$ is the average energy required to produce one electron-ion pair in the 
gas, (dE/dx)$_{mip}$ is the most probable energy-loss by a minimum ionizing particle and X$_0$ is the radiation length. 
For gas mixtures, the weighted average value has been taken.}
\label{tab:stt:des:gas}       
\begin{tabular*}{\textwidth}{lccccccccc}
\noalign{\smallskip\hrule\smallskip}
Gas or  & Z & A & E$_x$  & E$_i$ & W$_i$ &dE /dx  &N$_p$ & N$_t$ & X$_0$ \\
gas mixture& & & [eV]& [eV]& [eV]&  [keV/cm] & [cm$^{-1}$ ]& [cm$^{-1}$ ]&[m]\\
\hline
%\noalign{\smallskip}\hline
 He & 2& 4& 19.8& 24.5& 41& 0.32& 4.2& 8& 5299 \\
 Ar & 18 & 40& 11.6& 15.7& 26& 2.44& 23& 94& 110 \\
 CO$_2$& 22& 44& 5.2& 13.7& 33& 3.01& 35.5& 91& 183 \\
 i--C$_4$H$_{10}$& 34& 58& 6.5& 10.6& 23& 5.93& 84& 195& 169 \\
 Ar+10\,\%\,CO$_2$& -& -&  -&  -&  26.7&  2.5&  24.6&  93&  117 \\
 He+10\,\%\,i--C$_4$H$_{10}$& -&  -&  -&  -&  39.2&  0.88&  12.7&  26.7&  1313 \\
 He+20\,\%\,i--C$_4$H$_{10}$&  -&  -&  -&  -&  37.4&  1.44&  20.6&  45.4&  749 \\
\noalign{\smallskip}\hline
\end{tabular*}
\end{center}
\end{table*}
In order to select the most suited gas mixture for the STT detector, it is useful to consider two essentially 
different situations. Some gas mixtures, if a low electric field is used, can effectively quench the electron 
kinetic energy, preventing them to gain enough energy between collisions. In this case, electrons are in thermal 
equilibrium with the surrounding medium and the drift velocity is proportional to the electric field. 
Such gases are usually called ``cold'' for that given electric field strength.
\par
On the contrary, if the electron average kinetic energy differs from the thermal energy, the drift velocity behavior 
becomes more complicated. In many gas mixtures the drift velocity becomes saturated and does not depend strongly on the 
electric field strength. That makes the reconstruction of the track coordinates easier. However, it is difficult to get high 
spatial resolution in these ``hot'' gas mixtures, in principle due to the large diffusion.
The standard choice of many experiments is to have a ``hot'' or ``warm'' gas mixture, that has a weak dependence of the drift 
velocity on the applied electric field. In this case, the electric field inhomogeneities do not play a significant role, 
which makes the calibration simpler. An overpressure can be used in these cases to reduce the diffusion. 
\par
The main requirements, that should be taken into account for the choice of the most suited gas mixture, are:
\begin{itemize}
 \item good spatial resolution;
 \item rate capability;
 \item radiation hardness;
 \item radiation length;
 \item chemical inactivity;
 \item working voltage;
 \item working pressure; 
 \item accessibility on the market and price.
\end{itemize}

For the \PANDA CT the spatial resolution, the rate capability and the radiation hardness are the points of highest 
importance. Initially a ``cold'' gas mixture of He + 10\,\%\,i--C$_4$H$_{10}$ was proposed for the Conceptual Design Report 
\cite{bib:stt:des:CDR}. 
Although this gas mixture has one undoubted advantage, the long radiation length X$_0$, it provides a relatively low drift
velocity, which is a disadvantage more or less peculiar for all ``cold'' gases. As a result, a gas mixture based on Ar + 
10\%CO$_2$ has been suggested.
\par
\begin{figure}
  \centering
 \includegraphics[width=0.95\swidth]{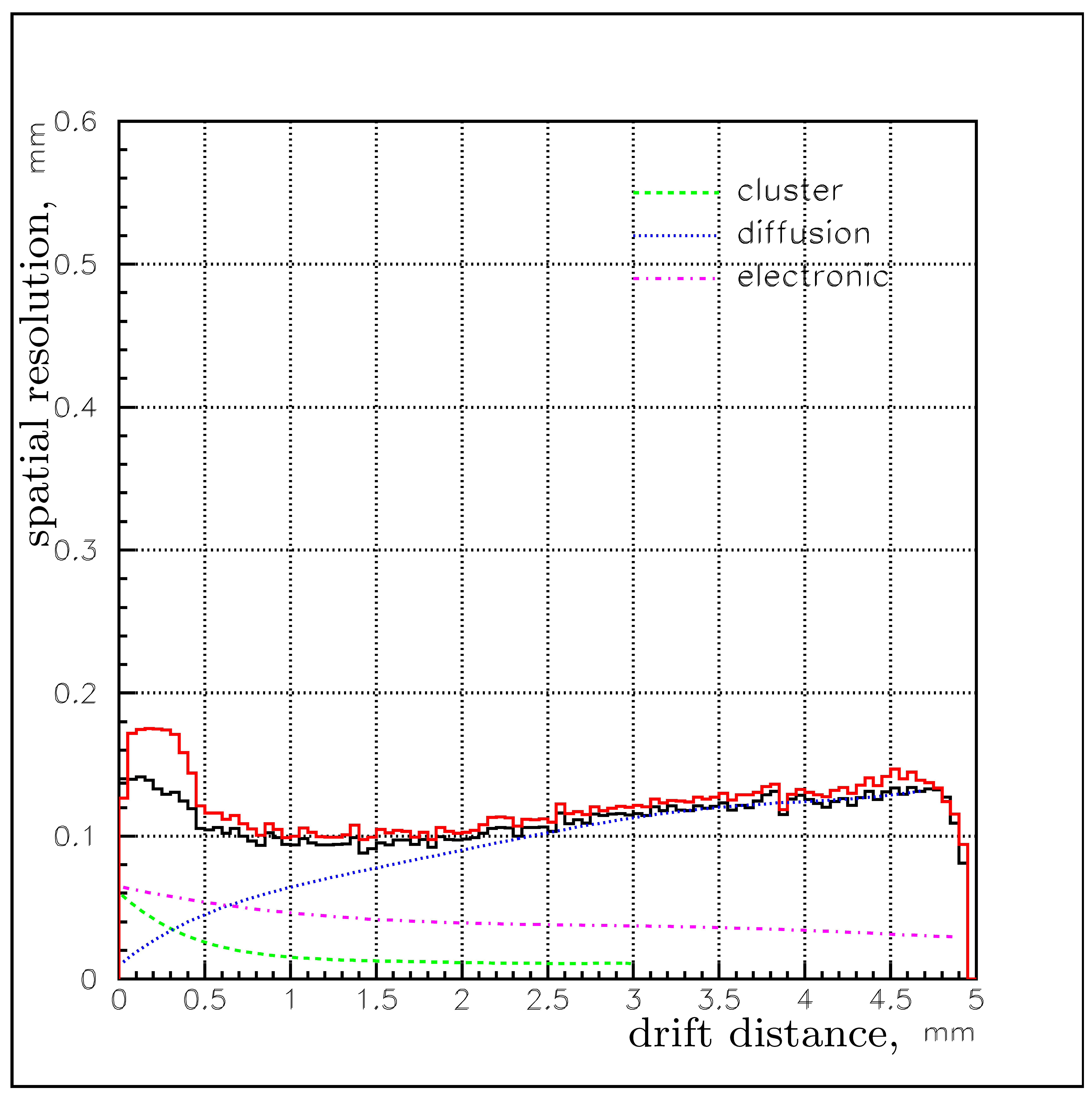}
 \includegraphics[width=0.95\swidth]{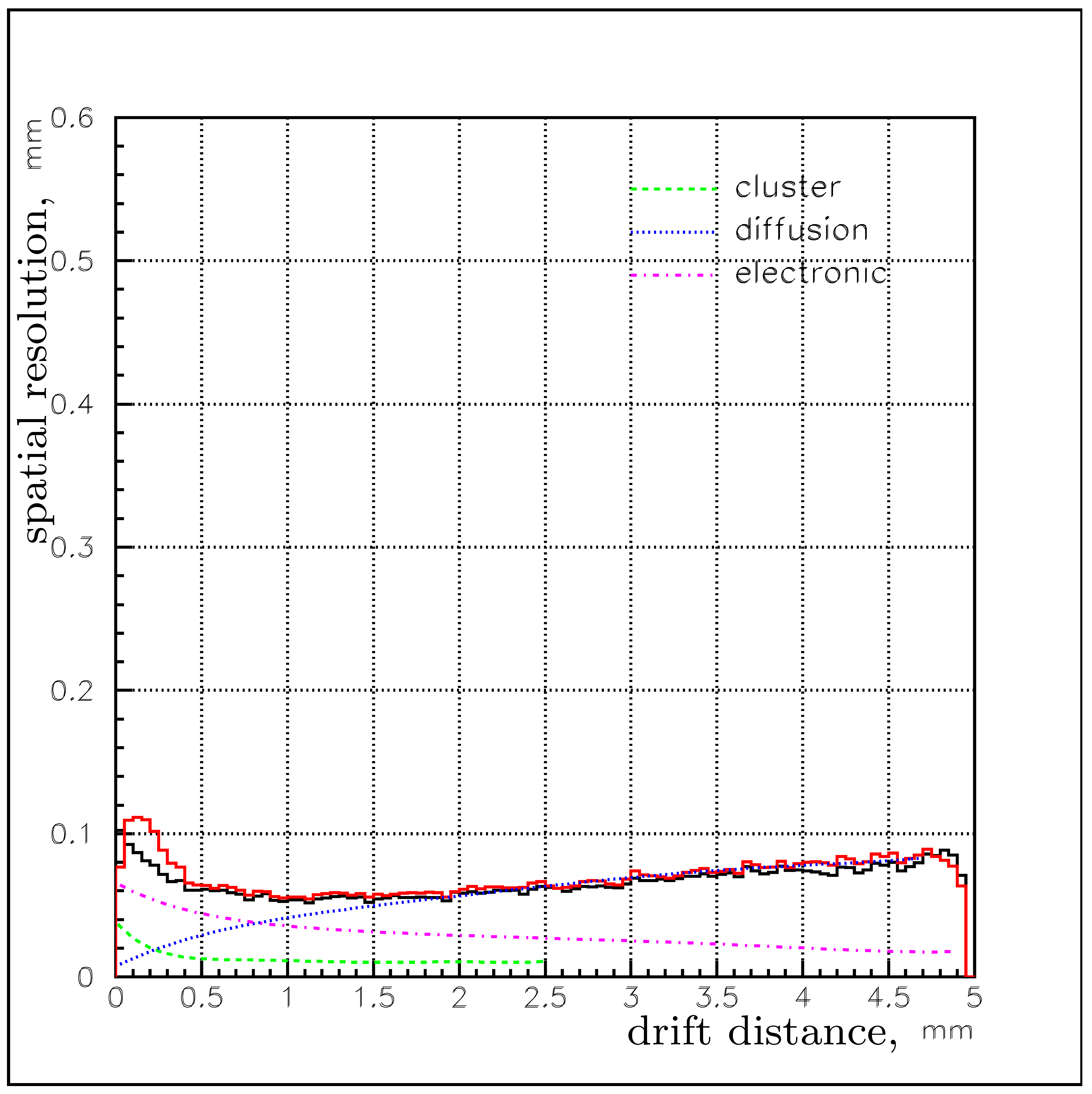}
 \caption[Spatial resolution for the Ar+10\,\%\,CO$_2$ gas mixture]{The spatial resolution for the Ar+10\,\%\,CO$_2$ gas mixture for 1 (top) and 2~atm
(bottom) pressures. The red line corresponds to an ideal r(t) relation, the black one to the measured. The main contributions to the resolution are also shown in different colors.}
  \label{fig:stt:des:gas:ArCO2}
\end{figure}
\begin{figure}
  \centering
 \includegraphics[width=0.95\swidth]{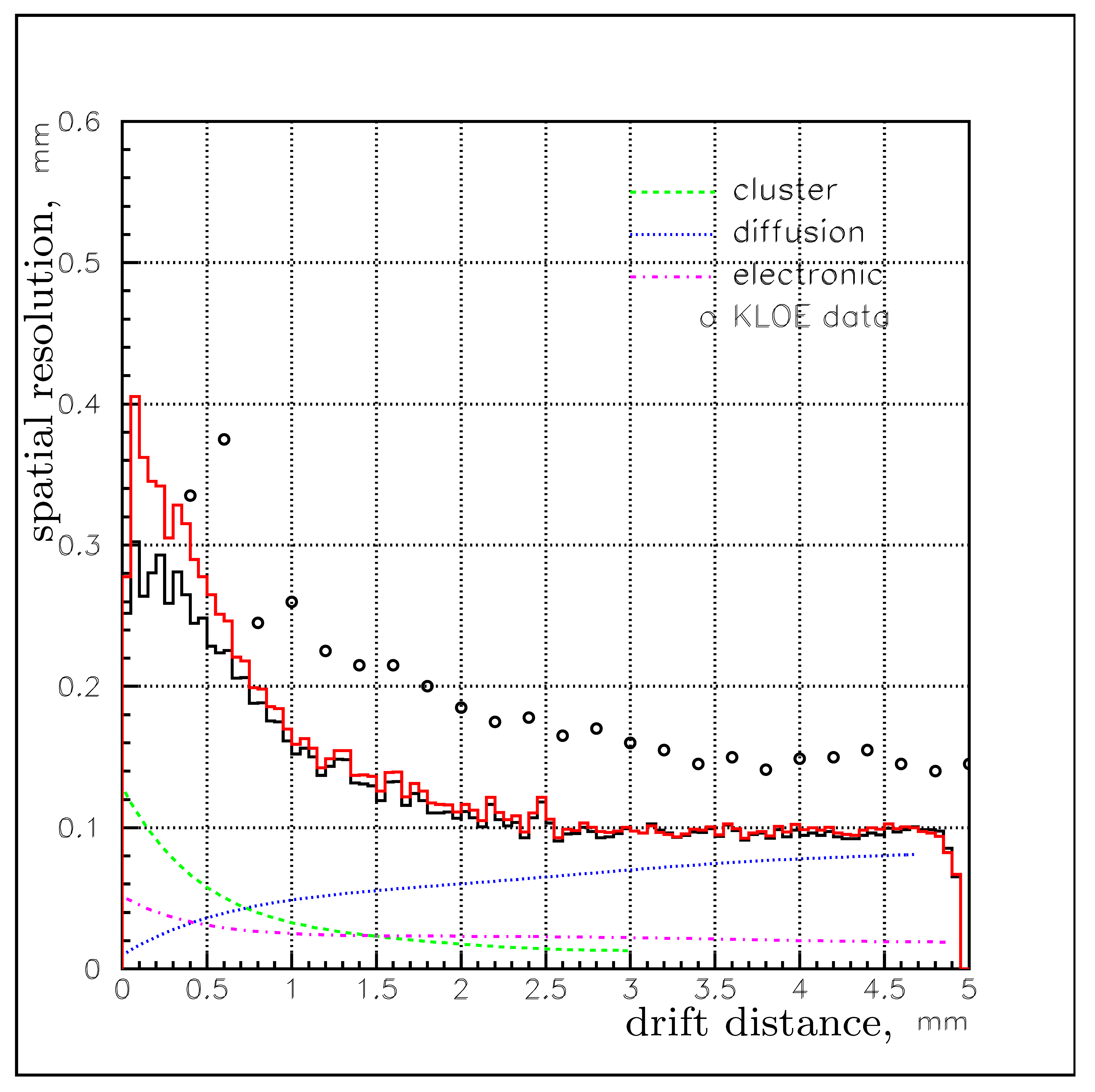}
 \includegraphics[width=0.95\swidth]{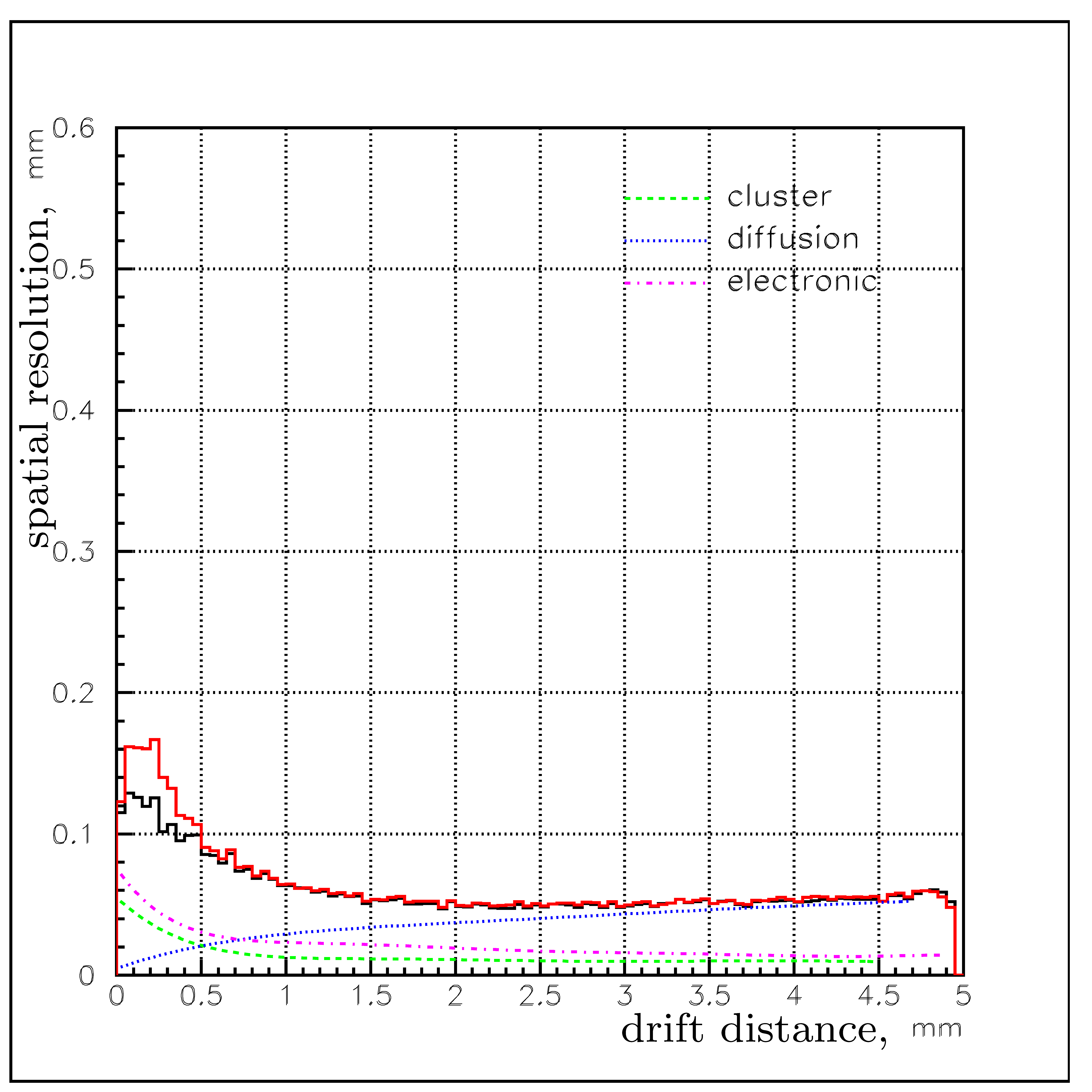}
 \caption[Spatial resolution in He+10\,\%\, i--C$_4$H$_{10}$]{Spatial resolution in He+10\,\%\, i--C$_4$H$_{10}$ with 1 (top) and 2~atm (bottom). The red line corresponds to an ideal r(t) relation, the black one 
to the measured. The main contributions to the resolution are also shown in different colors.
The experimental spatial resolution of the KLOE drift chamber, denoted by the open circles, is given for comparison \cite{bib:stt:des:Kloe}.}
  \label{fig:stt:des:gas:Ariso}
\end{figure}

\Reffig{fig:stt:des:gas:ArCO2} and \Reffig{fig:stt:des:gas:Ariso} show the results of the simulation for the spatial resolutions 
achievable for the Ar + 10\,\%\,CO$_2$ and He + 10\,\%\,i--C$_4$H$_{10}$ for 1 and 2~atm gas pressure. The simulations have been performed using
the GARFIELD program and the build-in MAGBOLTZ package \cite{bib:stt:cal:garfield}.
The good agreement of these simulation data with the experimental results obtained by the KLOE drift chamber prototype \cite{bib:stt:des:Kloe}, 
as shown in  \Reffig{fig:stt:des:gas:Ariso}, can be interpreted as a proof of the validity of the simulations of the straw tube parameters.

The spatial resolution of the Ar + 10\,\%\,CO$_2$ mixture is satisfactory even at 1~atm pressure, while the spatial resolution in 
the He + 10\,\%\,i--C$_4$H$_{10}$ is worse than the required 150\,$\mu$m, and only an increase of the pressure could improve this situation. 
The total drift time is also an important parameter.  The Ar+10\,\%\,CO$_2$ mixture has a drift time of 80 ns for a 4 mm drift path. 
The He+10\,\%\,i--C$_4$H$_{10}$ has double the drift time.
Since the average time between two events in PANDA will be $\sim$ 100~ns, when using the He+10\,\%\,i--C$_4$H$_{10}$ gas mixture, 
the information from consecutive events could be contained in the STT at any time. This event mixing in the tracker will 
result in a significant complication of the trigger logic and of the pattern recognition algorithm. By increasing the pressure two 
times, the drift time for the He+10\,\%\,i--C$_4$H$_{10}$ grows by 50~ns, while for the Ar + 10\,\%\,CO$_2$ only by 10~ns. That makes 
the situation with the event mixing even more difficult.
\par
The effect of the electronics threshold on the spatial resolution has also been studied. The average gas gain has been 
reduced by a factor two using the same electronic threshold. \Reffig{fig:stt:des:gas:Ariso} shows only a small deterioration of the 
Ar+10\,\%\,CO$_2$ resolution and a strong worsening in the case of the He+10\,\%\,i--C$_4$H$_{10}$ gas mixture. This is one more 
argument in favor of the Ar + 10\,\%\,CO$_2$ usage.

All these considerations show strong advantages for the Ar + 10\,\%\,CO$_2$ gas mixture for the \PANDA{} 
STT compared to the He + 10\,\%\,i--C$_4$H$_{10}$ gas composite. 
%Parameters of the Ar + CO$_2$ gas mixture can be modified in the 
%future, by changing pressure, CO$_2$ percentage or admixture of other gases, in order to sligthly tune STT performance.
\par
\begin{figure*}
  \centering
 \includegraphics[width=1.\swidth]{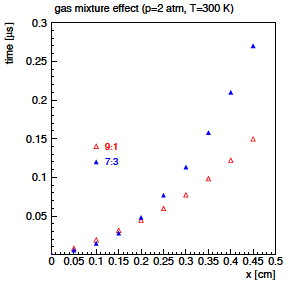}
 \includegraphics[width=.95\swidth]{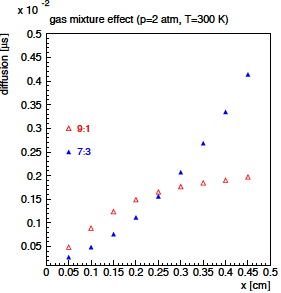}
 \caption[Drift and diffusion at different CO$_2$ admixtures]{The graphs refer to two gas mixtures with different CO$_2$ percentage. Red points correspond to a percentage of 
10\,\% of CO$_2$, blue to 30\,\%.}
  \label{fig:stt:des:gas:perc}
\end{figure*}

The possibility to use higher percentages of CO$_2$  has been investigated. \Reffig{fig:stt:des:gas:perc} shows the
space-time relation with two different CO$_2$ percentages: 10\,\% and 30\,\%, respectively.
A greater percentage of CO$_2$ produces an increase of the electron diffusion which worsens the achievable space resolution.
For completeness, we notice that a greater fraction of the quench gas will reduce the effect of the magnetic field
on the mixture (Lorentz angle). Therefore the final concentration of the CO$_2$ component can be defined only after tests with 
magnetic field.
\par
\begin{figure}
  \centering
  \includegraphics[width=\swidth]{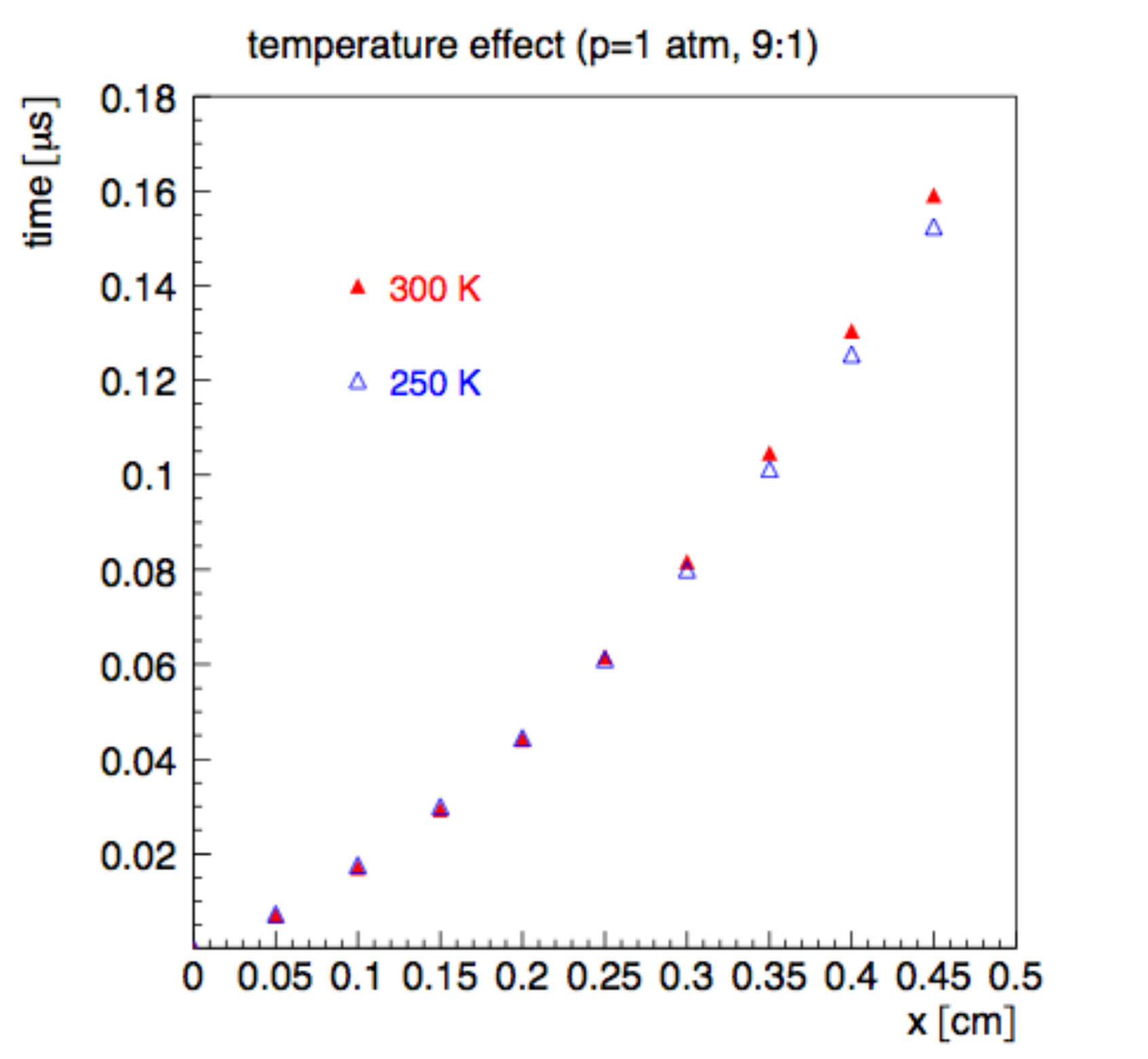}
 \caption[Space time relation in Ar+10\,\%\,CO$_2$ for different temperatures]{Space time relation for the Ar+10\,\%\,CO$_2$ mixture at 1~atm for two different temperatures.}
  \label{fig:stt:des:gas:temp}
\end{figure}

The variations of the gas mixture performance due to changes of the absolute temperature have been studied.
The space time relation for the Ar+10\,\%\,CO$_2$ mixture at 1~atm for two different temperatures,
250 and 300 K, is shown in \Reffig{fig:stt:des:gas:temp}.
No significant differences are present between the two curves. Therefore, it will not be necessary to control
the temperature variation very precisely.
\eject
%
%EOF: panda_tdr_stt_des.tex

% FILE: panda_tdr_stt_lay.tex
%
\section{The STT Detector}
\label{sec:stt:lay}
%\COM{Author(s): P. Gianotti/P.Wintz}
\vspace{-0.3cm}
The surrounding detector systems define the available space for the \PANDA-STT as a cylindrical 
volume with an inner radius of 150\,mm, outer radius of 420\,mm and length of 1650\,mm, at a 
position in z-direction relative to the target from about \mbox{z=--550\,mm} to z=+1100\,mm. 
The space for the target pipe of the pellet beam at the vertical axis cuts this volume into two 
semi-cylinders with a gap of 42\,mm in between. To facilitate access and maintenance the layout 
of the STT detector is split into two independent semi-cylindrical systems, with two separated 
mechanical frame structures, separated frontend electronic, gas and high voltage supply. The two 
systems are mounted at the opposite sides of the vertical Central Frame (CF) structure which also 
supports the inner MVD detector system and the beam-target cross-pipe. The electronic frontend readout 
cards, supply and other services of the STT are placed at the upstream end of the detector within a 
space of 150\,mm in z-direction. The remaining active detection volume with a length of 1500\,mm is 
filled by layers of straw tubes, each tube with a diameter of 10\,mm and a length of 1500\,mm. A few 
dedicated tubes have shorter lengths to fill some rest gaps in the volume.

\subsection{The Straw Layout in the STT}
\label{sec:stt:lay:geo} 
The solenoid magnetic field is parallel to the beam axis and forces charged particles to helical trajectories, 
which are described by the helix circle in the projection on the xy-axis and by the helix slope in the perpendicular 
projection in the z-direction. For the spatial reconstruction of the trajectory the STT consists of a number of straws 
precisely aligned parallel to the beam and magnetic field, which measure the helix circle. Additional straws which are 
skewed by a few degrees to the axial direction provide a stereo view of the track and measure the z-information of the 
track for reconstructing the helix slope.     

The \PANDA-STT uses the technique of pressurized straw tubes, closely packed and glued together to planar multi-layer modules. 
As discussed in the previous section such self-supporting straw modules show a high rigidity and mechanical precision and allow 
to reduce the weight and size of the mechanical frame structure to an absolute minimum. In addition, the close-packaging yields 
the highest straw density with a maximum number of straws per cross-sectional area. 
Therefore, the planar layer modules are arranged in a hexagonal layout which preserves the 60$^\circ$ position symmetry of close-packed,
 parallel straws.

\begin{figure}%[b] 
\begin{center}
\vspace{-3cm}
\includegraphics[width=\swidth]{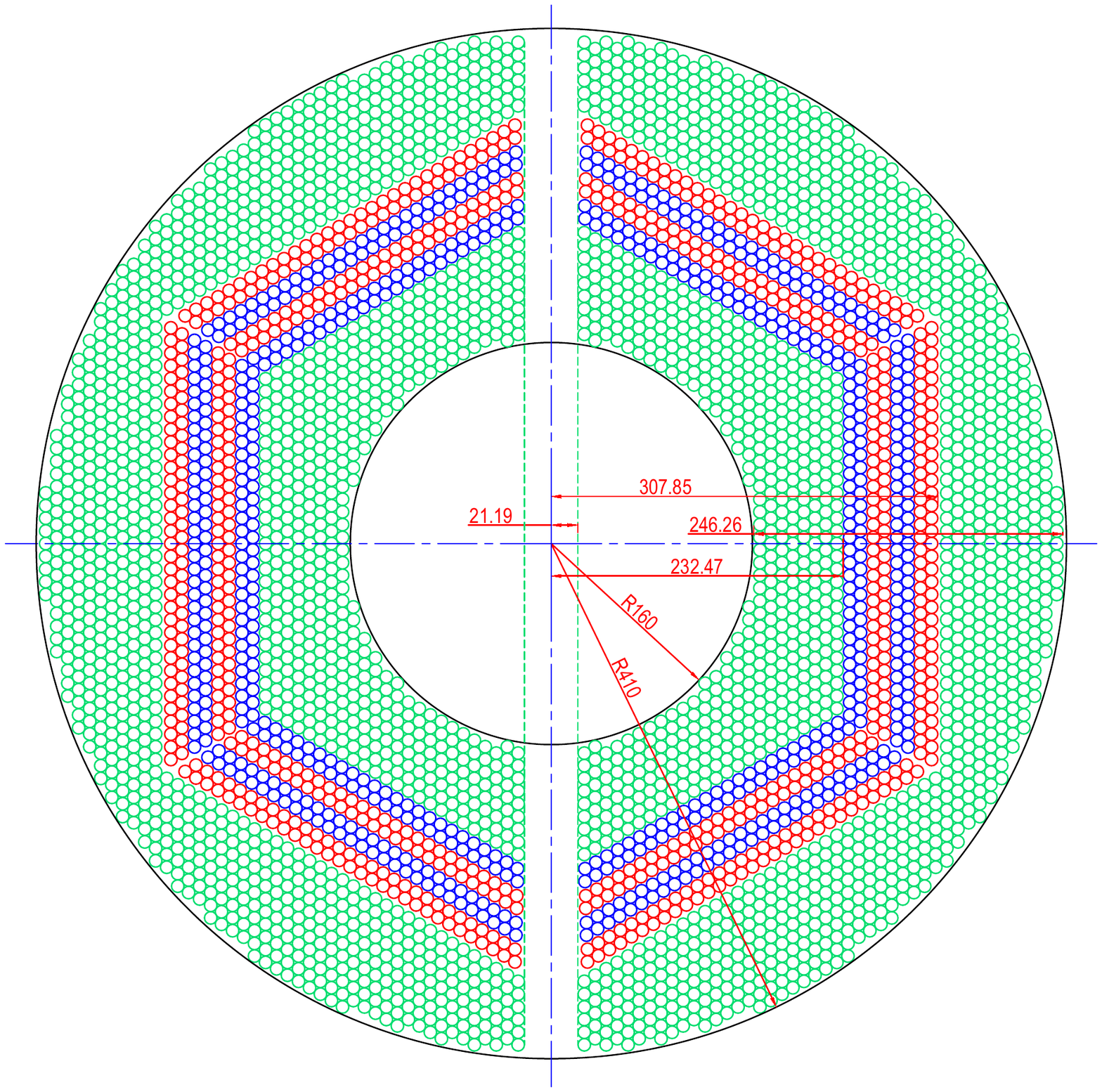}
\caption[Layout of the straw tubes in the STT]{Layout of the straw tubes in the STT in xy-view. The straws marked in green are 
parallel to the beam axis. The blue and red marked straw layers are skewed relative to the axially aligned straws in the same sector by a small angle of 
$+$2.9$^\circ$ and $-$2.9$^\circ$, respectively.}
\label{fig:stt:lay:geo:Detectorlayout}
\end{center}
\end{figure}

Each of the two semi-cylindrical \PANDA-STT volumes is filled by three sectors of straw tubes aligned in the z-direction and arranged 
in stacks of planar multi-layer modules. The hexagonal layout of both volumes together has an almost cylindrical shape with a 42\,mm 
gap for the target pipe (\Reffig{fig:stt:lay:geo:Detectorlayout}). 

The arrangement of the straw layers in each of the six hexagonal sectors is as follows. In radial direction and starting from the inner radius in a sector there are 8 straw layers parallel to the beam axis, followed by a block of 4 stereo double-layers, alternately skewed by $\pm2.9^\circ$ relative to the axially aligned straw layers, and again a block of 4 layers parallel to the beam axis. Then, there are another 7 layers aligned parallel to the beam with a decreasing number of straws per layer to achieve the outer cylindrical shape of the STT. The inner cylindrical shape is reached by placing a few axially aligned straws in the inner corner region of each hexagon sector (see \Reffig{fig:stt:lay:geo:Detectorlayout}).

In total, there are 4636 straws in the layout. All straws have the same inner diameter of 10\,mm and length of 1500\,mm, except a few outer straws in the border region of each skewed layer, which have different, reduced lengths
(see \Reffig{fig:stt:lay:mod:Strawmodule0}). The film wall thickness of all straws is 27\,$\mu$m Mylar, aluminized on the inner side and outer side of the tube. 

The close-packaging of the straws with less than 20\,$\mu$m gaps between adjacent tubes yields the highest straw density with up to 27 layers 
in radial direction for the 3-dimensional track reconstruction. Up to 19 layers with axial straws parallel to the beam measure the helix circle 
in the xy-projection with a single (mean) isochrone resolution of better than 150\,$\mu$m ($\sigma_{r}$). The association of the isochrone hits 
in the 8 stereo layers to the helix circle provides the z-coordinates of the track with a single hit resolution of slightly better than 3\,mm 
($\sigma_z$), which is determined by the isochrone resolution and the skew angle ($\alpha$) of $\pm2.9^\circ$ ($\sigma_z=\sigma_r/sin(\alpha)$). 
The tracking efficiency for a single layer is  98.5\,\% and only slightly reduced compared to the single tube radial efficiency (99.5\,\%) by 
the thin tube wall (27\,$\mu$m) and minimal spacing (20\,$\mu$m gaps) between adjacent tubes. 

Since the momentum resolution is dominated by the transverse momentum reconstruction and the stereo layers distort the close-packed cylindrical 
geometry their skew angle should be kept as small as possible to a few degrees. As can be seen in \Reffig{fig:stt:lay:geo:Detectorlayout} the 
chosen value of 2.9$^\circ$ creates only minor gaps between two hexagon sectors. 

\begin{figure}%[b] 
\begin{center}
\includegraphics[width=\swidth]{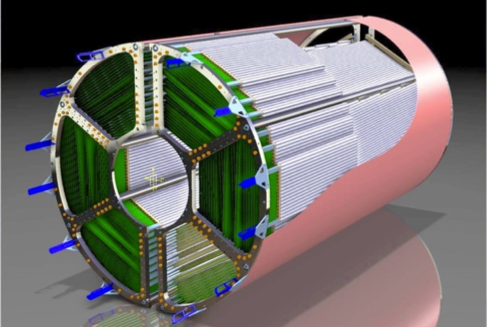}
\caption[Three-dimensional view of the STT]{Three-dimensional view of the STT with the mechanical frame consisting of light-weight profiles at 
both ends to attach and support the straw modules.}
\label{fig:stt:lay:geo:Detectorlayout1}
\end{center}
\end{figure}

Due to the technique of the self-supporting straw modules no support or reinforcement structures in the tracking volume are needed. 
\Reffig{fig:stt:lay:geo:Detectorlayout1} shows a three-dimensional view of the STT including the light-weight 
mechanical frame which consists of end flange profiles with precision holes to attach and support the straw modules.
The inner and outer semi-cylinder surfaces will be covered by a thin wall of a light-weight composite material, consisting of a 1\,mm Rohacell layer with a 0.17\,mm thin Carbon Fiber skin, to protect the straw film tubes against a mechanical hazard from outside.      

The low material budget of 1.23\,\% (X/X$_0$) in the radial direction is the sum of the 24 average straw layers in the STT (1.06\,\%) and the two protection walls (2$\times$0.084\,\%). It is dominated by the film wall thickness of the straw layers (0.72\,\%) and the gas (0.31\,\%). 
As discussed in the previous section the chosen film thickness of 27\,$\mu$m Mylar is at a minimum and can not be further reduced. 
The resolution of the reconstructed momentum from the spatial trajectories is about 1-2\,\% ($\sigma_p$/p) for simulated charged particles originating from the beam-target interaction point and including the track hits in the MVD. The material budgets of the straw layers and the walls have been taken into account in the simulation. 

At the 2\,Tesla solenoid magnetic field the minimum transverse momentum for charged particles to reach from the interaction point the innermost straw layer 
in the STT is about 50\,MeV/c. A minimum transverse momentum of about 100\,MeV/c is needed to reach enough straw layers for a complete three-dimensional 
reconstruction of the helical trajectory. 
The STT covers a polar angle range from about 10$^\circ$ to 140$^\circ$. The azimuthal coverage is only limited by the gap for the target pipe at $\pm$90$^\circ$.

The high number of up to 27 hit straws in radial direction is important to achieve a high resolution for the specific energy-loss measurement (dE/dx). The high sampling number per track allows to truncate such hits with large deviations from the mean energy-loss per tube. This so-called truncated mean method for the measured Landau-distributed energy-losses improves the resolution significantly. From prototype measurements an energy resolution of better than 8\,\% ($\sigma(E)/E$) is expected for the \Panda-STT (see \Refsec{sec:stt:pro:eloss}).

\subsection{Layout Considerations}
\label{sec:stt:lay:cons} 
The specific straw layout described in the previous section has been optimized to achieve highest geometrical efficiency and spatial resolution for 
the track reconstruction in the \Panda target spectrometer environment. 
By choosing a hexagonal geometry the straw tubes can be arranged in close-packed layers and the number of straws per cross-sectional area is largest. 
Then the main parameters for the STT layout determination are the straw diameter, number and position of the axial and skewed stereo layers, and the 
stereo angle which are discussed in the following. 

The inner straw diameter of 10\,mm is the same for all tubes which avoids different end plug designs, assembly tools and techniques. Therefore the cost and time for the mass production of the 4636 straws are strongly reduced.
The expected highest particle rates of the single straws in the innermost layers scale roughly with the straw diameter and are about 5-8\,kHz/cm, corresponding to 800\,kHz per tube. These rates are still tolerable concerning signal distortion and aging properties. 
In a closed-packed geometry smaller tube diameters would increase the number of readout channels and reduce the available cross-sectional space for the electronic readout and gas supply per channel. In addition the material budget would be higher. 
All these aspects together favor the 10\,mm tube diameter.

The tracking properties of the STT are mainly defined by the number and radial position of the axial and stereo straw layers. The axial straws are 
used for the measurement of the helix curvature and transverse momentum with high resolution. Then the hits in the stereo layers are associated to 
the found circular trajectory in the xy-projection and determine the helix slope in z-direction. Instead of choosing a layout with many alternating 
axial and stereo layers in radial direction, the specific requirements for a highly efficient and high resolution reconstruction of charged particle 
tracks in the \Panda environment favors a different layout. 

The chosen layout with a larger inner block of close-packed, axial straws, central block of stereo layers, followed again by an outer block of close-packed axial straws has the advantage of an almost continuous tracking, which is important for the 
particular \Panda tracking environment with a high $p\bar{p}$ interaction rate of 2$\times$10$^7$\,s$^{-1}$ and a mean particle multiplicity of about 4 charged tracks per event. The close-packing of many layers of axial straws yields the highest possible number of straw layers in the radial direction.

An important task is the recognition and reconstruction of the decay vertices of the $\Lambda$ ($\bar{\Lambda}$) by the tracks of the charged decay particles. 
Up to a few 10\,\% of the $\Lambda$ ($\bar{\Lambda}$) can decay inside or even outside the region of the outer MVD layers. Then the vertex 
finding and reconstruction can only be done by the STT and needs a larger number ($\ge$6) of inner axial straw layers for the precise track 
reconstruction in the xy-plane with a single hit resolution of about 150\,$\mu$m. Although a complete secondary track finder program
should combine the information of all the tracking detectors of the target spectrometer, the STT capability in this respect has been checked an 
preliminary results are described in \Refsec{sec:stt:sim:pr2}.
The resolution in z-direction by the skewed layers is about 
3\,mm for single hits. As discussed in the previous section larger stereo angles which would improve the z-resolution are not favorable because 
they distort the cylindrical geometry, cause larger gaps in the close-packed layout and have a higher material budget. For forward emitted decay tracks 
which hit only the inner axial layers and then leave the STT the hits in the vertical GEM-tracker are associated to the found trajectories and add the z-information. 

In general the large inner and outer blocks of axial layers in the STT provide a continuous tracking in the xy-plane with high resolution for tracks 
entering the STT from the target interaction point or for background tracks entering the STT from outside. This is important to recognize a distortion 
of the helical trajectory by interactions with the MVD material or secondary background production inside the MVD volume or the outer DIRC and EMC volumes.

In summary, the STT layout combines a large acceptance and high momentum resolution for charged particle tracks originating from the beam-target 
interaction point and a high efficiency for the reconstruction of displaced vertices, even outside the MVD. The detailed properties and performance 
results for the STT are described in the chapter about the physics analyses.

\subsection{Straw Layer Modules}
\label{sec:stt:lay:mod} 
\begin{figure}%[b] 
\begin{center}
\includegraphics[width=\swidth]{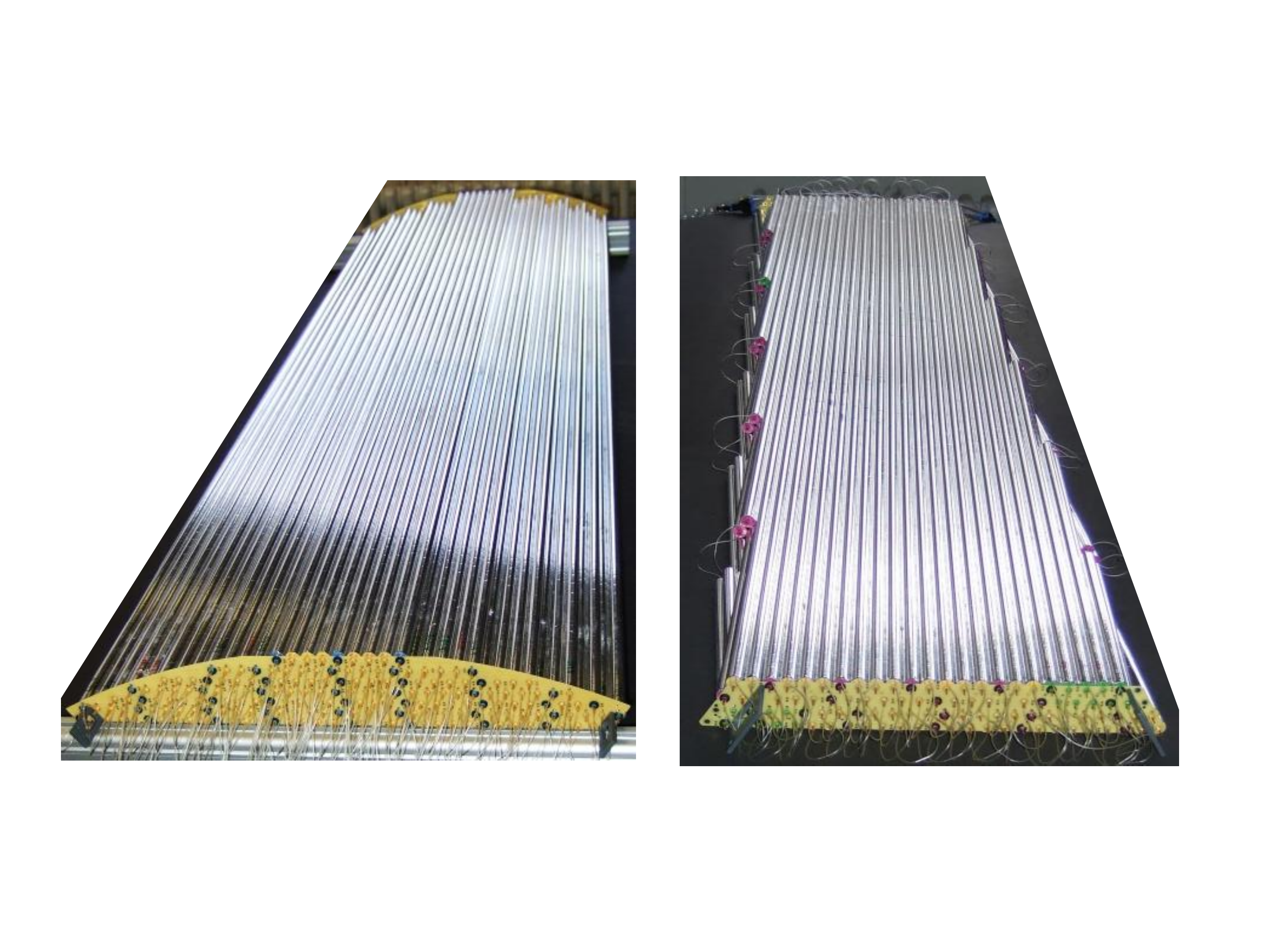}
\vspace{-1cm}
\caption[Photographs of axial and skewed straw layer modules]{Photograph of an axial straw layer module for the outer cylindrical shape and module with two double-layers of with opposite skew angle.}
\label{fig:stt:lay:mod:Strawmodule0}
\end{center}
\end{figure}

The layers of a sector are grouped into multi-layer modules, consisting of four close-packed axial layers or two close-packed double-layers with opposite skew angle. The outermost module in a sector consists of 7 close-packed axial layers with a varying number of straws per layer to reach an outer cylindrical shape (see \Reffig{fig:stt:lay:mod:Strawmodule0}). 
For the innermost straw module a few single straws are added in the corners to reach the inner cylindrical shape. 

\begin{figure}%[b] 
\begin{center}
\includegraphics[width=\swidth]{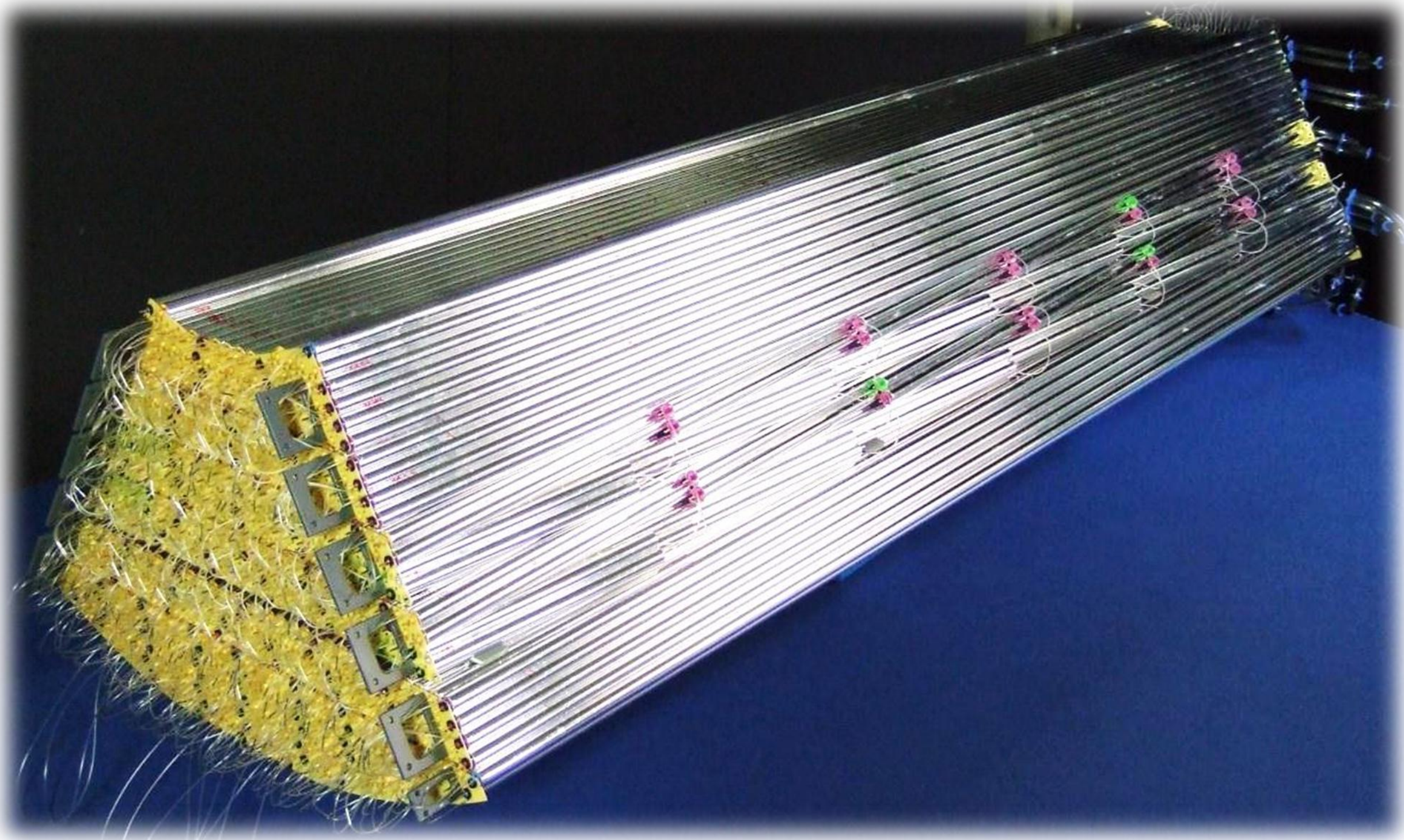}
%\vspace{-1cm}
\caption[Photograph of all straw modules of one STT hexagon sector]{Photograph of all straw modules of one STT hexagon sector. Two thermoplastic mounting brackets at both ends of a module are used for its support and positioning in the mechanical frame.}
\label{fig:stt:lay:mod:Strawmodule1}
\end{center}
\end{figure}

The close-packed layer modules show a strong rigidity when the straws are pressurized to the nominal overpressure of 1\,bar. 
No stretching from a mechanical frame structure to sustain the wire tension or reinforcements for the tube shape are needed.
Due to the high overpressure the thin-wall tubes have a perfect and strong cylindrical shape and the modules are self-supporting.
At both ends of a module dedicated strips made of 0.7\,mm thin glass fiber are attached and fixed to the end plugs by thermoplastic snap rings (see \Reffig{fig:stt:des:mat:strawtube}). The strips provide the electric grounding of the individual straws and the mechanical support and positioning of a module by two additional thermoplastic mounting brackets per strip. 
\Reffig{fig:stt:lay:mod:Strawmodule1} shows all modules of one full hexagon sector together.  

The modules consisting of two stereo double-layers with opposite $\pm2.9^\circ$ skew angle have several straws with different, shorter lengths at the corners to adopt the hexagonal sector shape. 
For the gas supply and the electric connection of these tubes the shorter straws in the lower double-layer are connected to the corresponding, at the same z-position attaching straws in the upper double-layer. About 2\,cm space in the z-direction is foreseen for connecting the gas tubes, sense wires and electric ground. The electric connection scheme was tested by 
illuminating two connected straws with an $^{55}$Fe radioactive source and comparing the shape of their analog signals. No obvious distortions of the signals were observed due to the short length of the electric connection.       
  
For all modules the electronic readout, gas and high voltage supply are at the upstream end of the detector to reduce the material budget for tracks going in forward direction through the downstream end of the STT. 

\subsection{Assembly of Straw Modules}
\label{sec:stt:lay:ass}
The construction of the straw modules consists of several assembly steps, starting with the production of the single straw tubes and ending with a final, self-supporting straw module, consisting of several straw layers. Such a module is then mounted in the mechanical frame structure which is discussed in detail in the next section.
In the following, the main steps of the assembly procedure of the single straws and modules are described.
\begin{itemize}
\item Mylar film tubes are cut to the nominal length of 1500\,mm and gas pipes are glued to the end-plugs using a plastic glue (Pattex plastik \cite{bib:stt:lay:glue1}).
\item The anode wire is fed through a crimp pin, end-plug, Mylar film tube, next end-plug and crimp pin. The crimp pins are then glued in the end-plugs, and afterwards the end-plugs are glued inside the Mylar film tube leaving a small \mbox{$\sim$ 1.5\,mm} 
film overlap at both ends. In the film overlap, later a dedicated spring with an outer snap-ring contact is inserted, which provides the electric grounding and can compensate the elastic elongation of the film tube under overpressure.
The glue used for gluing the end-plug and crimp pin is a 2-component epoxy adhesive (UHU endfest 300  
\cite{bib:stt:lay:glue2}) with 2\,h working time, and 12\,h setting time.
\item After glue hardening, a single straw is placed in a long v-shaped profile and at one end the wire is crimped. At the other end the wire is stretched by a weight of 50\,g. The straw is then connected to a gas supply and the gas pressure inside is 
raised smoothly to the nominal overpressure of 1\,bar. Then, the wire is crimped in the second end-plug.
\item After having produced a sufficient number of straws each of them is tested for gas leakage and correct wire tension. 
The wire tension is measured by placing the pressurized tube in a strong magnetic field and applying an AC-current to the wire.
The tension can be calculated by measuring the first harmonic of the oscillating wire. 
Tubes showing deviations from the nominal 50\,g wire tension or gas leakage and tubes with broken wires are rejected.
\begin{figure}%[b] 
\begin{center}
\includegraphics[width=0.95\swidth]{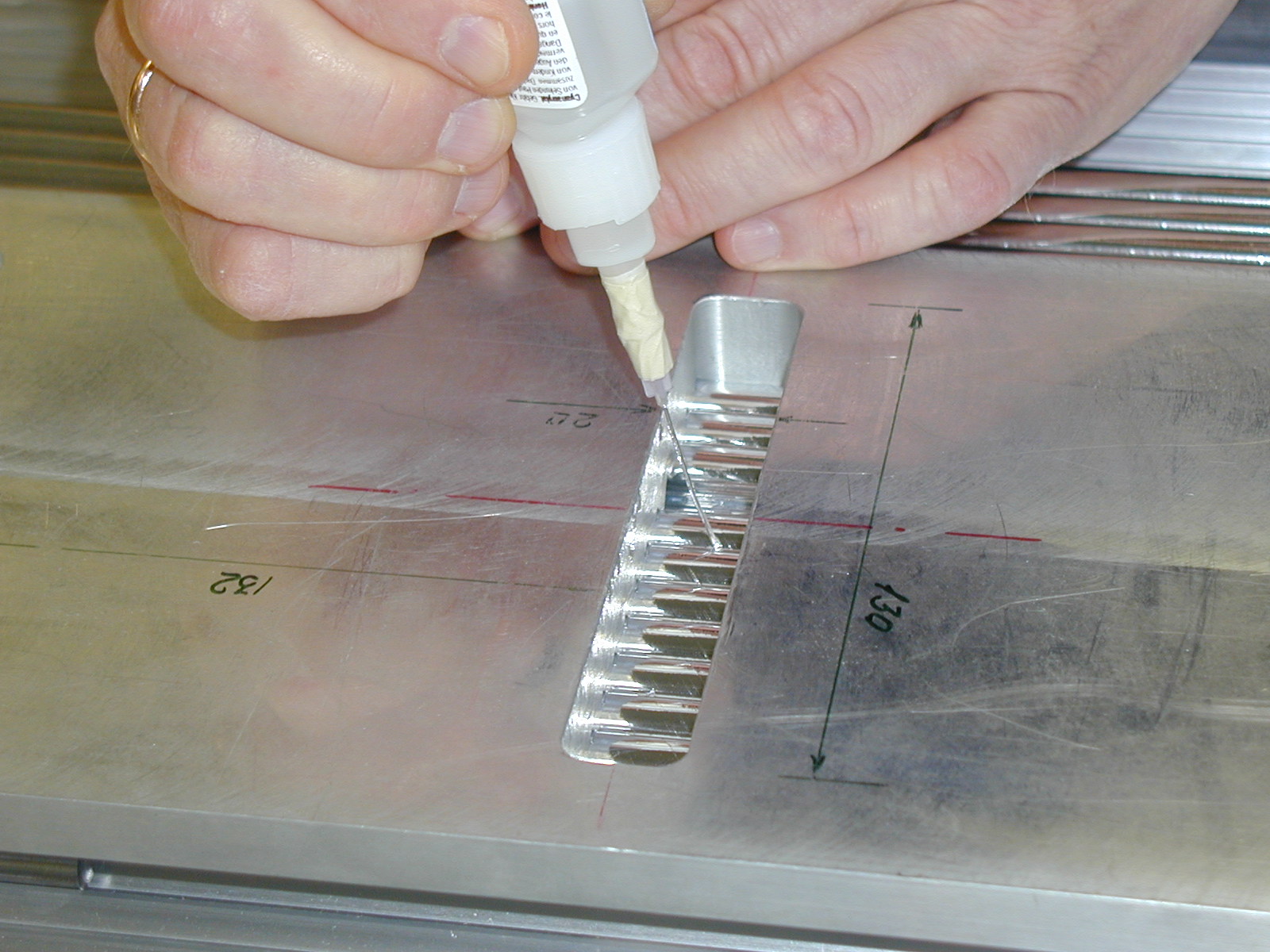}
\caption[Gluing of straw tubes to multi-layers]{Gluing of straw tubes to multi-layers.}
\label{fig:stt:lay:ass:glue}
\end{center}
\end{figure}
\begin{figure}%[b] 
\begin{center}
\includegraphics[width=0.95\swidth]{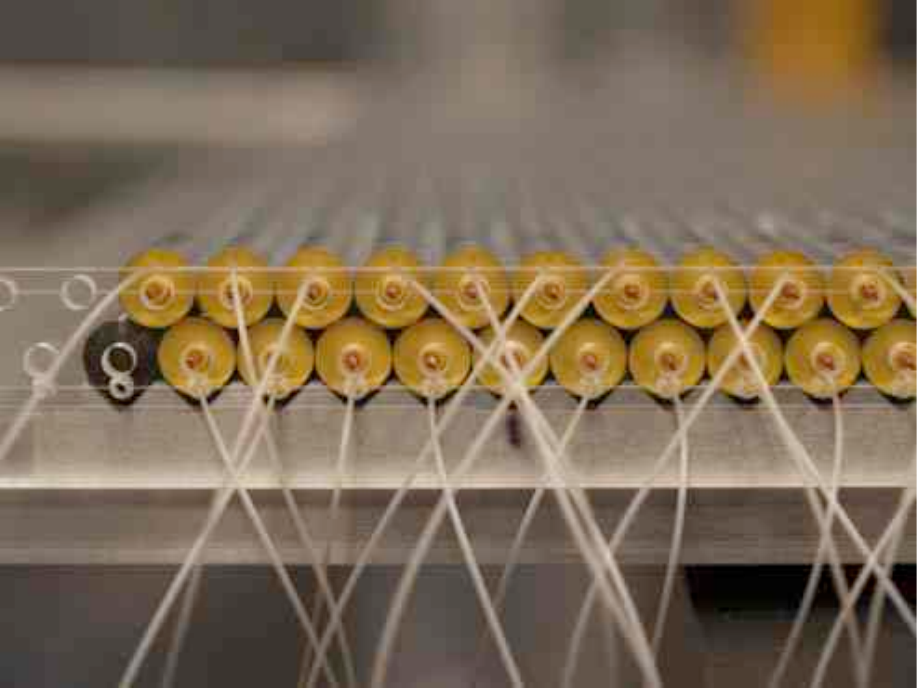}
\caption[A straw tube double-layer on the reference plate]{A straw tube double-layer on the reference plate.}
\label{fig:stt:lay:ass:DL}
\end{center}
\end{figure}
\item After this selection a number of straws is placed as a mono-layer onto a reference groove plate, connected to a gas supply and pressurized to the nominal pressure of 1\,bar. The individual tubes are aligned with high precision also from the top by smaller reference plates (see \Reffig{fig:stt:lay:ass:glue}). Then, each tube is glued to the two adjacent ones at several defined points along their length. The glue used here is an instant cyanoacrylate adhesive 
(Loctite 408 \cite{bib:stt:lay:glue3}). After that, the second layer of straws is precisely positioned on top of the first one, pressurized to the nominal pressure and the single tubes are then glued to the adjacent ones in the same layer and in the lower layer (see \Reffig{fig:stt:lay:ass:DL}).
\item This procedure is repeated depending on the number of layers in the straw module.
\item Springs at both straw ends are inserted and finally side-bands are fixed to both ends of the straw module
by thermoplastic snap-rings attached to the end-plugs (see \Reffig{fig:stt:lay:ass:stendplug} and \Reffig{fig:stt:des:mat:strawtube}).
\begin{figure}%[b] 
\begin{center}
\includegraphics[width=0.95\swidth]{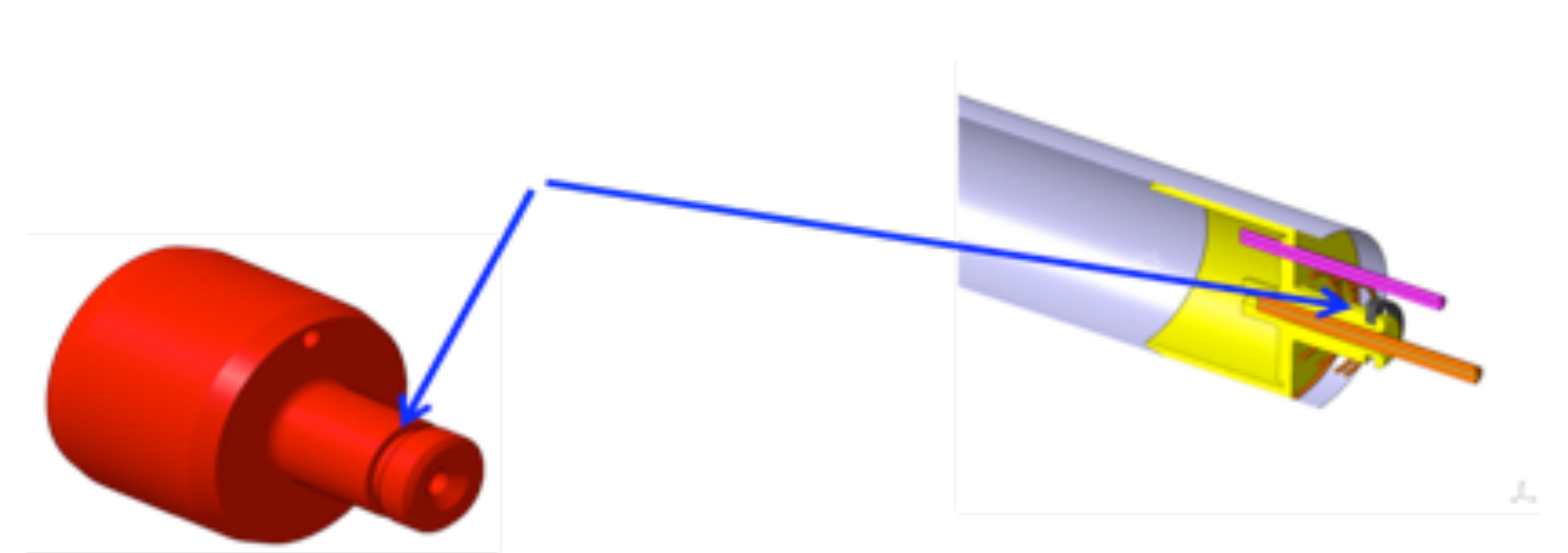}
\caption[Straw end-plug]{Straw end-plug with a groove (indicated by the arrows) for a thermoplastic snap-ring to attach it  
to the side-band.}
\label{fig:stt:lay:ass:stendplug}
\end{center}
\end{figure} 
\end{itemize}

%
%EOF: panda_tdr_stt_lay.tex

% FILE: panda_tdr_stt_mec.tex
%
\section{Mechanical Structure}
%\COM{Author(s): D. Orecchini, B. Dulach, P. Wintz}
\label{sec:stt:mech}
The STT layout is split into two independent semi-cylindrical detector volumes with two separated mechanical frames. 
The symmetrical, mirrored layout of the two detector volumes is also adopted for the frame system. 
The mechanical frame structure for each volume has to support and precisely position the straw layer modules
at both ends. In addition the structure has to support all the electronic readout and supply elements, which are connected and placed at the detector front-end: the electronic readout cards, all readout and supply cables, the gas manifolds and supply pipes. 
 
After the detector assembly the two mechanical frames are mounted at the opposite sides of a vertical mechanical frame structure, which has the additional task to support and align the beam-target cross-pipe and the MVD detector system.
This vertical central frame (CF) is mounted on rails to move the entire system in and out of the \Panda target spectrometer. 
The CF system is described in \Refsec{sec:stt:centralframe}.

\subsection{The STT Mechanical Frame}
\label{subec:stt:mech:sup}
\begin{figure}%[b] 
\begin{center}
\includegraphics[width=\swidth]{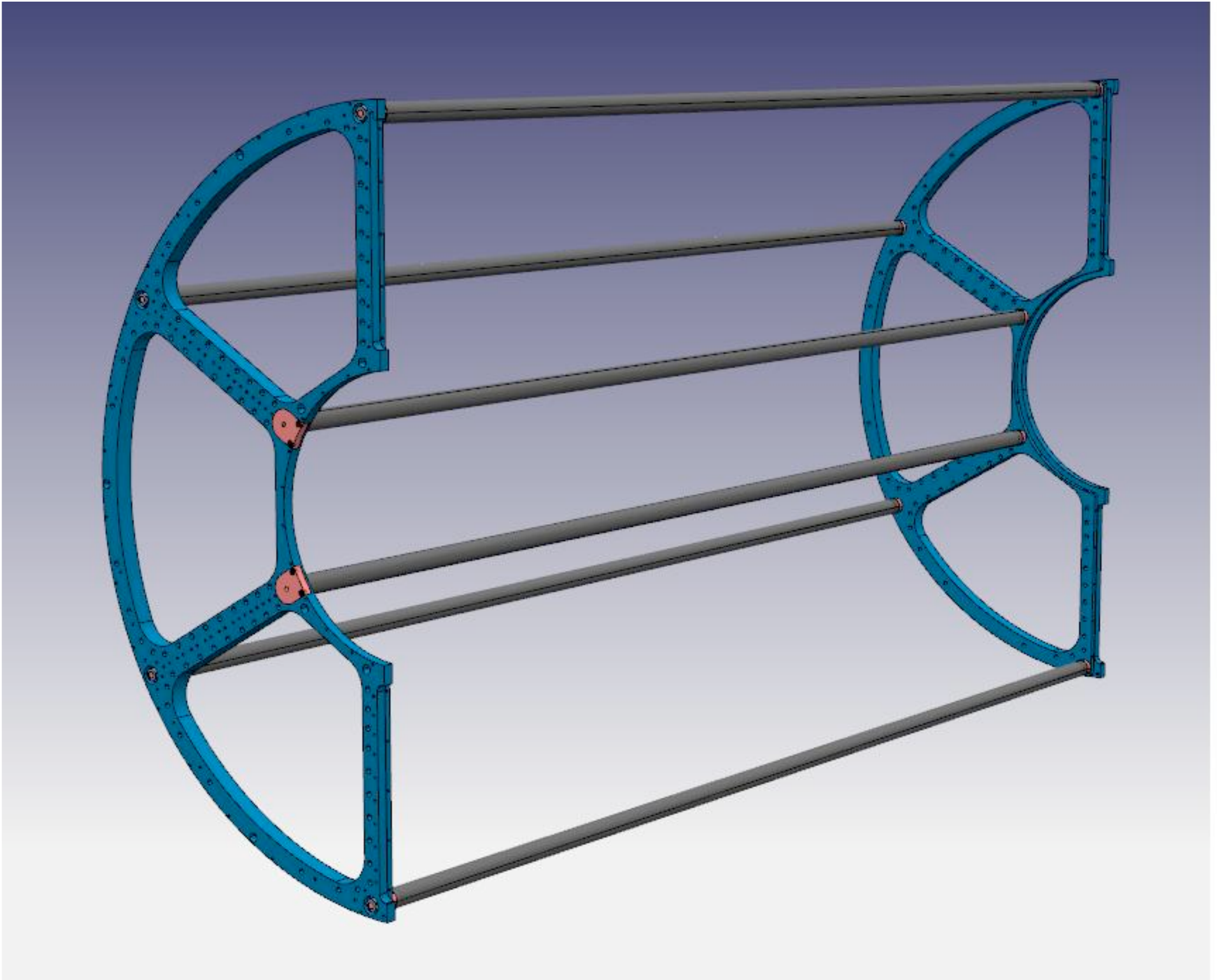}
\caption[Pictorial drawing of one of the two frame structures of the STT]{Pictorial drawing of one of the two frame structures of the STT.}
\label{fig:stt:mech:MechFrame}
\end{center}
\end{figure}

In order to obtain a structure with high mechanical accuracy and rigidity, but being extremely light-weight, we have conceived a simple and easy to realize solution using aluminum or carbon fiber. 
Following the experience gained in previous experiments \cite{bib:stt:mec:FIN,bib:stt:mec:BTeV}, 
the solution shown in \Reffig{fig:stt:mech:MechFrame} has been designed.

The structure consists of two identical flanges, shaped and drilled individually, connected by screws
to six precise tubular spacers. The two inner ones are only necessary during the mounting phase 
and could be removed after the installation of the straw modules (see \Reffig{fig:stt:mech:MechFrame}).
The two flanges with semi-cylindrical shape are mounted to the Central Frame structure and are separated by 42\,mm
to leave space for the target pipe.

\begin{figure}%[b] 
\begin{center}
\includegraphics[width=\swidth]{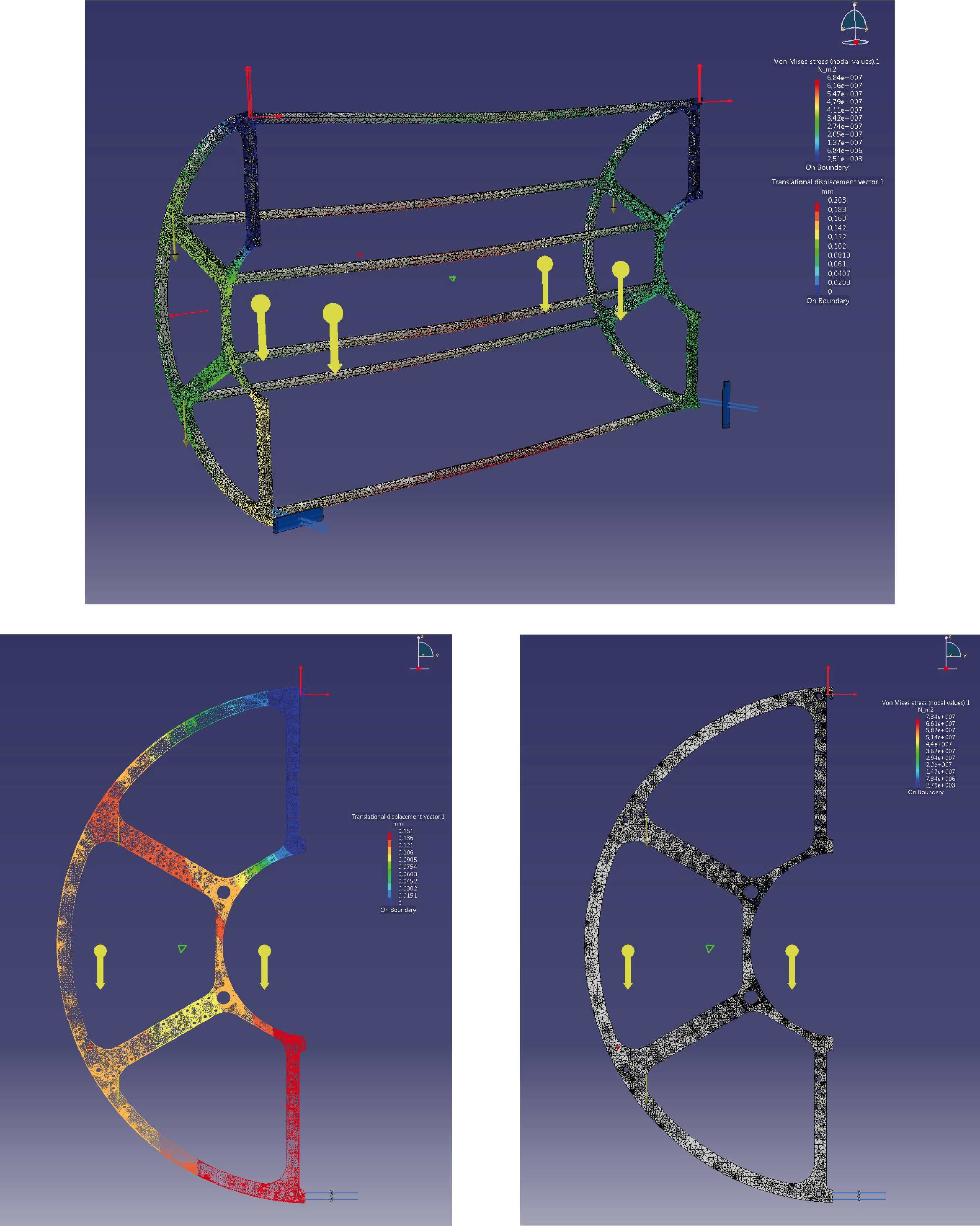}
\caption[Results of the finite element analysis of the STT support structure]{Results of the FEA of the STT support structure. The maximum deflection of the frame is 0.03~mm.}
\label{fig:stt:mech:FEA}
\end{center}
\end{figure}

High-precision boreholes  on the spokes of the flanges are used to mount the straw 
layer modules with the required position accuracy of $\pm$ 0.05~mm.
In order to check the solution, a thorough finite element analysis (FEA) of the whole structure 
has been performed. The results confirm the validity of the frame design, both from the functional and the structural point of view. 
The input parameters of the FEA-analysis are listed in \Reftbl{tab:stt:mec:FEAAn} for aluminum.
\begin{table}
\begin{center}
\caption[Data used for the FEA (Finite Element Analysis) of the support structure]{Data used for the FEA (KinonRisic@
Finite Element Analysis) of the support structure.}
\label{tab:stt:mec:FEAAn}
\begin{tabular}{|c|c|}
\hline\hline
2 End-plates & 60 N \\
6 Connecting bars & 30 N \\
2100 Straw Tubes & 80 N \\
Electronics, gas, services & 110 N \\
{\bf Total weight }& {\bf 280 N} \\
Density & 2.7 g/cm$^3$ \\
Young’s modulus & 70 GPa \\
\%% Radiation length (X$_0$)&  9 cm \\
Thermal expansion & 24 ppm/$^o$C \\
\hline\hline
\end{tabular}
\end{center}
\end{table} 
\Reffig{fig:stt:mech:FEA} shows the calculated stress on the structure. The maximum value of the 
deflection on the frame is 0.03~mm. The biggest stress is suffered at the backward flange, which 
also has to support the weight of the electronics and  
supply services of the detector (see \Reffig{fig:stt:mech:FEA} bottom left panel). 
%The two internal rods (see \Reffig{fig:stt:mech:MechFrame}) are needed only during the straw layers 
%mounting phase and then could be removed reducing further the material budget.

\subsection{Mounting of the Straw Layer Modules in the Support Structure}
\label{subec:stt:mech:inst}
\begin{figure}%[b] 
\begin{center}
\includegraphics[width=0.95\swidth]{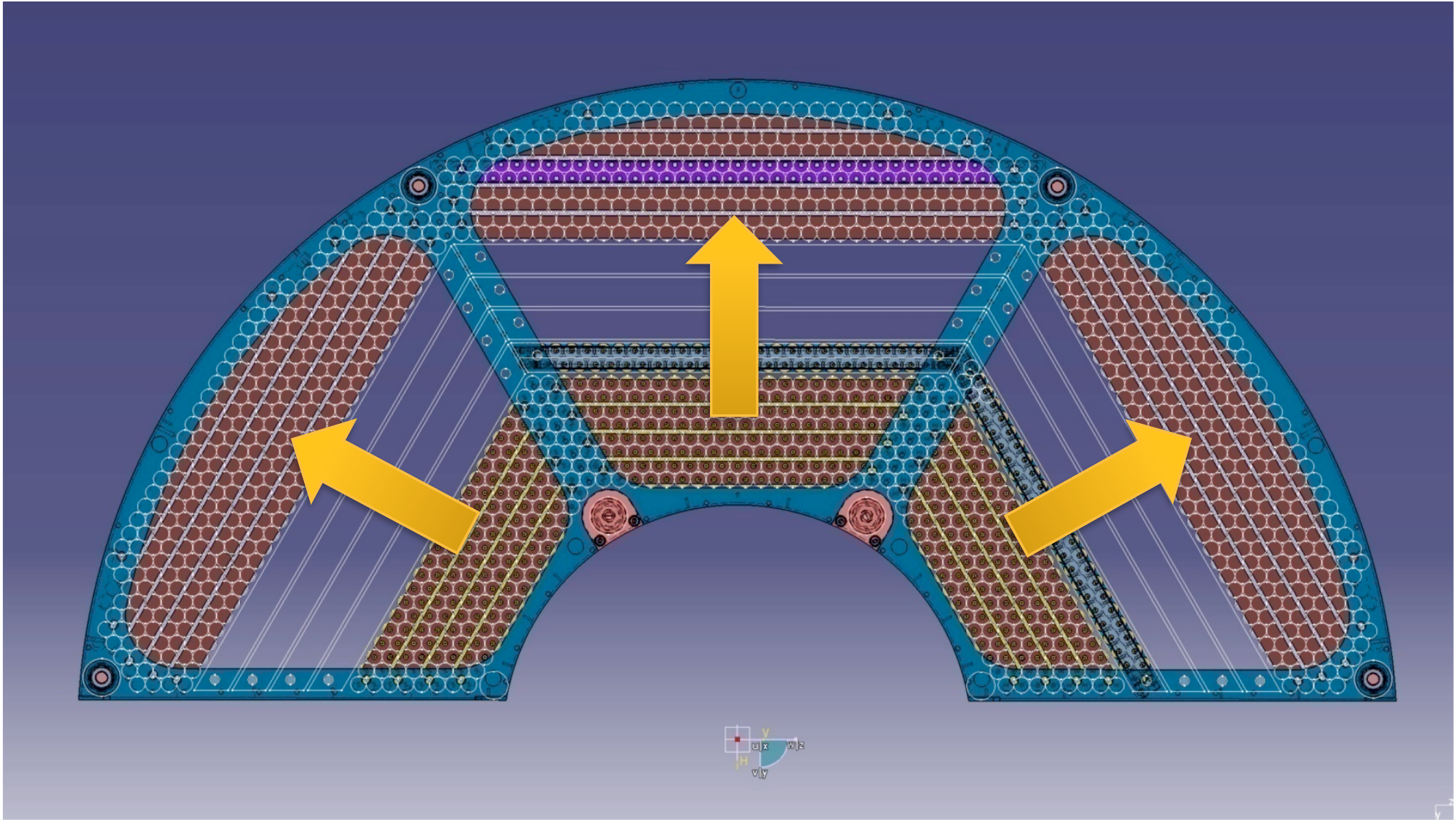}
\caption[Mounting scheme of the straw tube layer modules in the support structure]{Mounting scheme of the straw tube layer modules in the support structure. The free space in the 
middle is filled by the skewed layer modules.}
\label{fig:stt:mech:Moun}
\end{center}
\end{figure}

\begin{figure}%[b] 
\begin{center}
%\vspace{-2cm}
\includegraphics[width=0.95\swidth]{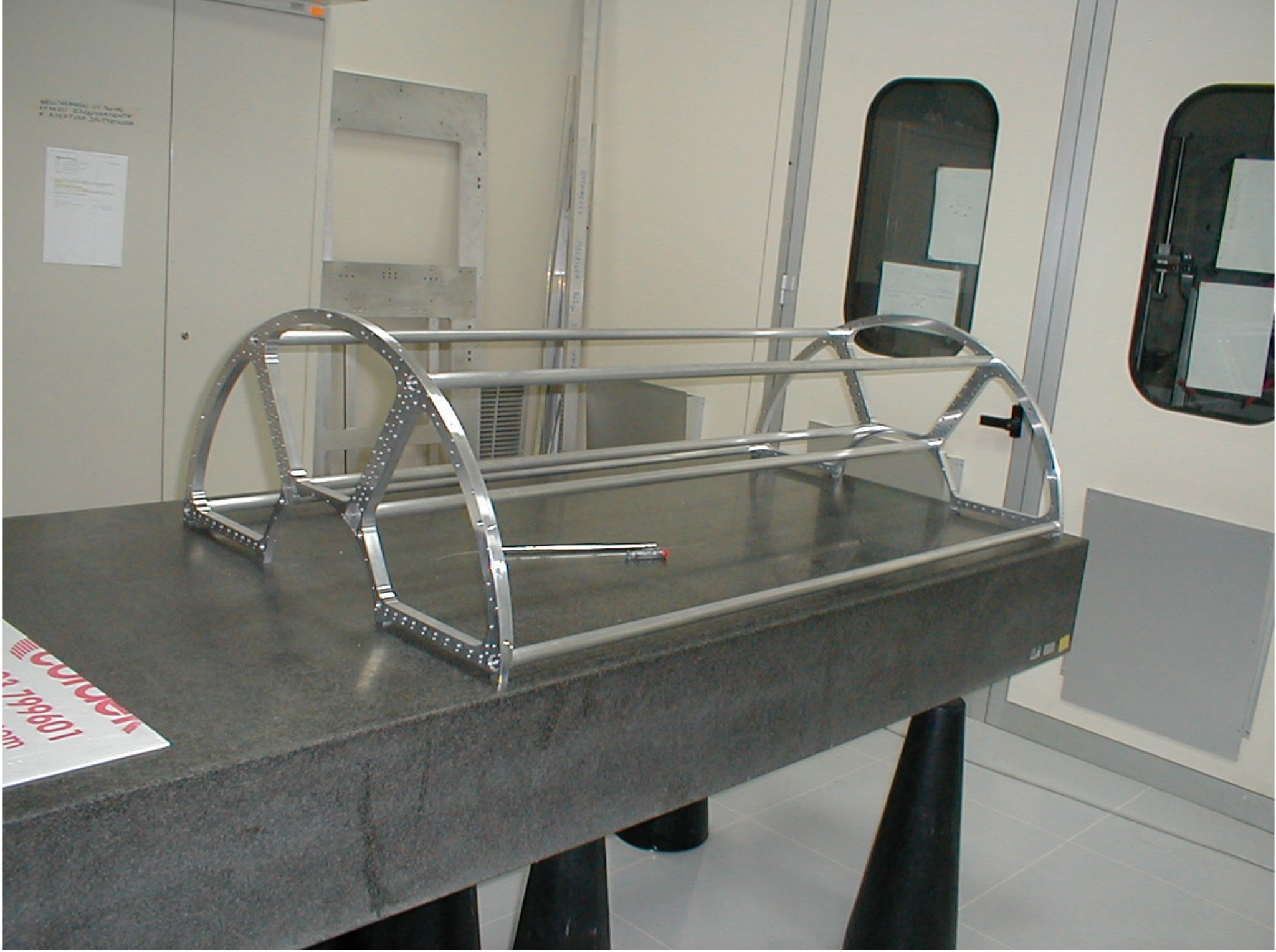}
%\vspace{-2cm}
\caption[Full-size prototype of the mechanical frame structure of the STT]{Full-size prototype of the mechanical frame structure of the STT.}
\label{fig:stt:mech:frame}
\end{center}
\end{figure}

The assembled straw layer modules will be mounted in the mechanical frame structure, starting with the inner axial ones.  \Reffig{fig:stt:mech:Moun} shows a scheme of the mounting procedure. As described in \Refsec{sec:stt:lay:mod} each module is precisely positioned by two mounting brackets at both ends, which are fixed by special pins to the reference boreholes on the spokes of the frame flanges.
A full-size prototype of the mechanical frame structure made of aluminum is shown in \Reffig{fig:stt:mech:frame} and is used to check the mechanical precision and complete mounting scheme of the straw layer modules. 

A very limited space of only 15\,cm in the z-direction upstream of the detector is foreseen for the readout electronics, high voltage supply, gas manifold lines, and distribution of all cables and supply pipes. The electronic readout scheme 
of the STT is split into the analog part, consisting of an Application Specific Integrated Circuit (ASIC) for each individual straw,
and the digital readout part, which is located close to the \Panda spectrometer and connected by 5-6\,m long cables.
The STT readout is described in detail in the next chapter.
\begin{figure}%[b] 
\begin{center}
\includegraphics[width=\swidth, height=5cm]{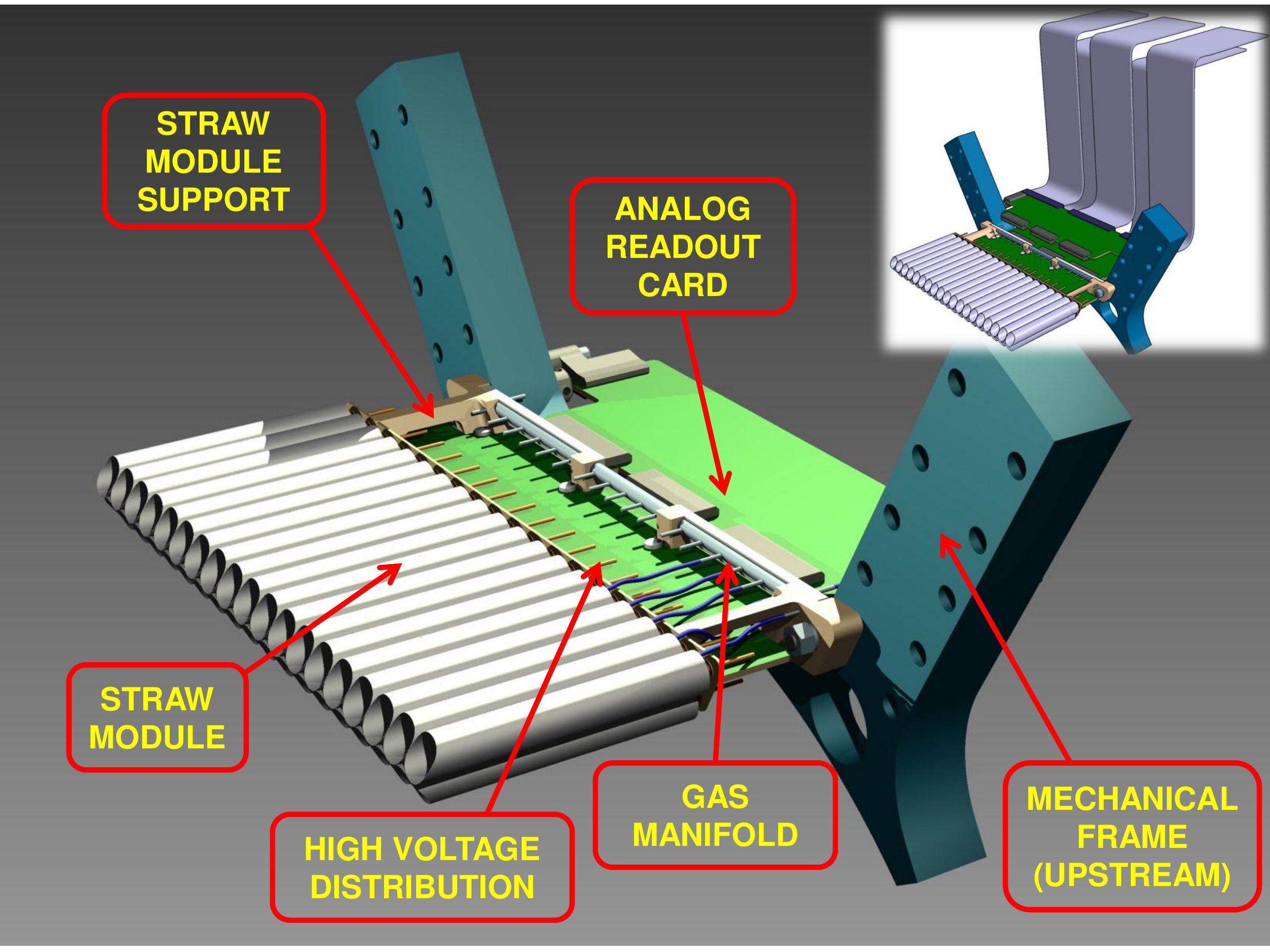}
\caption[CAD drawing of the STT front-end part]{CAD drawing of the STT front-end part with the analog readout boards, high voltage and gas supply. The cabling scheme is shown in the upper right corner.}
\label{fig:stt:mech:sttfront}
\end{center}
\end{figure}

As can be seen in \Reffig{fig:stt:mech:sttfront} the readout part, high voltage and gas supply are located at the same upstream end of the detector. Thus, the material budget at the downstream end of the STT can be kept very low, which is important for the particle tracking in the forward direction.

\subsection{The Central Frame Structure}
%\COM{Author(s): B. Dulach}
\label{sec:stt:centralframe}
The inner central region of the \Panda target spectrometer consists of the beam-target cross-pipe, the Micro Vertex Detector (MVD) and the STT detector, which are supported and precisely positioned by a common mechanical Central Frame (CF) structure. In addition, the CF structure has to support all services, readout components, cable routing, gas pipes, and cooling pipes in case of the MVD.

The mounting procedure of this system will start with connecting the beam-target cross-pipe to the CF structure. Next the MVD detector and all related services will be mounted to the CF and precisely positioned to the cross-pipe. Finally, the two semi-cylinders of the STT will be attached to both sides of the vertical CF structure, including the cable routing. Then the completed system will be inserted in the target spectrometer using two top and bottom rails, which are installed in the apparatus parallel to the spectrometer axis. Three skates on top of the CF structure and two skates at the CF bottom move the 
system on the two rails in and out of the target spectrometer.   

\begin{figure}%[b] 
\begin{center}
\includegraphics[width=0.95\swidth]{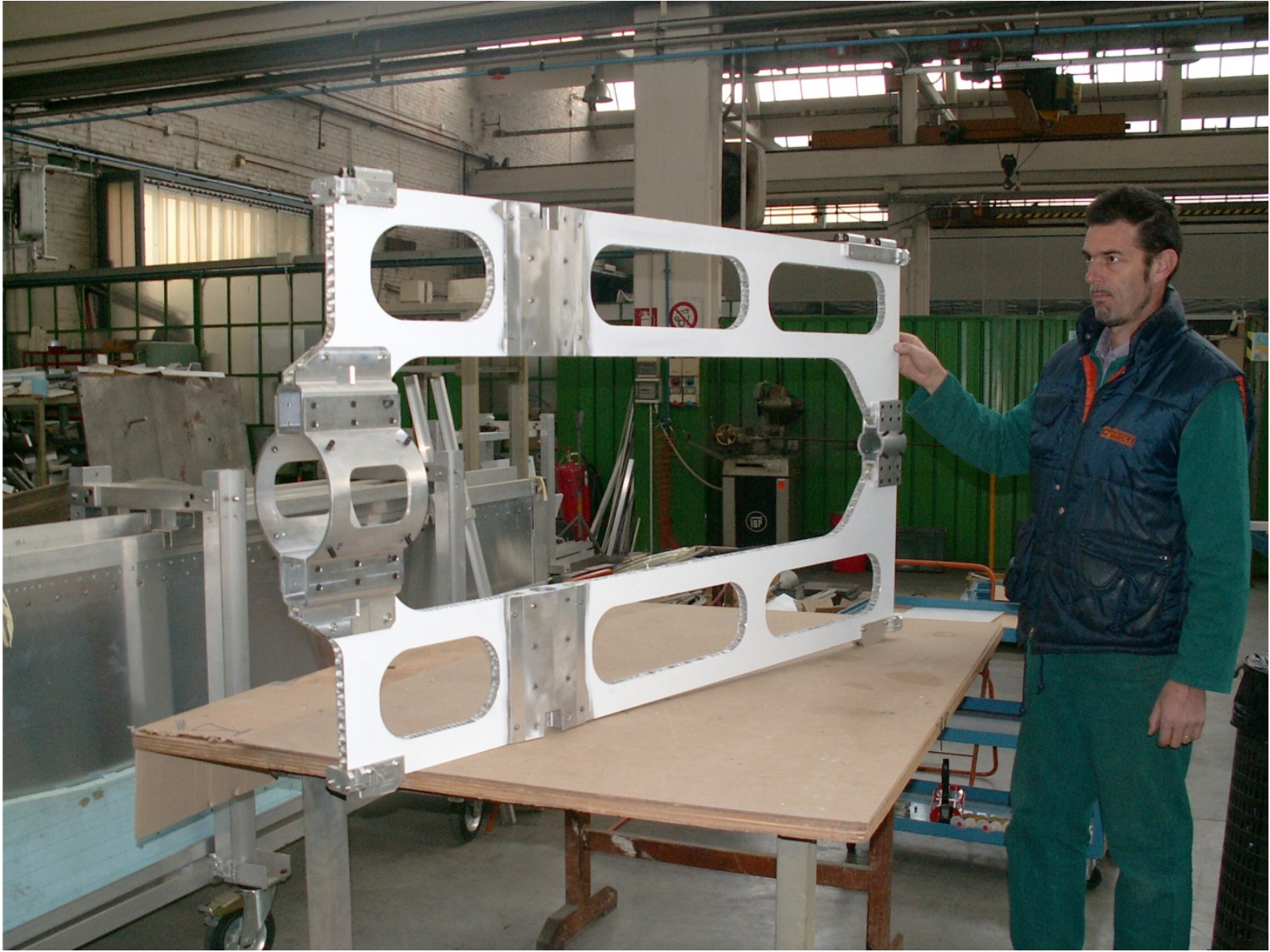}
\caption[First prototype of the central frame]{First prototype of the Central Frame (see text for more details).}
\label{fig:stt:mech:CF}
\end{center}
\end{figure}

\begin{figure}%[b] 
\begin{center}
\includegraphics[width=\swidth]{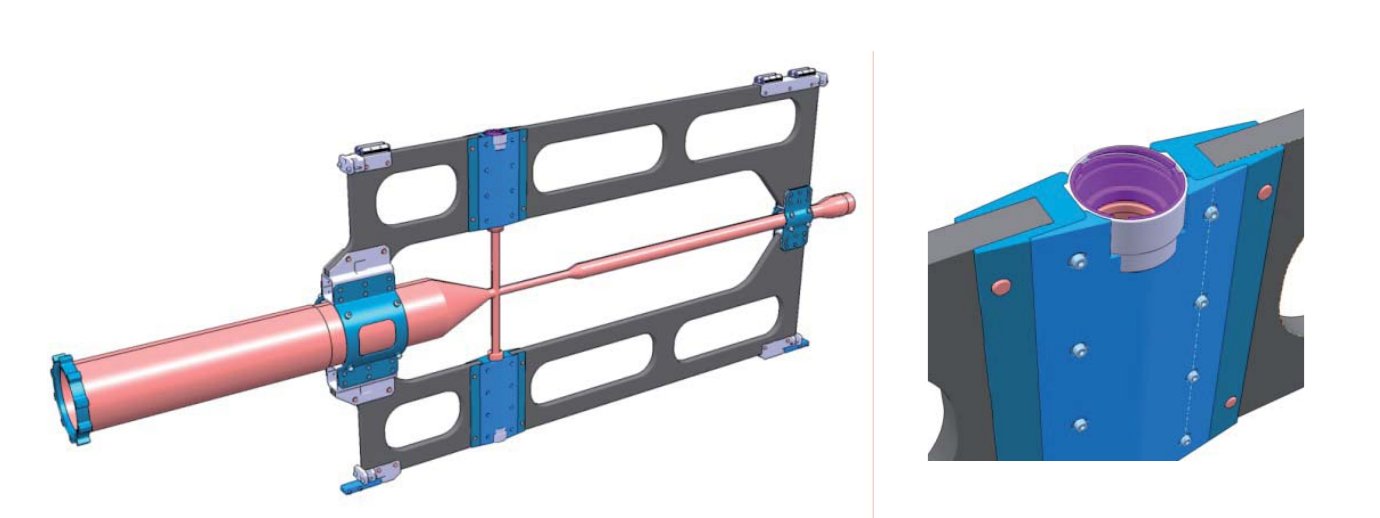}
\caption[CAD drawing of the central frame and target pipe support]{CAD drawing of the Central Frame (CF) structure with the mounted beam-target cross-pipe (left figure). The right figure shows an enlarged cut-out of the target pipe support in the CF.}
\label{fig:stt:mech:CR}
\end{center}
\end{figure}

The CF frame has a total thickness of 20\,mm and consists of a sandwich made with a central core of honeycomb 
covered on both sides with pasted skins of carbon fiber. By this the structure is very thin 
and light-weight. The most massive parts are those necessary to fix the beam-target cross-pipe. In fact, on top and bottom, the fixing pads must be able to support the torque of 1\,Nm that will be applied when the pipe is connected. Moreover, the 
backward support of the beam pipe must leave sufficient space for the cable routing of the MVD. Therefore, it has two 
rectangular housings, located top and bottom. The attachment point of the beam pipe in the forward region is less critical.
\Reffig{fig:stt:mech:CR} shows the CF after the mounting of the beam-target cross-pipe and an enlarged cut-out of the target pipe support at the top and bottom of the CF.
\par
\begin{figure}%[b] 
\begin{center}
\includegraphics[width=0.95\swidth]{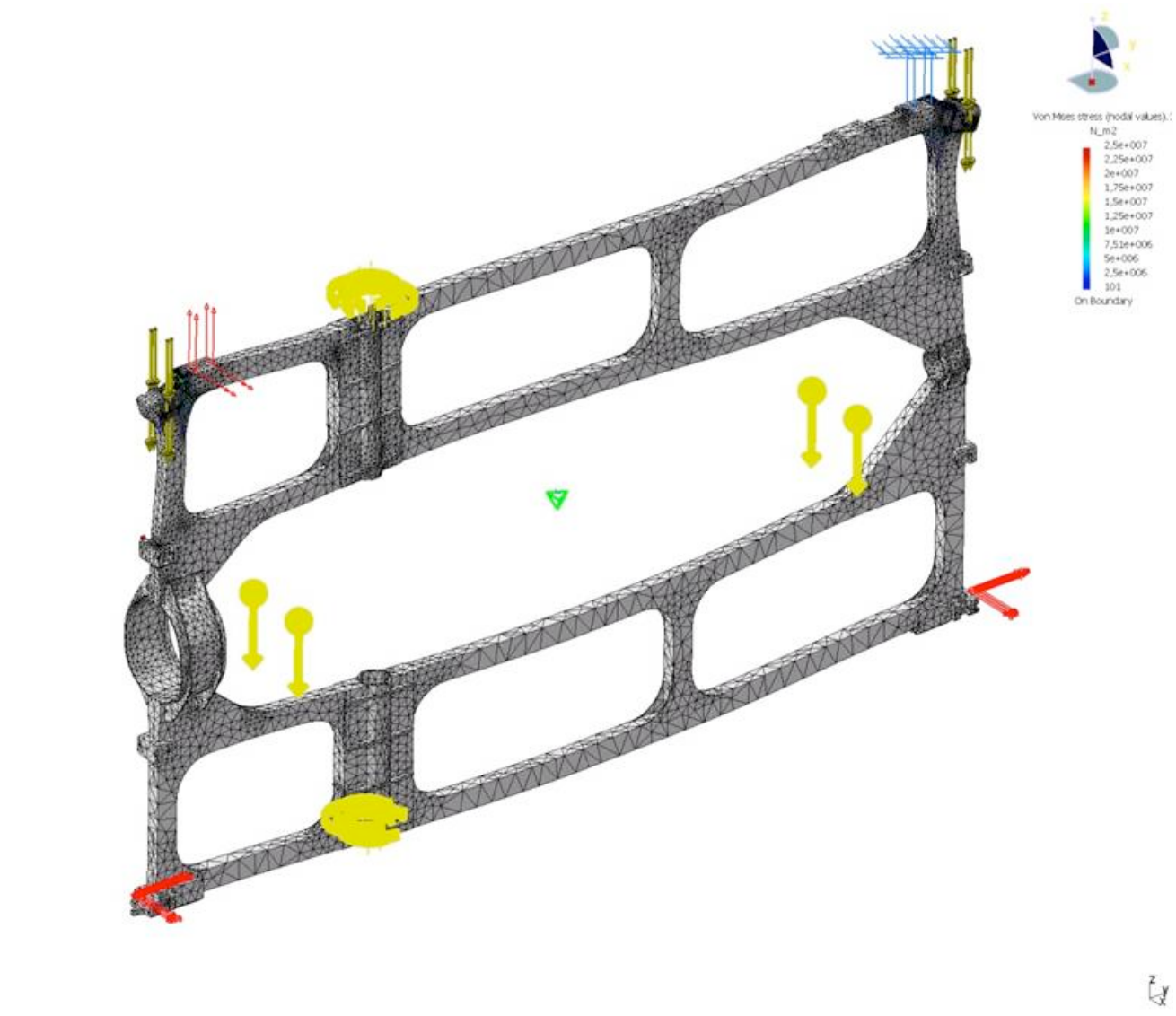}
\caption[Results of the finite element analysis of the central frame]{Results of the Finite Element Analysis (FEA) performed on the central frame.}
\label{fig:stt:mech:feaCF}
\end{center}
\end{figure}

\Reftbl{tab:stt:mec:CF} lists the input data for a FEA-analysis of the mechanical stress of the CF structure.
A maximum equivalent stress of 25\,MPa and maximum deflection (sag) of 0.5\,mm has been calculated (see \Reffig{fig:stt:mech:feaCF}), which is far below the stress limit of the structure of 95\,MPa. 
\begin{table}
\begin{center}
\caption[Input data used for the FEA-analysis of the central frame structure]{Input data used for the FEA-analysis of the central frame structure.}
\begin{tabular}{|c|c|}
\hline\hline
Straw tube tracker &  700.0 N\\
Beam-target cross-pipe &  45.0 N \\
Micro vertex detector& 300.0 N \\
Gas pipes, electronics, cables, others & 100.0 N \\
Safety load &  250.0 N \\
{\bf Total load}& {\bf 1395.0 N }\\
\hline\hline
\end{tabular}
\label{tab:stt:mec:CF}
\end{center}
\end{table} 
\par

\begin{figure}%[b] 
\begin{center}
\includegraphics[width=0.95\swidth]{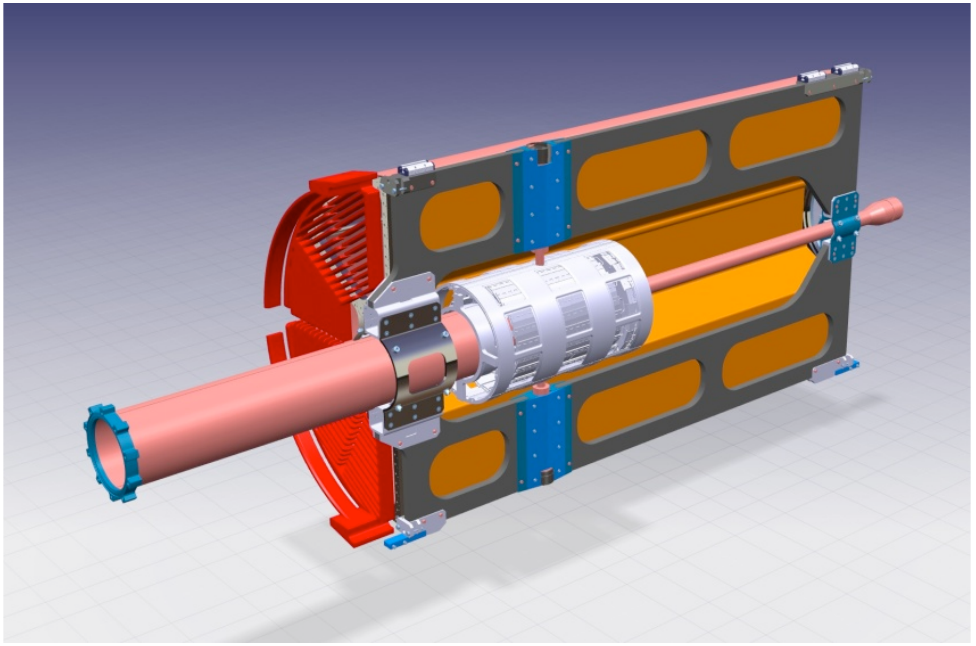}
\caption[CAD drawing of the central frame and MVD]{CAD drawing of the central frame with the target-beam cross-pipe, the MVD detector and one semi-cylinder of the STT at the backside of the CF.}
\label{fig:stt:mech:Global}
\end{center}
\end{figure}

In order to test the solution and the mounting sequence, a prototype of the CF has been constructed. 
\Reffig{fig:stt:mech:Global} shows a CAD drawing of the CF structure with the target-beam cross-pipe, the micro vertex detector and one semi-cylinder of the STT mounted at the backside of the CF.

\subsection{Positioning and Alignment}
\label{sec:stt:all}

As described in \Refsec{subec:stt:mech:sup} the position accuracy of the mounted straw layer-modules relative to the 
precision boreholes in the end flanges of the STT mechanical frame is within 50\,$\mu$m. Due to the close-packaging of the glued straws in a module
with a precise 10.1\,mm tube-to-tube distance, deviations in the position of a single straw larger than about 40\,$\mu$m are not possible.

After the complete assembling of each STT chamber, they will be attached to the CF by means of a set of precise positioning pins.
This solution ensures that the two semi-cylindrical STT volumes are placed with the correct relative position to each other.
It is worth to remind here that the CF not only holds the STT but also the MVD and the target-beam cross-pipe. Therefore it will house a 
complete set of reference marks so that the three systems can be aligned relative to each other with a precision of about 100\,$\mu$m. This 
is also the reason why the attach points on the CF for each detector are made as hardened titanium inserts, embedded in the carbon fiber structure. 
They will be precisely machined to allow a high dimensional and geometrical accuracy.
 
The relative alignment of the three components (cross-pipe, MVD, STT) of the \Panda target spectrometer will be  
done during the installation, outside of the solenoid magnet. Afterwards, only the Central Frame structure must be precisely positioned inside the 
magnet by means of the sliding rail system. 
Reference marks will be placed on the CF to allow an external survey to define its position with respect to the general \Panda
reference frame. 
%The correct positioning of the rails will be checked with a dummy frame, with the same characteristics as the real one, before
%the insertion of the whole detector system. 
The overall mechanical precision in the x,y--plane will be below 150\,$\mu$m.
%EOF: panda_tdr_stt_mec.tex

% FILE: panda_tdr_stt_gas.tex
%
\section{The Gas System}
%\COM{Author(s): V. Lucherini}
\label{sec:stt:gas}
\begin{figure*}[!ht]
\begin{center}
\includegraphics[height=9cm]{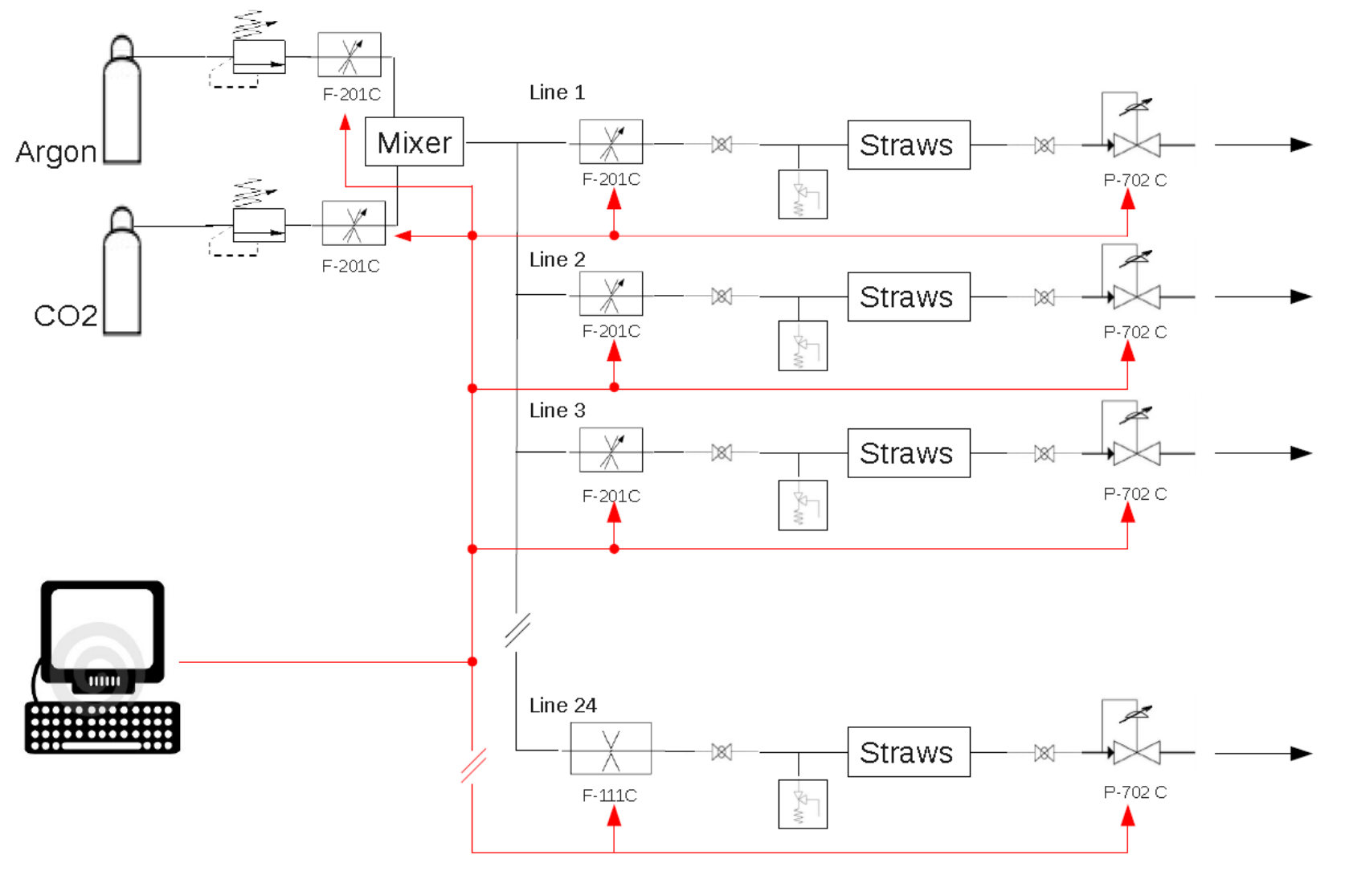}
\caption[Scheme of the gas system]{Scheme of the gas system.}
\label{fig:stt:gas:scheme}
\end{center}
\end{figure*} 
The preferred gas mixture for the \Panda-STT is argon with an admixture of 
about 10--20$\,\%$ CO$_2$ as the quenching component. The details that brought
to this choice are illustrated in \Refsec{sec:stt:des:mix}.
This gas mixture has also good capability 
to tolerate high irradiation levels (see \Refsec{sec:stt:pro:aging}) since
no deposits on the straw tube electrodes from polymerisation reactions occur, 
provided that there is a clean gas environment including all materials and parts of the 
detector and gas supply system in contact with the gas. For both gas components a 
high purity grade is required (argon with grade 5.0, CO$_2$ with 4.8). The supply 
lines consist of polished stainless steel pipes and thermoplast (PA) tubings where 
a higher flexibility is needed. Since argon and CO$_2$ are non-flammable, not 
expensive, and components of the atmosphere, no recirculation and containment of 
the gas mixture is needed, and the gas supply of the detector is done in flushing mode. 
The STT will be operated at a gas pressure of about 2\,bar (absolute) and preferably at 
room temperature. 
The total STT gas volume of about 1040\,~l %(520\,~l $\times$ 2\,bar) 
is exchanged typically 
every six hours with a flow rate of about 3\,~l per minute to refresh the gas mixture and 
to prevent an accumulation of contaminants in the detector and gas system.

The gas system of the STT consists of supply gas bottles for each mixture component, 
cleaning filters in the gas lines, a mixing section with ratio-based mass flow controllers, 
regulated by a pressure transducer inserted in the STT volume to set a constant absolute pressure 
of about 2\,bar in the detector, the supply lines in and out of the detector and outlet valves to 
a dedicated exhaust line at the \Panda experimental area. The scheme of the gas distribution system 
is shown in \Reffig{fig:stt:gas:scheme}. 
The mass flow controller and meter devices \cite{bib:stt:gas:Bronkhorst} are based on digital 
electronics. In these devices the analog sensor signal is sent directly to a micro processor. 
By doing so, optimum signal stability and accuracy is achieved. An integral alarm function 
continuously checks the difference between the set point and the measured value. If the supply 
pressure drops the instrument gives a warning. In addition the instrument runs a self diagnostics 
routine, and controller settings can be remotely adjusted with a hand terminal or a computer using 
an RS-485 busline.
For the Ar/CO$_2$ gas mixture the required accuracies of the settings and control have to be better 
than 0.3\,\% (absolute) for the mixture ratio, about 1-2\,mbar for the pressure, and 1\,K for the 
temperature.  

The \Panda straw tubes are arranged in six sectors forming a hexagon 
around the interaction point. Each sector can be further sub-divided in 
three regions in the radial direction: the inner and outer region ones, that are equipped with axially aligned
straws, and the middle section which houses stereo straw tubes, 
see \Reffig{fig:stt:lay:geo:Detectorlayout} 
(more details of this arrangement can be found in \Refsec{sec:stt:lay}).
To flush the straws with the required two-components gas mixture, the following 
guidelines have been adopted:
\begin{itemize}
\item reducing the redundancy of the system to lower the costs and the complexity of the system;
\item assure a minimal redundancy to guaranty, in case of failure, the operation of at least parts of 
any sector;
\item keeping the space needed for the system within reasonable boundaries ($\sim$ few cm);
\item automatization and remote control of the gas flow parameters, with the possibility to switch to 
manual/local operation for: 
\begin{itemize}
\item setting the ratio of the two components in the gas mixture;
\item quantifying the gas flow;
\item setting the gas pressure;
\item controlling the temperature.
\end{itemize}
\end{itemize}
Following these guidelines we decided to have 24 independent lines, organized as follows:
\begin{itemize}
\item a line per sector for each axial straw regions: $6 \times 2 = 12$;
\item two lines per sector for each of the stereo straw groups: $6 \times 2 = 12$.
\end{itemize}

\begin{figure}
\begin{center}
\includegraphics[width=\swidth]{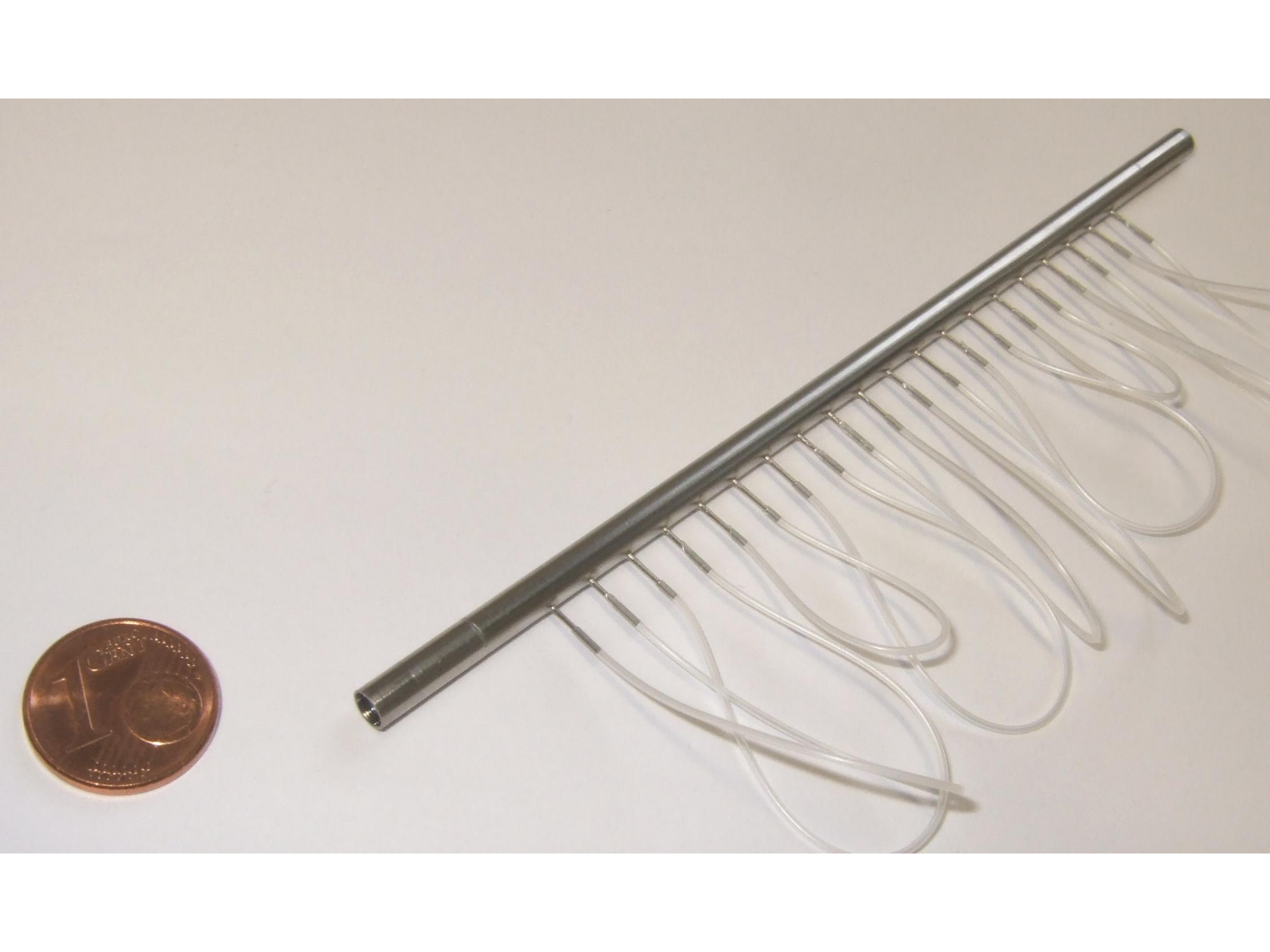}
\caption[Prototype of a gas distributor]{Prototype of a gas distributor consisting of a stainless steel pipe (4\,mm diameter, 0.1\,mm wall thickness) with small capillaries (0.55\,mm diameter) for connecting the individual straws.}
\label{fig:stt:gas:distributor}
\end{center}
\end{figure} 
With this selection of multiplicity and topology, in case of failure of one line, it is assured that 
at least one half of any sector remains operative. This is particularly important for the stereo 
tubes that allow to determine the z-coordinate of particle trajectories. 
The gas lines are connected to gas manifold pipes which supply each individual straw.
\Reffig{fig:stt:gas:distributor} shows a first prototype of such a gas manifold. It consist of a 
stainless steel pipe having a wall thickness of 0.1 mm and a diameter of 4 mm. On it, small 
capillaries (diam. 0.55 mm) are welded and will be connected to the individual straws. By connecting 
two straws in series it is possible to place the in- and outlet gas manifolds together at the upstream 
end of the detector.

%
%EOF: panda_tdr_stt_gas.tex

%
% Bibliography for this chapter (remove %)
%
\bibliographystyle{panda_tdr_lit}
\bibliography{./stt/lit_stt}
% EOF

%
% STT TDR
% File for chapter 2
\chapter{The Read Out and Control Systems}
% FILE: panda_tdr_stt_ele.tex
%
%
%
%\section{Requirements}
%\COM{Author(s): H.~Kuc, J.~Smyrski}
\label{sec:stt:ele:req}
The input characteristics of the front end electronics have to match
the electrical properties of the straw tubes, which are listed in \Reftbl{tab:stt:ele:req:straw_prop}.
\par
\begin{figure}[b]
\begin{center}
\includegraphics[width=0.95\swidth]{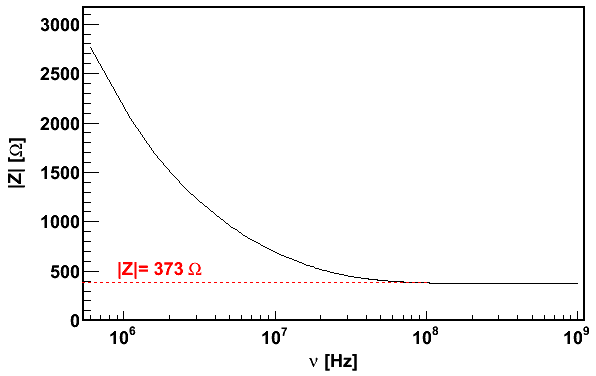}
\caption[Straw tube impedance as a function of frequency $\nu=\omega/2\pi$]{Straw tube impedance as a function of frequency $\nu=\omega/2\pi$
(solid black line). The high frequency limit is indicated with the red dashed
line.}
\label{fig:stt:ele:req:impedance}
\end{center}
\end{figure}

From the point of view of the pulse propagation, the straw tube
acts as a coaxial transmission line with loss, with an impedance given by the
formula:
\begin{equation}
Z = \sqrt\frac{R+i\omega L}{i\omega C},
\end{equation}
where $R$ is the electrical resistance, $L$ is the inductance,
$C$ is the capacitance and $\omega$ is the angular frequency.
For high frequencies ($>100$\,MHz),
the impedance of the straw tubes tends to the limit
$Z\rightarrow \sqrt\frac{L}{C} = 373\,\Omega$ (see \Reffig{fig:stt:ele:req:impedance}).
\begin{table}[h]
\caption[Straw tube electrical properties]{Straw tube electrical properties.}
\smallskip
\begin{center}
\begin{tabular}{|c|c|}
\hline\hline
Capacitance & 8.9\,pF/m \\
Sense wire resistance & 258\,$\Omega$/m  \\
Inductance & 1.24\,$\mu$H/m \\
Impedance & 373\,$\Omega$ \\
Analog cross talk & $<$ 1\,\% \\
\hline\hline
\end{tabular}
\label{tab:stt:ele:req:straw_prop}
\end{center}
\end{table}

The basic requirements for the straw tube front-end electronics are listed in
\Reftbl{tab:stt:ele:req:require}.
\begin{table}
\begin{center}
\caption[Front end electronics requirements]{Front end electronics requirements.}
\smallskip
\begin{tabular}{|c|c|}
\hline\hline
Peaking time & $\leq$ 20\,ns \\
Double pulse resolution & $\sim$ 100\,ns \\ %150
Intrinsic electronic noise & $<$1\,fC \\
Discrimination threshold & $\approx$5\,fC \\
Max. drift time & 200\,ns \\   %150
Max. occupancy & $<$ 10\% \\
TDC resolution & $\sim$ 1\,ns \\
%Neutron exposure (10 years) & ???\,1/cm$^2$\\
%Radiation dose (10 years) & ???\,krad  \\
\hline\hline
\end{tabular}
\label{tab:stt:ele:req:require}
\end{center}
\end{table}

\section{General Concept}
%\COM{Author(s): M. Idzik, P.~Salabura, M. Palka, M. Kajetanowicz}
\label{sec:stt:ele:gen}
The STT detector consists of 4636 tubes arranged in 27 layers (see sec. \ref{sec:stt:lay}).
The maximum counting rate of 800 kHz is expected for straws in the innermost layer for the $\bar{p}p$ annihilations at the highest energy and $2\times 10^7$ interactions/s. The maximum drift time of a straw is 200 ns which results in an average double-hit probability less than 10\,\% for the innermost layers. The requested electronic time resolution should be around 
1\,ns and the intrinsic noise level below 1\,fC. The maximum analog pulse duration should be comparable to the maximum drift time of 200 ns. Furthermore, an energy-loss measurement with all straws is requested for particle identification.

To fulfill all these requirements, the proposed straw tube read-out organization comprises 3 stages:
\begin{enumerate}
\item Analog Front End Electronics (FEE) cards hosting, depending on radial distance from the beam, 36-80 channels. FEE is composed of preamplifier, 
amplifier with analog signal shaping and discriminator unit with differential output.
\item Digital Board (DB) for time and amplitude (or charge) measurements, local logic resources for noise suppression, fast hit detection, memory buffer 
for hit storage, serial Gbit optical links for the data transmission and slow control.
\item Detector Concentrator Board (DCB) (optional) receiving and merging inputs from several DB in local memory buffer and sending it to the \Panda DAQ system.
\end{enumerate}

The data from the DCBs will be transferred via fast optical links to Compute Nodes for the on-line track reconstruction and subsequently, after merging 
with the information from other \Panda detector systems, for the event selection. It will be possible to perform some local correlations on the data inside the single DB to suppress noise and reduce the amount of data sent from the board.

\par
The \Panda data acquisition and filtering systems will implement a trigger-less architecture. Instead of having a hardware trigger signal, which indicates the occurrence of a valid event, each DB will receive a precise clock signal distributed centrally from a single source: the Syncronization 
Of Data Acquisition (SODA). The DB boards will continuously monitor the detector channels and will generate data packets whenever the number 
of the input signals exceeds programmed thresholds. These data will be tagged with time-stamps obtained from the SODA.

The data acquisition system will profit from the structured running mode of the \HESR operation.
Periods of $2 \mu$s with antiproton interactions will be interleaved with periods of 400 ns of idle time.
The information on the accelerator activity will be distributed to DCBs via the Clock and Timing Distribution System. The data recorded during 
the interaction intervals will be grouped together in DCB to form a burst which will be then uniquely tagged.
Grouping of data from many bursts into predefined epoques (e.g. 500\,$\mu$s) inside DB is also considered in order to reduce network traffic.
Data from all \Panda detectors tagged with the same burst identification number will be grouped together and will be made accessible to filtering 
algorithms implemented in the Compute Nodes (CN). Decisions produced by these algorithms will thus be based on the complete detector data with full 
granularity.

\section{Analog Front-End Electronics}
%\COM{Author(s): M. Idzik, D. Przyborski}
\label{sec:ft:ele:amp}
An Application Specific Integrated Circuit (ASIC) is being developed in order to read out the straw tube pulses.
The main specifications of this chip are summarized in \Reftbl{tab:stt:ele:tab_spec}.
\begin{table}%[htb!]
\caption[Main parameters of the new straw tube front-end readout chip]{Main parameters of the new straw tube front-end readout chip (see text for more details).}
\label{tab:stt:ele:tab_spec}
\begin{center}
\begin{tabular}{| l | l |}
\hline
%				& This work
%				\hline \hline		
Technology & 0.35 $\mu m $ CMOS \\	
\hline
Number of channels & 16 \\			
\hline
Input Resistance & $\sim$120~\,$\Omega$ \\		
\hline
Default gain & $\sim$10~mV/fC \\		
\hline
Peaking time (for delta) & 20~ns \\		
\hline
Timing resolution & 1-2~ns \\			
\hline
Equivalent (delta) input range & 0-200~fC \\			
\hline
Noise ENC & < 0.4~fC \\			
\hline
Output standard & LVDS and analog\\
\hline	
Power consumption & $\sim$\,30~mW \\			
\hline
\end{tabular}	
\end{center}
\end{table}
\par

\begin{figure*}%[htb!]
\begin{center}
\includegraphics[width=0.9\dwidth]{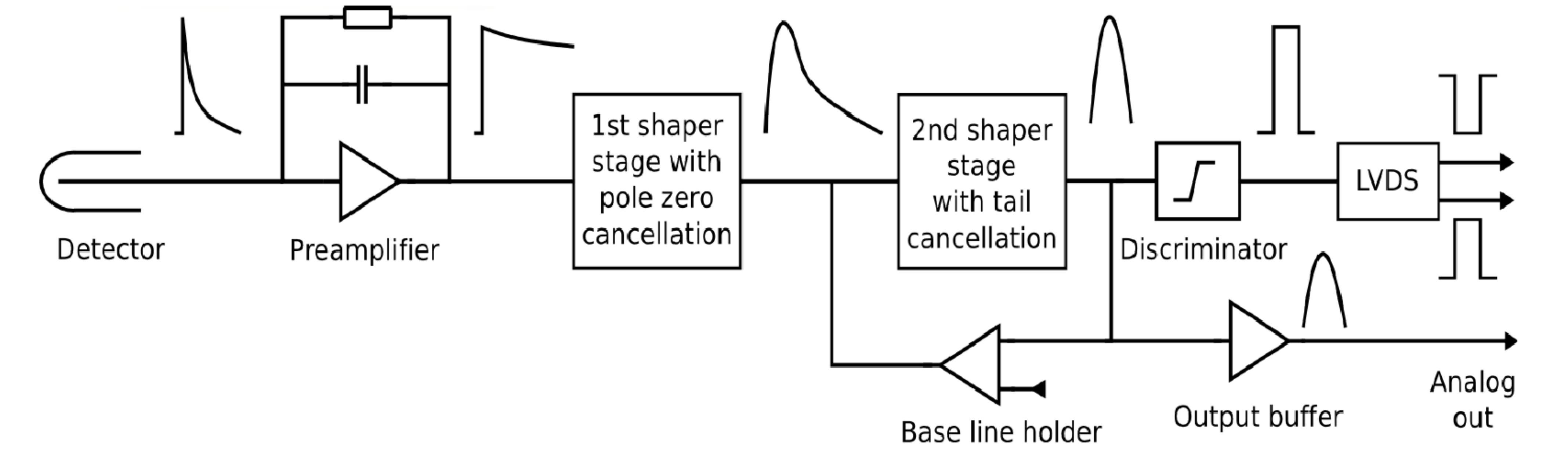}
\caption[Block diagram of ASIC proposed for straw-tube read-out]{Block diagram of the ASIC under development for the straw-tube read-out.}
\label{fig:stt:ele:amp:fe_block}
\end{center}
\end{figure*}

The ASIC's channel comprises a charge preamplifier stage, a pole-zero cancelation network (PZC), a shaper stage, a tail cancelation network, a 
discriminator circuit, a baseline holder (BLH), a fast differential LVDS output and an analog output.
The block diagram of the designed readout channel is shown in \Reffig{fig:stt:ele:amp:fe_block}.

The solution for the FEE should provide both, the timing and the amplitude information. Since it is still under study whether the Time Over Threshold (TOT) 
technique or the analog amplitude information will be used for the energy-loss measurement, the first ASIC prototype provides both the amplitude and TOT information.

\begin{figure}%[htb!]
\centering
\includegraphics[width=0.95\swidth]{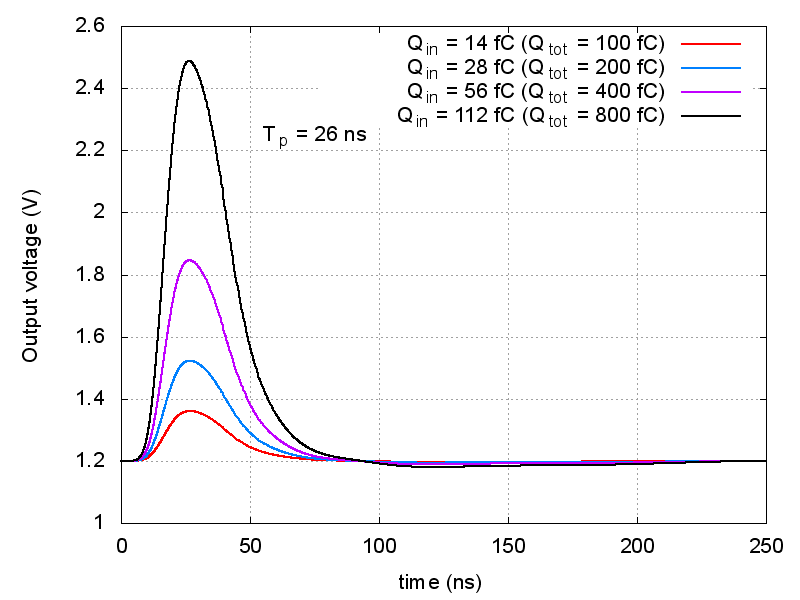}
\caption[Simulated analog responses for different input charges]{Examples of the simulated analog responses for different input charges.}
\label{fig:stt:ele:amp:tran}
\end{figure}

A typical simulated analog response of the amplifier for straw tube pulses  (generated with GARFIELD \cite{bib:stt:ele:gar}) for different charge depositions is shown in 
\Reffig{fig:stt:ele:amp:tran}. The charge depositions are expressed both as equivalent charges of ``delta-like'' pulses and as integrated charge carried by the pulses. 
The tail cancelation network assures that the pulse length is shorter than about 150\,ns.

\begin{figure}%[htb!]
\centering
\includegraphics[width=0.95\swidth]{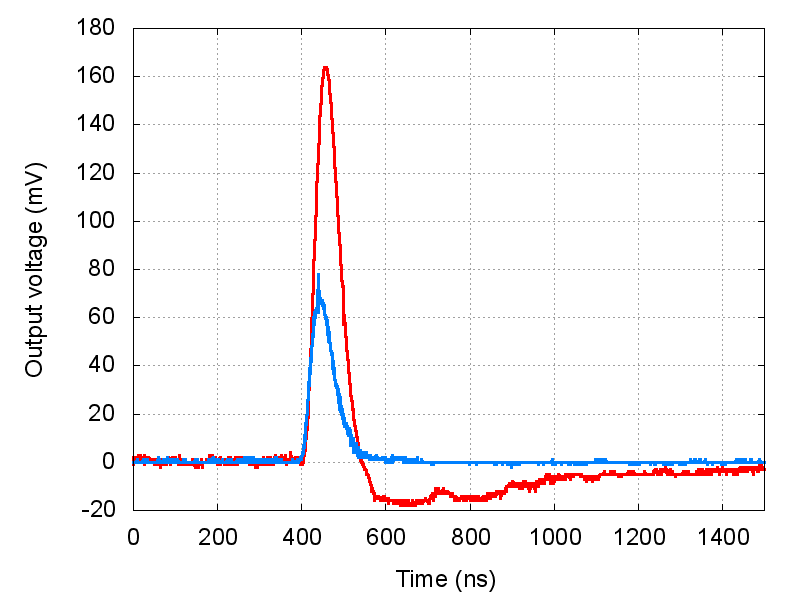}
\caption[Front-end pulses for different settings of the ion cancellation network]{Examples of front-end pulses for not optimized (red) and optimized (blue) settings of the ion cancellation network.}
\label{fig:stt:ele:amp:pulse}
\end{figure}

The design of the first version of the ASIC channel was completed and a first prototype containing 4 readout channels has been fabricated and delivered in 
the second part of 2011. Signals from an $^{55}$Fe source measured with the ASIC prototype connected to the illuminated straw tube, for different settings of the 
ion cancelation network, are shown in \Reffig{fig:stt:ele:amp:pulse}. It is seen that, with optimized parameters of the network, the long tail can be eliminated.

\begin{figure}%[htb!]
\centering
\includegraphics[width=0.95\swidth]{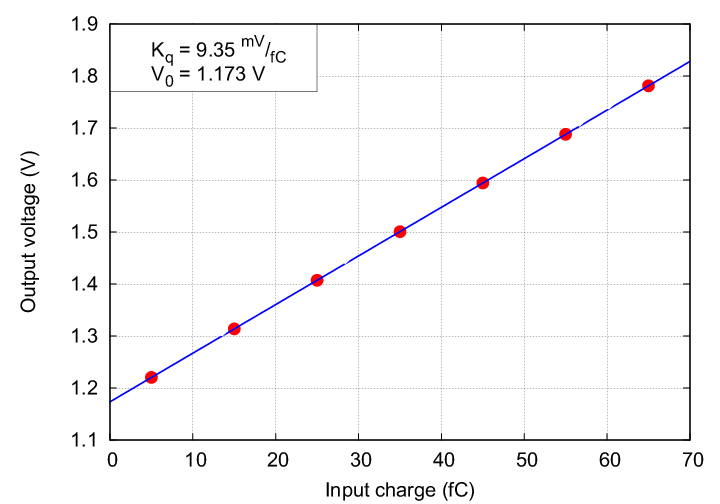}
\caption[Front-end gain measurement]{Examples of the front-end gain measurement for default settings with ``delta-like'' current pulses.}
\label{fig:stt:ele:amp:gain}
\end{figure}

\begin{figure}%[htb!]
\centering
\includegraphics[width=0.95\swidth]{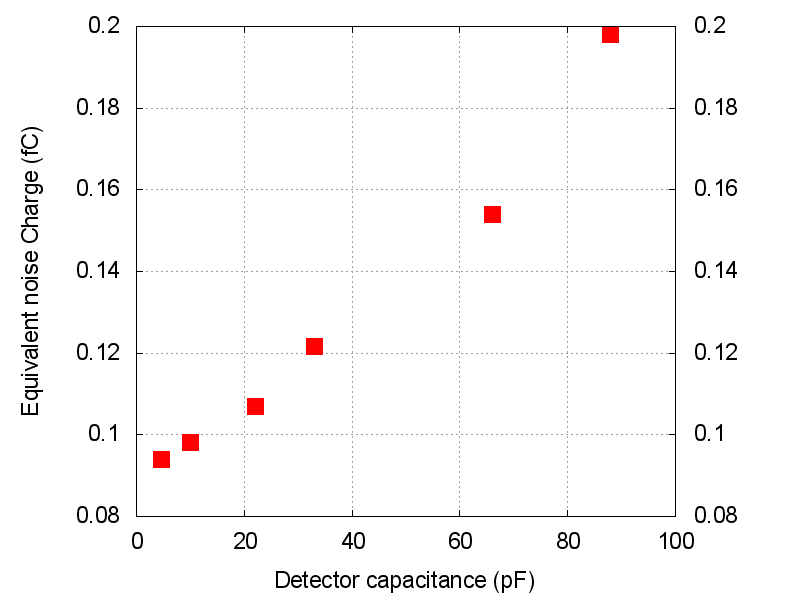}
\caption[Measurement of the front-end noise vs input capacitance]{Example measurement of the front-end noise vs input capacitance.}
\label{fig:stt:ele:amp:noise}
\end{figure}

Preliminary measurements of the front-end gain and noise are shown in \Reffig{fig:stt:ele:amp:gain} and \Reffig{fig:stt:ele:amp:noise}. 
The gain characteristics have been measured with a step-like voltage pulse injected into the ASIC channel via a capacitance (``delta-like'' pulse).

\begin{figure}%[htb!]
\centering
\includegraphics[width=0.95\swidth]{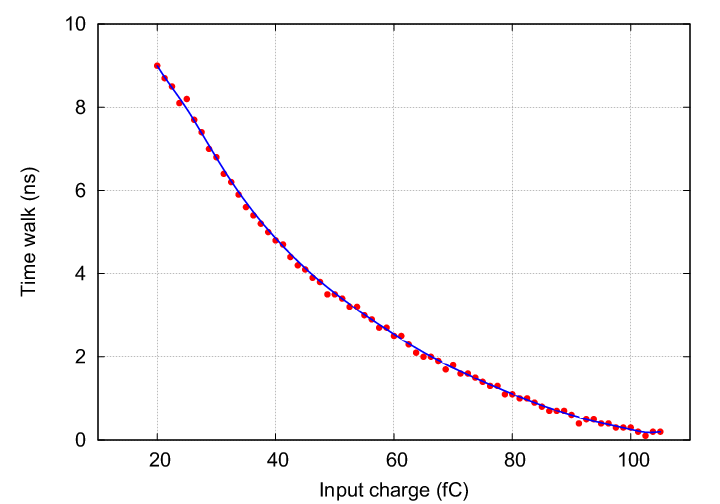}
\caption[Measurement of the discriminator time walk]{Example measurement of the discriminator time walk.}
\label{fig:stt:ele:amp:timewalk}
\end{figure}

Both results stay well within the requested specifications. The discriminator circuit uses a simple leading edge configuration. A preliminary 
measurement of the discriminator time-walk, shown in \Reffig{fig:stt:ele:amp:timewalk}, shows the typical leading edge behavior.

\begin{figure}%[htb!]
\centering
\includegraphics[width=0.95\swidth]{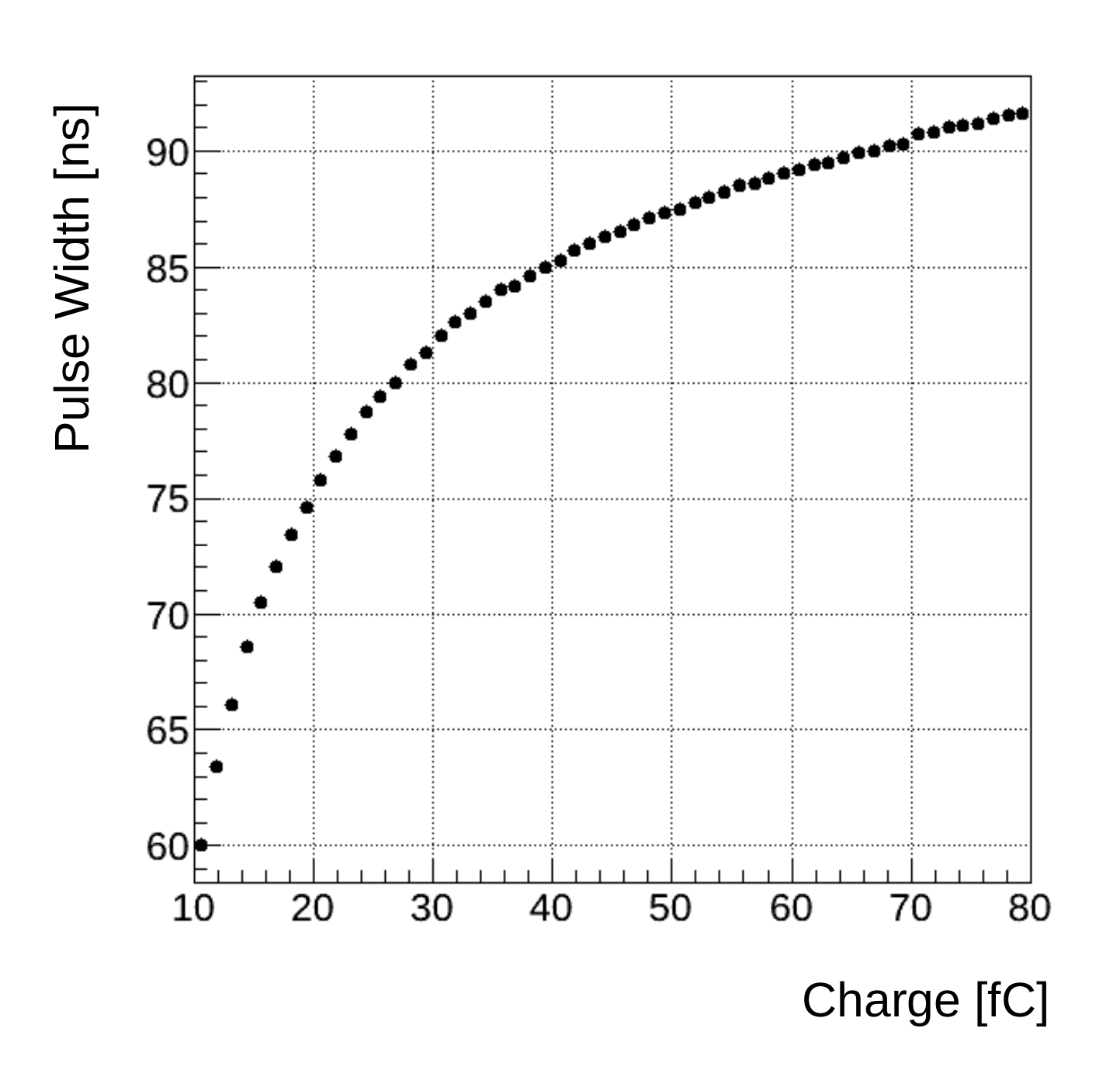}
\caption[Measurement of time-over-threshold vs charge]{Time-over-threshold vs charge measured with ``delta-like'' current pulses.}
\label{fig:stt:ele:amp:TOT}
\end{figure}

The charge vs Time Over Threshold (TOT) behavior of the ASIC is shown in \Reffig{fig:stt:ele:amp:TOT}. It has been measured with ``delta-like'' 
pulses for an input charge range of 10-80\,fC. It shows a non-linearity which is typical for gaussian like pulses.
It can be optimized in a future version by implementing a linear discharge of the front-end output capacitance. A discharging capacitance 
by a constant current provides a linear shape of the analog pulse and then the width of the discriminator response may be proportional to the collected charge.
A similar idea was successfully used in previously reported designs \cite{bib:stt:ele:peric,bib:stt:ele:kugathasan}.

\begin{figure}%[htb!]
\centering
\includegraphics[width=0.95\swidth]{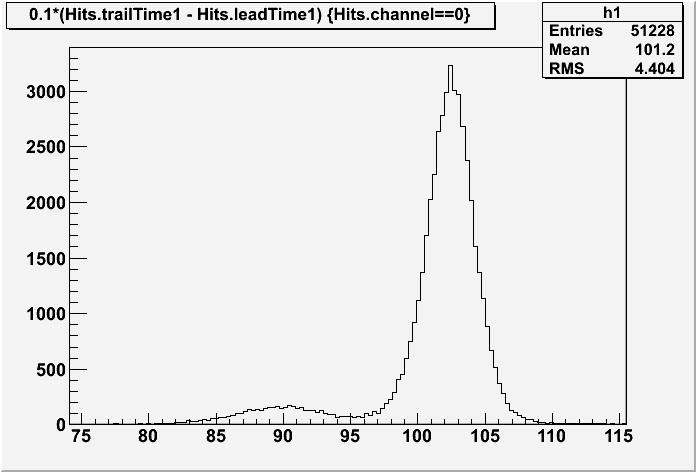}
\caption[Time-over-threshold spectrum measured with an $^{55}$Fe source]{Time-over-threshold spectrum measured with an $^{55}$Fe source and the straw tube at HV=1750\,V.}
\label{fig:stt:ele:amp:amplitude}
\end{figure}

However, it should be noted that already with the present design, the amplitude spectrum measured with an $^{55}$Fe source exhibits two clearly 
separated peaks corresponding to the characteristic 2.9\,keV and 5.8\,keV energy deposits of the source, as shown in \Reffig{fig:stt:ele:amp:amplitude}. Further simulations are needed to 
answer the question whether the present non-linearity is acceptable for particle identification without losing too much performance.

\begin{figure}%[htb!]
\centering
\includegraphics[angle=0,width=0.95\columnwidth]{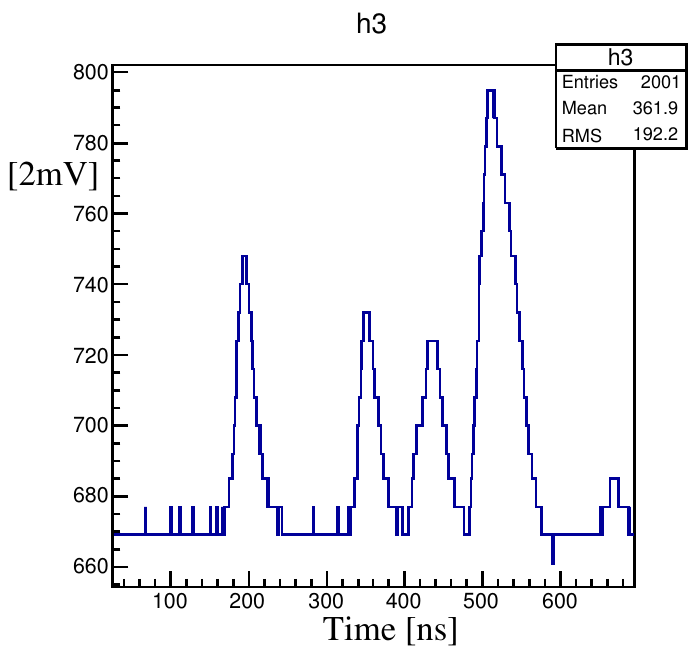}
\caption[ASIC analog signal output measured at a high hit rate]{Example of the ASIC analog signal output measured at a high (few MHZ) hit rate.}
\label{fig:stt:ele:amp:high_rate}
\end{figure}

Despite the expected large counting rate and long time constant related to the ion propagation it is very crucial to demonstrate that the ion tail cancelation and the base holder circuits work according to the design.
Recently, first measurements with a high-intensity proton beam were performed in J\"ulich in order to verify the signal readout at high rates. As an example, \Reffig{fig:stt:ele:amp:high_rate} shows the analog output of the ASIC 
recorded by an oscilloscope. No baseline distortion and a clear separation of the four individual signals can be seen within a time window of about 700\,ns, which corresponds roughly to a proton rate of 6\,MHz in the single straw.

A further optimization of the ASIC with a systematic study of the design 
parameters is still going on. In addition, a thorough analysis of the two 
different methods of the signal amplitude measurement will be done by comparing 
the TOT information with the analog signal shape, which are both provided by the 
ASIC. Depending on the result, the final architecture of the ASIC-chip including 
the specific method for the amplitude measurement will be determined. 

Apart from the not yet decided method for the amplitude measurement, some other design aspects have to be considered. 
In particular the DAC converters need to be designed and added to each channel in order to tune independently the discrimination 
threshold of each channel, the reference voltage source needs to be designed and added, the digital part of the ASIC needs to be 
implemented. Although the present ASIC was designed in 0.35\,$\mu$m technology, the final technology choice has not yet been done. The final ASIC will probably contain 16 channels.

The multi-layer printed circuit board (PCB) for the FEE will contain 2 to 5  ASICs (for 16 channels each), a dedicated logic for the ASIC configuration and connectors for flat twisted-pair cables with signal inputs (for slow control) and outputs to the DB. The 
inputs from the single straws will be provided by thin single-ended, coaxial cables.
%The power and the slow control connectors will be placed on the PCB
%together with a number of Digital to Analog Converters (DAC) for the threshold setting.
Since the mechanical frame structure of the STT has two support flanges with a six-fold symmetry, a back plane with the analog 
read-out will be composed out of six independent sectors, each serving 768 straws. Thus, 13 FEE boards with a varying number of channels, from 36 (innermost) to 80 (outermost), are needed for the read-out of one complete sector. 

\begin{table*}
\caption[Basic technical parameters of the CARIOCA-10 chip]{Basic technical parameters of the CARIOCA-10 chip.}
\smallskip
\begin{center}
\begin{tabular}{|l|c|c|}
\hline\hline
General parameters & Number of channels & 8 \\
& Radiation resistance & 20\,Mrad \\
& Technology & IBM CMOS 0.25 $\mu m $ \\
\hline
& Input impedance & 45\,$\Omega$ \\
& Range of input charge & $2.5 \div 300$\,fC \\
& Peaking time & 14\,ns\\
& Sensitivity with detector capacitance 220 pF for & \\
& positive input & 8.21\,mV/fC \\
& negative input & 7.7\,mV/fC \\
Input parameters  & Width of output pulse for charge $<$ 300\,fC at& \\
& positive input & 55\,ns \\
& negative input & 65\,ns \\
& Minimum charge & \\
& positive input & 2.4\,fC (rms 0.37\,fC) \\
&negative input & 2.4\,fC (rms 0.24\,fC)\\
\hline
Output parameters & Standard of pulses & LVDS \\
\hline\hline
\end{tabular}
\label{tab:stt:ele:amp:carioca}
\end{center}
\end{table*}

During the design phase of the new ASIC several FEE prototypes, based on CARIOCA chips, have also been tested with prototype chambers and showed 
satisfactory performance.
\par
The CARIOCA is an 8 channel, radiation hard (up to 20 Mrad dose) ASIC, featuring preamplifier,
shaper, base line restorer and discriminator.
One single FEE board consists of four CARIOCA chips for the read-out of 32 channels.
The FEE board provides an LVDS differential output which is connected by a flat cable to the DB.
The threshold for the CARIOCA's leading-edge discriminators is set by the on-board DAC, which is controlled by
dedicated lines in the cable connection to the DB. The total power consumption per channel of the CARIOCA chip is 25\,mW.

\begin{figure}
\begin{center}
\includegraphics[width=0.95\swidth]{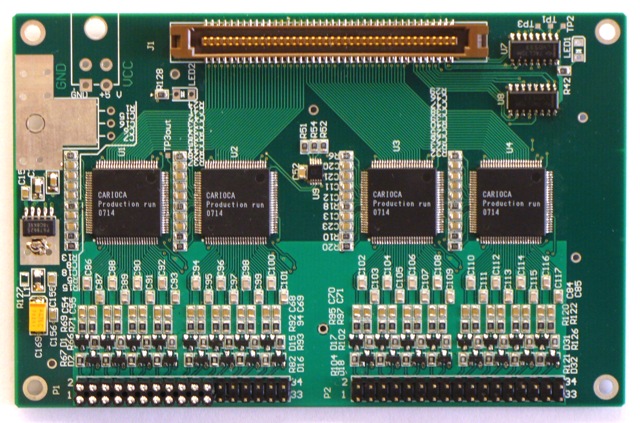}
\caption[Prototype board of a 32-channel preamplifier/discriminator based
on CARIOCA-10 chips]{Prototype board of a 32-channel preamplifier/discriminator based
on CARIOCA-10 chips.}
\label{fig:stt:ele:amp:preamp}
\end{center}
\end{figure}

The most recent, second version of the preamplifier/discriminator board
is shown in \Reffig{fig:stt:ele:amp:preamp}. Its basic parameters are given in
\Reftbl{tab:stt:ele:amp:param}.

\begin{table}
\caption[Technical characteristics of the prototype preamplifier/discriminator board - version-2]{Technical characteristics of the prototype preamplifier/discriminator board - version-2.}
\begin{center}
\begin{tabular}{|c|c|}
\hline\hline
Supply voltage & $+4.5 \div +12$\,V DC\\
Supply current & 560\,mA \\
Power consumption & 3.3\,W \\
Number of channels & 32 \\
Dimensions of board &  124\,mm x 80\,mm x 16\,mm\\
\hline\hline
\end{tabular}
\label{tab:stt:ele:amp:param}
\end{center}
\end{table}

The main limitation of the CARIOCA chip is the lack of the signal amplitude information which is crucial for the STT for the dE/dx measurement. However, it is still considered as a back-up solution for the \Panda Forward Tracker (FT) where a dE/dx measurement is not required.

\section{Digital Electronics}
%\COM{Author(s): P. Salabura, M. Kajetanowicz, M.Palka}
\label{sec:stt:ele:tdc}

The DB readout will be located outside the \Panda target spectrometer in a distance of 5-6 meters to the STT.  The DB will contain a multi-hit TDC measuring the signal arrival time with respect to the external clock provided to the DB by the \Panda SODA. For the amplitude measurement and depending on the test results, either the TDC will provide the sufficient pulse length information (TOT) or the DB will contain an additional fast sampling ADC for the analog signal.
\par
A time measurement system based on FPGA is foreseen for the STT. Recently, a time measurement board (TRBv3 see  below) based on the Lattice ECP3 family, has 
been developed at GSI, University of Frankfurt and Jagiellonian University. The implementation of a TDC in FPGA allows for a large flexibility in the selection of main measurement parameters like time range,
binning etc., and makes this approach very attractive for a broad range of applications. The implementation of the TDC functionality in FPGA is achieved by 
using its internal architecture elements - carry chains \cite{bib:stt:ele:TDC_A}, \cite{bib:stt:ele:TDC_B}, \cite{bib:stt:ele:TDC_C}.

\begin{figure}
\begin{center}
\includegraphics[width=0.95\swidth]{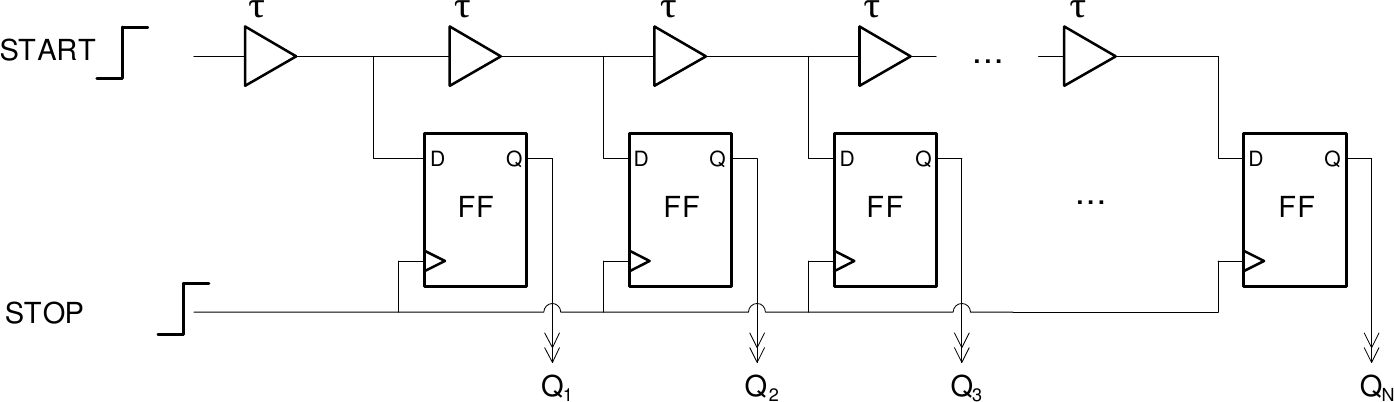}
\caption[Scheme of TDC-FPGA implementation with carry chain usage]{Scheme of TDC-FPGA implementation with carry chain usage.}
\label{fig:stt:ele:tdc:trb_tdc_carry}
\end{center}
\end{figure}

As presented in \Reffig{fig:stt:ele:tdc:trb_tdc_carry}, the time measurement is based on the information (from the carry chain - START signal in 
\Reffig{fig:stt:ele:tdc:trb_tdc_carry}) saved in the flip-flops ($Q_1-Q_n$) on the rising edge of the system clock (STOP signal in 
\Reffig{fig:stt:ele:tdc:trb_tdc_carry}). Each carry chain element delays the signal in average by 30 ps. Time measurements done at GSI demonstrate a 
$\sim 17$  ps resolution. For the STT detector, a TDC binning of 0.5 ns will be sufficient. To have all needed information about the signal from the detector 
it will be required either to measure the amplitude/charge of the straw signal via TOT or even, if it will turn-out necessary, sample its shape. This is 
possible since the FEE-ASIC can provide both, the digital (time and TOT) and analog signals.

\begin{figure}
\begin{center}
\includegraphics[width=0.95\swidth]{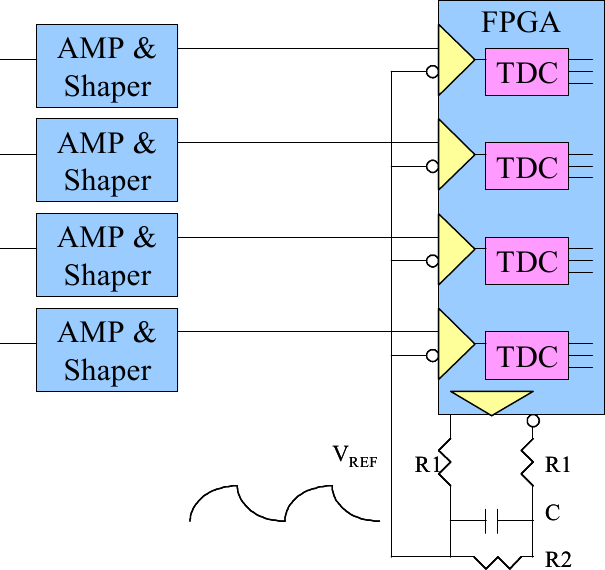}
\caption[ADC-FPGA implementation \cite{bib:stt:ele:adc_tdc}]{ADC-FPGA implementation \cite{bib:stt:ele:adc_tdc}.}
\label{fig:stt:ele:tdc:adc_tdc_fpga}
\end{center}
\end{figure}

\begin{figure}
\begin{center}
\includegraphics[width=0.95\swidth]{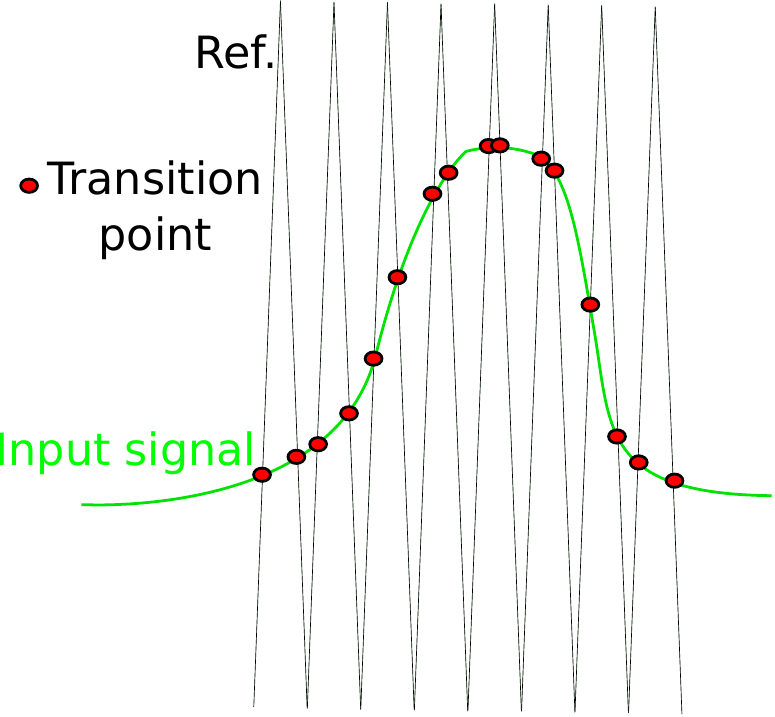}
\caption[Input and reference signal]{Example of the input and reference signals. The red points mark the transition point when the FPGA should see a change from the logical $0~(1)$ to the $1~(0)$ level. }
\label{fig:stt:ele:tdc:adc_fpga_ex}
\end{center}
\end{figure}

Therefore, along with the development of the TDC-FPGA measurement techniques the utilization of the ADC functionality into the FPGA is under investigation. 
The TDC implementation together with just a few components (resistors and capacitors) allows to perform an additional ADC measurement. \Reffig{fig:stt:ele:tdc:adc_tdc_fpga} shows a scheme of such an approach.
The differential input of the FPGA is used as a comparator. If a defined signal generated by the FPGA ($V_{ref}$ in 
\Reffig{fig:stt:ele:tdc:adc_tdc_fpga} and \Reffig{fig:stt:ele:tdc:adc_fpga_ex}) is larger than the input signal, the FPGA logic sees a $0$, otherwise a $1$ level. 
The transitions from the $0$ to the $1$ level are again saved in the flip-flop chain (see \Reffig{fig:stt:ele:tdc:trb_tdc_carry}). At the end the time measurement of the transition 
can be translated to a voltage. The advantages of this solution compared to the usage of a standard sampling ADC are the smaller power consumption and lower price. The decision about the final method of the ADC measurement will be taken after thorough tests. Consequently, appropriate mezzanine cards will be build. Such mezzanine cards can be added as an 
add-on to the basic Time Read-out Board (TRB, see below) containing the TDC.

\par

\begin{figure}
\begin{center}
\includegraphics[width=0.95\swidth]{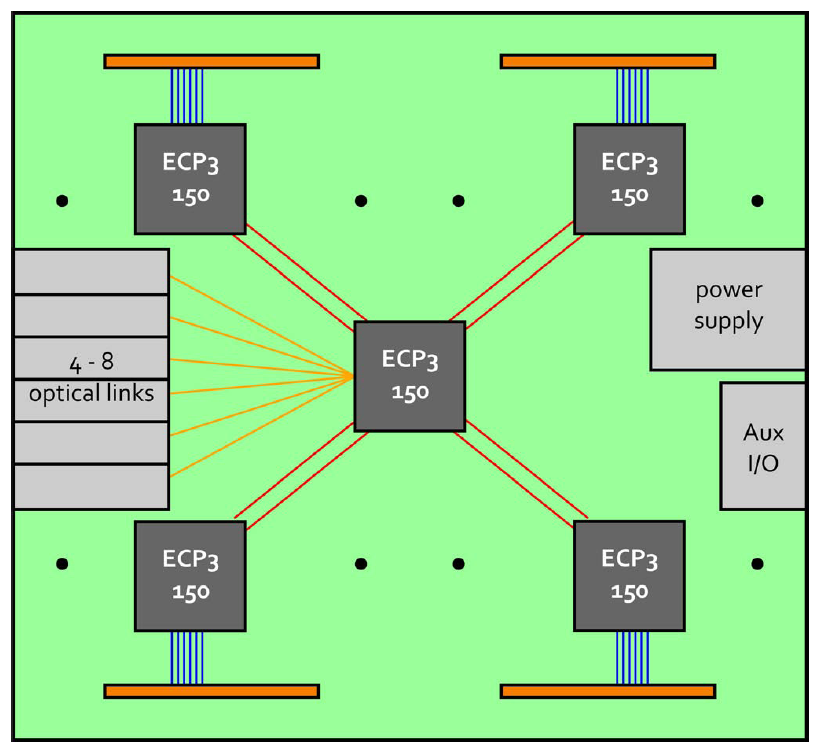}
\caption[Block diagram of the TRBv3]{Block diagram of the TRBv3.}
\label{fig:stt:ele:tdc:TRB_v3}
\end{center}
\end{figure}

\begin{figure}
\begin{center}
\includegraphics[width=0.95\swidth]{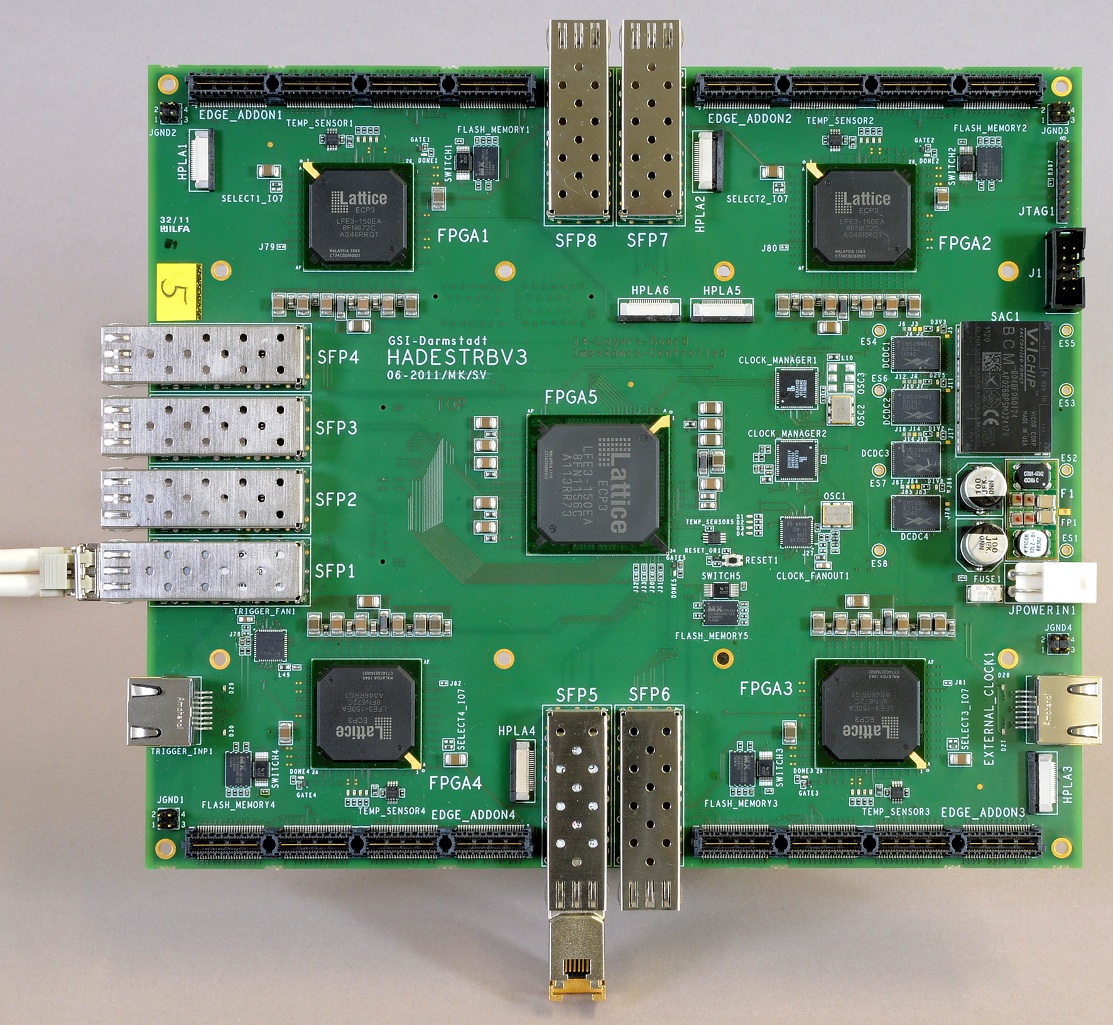}
\caption[Produced TRBv3 board]{Produced TRBv3 board.}
\label{fig:stt:ele:tdc:TRB_v3_PCB}
\end{center}
\end{figure}

The block diagram and a recently produced TRBv3 board are shown in \Reffig{fig:stt:ele:tdc:TRB_v3} and \Reffig{fig:stt:ele:tdc:TRB_v3_PCB}, respectively. Four out of five 
FPGAs on the TRBv3 are located along the edges of the board. Each of them has a 208-pin input/output connector assigned. The fifth FPGA is located in the center of 
the board and coordinates the work of the edge FPGAs as well as communicates with the data acquisition system. The TRBv3 board is equipped with eight optical SFP 
connectors. The maximum transmission speed of each optical connector is 3.2 Gbit/s. The input/output connectors are used to plug in mezzanine cards.
The connector contains 188 general purpose lines and 6 high speed serial connections between edge FPGA and a mezzanine board. Eight lines are connected to the 
central FPGA from each connector. The design of the mezzanine card is left for future TRBv3 users. It may range from a simple
flat cable adapter to a sophisticated board (e.g multichannel ADC). A set of two connectors is placed on the bottom side of the TRBv3 allowing for yet another
mezzanine card connection. All 160 general purpose lines from the bottom connectors are controlled by the central FPGA. Both top and bottom
connectors provide a power and ground for mezzanine cards.
\par
The edge FPGA may contain up to 64 time measurement channels. The DB based on TRBv3 for the STT will have 48 TDC channels in FPGA, giving the total number of 192 channels 
per board. Thus, four TRBv3 boards will be sufficient to collect data from one STT sector.
\par
\begin{figure}
\begin{center}
\includegraphics[width=0.95\swidth]{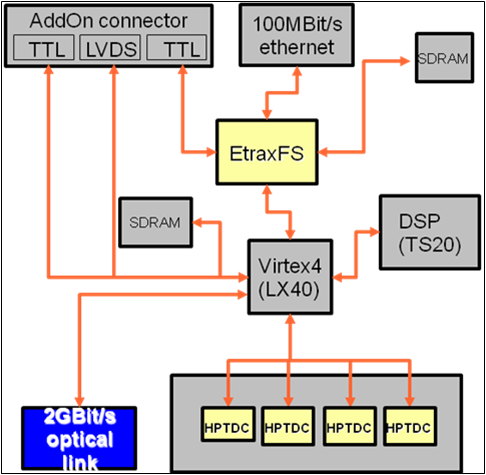}
\caption[Schematics of the HADES TRBv2]{Schematics of the HADES TRBv2 board used
for the time-of-flight measurements.}
\label{fig:stt:ele:tdc:TRB_scheme}
\end{center}
\end{figure}
\begin{figure}
\begin{center}
\includegraphics[width=0.95\swidth]{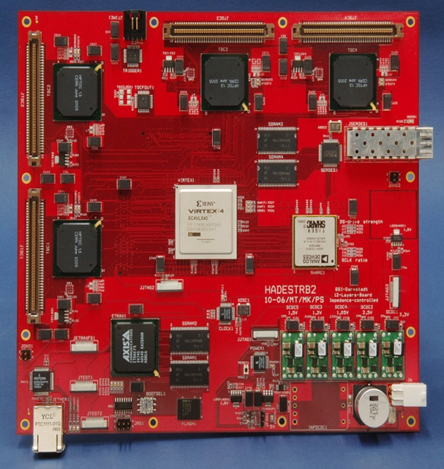}
\caption[HADES TRBv2 board]{HADES TRBv2 board.}
\label{fig:stt:ele:tdc:TRB_v2}
\end{center}
\end{figure}

The former version of the TRB (TRBv2), containing four HPTDC chips is presently used for straw detector tests (as for example shown above). It has been built 
for the HADES
experiment at GSI  \cite{bib:stt:ele:TDC} (schematics and photograph of TRBv2 are shown in
\Reffig{fig:stt:ele:tdc:TRB_scheme} and \Reffig{fig:stt:ele:tdc:TRB_v2}, respectively). The HPTDC chip (32 channels) has been developed at CERN for LHC experiments.
The HPTDC can operate with a maximum trigger rate of 1\,MHz and a maximum of
2\,MHz hit rate per channel. Four TDC binning widths (25, 100, 195 or 785\,ps) can be selected by software during the chip initialization. For the straw tubes, a binning of 
785\,ps has been selected. The measured hit times together with the trigger time stamp are stored in the local TDC memory (up to 256 hits/shared by 8 channels can be 
stored) and read-out from the TDC read-out FIFO (also 256 hits deep) with 40 MHz clock (8 bit parallel bus). Noise suppression and a fast hit detection on single wires 
is performed in FPGA located on the board.
The HPTDC allows also to measure the time over threshold which is used for a noise suppression.
The FPGA controls also the data flow. The data are transmitted from the DB via 8b/10 serial 2.5\,Gbit optical link driven by TLK2501 transceiver from Texas Instruments.
\par
For tests an external trigger (clock) was connected to the TRBv2 by a dedicated line, not shown in the schematics. The slow control is provided by the Etrax FS CPU 
running a LINUX OS and an EPICS client \cite{bib:stt:ele:epics}.

\section{Data Rate}
%\COM{Author(s): P. Salabura, M. Kajetanowicz}
\label{sec:stt:ele:data}

An average maximum hit rate of 800 kHz per channel is expected for the innermost straws when operating at an interaction 
rate of $2 \times 10^7$ proton-antiproton annihilations per second. One DB, with 196 channels, will provide on average 157\,Mhits/s.
The TDC will require 11 bits in order to measure a 1\,$\mu$s range with a 0.5\,ns binning with. The Time-Over-Threshold will require 8 bits to cover a range up to 200\,ns. 
The channel number (1-48) will require 6\,bits and the encoding of the time stamp (1-500) will require 9 bits.
The latter assumes that epoques of 500\,$\mu$s are stored in the DB buffer. Sufficient memory (400 kB) for one epoque can be easily implemented in the FPGA.
\par
Altogether 34 bits represent one TDC channel result in a given TDC on the DB. Two more bits are necessary to distinguish the FPGA. Thus a 5 byte word is
generated for each hit on the DB. Assuming 800 khits per second a data rate of 4 MB/s is generated in each TDC channel. This results in a 784 MB/s data rate from one DB. This data 
rate can be handled by four 3.2 GBit/s optical serial links. Twenty four 196-channel DBs will be necessary for the full CT read-out.
\par

It seems reasonable to merge the data from the straw layers belonging to one STT-sector and sent them to a common DCB.  Such a layout can be more favorable for a cluster 
search but is not mandatory since the DB have also features of the Detector Concentrator (grouping of bursts in epoques).

A board which could be considered as a prototype of the \Panda DCB has also been developed by the HADES DAQ group (optical hub module) and is currently 
installed in the Krakow straw tube test set-up. The prototype is equipped with several Small Form-factor Pluggable Transceivers (SFPT) serving as optical connectors and FPGAs controlling the data transfer. An attractive feature of this unit is the possibility to create groups of links (4 in the present prototype) 
into one protocol standard.  This is provided by the FPGA controlling the data flow ({\it i.e.} Lattice SCM 50 chip) which currently supports 8b/10b and  GBit Ethernet format. The optical links used on this prototype can send data with a maximal speed of 3.8 Gbit/s.

\begin{figure}
\begin{center}
\includegraphics[width=0.95\swidth]{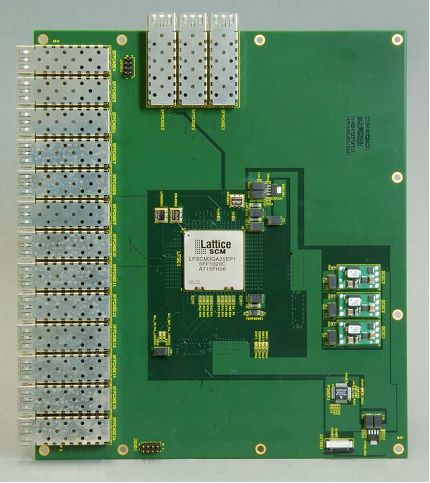}
\caption[HADES optical HUB board]{Optical HUB board for the HADES experiment.}
\label{fig:stt:ele:tdc:Concentrator}
\end{center}
\end{figure}

%
%EOF: panda_tdr_stt_ele.tex

% FILE: panda_tdr_stt_dcs.tex
%
\section{Detector Control System}
%\COM{Author(s): P. Gianotti}
\label{sec:stt:dcs}
The Detector Control System (DCS) of the STT is one branch of the general \PANDA Experiment Control System (ECS).
The control system has to continuously collect the actual parameters from the supply and electronic readout systems
and compare them with their set values and predefined tolerances. In addition, parameters of the detector environment, like for instance temperatures and humidities at several locations must be monitored. 

In case of certain deviations a specific alarm message should be generated to inform the detector operators.
Some of the data must be stored on disk and added to the experiment event data for possible offline corrections of the detector measurements. 
The DCS system will manage the database with the mapping of the physical channel source (voltage, current, temperature, aso.), the label and the relevant calibration constants.

The parameters which have to be considered for the STT are listed below:
\begin{itemize}
\item gas supply system: gas mixture composition, pressure and temperature at several locations in the in- and outlet lines, gas flow;
\item high voltage supply system:  high voltage, current and status (trip, ramp, on, off) values for every supply line, temperature of the supply boards;
\item electronic readout system: discriminator thresholds for every readout channel, supply and reference voltage of every readout board, temperature of the readout boards;
\item detector environment: temperature and humidity at several locations at the detector.
\end{itemize}

\begin{figure}[h]
\begin{center}
\includegraphics[width=\swidth]{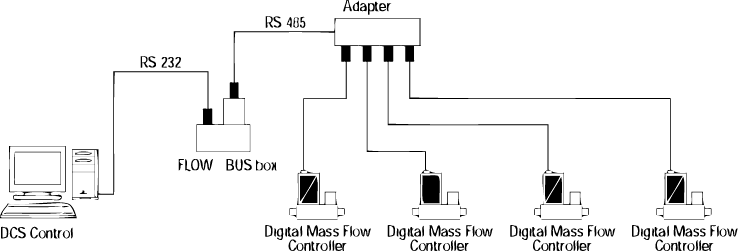}
\caption[Layout of the STT gas control system]{Layout of the STT gas control system.}
\label{fig:stt:dcs:control}
\end{center}
\end{figure}

For the Ar/CO$_2$ gas mixture in the STT the control of the fraction of the two components
will be done within a level of 0.3\,\% (i.e. within 10.0$\pm$0.3\,\%), while the mixture pressure and temperature will be 
kept stable at the level of 1 mbar and 1$^\circ$\,C, respectively. 
This will be done using components, for the gas system, with digital communication
capability controlled by the user via Graphical User Interfaces (e.g. LabVIEW).
An example of the DCS implementation for the gas system is shown in \Reffig{fig:stt:dcs:control}.
For the high and low voltage distribution, multichannel power supply systems like the CAEN SY1527
will be used. This system consists of a main frame allowing the housing of a wide range 
of boards providing high and low voltages.
For the high voltage, the straw tubes will require boards which are able to provide up to 3\,kV with a current 
per channel of about 100\,$\mu$A. The straws will be grouped to sectors, which will be fed in parallel
by a single system channel each. High voltage supplied, current, ramp up and down times, will be 
the parameters to be controlled by the DCS system using a dedicated bus or Ethernet connection.
\par
For the front end electronics a few channels of low voltages can be housed in the same main frame.
This system allows an easy communication with the mainframe through ethernet using an OPC server software which 
can be easily integrated in the user DCS.
\par
For a stable STT operation, remote access and monitoring of the front-end
electronics (FEE) cards is mandatory. Therefore, the DCS will be
connected to the STT FEE via a dedicated bus. The software
will permit adressing, via read-write operations, the status and control
registers of the FEE ASIC chips.
Moreover, through the slow control software interface, various test and/or
calibration pulses will be generated.
%
%EOF: panda_tdr_stt_dcs.tex

%
% Bibliography for this chapter (remove %)
%
\bibliographystyle{panda_tdr_lit}
\bibliography{./stt/lit_stt}
% EOF

%
% STT TDR
% File for chapter 3
\chapter{Calibration Method}
% FILE: panda_tdr_stt_cal.tex
%
%\COM{Author(s): S. Costanza/ P. Wintz}
The calibration of the STT includes a determination of the position in space of the straw tubes and the characteristic relation 
between the measured drift time and the isochrone radius. In principle, both calibrations have to be done for each single straw,
 but the layout and properties of the pressurized tubes simplify the calibration method to a large extend.

%Due to the close-packaging of the glued straws in a layer-module with a precise tube-to-tube distance of 10.1\,mm, clear deviations 
%of single straws larger than about 40\,$\mu$m are not possible. Thus, only the position of the whole module must be determined. As a 
%first input the precise mounting boreholes in the mechanical frame flanges will be used. Then, using reconstructed tracks an iterative 
%recalibration of the positions will be done by correcting systematic shifts. Since the position calibration must be carefully done only 
%once, after the installation of the STT in the \Panda target spectrometer, it is best to perform this task with the \Panda solenoid 
%magnetic field switched off to have straight cosmic tracks.

The calibration of the isochrone radius and drift time relation benefits from the mechanical properties of the pressurized, thin-wall 
(27\,$\mu$m) straw tubes, having a perfect cylindrical shape and precise diameter. The experience from the COSY-STT with 2700 straw tubes, 
also pressurized and with a similar thin film wall (32\,$\mu$m), showed that a global isochrone calibration for all straws together is sufficient. 
The isochrone relation only depends on the specific gas and electric field parameters. The individual time offsets from the electronic readout system
 have to be corrected only once.

In the following, algorithms for the isochrone calibration are described, which can be 
easily adapted for the \Panda-STT. The methods presented were checked with experimental data 
from different straw test systems. 
In particular the COSY-STT detector is considered as an ideal test system for the whole calibration method and 
the performance results can be extrapolated to the \Panda-STT due to the similar technique of close-packed straw layer-modules.
The test systems and measurements are described in detail in \Refsec{sec:stt:pro}.    

%\section{Straw tube calibration}
\label{sec:stt:calib}
\section{Drift Time Spectra}
%\subsection{Fit of TDC spectra}
\label{sec:stt:calibcurve}
\begin{figure}[ht!]
\begin{center}
\includegraphics[width=0.9\swidth]{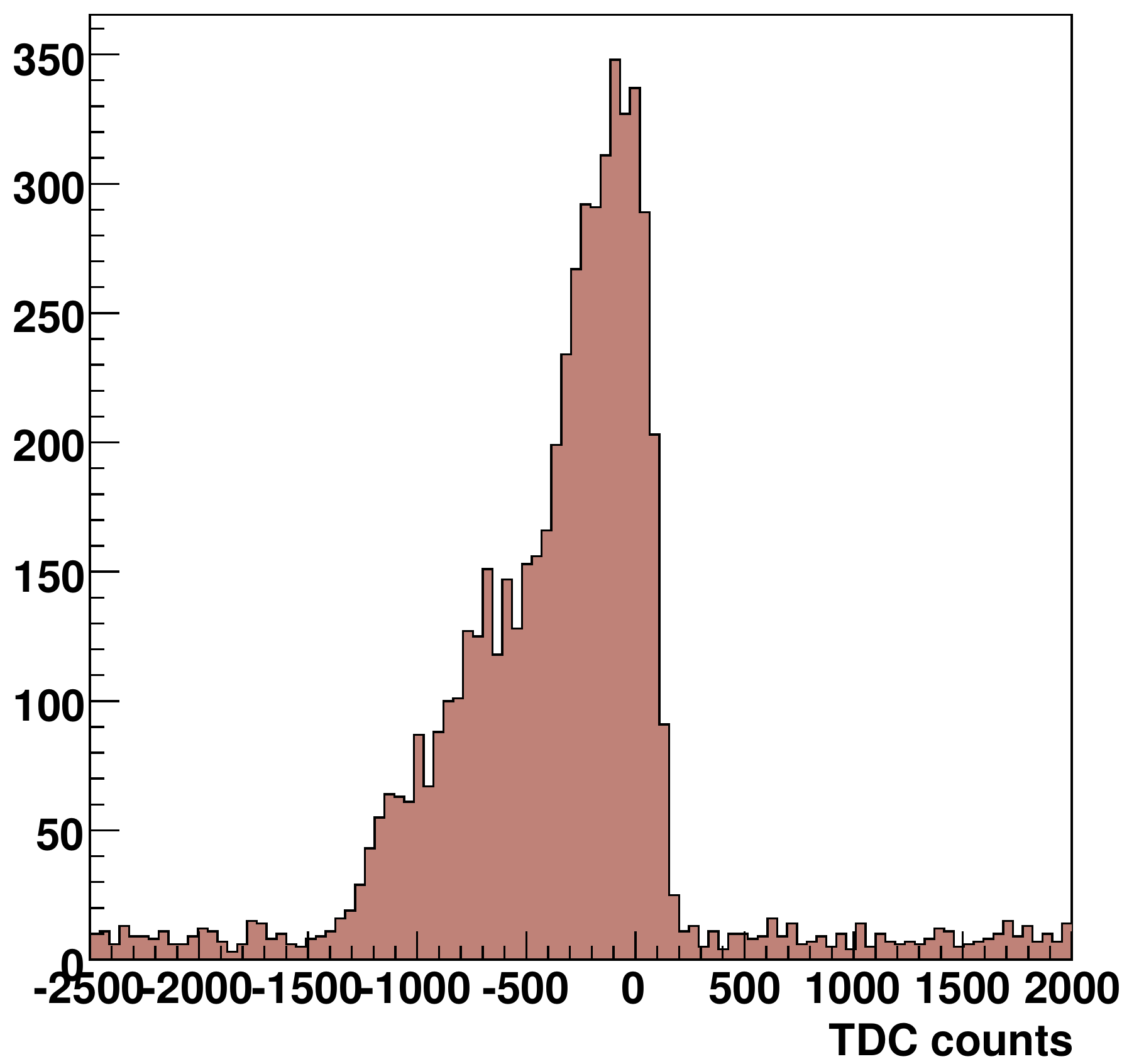}
\caption[Measured raw TDC spectrum]{Example of a measured raw TDC spectrum with some particular noise contribution from 
the electronic readout system. On the $x$ axis, the time is expressed 
in TDC counts (in this case, one TDC channel corresponds to 130 ps) 
and runs from right to left.}
\label{fig:stt:cal:tdcraw}
\end{center}
\end{figure}
\begin{figure}[ht!]
\begin{center}
\includegraphics[width=0.9\swidth]{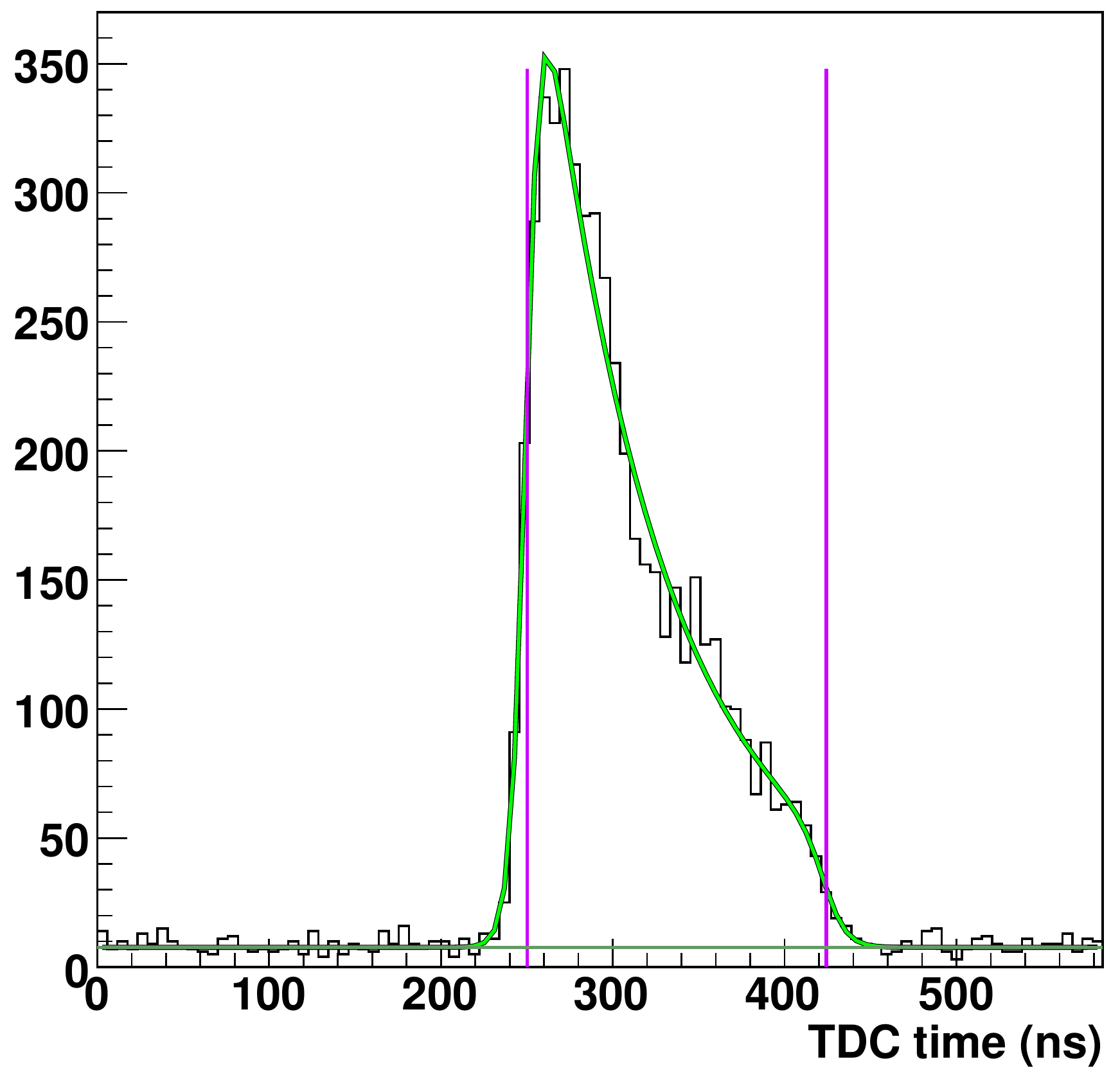}
\caption[Fitted TDC spectrum]{Example of a fitted TDC spectrum. The light green line is the fit of 
the distribution; the violet vertical lines correspond to the $t_0$ and 
$t_{max}$ values determined by the fit. The dark green horizontal line indicates
 the noise level.}
\label{fig:stt:cal:spettro}
\end{center}
\end{figure}
\Reffig{fig:stt:cal:tdcraw} shows an example of a measured time spectrum for a 
uniformly illuminated straw tube, with some particular noise contribution from the electronic readout system.  In this figure, the time is expressed in TDC counts 
and runs from the right to the left. In order to get the time spectrum, the TDC counts are 
converted into seconds and the time is reversed; finally, the spectrum shown in 
\Reffig{fig:stt:cal:spettro} is obtained.
The analysis of the time distributions of individual straws allows the monitoring of
 the data quality: the minimum and the maximum drift times, $t_0$ and $t_{max}$,
 correspond to a track traversing the tube close to the wire and close to the cathode
 wall, respectively. The value of $t_0$ depends on the signal cable length, discriminator 
threshold, high voltage setting and delays in the 
readout electronics.
 Nearby tubes sharing the same front--end electronics are expected to have a
similar value of $t_0$; on the contrary, the drift time $\Delta t=t_{max}-t_0$ 
depends only on the drift properties of the tubes. The number of events outside 
the drift time window gives an estimate of the random, constant noise level over
 time range (see \Reffig{fig:stt:cal:tdcraw}) \cite{bib:stt:cal:atlas2}.
For each tube, the parameters of the drift time distribution are derived from a 
fit performed with the following empirical function \cite{bib:stt:cal:atlas2,bib:stt:cal:atlas1,bib:stt:cal:atlas3}:
%\begin{align}\nonumber
%\frac{\mathrm{d} n}{\mathrm{d} t} =  P_1 + \\ \nonumber
% \frac{P_2\left[1+P_3\exp((P_5-t)/P_4)\right]}{\left[1+
%\exp((P_5-t)/P_7)\right]\left[1+\exp((t-P_6)/P_8)\right]}\\
%\label{eq:stt:cal:fitsingle}
%\end{align}
\begin{equation}
%\frac{\mathrm{d} n}{\mathrm{d} t} = 
\begin{array}{l l}
\frac{\mathrm{d} n}{\mathrm{d} t} = P_1 + & 
\frac{P_2\left[1+P_3\exp((P_5-t)/P_4)\right]}{\left[1+\exp((P_5-t)/P_7)\right]\left[1+\exp((t-P_6)/P_8)\right]}\\
\end{array}
\label{eq:stt:cal:fitsingle}
\end{equation}
where $P_1$ is the noise level, $P_2$ is a normalisation factor, $P_3$ and $P_4$
 are related to the shape of the distribution, $P_5$ and $P_6$ are the values of
 $t_0$ and $t_{max}$. $P_7$ and $P_8$ describe the slope of the leading and 
trailing edge of the distribution, so they are indicators of the drift tube 
resolution close to the wire and to the tube wall, respectively. The fit result of  
\Reffig{fig:stt:cal:spettro} shows as an example the fit of the function (green line) to 
a measured TDC spectrum.
In order to do a common calibration for all the tubes, their time spectra must 
have approximately the same shape and the same maximum drift time $\Delta t$. A quality
check on the uniformity of the tubes, as well as on the quality of the fit, can
be done by looking at the distributions of the fit parameters.
\begin{figure}%[ht!]
\begin{center}
\includegraphics[width=0.9\swidth]{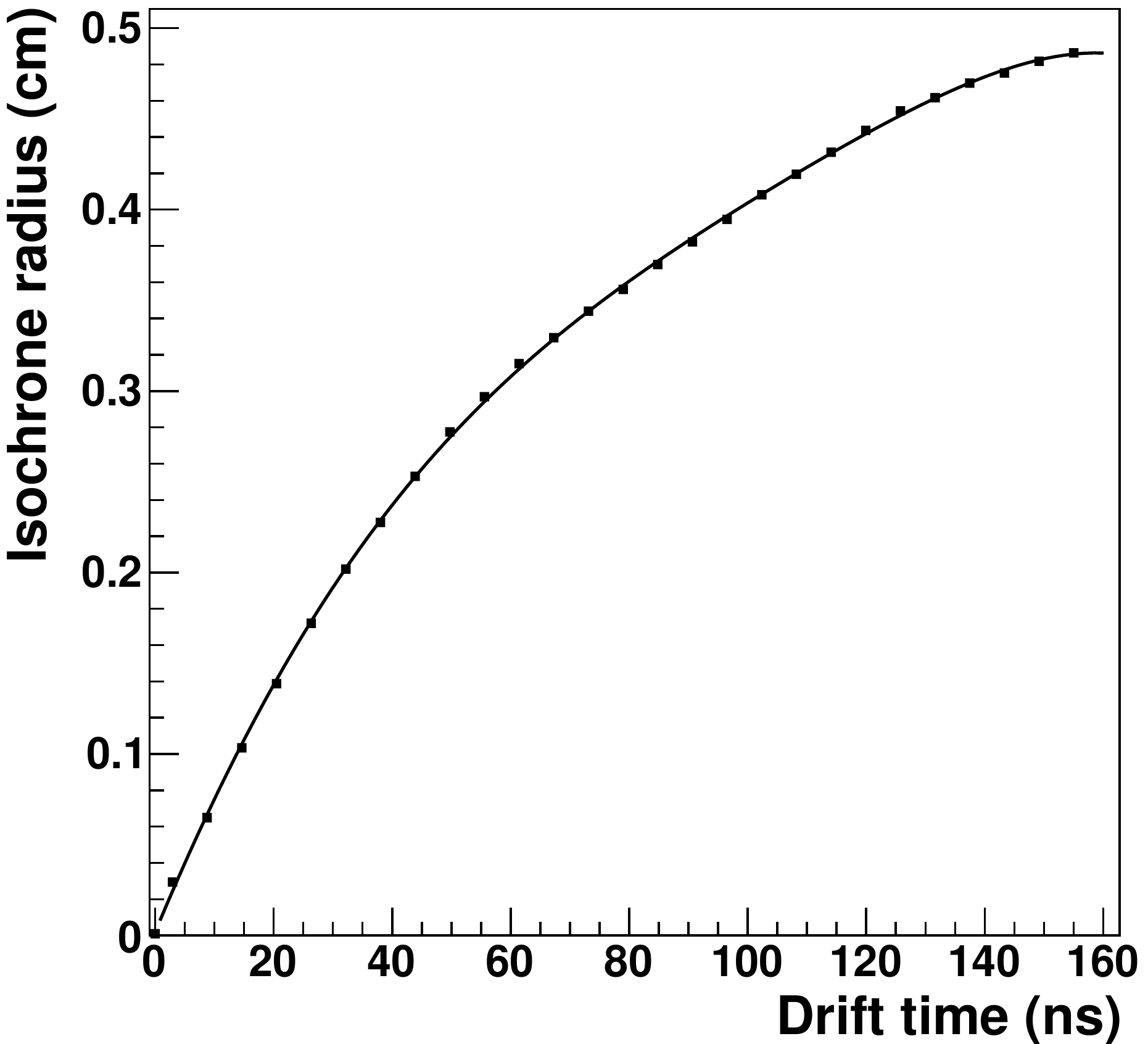}
\caption[Isochrones radius -- drift time relation]{Isochrones radius -- drift time relation ($r(t)$), parametrised using 
a combination of Chebyshev polynomials of the first kind, up to the fifth order.}
\label{fig:stt:cal:calibcurve}
\end{center}
\end{figure}

\section{{$r(t)$} Calibration Curve}
\label{sec:stt:rtcurve}
\begin{figure}
\begin{center}
\includegraphics[width=0.9\swidth, height=0.9\swidth]{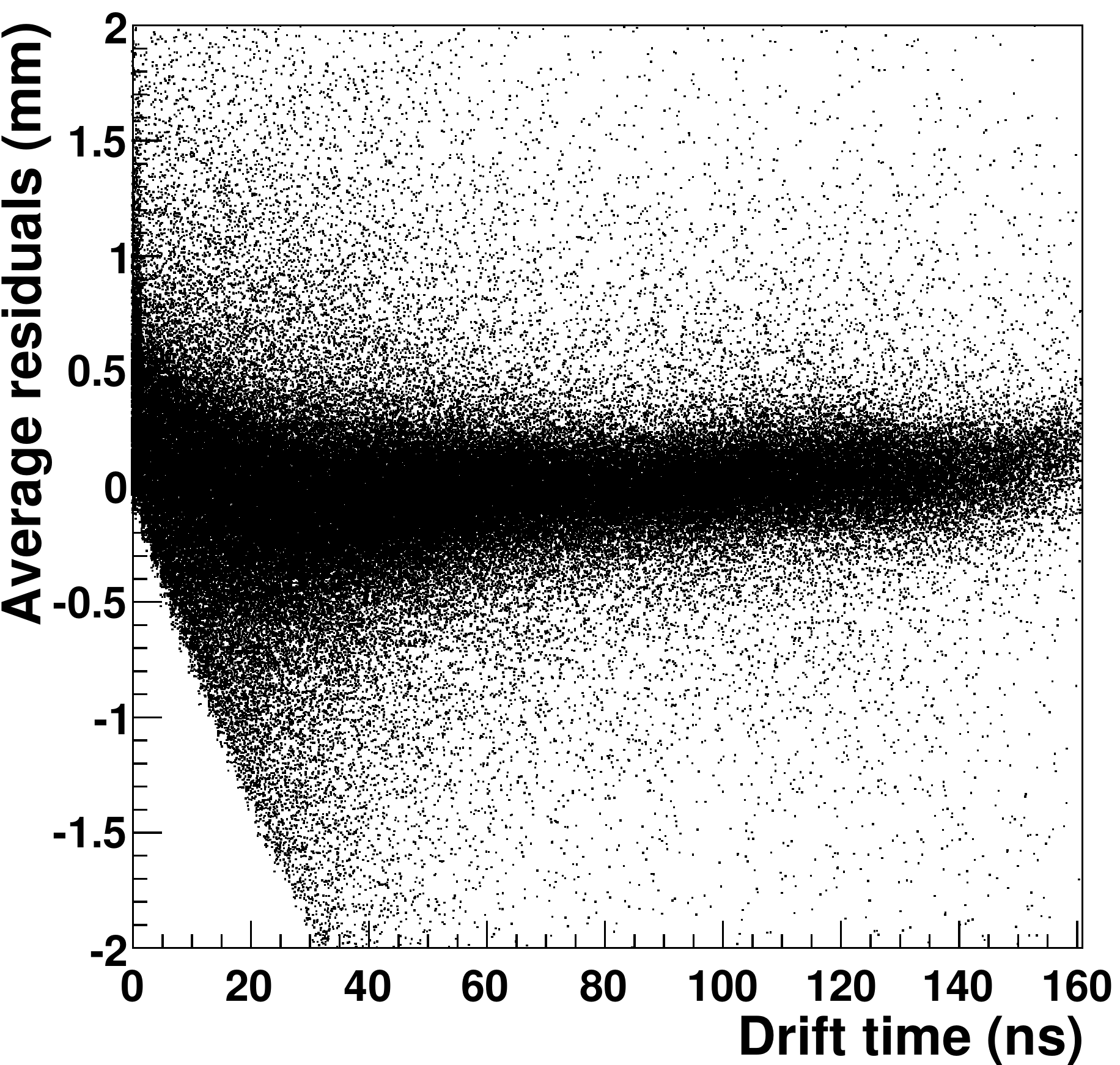}
\caption[Distribution of the average residuals as a function of the drift time]{Distribution of the average residuals as a function of the drift time at the first iteration.}
\label{fig:stt:cal:meanres1}
\end{center}
\end{figure}
%\noindent
After the selection of the similar spectra, their specific time offset $t_0$ is
corrected and their noise level is subtracted; then, they are added into a sum
spectrum, each in its $\Delta t$ range.
Under the hypothesis of a uniform illumination of the tube and a constant 
efficiency over the tube volume, the isochrone radius -- drift time relation 
($r(t)$ relation in the following) can be obtained by the following 
integration:
\begin{equation}
r(t)=\frac{R_{\rm tube}}{N_{\rm tot}}\int_0^t\frac{\mathrm{d}n}{\mathrm{d}t'}\mathrm{d}t'
\label{eq:stt:cal:scal2}
\end{equation}
where $n$ is the number of tracks, $N_{\rm tot}$ is the total number of tracks 
and $R_{\rm tube}$ the tube radius.\\
Taking into account the finite TDC resolution (bin size) and the wire radius 
$R_{wire}$, \Refeq{eq:stt:cal:scal2} becomes:
\begin{equation}
r(t_i)=\frac{\sum_{i=1}^{i_t}N_i}{N_{tot}}\cdot\left( R_{tube}-R_{wire}\right)+R_{wire}
\end{equation}
$R_{wire}$ is the wire radius and $N_{tot}$ is the sum of all bin entries 
$N_i$. 
The obtained space--time relation is then parametrised as a polynomial 
function. 

An example of the $r(t)$ curve is shown in \Reffig{fig:stt:cal:calibcurve}; 
in this case, the space--time relation has been parametrised with a combination
 of Chebyshev polynomials of the first kind up to the fifth order:\footnote{The Chebyshev polynomials of the first kind are defined by the recurrence
relation:
\begin{eqnarray} \nonumber
T_0(x) &=& 1, \\ \nonumber
T_1(x) &=& x, \\ \nonumber
T_{n+1}(x) &=& 2xT_n(x)-T_{n-1}(x).
\end{eqnarray}}
\begin{eqnarray}\nonumber
r(t)&=&p_0+p_1t+p_2(2t^2-1)+p_3(4t^3-3t)\\\nonumber
&+&p_4(8t^4-8t^2+1)+p_5(16t^5-20t^3+5t)\\
\label{eq:stt:cal:fitcalib}
\end{eqnarray}
Once the space--time relation is known, the isochrone radius of a certain tube 
is computed by substituting in \Refeq{eq:stt:cal:fitcalib} the measured drift 
time. This is calculated by subtracting from the measured drift ``raw'' time the
 time offset $t_0$ of that tube, obtained from the fit of 
\Refeq{eq:stt:cal:fitsingle}.
%=======================
\begin{figure}[t]
%[ht!]
\begin{center}
\includegraphics[width=0.5\swidth, height=0.5\swidth]{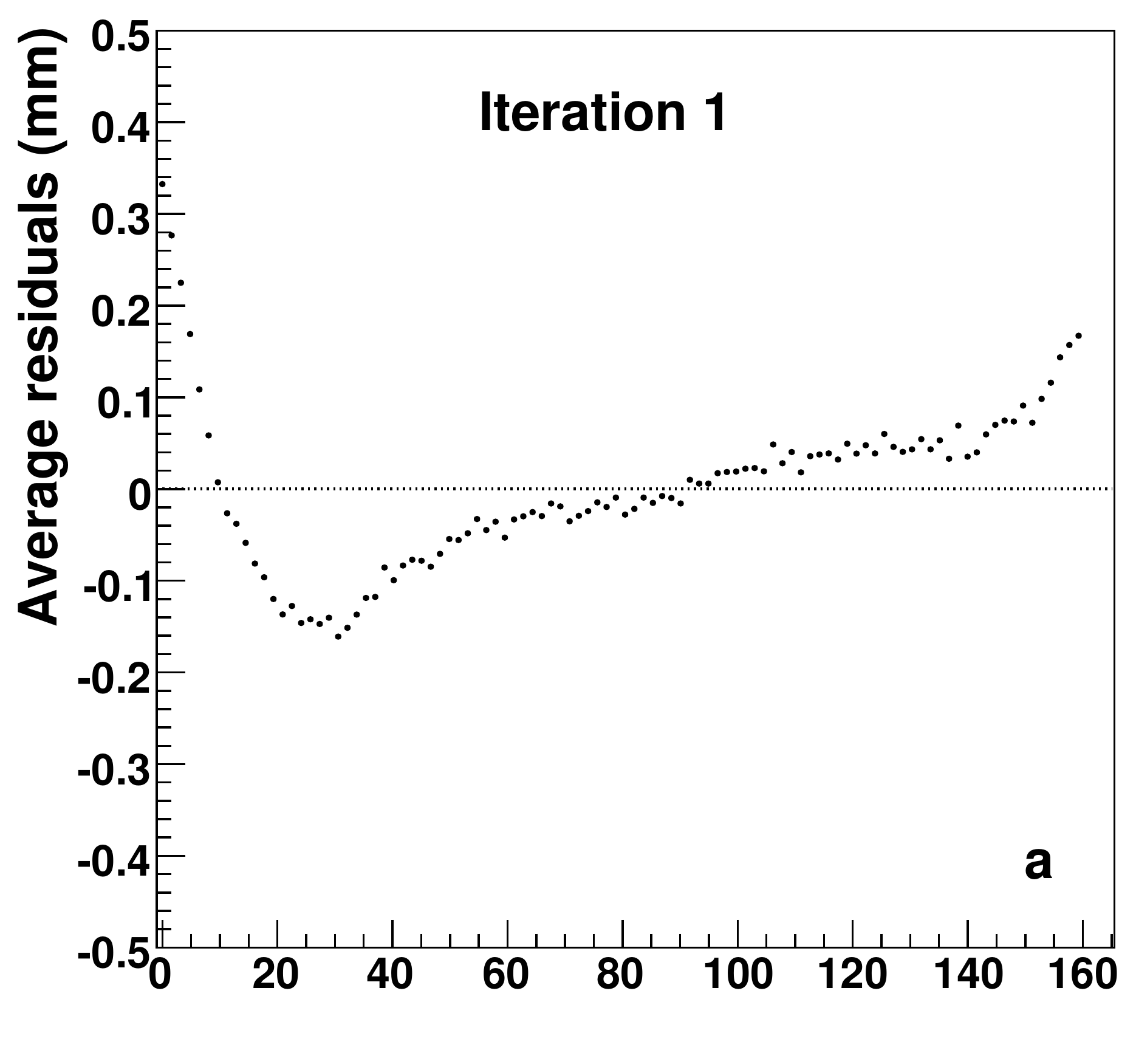}
\includegraphics[width=0.5\swidth, height=0.5\swidth]{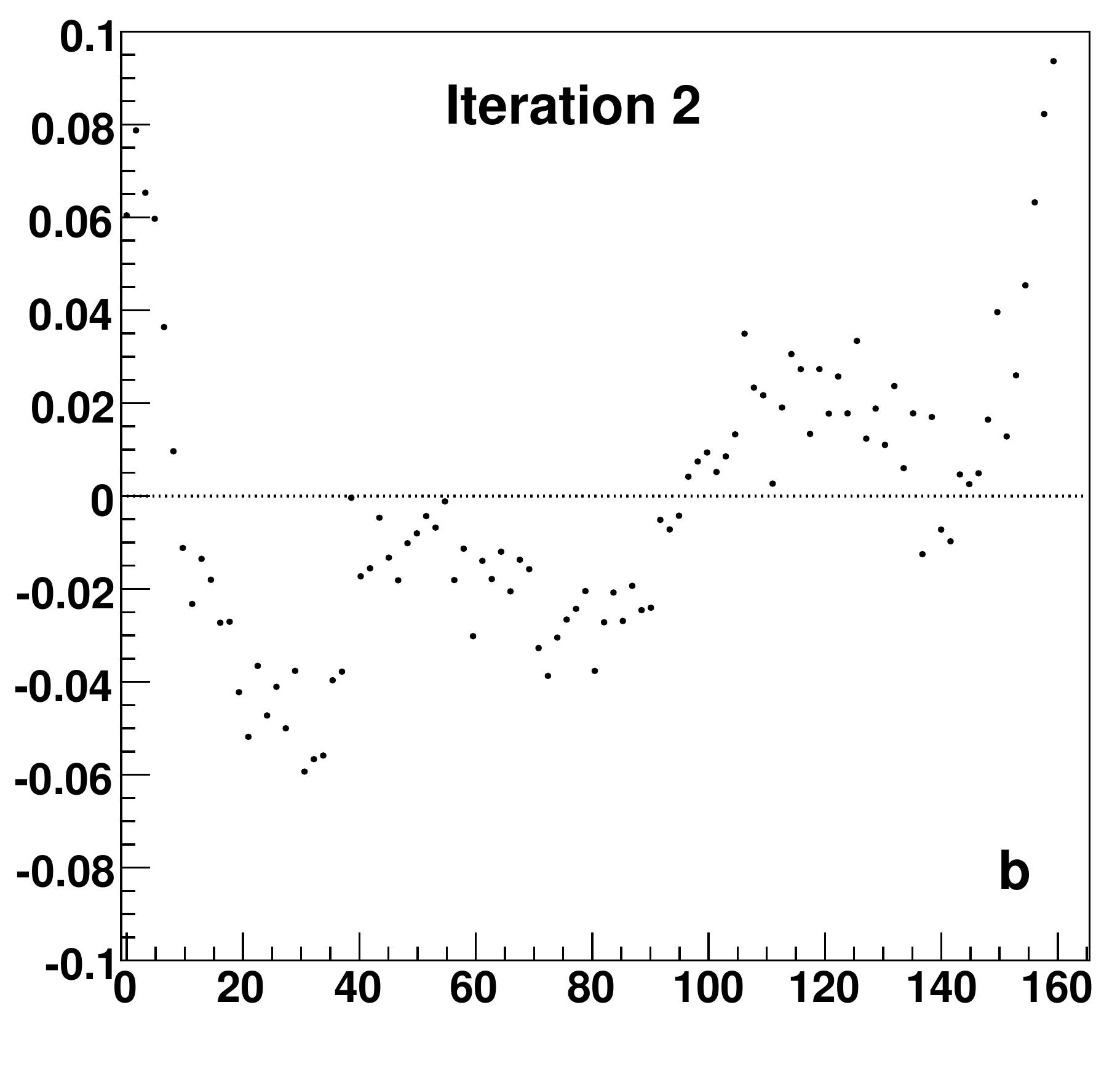}
\includegraphics[width=0.5\swidth, height=0.5\swidth]{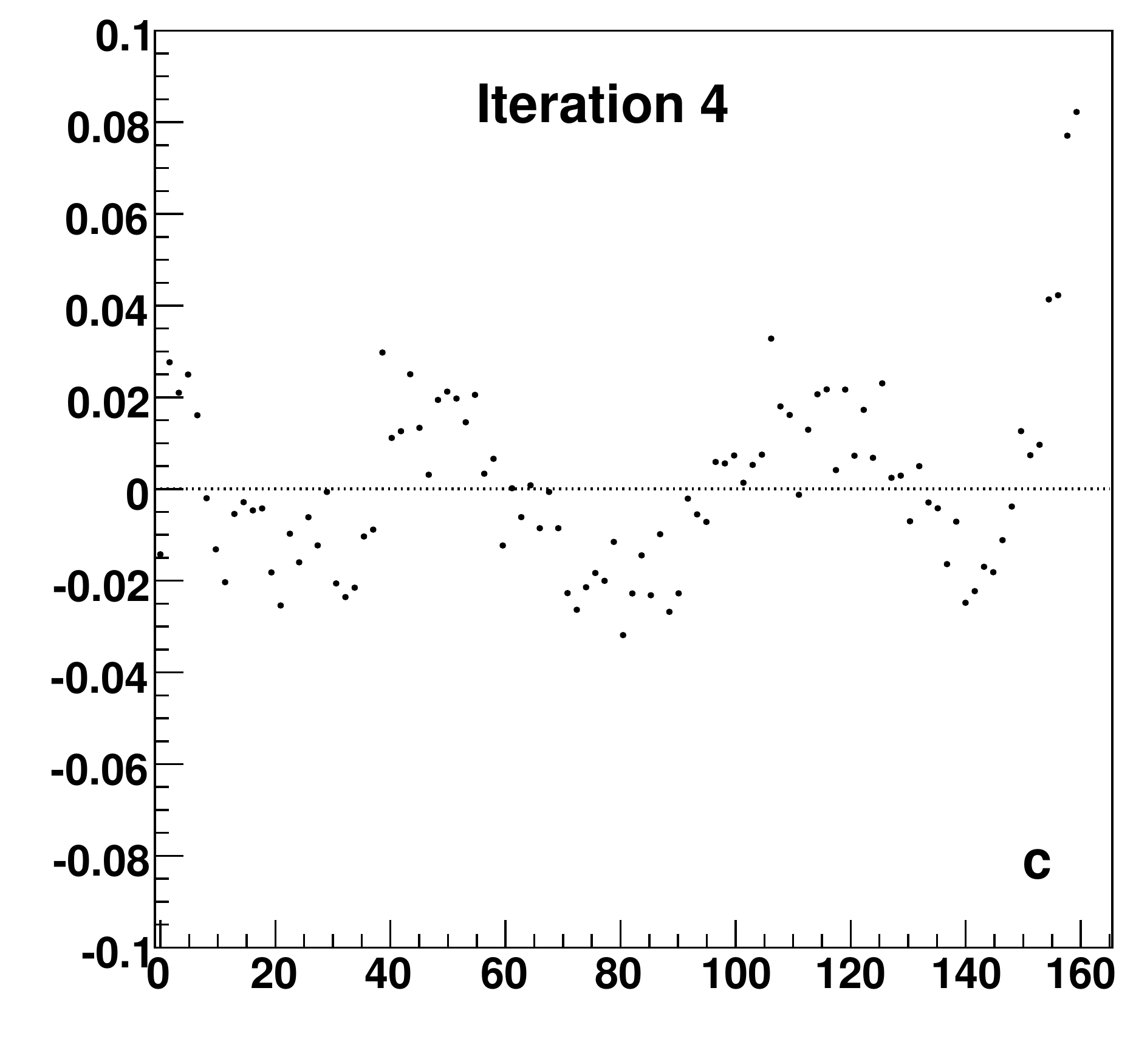}
\includegraphics[width=0.5\swidth, height=0.5\swidth]{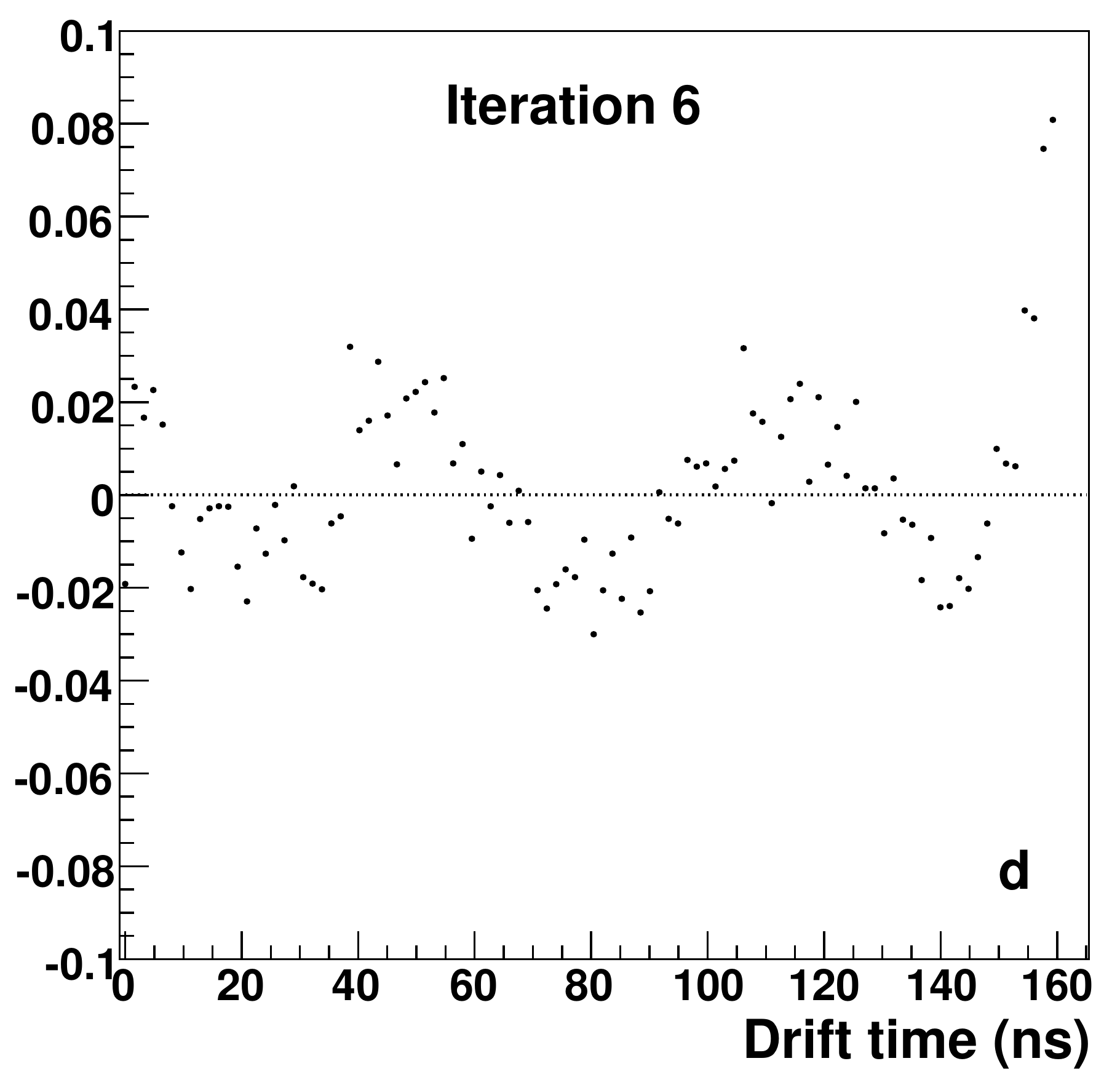}
\caption[Distribution of the average residuals after autocalibration]{Distribution of the average residuals as a function of the drift time after one (a), two (b), four (c) and six (d) iterations of the autocalibration procedure.}
\label{fig:stt:cal:meanres6}
\end{center}
\end{figure}
\section{Autocalibration}
\label{sec:stt:autocalib}
Once the calibration curve has been derived, it is possible to proceed with 
the track reconstruction.
In order to perform a good track fitting, it is necessary to know with high 
precision the relation between the measured drift time and the distance of 
closest approach of the particle trajectory to the wire. This implies an 
accurate knowledge of the $r(t)$ relation, that can be achieved with an 
iterative procedure called {\bf autocalibration}, since it makes use of just the
information from the tubes under investigation.
\par
The autocalibration works as follows: at each step of the procedure, the $r(t)$ 
relation derived in the previous iteration is used to convert the measured drift
 times into drift radii, which will be used in the track fitting. At the first 
step, the $r(t)$ relation obtained directly from the integration of the drift 
time spectra (\Refsec{sec:stt:rtcurve}) is used. 
Once a track candidate has been identified through dedicated pattern recognition
algorithms, the track is reconstructed by using a track fitting algorithm. It
allows to extract the $(x,y)$ hit coordinates from the drift times and the 
$(x,y)$ coordinates of the firing wires, which are the observables measured by
the straw tubes.
The fitting algorithms implemented for the \panda--STT will be described in 
detail in \Refsec{sec:stt:sim}; the tracking procedure used for the test systems
will be briefly presented in \Refsec{sec:stt:pro}. 
In this last case, in general tracks are reconstructed as straight lines 
$y=a+bx$, where the parameters $a$ and $b$ are obtained by a least squares fit
($\chi^2$) that minimizes the track residuals, i.e.~the difference of the 
distances of closest approach of the best fit lines to the centers of the 
firing tubes and the corresponding isochrones calculated from the measured
drift times using the $r(t)$ relation.
For each tube of the pattern associated with a track, the residuals are then 
computed and represented as a function of the $N$ bins the drift time interval 
is divided into. 
If the $r(t)$ relation was exact, the average residuals would be zero at all 
radii.
\par
An example of a residual distribution is shown in \Reffig{fig:stt:cal:meanres1}: at this 
step, the mean value of the residuals varies from a minimum of $\sim$ -160 $\mum$ to a 
maximum of $\sim$ 320 $\mum$ for small radii. 
These deviations from zero indicate a miscalibration in the $r(t)$ relation, which 
is then corrected by taking the average deviation.
The track reconstruction and the $r(t)$ calibration with the residuals as input 
are then repeated until the corrections become negligible and the
mean value of the residuals is close to 0, as 
shown in \Reffig{fig:stt:cal:meanres6}.
\begin{figure}
%[ht!]
\begin{center}
\includegraphics[width=0.95\swidth]{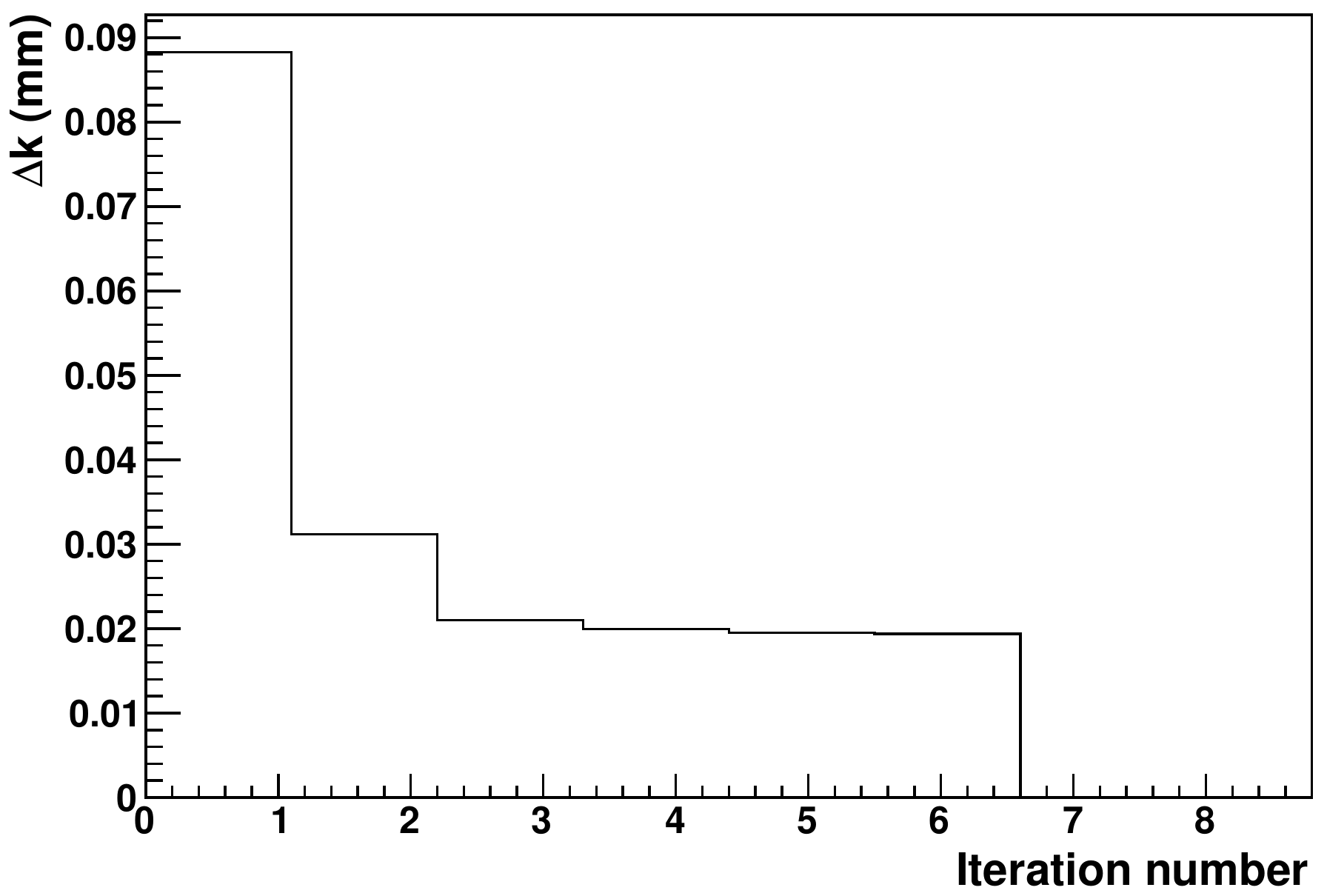}
\caption[Root mean square correction from the autocalibration procedure]{Root mean square correction from the autocalibration procedure.}
\label{fig:stt:cal:msc}
\end{center}
\end{figure}

To study the speed and stability of the convergence of the method, the following quantity
 (mean square correction):
\begin{equation}
\Delta^2_k=\frac{\sum_{i=0}^N\delta_{ik}^2}{N}
\end{equation}
where $\delta_{ik}$ is the mean value of the residuals in the $i^{th}$ time bin 
and $N$ is the total number of bins, can be used as figure of merit (\Reffig{fig:stt:cal:msc}).

The recalibration procedure is iterated until the mean square correction  has converged 
to a stable solution \cite{bib:stt:cal:susannathesis}. 

%
%EOF: panda_tdr_stt_cal.tex

%
% Bibliography for this chapter (remove %)
%
\bibliographystyle{panda_tdr_lit}
\bibliography{./stt/lit_stt}
% EOF

%
% STT TDR
% File for chapter 4
\chapter{Prototype Tests}
\label{sec:stt:pro}
% FILE: panda_tdr_stt_pro.tex
%
%\section{Small--scale setup}
%\label{sec:stt:setup}
%\COM{Author(s): S. Costanza}
\section{Test Systems}
\label{sec:stt:setup}
\begin{figure}[h]
\begin{center}
%\vspace {-3cm}
\includegraphics[width=\swidth]{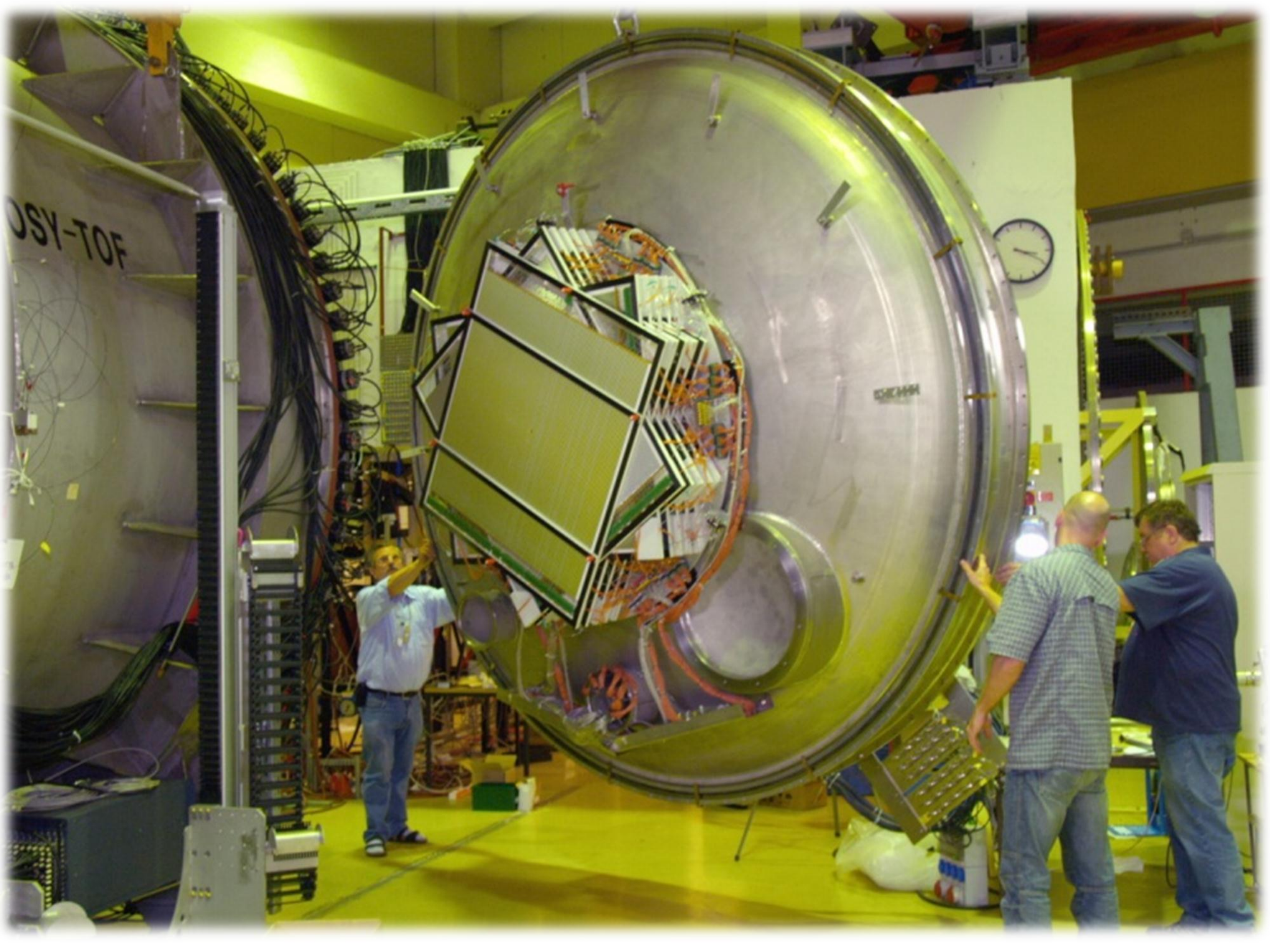}
\caption[Mounting of the COSY-STT in the time-of-flight barrel of the COSY-TOF spectrometer]{Mounting of the COSY-STT in the time-of-flight barrel of the COSY-TOF spectrometer.}
\label{fig:stt:pro:cosy_stt_mount}
\end{center}
\end{figure}

%\COM{Author(s): P. Wintz}
The experience from different straw test systems and dedicated test measurements
has been taken into account in the design of the \Panda-STT. 
The Straw Tube Tracker (COSY-STT) of the COSY-TOF experiment is considered to be 
a global test system, due to the similar mechanical layout of close-packed, self-supporting 
straw modules and its operation in the 
experimental environment of proton-proton collisions with a proton beam momentum around 3\,GeV/c.
COSY-TOF is a non-magnetic spectrometer (see \Reffig{fig:stt:pro:cosy_stt_mount}) and 
the tracks are reconstructed as straight lines
instead of the helical trajectories in the solenoid magnetic field at \Panda. 
Nevertheless, 
the calibration method of the COSY-STT, the obtained spatial resolution and the mechanical precision 
of the detector, which consists of 2700 straw tubes, are of interest for the \Panda-STT design and expected performance.
The COSY-STT is operated in the large (25\,m$^3$) evacuated time-of-flight barrel of the spectrometer since 
about 3 years. The surrounding vacuum is a strong test of all straw materials and assembly techniques, which are
the same or similar to the \Panda-STT. \Refsec{sec:stt:pro:cosystt} describes the properties and results of the
COSY-STT.

A dedicated straw system consisting of a close-packed double-layer of 32 tubes was used for specific aging tests. The setup was installed behind the COSY-TOF apparatus and exposed to the residual proton beam with a momentum of about 3\,GeV/c during about 2 weeks. Except of their shorter length of 1\,m the straw design and materials are similar to the \Panda type straws. 
The aging test and obtained results are summarized in \Refsec{sec:stt:pro:aging}. 

\begin{figure}[ht]
\begin{center}
%\vspace {-3cm}
\includegraphics[width=\swidth]{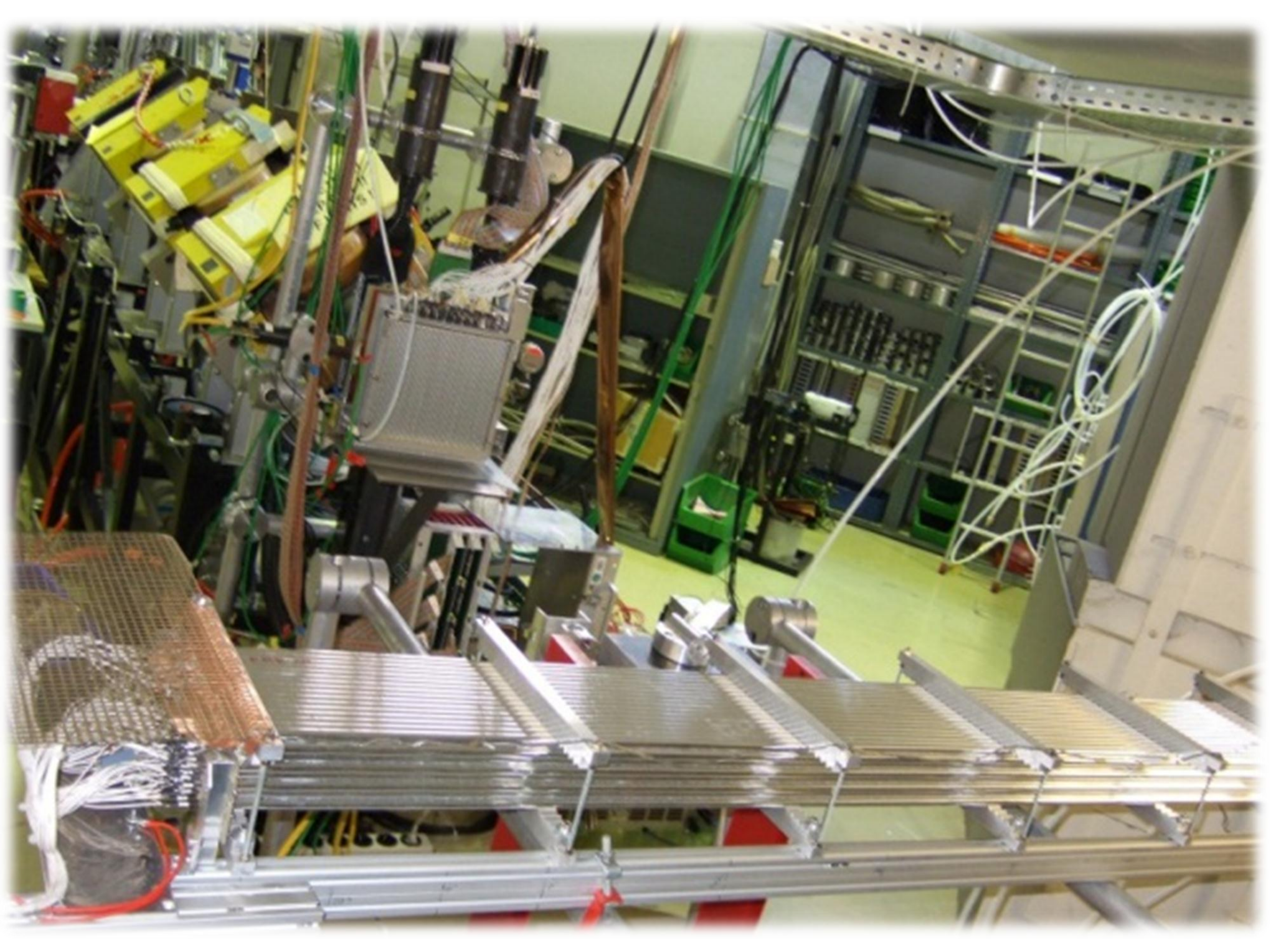}
\caption[Straw setup for the energy-loss test measurements]{Straw setup (in front) for the energy-loss test measurements. The proton beam is coming from the back.}
\label{fig:stt:pro:stt_prot_bigkarl}
\end{center}
\end{figure}

The measurement of the energy-loss of a charged particle with high resolution is a non-standard task for a straw detector. It requires a novel electronic readout design, as well as a dedicated layout of the straw tubes in the \Panda-STT. A specific test system consisting of 128 straws, arranged in 8 close-packed layers, has been set up and measurements with high intensity proton beams have been carried out (see \Reffig{fig:stt:pro:stt_prot_bigkarl}). The next section describes in detail the tests and developments concerning the energy-loss measurement for the \Panda-STT.

\section{Energy-loss Measurements}
%\subsection{energy-loss measurements}
%\COM{Author(s): K.Pysz}
\label{sec:stt:pro:eloss}
The STT has to measure, in addition to the trajectory reconstruction with high resolution, the specific energy-loss of the 
charged particle for an identification of the particle species. In particular, an efficient separation of pions, kaons and protons 
in the momentum region below 1\,GeV/c is needed. 
\Refsec{sec:stt:sim} shows the results from a simulation analysis of the energy-loss measurement with the STT and the expected separation power of the particle species.   

Charged particle identification with drift chambers of comparable size and channel numbers to the \Panda--\Stt has been performed in experiments like BABAR  \cite{bib:stt:pro:babar}, BESIII \cite{bib:stt:pro:bes3} and HADES \cite{bib:stt:pro:hades1} and 
are very encouraging. In all gaseous detectors with segmented readout, there are
factors which limit the precision of the energy-loss measurements:
\begin{itemize}
\item the statistical nature of the ionization process that results in an extended and asymmetric
  energy-loss distribution (Landau-curve shape);    
\item the limited numbers of sampling points used for the energy-loss estimation;
\item the long range of the drift time;
\item the compromises necessary on the readout electronics, which should provide, at the same time, a 
  ``fast'' time-information for an efficient tracking, and which have to integrate for 
  a sufficiently ``extended'' time interval the charge necessary for energy-loss measurements. 
\end{itemize}

%KP: what is below:
%In order to evaluate the detector energy resolution we had to study, for the \Panda-STT, the 
%above mentioned factors. This was the aim and the strategy followed in the studies performed on the 
%prototypes:
%\begin{itemize}
%\item to fix the requirements for the readout electronics;
%\item to select and optimize the method of signal processing and data treatment;
%\item to optimize the detector performance, the electronics coupling, and the noise
%  suppression.
%\end{itemize}
%KP: might be replaced by:
The experimental investigations to evaluate the achievable energy
resolution of the \Panda--\Stt are described in the following sections and aim at:
\begin{itemize}
\item fixing of the requirements for the readout electronics;
\item selecting and optimizing the method of signal processing and data treatment;
\item optimizing the detector performance, the electronics coupling and the noise suppression.
\end{itemize}

\subsection{Experimental Setup}
The test setup consists of 128 \panda--type straw tubes which are arranged 
in four double--layers of 32 straws each (see \Reffig{fig:stt:pro:stt_prot_bigkarl}). 
The tubes, with an aluminized Mylar wall of 30 $\mum$ thickness, are 150 cm long and 
have an inner diameter of 10\,mm. Several scintillators are placed in front and after the straw setup in the 
proton beam and are used to trigger on a coincident event and start the data-acquisition.

The electronic readout of the straw signals consists of front-end transresistance amplifiers with about 8\,ns rise time
and a gain factor of about 
% 20,
360, 
and flash--analog--to--digital converters (FADC) which sample the analog signal amplitude with a frequency of 240\,MHz. 
FPGAs (Field Programmable Gate Array) controlling the readout of an FADC module are 
programmed for high flexibility to permit also the total readout in the ``oscilloscope mode'' and to 
record single spectra in a self-triggering mode for calibration measurements with an $^{55}$Fe $\beta^+$ 
source.
% KP !!! what is below has to be removed:
%In view of the limited number of oscilloscope channels several amplifiers were fed via 
%analog OR into the oscilloscope and then the data were transferred to a computer.
%The actual number of firing straws in each event can be deduced from a multiplicity signal delivered 
%by a 8-channel discriminator \cite{bib:stt:cal:valeryi}.
% KP !!!
%\begin{figure}[ht!]
%\begin{center}
%\includegraphics[width=0.7\swidth]{./stt/fig/prototype_detail.png}
%\caption{Detail of the prototype shown in Fig.~\ref{fig:prototype}.}
%\label{fig:prot_detail}
%\end{center}
%\end{figure}

\begin{figure}[h]
\begin{center}
\includegraphics[width=\swidth,height=0.4\dwidth]{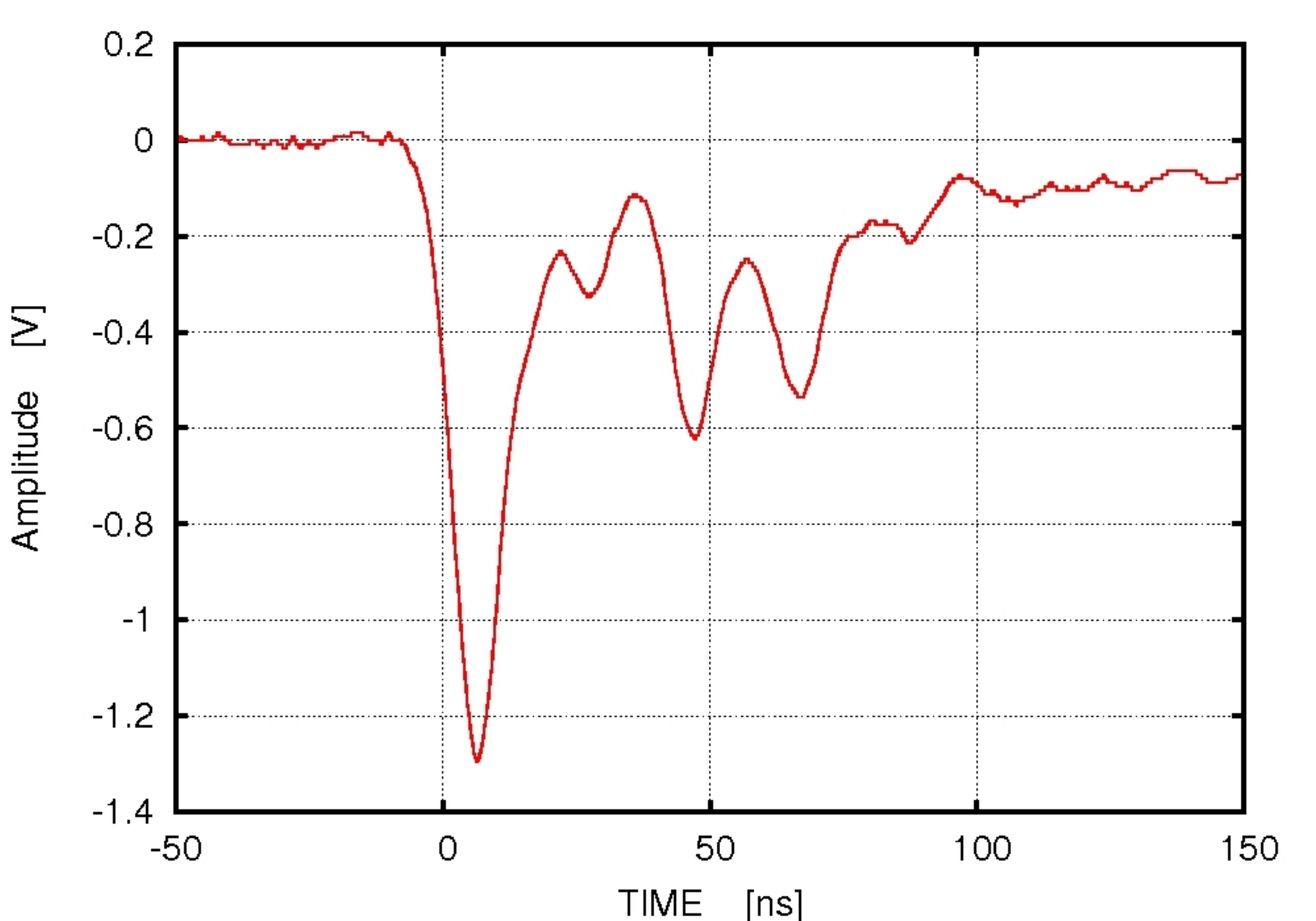}
\caption[Example of an analog output signal of the amplifier from a straw irradiated by a $^{90}$Sr $\beta$-source]{Example of an analog output signal of the amplifier from a straw irradiated by a $^{90}$Sr $\beta$-source.} 
\label{fig:stt:pro:signal1}
\end{center}
\end{figure}

In order to test different analysis methods of the signal shape, 
the analog output signals were recorded as a sampled waveform.
%The front-end transresistance amplifier of 8 $ns$ rise-time and
%gain factor of 10 has been followed by a booster amplifier (gain factor 2)
%translating the differential outputs of transresistance amplifier into
%single-ended (LEMO) signals.  
%The total integrating time-constant of both amplifiers was about 7 $ns$. 
\Reffig{fig:stt:pro:signal1} shows an example of a straw tube signal from a $^{90}$Y/$^{90}$Sr
$\beta$-particle processed by the transresistance amplifier. 
Due to the fast risetime of the amplifier several ionization clusters are resolved as distinct peaks in the 
output signal.  
\noindent

\begin{figure}[h]
\begin{center}
\includegraphics[width=\swidth,height=0.4\dwidth]{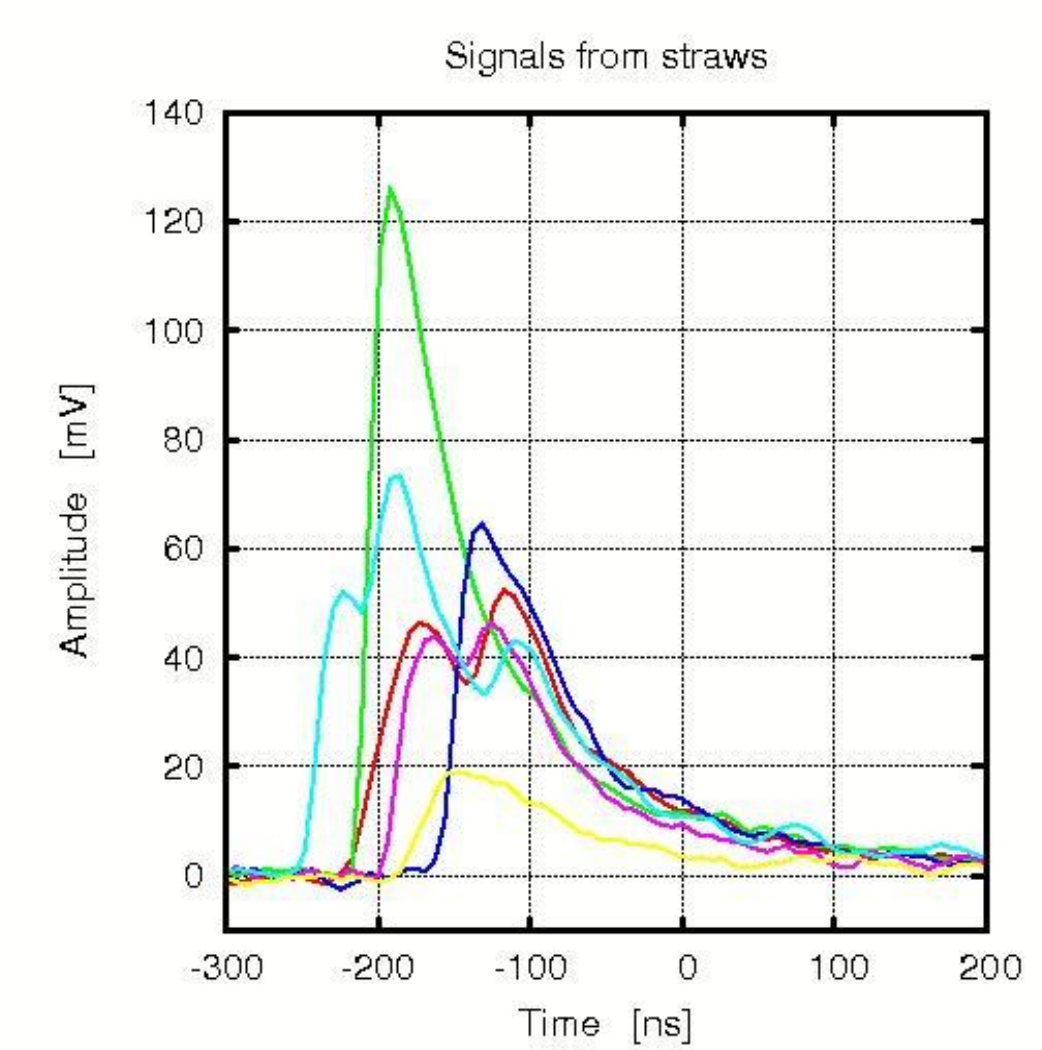}
\caption[Analog signals from the straw tubes as recorded by the 240 MHz FADC]{Analog signals from the straw tubes as recorded by the 240 MHz FADC.}
\label{fig:stt:pro:signal2}
\end{center}
\end{figure}

The amplified signals are fed into the FADCs with a sampling time interval of 4.17\,ns (240\,MHz). An example of the
recorded straw signals is shown in \Reffig{fig:stt:pro:signal2}. 
As can be seen the limited precision of the sampling and the additional integration 
deteriorates the shape of the initial signals to some extent. Nevertheless, the envelopes of the groups
of clusters are still visible. Since the straws are operated in proportional mode
and the response of the electronics is almost linear, the area of the signal 
is directly proportional to the primary ionization in the straw tube, and therefore 
proportional to the energy-loss of the traversing particle. 
\noindent

\begin{figure}[h]
\begin{center}
\includegraphics[width=\swidth,height=0.4\dwidth]{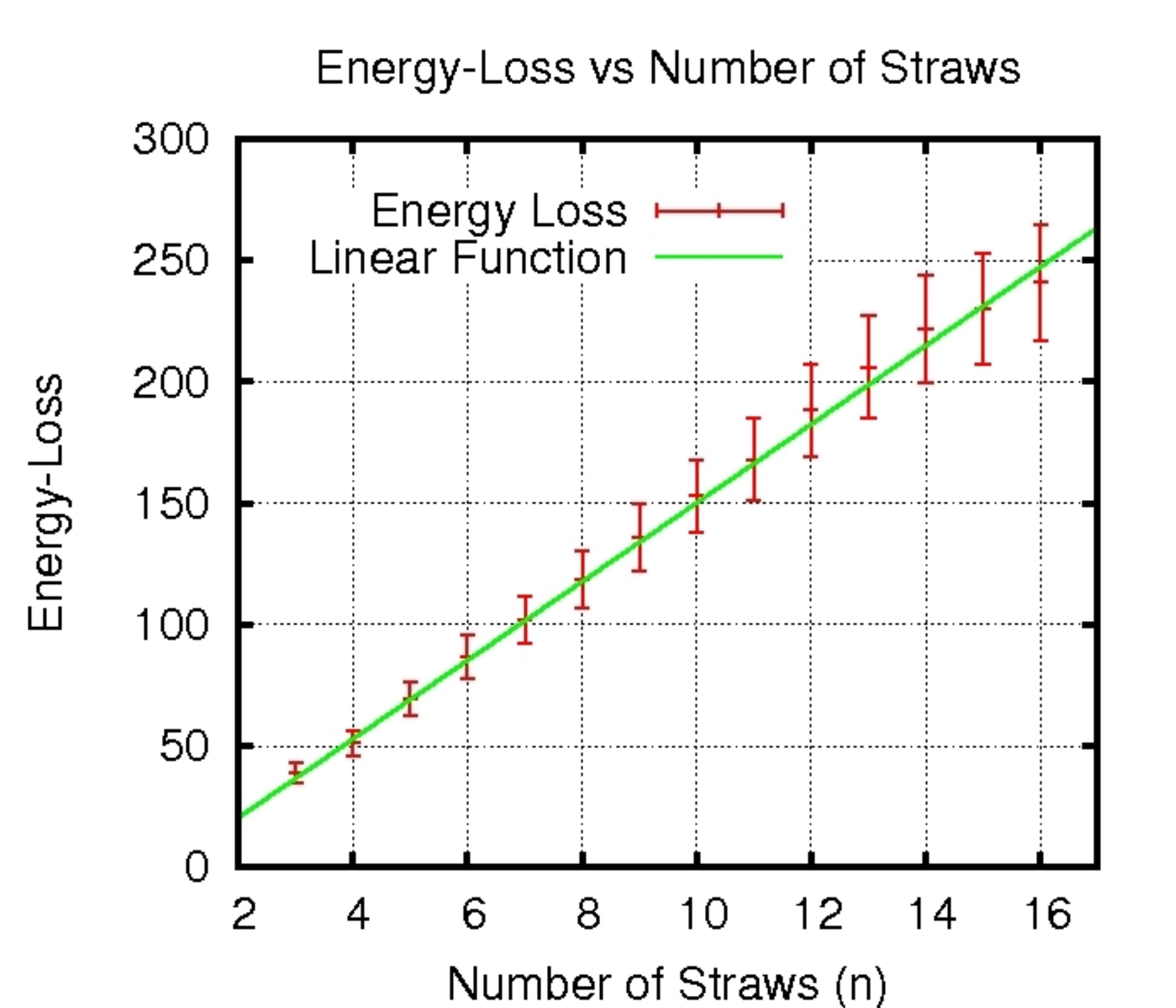}
\caption[Dependence of the energy-loss for minimum ionizing $\beta$-particles on the number of the traversed straws]{Dependence of the energy-loss for minimum ionizing $\beta$-particles on the number of the traversed straws. The energy-loss is given in arbitrary units.}
\label{fig:stt:pro:eloss_vs_n}
\end{center}
\end{figure}

\begin{figure}[h]
\begin{center}
\includegraphics[width=\swidth,height=0.4\dwidth]{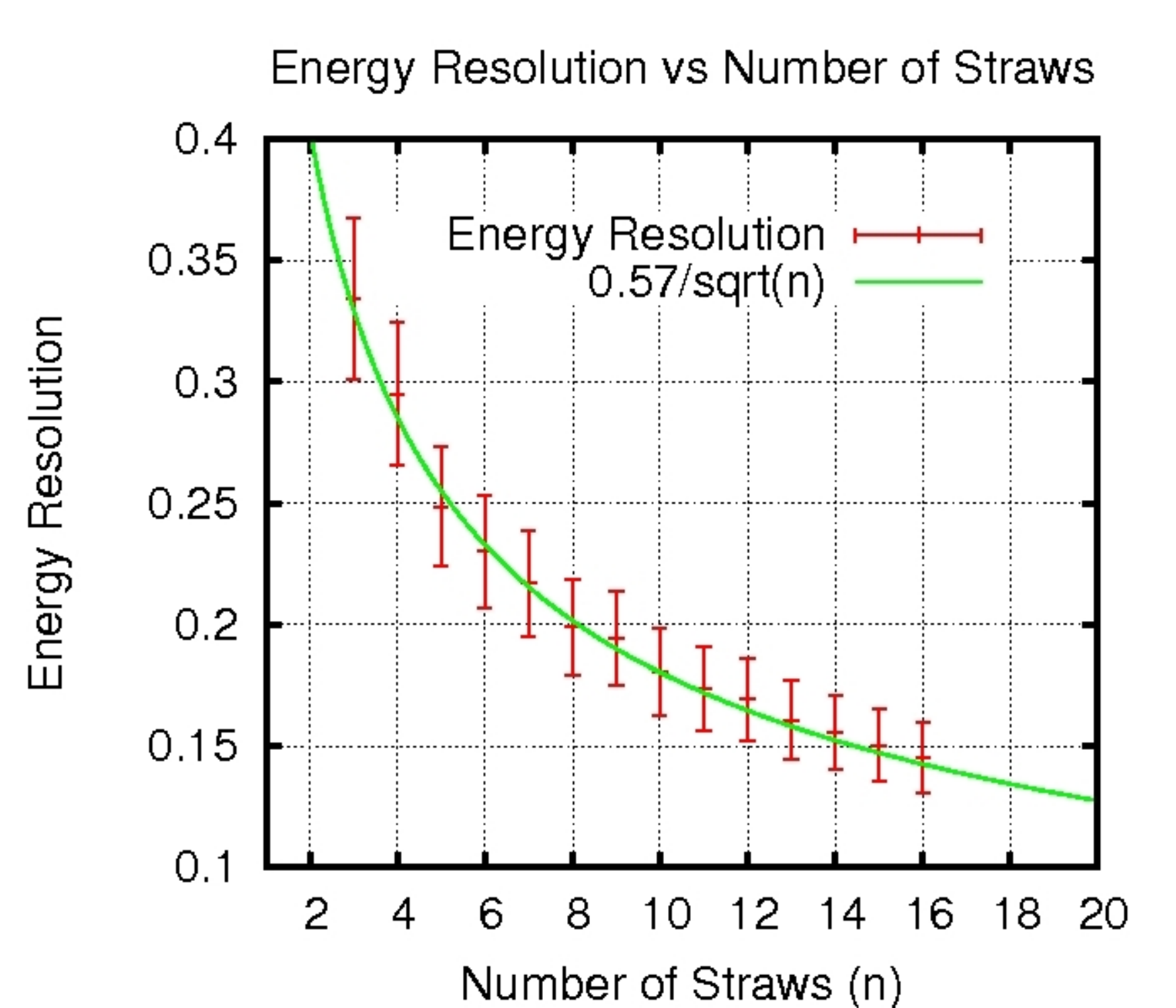}
\caption[Dependence of the energy resolution for minimum ionizing $\beta$-particles on the number of traversed straws]{Dependence of the energy resolution for minimum ionizing $\beta$-particles on the number of traversed straws.}
\label{fig:stt:pro:eres_vs_n}
\end{center}
\end{figure}

Initial checks of the detector response were performed by using 
$\beta$-particles from a $^{90}$Y/$^{90}$Sr source. The geometry and the trigger conditions
were optimized in order to select only the highest energy fraction of the $\beta$-decay 
spectrum containing minimum ionizing electrons.
\Reffig{fig:stt:pro:eloss_vs_n} shows the dependence of the energy-loss of the $\beta$-particles
on the number of traversed straws.
The relation between the energy resolution and the number of traversed straws has also been
tested and the result is presented in \Reffig{fig:stt:pro:eres_vs_n}. 
\noindent
\noindent
In this case the energy-loss spectra have been built integrating the area of the recorded signals. 
No further cuts were applied. The resolution is calculated as the ratio between the width 
and the mean value derived from the fit with a Landau curve. As expected, the energy dependence 
is perfectly linear and the resolution follows an inverse square-root relation 
to the number of fired straws. 

Further tests were performed using a monoenergetic proton beam of the COSY-accelerator. 
A low intensity beam (up to 1 $\cdot$ 10$^{4}$ protons/$s$) was selected and focussed 
directly onto the straw prototype. 

\begin{figure*}[h]
\begin{center}
\includegraphics[height=0.2\dwidth]{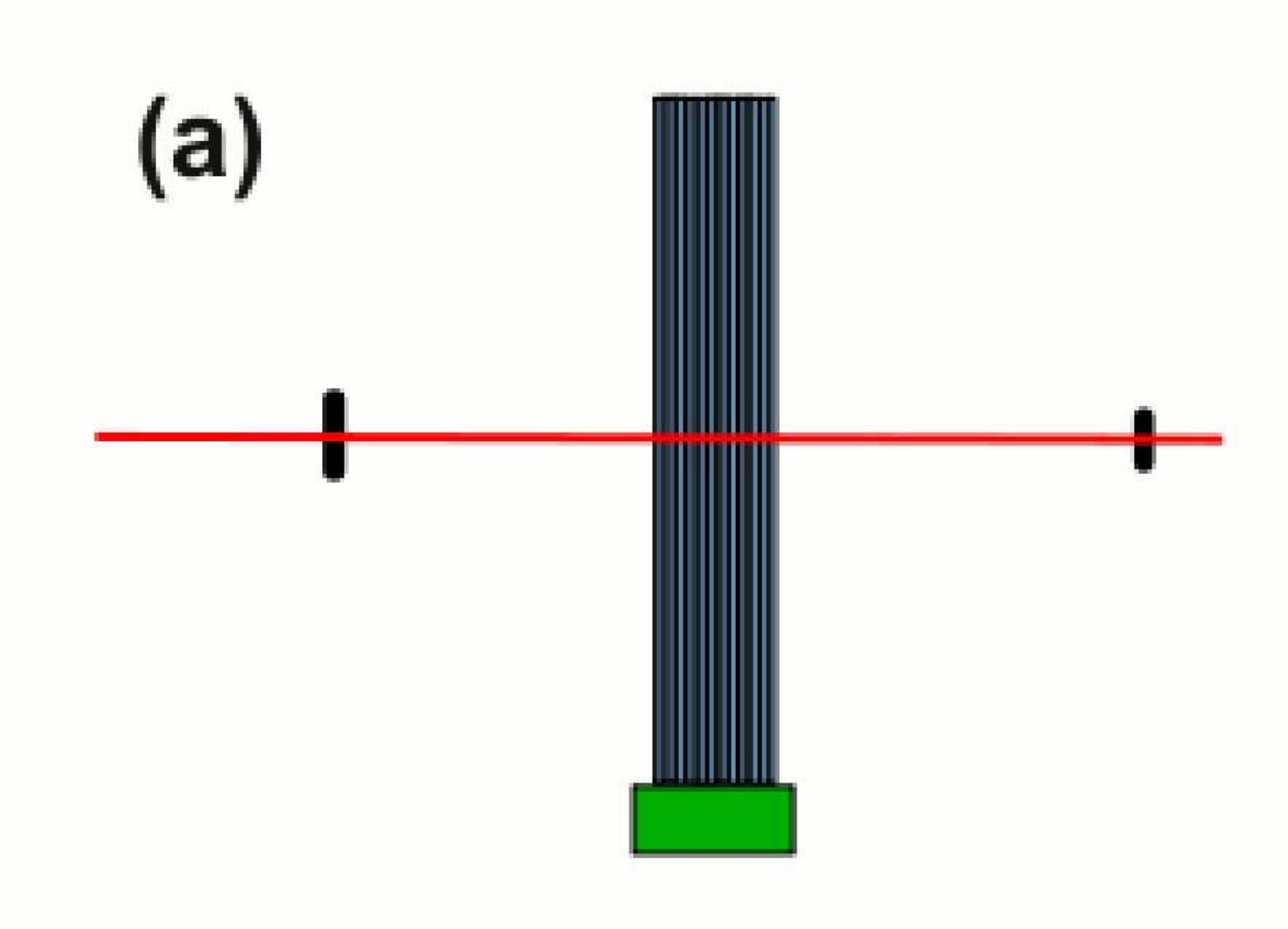}
\includegraphics[height=0.2\dwidth]{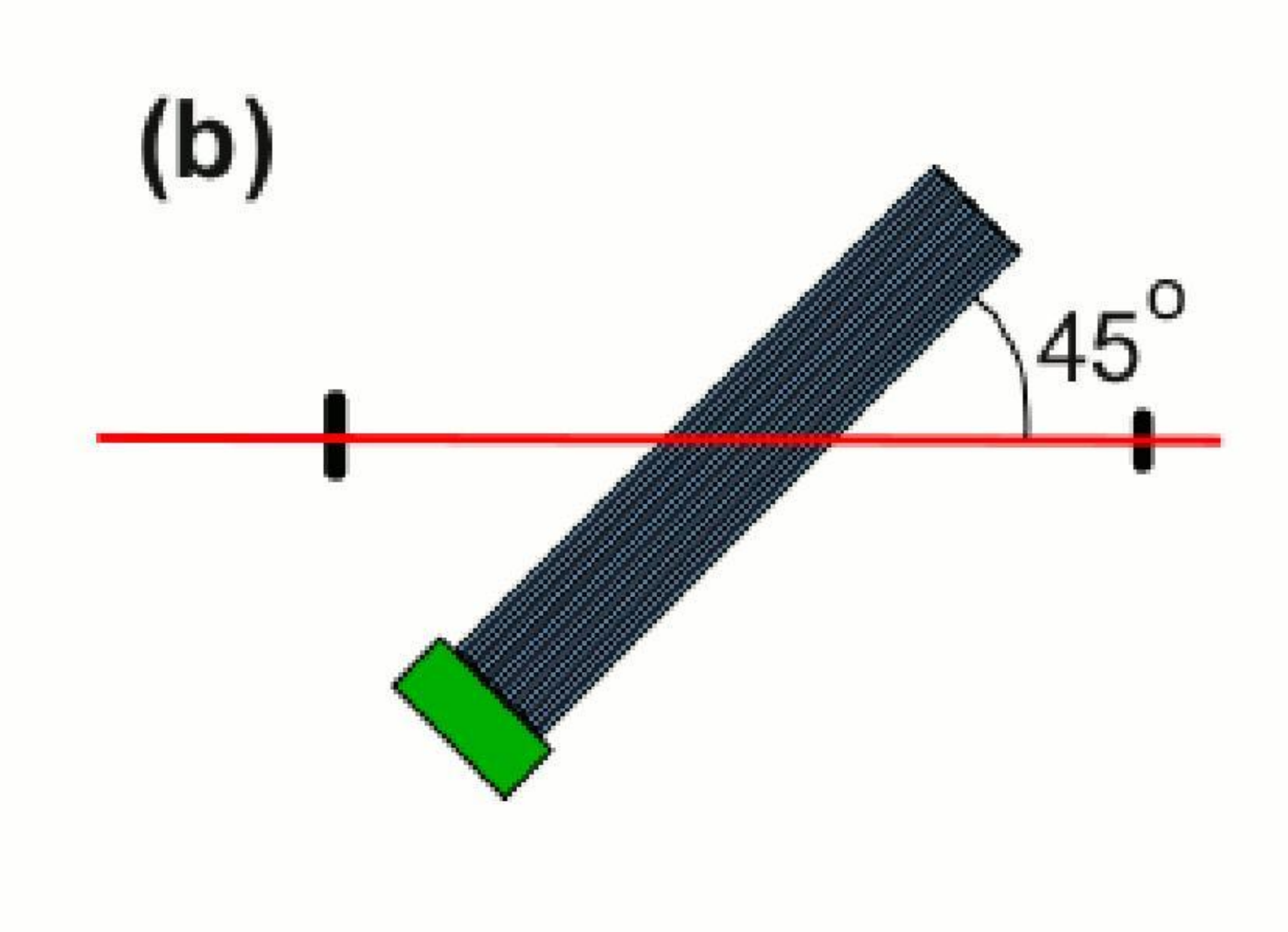}
\includegraphics[height=0.2\dwidth]{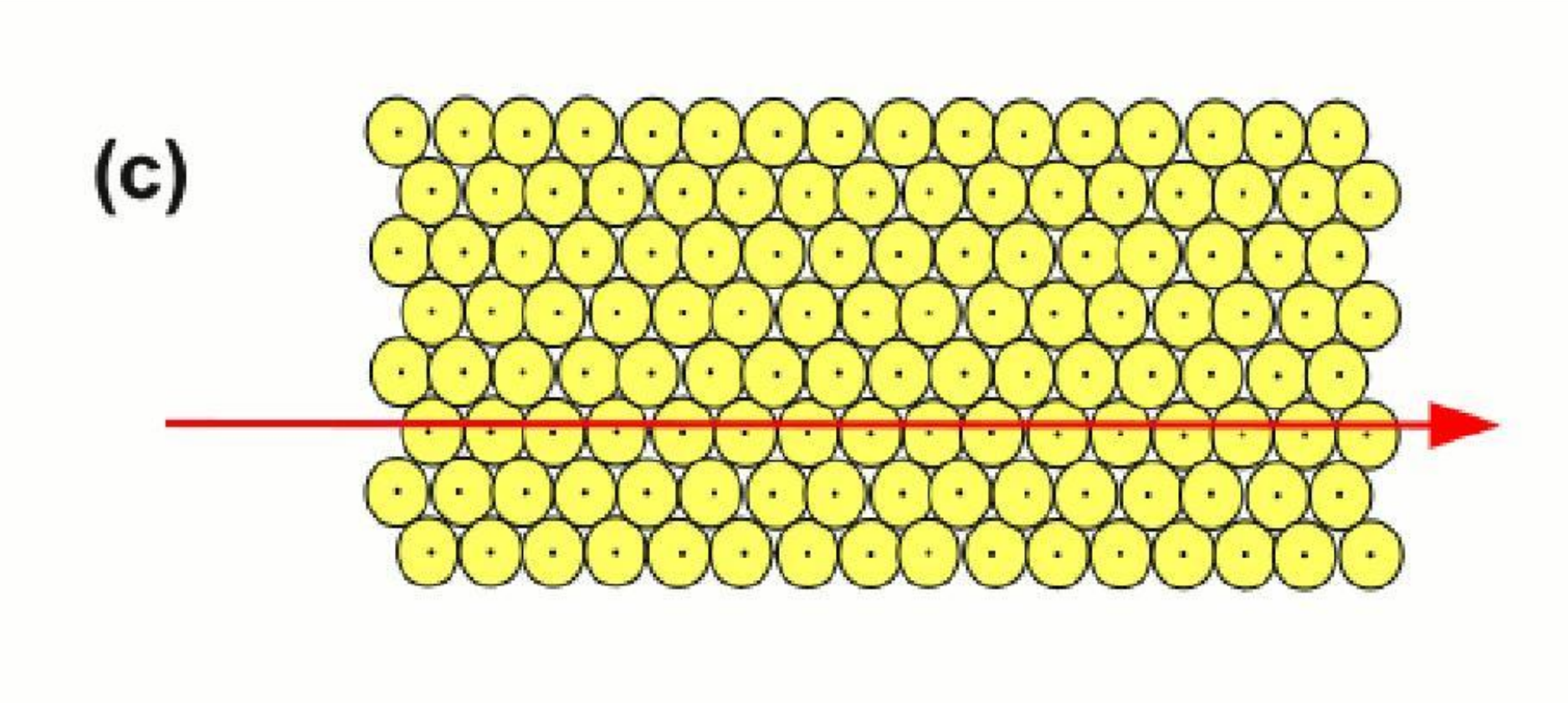}
\includegraphics[height=0.2\dwidth]{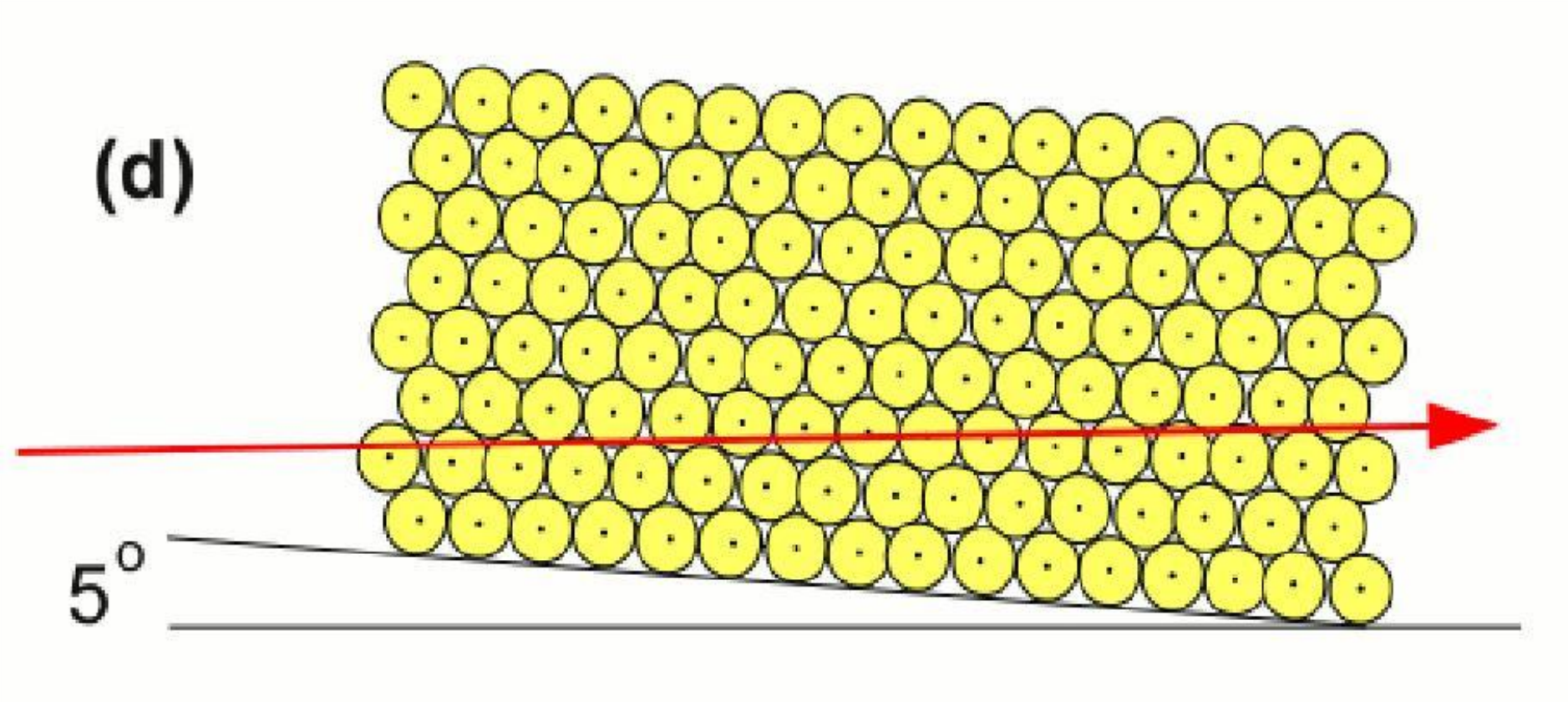}
\includegraphics[height=0.2\dwidth]{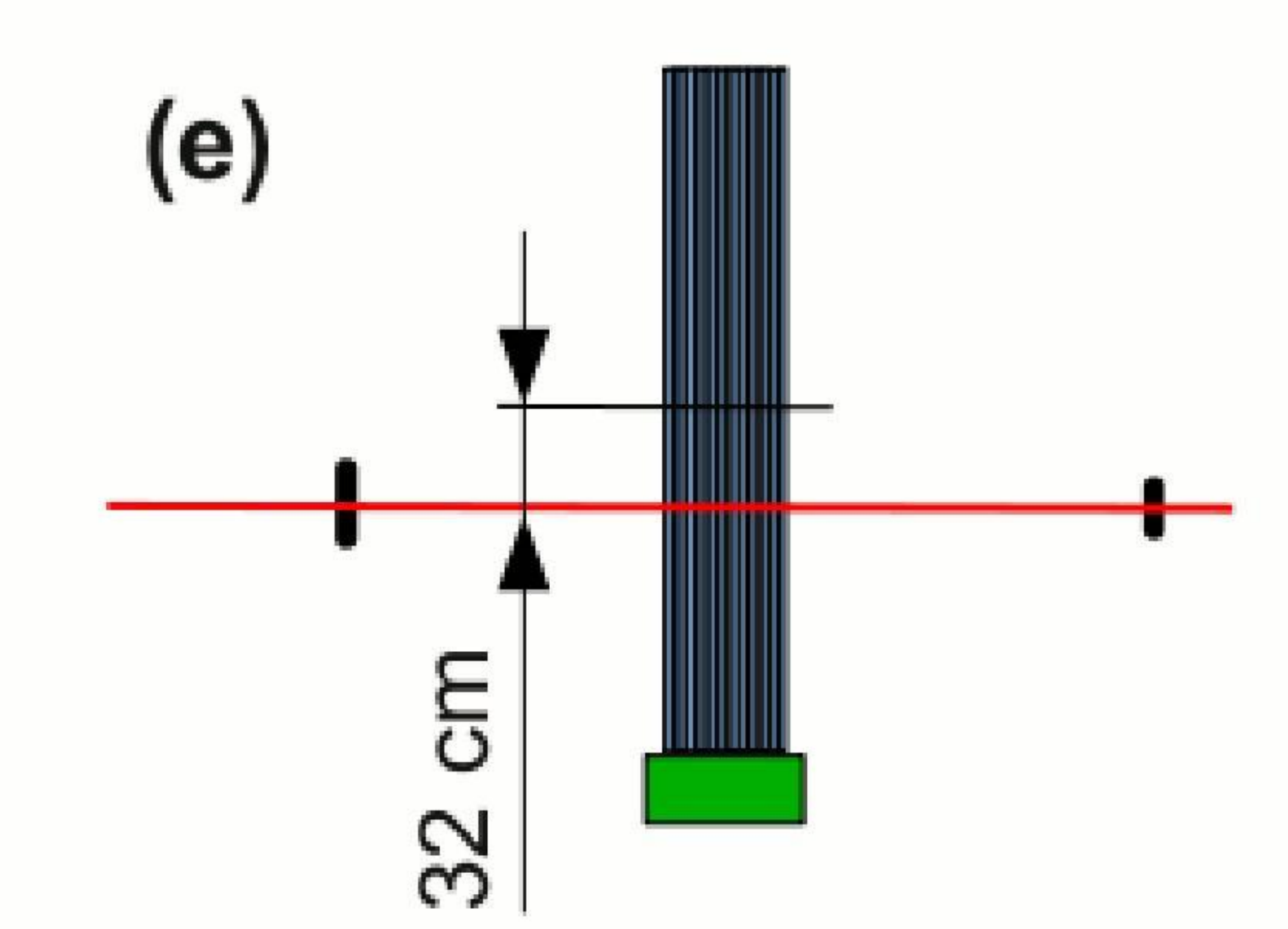}
\caption[Configurations of the straw prototype detector during the tests with the proton beam]{Configurations of the straw prototype detector during the tests with the proton beam:
(a) protons hit the straw tubes perpendicularly; 
(b) protons hit at 45$^{\circ}$ with respect to the straw axis; 
(c) the beam direction is parallel to the layers of straw tubes; 
(d) the prototype is rotated by 5$^{\circ}$;
(e) the prototype is displaced horizontally by 32 cm. 
The red lines indicate the beam.
}
\label{fig:stt:pro:det_pos}
\end{center}
\end{figure*}

During the measurements the detector was first oriented perpendicular to the beam axis (see \Reffig{fig:stt:pro:det_pos}a) to 
have the protons impinging at 90$^{\circ}$ with respect to the straw axis.
Then, it was turned by 45$^{\circ}$ as it is shown in \Reffig{fig:stt:pro:det_pos}b.
%KP: what is below:
%Later on,
%KP: has to be repalced by:
In vertical direction 
the straw tracker was positioned allowing the beam to pass parallel to the 
layer structure or with an angle of 5$^{\circ}$ in order to increase the number 
of crossed straw tubes and to enhance the merit of the isochrone calibration method:
\Reffig{fig:stt:pro:det_pos}c and \Reffig{fig:stt:pro:det_pos}d, respectively.
All these measurements were done with the beam hitting the straw in the middle. 
In order to test possible amplitude differences along the straw length, the prototype was finally displaced
by 32~cm. In this case, the beam entered the detector closer to the straw end equipped with the readout 
electronics (\Reffig{fig:stt:pro:det_pos}e).

The trigger signal was generated by the coincidence of the signals of
two small ($\sim$ 10 $\times$ 10 cm) and thin (5-10 mm)
scintillation detectors situated upstream and downstream the straw prototype. The beam 
was defocused in the vertical direction in order to
cover a broad range of straws. 
The small size of the triggering scintillators assured a negligible horizontal angular spread of the 
beam. 

The detector was filled with an Ar/CO$_2$ (9/1) mixture at 1\,bar overpressure.
The high voltage was set to keep the gas gain factor at a moderate level of about 
5 $\times$ 10$^{4}$. 
%(cf. \Refsec{sec:stt:sim}, \Reffig{fig:stt:sim:ExpGain}).

%\subsubsection{Analysis}
\subsection{Analysis Method}
The shapes of the recorded signals (see \Reffig{fig:stt:pro:signal1} and \Reffig{fig:stt:pro:signal2}) 
do not allow to estimate the particle energy-losses 
%KP: this has to be used : 
from the numbers and 
%KP: this has to be removed:   due to the variations and 
the distributions of the initial ionization clusters. 
The energy-losses can be deduced only from the integrated charge of the output signal, or from any 
other parameter which is linear or related through an other known function to the collected charge.
Two different methods have been tested: the so called \emph{Truncated Mean} and the
\emph{Time over Threshold}. The methods and their results are described in the following.

{\bf Selection of Events}

\begin{figure}[h]
\begin{center}
\includegraphics[width=\swidth,height=0.4\dwidth]{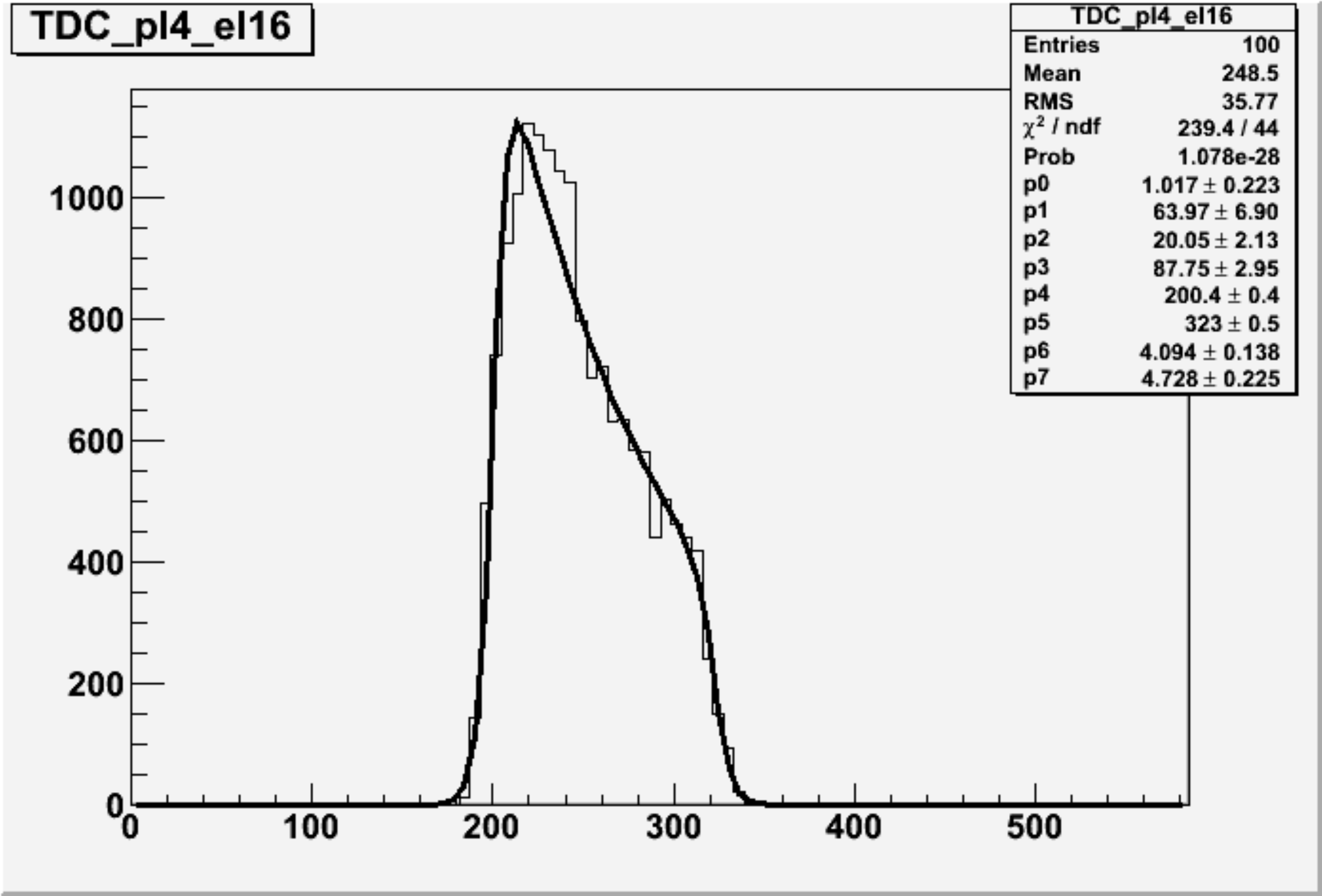}
\caption[Drift time distribution for a straw irradiated by a 2.95\,\gevc proton beam]{Drift time distribution for a straw irradiated by the proton beam with a momentum of 2.95\,\gevc.
The fit function to the time distribution is described in \Refsec{sec:stt:calib}.}
\label{fig:stt:pro:drift_time_distr}
\end{center}
\end{figure}

\begin{figure}[h]
\begin{center}
\includegraphics[width=\swidth,height=0.4\dwidth]{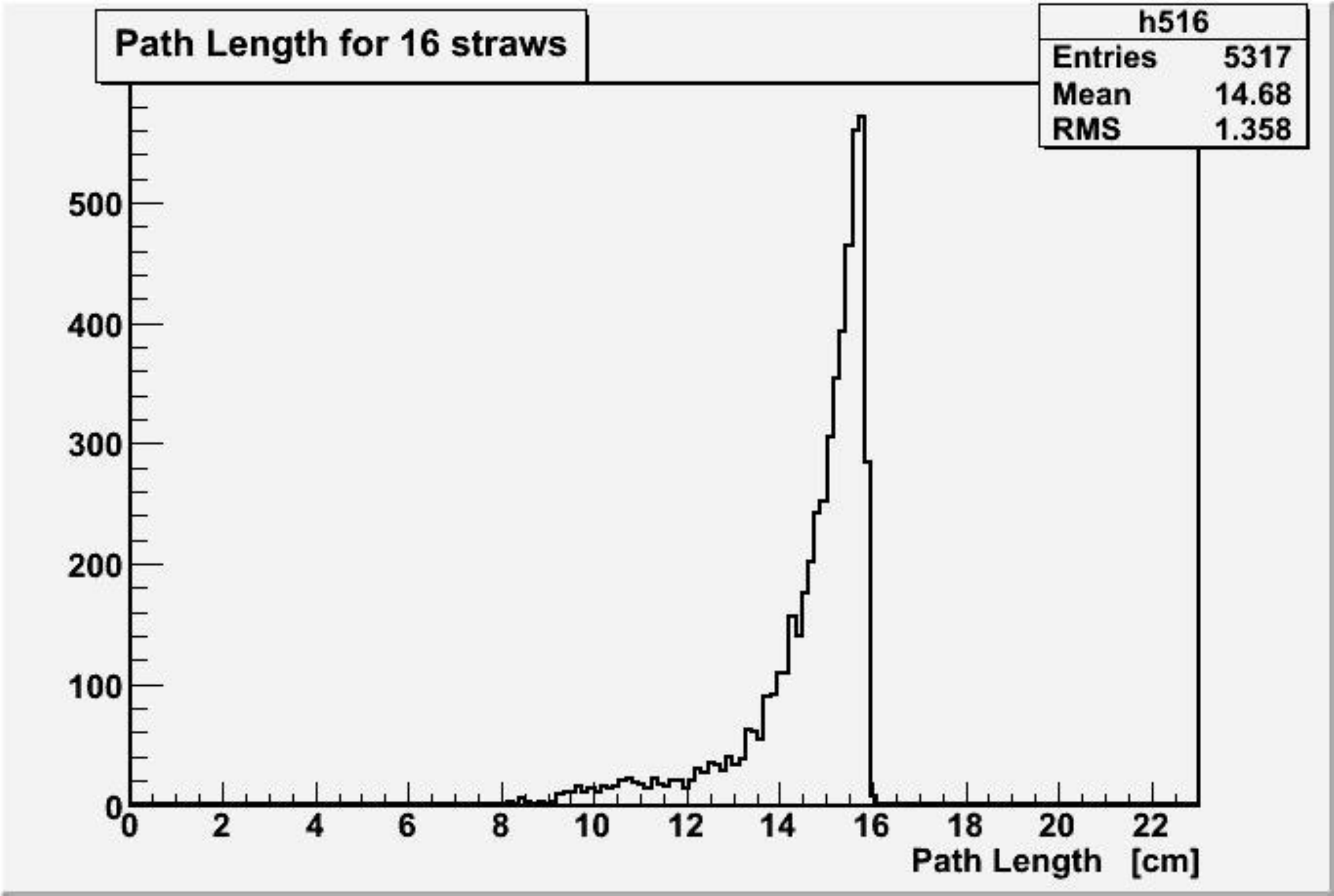}
\caption[Path length distribution for reconstructed tracks where 16 straw tubes were hit]{Path length distribution for reconstructed tracks where 16 straw tubes were hit.}
\label{fig:stt:pro:path_length}
\end{center}
\end{figure}

\begin{figure}[h]
\begin{center}
\includegraphics[width=\swidth,height=0.4\dwidth]{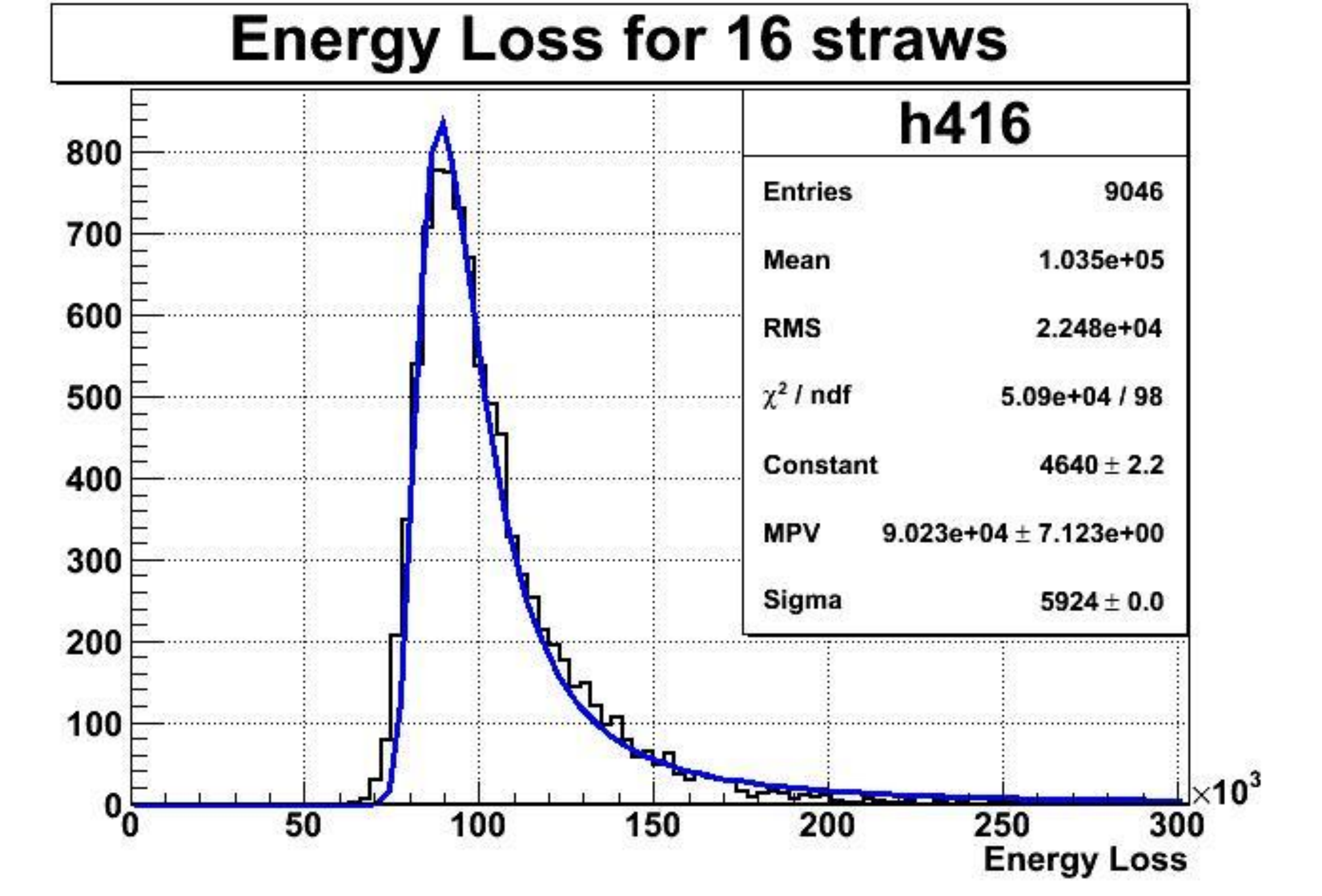}
\caption[Energy-loss distribution for 2.95\,\gevc protons for reconstructed tracks hitting 16 straw tubes]{Energy-loss distribution for 2.95 \gevc protons for reconstructed tracks hitting 16 straw tubes. 
The energy-loss is given in arbitrary units, and the distribution is fitted with a Landau function.}
\label{fig:stt:pro:de_16}
\end{center}
\end{figure}

The sampling frequency of the used FADC provides a drift time precision of 4.17\,ns. \Reffig{fig:stt:pro:drift_time_distr}
shows a typical drift time spectrum. The isochrone calibration was performed as it is described in \Refsec{sec:stt:calib}.
The gas mixture in the straws was Ar/CO2 (9/1) at a pressure of 2\,bar.
\noindent
The tracking procedure allowed the selection of the fired straws belonging to an
event and the calculation of the particle path length (\Reffig{fig:stt:pro:path_length}).
\noindent
The signals from the fired straws were then used to build the energy-loss distribution. 
The energy-loss distribution for 2.95 \gevc momentum protons 
is shown in \Reffig{fig:stt:pro:de_16}. As expected, it shows a Landau distribution shape.
% KP: what is below
%The tail of the distribution extend to higher energies and overlap with the 
%ionization curves of higher energy particles and deteriorate the separation between them. 
% KP: might be replaced by following:
The tail of the distribution extends to higher energies deteriorating the separation between 
the neighbouring ionization curves.
In order to avoid the tail of the energy distribution 
a conversion of the Landau function into a symmetric Gaussian-like function by means of the 
so called \emph{Truncated Mean} is performed.

{\bf Truncated Mean method}

\begin{figure}[h]
\begin{center}
\includegraphics[height=\swidth,width=0.4\dwidth,angle=90]{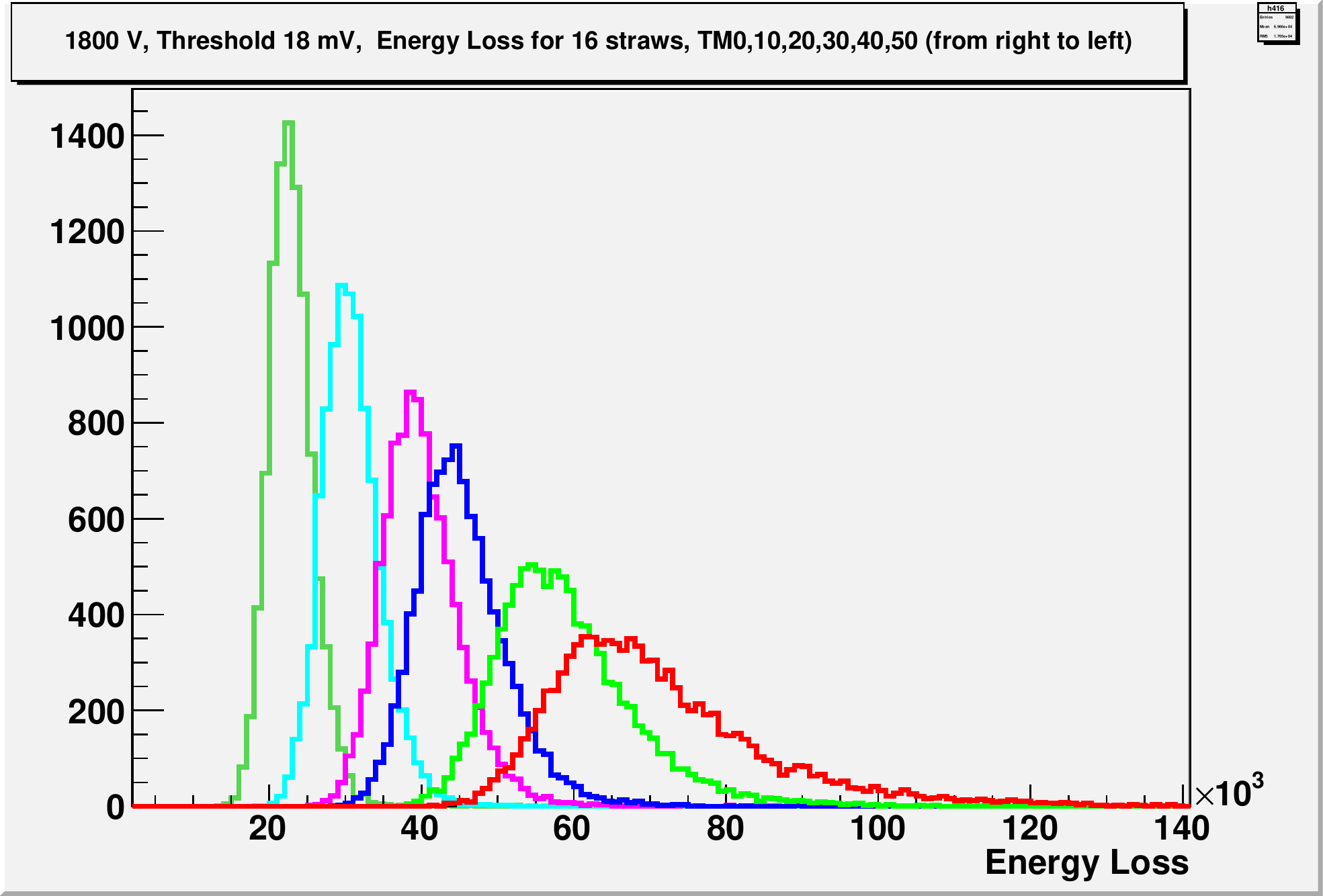}
\caption[Modification of the experimental Landau distribution by the Truncated Mean method]{Modification of the experimental Landau distribution by the so called \emph{Truncated Mean} method. 
From right to left: original distribution, truncated by 10\%, 20\%, 30\%, 
40\% and 50\%. The energy-loss is given in arbitrary units.}
\label{fig:stt:pro:tm}
\end{center}
\end{figure}

The probability of a certain mass assignment to a charged track may be determined
by means of the observed ionization values.
The average of all $n$ measurements performed in a gaseous detector, is a bad estimator of the particle energy 
since it fluctuates a lot from track to track, because the underlying mathematical ionization
distribution has no finite average and no finite variance. A good estimator
is either derived from a fit to the shape of the measured distribution or from a
subsample excluding the very high measured values \cite{bib:stt:tub:blum}.
By taking a fixed fraction $r$ of the signals with the smallest amplitudes and evaluating
their mean one finds a shallow minimum of this quantity as a function of $r$ for values between about 0.35 and 0.75. 
This procedure has been deeply studied in the field \cite{bib:stt:pro:tm1,bib:stt:pro:tm,bib:stt:tub:blum} finding that 
in this range of $r$ it is an empirical fact that the truncated mean values are distributed almost like a Gaussian.
\Reffig{fig:stt:pro:tm} shows energy-loss distributions obtained with the straw tube signals when different truncation factors f (f=1-r) are 
applied. 
\noindent
The most suitable truncation factor has been determined by optimizing the resolution for the truncated mean distribution of the specific 
energy-loss. For the analyzed data the best truncation fraction is 30\,\% (see \Reffig{fig:stt:pro:fractions_TM}).

\begin{figure}[h]
\begin{center}
\includegraphics[width=\swidth,height=0.4\dwidth]{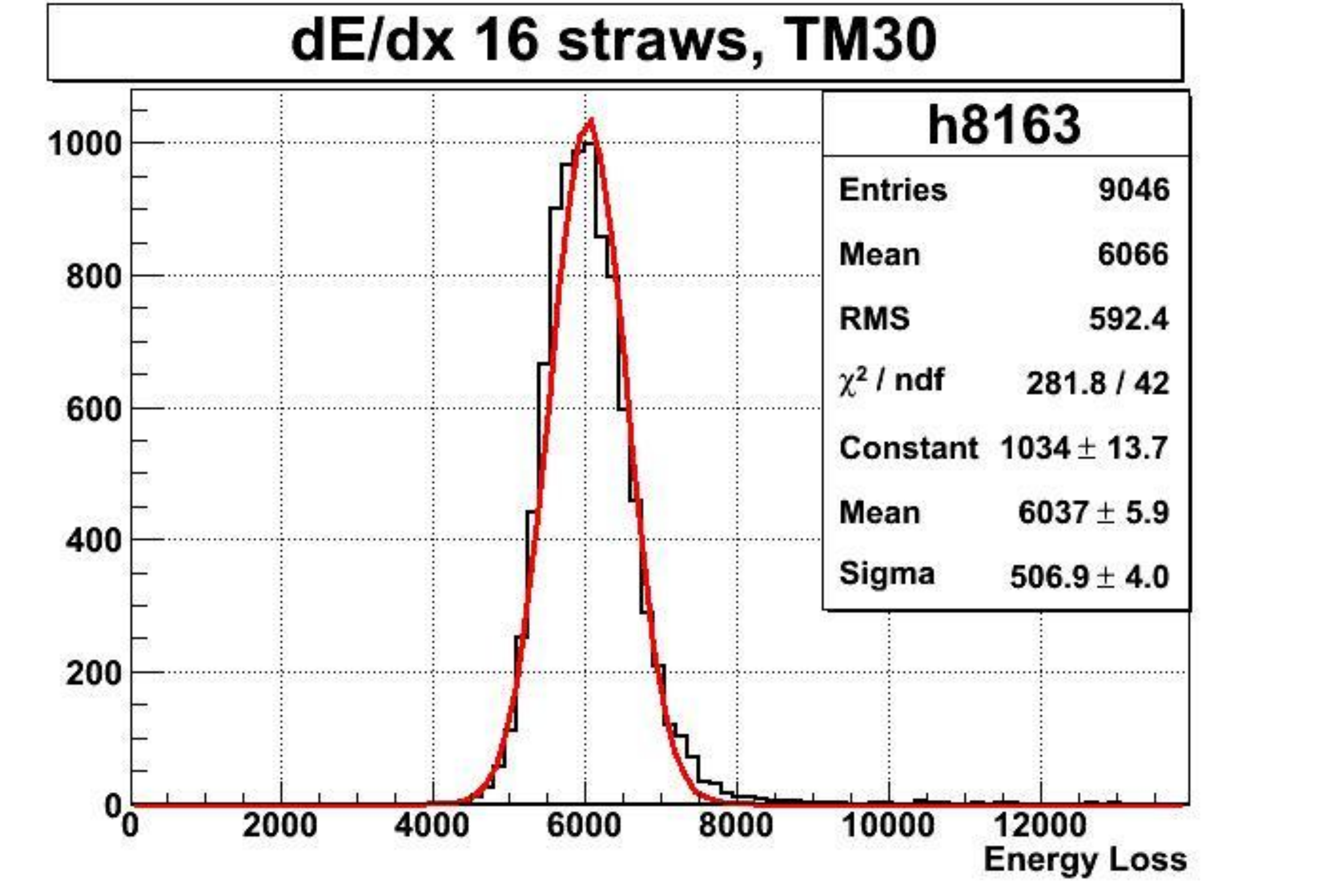}
\caption[dE/dx distribution for protons with a momentum of 2.95 \gevc]{dE/dx distribution for 2.95 \gevc protons fitted with a Gaussian curve. 
The protons hit 16 straw tubes, and a truncation of 30\% has been applied.}
\label{fig:stt:pro:de_dx_16}
\end{center}
\end{figure}

The energies of the truncated distributions have then to be divided by the appropriate reconstructed 
path lengths.
The energy-loss distribution 30\,\% truncated, corrected for the path length for protons of 
2.95 \gevc momentum, is shown in \Reffig{fig:stt:pro:de_dx_16}.
The distribution has a shape well resembling the normal distribution 
and the fit with a Gaussian curve permits to derive the parameters of 
the distribution biased only with minimal uncertainties.
\noindent  
\newpage

{\bf Time over Threshold}

Another technique that can be used to determine particle energy-loss, avoiding two electronic-readout 
branches (time and amplitude), is the so called \emph{Time over Threshold (ToT)}. This method postulates that
the energy deposit inside the drift cell could be related to the time duration of the output signal and has 
been exploited by the ATLAS experiment \cite{bib:stt:pro:atlas1,bib:stt:pro:atlas2,bib:stt:pro:atlas3}.

Utilizing the recorded signal shapes from the straw prototype, it was checked 
that the direct measurement of the signals duration does not give a satisfactory relation with the specific 
energy-loss for any reasonable threshold value.
Only a coarse relation between the time width and the deposited charge is observed. Moreover, for high 
counting rates it is not affordable to follow the output signal over the whole drift time.

A most successful compromise solution, utilizing only a sensible fraction
of the output signals, has been worked out and applied in the HADES experiment \cite{bib:stt:pro:hades1}. 
Unfortunately, due to the lack of a proper readout electronics, this method could not be
checked in the tests we performed. However, since the characteristics of the signals of the HADES drift 
chambers are similar to those of the \Panda--\Stt, the method and the results obtained by the HADES 
collaboration are recalled here as an example of what could be achieved with the use of timing electronics 
only. 

In HADES the ToT-method was used for particle discrimination by means of dE/dx measurements 
in Mini-Drift Chambers (MDC), a four sections gaseous tracker.
Each section of the detector consists of 6 separate drift chambers forming in total a 24 layer structure.
Hence, particles traversing the tracking system may induce up to 24 individual output 
signals per track.
The drift cells of each section have different 
dimensions and alternates from 5 $\times$ 5 \,mm for the first section, up to 
14 $\times$ 10 \,mm for the chambers in section IV.
The chamber windows are made of 12 $\mum$ aluminized Mylar foils. Aluminum, tungsten, 
bare and gold-plated wires are used for anodes, cathodes and field-shaping. Diameters of the wires 
vary from 20 to 100 $\mum$. The detector is filled with an Helium/Isobutane gas mixture (6/4) 
at atmospheric pressure.

\begin{figure}[h]
\begin{center}
\includegraphics[width=\swidth,height=0.4\dwidth]{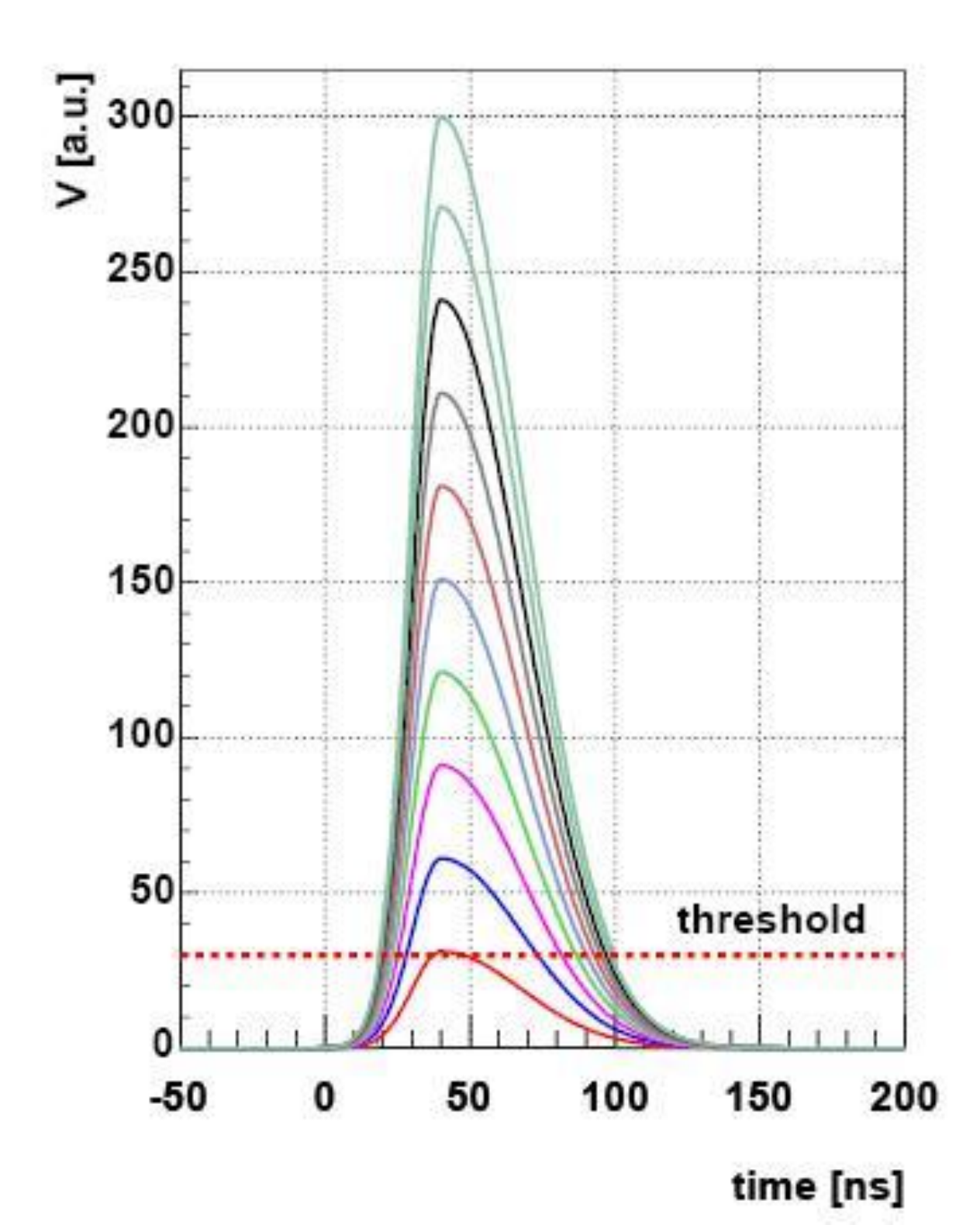}
\caption[Time over Threshold dependence on the initial signal \cite{bib:stt:pro:hades1}]{Illustration of the \emph{Time over Threshold} dependence on a preshaped fraction of the 
initial signals from the MDC-drift cells of the HADES experiment \cite{bib:stt:pro:hades1}.
}
\label{fig:stt:pro:hades_tot_1}
\end{center}
\end{figure}

The outputs of the sense wires are connected to analog boards \cite{bib:stt:pro:hades2} 
allowing for differential amplification, shaping and discrimination.  
The signals are digitized by means of an ASD8-B chip \cite{bib:stt:pro:hades3}, 
which delivers a logical (LVDS) signal whose width is proportional to the time that 
the shaped signal remain above a fixed threshold value. 
Logical signals are then fed to multi-hit TDCs allowing both the 
time stamping with a 0.5 ns precision (from the leading edge of the signal), as well as the ToT
evaluation (from the signal width). The method is illustrated in \Reffig{fig:stt:pro:hades_tot_1}.
\noindent 
In order to extract the energy-loss from the measured ToT a careful calibration has been performed.
For each particle species the non-linear correlation function has been determined by means of a fitting
procedure taking into account various incident angles and drift distances 
(in bins of 5$^{\circ}$ and 100 $\mum$, respectively). Finally, the Truncated Mean method, similar to that
already described, has been applied.

Eventually, an energy resolution of the order of 7\,\% has been obtained for minimum ionizing 
particles, whereas for higher ionizing particles the achieved resolution 
%KP: what is below:
%of about 4$\percent$ \cite{bib:stt:pro:hades1} has been obtained. 
%KP: might be repalced by:
was about 4\,\% \cite{bib:stt:pro:hades1}.

%\subsubsection{Results}
\subsection{Results}

In this subsection the results obtained with the use of the Truncated Mean method are reported. 

\begin{figure}[h]
\begin{center}
\includegraphics[width=\swidth,height=0.4\dwidth]{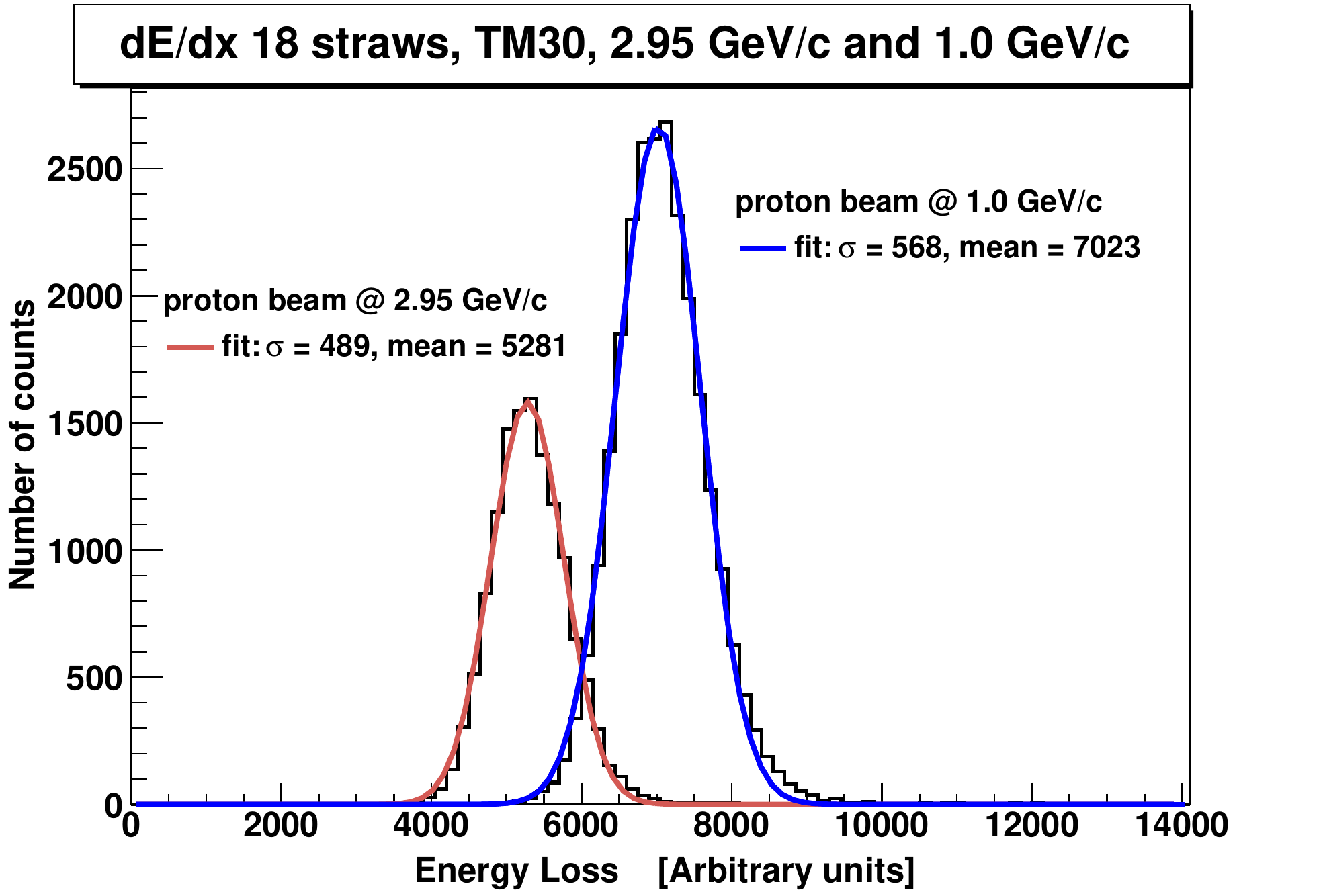}
\caption[d$E$/d$x$ distributions for monoenergetic protons of 2.95\,\gevc and 1.0\,\gevc]{
d$E$/d$x$ distributions for monoenergetic protons of 2.95 \gevc (left one) 
and 1.0 \gevc (right one) with Gaussian fits. The proton beam hit a maximum of 18 tubes for the 5$^\circ$ inclined straw setup.}
\label{fig:stt:pro:de_dx_18_TM30_2_95_1_0_fit}
\end{center}
\end{figure}

\begin{figure}[h]
\begin{center}
\includegraphics[width=\swidth,height=0.4\dwidth]{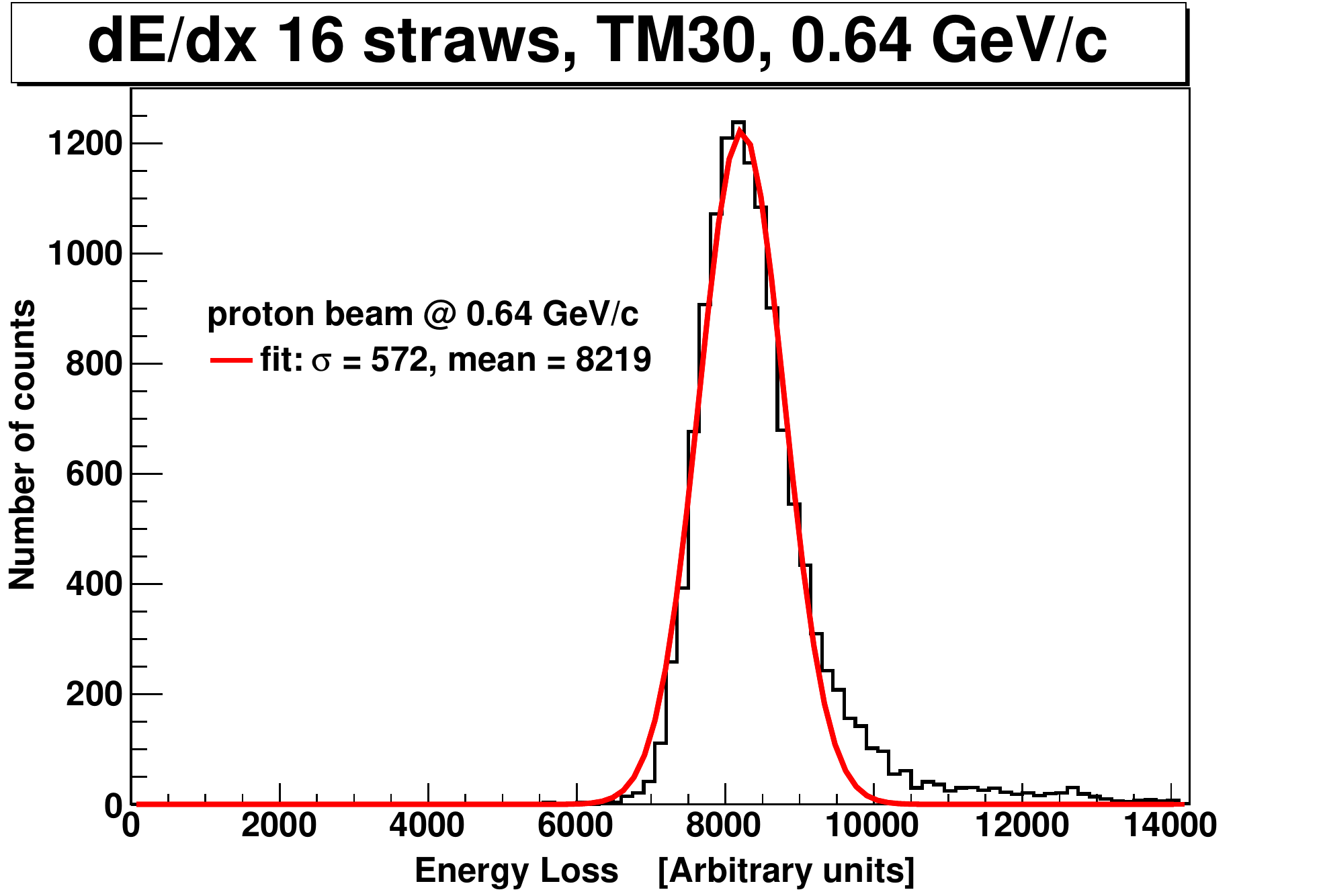}
\caption[d$E$/d$x$ distribution for monoenergetic protons of 0.64 \gevc]{d$E$/d$x$ distribution for monoenergetic protons of 0.64 \gevc with Gaussian fit. 
The protons hit a maximum of 16 straw tubes.}
\label{fig:stt:pro:de_dx_16_TM30_0_64_fit}
\end{center}
\end{figure}

Truncated energy-loss distributions for different proton beam momenta and for selected tracks of high 
statistics are shown in \Reffig{fig:stt:pro:de_dx_18_TM30_2_95_1_0_fit}, 
for 2.95 and 1.0 \gevc and in \Reffig{fig:stt:pro:de_dx_16_TM30_0_64_fit} for 0.64 \gevc. 
A Truncated Mean cut of 30\,\% of the hits has been applied. The Gaussian fits are superimposed and the fit parameters are given in the figures. 
The results for the 0.64 \gevc protons cannot be presented on the same energy scale of the results 
of 1.0 and 2.95 \gevc due to a slightly different orientation of the setup (5$^\circ$ inclination and 18 hit straws instead of 16 for the latter two momenta). Also a higher threshold was used during the analysis of this data set. 
This was necessary since a higher pick-up noise was observed during this measurement 
created by an insufficient shielding of the scintillator bases. The selected threshold was two times
higher than that set for the other tests. 
In all cases the applied HV was the same: 1800 V (gas gain of about 5 $\cdot$ 10$^{4}$).
\noindent
\noindent
%For the sake of a most effective comparison of the results obtained for 
%different beam momenta, data sets collected and analyzed in the same conditions 
%are presented in this summary. 
%It means that the results presented here are not always the best ones. 
%This is e.g. for the case of 2.95 \gevc data. 
%In \Reffig{fig:stt:pro:de_dx_16} result obtained 
%for another measurement where only 2.95 \gevc momentum was available is given. 
%At conditions of that measurement both the threshold and time calibration were different. 
%The energy resolution derived from these data is better than for 2.95 \gevc data shown in this subsection.   
 
\begin{figure}[h]
\begin{center}
\includegraphics[width=\swidth,height=0.4\dwidth]{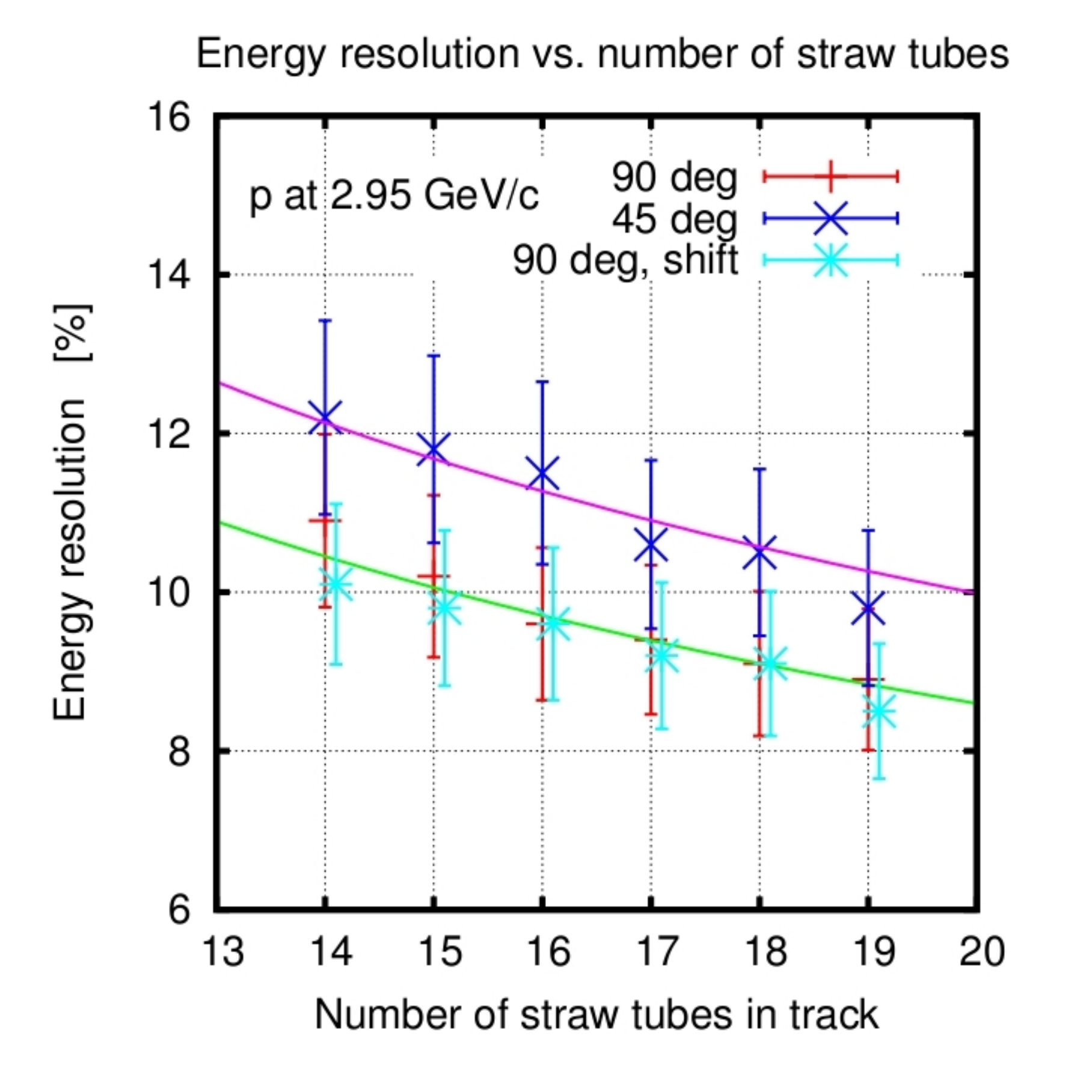}
\caption[Dependence of the energy resolution on the number of straw hits (2.95\,\gevc protons)]{Dependence of the energy resolution on the number of hit straws
for protons with 2.95 \gevc momentum. The different measurement setups are marked by different colors 
(see also \Reffig{fig:stt:pro:det_pos}). Red: straw setup perpendicular to the proton beam; 
Dark blue: setup skewed horizontally by 45$^\circ$; Light blue: setup shifted by 32\,cm in straw direction. 
The superimposed curves are functions $\propto(n)^{-1/2}$, where $n$ is the number of hits.}
\label{fig:stt:pro:de_n_2_95_1800_TM30_th75_90_45_shift}
\end{center}
\end{figure}

\begin{figure}[h]
\begin{center}
\includegraphics[width=\swidth,height=0.4\dwidth]{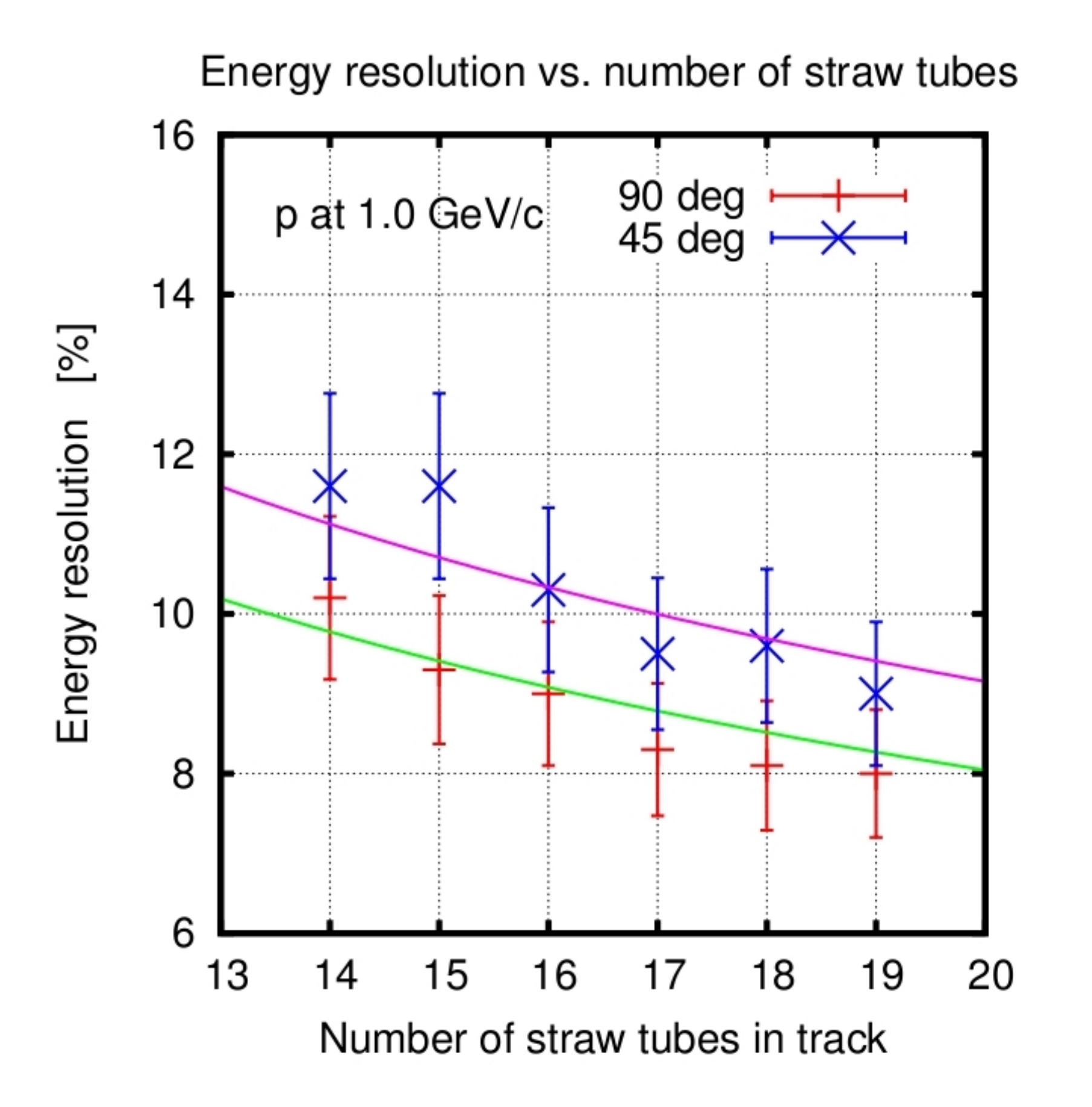}
\caption[Dependence of the energy resolution on the number of hit straw (1.0\,GeV/c protons)]{Dependence of the energy resolution on the number of hit straws for protons with 1.0\,GeV/c momentum. 
For the setup inclined by 45$^{\circ}$ the resolution is slightly worse, from about 8\,\% to 9\,\% at 19 hit tubes.}
\label{fig:stt:pro:de_n_1_0_1800_TM30_th75_90_45}
\end{center}
\end{figure}

\begin{figure}[h]
\begin{center}
\includegraphics[width=\swidth,height=0.4\dwidth]{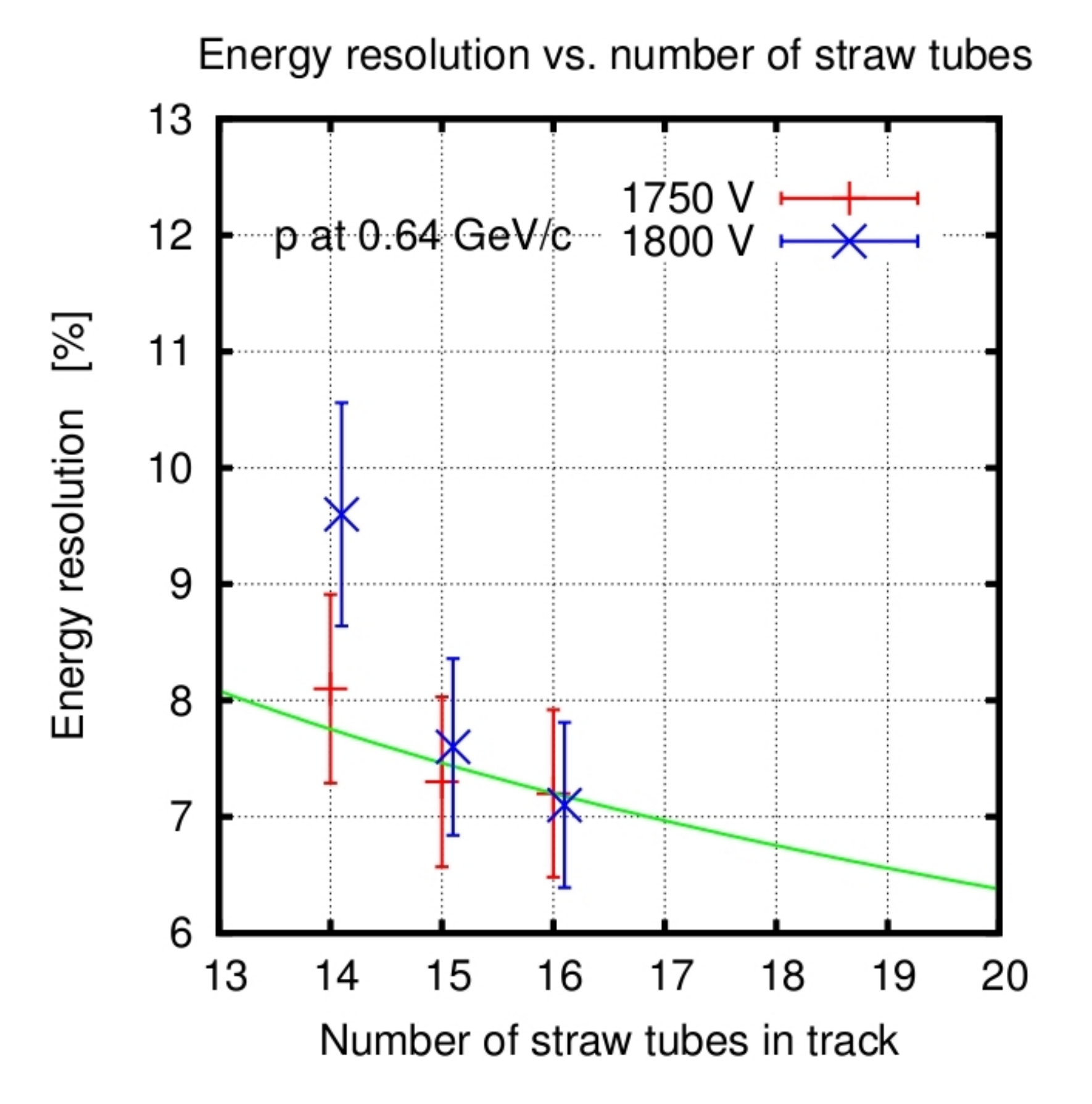}
\caption[The dependence of the d$E$/d$x$ on the number of hit straws (0.64\,\gevc protons)]{Dependence of the d$E$/d$x$ on the number of hit straws for protons at 0.64 \gevc 
momentum for two different high voltage settings. 
}
\label{fig:stt:pro:de_n_0_64_1800_TM30_th150_1750_1800}
\end{center}
\end{figure}

\Reffig{fig:stt:pro:de_n_2_95_1800_TM30_th75_90_45_shift} and 
\Reffig{fig:stt:pro:de_n_1_0_1800_TM30_th75_90_45} 
present the energy resolution dependence on the number of
straws used to reconstruct the track for various geometrical configurations of the detector  
for 2.95 and 1.0 \gevc proton momentum, respectively. 
\noindent
The dependence of d$E$/d$x$ on the number of hit straws for protons at 0.64 \gevc momentum 
is shown in \Reffig{fig:stt:pro:de_n_0_64_1800_TM30_th150_1750_1800}. Due to technical problems, for
this beam momentum the inclined tracks could not be registered. In this case the beam passed parallel to the
detector layers and the highest statistics was obtained for reconstructed tracks with only 16 hits.
Results for two different values of HV: normal one (1800 V) and decreased by 50 V (1750 V), are given. 
No significant difference in resolution for the lower voltage is observed.     
\noindent
The presented results show that within the geometrical constrains of the laboratory test 
and in the kinematical momentum range of interest, the achievable energy resolution
is equal to 8\,\% for 1.0 \gevc proton momentum and improves at lower momenta with the 
increase of the particle energy deposit. At 0.64 $\gevc$, with only 16 
straws in the track, the resolution is equal to 7\,\%. For tracks inclined by 45$^{\circ}$ 
a systematical deterioration of the resolution of 1\,\% is observed. 
For minimum ionizing protons of 2.95 \gevc the energy resolution is about 9\,\%, and 
for tracks at 45$^{\circ}$ it is worse of 1.5\,\%. 
No significant effect on the energy resolution is observed for tracks hitting the straws at different 
longitudinal positions.

In order to check the possibility to perform d$E$/d$x$ measurements also for high particle rates, which are 
expected in the innermost layers of the \Panda--\Stt, and the possibility to use the ToT-method for the
energy-loss determination, we explored the possibility of shortening the signal integration time.

\begin{figure}[h]
\begin{center}
\includegraphics[width=\swidth,height=0.4\dwidth]{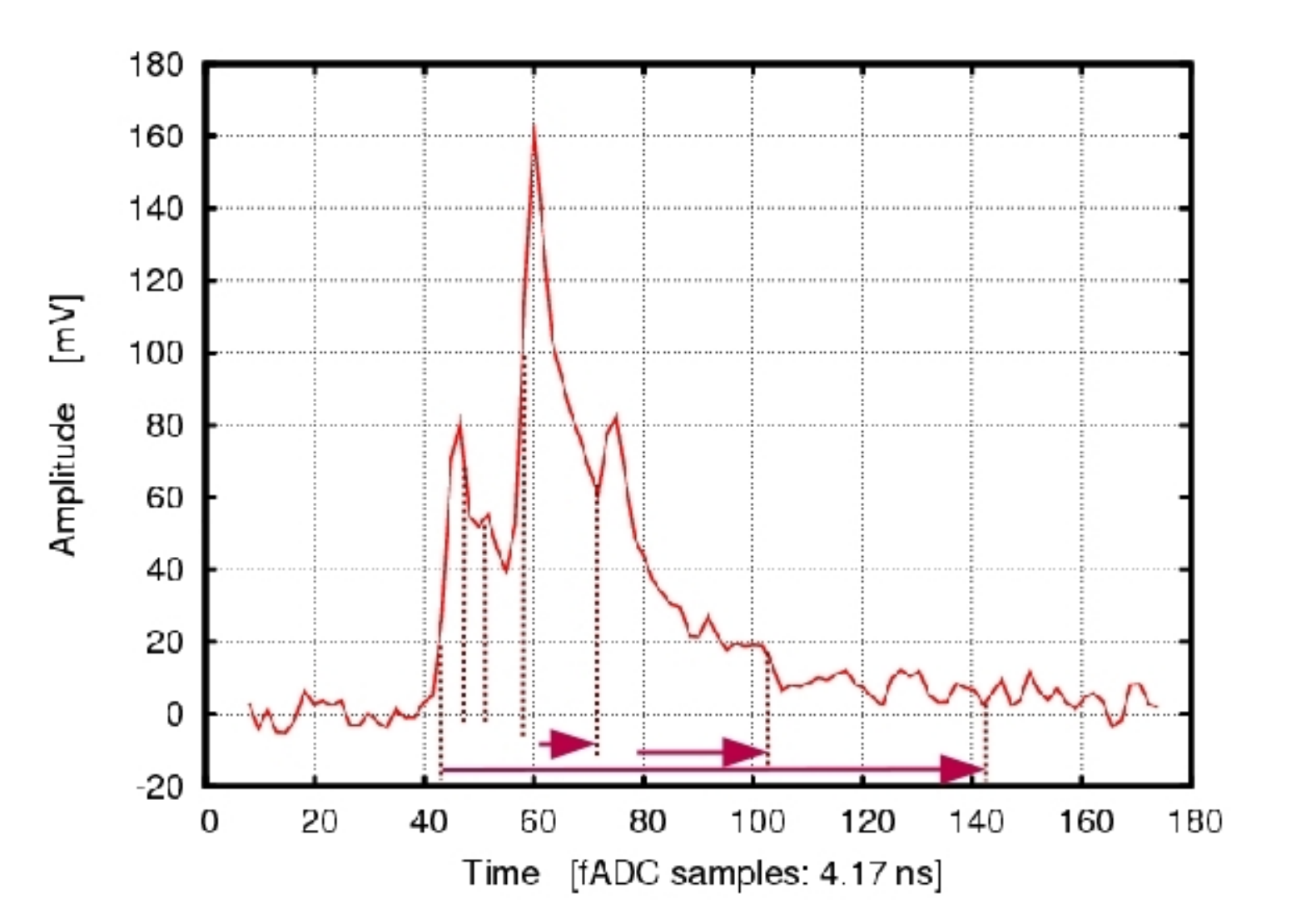}
\caption[Straw tube signal recorded by the FADC]{Straw tube signal recorded by the FADC. The integration range indicated by the lines are 4, 8, 16, 30, 60, and 100
times the 4.17\,ns sampling time. 
}
\label{fig:stt:pro:fractions}
\end{center}
\end{figure}

\begin{figure}[h]
\begin{center}
\includegraphics[width=\swidth,height=0.4\dwidth]{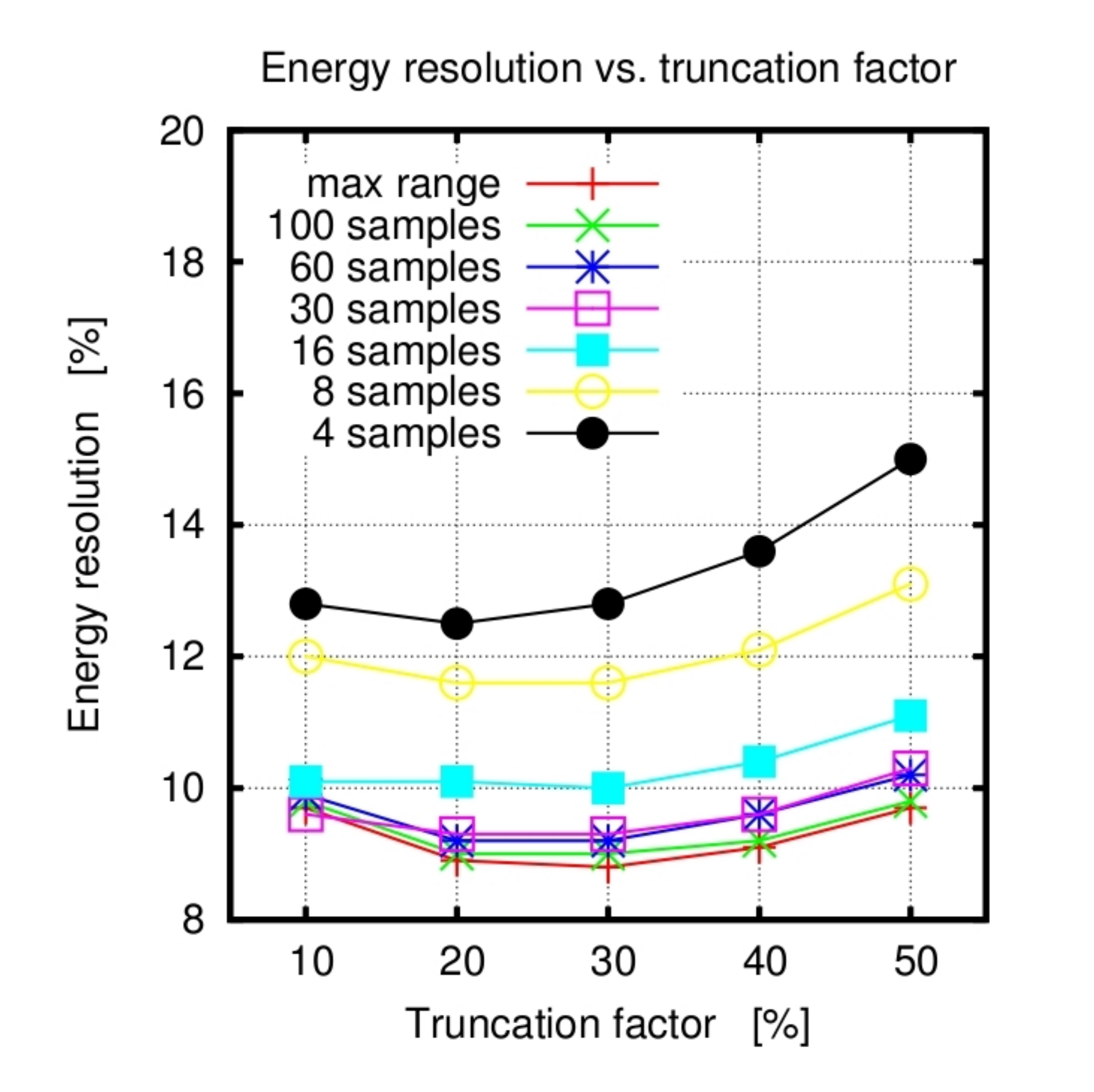}
\caption[Energy resolution as a function of the truncation factor]{Energy resolution as a function of the truncation factor for various 
fractions of the deposited charge.}
\label{fig:stt:pro:fractions_TM}
\end{center}
\end{figure}

By using the recorded analog signals of protons of 1.0 \gevc momentum, an
analysis changing the fractions of the integrated signals has been performed. 
The concept of this analysis is shown in \Reffig{fig:stt:pro:fractions}.
The signals have been integrated over 4, 8, 16, 30, 60 and 100 FADC samples
(sample width is equal to 4.17 ns). 
\noindent
The resulting energy-loss distributions have been truncated and a Gaussian fit has been performed.
The results for events with 18 hit straws per track are shown in \Reffig{fig:stt:pro:fractions_TM}.
For this analysis, the path length correction has been applied before making the 
tail truncation (this was not done for the results shown above).  A significant deterioration 
of the energy resolution is observed when the integration range is reduced to less then 16 samples (time interval 16$\times$4.17\,ns). 
``max range'' means that the integration is done over the whole time window
of the FADC.  
% what due to the limited tracking 
%precision caused a cumulation of individual errors and 
A small deterioration of the resolutions is obtained in comparison with that presented in 
\Reffig{fig:stt:pro:de_n_1_0_1800_TM30_th75_90_45}.

\begin{figure}[h]
\begin{center}
\includegraphics[width=\swidth,height=0.4\dwidth]{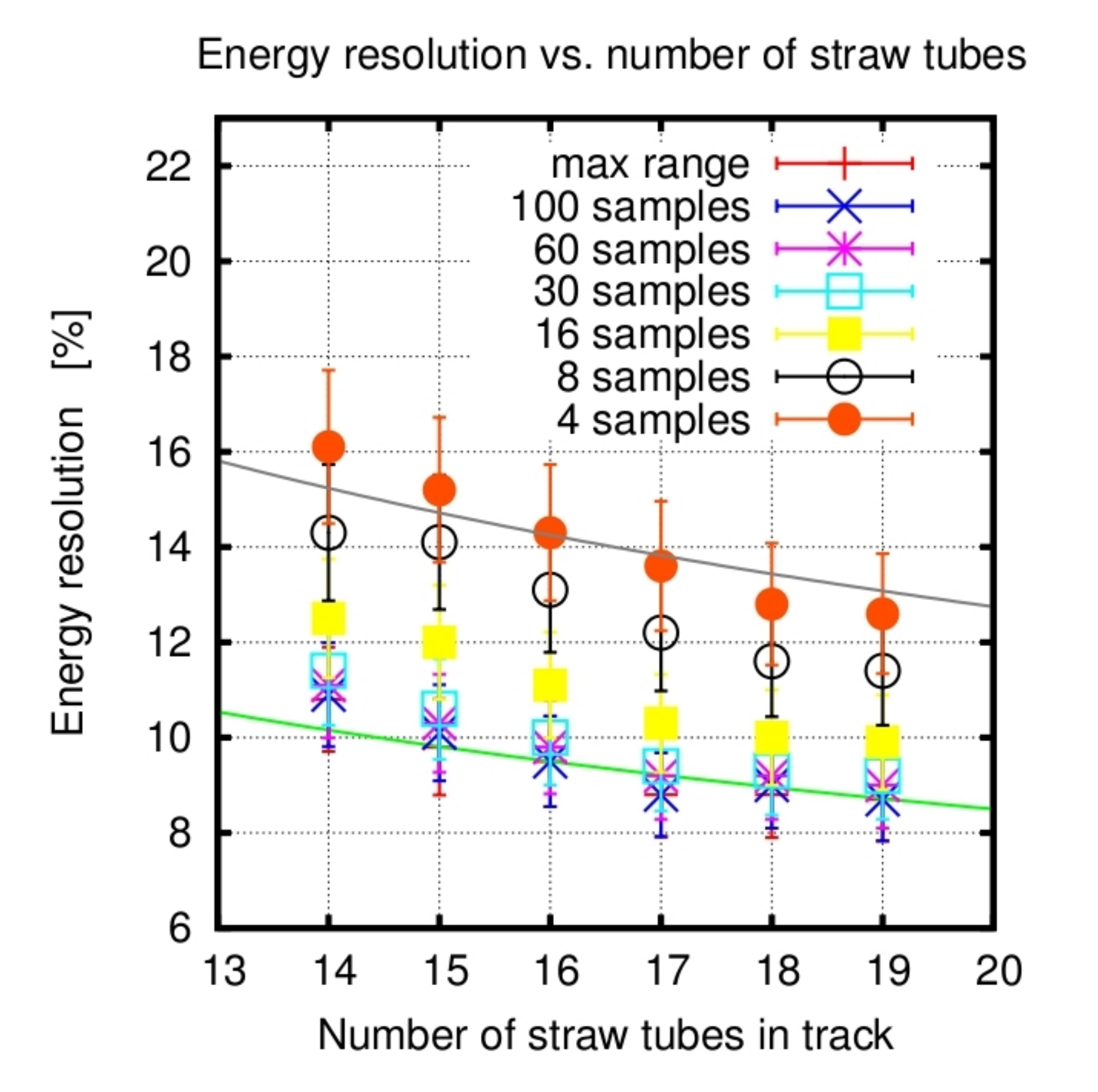}
\caption[Dependence of the energy resolution on the fraction of the integrated charge]{Deterioration of the energy resolution by decreasing the fraction of the integrated charge. 
For the superimposed curves refer to \Reffig{fig:stt:pro:de_n_2_95_1800_TM30_th75_90_45_shift}.
}
\label{fig:stt:pro:fractions_n_TM}
\end{center}
\end{figure}

Since almost the whole charge of the signals is contained within the first 40--60 samples, there 
is not a significant decrease of the energy resolution measured, when the integration range extends 
beyond 30 samples. On the other hand, the 16 sample resolution is worse by 1\,\% with respect to that 
obtained with the whole integration, and any further reduction of the integration range causes big worsening
of the resolution. For the shortest integration time, only 4 samples, the energy resolution increases by 
an additional 3--4\,\%.
This result is shown in \Reffig{fig:stt:pro:fractions_n_TM}.
Here the energy resolution and its dependence on the number of hit straws is presented.
The energy resolution is evaluated with a truncation factor of 30\,\% only, which gives the best 
results.

The presented analysis, 
%although charged with the uncertainty caused by an unknown response function of the 
%whole electronic chain used for creating and recording of the signal shapes, 
indicates that in order to obtain a satisfactory energy resolution with the \Panda--\Stt, there is no need to 
integrate the signal charges over the whole drift time. A preshaping over 65\,ns would be sufficient
in order to keep the energy resolution below 10\,\%, if the amplitude is used to  
measure the energy-loss. On the other hand, if ToT is used, in the version developed in the HADES 
experiment, 
%where
%extracting of the energy resolution is supported by an extensive multiparameter calibration, 
the integration time could be even shorter. In HADES the charge is integrated over few tens of ns. 
Further test exploring the possibility of using this technique will be done.

\subsection{Detector Performance at High Counting Rates}
%\subsubsection{Detector performance at high counting rates}

\begin{figure}[h]
\begin{center}
\includegraphics[width=\swidth,height=0.4\dwidth]{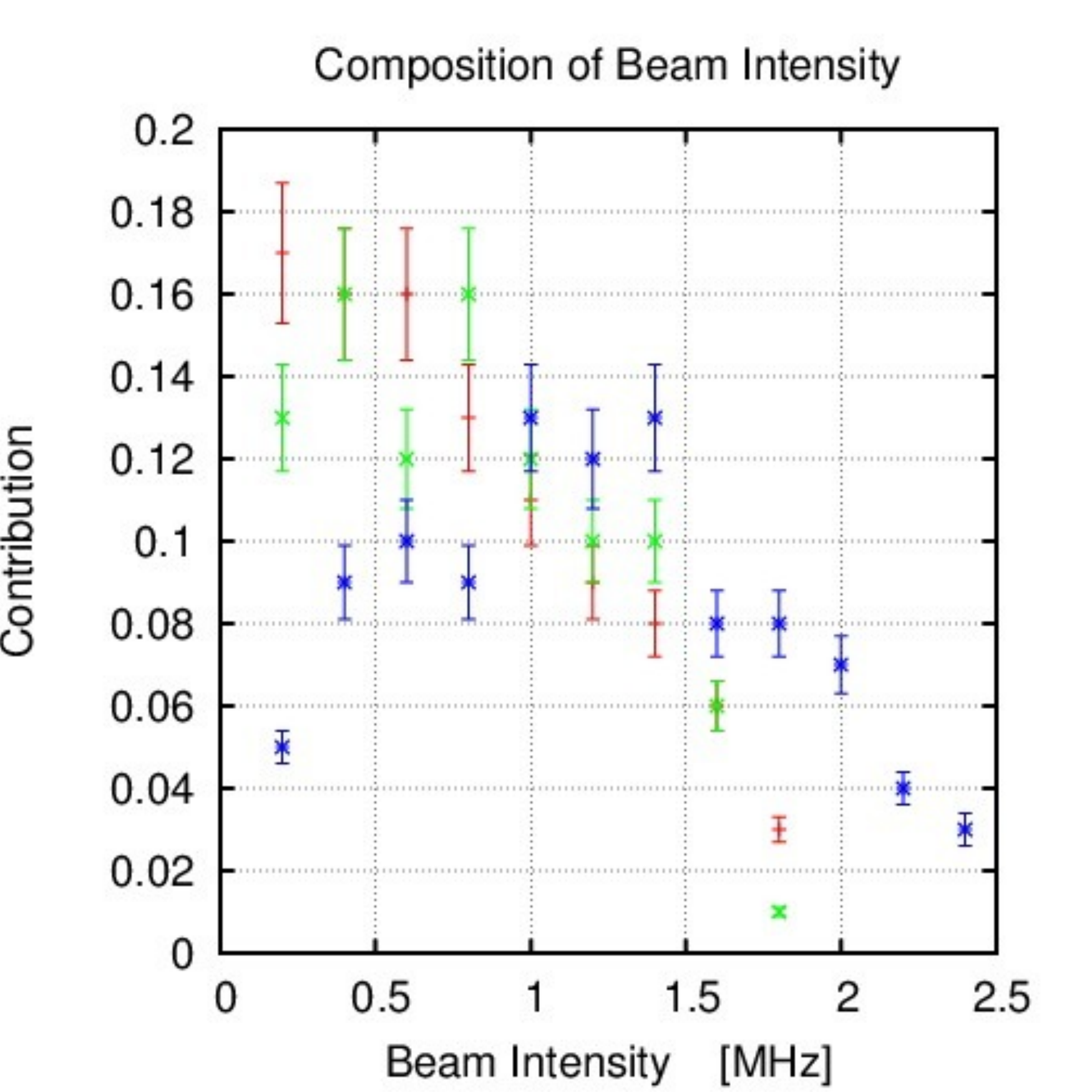}
\caption[Beam intensity variation during the high-rate test of the straw prototype]{Examples of the beam intensity variation during the high-rate test of the straw prototype. 
The different colors give the distributions of the measured beam intensity in different straws.}
\label{fig:stt:pro:beam_intensity}
\end{center}
\end{figure}

The results described above were obtained with monoenergetic beams of protons of intensity up to 10$^{4}$/s. 
At \Panda the experimental conditions foreseen predict for the innermost layer of \Panda--\Stt a particle rate of up to 0.8 MHz/straw. 
The rather long ion tail of the signals with this heavy particle flux calls for very efficient baseline restoration circuits,
furthermor it could produce space charge distortions that can cause gas gain reduction with loss in resolution.
In order to start addressing these problems
% Therefore, the problem of stability 
%and quality of the output signals at such extreme conditions has been addressed as well.
the experimental setup was exposed to a proton beam of a momentum of 2.7\,GeV/c of intensity up to 2.4\,MHz.
The actual beam intensity was monitored by counting the signals of each of the first straw in the layer.
Due to the high beam divergence a stable high intensity beam could not be kept on the whole detector. 
Therefore, a variation of the instantaneous beam intensity was observed, that can even better simulate the 
experimental conditions at \Panda. Examples of the spreads of the beam intensity are given in 
\Reffig{fig:stt:pro:beam_intensity}. 
\noindent

\begin{figure*}[h]
\begin{center}
\includegraphics[height=0.35\dwidth]{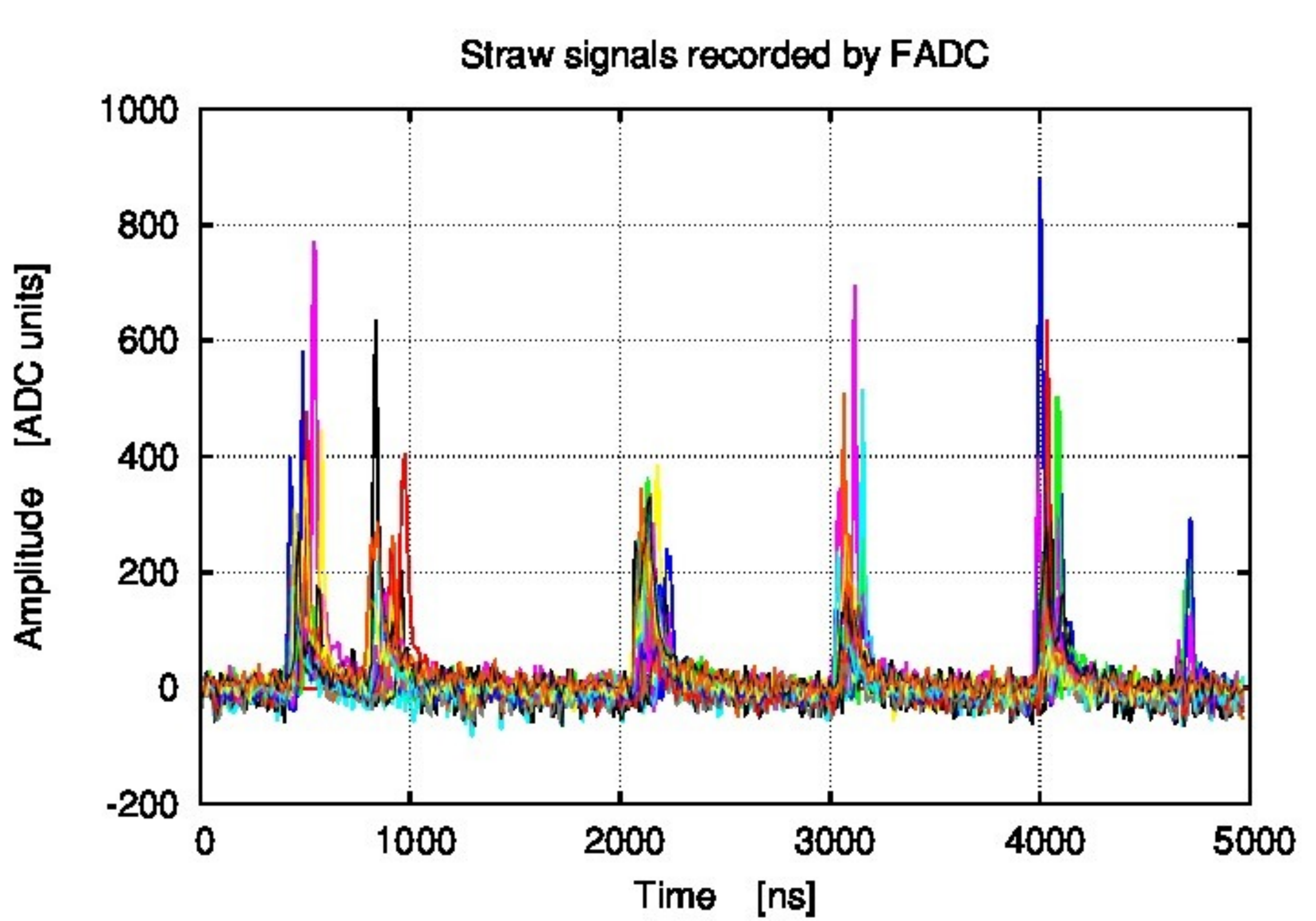}
\includegraphics[height=0.35\dwidth]{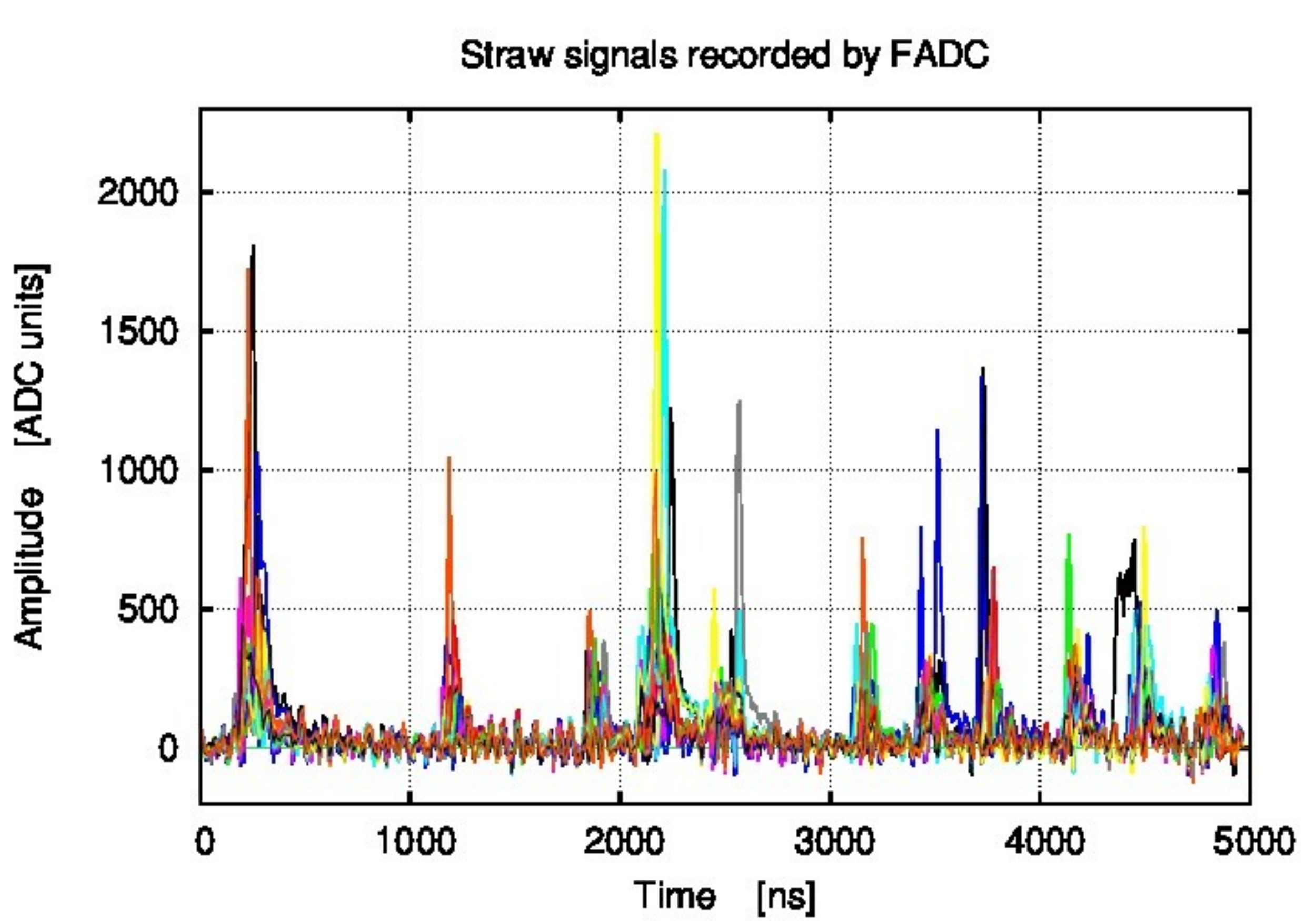}
\includegraphics[height=0.35\dwidth]{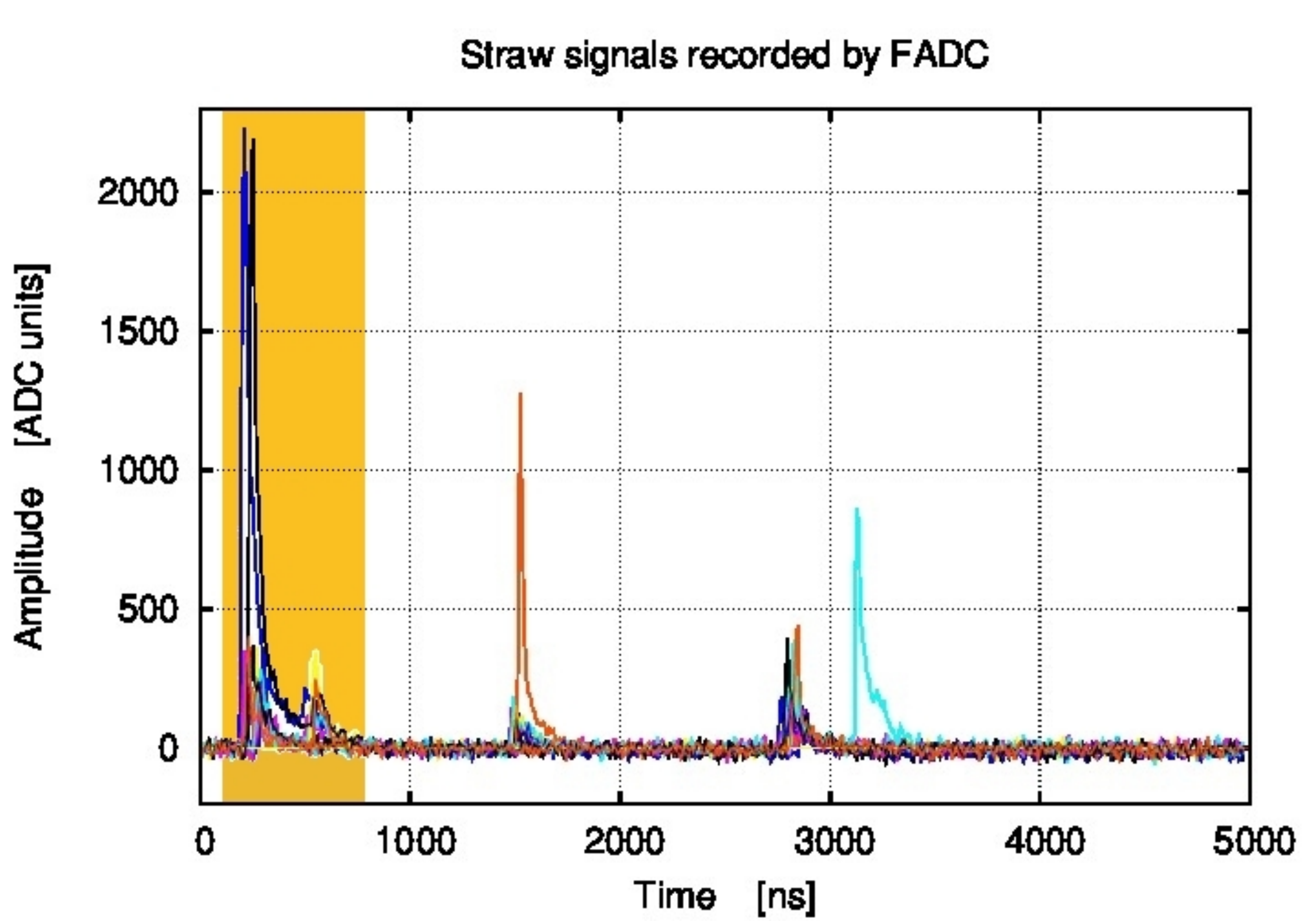}
\includegraphics[height=0.35\dwidth]{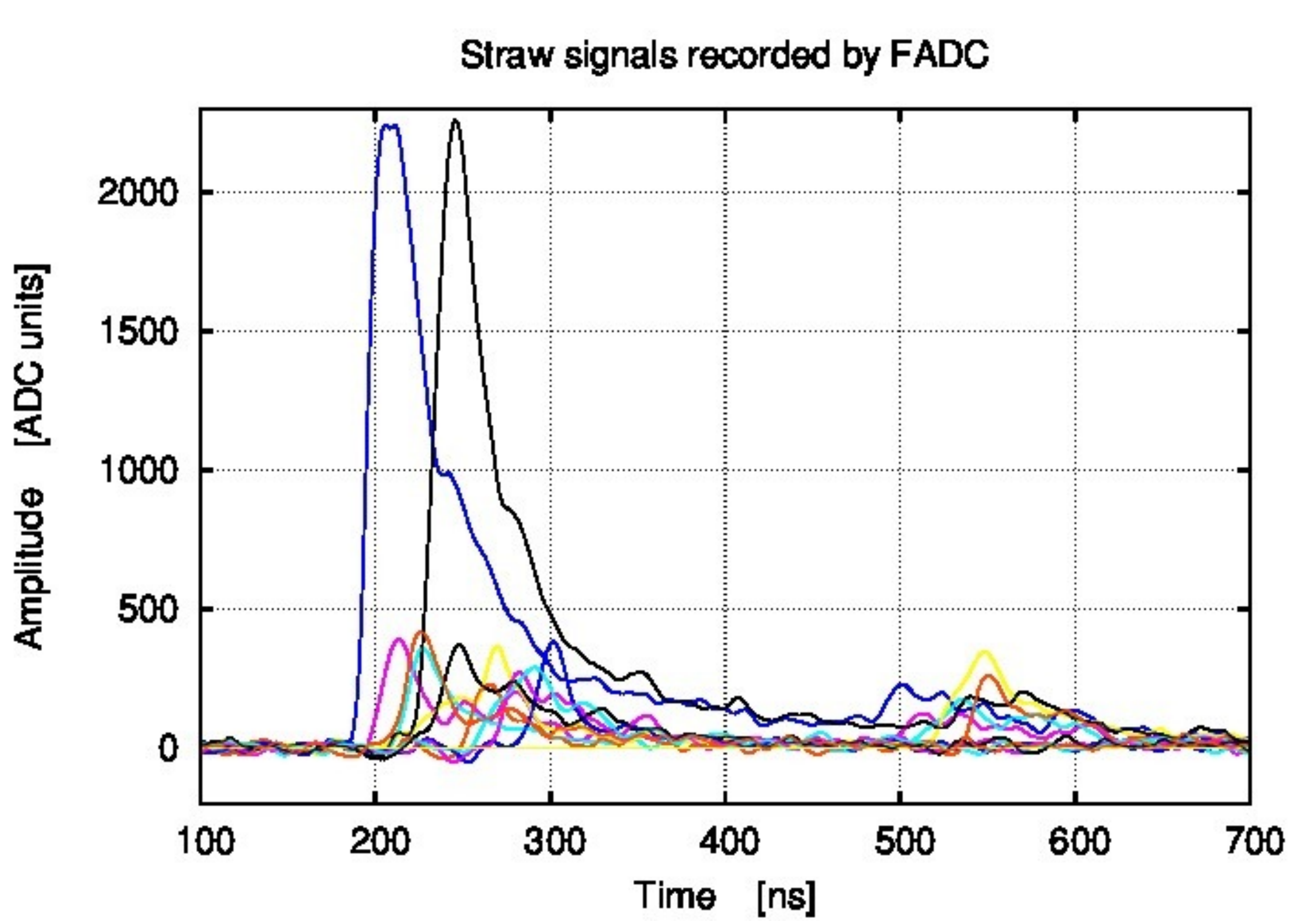}
\caption[Signals from different straw tubes recorded by the 240 MHz FADC]{Signals from different straw tubes recorded in 5 $\mus$ time windows by the 240 MHz FADC 
at high beam intensities. Top left: 6 particles crossings within the time window are visible, equiv. to $\simeq$1.2\,MHz rate. Top right:
Beam intensity of about 2.2 MHz. Bottom left: Example of signals pileup. The colored region is shown enlarged in the bottom right panel of the figure. No baseline shift or signal shape deterioration is visible.}
\label{fig:stt:pro:signals_high_intensity}
\end{center}
\end{figure*}

The short integration constant of the front-end electronics permits to observe an almost 
undistorted shape of the anode current signal of the straws. These shapes were recorded by the
FADC in time windows of 5\,$\mu$s length. 
\Reffig{fig:stt:pro:signals_high_intensity} shows the signals recorded for one complete layer 
(16 straws) of the straw prototype. The individual groups of the signals may consist of up to 16 components.
\noindent

%During the high intensity test none of the unfavorable phenomena previously mentioned 
%, expected when some ionization density 
%in the detector volume is exceeded, were observed. 
Even at a very high beam intensity of 2.2\,MHz, which is 
significantly beyond the expected rate during the operation of the \Panda--\Stt, the signal's baseline was stable and 
no onset of space-charge effects were recognized as well. At the normal operational voltage with particle fluxes 
of the order of 0.8 MHz/straw, both space- as well as energy resolution of the tracker will not be deteriorated.

%\subsection{Aging tests}
\section{Aging Tests}
%\COM{Author(s): P. Wintz}
\label{sec:stt:pro:aging}
A degradation of the straw tube properties like a specific gas gain reduction or high voltage instabilities during operation caused by 
irradiation is expressed as aging. In general, aging is induced by the plasma-chemical processes during the gas amplification processes with a 
high density of ions, electrons, excitation photons, free radicals and possible molecular dissociations and polymerizations. A complete 
overview and description of the aging phenomena in gaseous detectors can be found in \cite{bib:stt:des:hohlmann} which is a summary of a 
dedicated workshop with about 100 detector experts, held at DESY (Hamburg, Germany) in 2001. In the following, the main aspects relevant for 
the \PANDA-STT are discussed. 

\par
Two main sources of aging have been identified in wire chambers. A growth of polymeric deposits on the electrodes which can change the 
electric field, create sparking, produce dark- or even self-sustaining (Malter) currents. At high irradiation densities and high gas gains 
already trace contaminations on the sub-ppm level in the gas can lead to such deposits. Another aging source is a possible oxidation of the 
sense wire. Usually the wire is protected by an outer gold-plating layer which makes the wire highly inert to chemical reactions. If oxygen 
produced in the amplification avalanche penetrates through the gold-layer to the inner wire by permeation or at imperfection spots (holes) it 
can oxidize the wire with a swelling of the inner wire diameter and a cracking of the gold-plating layer \cite{bib:stt:des:ferguson}. The 
increased wire diameter reduces the gas gain at a given voltage by the lower electric field strength on the wire surface.      
A quantitative description of the aging process is difficult due to the high complexity with an influence for instance of the gas mixture and 
purity, trace contaminations, construction materials, gas flow, irradiation area and intensity, ionization density, high voltage setting, 
particle type and energy.

\par
The proposed Ar/CO$_2$ gas mixture is known as being one of the best gas mixtures for high-rate hadronic environments due to the absence of 
polymerization reactions of the components. Contaminations of the gas or detector materials with silicone, e.g. from lubricants must be 
avoided, since they produce a strong growth of non-volatile SiO$_2$ crystals on the wire.  An admixture of CF$_4$ to the gas can remove such 
SiO$_2$ deposits, but due to its high additional wire etching capability special care is needed. Hydrocarbons are better quenching agents 
compared to CO$_2$, but not considered for the \Panda--\Stt because of their high polymerization rate, which can lead to deposits on the 
electrodes. In particular deposits on the cathode can produce 
self-sustaining currents with a possible high voltage breakdown (Malter-effect) \cite{bib:stt:des:hohlmann}. 
In general a moderate gas gain of about 5$\times$10$^4$ is recommended which reduces the occurrence of limited streamer mode pulses with an 
increased avalanche size and possible accelerated aging \cite{bib:stt:des:titov}.    

\par
The behaviour of the straw tubes under very high irradiation was studied at COSY with a proton beam. The goal
was to check the influence of the beam exposure and charge deposition on the
straw gas gain, high voltage operation stability and to verify that all assembled materials
including the gas system do not create harmful pollution, e.g. by out-gassing. 
Within the short time of about 10 days beam irradiation it was possible to
collect a charge deposition in single tubes up to about 1.2\,C/cm equivalent to more than 5 years in 99.7\% of the STT volume
when operated in the \Panda detector at full luminosity. 

\par
The straw setup consisted of a planar
double-layer of 32 close-packed tubes installed behind
the COSY-TOF apparatus and exposed to the residual proton beam with a momentum
of about 3\,GeV/c. The straw design and all materials were the same as used for the 
COSY-TOF straw tracker assembly, i.e. 
30\,$\mu$m thick Mylar film tubes with 10\,mm diameter 
and a length of 105\,cm. For the \Panda detector the same straw tube design is
proposed, but with a length of 150\,cm. 
Due to the horizontal placement of the double-layer and a beam spot of about 2$\times $2\,cm$^2$ 
the particle rate through all tubes was almost
the same. The surrounding alignment frame consisted of sandwich bars with a
Rohacell core reinforced by Carbon fiber skins \cite{bib:stt:des:wintzaip}. Therefore, interaction
of the beam with this low-density foam material ($\rho$=0.05\,g/cm$^3$) was negligible.

\par
The gas supply was divided into four parallel gas lines, each serving eight straws.  
Thus, it was possible to test at the same time straws filled with
four different gas mixtures and gas gains with the same particle rates. 
The chosen gas mixtures were argon based, with
different fractions of CO$_2$ (10\,\% and 30\,\%) and one mixture with 10\,\%
ethane. The gas pressure for all mixtures was 1650\,mbar. 
The typical gas flow was one volume exchange per hour. 
In total, 16 high voltage supply channels (one channel per two straws) 
allowed to operate the straws at different voltage levels and gas gains. The current of every
voltage channel was monitored with a resolution of 2\,nA.
All straws were equipped with preamplifiers and 30\,m long signal cables
ending in the counting room. Therefore, it was possible to check analog
signal shapes and signal rates during beam irradiation for every straw.
\Reftbl{tab:stt:pro:tab2} lists the straw settings during the beam test.

\par
\begin{figure}%[ht]
\begin{center}
%\vskip 9cm
%\includegraphics[scale=0.4]{./stt/fig/Nhits_2atm.pdf}
%\includegraphics[width=\swidth]{./stt/fig/Nhits_2atm.pdf}
\includegraphics[width=\swidth]{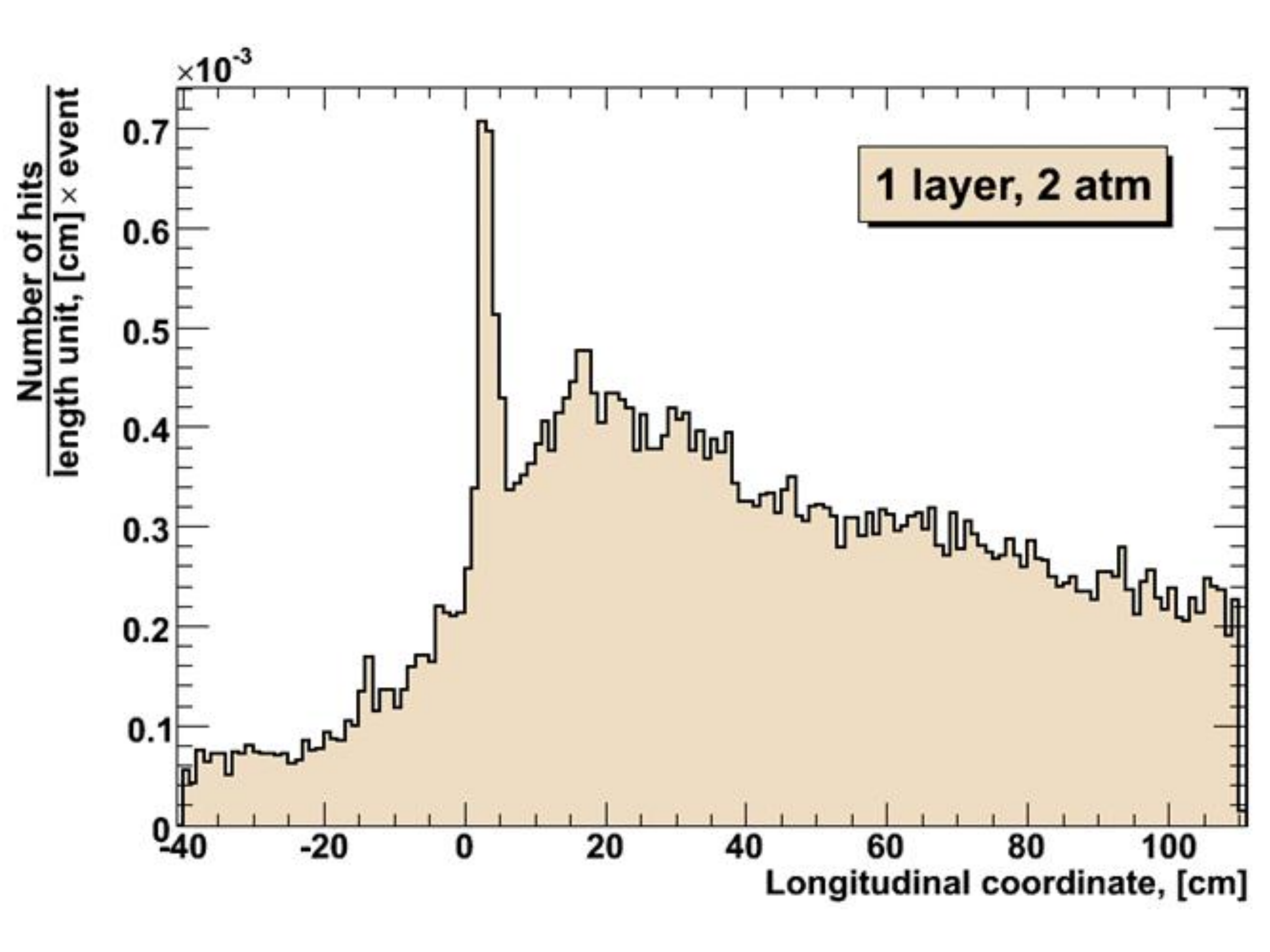}
\caption[Number of hits per event and cm along the tubes in the innermost straw layer]{Simulation of $\bar{p}p$ reactions giving the number of hits per
  event and per cm along the tubes in the innermost layer of the 
 \PANDA straw tube tracker. The
  target position is at z=0\,cm.}
\label{fig:stt:pro:ppsimu}
\end{center}
\end{figure}

The expected particle rates for the individual tubes in the \PANDA central
tracker volume were derived from a simulation of $\bar{p}
p$ interactions and assuming an event rate of
2$\times$10$^7$~s$^{-1}$ (see \Reffig{fig:stt:pro:ppsimu}). 
The mean particle flux for straws in the innermost layer 
was $\simeq$800\,kHz per 1500\,mm long tube and about $\simeq$7\,kHz/cm in
the forward region (z$>$0\,cm).
The maximum flux of $\simeq$14\,kHz/cm in the tube was concentrated within
z=2$\pm$1\,cm (target position at z=0\,cm) coming from $\bar{p}
p$ elastic interactions with a laboratory scattering angle $\theta \simeq $90$^o$ and
relatively low momentum. These particles crossing the tubes around z=2$\pm$1\,cm were 
highly ionizing and produced a high charge load of $\simeq$1\,C/cm, if one assumed a 
typical gas gain inside the tubes of 5$\times$10$^4$. 
At all other positions, which represent 99.7\,\% of the STT volume, the mean charge load was about 0.2\,C/cm.
All quoted charge loads were equivalent to an expected typical beam time for 
\Panda of one year with 50\,\% live-time.
\par

\par
The total live-time with
beam on the straws was 199 hours after correcting the COSY spill time
structure and beam breaks. 
All straws were exposed to the 
proton beam at the same longitudinal position, in the middle of each tube.  
The beam rate and cross section profile was measured by 
a scintillating fiber 
hodoscope placed behind the COSY-TOF apparatus and in
front of the straw setup. The derived proton intensity per 
straw diameter during extraction was about 2.3$\times$10$^6\,s^{-1}$.
The slightly lower pulse rate of $\simeq$2.0$\times$10$^6\,s^{-1}$ 
measured for the single straws could be explained by losses of low amplitude
signals due to the damping inside the 30\,m long cables.

\par
During the beam time no high voltage failures, dark currents or broken wires 
due to the high charge load were observed.
A high maximum current of a single straw wire of up to 2.3\,$\mu$A was measured.

\par
\begin{figure}%[ht]
\begin{center}
%\vskip 6cm
\includegraphics[width=\swidth]{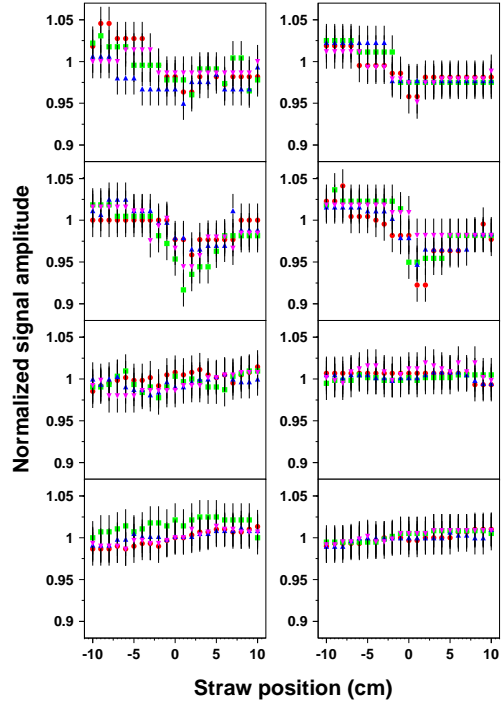}
\caption[Measured normalized gas gain reduction along the tube]{Measured normalized gas gain reduction ($\Delta$G/G$_0$) along the tube for all 32 straws, shown in groups of 4 straws. Straw no. 1--4 (upper left), straw no 29--32 (lower right). The beam hits all tubes around 0\,cm longitudinal position.}
\label{fig:stt:pro:amplitudes32}
\end{center}
\end{figure}

A possible gas gain reduction due to the proton beam irradiation 
was checked after the beam time by exposing all
straw tubes to a $^{55}$Fe radioactive source with 5.9\,keV $\gamma$-emission. 
In the argon-based gas mixtures the photo-absorption produces a localized
ionization spot with a characteristic number of about 220 electrons. 
Therefore, the recorded signal amplitude height was a direct measure
of the gas gain. The amplitude heights were checked for each straw at different
longitudinal positions around the beam irradiation spot and normalized to the amplitude heights far from the 
irradiation spot (see \Reffig{fig:stt:pro:amplitudes32}). A lower amplitude height indicates a reduction of the gas gain 
($\Delta$A/A$_0$=$\Delta$G/G$_0$). The estimated resolution error of the measurement was about 2\,\% of amplitude height.
\par

\begin{table*}%[ht]
\caption[List of straw settings and charge load during the beam test]{
List of straw settings and charge load during the beam test. The last column shows the normalized gas gain reduction in the irradiated straw region with a measurement resolution of about 2\,\%. The aging intervals give the minimum and maximum gain reductions, e.g. 0--7\,\% means that at least one straw showed no gain reduction and one a maximum of 7\,\%.}
\label{tab:stt:pro:tab2}
\par\bigskip

\hfill\vbox{{\halign
{\hfil#\quad\hfil&
\hfil#\quad\hfil&
\hfil#\quad\hfil&
\hfil#\quad\hfil&
\hfil#\quad\hfil\cr
\noalign{\smallskip\hrule\smallskip}
 
\hfil Straw \hfil & 
\hfil Gas mixture \hfil & 
\hfil Voltage \hfil & 
\hfil $\sum Q$ \hfil &
\hfil Aging \hfil \cr
\hfil no. \hfil & 
\hfil  \hfil & 
\hfil (V) \hfil & 
\hfil (C/cm) \hfil &
\hfil  $\Delta$G/G$_0$\hfil \cr
\noalign{\smallskip\hrule\smallskip}
 1--8 & Ar/CO$_2$~(10~$\%$) & 1750 &  0.72 & 0--3\,\% \cr
 9--16 & Ar/CO$_2$~(10~$\%$) & 1700 &  0.58 & 0--7\,\% \cr
 17--20 & Ar/CO$_2$~(30~$\%$) & 2200 &  1.23 & no \cr
 21--24 & Ar/CO$_2$~(30~$\%$) & 2100 &  0.79 & no \cr
 25--32 & Ar/C$_2$H$_6$~(10~$\%$) & 1550 & 0.87 & no \cr
\noalign{\smallskip\hrule\smallskip}
}}}
\hfill\break\par
\end{table*} 

\par
It can be seen that for all straws filled with 30\,\% CO$_2$ or 10\,\% ethane in argon no gas gain reduction was measured, even for the highest charge loads up to 1.2\,C/cm. Some but not all straws filled with 10\,\% CO$_2$ in argon showed a small gas gain loss of up to 7\,\% at the beam irradiation spot. A clean spatial correlation between the reduced gas gain and beam intensity distribution, measured by the scintillating fiber hodoscope in front of the straws, was observed. 
The results of the gas gain measurement together with the total charge loads for all 32 straws are summarized in \Reftbl{tab:stt:pro:tab2}.    

\par
The absence of any aging in the straws filled with ethane or the higher CO$_2$ percentage in argon indicated no general problem with the gas purity, and a pollution by the used straw materials or gas system could be excluded. The small gas gain reduction observed only for some of the straws operated with the lower 10\,\% CO$_2$ admixture might be explained by the known poor quenching capabilities of CO$_2$, together with the very high irradiation perpendicular to the wire and concentrated at a small spot of about 2\,cm along the wire during the beam test. Due to the incomplete avalanche quenching the occurrence of limited streamer mode pulses, with the characteristic double-peak signal shape, was higher for that gas mixture. The high ionization density with a large number of produced oxygen ions and radicals increased the probability of oxygen permeation through the gold-layer to the inner wire. The oxidation of the inner tungsten-rhenium wire caused a swelling of the wire diameter, and as a result the electric field strength at the wire surface was reduced (E\,$\propto$\,1/r) which lowered the gas gain at the same high voltage setting. 
Since the observed gas gain reduction was very small the occurring aging processes were rather weak. To clearly identify the sources of aging, dedicated investigations with a higher charge load over a much longer time period would be needed.

\par
Ar/CO$_2$ is the preferred gas mixture for the \Panda--\Stt since it is highly tolerant to highest irradiation, not expensive, and non-flammable. The measurements confirm that the straw design and all used materials are suited and will not limit the life time of the detector. No aging in the straws is expected at moderate gas gains of about 5$\times$10$^4$ for 99.7\,\% of the STT volume during more than 5 years of \Panda operation at full luminosity. A small aging on the low percent level may start first in the region at z=2$\pm$1\,cm (= 0.3\,\% of the STT volume) after about 2 years of operation, caused by low energy protons from elastic scattering. The modular mechanical design of the \Panda--\Stt allows to replace even single straws showing aging or other failures inside the layer modules after some years of operation during the \Panda maintenance time.

\section{The COSY-TOF Straw Tube Tracker}
\label{sec:stt:pro:cosystt}
%\COM{Author(s): P. Wintz}
\par
The technique of pressurized, self-supporting straw tube layers was first developed for the Straw Tube Tracker of the COSY-TOF 
experiment (COSY-STT) at the COSY-accelerator (J\"ulich, Germany). The used straw tube materials and dimensions, and the geometry of planar, close-packed multi-layers are the same or quite similar as for the \Panda--\Stt. Although the COSY-STT is a non-magnetic spectrometer, the calibration methods for the straw tube positions and isochrone radius - drift time relation are similar for both detectors. The operation of the COSY-STT with about 275\,l gas volume in surrounding vacuum is an outstanding technical challenge. The required minimal leakage of the detector in vacuum is a strong and sensitive proof of all straw materials, glueing and assembly techniques, which are also crucial for the \Panda--\Stt. The COSY-STT is considered to be a global test system for the \Panda--\Stt and its properties and performance results are summarized in the following. 
\par
The COSY-STT was installed in 2009 as an upgrade of the COSY-TOF spectrometer, which consists of a large 25\,m$^3$ vacuum barrel with a liquid hydrogen target cell at the entrance, followed by a start detector, silicon-microstrip detector, the straw tube tracker (STT), and scintillator hodoscopes covering the barrel walls and end cap. The apparatus allows to measure kinematically complete the time-of-flight and space directions of the reaction particles of hyperon production in proton-proton and proton-deuteron collisions with polarized proton beam. The vacuum ensures lowest background produced by beam and reaction particles with up to 3.5\,m track lengths. More details about the experimental program and the STT installation can be found in \cite{bib:stt:pro:cosy-tof-prop} and \cite{bib:stt:pro:cosy-stt-prop}. A first experiment beam time of hyperon production with polarized proton beam was carried out in 2010. 
\par
\begin{figure}%[ht]
\begin{center}
\includegraphics[width=\swidth]{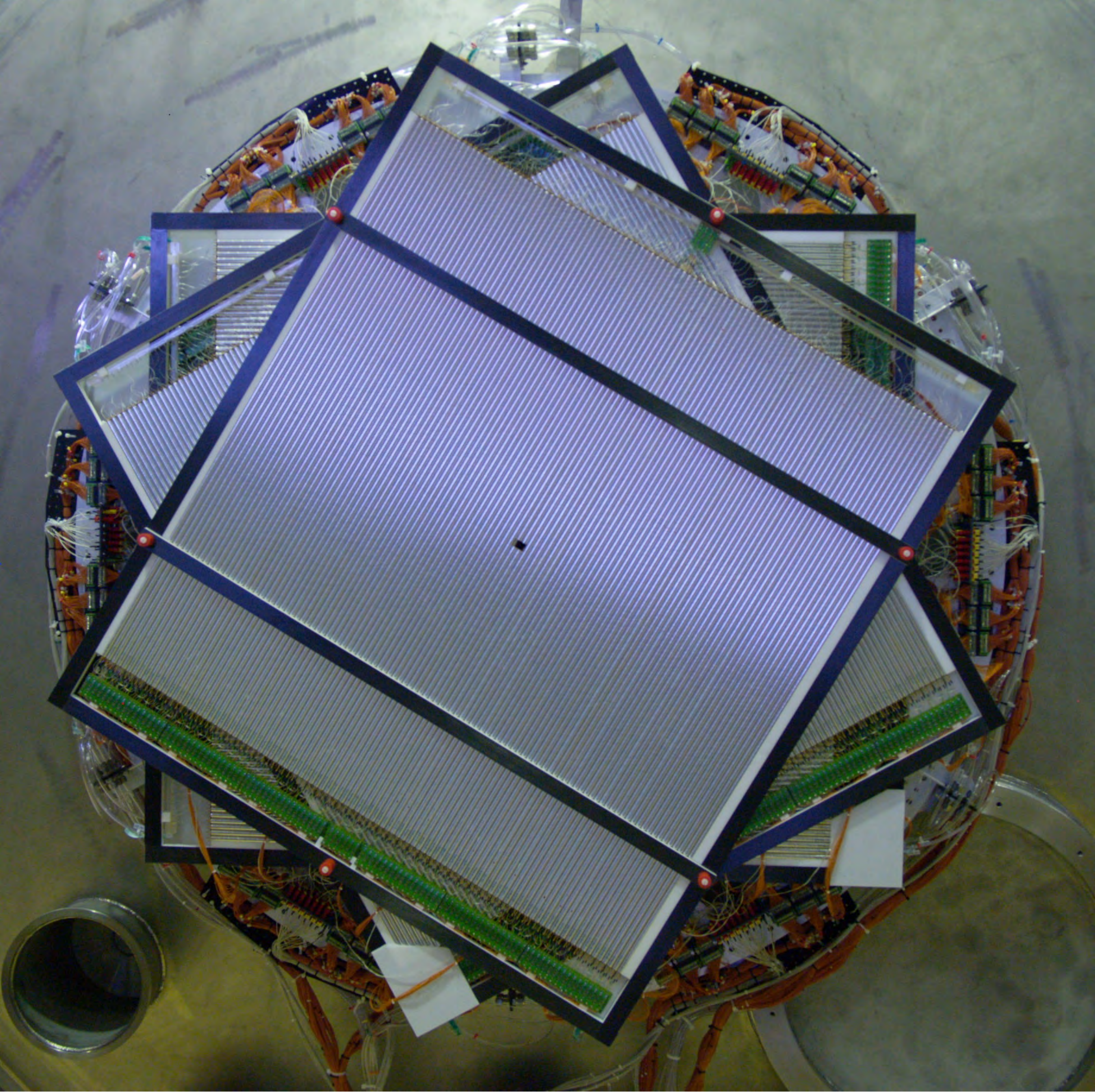}
%\vspace {-1.5cm}
\caption[The COSY-STT mounted at the front cap of the COSY-TOF spectrometer]{The COSY-STT mounted at the front cap of the COSY-TOF spectrometer. The detector consists of 2704 straw tubes of 1\,m length and 10\,mm diameter, arranged as a vertical stack of 13 close-packed double-layers at three different orientations.}
\label{fig:stt:pro:cosy_stt_1}
\end{center}
\end{figure}

The COSY-STT consists of 2704 straw tubes, each with a length of 1050\,mm, inner diameter of 10\,mm, and 32\,$\mu$m wall thickness of aluminized Mylar film. The tubes are arranged as a vertical stack of 13 close-packed double-layers with three different orientations ($\phi$=0$^\circ$, 60$^\circ$, 120$^\circ$) for a 3-dimensional track reconstruction. A 15$\times$15\,mm$^2$ beam hole in the center of every double-layer is realized by splitting the 4 central straws into 
8 straws with about half length (see \Reffig{fig:stt:pro:cosy_stt_1}).
The straws are filled with a gas mixture of Ar/CO$_2$ (80/20\,\%) at a pressure of 1.25\,bar. The typical operation voltage is 1840\,V. The electronic readout consists of low-power trans-impedance preamplifiers directly connected to each straw in vacuum and feeding the signals through 13\,m coaxial signal cables to ASD8B-discriminators and TDCs, which are located outside the 
vacuum barrel. 

\par
\begin{figure}%[ht]
\begin{center}
\vspace {-3cm}
\includegraphics[width=\swidth]{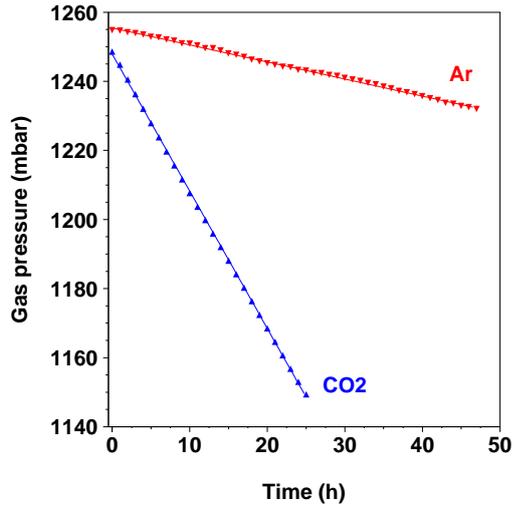}
\caption[Gas leakage of the COSY-STT]{Gas leakage of the COSY-STT filled with pure argon (red) and pure CO$_2$ (blue), measured by the gas pressure drop of the straws in surrounding vacuum.}
\label{fig:stt:pro:cosy_stt_2}
\end{center}
\end{figure}

The COSY-STT is now since about three years in surrounding vacuum and no real leakage sources of the detector, caused by dissolving glue spots, brittle materials, or loose gas connections, have been observed. The gas leakage stays on the permeation level, which is caused by the flow of the gas molecules inside the straws through the thin Mylar film wall to outside vacuum. \Reffig{fig:stt:pro:cosy_stt_2} shows the gas loss by measuring the pressure drop inside the straws in surrounding vacuum if the STT is filled with pure argon and pure CO$_2$. The difference in the gas loss rate for argon and CO$_2$ of about a factor of 10 is characteristic for the different permeation of the specific gas molecules through the Mylar film and in accordance with reference measurements by the manufacturer (DuPont Teijin Films, USA).    
For the used gas mixture of Ar/CO$_2$ (80/20\,\%) the total leakage is about 2\,\% of the STT volume per day. The typical gas flow during the high voltage operation is about four times the STT volume per day (=1000\,l/day). 

\par
\begin{figure}%[ht]
\begin{center}
\vspace {-3cm}
\includegraphics[width=\swidth]{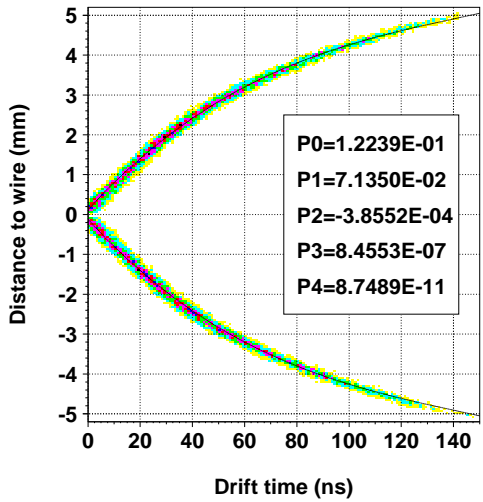}
\caption[Track to wire distances and measured drift times for the reconstructed tracks]{Track to wire distances and measured drift times for the reconstructed tracks. The r$_{iso}$(t)-relation (black line) is parametrized as a polynomial function of 4th order with the parameters P$_0$--P$_4$.}
\label{fig:stt:pro:cosy_stt_3}
\end{center}
\end{figure}

The calibration of the STT consists of the determination of the isochrone radius - drift time relation and the adjustment of the straw positions and is performed as an iterative procedure. At first, the isochrone - drift time relation (r$_{iso}$(t) in the following) is parametrized as a polynomial function of 4th order and obtained by an integration of the time offset corrected drift time spectrum (see  \Refsec{sec:stt:rtcurve}).
\begin{equation}
r_{iso}(t) = \sum_{i=0}^{4} P_i \times t^i 
\end{equation}
Then, tracks are reconstructed as straight lines with a least squares fit ($\chi^2$) to the isochrones calculated from the measured drift times using the defined r$_{iso}$(t)-relation. \Reffig{fig:stt:pro:cosy_stt_3} shows the distances to the fired straw wires versus the measured drift times for all reconstructed tracks. A systematic deviation in the track distance for single straws or straw groups from the expected r$_{iso}$(t)-relation is corrected by adjusting the straw position accordingly. Here, the assembly technique of the STT simplifies the position calibration to a large extent. Individual deviations of single tubes in the close-packed double-layers are not possible and only the vertical position of the 13 double-layers have to be adjusted. The track reconstruction is repeated using the new straw layer positions, the distances are checked and the positions are corrected again until the systematic deviations vanish. Finally, also the r$_{iso}$(t)-relation is verified by a new parameter fit of the reconstructed track to wire distances to the measured drift times.

\par
\begin{figure}%[ht]
\begin{center}
\vspace {-3cm}
\includegraphics[width=\swidth]{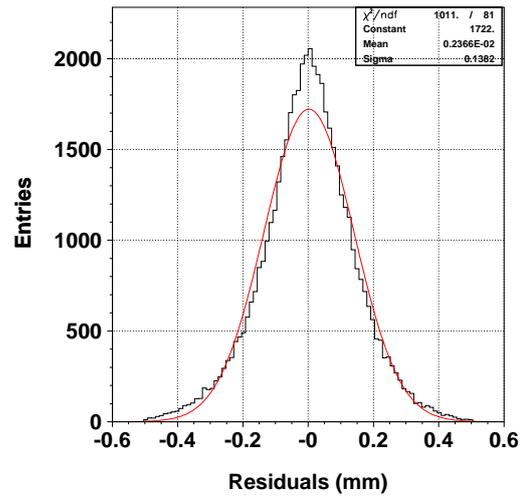}
\caption[Distribution of the residuals of all reconstructed tracks]{Distribution of the residuals of all reconstructed tracks as a measure of the COSY-STT spatial resolution. The width of 138\,$\mu$m ($\sigma$) and mean of 2\,$\mu$m are the results from the Gaussian fit (red line).}
\label{fig:stt:pro:cosy_stt_4}
\end{center}
\end{figure}

The distribution of the finally obtained residuals of the reconstructed tracks to the isochrones is a measure of the spatial resolution 
of the STT and is shown in \Reffig{fig:stt:pro:cosy_stt_4}. Only a simple filter for single hits from delta-electrons with large distortions to the fitted track has been applied. No drift time correction due to the signal propagation time along the wire and the particle time-of-flight have been made. The estimated drift time error is about $\Delta$t=2\,ns. Also the reconstruction of a straight line track does not take into account multiple-scattering inside the STT which contributes to a maximum of about 100$\,\mu$m for the first and last layers. The spatial resolution of the STT is given by the width of the residual distribution, which is 138$\,\mu$m ($\sigma$) for the gas mixture of Ar/CO$_2$ (80/20\%) at an absolute pressure of 1.25\,bar. The shape of the distribution is nicely symmetric with a low mean of 2\,$\mu$m, showing no distortion by additional systematic errors. 
\par
\begin{figure}%[ht]
\begin{center}
%\vspace {-3cm}
\includegraphics[width=\swidth]{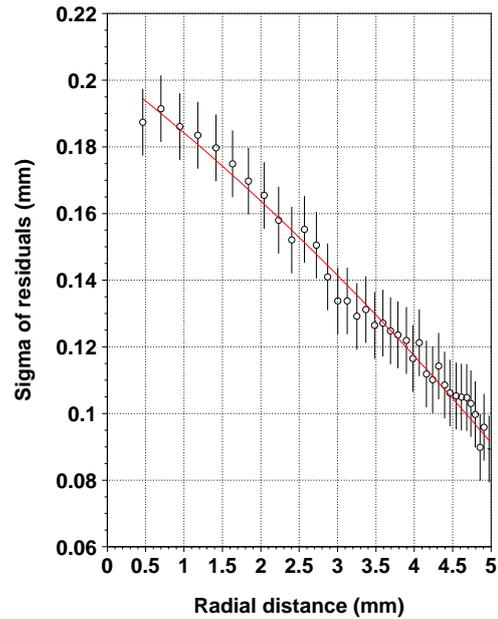}
\vspace {-1cm}
\caption[Width of the residual distributions for different intervals of the radial distances to the wire]{Width (sigma) of the residual distributions for different intervals of the radial distances to the wire.}
\label{fig:stt:pro:cosy_stt_5}
\end{center}
\end{figure}

The variation of the spatial resolution depending on the radial distance to the wire is shown in \Reffig{fig:stt:pro:cosy_stt_5}. 
Close to the wire the resolution is about 190\,$\mu$m, dominated by the primary ionization cluster spacing and time jitter together with higher drift velocities. Both effects are reduced more and more for larger distances to the wire and the resolution improves to about 100\,$\mu$m close to the straw cathode, where the electron diffusion during their drift to the anode is the limiting factor.
\par
\begin{figure}%[ht]
\begin{center}
\includegraphics[width=\swidth]{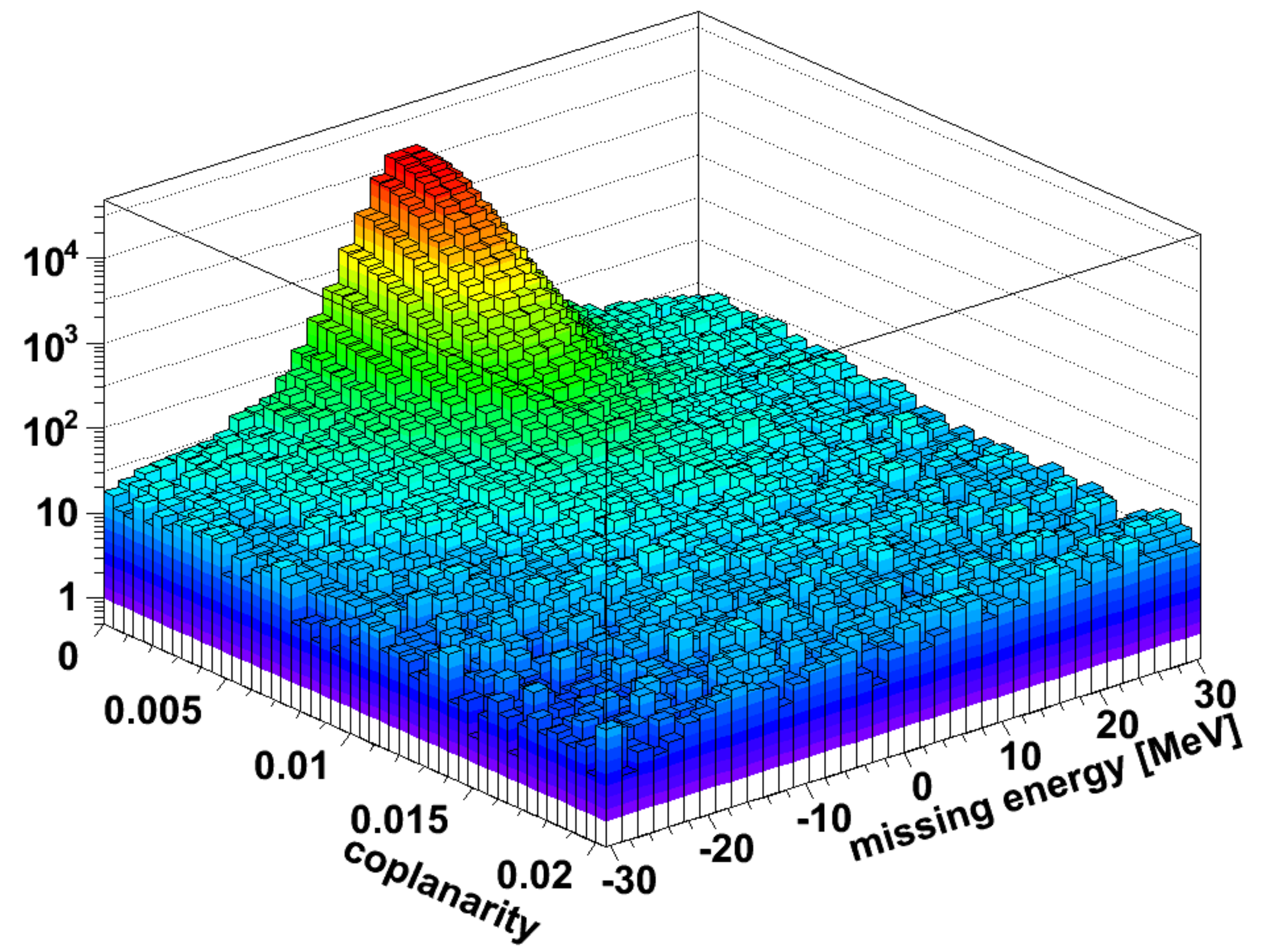}
\caption[The distribution of the missing energy in the primary vertex and the coplanarity (C) shows a clear peak from pp elastic scattering events]{The distribution of the missing energy in the primary vertex and the coplanarity (C) shows a clear peak from pp elastic scattering events.}
\label{fig:stt:pro:ppelastic}
\end{center}
\end{figure}

Since the installation of the STT in the COSY-TOF spectrometer several experiments with different momenta of the polarized proton beam have been carried out to study the hyperon production in proton-proton collisions. The performance of the track reconstruction with the STT can be studied by the analysis of the $pp\rightarrow pp$ elastic scattering process, which is used for the calibration of the detectors and for a determination of the luminosity in the experiment. The $pp\rightarrow pp$ elastic scattering event kinematics can be calculated from the measured direction of the two reconstructed tracks ($\hat p_{1,2}$)
and the known beam momentum ($p_{\text{beam}}$).
Requiring momentum conservation, the violation of the energy conservation
in the process is a strong selection criterion against physical background processes like $pp \to d\pi^{+}$ or final states with
higher track multiplicities (e.g. uncharged $\pi$). The latter are
also tested by the coplanarity (C) defined as
\begin{equation*}
  \label{eq:30}
C= \left|\left(\hat p_{1} \times \hat p_{2} \right) \cdot \hat
  p_{\text{beam}} \right|.
\end{equation*}
which measures how well the final state particles are reconstructed back to
back in the center of mass system.
\Reffig{fig:stt:pro:ppelastic} shows the coplanarity (C) versus the missing energy for an event sample
of 7.6 million triggered events. Due to the high reconstruction precision of the STT the signal events are
strongly peaked around (0, 0). A sample of 420\,000 elastic
scattering events can be selected with a circular cut around the peak
center. From the surrounding bins a background contamination of the
event sample from inelastic scattering can be determined to be lower than
$0.45\%$

\par
The results obtained from the COSY-STT can be extrapolated to the \PANDA-STT. Both detectors have a similar material budget and number of straw layers for the tracking. The main differences are the operation of the \PANDA-STT inside a solenoid field and at a higher straw gas pressure of about 2\,bar. The additional Lorentz force will change the radial drift path for the electrons inside a straw to a longer, spiral drift path and increased drift times. Still the isochrones have a cylindrical shape, only the r$_{iso}$(t)-relation will be different. The higher gas pressure will increase the maximum drift times and the ionization density which improves the spatial resolution. Therefore, assuming a comparable resolution of the drift time measurement of about $\Delta$t=2\,ns the spatial resolution of the \PANDA-STT is expected to be better than 140\,$\mu$m.

%
%EOF: panda_tdr_stt_pro.tex

%
% Bibliography for this chapter (remove %)
%
\bibliographystyle{panda_tdr_lit}
\bibliography{./stt/lit_stt}
% EOF

%
% STT TDR
% File for chapter 5
\chapter{Simulations}
% FILE: panda_tdr_stt_tub.tex
%
\section{The Single Straw Tube Simulation}
%\COM{Author(s): A. Rotondi}
\label{sec:stt:tub}

\subsection{The Charge Released into the Tube}

 We have performed a detailed simulation of the charge generation
 and collection in a single straw tube.

\begin{figure}[h!]  
\begin{center}
\includegraphics[width=\hwidth]{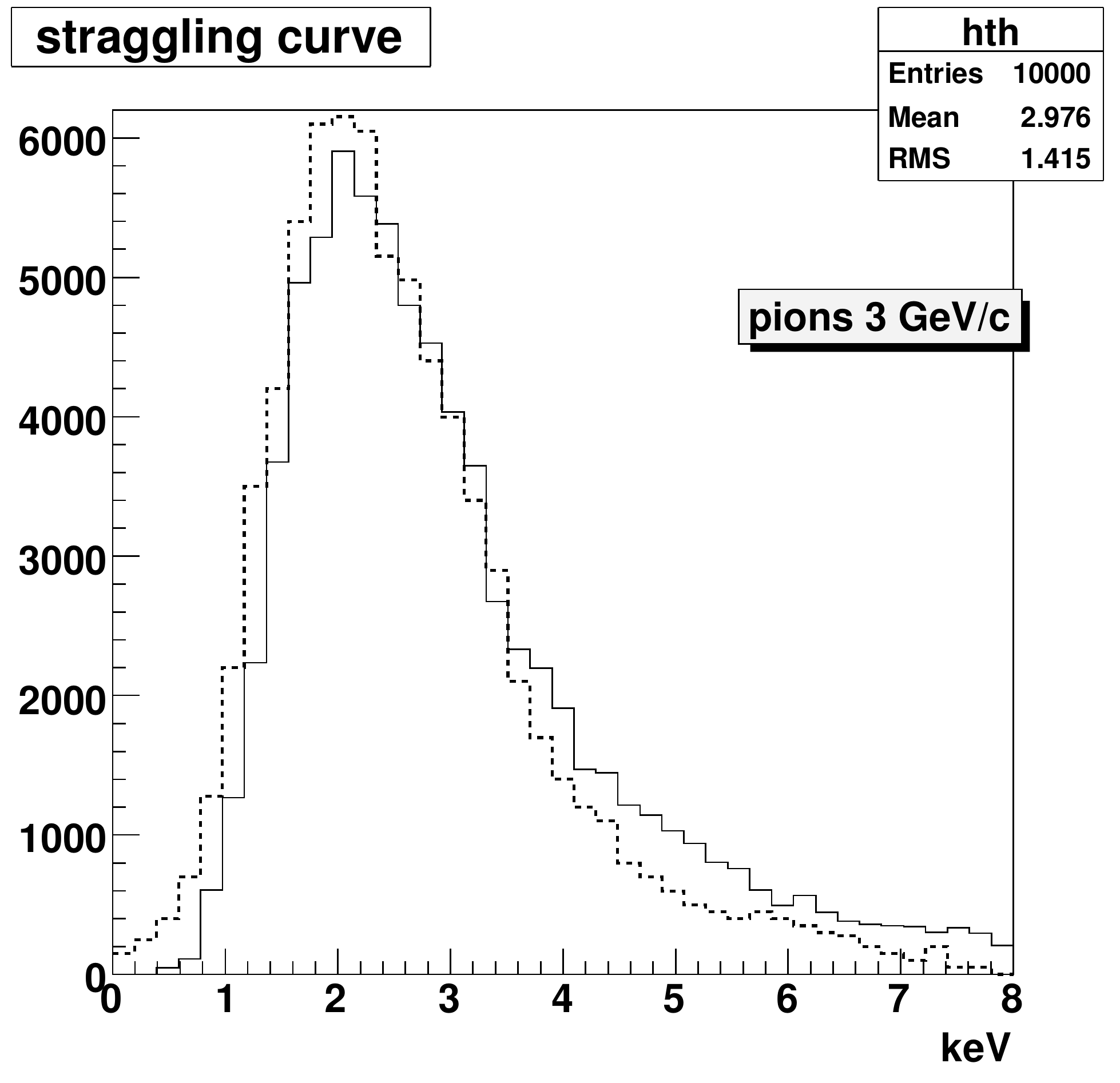}
\includegraphics[width=\hwidth]{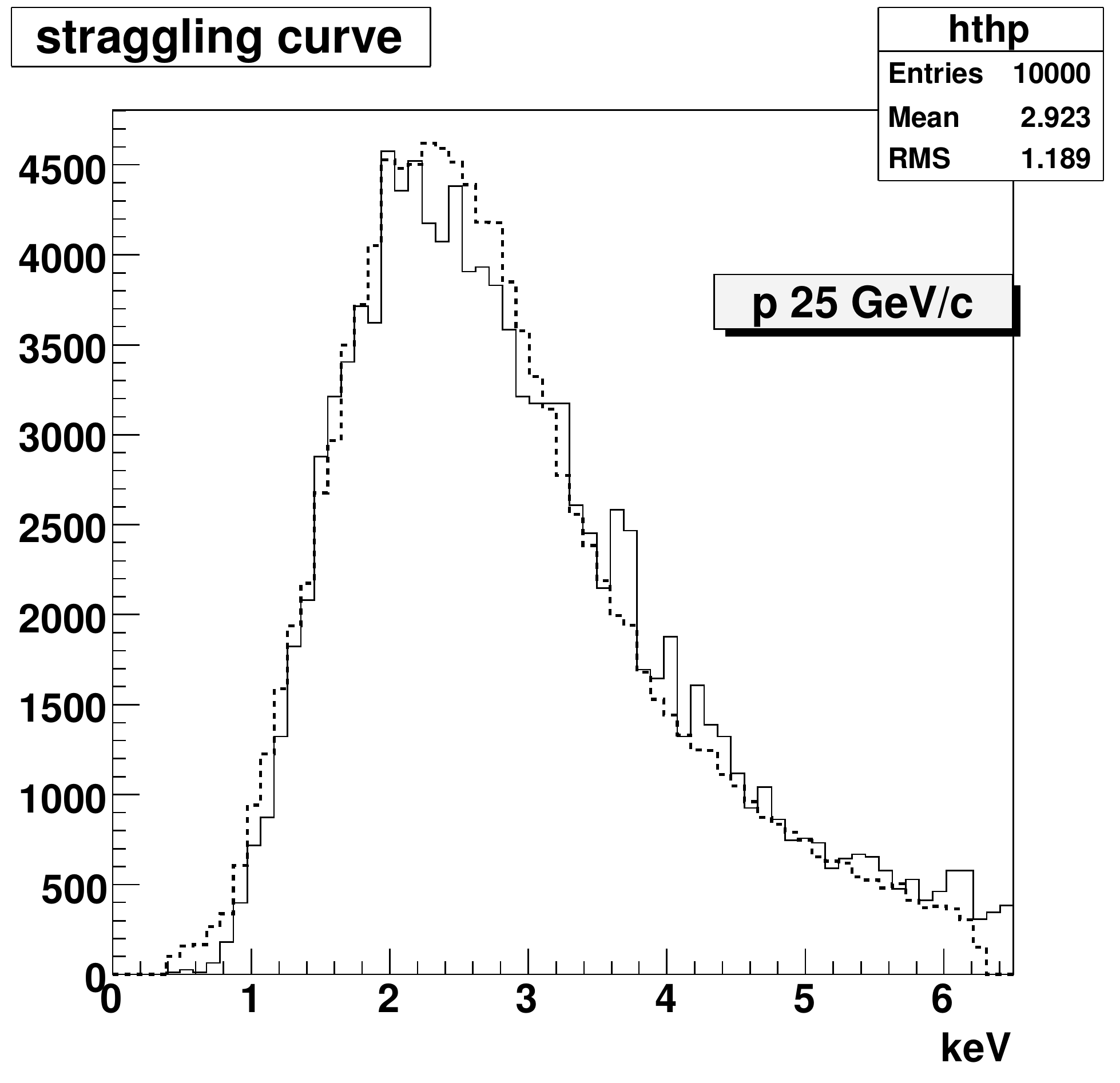}
\end{center}
\vspace*{-0.6cm}
\caption[Comparison between the simulation of the energy lost in a $1.5\,\cm$ Ar/CO$_2$ layer and the experimental values of Allison et al. \cite{bib:stt:tub:allison}]{Comparison between the simulated energy loss
  in a $1.5\,\cm$ Ar/CO$_2$ layer (line) and
  the experimental values of Allison et al.
 \cite{bib:stt:tub:allison} (dotted line).} \label{fig:stt:tub:onept}   
\end{figure}

 In correspondence of an incident charged particle,
 we sample from the exponential distribution 
 the point where an electron cluster is generated and from
 the proper distribution (see below)  the 
 number of electrons in the cluster. By stopping when the
 particle leaves the tube, we
 have the number of free electrons generated from a 
 poissonian number of clusters. 
 The mean number of clusters/cm is taken from \cite{bib:stt:tub:sauli} 
 (25 for Ar and 35.5 for CO$_2$).
 For the reliability of the simulation,
 it is crucial to know  the cluster size distribution, i.e. the 
 number of  electrons per cluster. We use
 the theoretical calculations of  
 \cite{bib:stt:tub:lapique} for Ar and the experimental data 
 on Ar and CO$_2$ from \cite{bib:stt:tub:fischle}. The comparison 
   with some available results in gas has shown that this
 choice is in reasonable agreement with the data (see \Reffig{fig:stt:tub:onept}).
%
%%%%%%%%%%%%%%%%%%%%%%%%%%%%%%%%%%%%%%%%%%%%%%%%%%%%%%%%%%%%%%%%%%%%%%
%%%%%%%%%%%%%%%%%%%%%%%%%%%%%%%%%%%%%%%%%%%%%%%%%%%%%%%%%%%%%%%%%%%%%%%
%
By knowing the
 mean value of the energy spent per free electron (i. e. to create an electron-ion pair),
 the overall
 energy loss of the projectile on the whole path can be calculated. 
 The assumed values  are 27\,eV for Ar and 33.5\,eV for
 CO$_2$ \cite{bib:stt:tub:sauli}.

%
%%%%%%%%%%%%%%%%%%%%%%%%%%%%%%%%%%%%%%%%%%%%%%%%%%%%%%%%%%%%%%%%%%%%%%
\begin{figure*}[h!]  
\begin{center}
\includegraphics[width=0.7\dwidth]{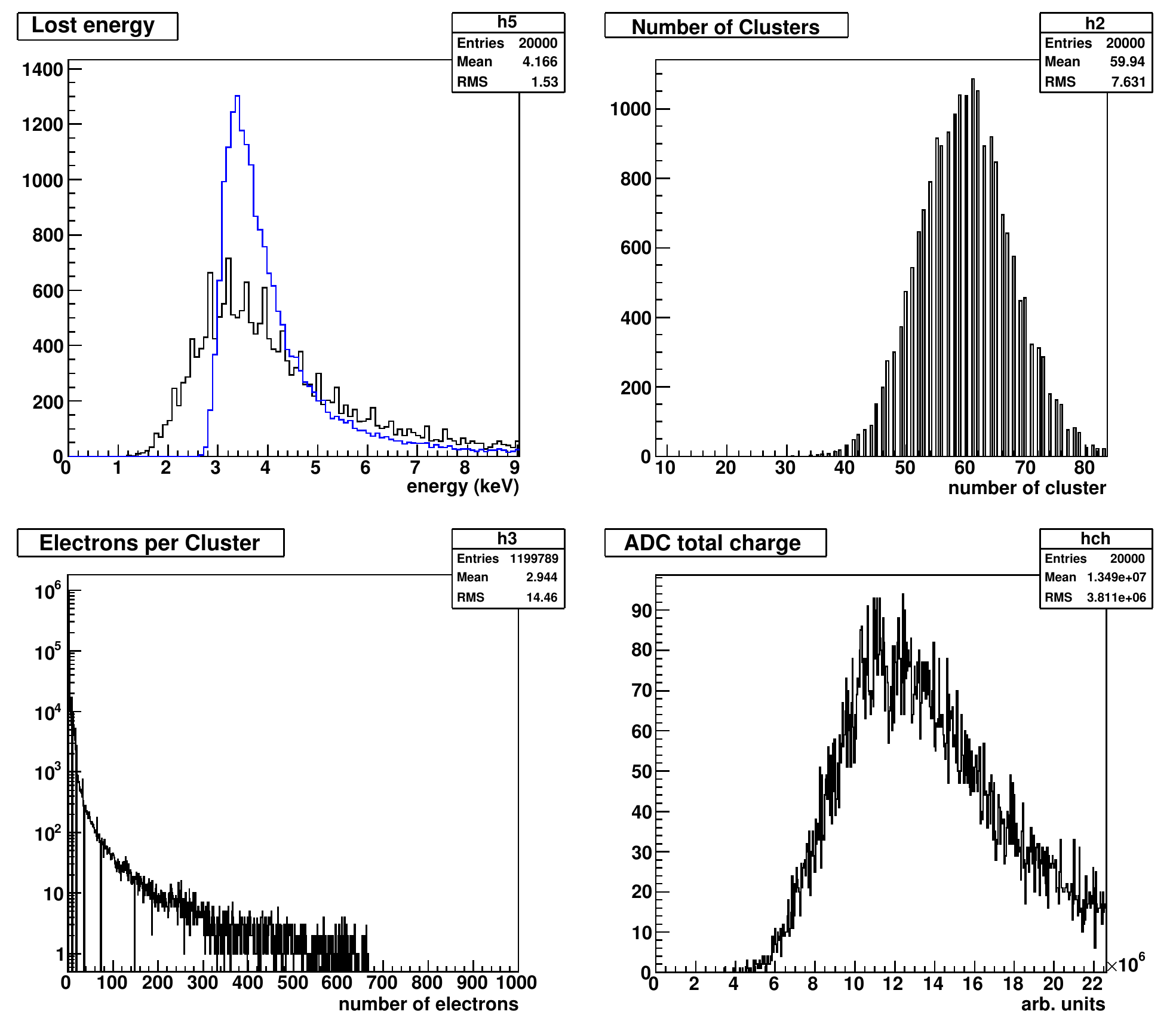}
\end{center}
\vspace*{-0.4cm}
\caption[Results of the single tube simulation]{Results of the single tube simulation for a $1\,\gev$ pion in a
 $2\,\atm$ pressure straw tube with a 90/10 Ar/CO$_2$ gas mixture.
 Upper left: energy lost in a tube compared with the 
 sharper Landau distribution; upper right: poissonian 
 distribution of the number of clusters; bottom left: cluster size
 distribution calculated as discussed in the text; bottom right:
 charge collected on the wire assuming a multiplication mechanism
 from the Polya distribution. By multiplying the
 number of clusters with the mean number of electrons per cluster,
 a mean
  number of primary electrons of about 200 is obtained.} \label{fig:stt:tub:genfig}   
\end{figure*} 
%%%
%

\begin{figure}[h!]  
\begin{center}  
\includegraphics[width=0.75\swidth]{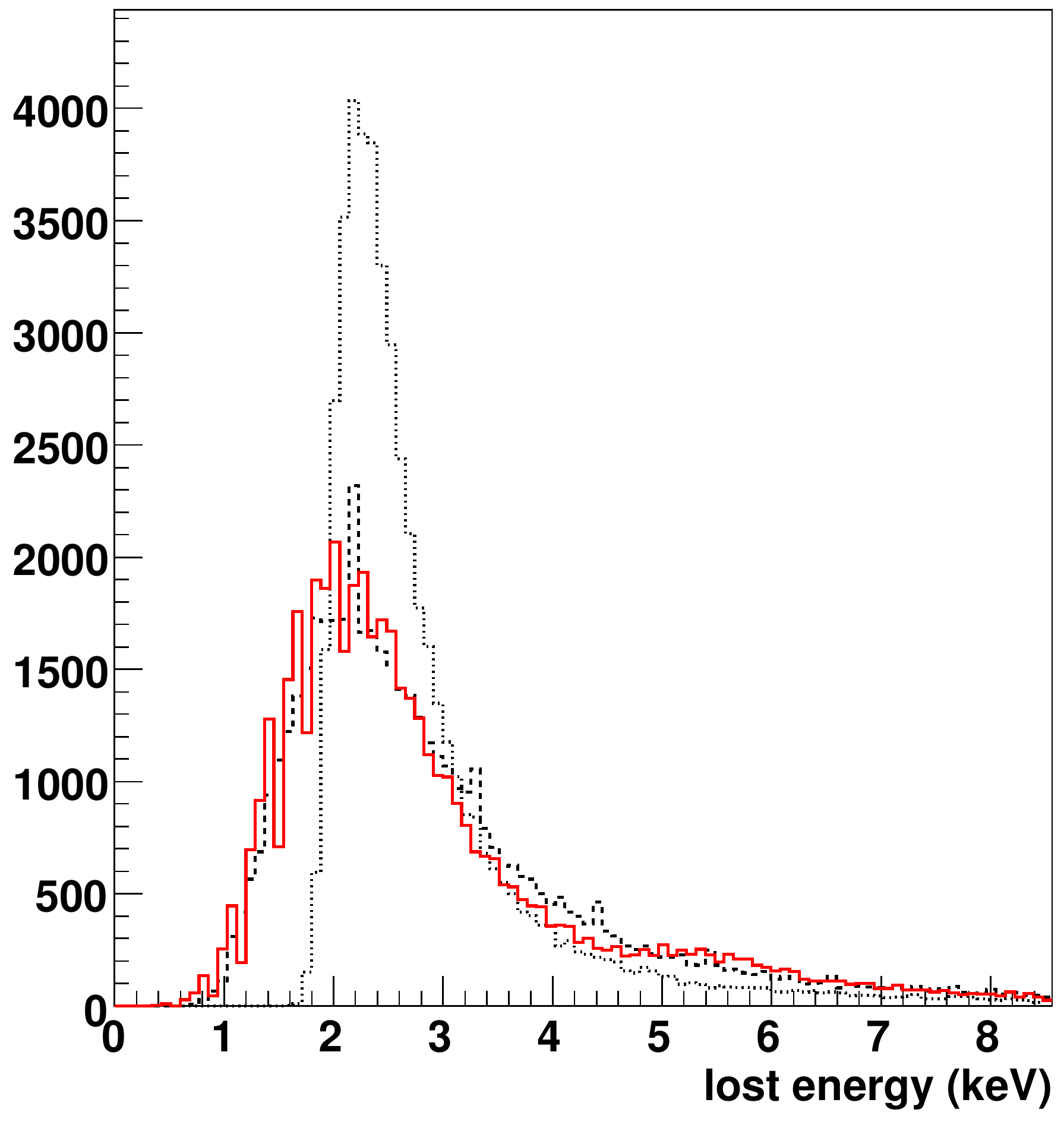}
\end{center}
\vspace*{-0.5cm}
\caption[Energy loss of a $1\,\gev$ pion in $1\,\cm$ ArCO$_2$ (9/1) gas mixture]{Energy loss of $1\,\gev$ pion
 traversing a $1\,\cm$ of 90\percent Ar, 10\percent CO$_2$ gas mixture at NTP.
 Solid line: Urban distribution; dashed line: specific simulation 
 model; dotted line: Landau distribution.} \label{fig:stt:tub:onept1}  
\end{figure}

 As a further check, we compared the energy lost in the tube, for a variety
 of projectiles and energies, with the 
 Urban model \cite{bib:stt:tub:urban}, which is used in GEANT3 and GEANT4 
 in the case of gaseous thin absorbers \cite{bib:stt:tub:geant3,bib:stt:tub:geant4}.
 The results, reported in \Reffig{fig:stt:tub:onept1},
 show good agreement with our simulation.
%
%%%%%%%%%%%%%%%%%%%%%%%%%%%%%%%%%%%%%%%%%%%%%%%%%%%%%%%%%%%%%%%%%%%%%%
%%%%%%%%%%%%%%%%%%%%%%%%%%%%%%%%%%%%%%%%%%%%%%%%%%%%%%%%%%%%%%%%%%%%%%%
%
%
\subsection{The Drift Process from GARFIELD} \label{sectgarf}
The tube response has been studied in detail giving the tube size, wire radius, high-voltage, gas mixture and magnetic field as input to the GARFIELD \cite{bib:stt:ele:gar} code.
 
 The mixture and the high voltage determine the behavior of a gas. In a weak electric field
 or in a mixture with high quenching, the electrons are in thermal 
 equilibrium with the surrounding medium and the drift velocity is 
 proportional to the electric field intensity. Such gases
  are usually called ``cold''.

 On the contrary, if the electron average kinetic energy differs
 from the thermal energy, the drift velocity behaviour becomes
 saturated and tends to be constant and 
 independent of the electric field strength, 
 that is of the distance from the wire anode.
 In this way the main sources of systematic errors
 are removed and the track reconstruction is easier.
  Such gases are called ``hot''.
 However, the spatial resolution in hot gas mixtures is limited 
 by the large diffusion and cannot be better than $50\,\mum$.

\begin{figure}[h!]  
\begin{center}  
 \includegraphics[width=0.75\swidth]{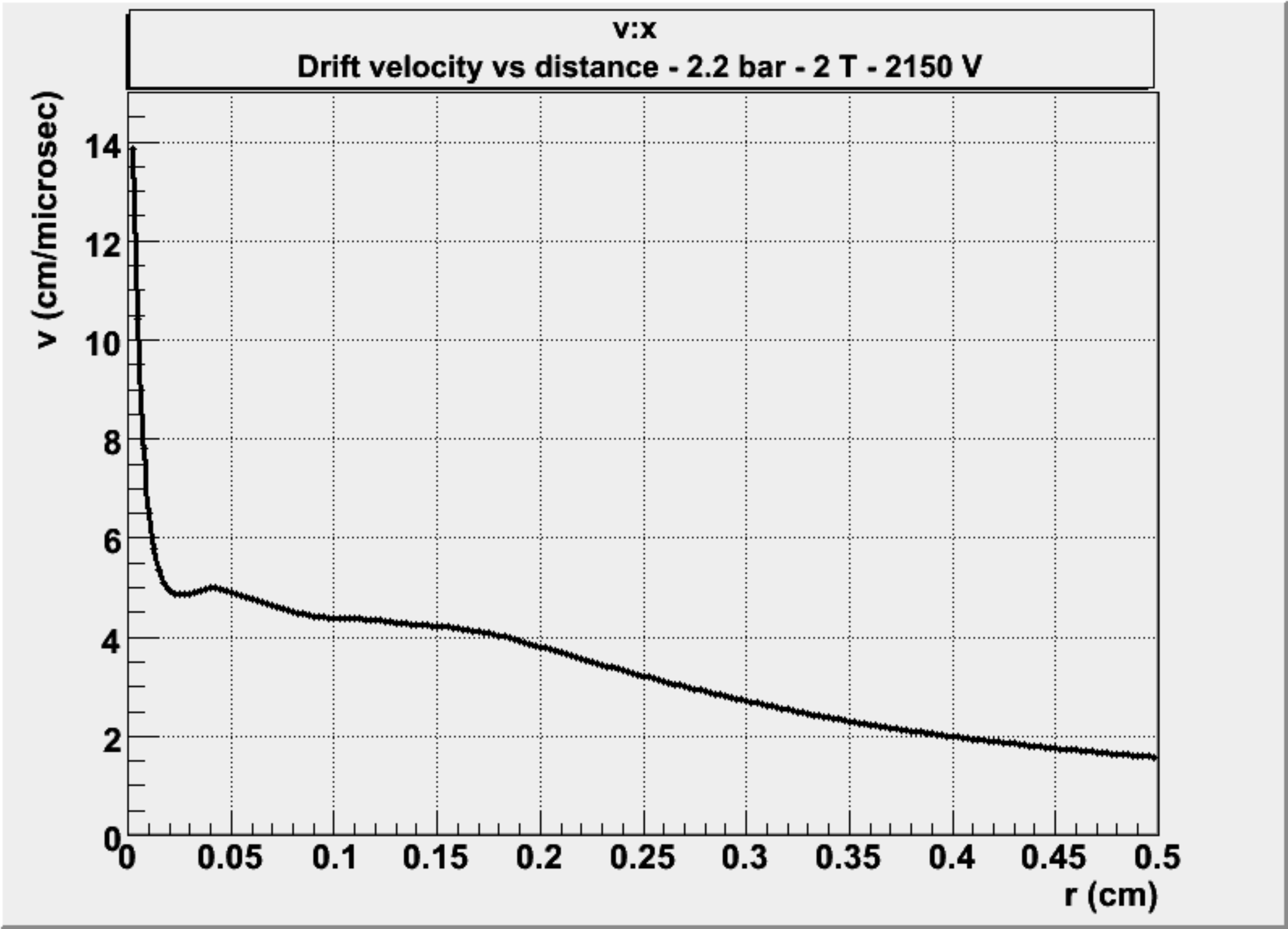}
 \includegraphics[width=0.75\swidth]{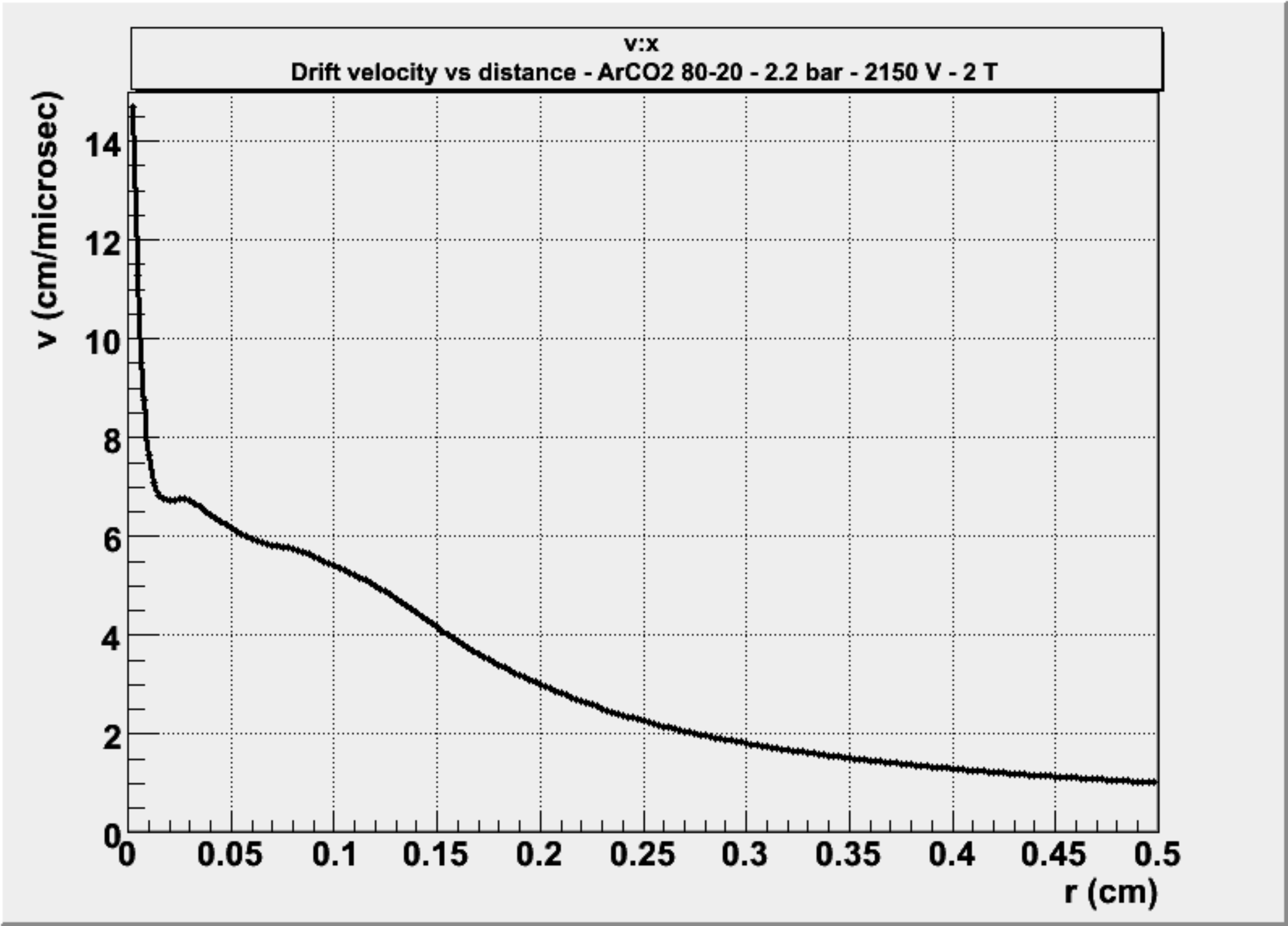}
\end{center}
\vspace*{-0.5cm}
\caption[Drift velocity vs wire distance in a straw tube]{Drift velocity vs wire distance in a straw tube 
of $0.5\,\cm$ radius, $1850\, V$ voltage,
$2.2\,$bar pressure and $2\,$T magnetic field for different 
gas mixtures: {\it (top)} $90/10\percent$ Ar/CO$_2$, 
{\it (bottom)} $80/20\percent$ Ar/CO$_2$.} 
\label{fig:stt:tub:Garf1}  
\end{figure}

 The drift velocity as a function of the wire distance is 
 reported in \Reffig{fig:stt:tub:Garf1} showing that the increase of the CO$_2$
 percentage tends to cool the gas, with a corresponding stronger
 dependence of the velocity from the wire distance. This effect
 could be recovered by an accurate self-calibration (see below), but makes the 
 tube stability more critical, requiring a precision control of
 temperature and pressure.

%%%%%%%%%%%%%%%%%%%%%%%%%%%%%%%%%%%%%%%%%%%%%%%%%%%%%%%%%%%%%%%%%%%%%%
%%%%%%%%%%%%%%%%%%%%%%%%%%%%%%%%%%%%%%%%%%%%%%%%%%%%%%%%%%%%%%%%%%%%%%%

 The effect of the magnetic field transforms the path
 between two collisions of a moving charge into circular trajectories.
 With obvious notation, the electron Lorentz angle is \cite{bib:stt:tub:blum}:
 $$ \tan \alpha =\tan \omega \tau =  \frac{e B}{m_e} \, \tau $$ \noindent
 where $\tau$ is the average time between collisions and $\omega$ is the Larmor
 frequency of the electron. In cold gases the drift velocity
 tends to be linear with the electric field $E$ and $\tau$ is almost constant, whereas
 in hot gases, where the drift velocity is more constant, $\tau$ is inversely
 proportional to $E$. Due to the much lower elastic cross section, $\tau$ in hot gases
 is about one order of magnitude higher. Estimations from experimental data
 show that for a 2\,T magnetic field and a 5 mm drift distance, the drift time
 for a CO$_2$/C$_4$H$_{10}$ (90/10) mixture increases by 15\percent in a magnetic field, that
 for an Ar/CO$_2$ (90/10) mixture increases up to 50\percent \cite{bib:stt:tub:magboltz}.

 All these effects are reproduced in the GARFIELD results.

\begin{figure}[h!]  
\begin{center}  
\includegraphics[width=0.75\swidth]{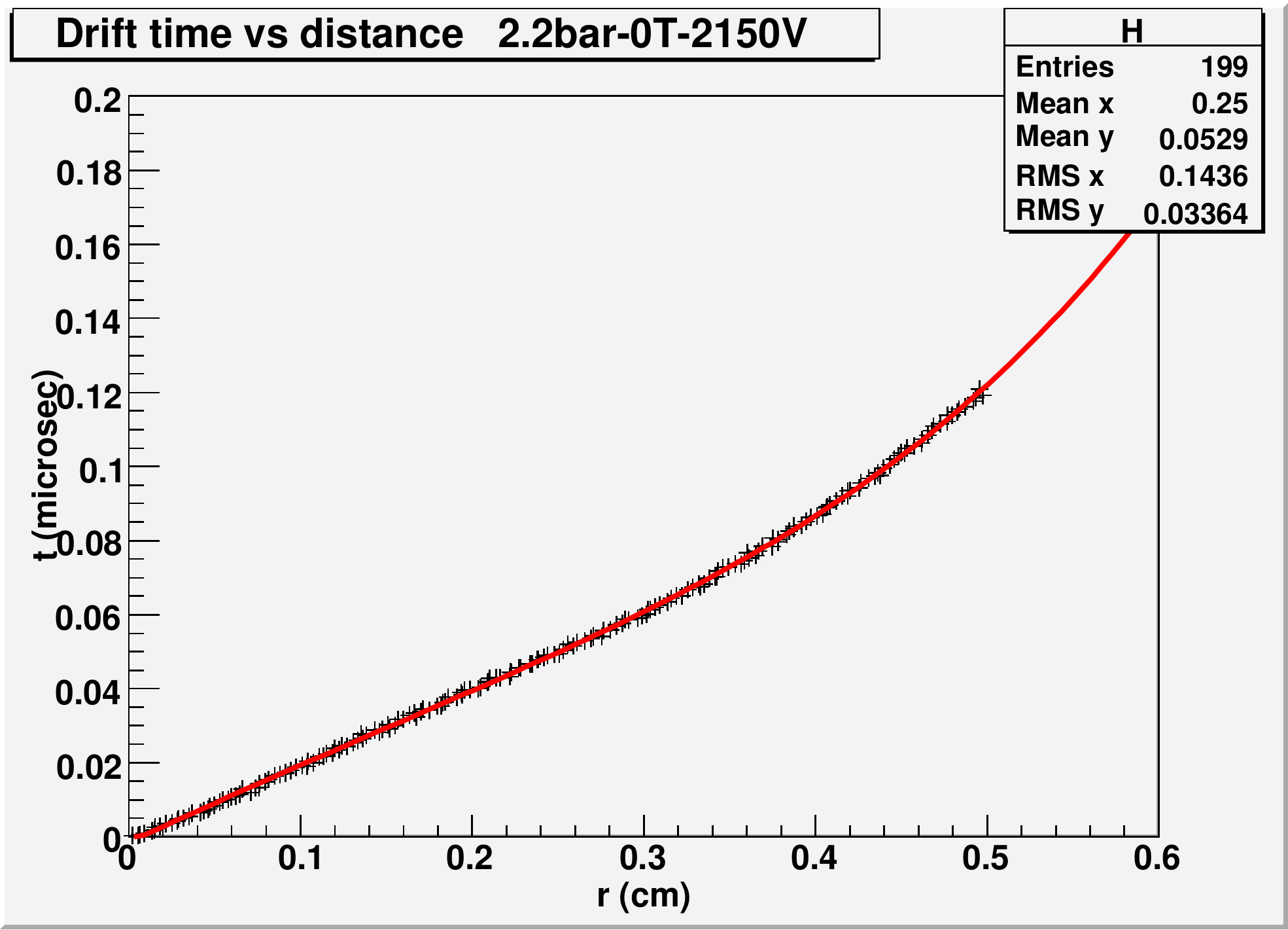}
\includegraphics[width=0.75\swidth]{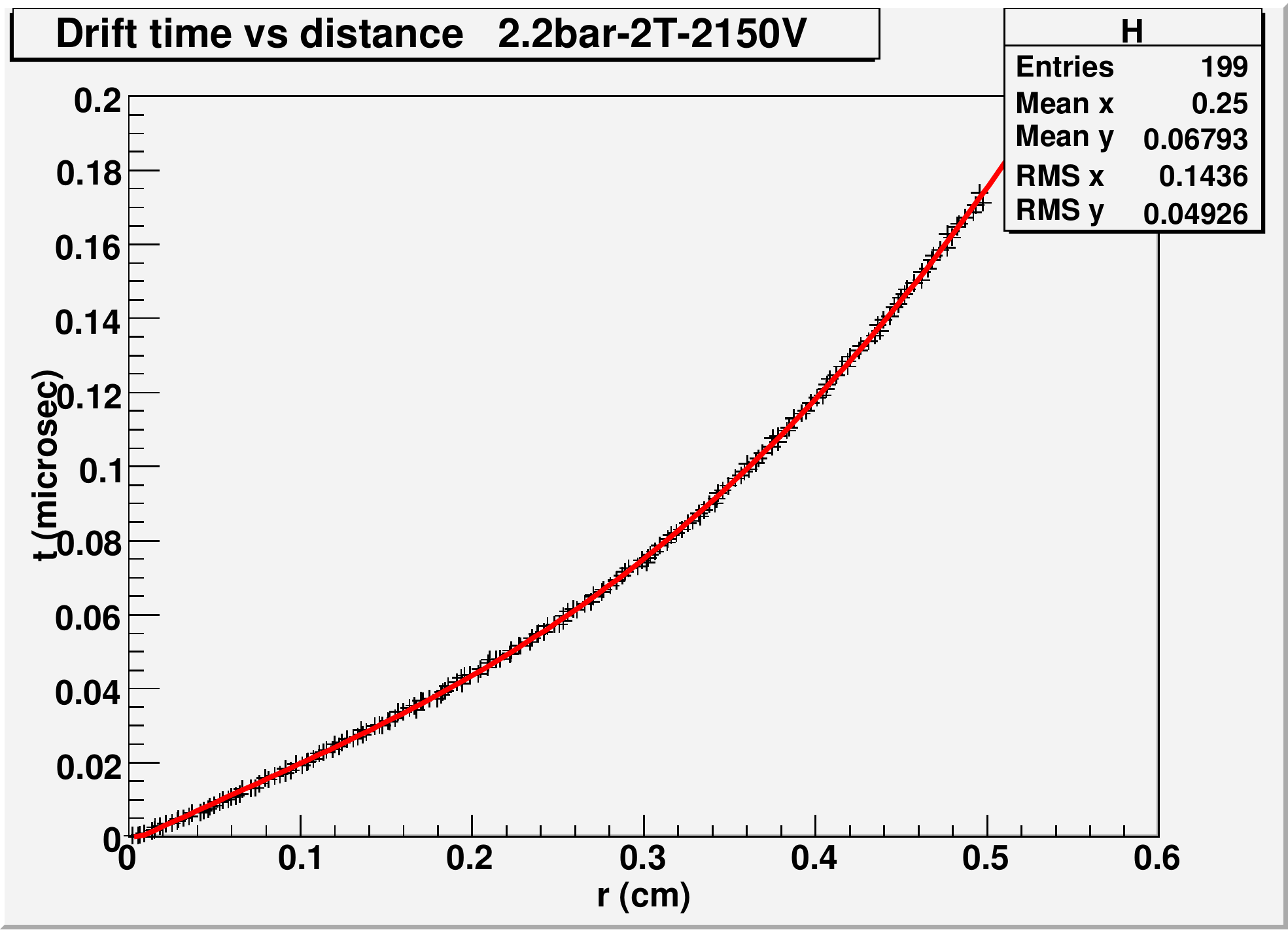}
\end{center}
\vspace*{-0.5cm}
\caption[Drift time vs wire distance in a straw tube]{Drift time vs wire distance in a $90/10\percent$ Ar/CO$_2$ 
straw tube of $0.5\,\cm$ radius  and 
2.2 bar pressure: up, without magnetic field; down, with magnetic field of 2 T
 (from GARFIELD).} 
\label{fig:stt:tub:Garf2}  
\end{figure}

 Typical  time vs distance curves for a hot gas mixture
 like Ar/CO$_2$ (90/10), with and without magnetic field, are
 reported in \Reffig{fig:stt:tub:Garf2}, where the increase of the drift time
 due to the  field is clearly visible. 
%
%%%%%%%%%%%%%%%%%%%%%%%%%%%%%%%%%%%%%%%%%%%%%%%%%%%%%%%%%%%%%%%%%%%%%%
%%%%%%%%%%%%%%%%%%%%%%%%%%%%%%%%%%%%%%%%%%%%%%%%%%%%%%%%%%%%%%%%%%%%%%%

\begin{figure}[h!]  
\begin{center}  
\includegraphics[width=0.75\swidth, angle=90]{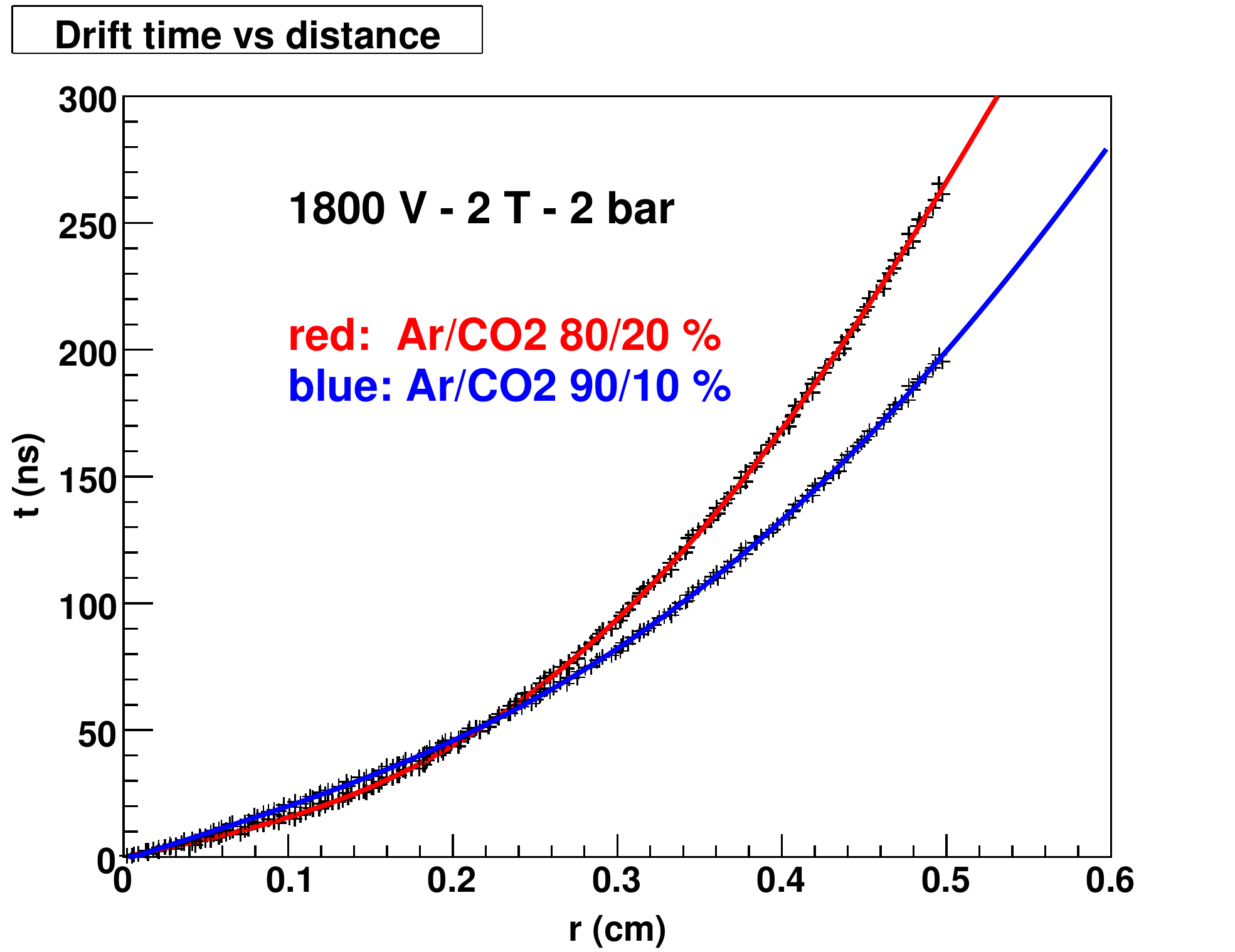}
\end{center}
\vspace*{-0.5cm}
\caption[Time vs wire distance for two different Ar/CO$_2$ mixtures in the presence of magnetic field]{Time vs wire distance for two different Ar/CO$_2$ mixtures
 in the presence of magnetic field (from GARFIELD).} 
\label{fig:stt:tub:Mixt}  
\end{figure}

The increase in the drift time while increasing the CO$_2$ percentage
 is also clearly shown  in \Reffig{fig:stt:tub:Mixt}.
%%%%%%%%%%%%%%%%%%%%%%%%%%%%%%%%%%%%%%%%%%%%%%%%%%%%%%%%%%%%%%%%%%%%%%
%%%%%%%%%%%%%%%%%%%%%%%%%%%%%%%%%%%%%%%%%%%%%%%%%%%%%%%%%%%%%%%%%%%%%%%

 Another important input to the simulation are the transverse and longitudinal
 diffusion curves, due to the thermal spreading of the electron clouds
 during the drift. 
 The GARFIELD results show that the high diffusion 
values of the hot gas (Ar/CO$_2$~=~90/10) are partially compensated by increasing
 the pressure. At 2 atm pressure the longitudinal and transverse diffusion coefficients,
 at 5~mm distance from the wire, are 100 and 140~$\mu m$, 
 whereas at 1~atm pressure the same coefficients are 120 and 220~$\mu m$, respectively.
%The GARFIELD curves for one of 
% the mixture of interest are reported in 
% \Reffig{fig:stt:tub:Garf3} a 1 and 2 bar pressure. We see that the 
% high diffusion values of the hot gas are partially compensated by increasing
% the pressure.
%
%%%%%%%%%%%%%%%%%%%%%%%%%%%%%%%%%%%%%%%%%%%%%%%%%%%%%%%%%%%%%%%%%%%%%%
% \begin{figure}[h!]  
%\begin{center}  
%\includegraphics[width=\hwidth]{./stt/fig/}
%\includegraphics[width=\hwidth]{./stt/fig/}
%\end{center}
%\vspace*{-0.5cm}
%\caption{Diffusion radius (standard deviation) vs distance
% for a $90-10\percent$ $Ar-CO_2$ mixture at 1 (dotted line)
% and 2 (full line) bar pressure} \label{fig:stt:tub:Garf3}  
%\end{figure} 
%%%%%%%%%%%%%%%%%%%%%%%%%%%%%%%%%%%%%%%%%%%%%%%%%%%%%%%%%%%%%%%%%%%%%%%

\begin{figure}[h!]  
\begin{center}  
\includegraphics[width=2.2\hwidth]{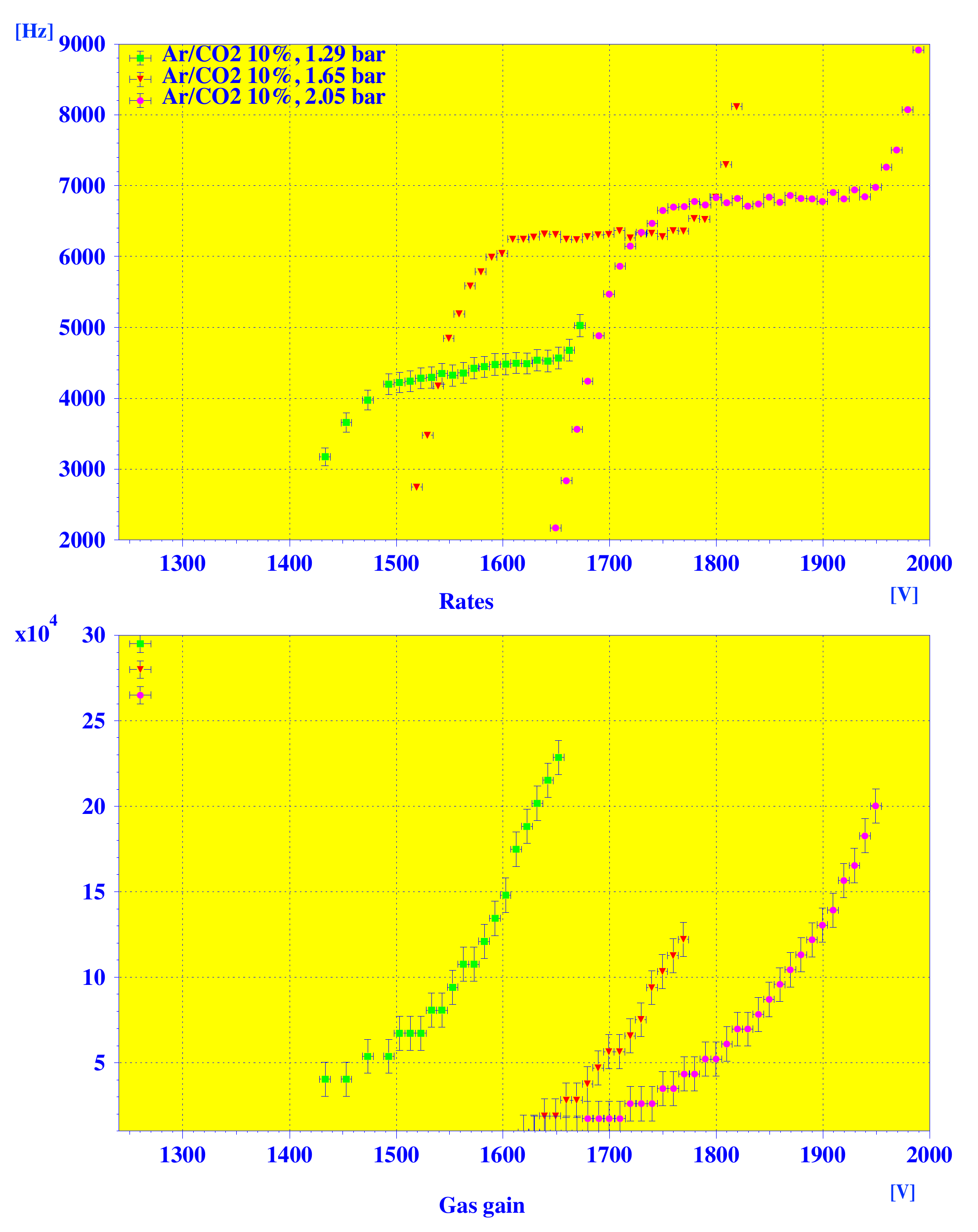}
\end{center}
\vspace*{-0.5cm}
\caption[Experimental plots of the tube rate and gas gain]{Experimental plots of the tube rate and gas gain relative to 
  a $90/10\percent$ Ar/CO$_2$ mixture.} \label{fig:stt:tub:ExpGain}  
\end{figure}

 Finally, the necessary input to the simulation  is the
 gas amplification, i.e. the multiplication factor of the avalanche 
 which is formed in the last tens of microns of the primary
 electron path in its drift to  the anode wire. This multiplication factor
 is given by \cite{bib:stt:tub:blum}:
 $$ G = \exp \left( \int_a^x \alpha(x) {\rm d}x \right) $$
 \noindent
 where $ \alpha(x)$ is the Townsend
 coefficient (inverse of the mean free
 path for ionization), $a$ is the anode wire radius and the integral is taken 
 along the whole drift path. A typical behaviour of the gas gain, measured for
 our mixtures of interest is  shown in \Reffig{fig:stt:tub:ExpGain}, where one sees that 
 in our case the tube remains in the region of direct proportionality. 
%
%
%
%%%%%%%%%%%%%%%%%%%%%%%%%%%%%%%%%%%%%%%%%%%%%%%%%%%%%%%%%%%%%%%%%%%%%%
%%%%%%%%%%%%%%%%%%%%%%%%%%%%%%%%%%%%%%%%%%%%%%%%%%%%%%%%%%%%%%%%%%%%%%%
%
%
\subsection{Simulation of the Drift Process}
 Once the free electrons have been created in some points of the 
 tube, their position is dispersed both longitudinally and 
 transversally according to the GARFIELD diffusion curves and
 the time of arrival on the wire is calculated from the GARFIELD
 distance-time curves.
% The GARFIELD outputs are described in more
% detail in sect.~\ref{sectgarf} below. 

 The movement of each electron gives rise to a charge, which is
 obtained by sampling from a Polya distribution \cite{bib:stt:tub:blum}
 having as a mean value the gain or multiplication
 factor (around $5 \cdot 10^4$).
 Then, by summing this signal over the number of electrons
 we obtain the total charge, as  shown in \Reffig{fig:stt:tub:genfig}.
\subsection{The Electrical Signal}
 By taking into account the 
 arrival time of each electron and 
 assigning a Gaussian-shaped electrical response to 
 each charge multiplication, we can reproduce also the
 shape of the electrical signal. 
 We added also a white noise component equal
 to the 3\percent of the primary signal peak value.

\begin{figure}[h!]  
\begin{center}  
\includegraphics[width=\hwidth]{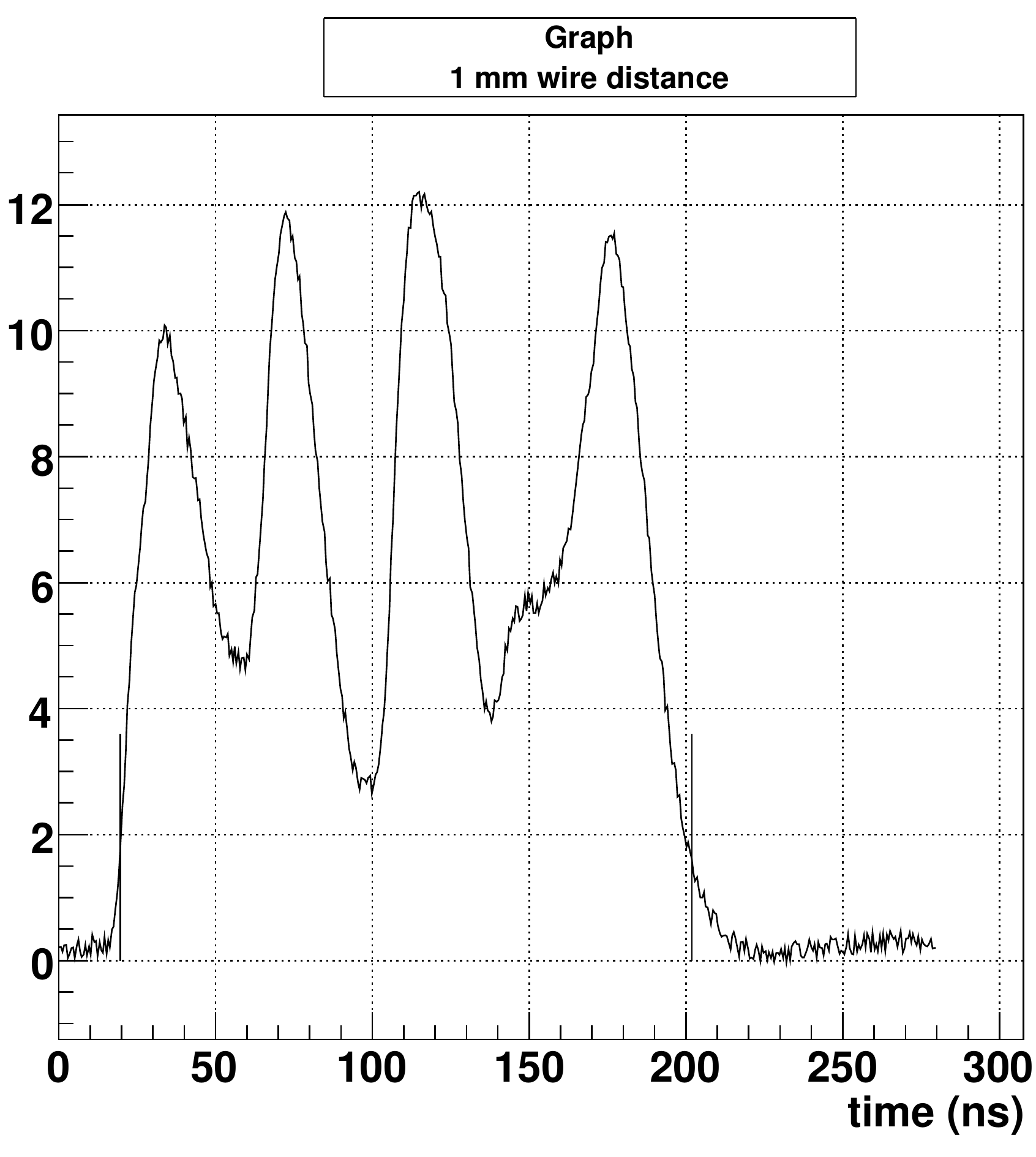}
\includegraphics[width=\hwidth]{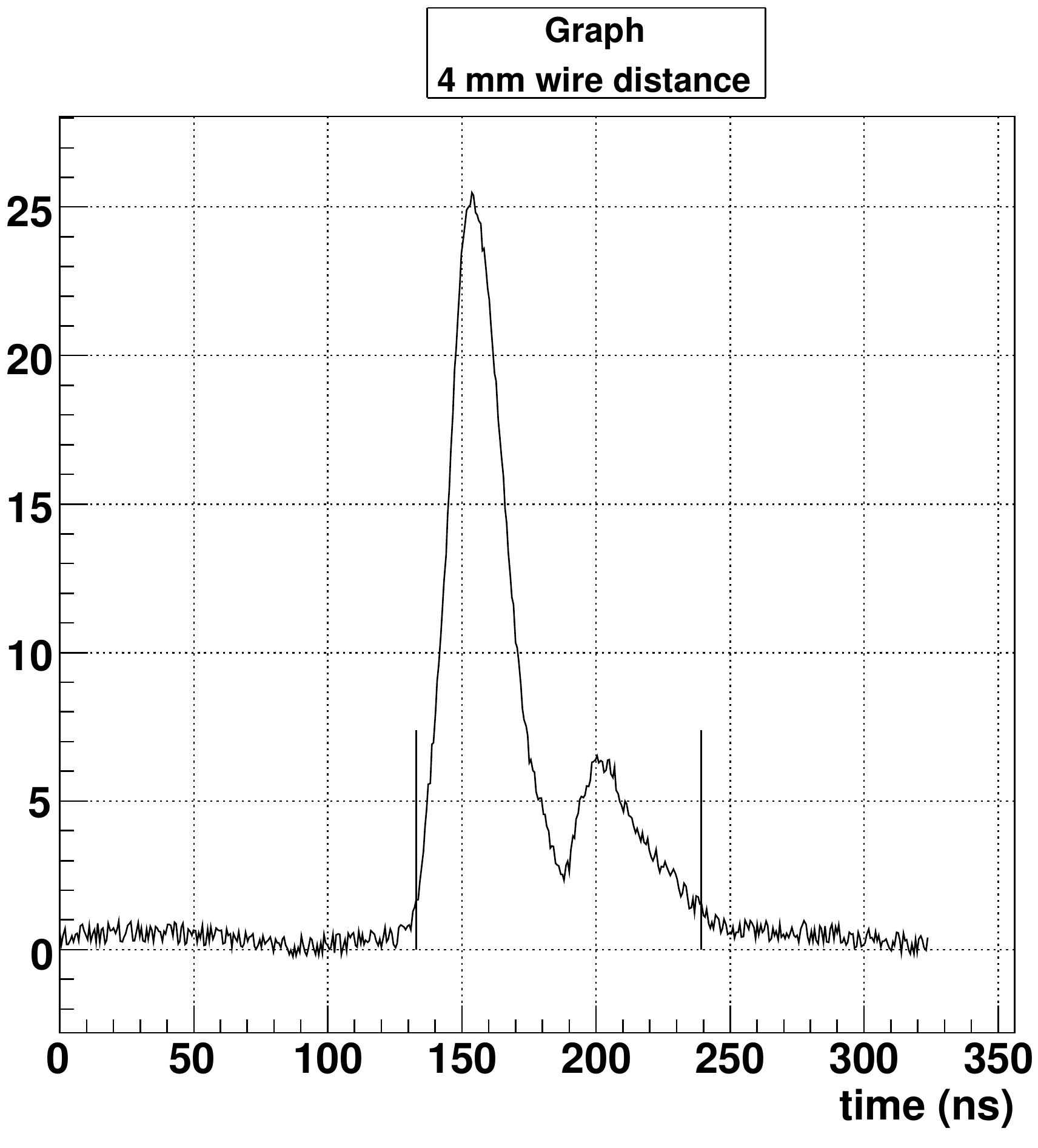}
\end{center}
\vspace*{-0.5cm}
\caption[Straw tube simulated signals for a track close to and far from the wire]{Straw tube simulated signals for a track close to (left) and far
  from (right) the wire.} \label{fig:stt:tub:signalf}  
\end{figure}

 Some examples are shown in \Reffig{fig:stt:tub:signalf}, where
 two typical signals are shown: the first one is generated from a track 1 mm near to
 the wire, the second one  from a track 4 mm far from the wire. 
 In the first case the
 clusters arrive dispersed in time, giving rise to an irregular structure of
 the signal. In this case the discrimination technique is crucial 
 for a good time resolution. In the second case the cluster arrival
 is more concentrated and the signal structure appears more regular.
 These example show the importance of the electronic treatment of the
 signal and of the discrimination technique to be used
 for obtaining the  drift time.

 We consider two discrimination techniques: fixed (F) and constant fraction (CF)
 thresholds. The F threshold is set to about 5\percent of the mean primary electron
 value, that is to 10 primary electrons in the 2 atm case (see \Reffig{fig:stt:tub:genfig}).
 This is compatible with previous studies \cite{bib:stt:tub:Branchini,bib:stt:tub:Riegler}.
 The CF threshold is set to 5\percent of the peak value of the current signal.

 In the following, unless specified otherwise, the displayed 
 results are obtained with the standard F threshold.
 
% These aspect are
% studied in sect.~\ref{subresolution}.
%
%%%%%%%%%%%%%%%%%%%%%%%%%%%%%%%%%%%%%%%%%%%%%%%%%%%%%%%%%%%%%%%%%%%%%%
%%%%%%%%%%%%%%%%%%%%%%%%%%%%%%%%%%%%%%%%%%%%%%%%%%%%%%%%%%%%%%%%%%%%%%%
\subsection{Simulation of the Self-Calibration Procedure}
\begin{figure}[h!]  
\begin{center}  
\includegraphics[width=\swidth]{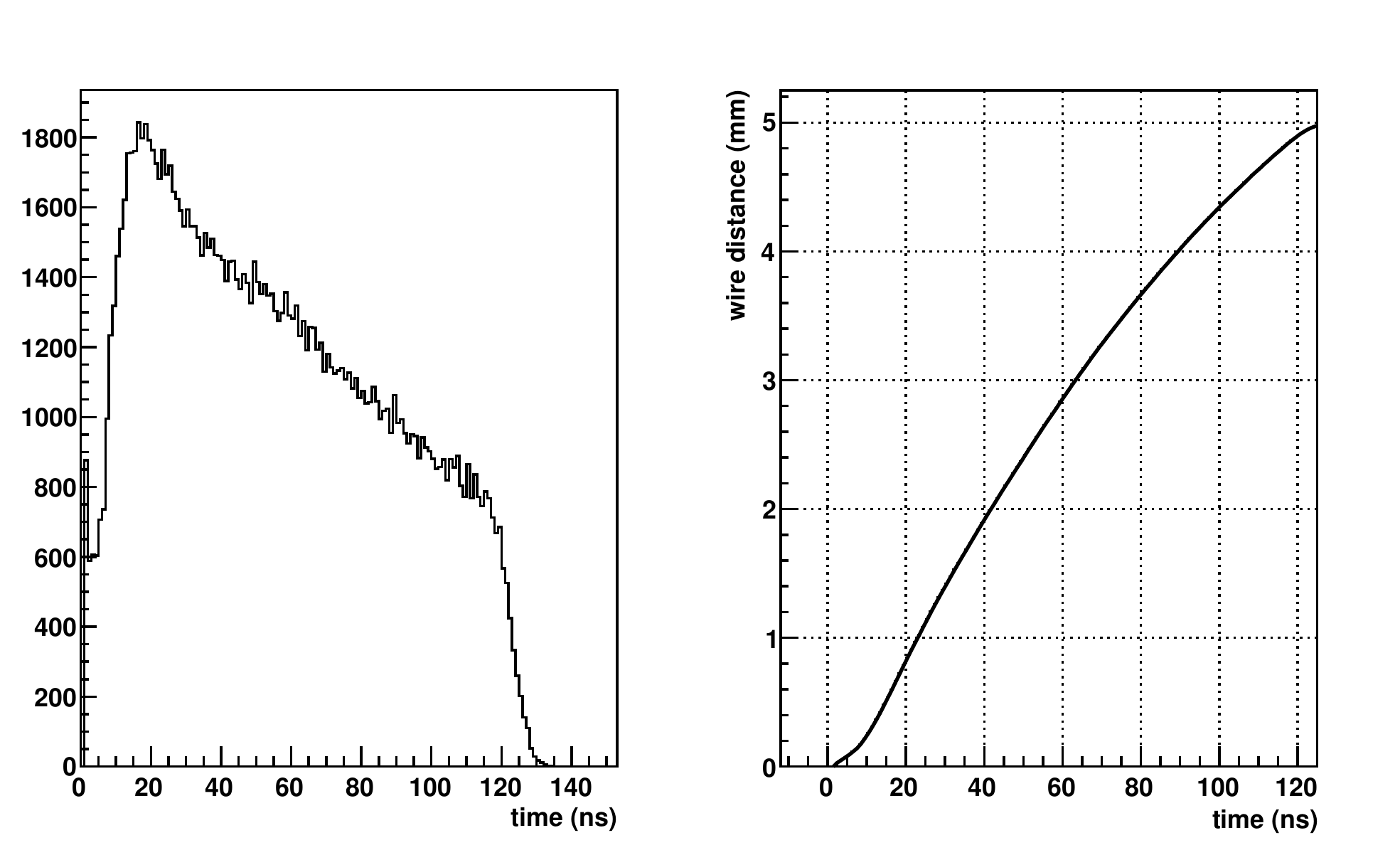}
\end{center}
\vspace*{-0.5cm}
\caption[Simulated TDC spectrum and space-time relation (no magnetic field)]{Simulated TDC spectrum without magnetic field
 for a single tube uniformly illuminated
 (left) and space-time relation obtained with the
 self-calibration method of \Refeq{eq:stt:tub:scal3}.} \label{fig:stt:tub:nofspacefig}  
\end{figure}

\begin{figure}[h!]  
\begin{center}  
\includegraphics[width=\swidth]{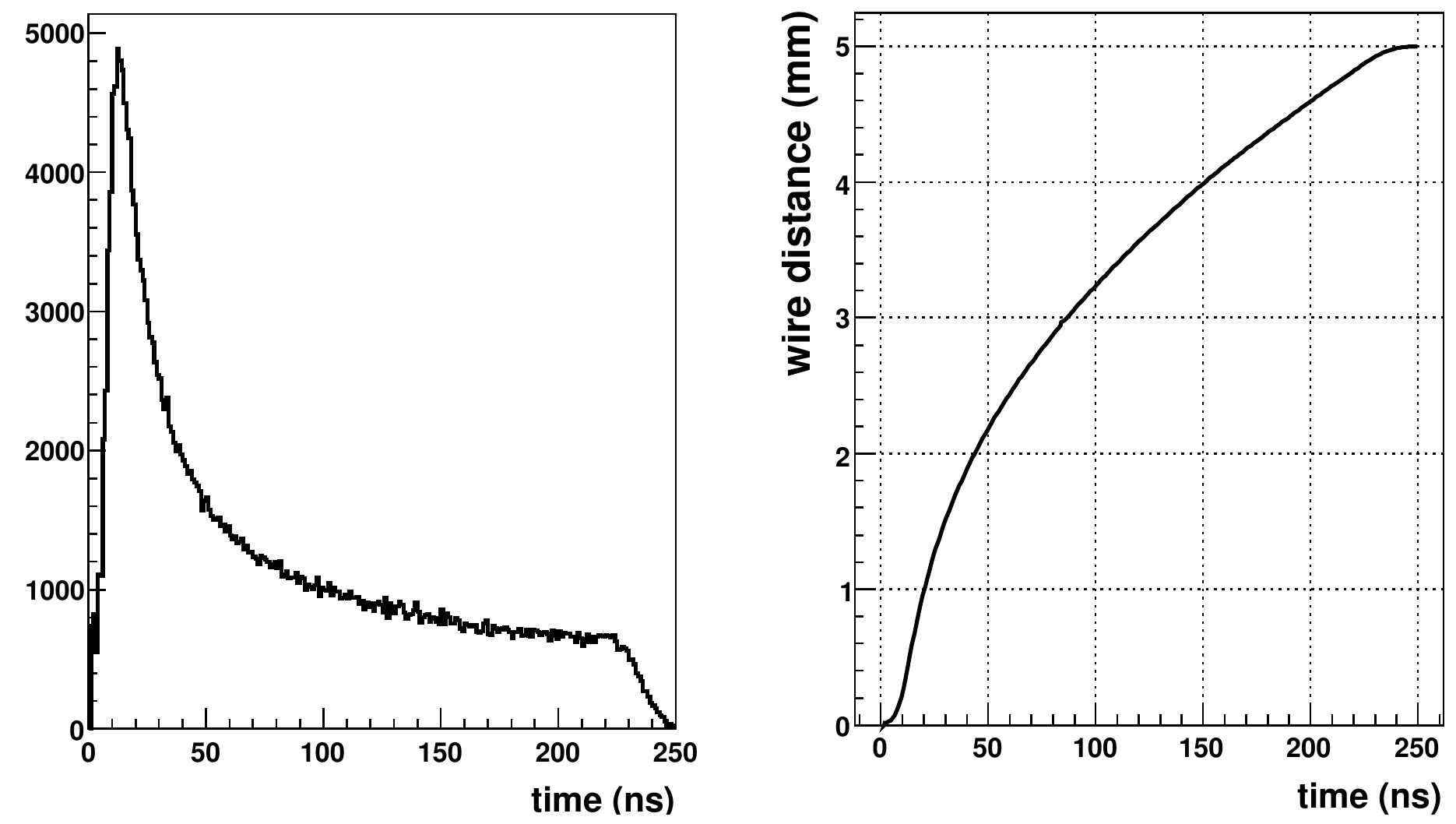}
\end{center}
\vspace*{-0.5cm}
\caption[Simulated TDC spectrum and space-time relation (2~T magnetic field)]{Simulated TDC spectrum for a 2\,T magnetic field
  for a single tube uniformly illuminated
 (left) and space-time relation obtained with the
 self-calibration method of \Refeq{eq:stt:tub:scal3}.} \label{fig:stt:tub:spacefig}  
\end{figure}

The primary information from the tube is the drift time distribution
 of the arriving signals, that
  is the number of tracks ${\rm d}N$ within the time interval ${\rm d} t$.
A typical distribution of this quantity, in the case of a parallel and
 uniform illumination of the tube is shown in \Reffig{fig:stt:tub:nofspacefig} 
 and  in  \Reffig{fig:stt:tub:spacefig} (left) in the case of the absence and the 
 presence of the magnetic field, respectively.

\begin{figure*}[h!]  
\begin{center}  
\includegraphics[width=0.7\dwidth]{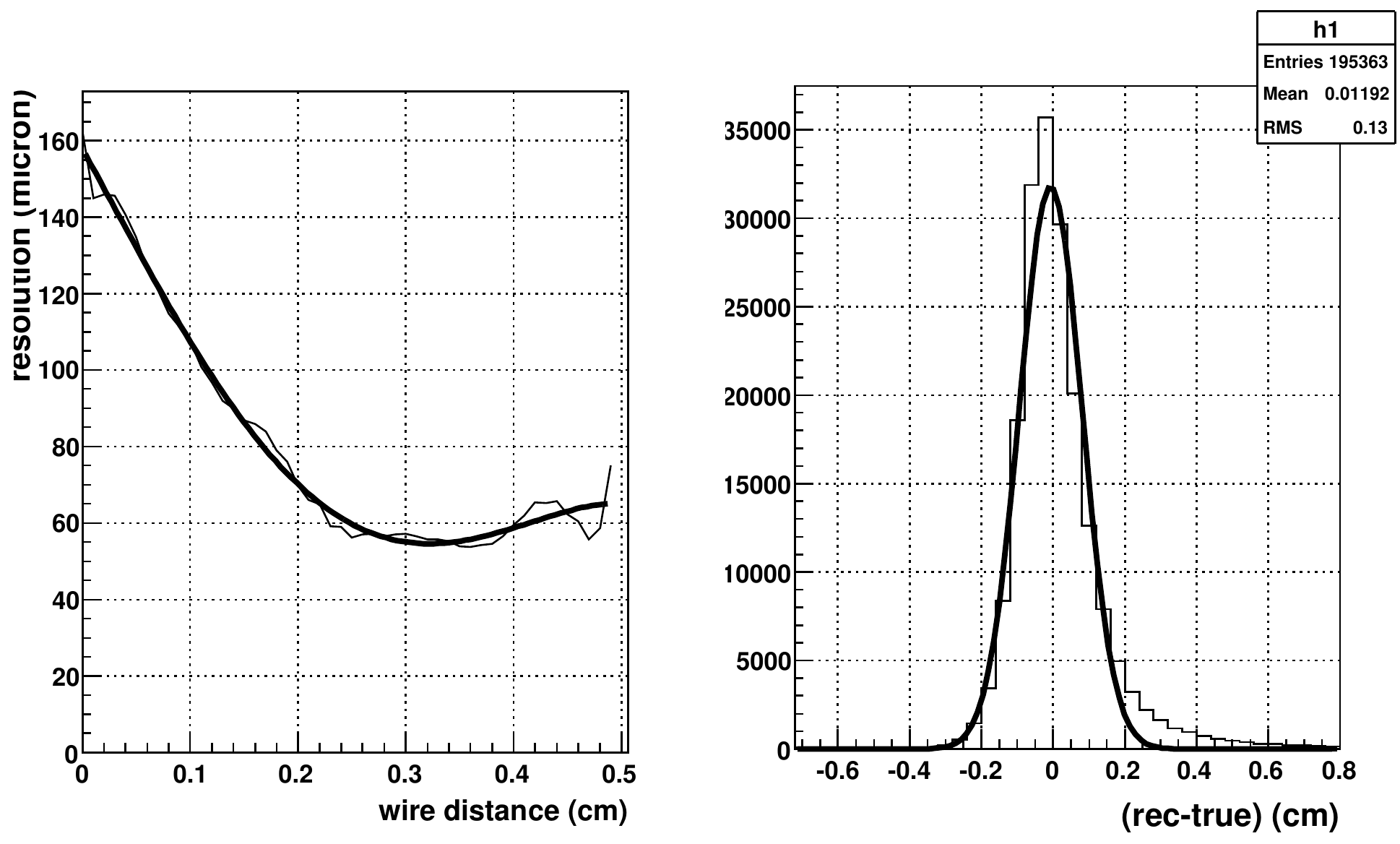}
\end{center}
\vspace*{-0.5cm} 
\caption[Simulated average residual width as a function of the track distance from the wire and residual distribution of (reconstructed-true) wire distance]{Simulated average residual width as 
a function of the track distance from the wire (left) and residual 
 distribution of (reconstructed-true) wire distance. 
  The bold line is the smoothing polynomial.} \label{fig:stt:tub:parfig2}  
\end{figure*}

The self-calibration method exploits the properties of this distribution. 
 Since the track density is constant over the tube diameter, one can write
\begin{equation}  \label{scal1}
     \frac{{\rm d}N}{{\rm d}r} = \frac{N_{\rm tot}}{R}
\end{equation}  \noindent
where $N$ is the number of tracks,
$r$ is the wire distance, 
 $N_{\rm tot}$ is the total number of tracks and $R$ the tube radius.
 The number of tracks in a time interval can be obtained directly
 from the above relation:
\begin{equation}  \label{scal2}
     \frac{{\rm d}N}{{\rm d}t} =  \frac{{\rm d}N}{{\rm d}r} \frac{{\rm d}r}{{\rm d}t}
 =  \frac{{\rm d}r}{{\rm d}t}  \frac{N_{\rm tot}}{R}
\end{equation}  \noindent
After integration, one obtains the desired space-time relation $r(t)$ 
  by integration of the time spectrum up to $t$:
\begin{equation}  \label{eq:stt:tub:scal3}
     r(t) = \frac{R}{N_{\rm tot}} \, \int_0^t \frac{{\rm d}N}{{\rm d}t} 
 \, {\rm d}t
\end{equation}  \noindent
The time spectrum and the space time relation $r(t)$ are shown in
  \Reffig{fig:stt:tub:nofspacefig} 
 (without magnetic field) and in \Reffig{fig:stt:tub:spacefig} (with magnetic field).
%
%%%%%%%%%%%%%%%%%%%%%%%%%%%%%%%%%%%%%%%%%%%%%%%%%%%%%%%%%%%%%%%%%%%%%%
%%%%%%%%%%%%%%%%%%%%%%%%%%%%%%%%%%%%%%%%%%%%%%%%%%%%%%%%%%%%%%%%
%
%%%%%%%%%%%%%%%%%%%%%%%%%%%%%%%%%%%%%%%%%%%%%%%%%%%%%%%%%%%%%%%%%%%%%%
%%%%%%%%%%%%%%%%%%%%%%%%%%%%%%%%%%%%%%%%%%%%%%%%%%%%%%%%%%%%%%%%
% After this step, the method requires the systematic error correction due to the
% threshold level, that appears as an offset  in the histogram of
% the residuals of the reconstructed distance, as in \Reffig{fig:stt:tub:parfig} (right).
%In the simulation, we adjust a time offset $t_0$ until the mean value
% of this distribution approaches zero.%
%
% After the correction $t-t_0$ to all the measured times of the spectrum of
% \Reffig{fig:stt:tub:spacefig}(left), the $r(t)$ curve is corrected
% and one can proceed to the derivation by
% simulation  of the resolution  curve. The improvement obtained
% with this method  is shown in  \Reffig{fig:stt:tub:parfig2}.
%%%%%%%%%%%%%%%%%%%%%%%%%%%%%%%%%%%%%%%%%%%%%%%%%%%%%%%%%%%%%%%%%%%%%%%
% \begin{figure*}[h!]  
%\begin{center}  
%\includegraphics[width=0.7\dwidth]{./stt/fig/restot0FT.pdf}
%\end{center}
%\vspace*{-0.5cm}
%\caption{Average residual width as 
% a function of the track distance from the wire (left) and residual 
% distribution of (reconstructed-true) wire distance.} \label{fig:stt:tub:parfig}  
%\end{figure*} 
%
%%%%%%%%%%%%%%%%%%%%%%%%%%%%%%%%%%%%%%%%%%%%%%%%%%%%%%%%%%%%%%%%%%%%%%
 The result of this  method  
 of calibration is shown in  \Reffig{fig:stt:tub:parfig2}.

%%%%%%%%%%%%%%%%%%%%%%%%%%%%%%%%%%%%%%%%%%%%%%%%%%%%%%%%%%%%%%%%
This simulated procedure corresponds, during the real calibration, to have an
 accurate knowledge of the relationship between the measured 
 drift time and the minimum approach distance of the particle trajectory 
 to the wire. The mean value of the residuals of tracks is then used to correct 
 the measured drift times until the residual distribution is symmetric about zero.
%
% The experimental set-up should be of the type sketched in Ref.~\cite{bib:stt:tub:Riegler},
% with the straw tube module sandwitched into a silicon strip detector assembly
% with a spatial accuracy determination of about 10 $\mu m$.

To explore the effect of the electronic threshold, we also simulate the
 resolution obtained by applying the constant fraction 
 discrimination technique, simulated as a fixed percentage (5\percent) of the 
 peak of the current signal.
%%%%%%%%%%%%%%%%%%%%%%%%%%%%%%%%%%%%%%%%%%%%%%%%%%%%%%%%%%%%%%%%%%%%%%
\begin{figure}[h!]  
\begin{center}  
\includegraphics[width=0.75\swidth]{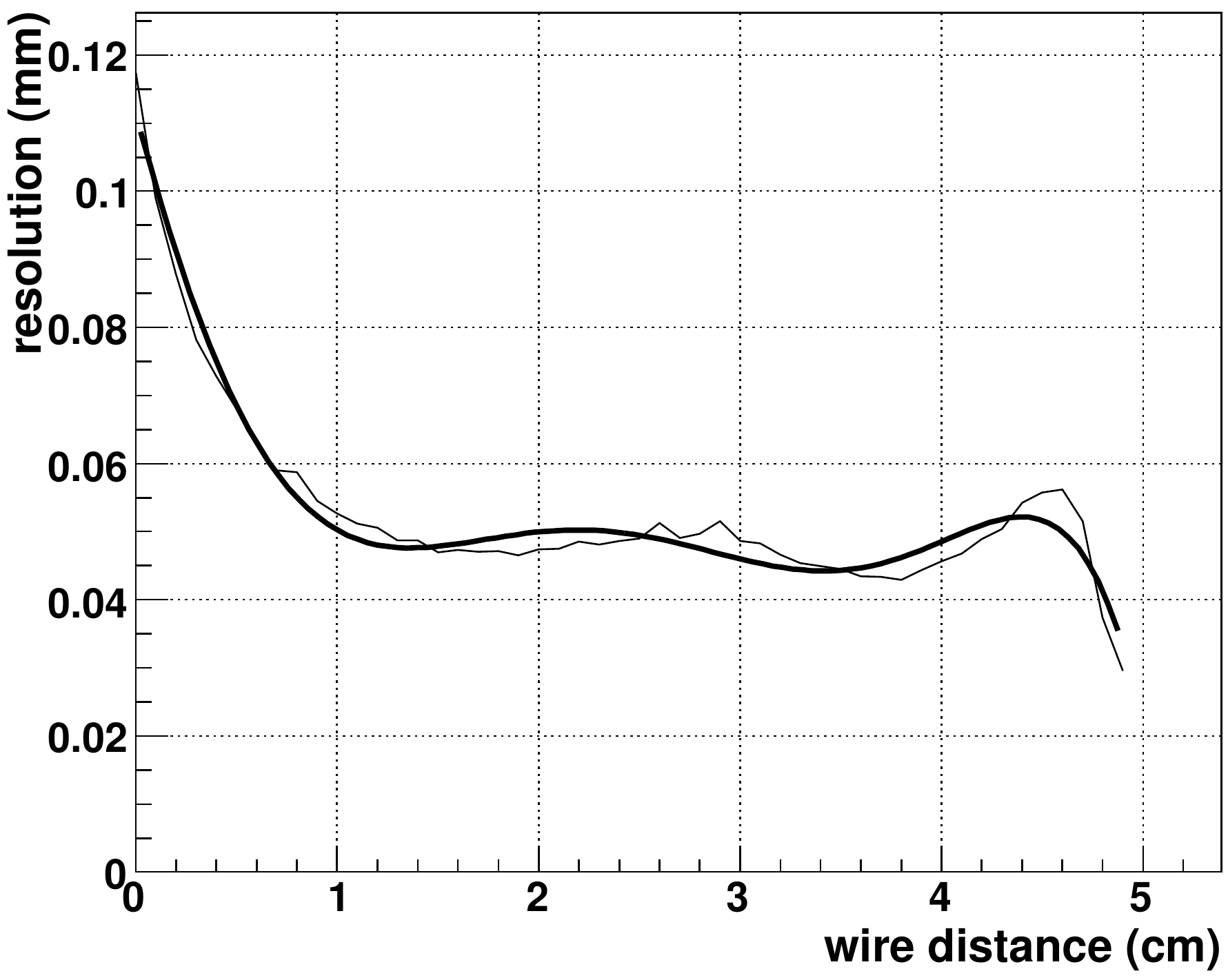}
\end{center}
\vspace*{-0.5cm}
\caption[Simulated average residual width as a function of the track distance from the wire with constant fraction discrimination]{As in \Reffig{fig:stt:tub:parfig2} with a constant fraction
 electronic discrimination.} \label{fig:stt:tub:parfig3}  
\end{figure} 
%%%%%%%%%%%%%%%%%%%%%%%%%%%%%%%%%%%%%%%%%%%%%%%%%%%%%%%%%%%%%%%%
 
 The improvement in the resolution, as shown in \Reffig{fig:stt:tub:parfig3},
 demonstrates the importance of the discrimination
 of the tube signals.

\subsection{Full and Fast Simulation}
%
%
%%%%%%%%%%%%%%%%%%%%%%%%%%%%%%%%%%%%%%%%%%%%%%%%%%%%%%%%%%%%%%%%%%%%%%
\begin{figure}[h!]  
\begin{center}  
\includegraphics[width=0.75\swidth]{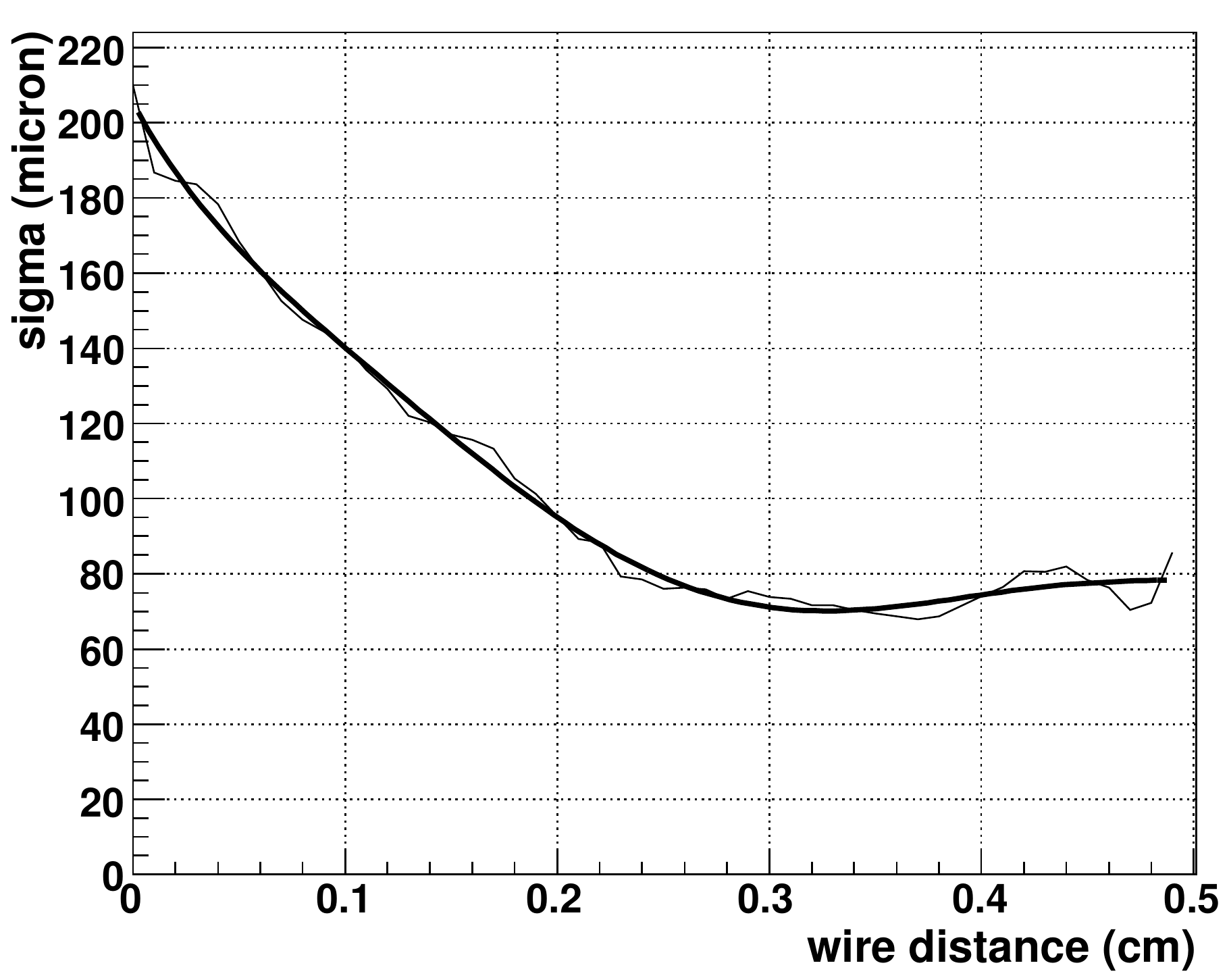}
\end{center}
\vspace*{-0.5cm}
\caption[Standard deviation corresponding to the residual width resolution of \Reffig{fig:stt:tub:parfig2} (left)]{Standard deviation corresponding to the 
 residual width resolution of \Reffig{fig:stt:tub:parfig2} (left).} \label{fig:stt:tub:parfig4}  
\end{figure} 
%%%%%%%%%%%%%%%%%%%%%%%%%%%%%%%%%%%%%%%%%%%%%%%%%%%%%%%%%%%%%%%%
 The full simulation reproduces the time output from the drift tube
and the ADC response on the charge collected starting from the primary
 cluster formation as discussed in the sections above. Since the time required 
 for each event is long, we also implemented into the simulation software
 a fast simulation option.

The spatial resolution is simply obtained through  
 the MC truth  for the true wire distance, which is used as the abscissa in 
 \Reffig{fig:stt:tub:parfig4} to extract the $\sigma$ for the Gaussian smearing to
 obtain a realistic position determination of the tube.

 The second important quantity, the charge collected on the wire, is 
 simulated in a fast manner by sampling the energy lost from the Urban 
 distribution as in \Reffig{fig:stt:tub:onept1},
 avoiding in this case the charged cluster generation.

 In this way the time spent in the tube response simulation results
 to be negligible when compared with the other part of the software.
%
%EOF: panda_tdr_stt_tub.tex

% FILE: panda_tdr_stt_sim.tex
%
\section{Simulation and Reconstruction Software}
%\COM{Author(s): A. Rotondi/L. Lavezzi/G. Boca}
\label{sec:stt:sim}

The simulation and reconstruction code for the STT is fully integrated in the \PANDA code framework \pandaroot \cite{bib:stt:sim:pr0}. 
\PANDA shares the base classes of a wider framework called \fairroot \cite{bib:stt:sim:fairroot} with other \FAIR experiments 
(CBM \cite{bib:stt:cbm}, HADES \cite{bib:stt:hades}, R3B \cite{bib:stt:r3b}) and adds its own specific tasks. In this section, a quick overview of the software framework will be given, 
with particular attention to the STT related code. The software and the procedure used to perform the tests which will be 
reported in \Refsec{sec:stt:stt:per} will be addressed.
% Different tests have been performed and will be discussed. At the beginning of each section, the specific features of the code used in each test will be recalled.

% THE VIRTUAL MC ************************
\subsection{The Framework}

The \fairroot framework is based on the Virtual Monte Carlo (VMC) \cite{bib:stt:sim:vmc1, bib:stt:sim:vmc2}, a tool developed at 
CERN by the ALICE collaboration, which allows the user to change the engine for the transport of particles in matter (\texttt{geant3}, 
\texttt{geant4}) at run-time without the need to change the input/output structure and to adapt the geometry description of the detector. 
The VMC classes decouple the user classes from the Monte Carlo classes and act as an interface allowing the interchange of the 
Monte Carlo codes. This grants high flexibility: the user can change the implementation of the detector and the algorithm of reconstruction 
and/or analysis independently from the core code.

\begin{figure*}%[!h]
\begin{center}
\includegraphics[width=\dwidth]{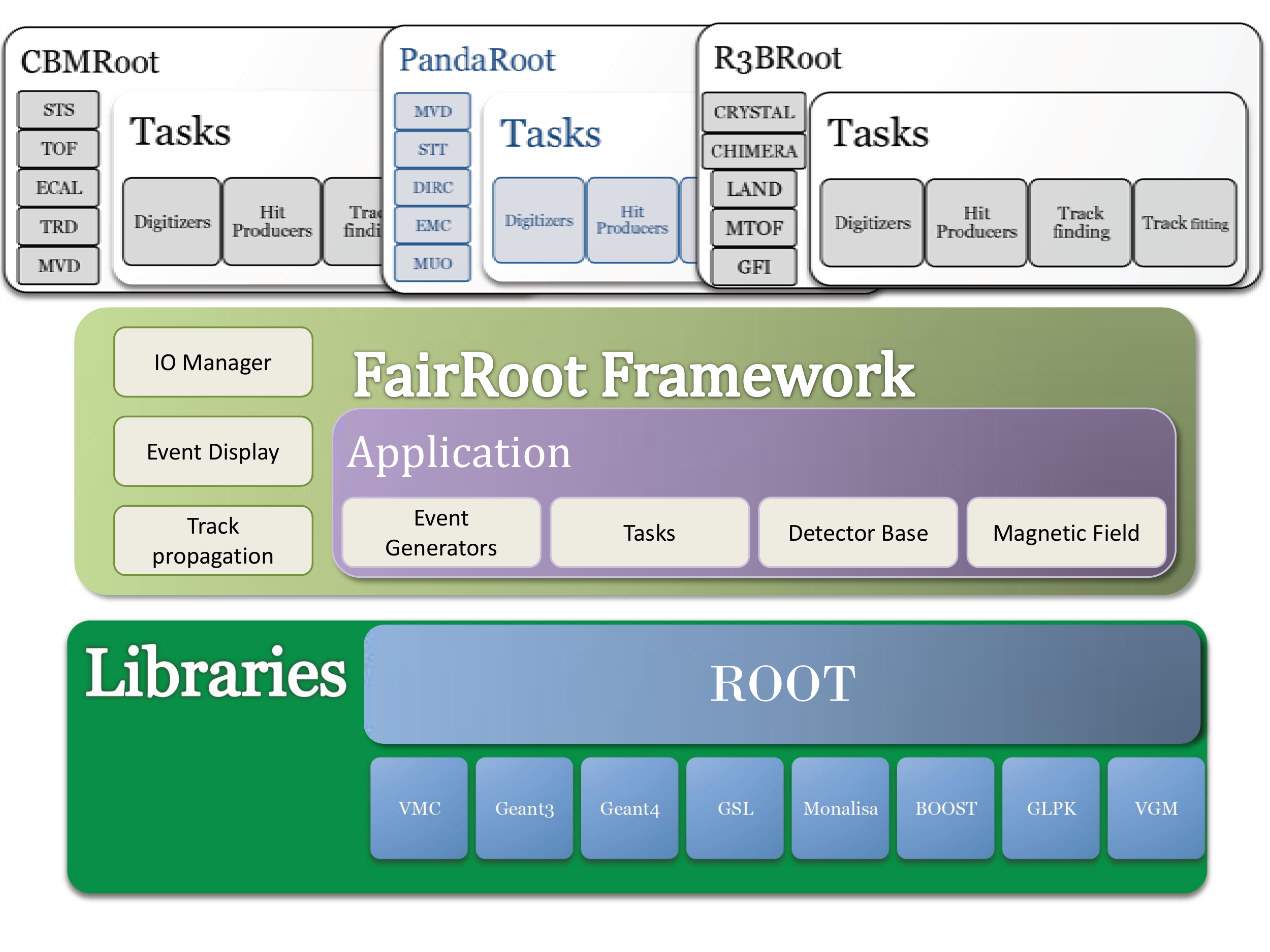}
\caption[Structure of the \fairroot framework with classes and applications]{Structure of the \fairroot framework with classes and applications.}
\label{fig:stt:sim:framework}
\end{center}
\end{figure*}

The following tools are made available to the user by the framework, in addition to the VMC and ROOT specific tasks (see \Reffig{fig:stt:sim:framework}):
\begin{itemize}
\item specific simulation and reconstruction classes for the detectors;
\item I/O Manager based on ROOT TFolder, TTree and TChain;
\item geometry readers: ASCII and ROOT (also CAD files converted to ROOT are usable);
\item track follower (GEANE\footnote{GEANE is a track follower distributed 
within the \texttt{geant3} package. It is written in FORTRAN and a C++ interface has been developed in \fairroot 
and is used also in the extrapolation step of the Kalman fit.} \cite{bib:stt:sim:geane});
\item event display based on TEve;
\item database for geometry and parameters handling.
\end{itemize}
% *****************************************************************
% STT CLASSES
\subsection{The STT Simulation and Reconstruction}
A full simulation chain can be characterized by four main steps: simulation, digitization, reconstruction and analysis. 
In this section only the simulation and reconstruction code, which provides the tracks used for the analysis, will be addressed exhaustively, while 
the digitization will only be mentioned since it has already been described in \Refsec{sec:stt:tub}.
The STT specific classes are all contained in the \texttt{stt} directory of \pandaroot.
\subsubsection{Simulation and Digitization}
During this step, realistic data, resembling the ones that will be available from the operating system, are generated, 
ready for the reconstruction. 
It can be divided in two parts, concerning, respectively, the tracker setup and its response to the passage of the particles. 
\par
The {\bf detector description} is contained in the \texttt{PndStt} class, where the geometry is loaded and the sensitive 
material is set to collect signals from charged particle transversing it. 
The straw tubes are built and positioned in an ASCII geometry file; for each tube, the coating, the filling gas mixture and the 
wire materials are implemented. The passive elements of the tube such as the plugs have not been implemented yet. On the other side, 
the passive support elements, which surround the central tracker, are present. It is however foreseen to insert all the information on 
passive elements in the future. Moreover, since the presence of many details will slow down the simulation, it is foreseen that the 
final geometry description will contain only {\it average} materials to take into account the correct material budget but be fast 
enough to grant good time performances. 
At the simulation stage the geometrical parameters of the tube are saved in the parameter file in order to be retrievable at any 
stage of the reconstruction. 
\par
After the collection of MC points from charged particles, the detector response of the STT is simulated as described in 
\Refsec{sec:stt:tub} during the {\bf digitization} step which provides the collection of realistic hits.
 These hits contain the information on the drift radius and the energy deposit: it must be pointed out that actual hits coming from the
  detector will contain only the time information together with the deposited energy, but in the present code the conversion from time
   to drift radius, which will be later part
   of the reconstruction, is inserted directly in the simulation of the single straw response (i.e. there is no separation between a
   ``digi'', with the time information, and a``hit'' with the reconstructed drift radius).
\subsubsection{Reconstruction} 
In a tracking detector, the aim of the reconstruction is to collect the hits, assign them to the different track candidates and then fit the
 obtained track candidates to get the momentum of each particle. 
% *****************************************************************
The STT does not provide the $x, y, z$ spatial coordinates of the point where the particle passed. When a tube is hit by a
 particle, the only available information for the track reconstruction is the measured drift radius, together with the position and
  orientation in space of the tube itself. A specific track finding (described in \Refsec{sec:stt:sim:pr} and \Refsec{sec:stt:sim:pr2}) and
   fitting (described in \Refsec{sec:stt:sim:kalman}) procedure has been developed relying only on this information. 
This procedure takes place through a chain of tasks, each one performing operations at event stage. Different packages devoted to the global
 tracking are available in \pandaroot \cite{bib:stt:sim:pr0,bib:stt:sim:pr1,bib:stt:sim:pr2,bib:stt:sim:pr3}: only the procedure and the code
  used to obtain the results presented in \Refsec{sec:stt:stt:per} will be described in \Refsec{sec:stt:sim:pr} and \Refsec{sec:stt:sim:kalman}. 
% *******************************
\, \\
A dedicated pattern recognition for secondary tracks, i.e. tracks whose origin is far from the interaction point, is under development and will
 be described in \Refsec{sec:stt:sim:pr2}.
\subsection{The Pattern Recognition for Primary Tracks} \label{sec:stt:sim:pr}
The track finder procedure for primary tracks crossing the STT detector is divided in several steps:
\begin{itemize}
\item track finding of the MVD stand-alone;
\item track finding starting from the STT hits only;
\item extension using also the MVD hits;
\item extension in the forward region using the GEM hits;
\item ``cleanup'' procedure to remove spurious tracks produced by the
high interaction rate of \PANDA.
\end{itemize}
\subsubsection{MVD Local Track Finding}
The MVD stand-alone pattern recognition divides the problem in a circle fit (in the xy plane) and a linear fit (in the arc length {\it vs} 
z coordinate plane). The circle fit is performed using the projection of the MVD hits to a Riemann sphere and fitting a plane through them. 
After this, the parameters describing the plane can be translated into the track parameters in the xy plane \cite{bib:stt:sim:riemann}.
\subsubsection{Track Finding Starting from the STT Hits}
The pattern recognition for the track identification proceeds in two steps, using at first the axial straws, then the skewed straws.
\par
\begin{figure}%[h!]  
\begin{center}
\includegraphics[width=\swidth]{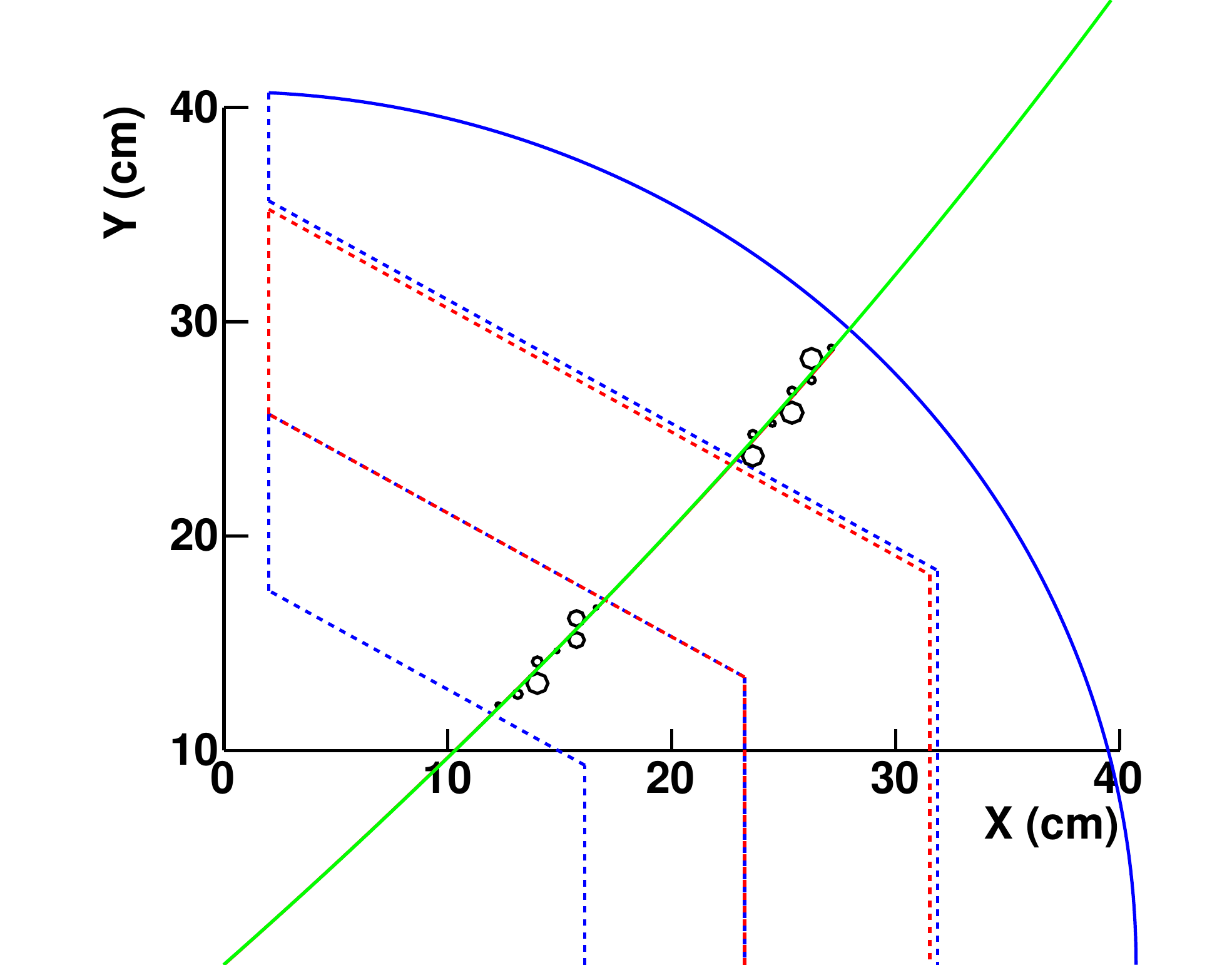}
\end{center}
%\vspace*{-0.6cm}
\caption[Simulated track found by the pattern recognition]{Track generated with Monte Carlo at the interaction vertex;
the small circles are the isochrone circles of the STT axial straws in the XY projection;
the track is the circle tangent to all drift circles.
The green curve is the Monte Carlo truth,
the red curve (almost not visible because essentially it coincides)
is found by the pattern recognition.
} \label{fig:stt:sim:pr1}
\end{figure}
\begin{figure}%[h!]  
\begin{center}
\includegraphics[width=\swidth]{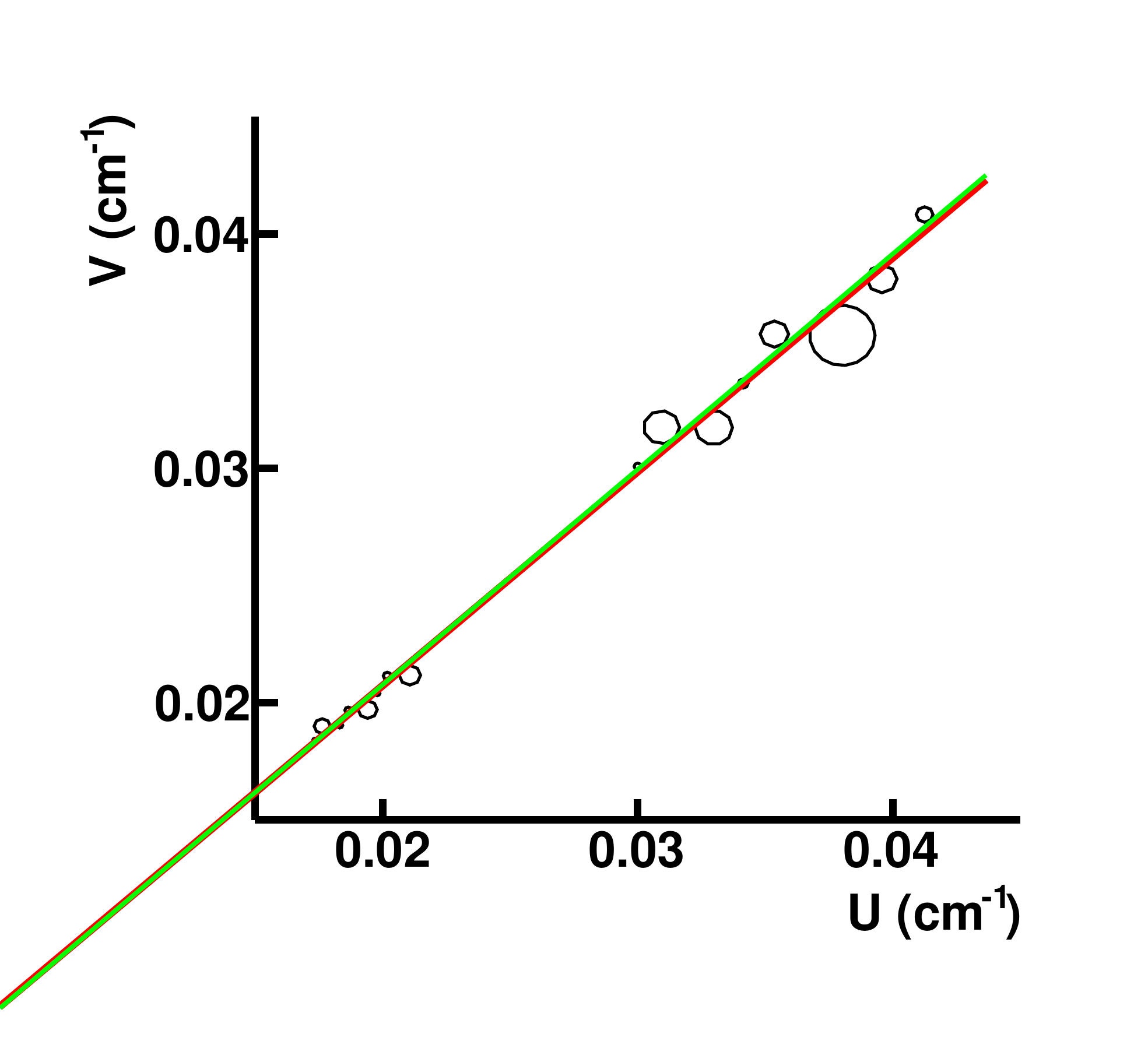}
\end{center}
%\vspace*{-0.6cm}
\caption[Simulated track after conformal transformation]{The same track of \Reffig{fig:stt:sim:pr1} plotted
after the conformal transformation. The track circle
transforms into a straight line, while the drift circumferences
transform into circumferences. The track straight line is still
tangent to all drift circles. The green line is the Monte Carlo truth,
the red line (almost not visible because essentially it coincides)
is found by the pattern recognition.
} \label{fig:stt:sim:pr2}
\end{figure}

In the first step only the hits of axial wires are used. The X and Y position
of the wires and the drift radius
define a small circumference in the XY plane (drift circles) to which the particle trajectory
is tangent (see \Reffig{fig:stt:sim:pr1}).
%%%%%%%%%%%%%%%%%%%%%%%%%%%%%%%%%%%%%%%%%%%%%%%%%%%%%%%%%%%%%%%%%%%%%%
%%%%%%%%%%%%%%%%%%%%%%%%%%%%%%%%%%%%%%%%%%%%%%%%%%%%%%%%%%%%%%%%%%%%%%%
The following conformal transformation~:
%\begin{eqnarray}
% U \equiv \frac{X}{X^2 + Y^2}\nonumber \\
% V \equiv \frac{Y}{X^2 + Y^2}\nonumber
%\end{eqnarray}
$$
 U \equiv \frac{X}{X^2 + Y^2}\qquad
 V \equiv \frac{Y}{X^2 + Y^2}
$$
is applied to the hit drift circles. New drift circumferences are obtained in the UV space.
The particle trajectory, a circle passing through the origin in the XY plane, transforms into a straight
line in the UV plane. A considerable mathematical simplification is obtained in this way since
the problem reduces to finding straight line trajectories tangential
to drift circles (see \Reffig{fig:stt:sim:pr2}).
%%%%%%%%%%%%%%%%%%%%%%%%%%%%%%%%%%%%%%%%%%%%%%%%%%%%%%%%%%%%%%%%%%%%%%
%%%%%%%%%%%%%%%%%%%%%%%%%%%%%%%%%%%%%%%%%%%%%%%%%%%%%%%%%%%%%%%%%%%%%%%
The Pattern Recognition proceeds by finding clusters of hits in the UV plane belonging to a
straight line. The search starts from hits belonging to the more external STT axial layers
where the hit density is lower.
A  classical ``road finding'' technique with a simple proximity criterion
is used and a first fit to
a straight line is attempted as soon as the cluster contains a minimum number of hits.
The fit is performed minimizing a ``cost function'' which is
the sum of the absolute values of the residuals (in the usual $\chi^2$
it is the sum of the squares of them). This minimization is performed
using a Mixed Integer Linear Programming (MILP) algorithm that is usually much faster than
the normal $\chi^2$ minimization. If a straight line is successfully fitted, a search
among all unused STT axial hits is performed and hits close
enough to it are
associated and a new candidate track is formed.
After this stage three of the five parameters of the track helix are known (the radius R,
the position of the helix center in the XY plane).
\par
\begin{figure}%[h!]  
\begin{center}
\includegraphics[width=\swidth]{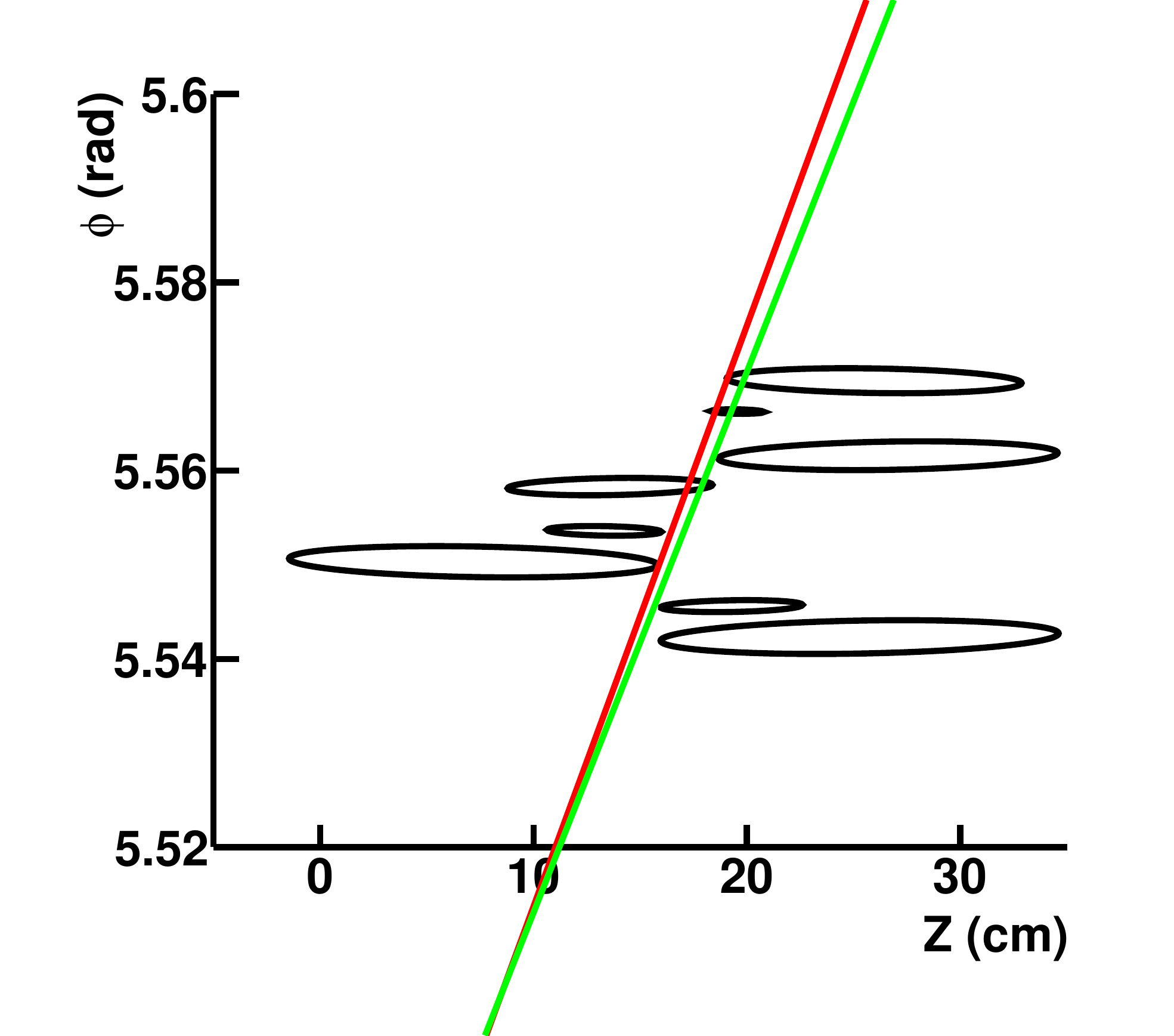}
\end{center}
%\vspace*{-0.6cm}
\caption[Simulated track plotted in the $\phi-Z$ plane]{The same track of \Reffig{fig:stt:sim:pr1} plotted
in the $\phi-Z$ plane.
The approximate ellipses are the intersection of the skewed
straws with the cylinder on which the helix trajectory
lies. The track straight line is
tangent to all skewed straw drift ellipses.
 The green line is the Monte Carlo truth,
the red line
is found by the pattern recognition.
} \label{fig:stt:sim:pr3}
\end{figure}

In the second step the remaining two parameters of the helix are determined by using the 
hits of the skewed STT straws. The ``drift cylinder'' is defined as an imaginary
cylinder coaxial to the straw wire and with radius equal to the drift radius.
Only the hits of those skewed straws are considered whose drift cylinder intersects the cylinder
on which the helix lies (see \Reffig{fig:stt:sim:pr3}).
%%%%%%%%%%%%%%%%%%%%%%%%%%%%%%%%%%%%%%%%%%%%%%%%%%%%%%%%%%%%%%%%%%%%%%
%%%%%%%%%%%%%%%%%%%%%%%%%%%%%%%%%%%%%%%%%%%%%%%%%%%%%%%%%%%%%%%%%%%%%%%
This intersection is approximately an ellipse.
The helix trajectory is a straight line
on the lateral surface of the helix cylinder ($\equiv \phi-Z$ plane)
with equation~:
$$ \phi = K Z + \phi_0$$ and tangent to the ellipses of the skewed straws.
\noindent
$K$ and $\phi_0$ are the remaining two parameters of the helix. At this stage
in the algorithm
$\phi_0$ is constrained with the requirement that the track originates from (0,0,0) in the
XYZ reference frame.
A fit with with a MILP algorithm gives $K$. Possible spurious skewed
 straw hits are rejected
if their distance from the fitted
 straight line exceeds a certain limit.
Then a track candidate is constructed, consisting of all STT associated
 axial and skewed hits.

It should be mentioned here that an extension of this scheme is being written presently and
although it is only in a preliminary stage it looks very promising
and it will be included in the future pattern recognition. The peculiarity of the new scheme
consists in the use of the SciTil hits at the very first stage of the algorithm.
Since SciTil hits are very fast ($\approx 100$ ps) the jitter of such pulses will be dominated
essentially by the length of the trajectory of the charged particles from the interaction
vertex to the SciTil detectors. This has been estimated to be of the order of 1 ns.
Consequently such hits will be only very marginally affected by pileup of previous events
and they will be very useful both in giving the time of production of an event and in
signalling that the charged track associated to them is not spurious.
That is why in the new scheme the pattern recognition algorithm starts clusterizing
hits from the SciTil hits, in the conformal space. Significant CPU time gains and spurious
track rejection are expected with this strategy.
\subsubsection{STT + MVD Track Finding}
In this stage all
the track candidates of the previous step are considered.
First the trajectory circle in the XY plane, found as described above,
is used to associate hits of MVD tracklets close to it. The MVD tracklets
were previously found by the MVD standalone pattern recognition. 
The fit of the trajectory circle is performed again,
including also the newly associated MVD hits and releasing the constraint that the
trajectory goes necessarily through (0,0,0). By
using the improved helix parameters in the XY plane (better radius and center of the trajectory cylinder),
skewed straw hits are associated to the candidate track in a similar way as in the previous step.
The fit in the $\phi-Z$ plane is performed again using these skewed straw hits plus the MVD hits.
The new trajectory parameters are used to eliminate some spurious skewed straw
hits and/or to include
some new axial straw
hits not yet included before. In this way the final track candidates are obtained.
Finally an attempt is made to find tracks starting from the MVD tracklets found by
the MVD stand-alone pattern recognition. MVD tracklets not yet used in the previous steps and
containing at least three MVD hits, are fitted with a straight line
in the conformal UV space first and then in the $\phi-Z$ plane with the fast MILP minimizer.
A new track candidate is added
only if the helix trajectory intersects the STT region.
The found trajectory is used
to collect straw hits, both axial and skewed, lying close to it.

\subsubsection{The GEM Extension}
Once the MVD + STT track finder has been run and a track hypothesis is available,
the GEM hit contribution can be exploited in the angular region where the tracks can cross both MVD/STT and the GEM
 detectors ($7^{o} < \theta < 21^{o}$). A simple extrapolation of the tracks using the track follower GEANE from the last point of the central 
tracker on each plane of the GEM detector is the starting point for adding
 the GEM hits to the tracks. 
For each extrapolation the distance between the propagated point and the error associated to it are calculated. 
The hit is associated to the track if the distance is within $5\, \sigma$. 
The GEM detector is composed of three stations, with two sensors each. Every sensor has two views. When more than one 
track hits the GEM stations, combinatorial background is present and has to be suppressed. A specific test has been 
written to take care of this: the two sensors in each station
 are overlapped and only when a hit has its counterpart 
on the other sensor, within $1$ cm, is considered {\it true}, otherwise it is flagged as {\it fake}.
Once a true hit pair
 has been found, the GEM channels of such hits are excluded by further combinations, namely
 all the hits on the same sensors defined by the same channels
are considered 
as combinatorial background. Only true hits are used in the next steps and possibly
assigned to tracks.  
The tracks are eventually refined by requiring that each GEM hit is associated to at most one track and each track is 
associated to at most one hit on every measurement plane.   % CHECK (performance?) AND MVD + GEM alone?
Once the hits have been attached to a track, a dedicated Kalman filter,
 specifically implemented inside the GEM extension code,
  is applied on that track, using the measured GEM hits: this 
is necessary because the extrapolation with GEANE is always performed with the mass
 hypothesis of the muon and this could lead 
to an underestimation or overestimation of the energy loss between the GEM stations,
 causing the propagated point to be too far 
from the measured one. The application of the Kalman filter forces the track to stick
 to the measured hits and allows to retrieve 
some hits formerly missed due to the wrong mass hypothesis. 
%
%Concerning the helix parameters found in the previous step, they are not recalculated
% since the GEM planes lie in the region between 
%the solenoidal an dipolar magnetic field, which is highly inhomogeneous.
%

\subsubsection{``Cleanup'' Procedure to Remove Spurious Tracks}
The average interaction rate of 20 MHz of \PANDA and the maximum drift time of the STT straws of
200 ns pose the problem of the presence of a large number of spurious hits in the
STT system (spurious $\equiv$ real hit belonging to a different event).
To every interesting physics event there will be superimposed some STT hits belonging to previous or later
events produced by the overwhelmingly large p$\overline{\rm p}$ total interaction.
This will cause an increase of
the number of spurious tracks found by the pattern recognition. A spurious track is formed by
spurious hits and its characteristic, most of the times, is the absence of MVD hits and
also gaps in the continuity of STT hits. The former happens because the time duration of a MVD
hit is typically 10 ns and so those hits disappear when the spurious event is late or early by more
than 10 ns. The latter happens because some STT (early) spurious hits have too small drift
time or (for the late spurious hits) too large drift time and they fall out of the time window
of the physics event leaving ``holes'' or gaps along the spurious track.

This section briefly describes
the ``cleanup'' algorithm applied after the pattern recognition.
This procedure is also useful to reject the (low percentage)
ghost tracks inevitably produced by the pattern recognition even in the absence of spurious hits.
%However it should be mentioned that the single track tests shown in \Refsec{sec:stt:stt:per} were produced
%without using this cleanup procedure (only for historical reasons). 
The ``cleanup'' procedures has been used in the studies of the physics channels described in \Refsec{sec:stt:ben}.

The algorithm begins using the helix track parameters
found by the pattern recognition.
If there are no MVD hits associated
to a track candidate going through the
MVD system, this is a typical spurious or ghost track and the candidate is rejected.
If gaps with more than 1 hit missing in the STT region are found,
the track candidate is rejected.

In order to check the effectiveness of the cleanup procedure,
a dedicated version of the \PANDA Monte Carlo was implemented, with the spurious hits superimposed
at an average rate of 20 MHz.
Presently the cleanup procedure is not in its final version yet.
The geometric accuracy in the determination of how many hits should be in a track is still
not refined and consequently
it ``cleans'' too much, lowering the
detection  efficiency of the true tracks down to around 90 \%.
In the near future this task will be refined and brought to conclusion.

\subsection{The Pattern Recognition for Secondary Tracks} 
\label{sec:stt:sim:pr2}
Events generated from the antiproton proton annihilation may produce neutral, long living particles, like $\Lambda$ or $K^0$, which travel 
a while before decaying (for example the $c\tau$ value for the $\Lambda$ is $\sim 7.9$ cm). The peculiarity of these decays is that the charged 
particles coming from them do not originate from the interaction point, but from a displaced position. Even though most of the secondary 
vertices will fall within the MVD, the neutral particles can as well decay outside it, thus demanding for good reconstruction capabilities of 
the other tracking detectors (STT, GEM). 
The secondary track finder is still under development. 

\subsubsection{General  Concept}
At present, the secondary track finder uses only the STT hits, but it will be extended to include also the MVD and GEM information. The final 
version of the global pattern recognition of secondary tracks in the target spectrometer will resemble the one for primary tracks and will 
consist in the following steps:
\begin{itemize}
\item MVD local track finding, 
\item STT local track finding for secondary tracks,
\item MVD + STT track finding, 
\item GEM extension. 
\end{itemize}
Since the MVD  + STT pattern recognition and the GEM extension for the primary tracks do not require that the particle comes from the interaction 
point they can also be used for secondary tracks, with minor modifications.
The GEM extension code can be used also for tracks hitting only the STT and GEM, without MVD hits. 
Moreover, the secondary track finder will run over the hits which are left unassigned by the primary track finder and which most likely will 
belong either to secondary tracks or to background tracks. 
The main difference between the primary track finder and the secondary track finder is the inclusion of the interaction point as a constraint 
for the primary which cannot be used for the secondary tracks.

\subsubsection{Secondary Track Finding Procedure}

Currently, the procedure starts from the xy plane to find the coordinates of the center of the track and its radius ($x_c, y_c, R$), and 
subsequently finds the remaining two parameters of the helix, $\tan(\lambda)$ and $z_0$. It is divided in several steps:
\begin{itemize}
\item {\bf clustering:} the STT hits are assigned to different sectors of the detector depending on their 
position; a first division is between left and right hits (which refers to the two half cylinders of the STT) 
and a second separation is made among inner parallel tubes, skewed tubes and outer parallel tubes, as shown in 
\Reffig{fig:stt:sim:sectors}. 

\begin{figure}%[!h]
\begin{center}
\includegraphics[width=\swidth]{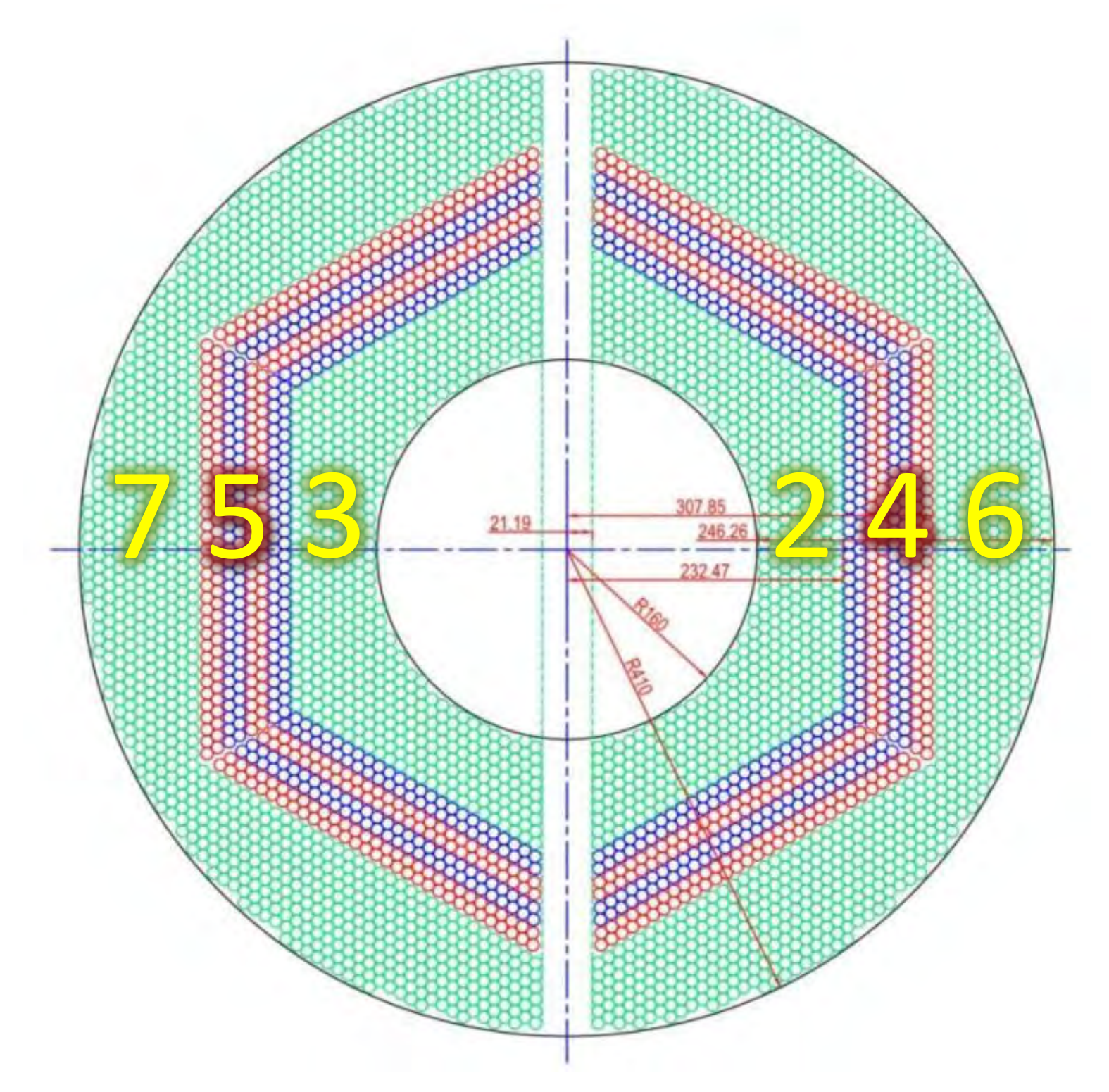}
\caption[Sector numbering of the STT]{The numbers correspond to the different sectors in which the STT is divided. $2$ and $3$ are inner 
parallel tubes, $4$ and $5$ are skewed tubes and $6$ and $7$ are outer parallel tubes.}
\label{fig:stt:sim:sectors}
\end{center}
\end{figure}

The clusterization is performed directly on the tubes without taking into account the drift radius. If the xy 
distance between the centers of two tubes is less than 1.2 cm the hits are collected in the same cluster.
\item {\bf xy fitting:} when possible, the clusters are coupled in order to have one inner cluster and one outer 
cluster of parallel tubes and they are fitted in the xy plane. The fit makes use of the conformal transformation, and considers 
the drift radii of the hits. Since the track cannot be assumed to  go through the interaction point, the positions of the tubes 
are translated by the coordinates of the hit tube with the smallest drift radius. This is necessary since the conformal 
transformation maps circular tracks into straight lines only if they come from the origin and with the 
translation it is assumed that the track is passing through the center of the tube. When the association 
inner/outer cluster is impossible, the single cluster is fitted alone.
\item {\bf z fitting:} Once the xy parameters of the track are found, the wires of the skewed tube are projected 
onto the xy plane. The ones which cross the trajectory circle are associated to the track. Their intersections 
are computed considering the drift radius and the left/right ambiguity: for this reason two solutions for each 
skewed tube are possible. For all the solutions the track length and the corresponding z coordinate are computed 
and are plotted on a z vs track length plane. Only the true intersections lie on a straight line. They are 
identified with a Hough transformation \cite{bib:stt:sim:hough} and the two remaining helix parameters are found by means of a straight line fit.
\end{itemize}

% ======================================
\subsubsection{First Test and Future Developments}
A test of this procedure has been made on $21500$ $p \overline{p} \rightarrow \Lambda \overline{\Lambda}$ events. 
They were generated with the \texttt{EvtGen} \cite{bib:stt:sim:evtgen} generator with an antiproton momentum of $4\, \gevc$. 
All the $\Lambda$ were forced to decay into $p \pi^-$ and the $\overline{\Lambda}$ into $\overline{p} \pi^+$. No forward 
peaking in the angular distribution was taken into account, but phase space was 
used as decay model. The tracks were simulated, digitized and finally reconstructed with the secondary track 
finder. Since this pattern recognition is still in its early stage, a full analysis of the channel was not 
possible. All the results were obtained without the MVD and GEM detectors, no Kalman filter was applied to the 
tracks and no kinematic fit was used. Further developments in the track finding and the use of the surrounding 
detectors will improve significantly the results. Once the information of these detectors will be used, a real 
event generator with the proper angular distributions will be adopted to evaluate the final event reconstruction efficiency. 

\begin{figure}[!t]
\begin{center}
\includegraphics[width=\swidth]{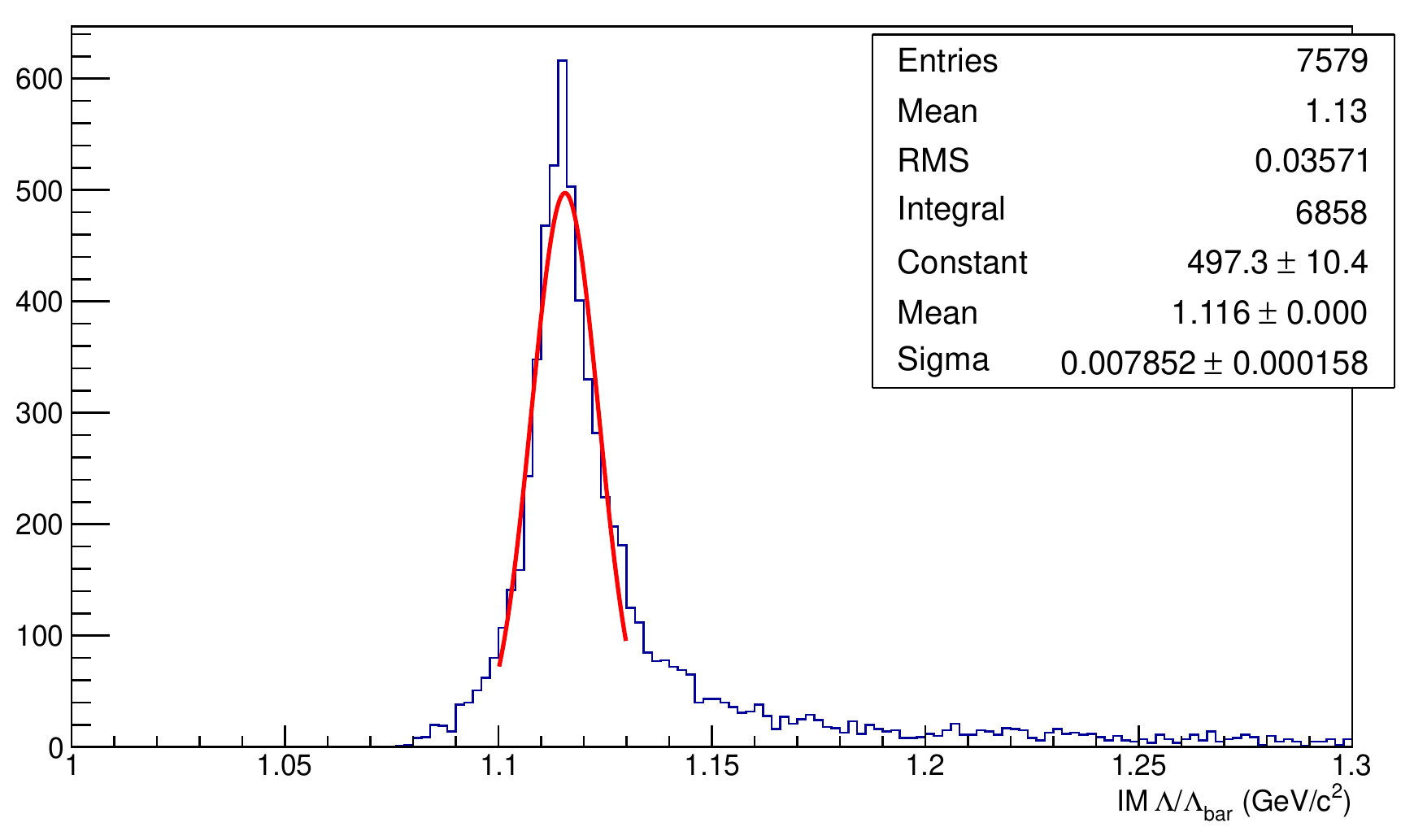}
\caption[$\Lambda$ and $\overline{\Lambda}$ invariant mass distribution]{$\Lambda$ and $\overline{\Lambda}$ invariant mass for tracks which cover the inner parallel tube sector, the skewed sector and the outer parallel sector.}
\label{fig:stt:sim:llbarim}
\end{center}
\end{figure}

\begin{figure*}
\begin{center}
\includegraphics[width=\dwidth]{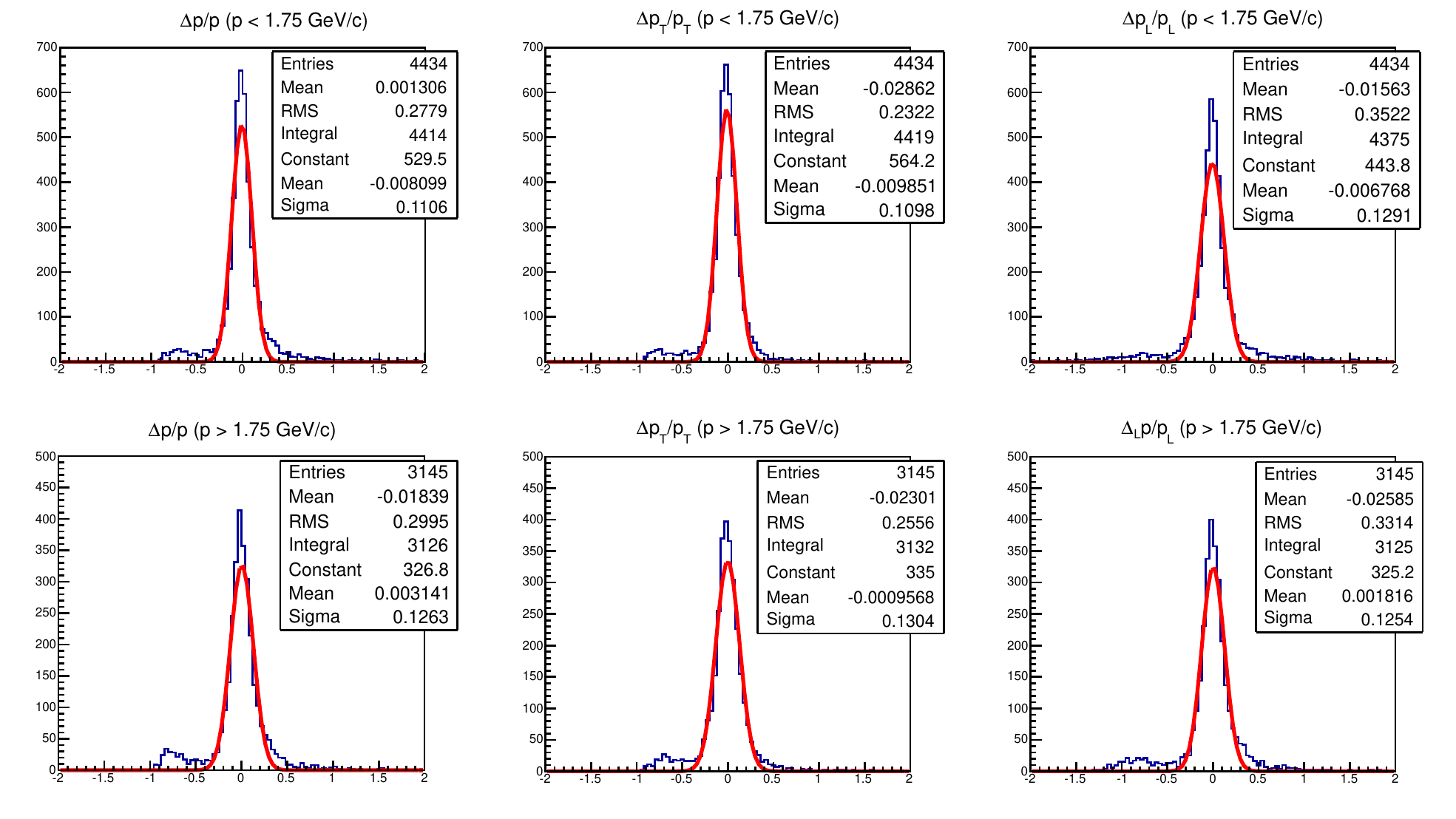}
\caption[Proton and antiproton momentum resolution]{$\frac{\Delta p}{p} = \frac{RECO_{p} - MC_p}{MC_p}$ for the protons and antiprotons which form the 
$\Lambda$ and $\overline{\Lambda}$ of \Reffig{fig:stt:sim:llbarim}. The first column plots the total momentum, 
the second column the transverse momentum and the third column the longitudinal momentum. The two rows represent 
the low momentum (top) and high momentum (bottom) particles.}
\label{fig:stt:sim:Dpoverpprot}
\end{center}
\end{figure*}

\begin{figure*}
\begin{center}
\includegraphics[width=\dwidth]{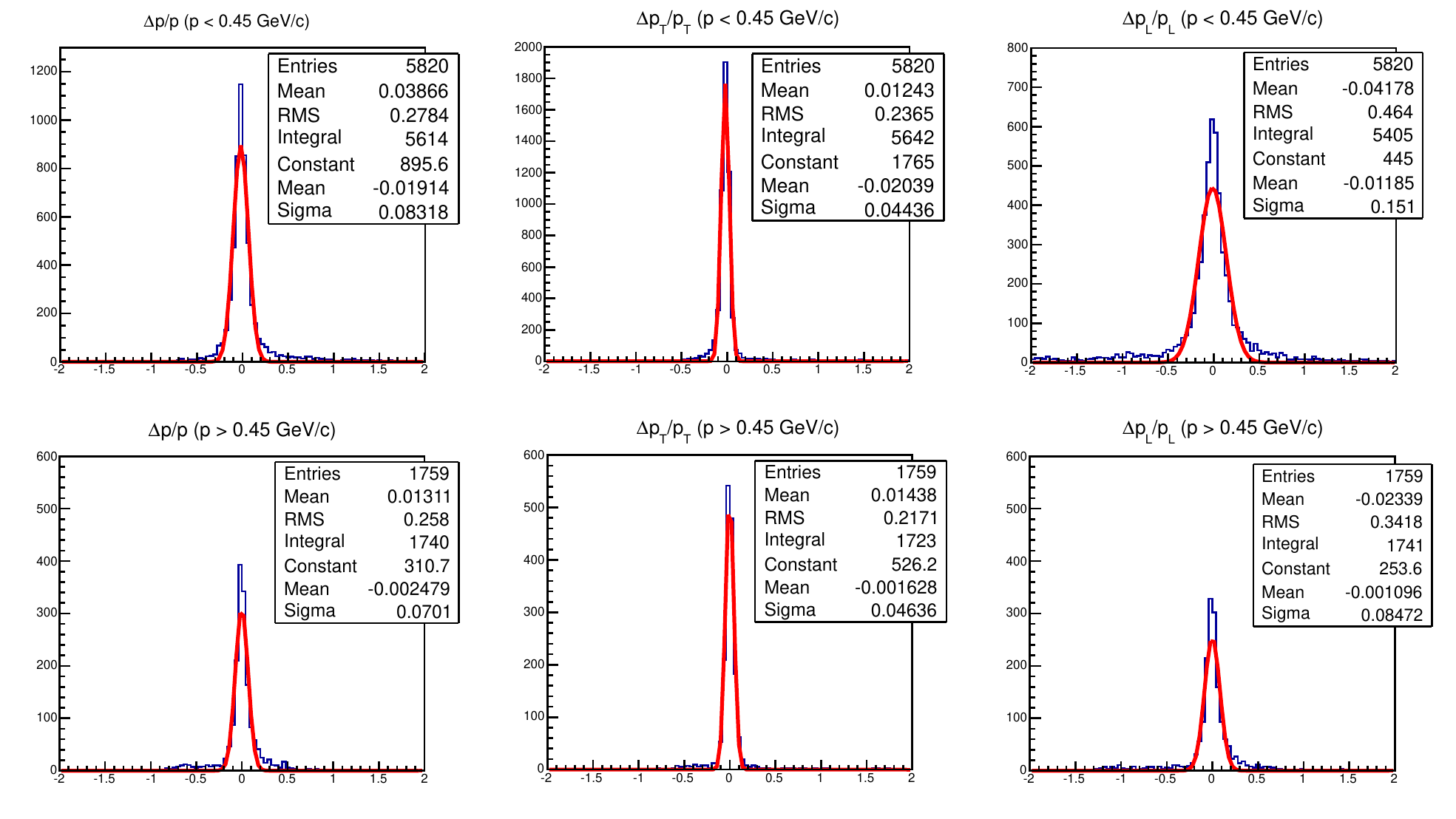}
\caption[$\pi^+$ and $\pi^-$ momentum resolution]{$\frac{\Delta p}{p} = \frac{RECO_{p} - MC_p}{MC_p}$ for the negative and positive pions which form the 
$\Lambda$ and $\overline{\Lambda}$ of \Reffig{fig:stt:sim:llbarim}. The first column plots the total momentum, 
the second column the transverse momentum and the third column the longitudinal momentum. The two rows represent 
the low momentum (top) and high momentum (bottom) particles.}
\label{fig:stt:sim:Dpoverppion}
\end{center}
\end{figure*}

The tracks reconstructed with the secondary pattern recognition were associated to pion and proton tracks using 
the MC particle identification. An algorithm to compute the distance among the two helices was then applied to 
each couple of $\overline{p} \pi^+$ and $p \pi^-$ tracks in order to find the point of closest approach. This 
point is used as the reconstructed vertex and the tracks are backtracked there. The invariant mass for the tracks 
traversing the inner parallel sector, the skewed sector and the outer parallel sector is shown in \Reffig{fig:stt:sim:llbarim}.
The corresponding $\Delta p/p$ for total, transverse and longitudinal momenta for both the protons and pions are
plotted in \Reffig{fig:stt:sim:Dpoverpprot} and \Reffig{fig:stt:sim:Dpoverppion}.

% ======================================

\begin{figure*}
\begin{center}
\includegraphics[width=\dwidth]{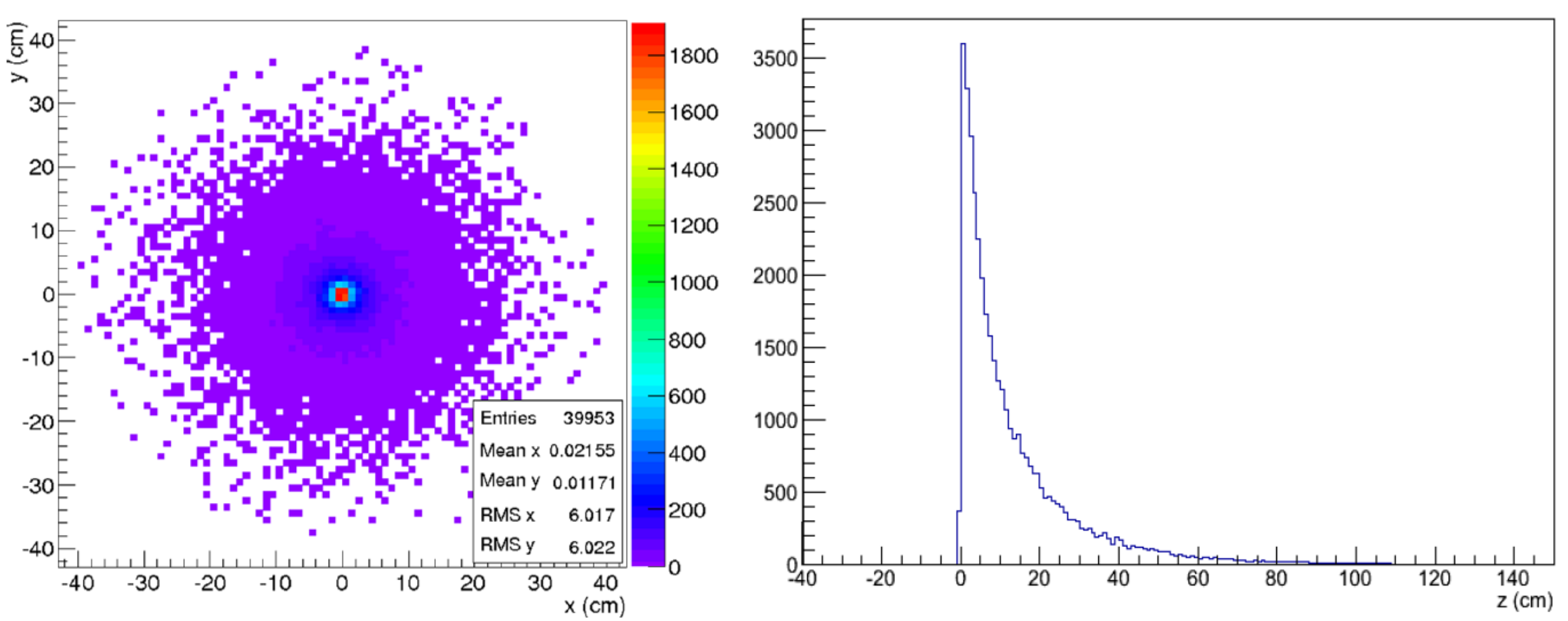}
\caption[Radial and $z$ distribution of the MC vertices leaving at least one track in the STT]{Radial (left) and $z$ (right) distribution of the MC vertices leaving at least one track in the STT.} 
\label{fig:stt:sim:vtx_distr}
\end{center}
\end{figure*}

As already mentioned, the $p \overline{p} \rightarrow \Lambda \overline{\Lambda}$ events have been generated a phase space distribution. 
The following considerations rely on this and thus the efficiencies and resolutions are not the ultimate ones. The number of Monte Carlo 
vertices which leave at least one track in the STT is $39953$ out of $43000$ generated, with a geometrical distribution shown in 
\Reffig{fig:stt:sim:vtx_distr}. Among these, $28303$ vertices leave two tracks in the STT and thus their invariant mass can be reconstructed 
in the STT alone. The fraction of events where both $\Lambda$ and $\overline{\Lambda}$ are reconstructable inside the STT is not considered 
here since the model used for the event generation does not correctly populate the phase space.

In \Reffig{fig:stt:sim:llbarim} $5433$ entries fall within $3 \sigma$ (corresponds to 19\,\% of the $\overline{\Lambda}/\Lambda$ which leave 
two tracks in the STT). The use of {\it all sectors} tracks will increase the efficiency. 
It is clear that there is room for further improvements: starting from the {\it three sectors} tracks, for which a fine tuning of the cuts and the 
implementation of additional functions to assign the hits which are left unassigned from the present procedure will improve the efficiency. 
In addition to this, a lot of events also leave hits in the MVD and in the GEM. As a result, their contribution must be taken into consideration to refine the results. 
In particular the GEM will have great importance when the realistic events will be considered. The momentum resolution can be improved with the application 
of the Kalman filter and of the kinematic fit. The invariant mass resolution itself is not yet the best achievable: For example, the use of a kinematic fit 
will lower the tail in the invariant mass distribution. 
In conclusion: Improvements in both efficiency and resolution are expected for the secondary track finder from the inclusion of the contribution of the 
surrounding detectors and from the application of special fitting techniques.
\clearpage

% ======================================
\subsection{The Kalman Filter} 
\label{sec:stt:sim:kalman}
The track fitting step is performed through the Kalman filter procedure, using the hits coming from MVD, STT 
and GEM where available and, as starting position and momentum, the values inferred by the pattern recognition
backtracked to the point of closest approach to the interaction point in case of primary particles.
In this section a short summary of the Kalman fit procedure \cite{bib:stt:sim:kalman,bib:stt:sim:fruh2} is 
reported. A more detailed description of this topic can be found in \cite{bib:stt:sim:pvreport} and 
\cite{bib:stt:sim:liathesis} and references quoted therein. The package devoted to the Kalman fit procedure is 
\texttt{\hyphenchar\font45\relax genfit} \cite{bib:stt:sim:genfit}.
\par
The Kalman fit is an iterative procedure which, unlike global methods such as the helix fit, takes into account 
the energy loss, the magnetic field inhomogeneities and the multiple scattering. 
The aim of the Kalman filter is to find the best estimation of the true track point $f_i$ on the $i$-th detector 
plane by minimizing the $\chi^2$:
\begin{eqnarray}  \label{eq:stt:sim:chisq4}
\chi^2(\bm{f}) &=& \sum_i [(\bm{e}_i[\bm{f}_{i-1}] - \bm{f}_i) {\bm W}_{i-1}
                   (\bm{e}_i[\bm{f}_{i-1}] - \bm{f}_i)]  \nonumber \\
   & \ &         + (\bm{x}_i- \bm{f}_i) {\bm V}_i (\bm{x}_i- \bm{f}_i) 
\end{eqnarray}
where $\bm{e}_i[\bm{f}_{i-1}]$ is the extrapolated point on the $i$-th detector plane starting from the true point on the $(i - 1)$-th plane and $\bm{x}_i$ 
is the measured one; $\bm{W}$ and $\bm{V}$ are the weight 
matrices containing respectively the tracking and the measurement errors.
The Kalman filter is a method to minimize the $\chi^2$ of \Refeq{eq:stt:sim:chisq4} to find the true points $\bm{f}_i$. Usually this is done through three steps 
\cite{bib:stt:sim:fruh,bib:stt:sim:innocente}:
\begin{itemize}
\item {\em extrapolation}: the status vector on the $i$-th plane is predicted starting from the knowledge gained up to the $(i-1)$-th plane; 
\item {\em filtering:} this is a preliminary evaluation of the track parameters on plane $i$, making a ``weighted mean'' between the measured and the predicted 
values on the same plane;
\item {\em smoothing:} on each plane, the Kalman point solution of the second step is refined to get the final estimate of its value. This last step is often 
substituted by an alternative option: the so-called ``backtracking'', which consists in repeating the first two steps while extrapolating in backward direction, 
from the last point of the track to the first one. 
\end{itemize}
The Kalman filter algorithm is a standard tracking procedure, but its use in the case of the STT has some peculiarities: the extrapolation method, the use of 
virtual detector planes and the z-reconstruction. 

\begin{figure}[h]
\begin{center}
\includegraphics[width=\swidth]{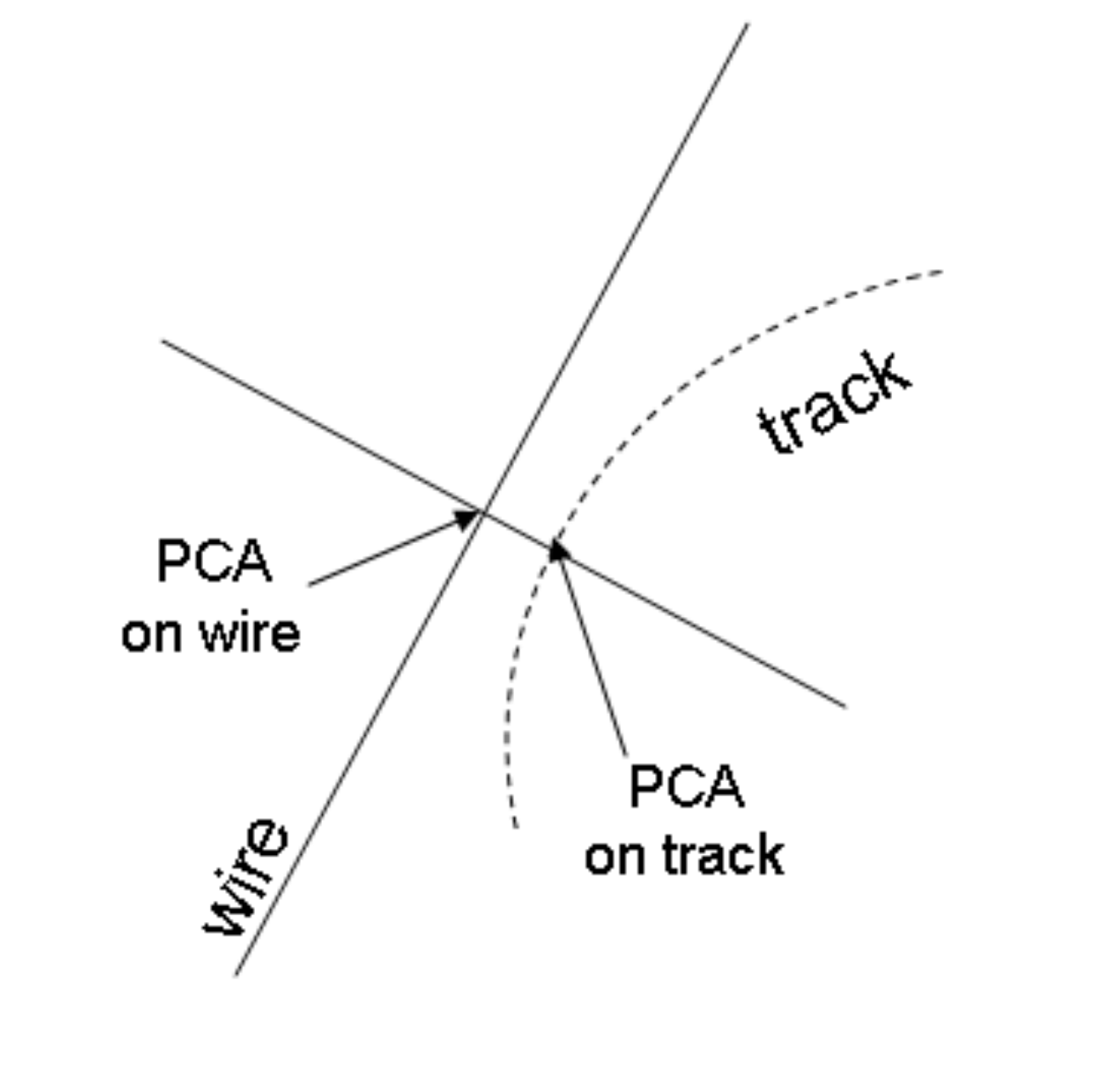}
\caption[Sketch of the positions of the point of closest approach (PCA) on the track and on the wire]{Sketch of the positions of the point of closest approach (PCA) on the track and on the wire.} 
\label{fig:stt:sim:poca}
\end{center}
\end{figure}

\begin{figure}[h]
\begin{center}
\includegraphics[width=\swidth]{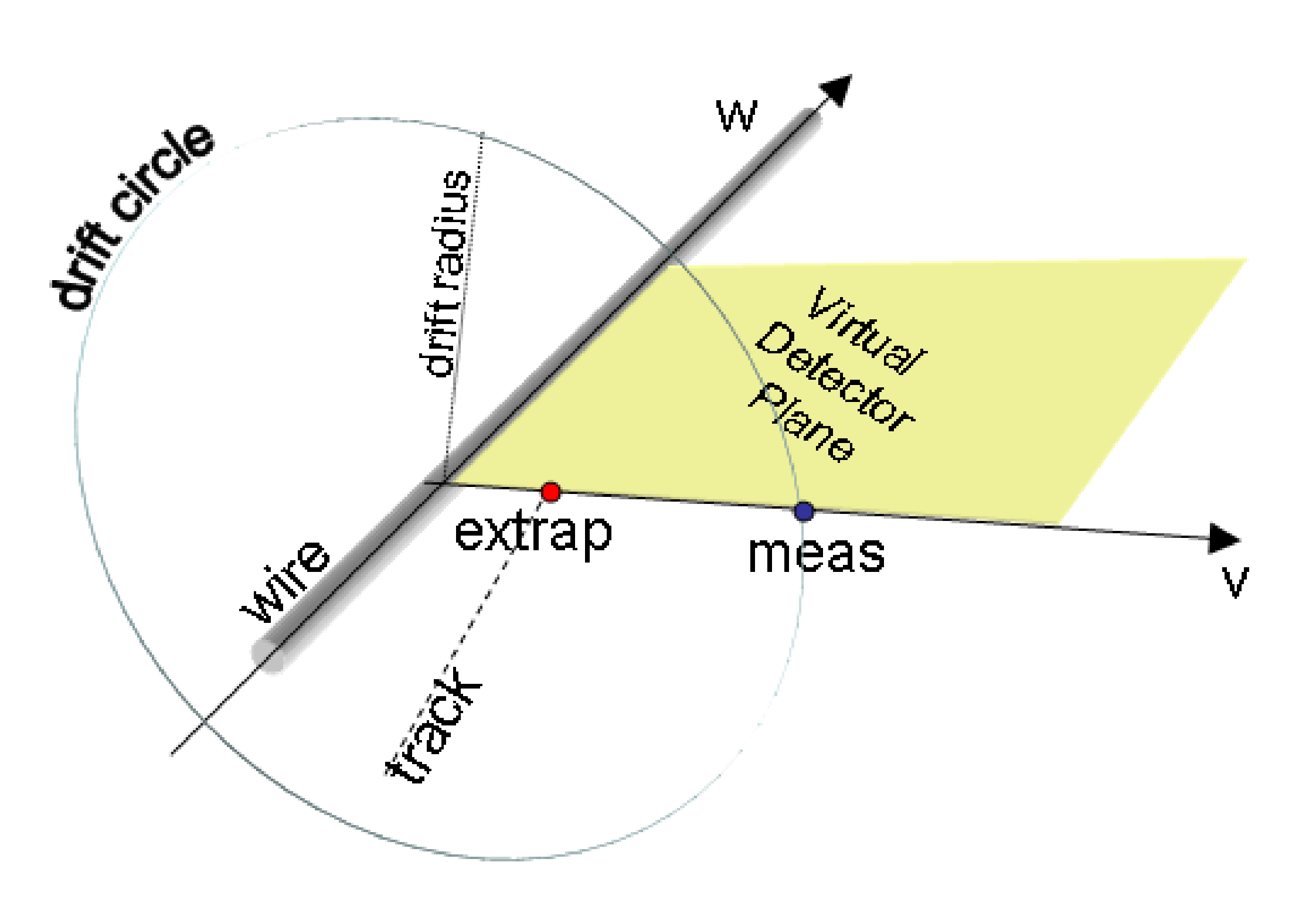}
\caption[Sketch of the virtual detector plane at the point of closest approach to the wire]{Sketch of the virtual detector plane at the point of closest 
approach to the wire.}
\label{fig:stt:sim:virtdetplane}
\end{center}
\end{figure}

The track follower used during the extrapolation step is GEANE. The standard ways of extrapolation made available by this tool are the one to a volume, to a 
plane and to a track length. Though the extrapolation to a plane was useful and easily applied to planar detectors as MVD and GEM, it was not suitable for the 
STT. None of the standard functions was and so a fourth method has been developed: The propagation to the point of closest approach to a line or to a space point. 
In particular, for the STT, the propagation to the point of closest approach to the straw wire is used (see \Reffig{fig:stt:sim:poca}). This method combines the 
propagation to a track length and to a plane: An extrapolation is performed, calculating the distance from the wire step by step, the minimum is found and a plane 
containing the wire and the point of closest approach is built there (see \Reffig{fig:stt:sim:virtdetplane}). Eventually a standard extrapolation onto this plane is made. 
As already pointed out, the straw tube is not a planar device, thus no real measurement plane can be identified. The chosen detector planes are virtual
and are built during the extrapolation step. Each plane is spanned by the axes $v$ and $w$ as shown in \Reffig{fig:stt:sim:virtdetplane}: The $w$ axis 
is along the wire and the $v$ axis orthogonal to it, through the found point of closest approach. 
The z coordinate of each single hit is unknown at this stage since no reconstruction of this coordinate has been done so far and it is not measured directly. 
It is reconstructed by the Kalman filter using the skewed tubes. In fact, given a starting position and direction (which contains also the z information) 
the extrapolation to the point of closest approach to the skewed tubes takes into account their position in tridimensional space and thus, indirectly, the 
z coordinate. When performing the filtering step on the virtual planes associated to the skewed tubes, all the track parameters are modified at the same time, 
taking into account the z information provided by the skewed tubes themselves.

\subsection{The dE/dx Simulation}
\begin{figure}[h]
\begin{center}
\includegraphics[width=\swidth]{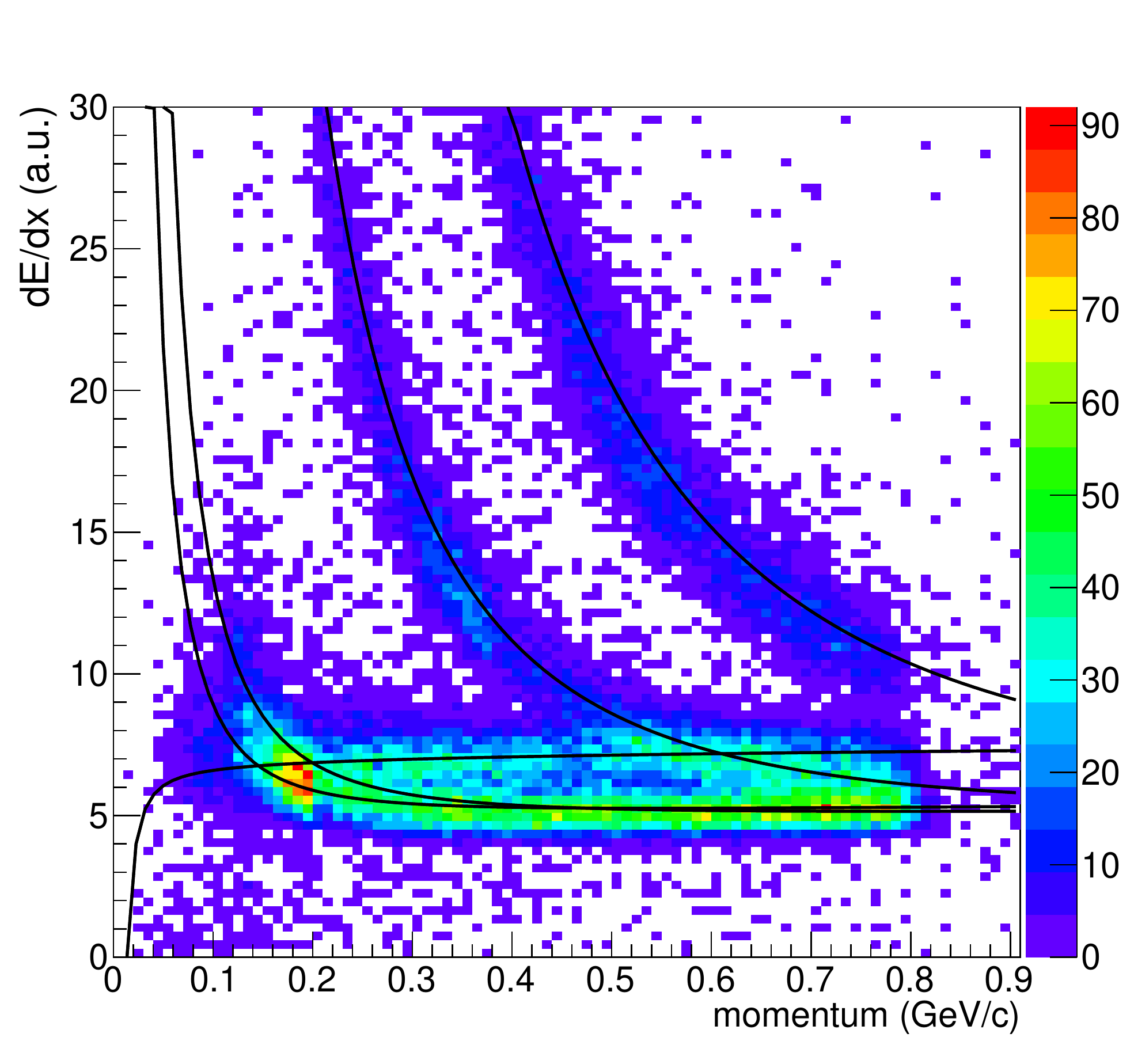}
\caption[Distribution of d$E$/d$x$ truncated mean values vs reconstructed momentum for electrons, muons, pions, 
kaons and protons]{Distribution of d$E$/d$x$ truncated mean values vs reconstructed momentum for electrons, muons, pions, 
kaons and protons. The superimposed lines are the mean value of the bands of \Reffig{fig:stt:sim:dedxbands}. The procedure to find them is described 
in the text.}
\label{fig:stt:sim:dedxvsp}
\end{center}
\end{figure}

\begin{figure}[h]
\begin{center}
\includegraphics[width=\swidth]{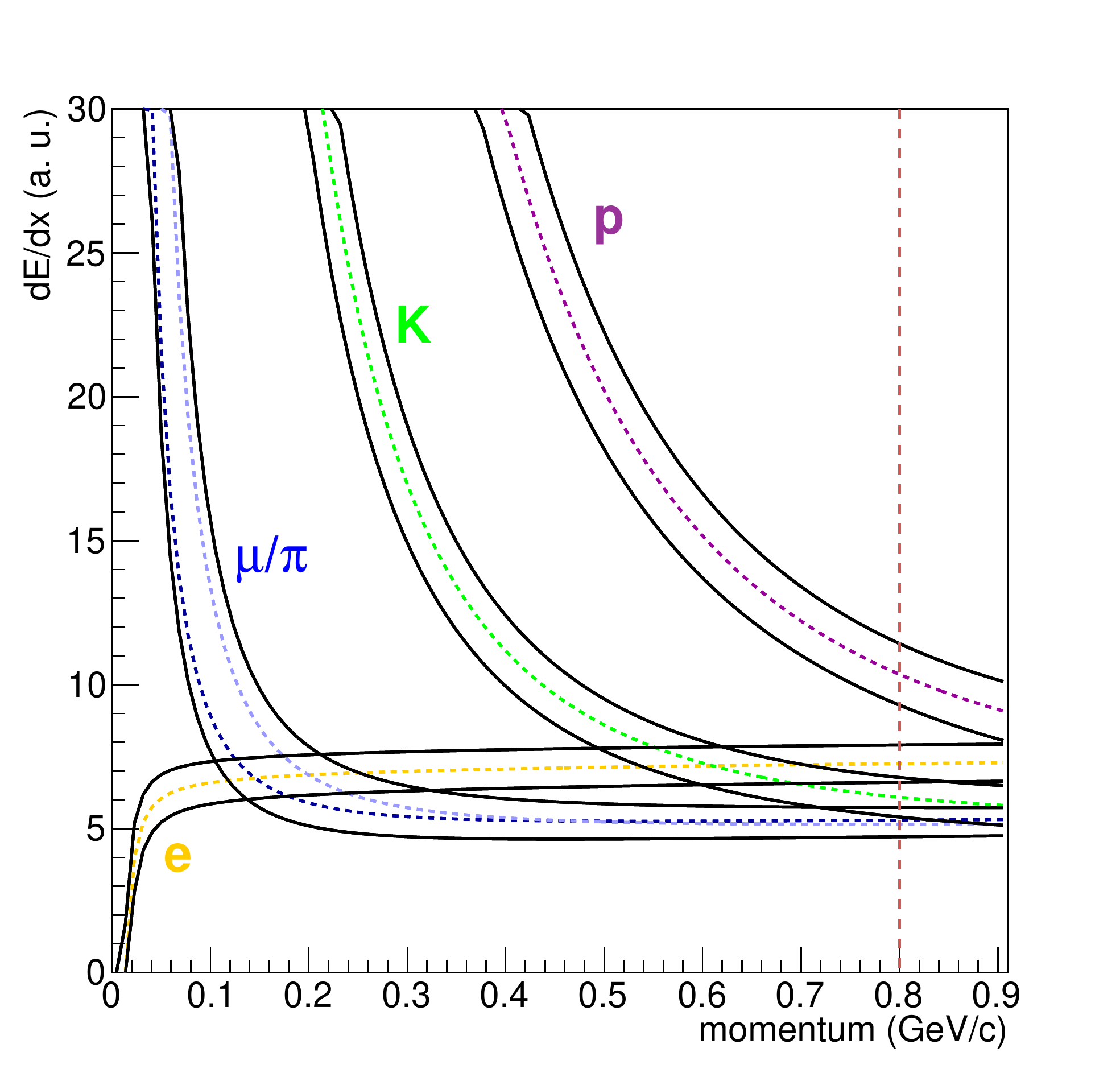} 
\caption[The bands identifying the regions of d$E$/d$x$ truncated mean values vs momentum for different particles]{The bands identifying the regions of d$E$/d$x$ truncated mean values vs momentum found for the different 
particles with the procedure described in the text are drawn (the tracks have been fitted by the Kalman filter with the mass hypothesis 
of muon). 
The muon and pion bands are highly overlapped due to the similarity in their masses: it is not possible to 
distinguish between these two particles in a reliable way with this method. The vertical line shows the chosen threshold 
value of $0.8\, \gevc$.}
\label{fig:stt:sim:dedxbands}
\end{center}
\end{figure}

The STT can also contribute to the particle identification in the low energy region, by means of the specific 
energy loss measurements. 
For gaseous detectors the particle identification is obtained from the simultaneous measurement of the d$E$/d$x$ 
and the momentum. 
For a $1\, \gevc$ track, the STT detector allows about $25$ energy loss measurements. Although this is usually 
considered rather low for a good particle identification, some capability exists in the low energy range. 
\Reffig{fig:stt:sim:dedxvsp} shows the distribution of specific energy loss for different particles plotted 
versus the momentum. 
The various regions have been identified as bands, with a mean value and an amplitude, as shown in \Reffig{fig:stt:sim:dedxbands}, 
using a sample composed of five types of charged particles (electrons, muons, pions, kaons and protons). They have 
been simulated, digitized and fully reconstructed in a momentum range between $0.05\, \gevc$ and $0.8\, \gevc$. 
% The sample was composed by three sets ($10000$ tracks each) divided according to the simulated momentum in the following way:
%\begin{itemize}  
%\item $0.05 - 1.5\, \gevc$
%\item $0.05 - 0.8\, \gevc$
%\item to gather more statistics around the bands in the region where there is the higher energy loss:
%	\begin{itemize}
%	\item e   $0.05 - 0.2\, \gevc$
%	\item \mu $0.05 - 0.2\, \gevc$
%	\item \pi $0.05 - 0.2\, \gevc$
%	\item k   $0.15 - 0.4\, \gevc$
%	\item p   $0.30 - 0.6\, \gevc$
%	\end{itemize}
%\end{itemize}
\par
\begin{figure}%[!h]
\begin{center}
\includegraphics[width=\swidth]{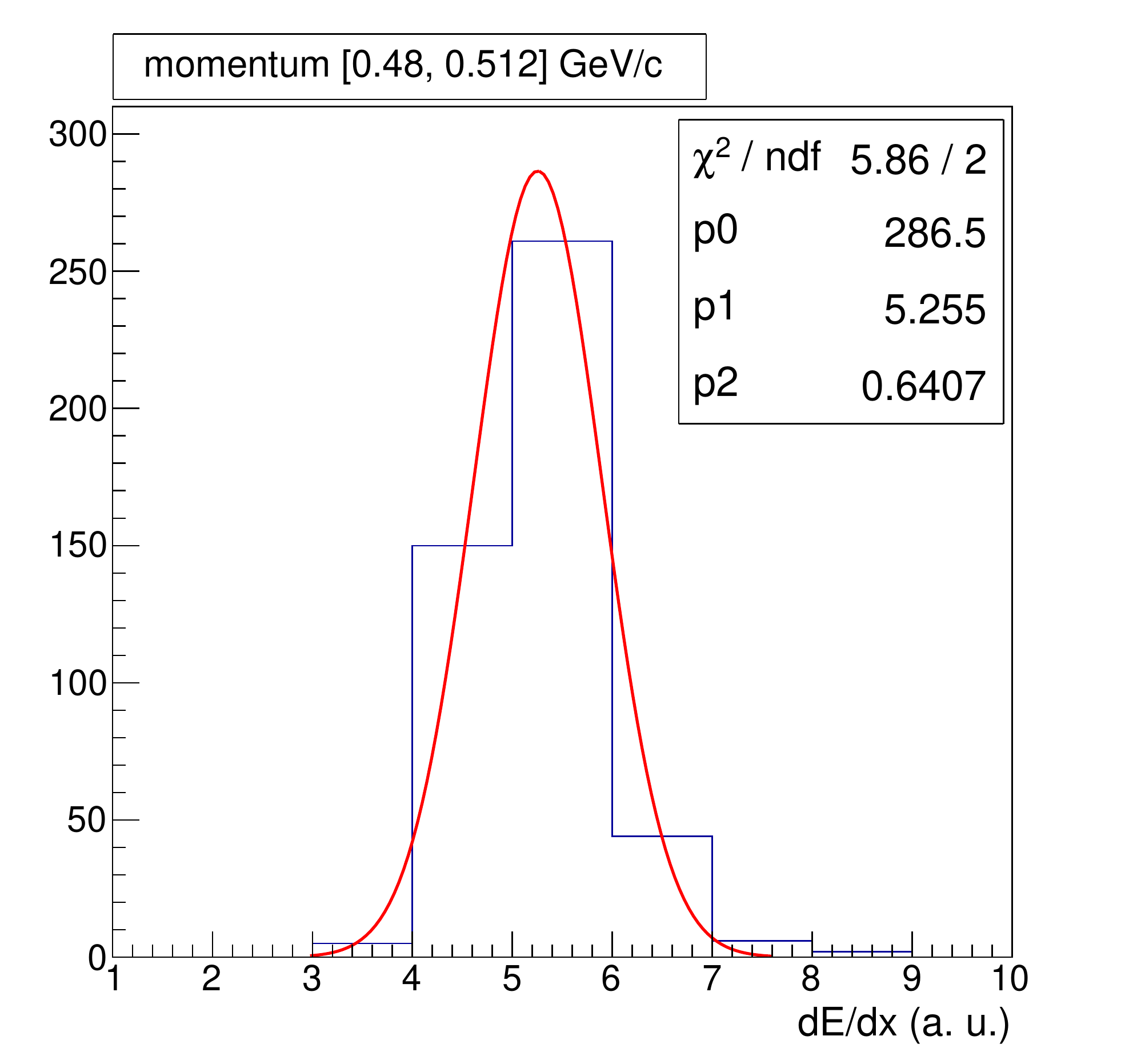} 
\caption[Example of a d$E$/d$x$ truncated mean distribution for muon tracks]{Example of a d$E$/d$x$ truncated mean distribution for muon tracks with momentum in the range [$0.48, 0.512$] $\gevc$: it 
shows a Gaussian shape as expected.} \label{fig:stt:sim:dedxgaussian}
\end{center}
\end{figure}
\begin{figure*}%[!h]
\begin{center}
\includegraphics[width=\dwidth]{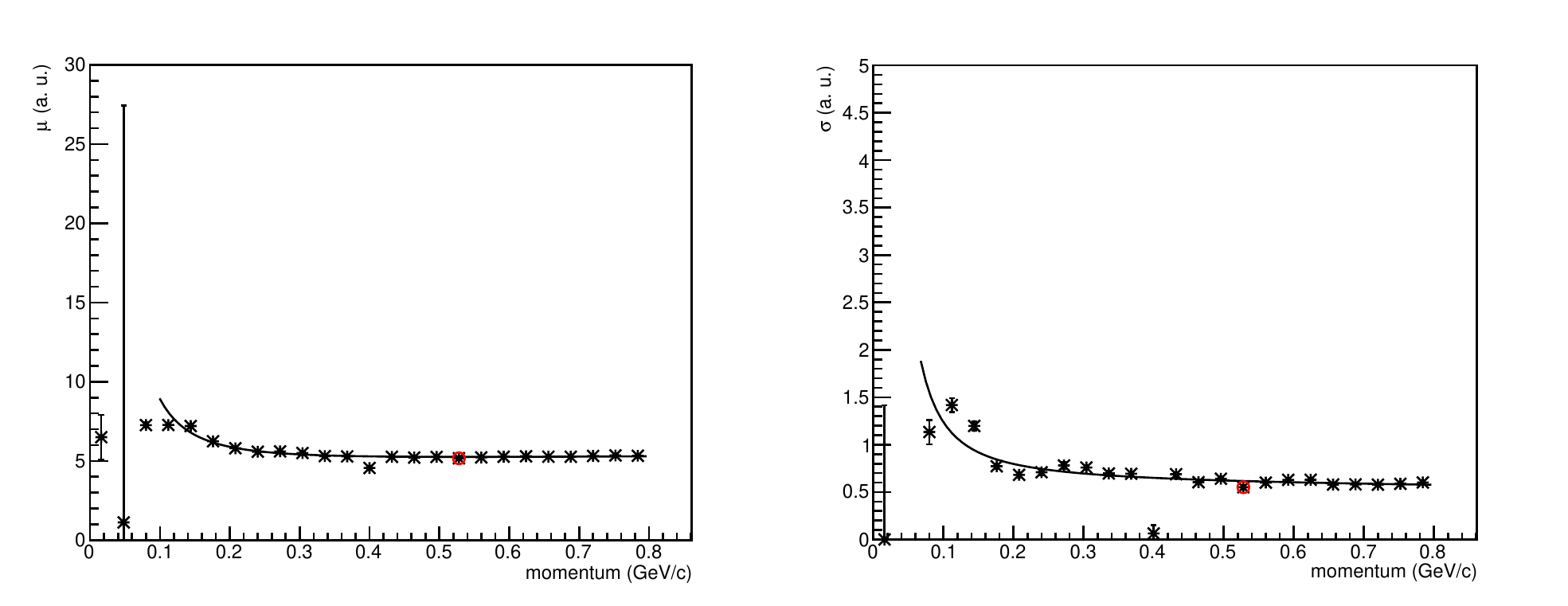} 
\caption[Plot of the mean and sigma values of the Gaussians as a function of the momentum]{Plot of the mean and sigma values of the Gaussians as a function of the momentum. The red circles correspond 
to the values for the momentum interval of figure \Reffig{fig:stt:sim:dedxgaussian}. From the fitting of these graphs 
the bands of d$E$/d$x$ truncated mean vs momentum are obtained. The points at the left extremity are not used since the 
statistics there is too low to perform a reliable fit.} \label{fig:stt:sim:dedxgraphs}
\end{center}
\end{figure*}

In each tube, the deposited energy was reproduced with a fast simulation tuned to the real data results.
The radial path has been reconstructed by the measured drift radius
and by the dip angle resulting from the fit.
The d$E$/d$x$ truncated mean value has been calculated with the truncation at $30\, \%$ in order to cut off the higher 
d$E$/d$x$ tail. 
The momentum has been obtained by fitting the hits from MVD, STT and GEM detectors with the Kalman filter procedure. 
Since at fixed momentum the d$E$/d$x$ truncated mean values are nearly Gaussian, the momentum range has been divided in 
many intervals $\sim 30\, \mevc$ large and for each interval the d$E$/d$x$ distribution has been fitted 
(\Reffig{fig:stt:sim:dedxgaussian}): the obtained mean and sigma values as a function of the momentum, whose 
graphs are shown in \Reffig{fig:stt:sim:dedxgraphs}, have been fitted to obtain the bands. 
Actually the d$E$/d$x$ truncated mean distribution at fixed momentum is not purely Gaussian, but has a small tail on the 
right side. To cope with this, the correct way of handling the fit procedure would be to consider the sum of two Gaussians,
each one with its mean and sigma, conveniently weighted. To do so, the following function should be used:
\begin{equation}
  p(x) = a \cdot p_1(x) + b \cdot p_2(x)
\end{equation}
with $a + b = 1$, once $p(x)$ has been normalized. All the parameters $a, b, \mu_1, \mu_2, \sigma_1, \sigma_2$ must be 
fitted by functions similar to the ones in \Reffig{fig:stt:sim:dedxgraphs}. In the following only the results with the 
single Gaussian will be shown.

\par
\begin{figure}%[!h]
\begin{center}
\includegraphics[width=\swidth]{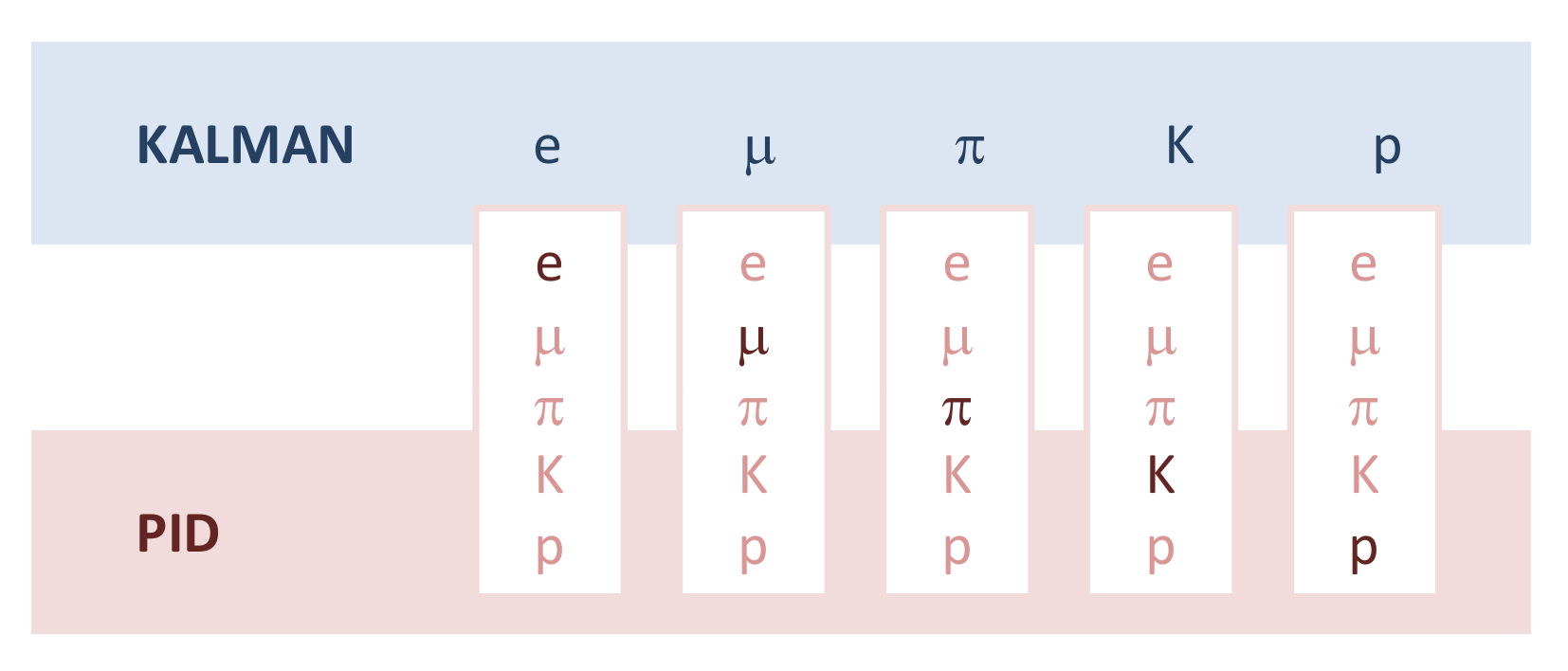}
\caption[Sketch of the association between the p.d.f. value from the particle identification (PID) procedure and the Kalman 
fitted track]{Sketch of the association between the p.d.f. value from the particle identification (PID) procedure and the Kalman 
fitted track.}
\label{fig:stt:sim:kalpid}
\end{center}
\end{figure}

When a particle of unknown mass has to be identified with this method, the d$E$/d$x$ and momentum are reconstructed and a 
point is identified in the plot of energy loss as a function of the momentum, where the bands are known. Then, for every 
particle hypothesis, the Gaussian corresponding to the reconstructed momentum of the track is chosen and it is evaluated at 
the measured track d$E$/d$x$ truncated mean. The resulting value, which comes from a standard normalized Gaussian, is the value of the
probability density function (p.d.f.) for that hypothesis. Since the momentum is the outcome of the Kalman filter 
procedure, for which a mass hypothesis has been used, two strategies can be followed: Either the Kalman fit is run on a 
track with all the mass hypotheses in parallel or the track is fitted with a unique mass hypothesis. In the first case, the particle 
identification from d$E$/d$x$ is used just to give the probability that the Kalman mass hypothesis was correct: Only the 
couples of reconstructed track and particle identification output with the same mass hypothesis are then taken into account 
(e.g., p.d.f. of the electron hypothesis for the momentum reconstructed with Kalman as an electron, p.d.f. of the muon 
hypothesis for the momentum reconstructed as a muon and so on as shown in \Reffig{fig:stt:sim:kalpid}).
They are normalized to their sum and the highest value gives the conclusive hypothesis. 
In the second case the track has only one reconstructed momentum and the particle identification is used to determine the 
mass hypothesis. After this the track should be refitted with the correct mass. 
Two different sets of bands have been identified. 
\par
To evaluate the performance of such a particle identification technique, a sample of particles of each kind, simulated 
with momenta between $0.05\,\gevc$ and $0.8\,\gevc$, has been used. For higher momenta the p.d.f value is set 
equal to $1$ for all the hypotheses, i.e. the procedure is not able to identify the particle mass. 
Each track has been reconstructed with the muon mass hypothesis (default in the code). For each reconstructed momentum, the 
p.d.f. values for all the five particle hypotheses are extracted from the d$E$/d$x$ truncated mean vs momentum. The one with 
the highest value determines the identified particle: the obtained results are shown in table \ref{tab:stt:sim:pidresults}. 
%%%%%%%%%%%
\definecolor{darkorange}{rgb}{0.77, 0.14, 0.09}
\definecolor{lightorange}{rgb}{0.87, 0.23, 0.0}
\begin{table}
\begin{center}
\caption[Results of the performance test of particle identification]{Results of the performance test of particle identification: for each row the frequency, 
in percentages, with which the simulated particle is recognized as electron, muon, pion, kaon and proton is written. Each row 
percentages sum up to $100\, \%$. The correct association is the one on the diagonal. The muon and pion frequencies 
must be summed, since with this method muons and pions can be hardly distinguished.}
\vspace{3mm}
\begin{tabular}{ccccccc}
& & \multicolumn{5}{|c}{frequencies of p.i.d. (\%)} \\ 
 \cline{3-7}
& & \multicolumn{1}{|c}{e}  & $\mu$ & $\pi$ & K & p \\
\hline 
\hline 
\multirow{5}{*}{\rotatebox{90}{{\it true} part.}} 
& \multicolumn{1}{|c|}{e}   & \textcolor{darkorange}{78.9} &  5.2  &  5.6 &  10.1 &  0.2 \\
\cline{2-7} 
& \multicolumn{1}{|c|}{$\pi$} &  9.0 & \textcolor{lightorange}{47.2}  & \textcolor{darkorange}{40.7} &  2.9 & 0.2 \\
\cline{2-7} 
& \multicolumn{1}{|c|}{K}   & 22.3 &  8.0  & 1.6 & \textcolor{darkorange}{65.1} &  3.0 \\ 
\cline{2-7} 
& \multicolumn{1}{|c|}{p}   & 0.1 & [0.01] &  0.1 &  1.0 & \textcolor{darkorange}{98.8} \\
\end{tabular}
\label{tab:stt:sim:pidresults}
\end{center}
\end{table}
%%%%%%%%%%%

\begin{figure}%[!h]
\begin{center}
\includegraphics[width=\swidth]{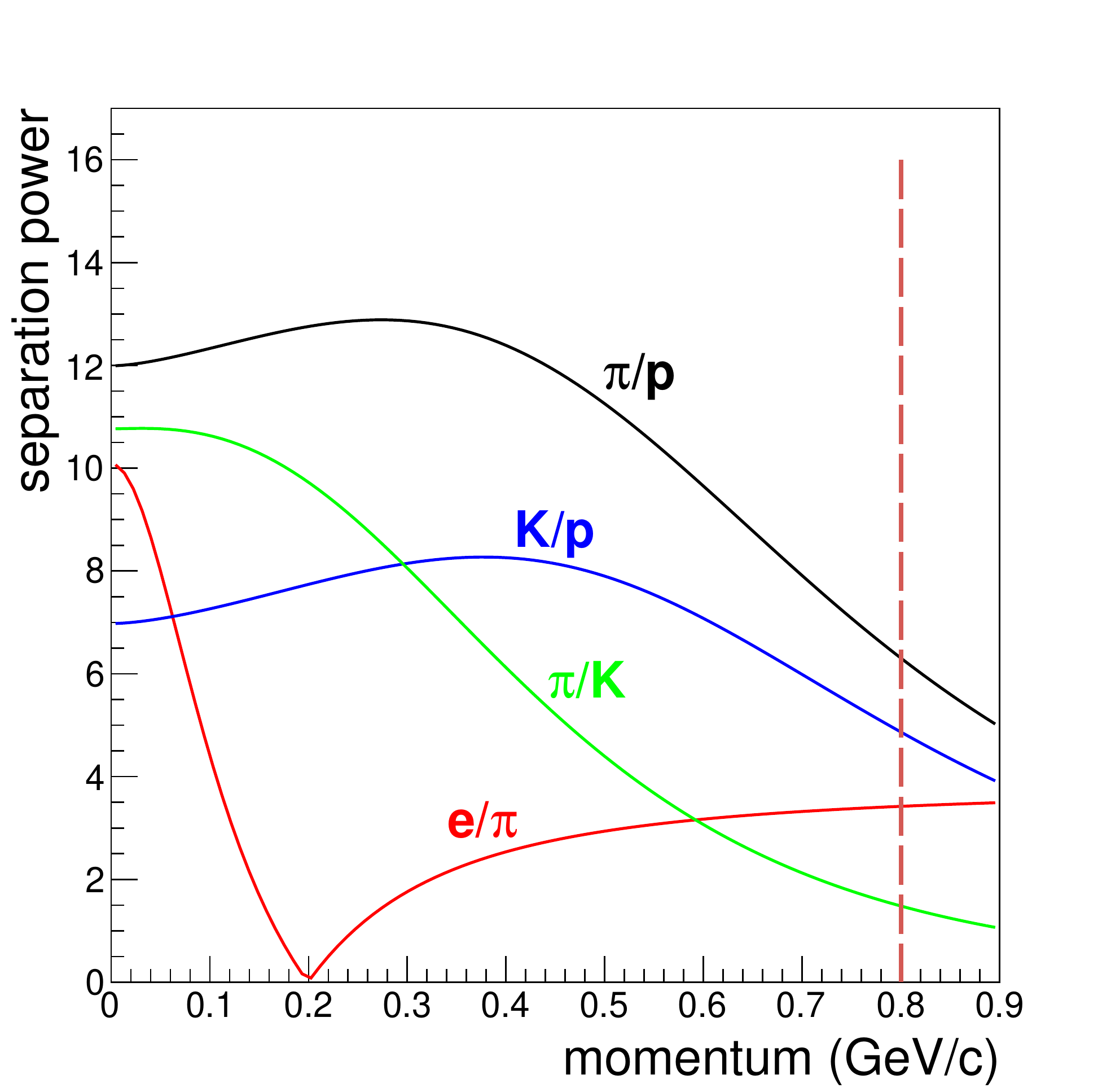} 
\caption[Separation power for particle identification in the STT detector]{Separation power in the STT detector for the bands built with particles all tracked with the 
same muon mass hypothesis. The vertical line at $0.8\, \gevc$ is the chosen threshold for the momentum 
to perform the particle identification in the STT.} 
\label{fig:stt:sim:dedx_separationpow}
\end{center}
\end{figure}

 The separation power $S =2 \, \Delta E$ between two particles is defined as the 
distance between the centres of the  two bands  
 $\left<E_1 \right>$ and  $\left<E_2 \right>$, measured in terms of the standard
 deviations $\sigma_1$ and $\sigma_2$ \cite{bib:stt:sim:PID}:
\begin{equation}
  \Delta E = \frac{E - \left< E_1 \right> }{\sigma_1} = 
   \frac{ \left< E_2 \right> - E}{\sigma_2}
\end{equation} \noindent
Eliminating $E$ from the previous equation and recalculating $S$
  the following separation power is obtained:
\begin{equation}   \label{eq:stt:sim:spow1}
   S = \frac{\left< E_2 \right> - \left< E_1 \right>}{\sigma_1/2 + \sigma_2/2}
\end{equation} \noindent
when $\left< E_2 \right> > \left< E_1 \right>$ or
\begin{equation}   \label{eq:stt:sim:spow2}
   S = \frac{|\left< E_2 \right> - \left< E_1 \right>|}{\sigma_1/2 + \sigma_2/2}
\end{equation} \noindent
in general, which  is shown in \Reffig{fig:stt:sim:dedx_separationpow}. This plot demonstrates clearly the 
 capability of the STT detector in the low energy PID.
\subsubsection{Effect of Pressure and Gas Mixture}
\begin{figure}%[!h]
\begin{center}
\includegraphics[width=\swidth]{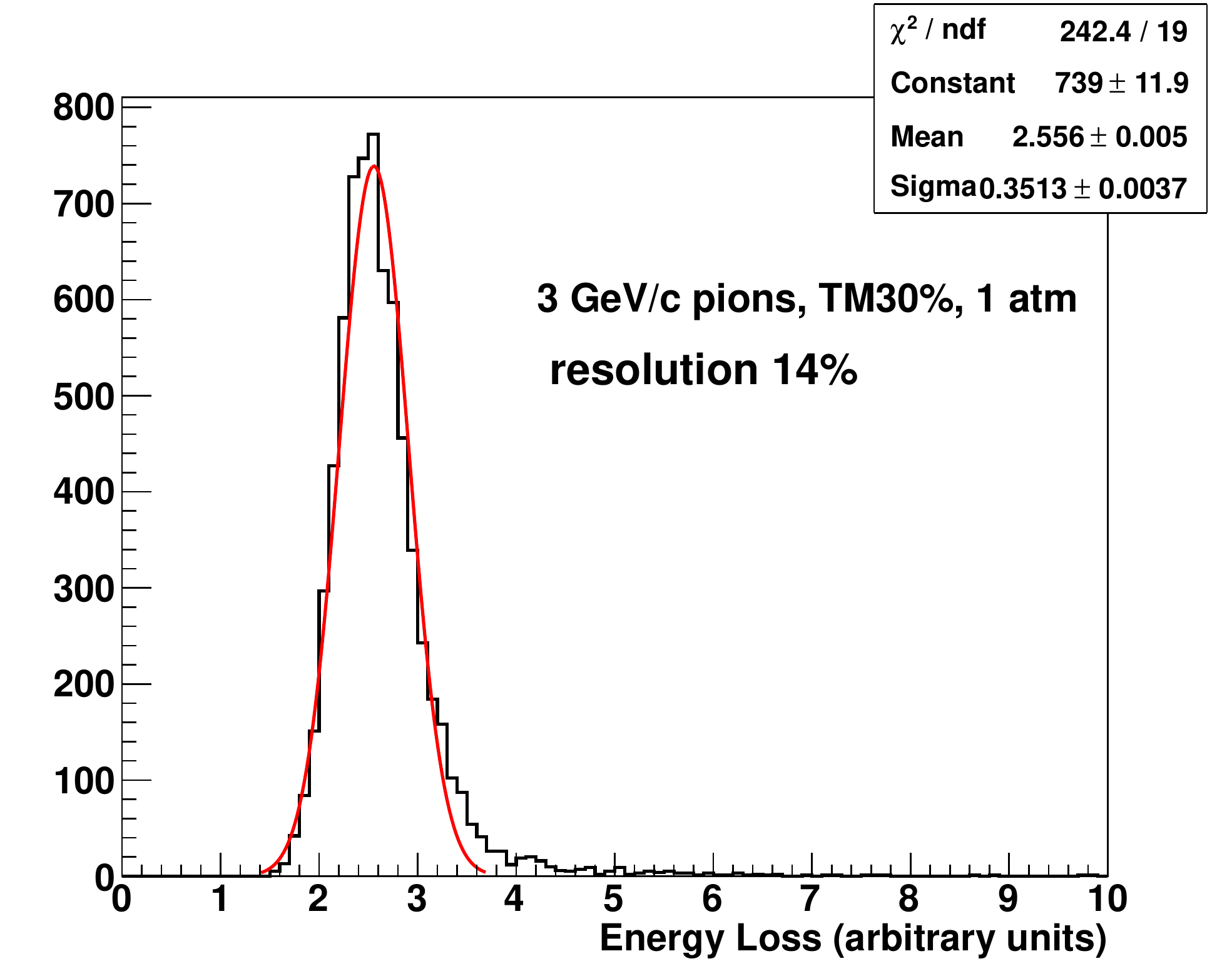}
\includegraphics[width=\swidth]{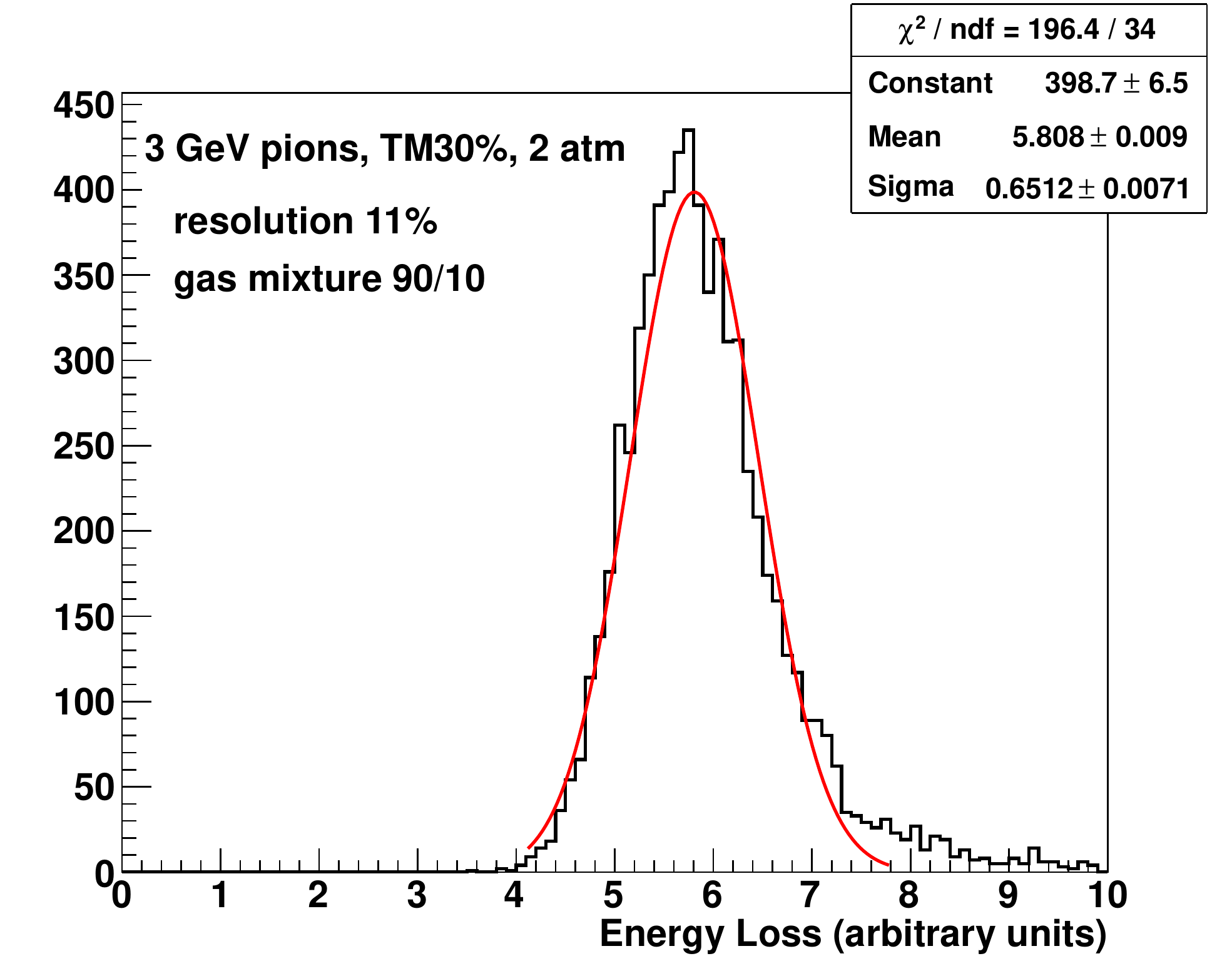}
\caption[Energy loss distributions for $3\, \gevc$ pions in $22$ straw tubes (ArCO$_2$ (9/1) gas mixture)]{Energy loss 
distributions for $3\, \gevc$ pions in $22$ straw tubes. Truncated mean of $30\,\%$ is applied. Upper figure shows the 
d$E$/d$x$ resolution at 1 $\atm$ absolute pressure, lower figure at 2 $\atm$: increasing the gas pressure a gain in resolution 
of about $20 \%$ is obtained. Both resolutions have been obtained with a gas mixture 
Ar~(90\,\%)/CO$_2$~(10\,\%).} \label{fig:stt:sim:dedxpressure}
\end{center}
\end{figure}

\begin{figure}%[!h]
\begin{center}
\includegraphics[width=\swidth]{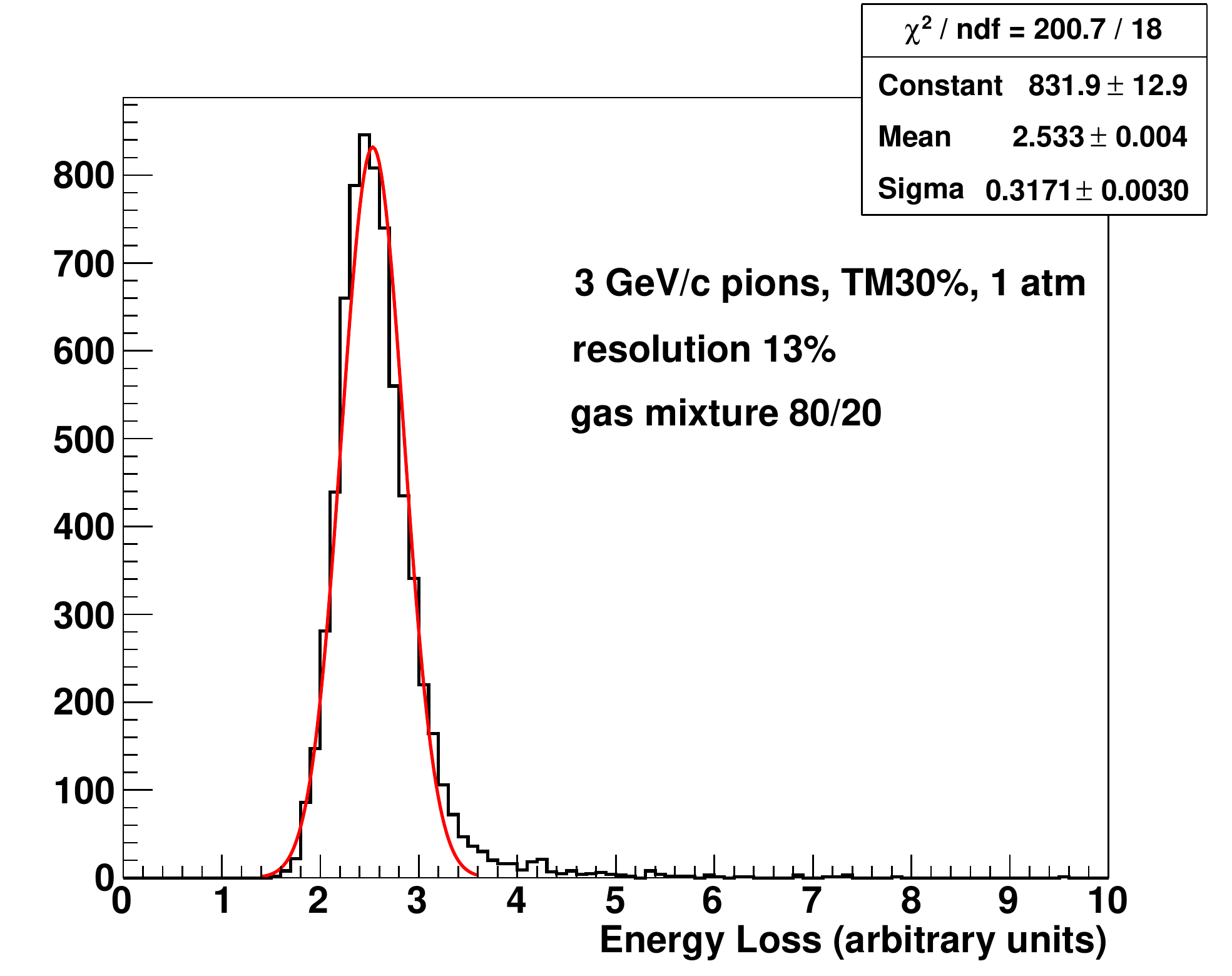}
\includegraphics[width=\swidth]{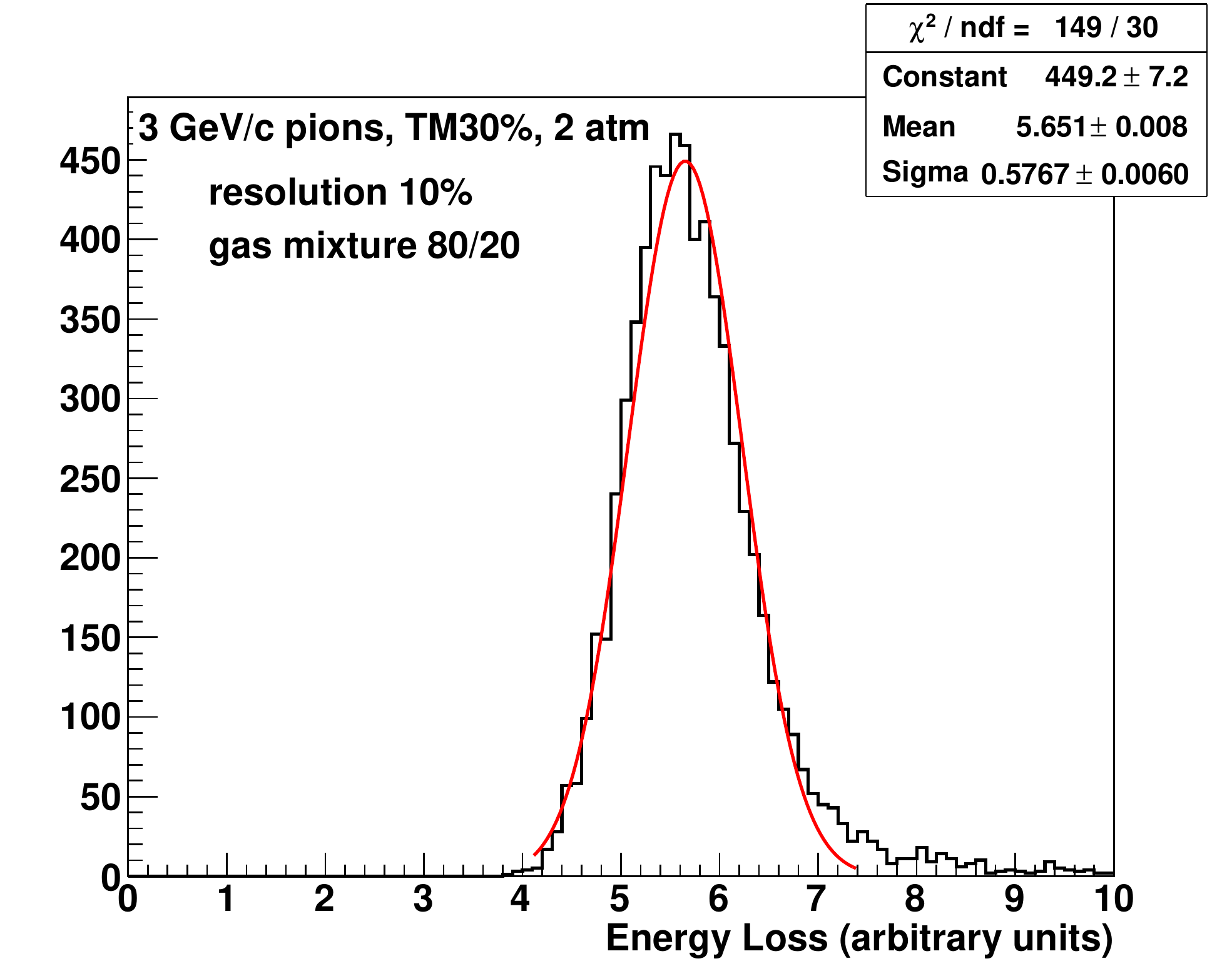}
\caption[Energy loss distributions for $3\, \gevc$ pions in $22$ straw tubes (ArCO$_2$ (8/2) gas mixture)]{The same as 
\Reffig{fig:stt:sim:dedxpressure}, but with a gas mixture Ar~(80\,\%)/CO$_2$~(20\,\%).} \label{fig:stt:sim:dedxpressure2}
\end{center}
\end{figure}

A simulation has been performed to investigate the relationship between gas pressure and d$E$/d$x$ resolution, for two 
different gas mixtures. The test has been performed with a simple setup, with $22$ samplings of $3\, \gevc$ pions from a 
single straw tube and using the truncated mean at $30\,\%$.
\Reffig{fig:stt:sim:dedxpressure} and \Reffig{fig:stt:sim:dedxpressure2} show the different d$E$/d$x$ distributions with 
a gas mixture of Ar and CO$_2$ in different ratios and at different pressures. The change in CO$_2$ percentages does not 
produce observable effects, while the change in operating pressure improves the specific energy resolution by about $20\,\%$.
% *******************************

%EOF: panda_tdr_stt_sim.tex

%
% Bibliography for this chapter (remove %)
%
\bibliographystyle{panda_tdr_lit}
\bibliography{./stt/lit_stt}
% EOF

%
% STT TDR
% File for chapter 6
\chapter{Straw Tube Tracker Performance}
% FILE: panda_tdr_stt_per.tex
%
%\COM{Author(s): G. Boca/L. Lavezzi/S. Costanza}
\section{Performance Studies with Single Tracks}
\label{sec:stt:stt:per}

In order to study the performances of the designed \panda Straw Tube Tracker in
 terms of geometrical acceptance of the layout, momentum resolution and 
reconstruction efficiency, systematic Monte Carlo studies have been performed with single track events.
 \subsection{Simulation Environment}
A summary of the choices made to perform the tests is given here. 
\par
The target spectrometer was simulated to have a realistic material budget. Specifically, the list of the simulated subdetectors 
contains: MVD, STT, Electromagnetic Calorimeter, TOF detector, Muon Chambers, Cherenkov detectors and 
forward GEM stations. In addition also the passive elements have been placed in order to take the correct amount of material into account: 
The Solenoid Magnet, the Target and Beam Pipes. 
\par
 The full magnetic field map has been used to account for magnetic inhomogeneities. 
%%%
\par
 % CHECK this parag
Different event generators are available in \pandaroot. For the single track tests, the \texttt{BoxGenerator} has been used, with the possibility to select ranges of momentum, both magnitude and direction, in addition to particle type and multiplicity.
\par
 % The digitization step was performed only for the tracker devices (MVD, STT and GEM) in order to save computation time, since the reconstruction studies would have been dedicated only to these subdetectors. % CHECK signle tracks?
The digitization step has been performed only for MVD, STT and GEM in order to save computation time, since the studies would have been dedicated only to the Central Tracker. It was performed in a realistic way to get a reliable detector response and the hits for the reconstruction.
\par
All realistic pattern recognitions were used, with no information taken from the Monte Carlo truth. The full chain of track finders was adopted. 
% following measured hits were used:
% \begin{itemize}
% \item for the {\bf MVD}, a plane coincident with the sensor one was chosen. The covariance values for the measured coordinates on this plane were $\sigma_{xy}^2 = (50\,\mum)^2$ for both pixel and strip hits. 
% \item for the {\bf STT}, virtual detector planes have been used, whose origin is the center of the firing wire and whose axes are one along the wire and the other passing through the point of closest approach to the wire itself; the covariances of the measured point were set in order to reproduce the real resolution of the drift radius: $(100\,\mum)^2$.
% \item for the {\bf GEM}, a plane on the sensor one was chosen, as in MVD, but rotated to have one axis passing through the center of the plane and the hit measured on it. The covariance values have been computed starting from the measurement errors of the hit distance from the center of the plane, {\it dr}, and on the orthogonal direction, {\it dp}.
% \end{itemize}
After the track finding, the Kalman filter was applied to the tracks, using the package \texttt{\hyphenchar\font45\relax genfit} \cite{bib:stt:sim:genfit} (see \Refsec{sec:stt:sim:kalman})
The starting point for the Kalman procedure was chosen by extrapolating the tracks fitted with the helix to the point of closest approach 
to the interaction point. The $xy$ plane was chosen as starting plane and only one iteration was performed in the fit procedure; this means that the filter step was performed on the plane corresponding to each measurement, both in the forward and in the backward direction. 

\subsection{Studies on the Number of Hits per Track}
In order to check the geometrical acceptance of the layout, the distributions of
 the number of hits coming from axial, skewed and short straws have been studied.
$10^5$ $\mu^-$ single track events have been generated in the interaction point 
I.P.~($x=y=z=0$), with random azimuthal angle $\phi$ 
($\phi \in [0^\circ, 360^\circ]$) and $\theta \in [7^\circ, \ 160^\circ]$, at fixed
total momentum (1 \gevc).
\begin{figure*}[h!]
\begin{center}
\includegraphics[width=0.45\dwidth, height=0.45\dwidth]{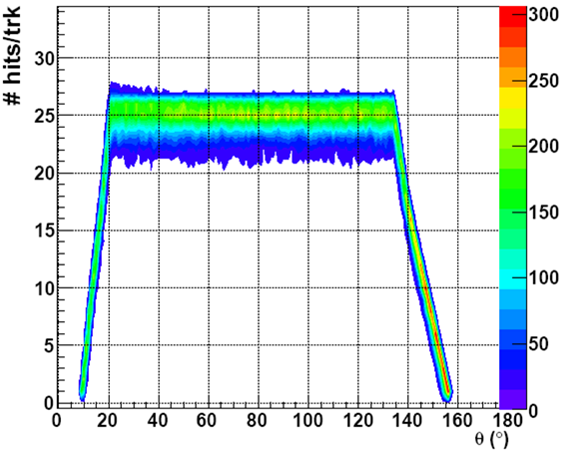}
\includegraphics[width=0.45\dwidth, height=0.45\dwidth]{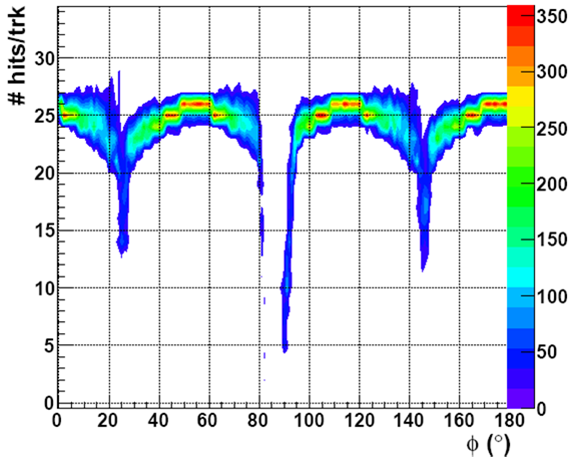}
\caption[Distribution of the number of hit straws as a function of the $\theta$ and $\phi$ angle]{Distribution of the number of hit straws as a 
function of $\theta$ (left) and $\phi$ (right) angles for 10000 $\mu^-$ generated
with a momentum of 1 \gevc.}
\label{fig:stt:per:hits_all}
\end{center}
\end{figure*}

\begin{figure*}[ht!]
\begin{center}
\includegraphics[width=0.45\dwidth, height=0.45\dwidth]{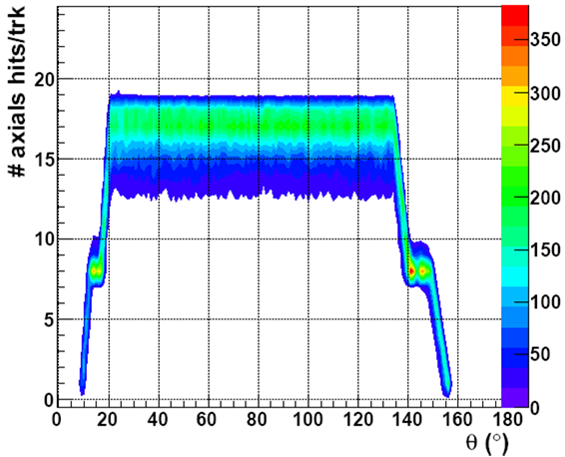}
\includegraphics[width=0.45\dwidth, height=0.45\dwidth]{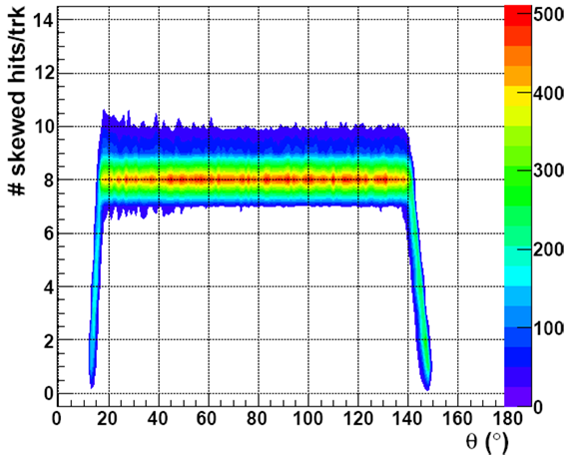}
\caption[Distribution of the number of hits per track as a function of $\theta$]{Distribution of the number of hits per track as a function of $\theta$ 
 angle for $10^5$ $\mu^-$ generated with a momentum of 1 \gevc, in case of axial (left) and 
skewed (right) hit straws.}
\label{fig:stt:per:hits_theta}
\end{center}
\end{figure*}

\begin{figure*}[ht!]
\begin{center}
\includegraphics[width=0.45\dwidth, height=0.45\dwidth]{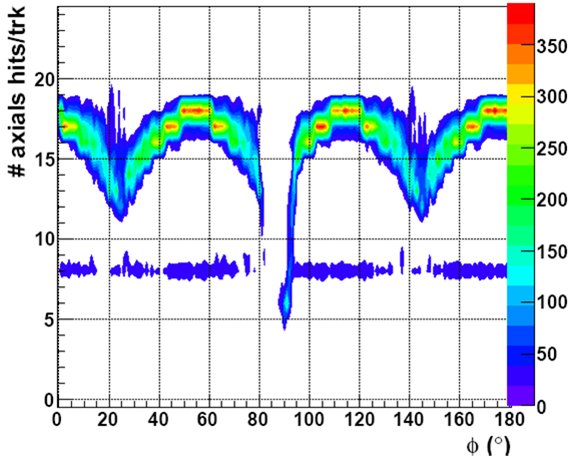}
\includegraphics[width=0.45\dwidth, height=0.45\dwidth]{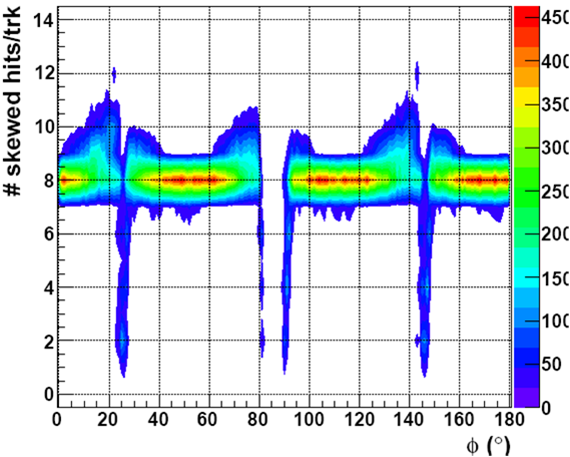}
\caption[Distribution of the number of hits per track as a function of $\phi$]{Distribution of the number of hits per track as a function of $\phi$ 
angle for $10^5$ $\mu^-$ generated with a momentum of 1 \gevc , in case of axial (left) and 
skewed (right) hit straws.}
\label{fig:stt:per:hits_phi}
\end{center}
\end{figure*}

The plots in \Reffig{fig:stt:per:hits_all} show the distributions of the hit numbers 
as a function of $\theta$ and $\phi$. Moreover, if we distinguish between the 
contribution of the axial and the 
skewed straws, more detailed considerations can be drawn. In particular, the plots 
in \Reffig{fig:stt:per:hits_theta} show the number of hits per track in case of 
axial (left) and skewed (right) hit straws as a function of $\theta$, and the ones
in \Reffig{fig:stt:per:hits_phi} are the analogous as a function of $\phi$.

As shown in the left plot of \Reffig{fig:stt:per:hits_theta}, the minimum number corresponds to 
the STT edge at $\theta = 7.8^\circ$; 
then the number of hits increases up to $\sim$ 8 around $\theta = 11.6^\circ$ and 
stays constant in the angular region where the skewed layers are placed. For larger
values of $\theta$, the number of hits for axial straws increases again, up to 
about $17-18$, corresponding to the region where the tracks with $\theta
\in [20.9^\circ, 133.6^\circ]$ hit all the straw layers.
For tracks with a bigger $\theta$ value, the number of hits decreases down to 8 hits 
and again, after the plateau, down to 0 at $\theta = 159.5^\circ$, corresponding to
the backward lower edge of the STT.

The number of hits from skewed straws (right plot of 
\Reffig{fig:stt:per:hits_theta}) increases, starting from $11.6^\circ$, since
below this $\theta$ value only axial double layers are placed. The maximum number of 
skewed hits is 8, according to the fact that the STT layout foresees four double 
layers of tilted tubes. A higher number of hits from skewed straws can be due to the
fact that, along their path, tracks may hit also the shorter tilted tubes placed in 
the corners of the hexagonal STT layout or more tubes due to the bending of
their trajectory.

The hit distributions vs $\phi$ are reported in \Reffig{fig:stt:per:hits_phi}. 
The results are in agreement with the ones as a function of $\theta$, showing that
the maximum number of hits from axial straws is about $17-18$ and that most tracks 
hit 8 skewed straws. 

In addition, the left plot of \Reffig{fig:stt:per:hits_phi} shows a structure around 8 
hits: This is due to the tracks which exit the front of the STT in the angular region of
the skewed layers. This prevents them from reaching the outer axial straw layers: Thus, the
maximum number of axial hits of these tracks is 8.

The hole at $\phi = 90^\circ$ and the low number of hits around this $\phi$
value are due to the gap for the target pipe. The losses at $\phi = 30^\circ$ and 
$\phi = 60^\circ$ are caused by the fact that the short tubes placed in 
the hexagon corners do  not completely fill the volume, leaving empty spaces. 
These losses are negligible: 
Only a small percentage of the total number of events hits less than 5 skewed straws.
Nevertheless, there is a gain in efficiency when including in the reconstruction 
procedure also the information of the hits from the MVD and the GEM chambers.

\begin{figure*}[ht!]
\begin{center}
\includegraphics[width=0.45\dwidth, height=0.45\dwidth]{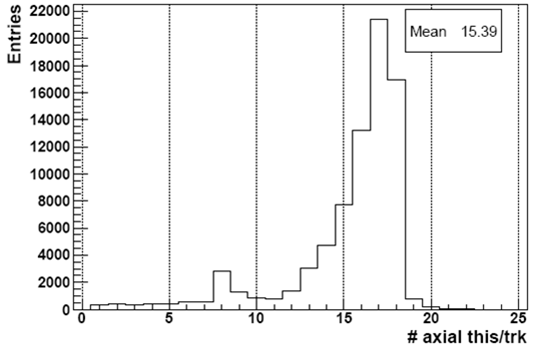}
\includegraphics[width=0.45\dwidth, height=0.45\dwidth]{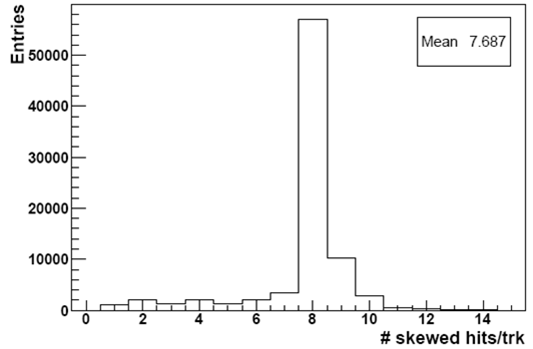}
\caption[Distribution of the mean number of axial and skewed hit straws per track]{Distribution of the mean number of axial (left) and skewed (right) 
hit straws per track.}
\label{fig:stt:per:summary}
\end{center}
\end{figure*}

As a summary, the distributions of the mean number of axial and skewed hit straws per
track are shown in \Reffig{fig:stt:per:summary}.

\subsection{Studies on Momentum Resolution and Reconstruction Efficiency}
\label{sec:stt:resolstudies}
\subsubsection{Studies with Uniform $\cos{\theta}$}
\label{sec:stt:resolstudies-uni}
$10^4$ $\mu^-$ single track events have been generated in the interaction point 
I.P.~($x=y=z=0$), with uniform azimuthal angle $\phi \in [0 ^\circ, 360 ^\circ]$ 
and uniform $\cos\theta$ ($\theta \in [7.8^\circ, 159.5^\circ]$) at fixed values of
 total momentum (0.3, 1, 5 \gevc). 

\begin{figure*}
\centering
\includegraphics[width=\swidth, height=\swidth]{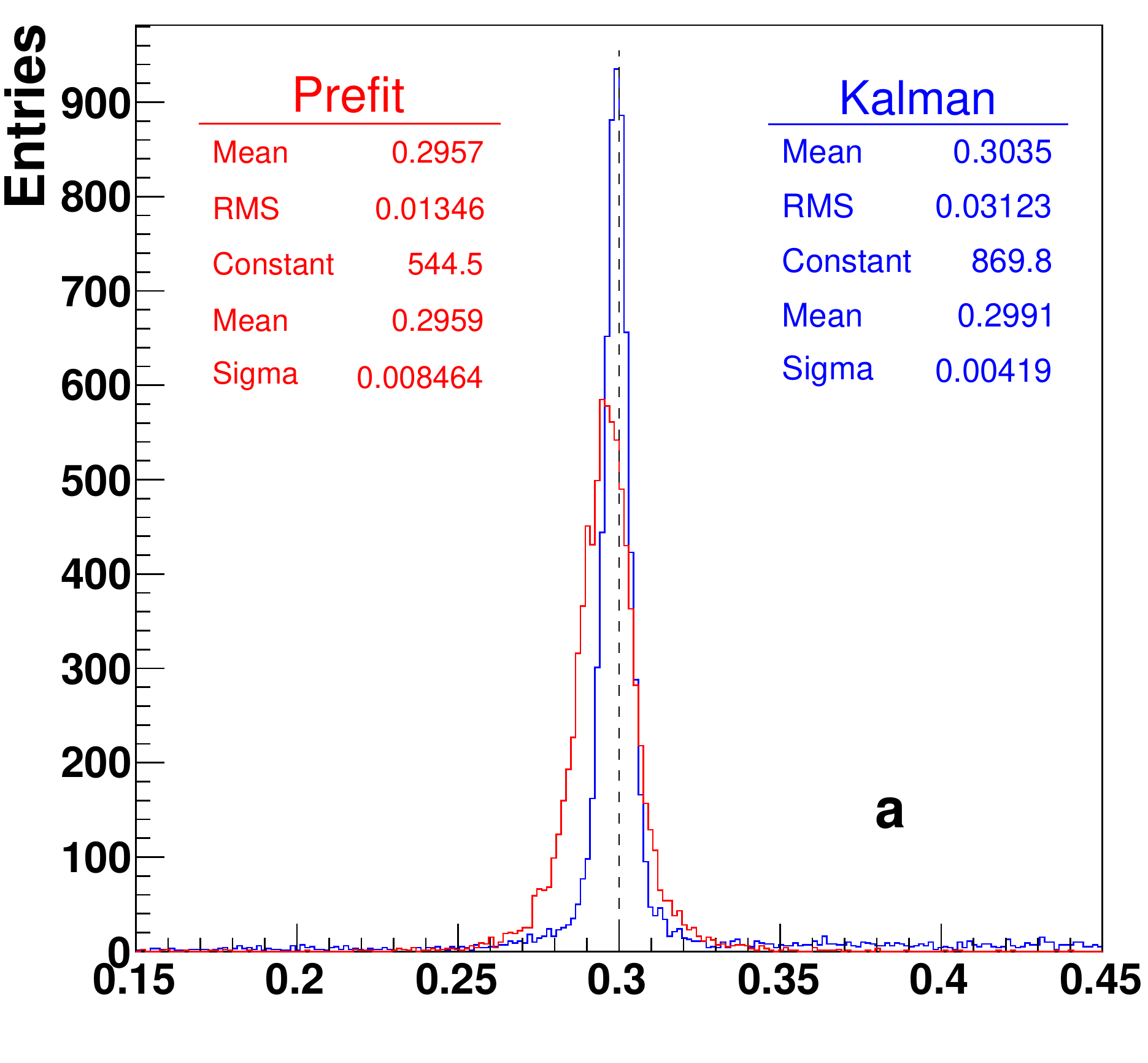}
\includegraphics[width=\swidth, height=\swidth]{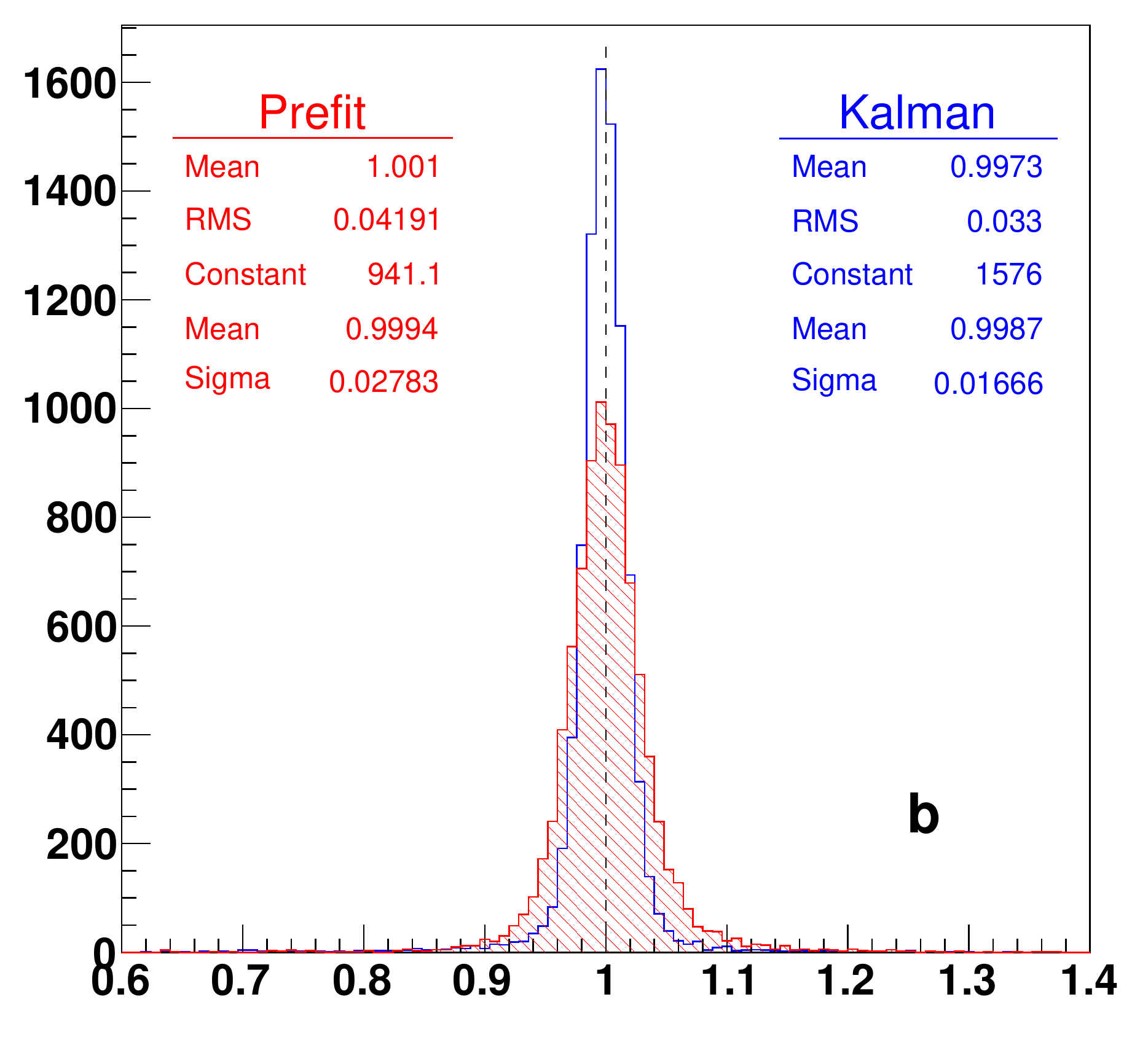}
\includegraphics[width=\swidth, height=\swidth]{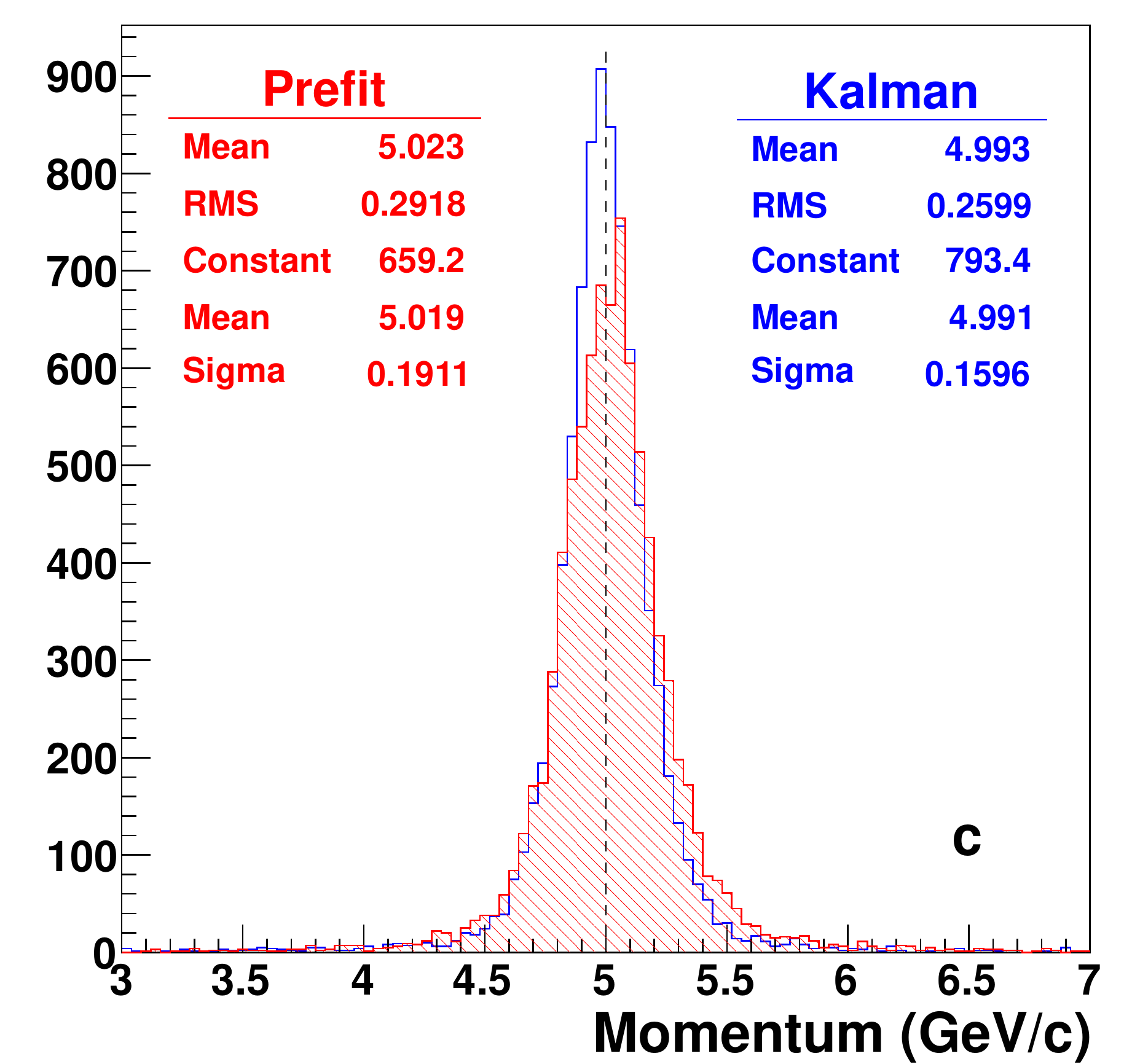}
\caption[Momentum distributions for $\mu^-$]{Momentum distributions for (a) 0.3, (b) 1 and (c) 5 \gevc $\mu^-$,
 reconstructed with helix (red dashed) and Kalman (blue) fits. The statistic boxes report the mean values 
and RMS of the non fitted histograms, as well as mean and sigma values of the 
Gaussian fits, before and after the Kalman fit.} \label{fig:stt:per:totalmom}
\end{figure*}

The reconstructed momentum distributions are shown in 
\Reffig{fig:stt:per:totalmom} for particles at (a) 0.3, (b) 1 and (c) 5 \gevc.
The red dashed histograms show the prefit results (the outcome of the pattern 
recognition, \Refsec{sec:stt:sim:pr}), while the blue histograms reproduce the Kalman fit  result. 

Each histogram has been fitted with a Gauss function in the range $[\mu - 3 
\sigma, \mu + 3 \sigma]$, where $\mu$ is the mean value of the momentum 
distribution and $\sigma$ has been calculated by dividing the FWHM of the 
histogram by 2.35. 

\Reftbl{tab:stt:per:totalmom} summarizes the obtained values of 
momentum resolution and efficiency. The 
resolution is calculated as $\sigma/\mu$, using the $\mu$ and $\sigma$ values 
from the Gaussian fit; it is then reported as relative resolution in percent. The efficiency is 
defined by the histogram integral divided by the number
 of generated tracks. In addition, the efficiency ``in peak'' is reported: it is
 the number of tracks in the fitted range ($\mu\pm 3\sigma$) with respect to the
 total number of tracks.

In all cases the Kalman fit results are better than the prefit ones (as 
expected), both in terms of mean value and sigma of the distributions. In fact 
the Kalman fit improves the helix fit results both reducing the width of the 
distribution (i.e.~improving the resolution) and shifting the distribution mean
 value towards a more correct value. On the other hand, the helix fit introduces
 a systematic offset in the momentum determination giving an underestimated 
value.

An efficiency loss of about 13\,\% from the prefit (78\,\%) to the kalman fit (65\,\%) in the helix reconstruction is observed for the tracks with 0.3\,\gevc total momentum (see \Reftbl{tab:stt:per:totalmom}, efficiency in peak values). This indicates a problem in the kalman fit algorithm for these low momentum tracks, which has to be investigated in detail. For the tracks with higher momenta, the efficiencies of the prefit and kalman fit are comparable and the differences are less than 2\,\%.

\begin{table*}[h!]
\begin{center}
\caption[Momentum resolution and reconstruction efficiency for 10$^4$ $\mu^-$]{Momentum resolution and reconstruction efficiency for 10$^4$ $\mu^-$ (\Reffig{fig:stt:per:totalmom}). 
The resolution is calculated as $\sigma/\mu$ (with $\mu$ and $\sigma$ values from the Gaussian fit); the efficiency is 
obtained as integral divided by the number of generated tracks and the efficiency ``in peak'' is the number of tracks in 
($\mu\pm 3\sigma$) divided by the total number of tracks (\Refsec{sec:stt:resolstudies-uni}).}
\smallskip
\begin{tabular}{c|c|c|c|c|c|c}
\hline \hline
Momentum    & \multicolumn{2}{c|}{Resolution (\%)}   &   \multicolumn{2}{c|}{Efficiency (\%)} &   \multicolumn{2}{c}{Eff. in peak (\%)}\\ \hline
(\gevc)     &       Prefit         &        Kalman       &   Prefit    &   Kalman  &   Prefit    &   Kalman    \\\hline\hline
$0.3$   & $2.86 \pm 0.03$ & $1.40 \pm 0.02$ & $82.75 \pm 0.38$ & $74.60 \pm 0.44$ & $77.64 \pm 0.42$ & $65.20 \pm 0.48$\\
$1.0$   & $2.78 \pm 0.03$ & $1.67 \pm 0.02$ & $86.89 \pm 0.34$ & $86.81 \pm 0.34$ & $81.64 \pm 0.39$ & $80.18 \pm 0.40$\\
$5.0$   & $3.81 \pm 0.05$ & $3.19 \pm 0.04$ & $84.91 \pm 0.36$ & $84.68 \pm 0.36$ & $79.07 \pm 0.41$ & $80.87 \pm 0.39$\\
\end{tabular}
\label{tab:stt:per:totalmom}
\end{center}
\end{table*}

\subsubsection{Studies at Fixed $\theta$ Values} 
A systematic scan of the momentum resolutions and efficiencies has been performed with fixed angle generated 
particles. $10^4$ $\mu^-$ single track events have been generated at the interaction point with fixed total 
momentum ($0.3$, $1$, $2$ and $5$ \gevc) and random $\phi$ ($\phi \in [0^\circ, 360^\circ]$). The $\theta$ 
angular range has been scanned as follows: 
\begin{itemize}
\item[{\it i.}] $\theta = 10^\circ,\ 12^\circ,$ \ldots,$\ 24^\circ$ in steps of 
$2^\circ$ ($\pm 1^\circ$);
\item[{\it ii.}] $\theta = 30^\circ,\ 40^\circ,$ \ldots, $150^\circ$ in steps of
 $10^\circ \ (\pm \ 5^\circ)$.
\end{itemize}\noindent
Finally, the events have been reconstructed and the Kalman fit has been 
performed.
\par
\begin{figure*}
\centering
\includegraphics[width=\swidth, height=\swidth]{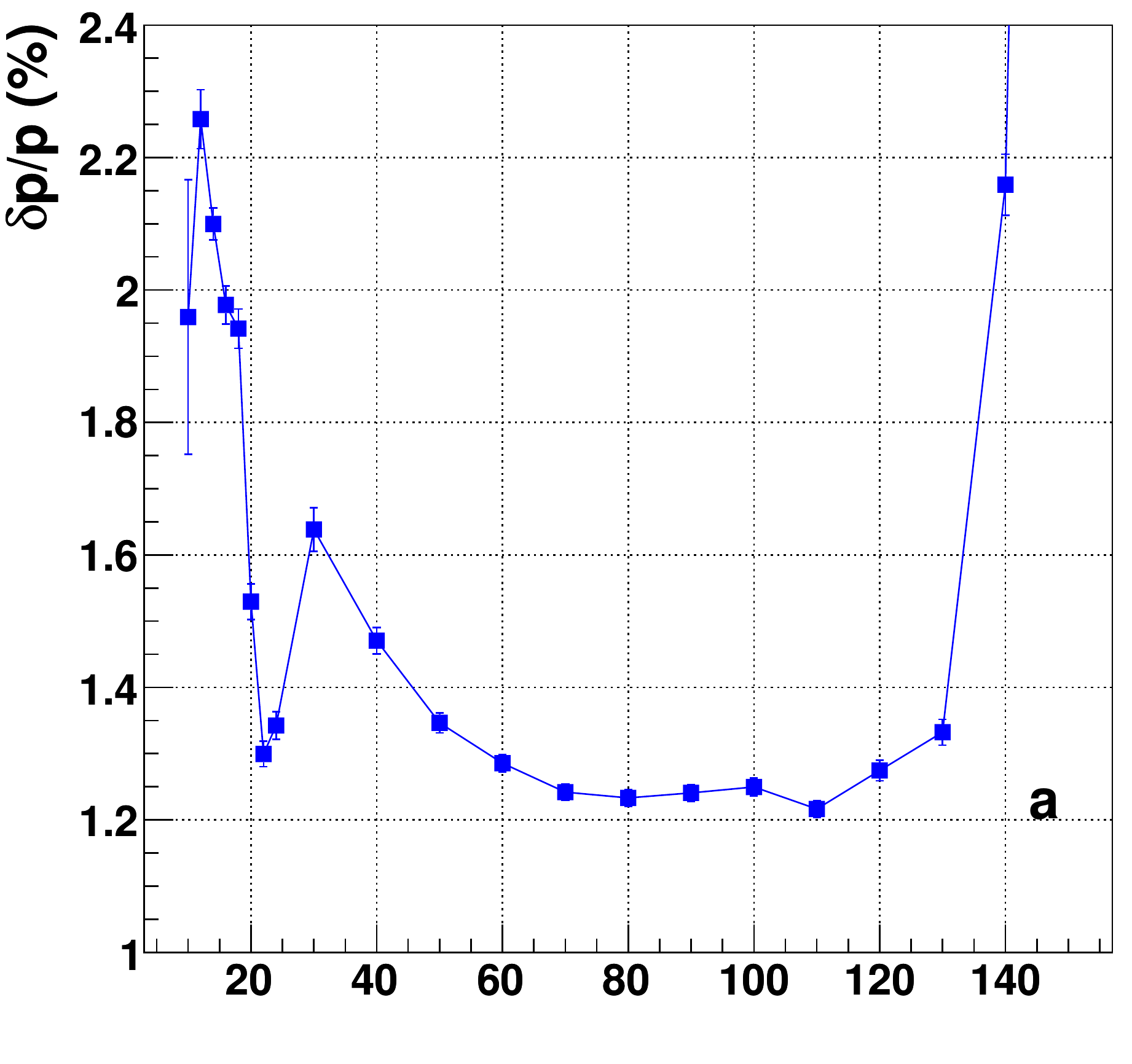}
\includegraphics[width=\swidth, height=\swidth]{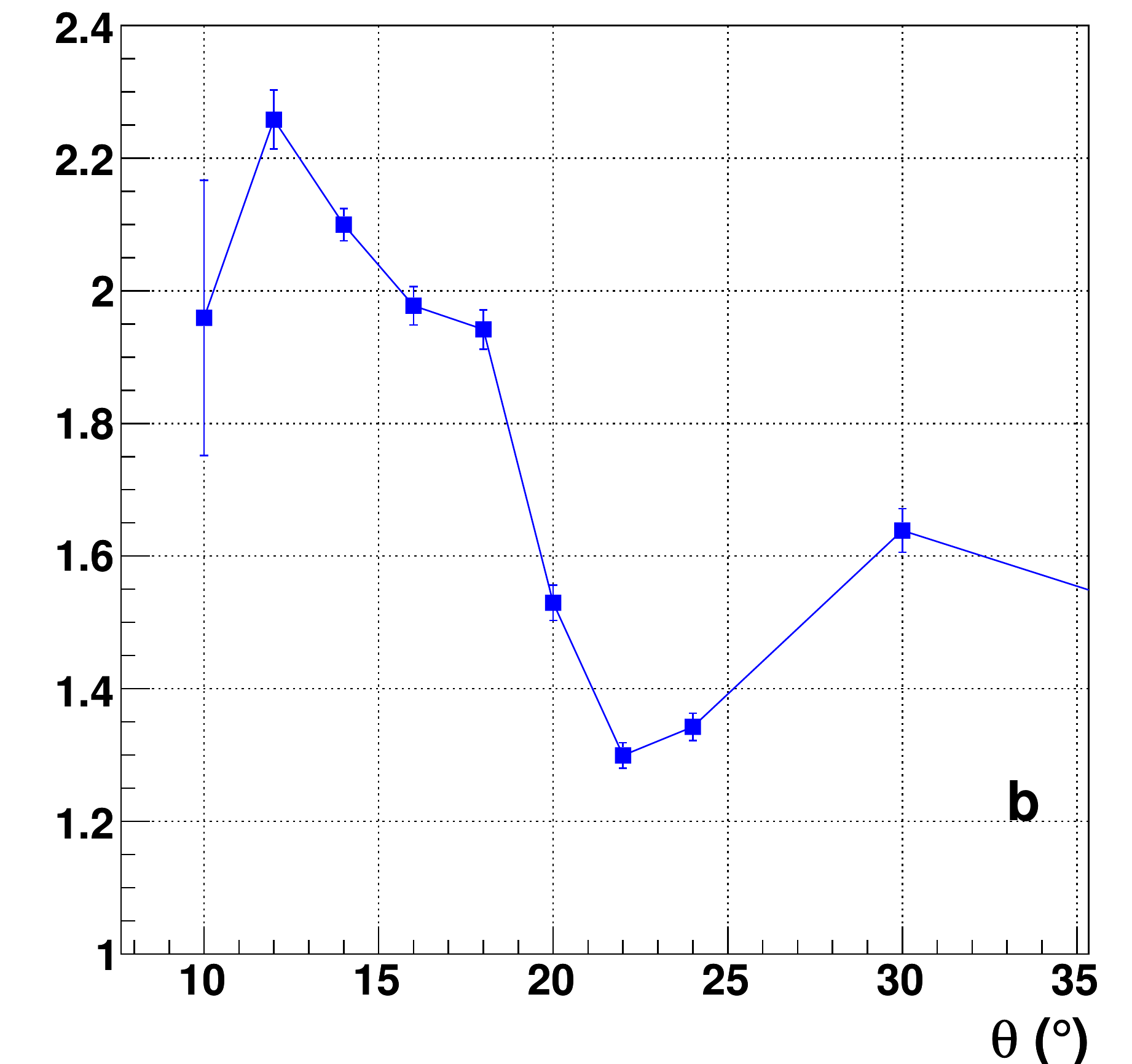}
\caption[Momentum resolution vs.~$\theta$ starting angle (0.3\,\gevc $mu^-$ single track events)]{Momentum resolution vs.~$\theta$ starting angle for {\bf 0.3 \gevc} 
{\boldmath $\mu^-$} single track events, in the full angular range {\boldmath $\theta \in [9^\circ, 160^\circ]$} (a) and in the forward region {\boldmath $\theta \in [9^\circ, 35^\circ]$}
(b) (see \Reftbl{tab:stt:per:perf_03}).} \label{fig:stt:per:resolution_03}
\end{figure*}
\begin{figure*}
\centering
\includegraphics[width=\swidth, height=\swidth]{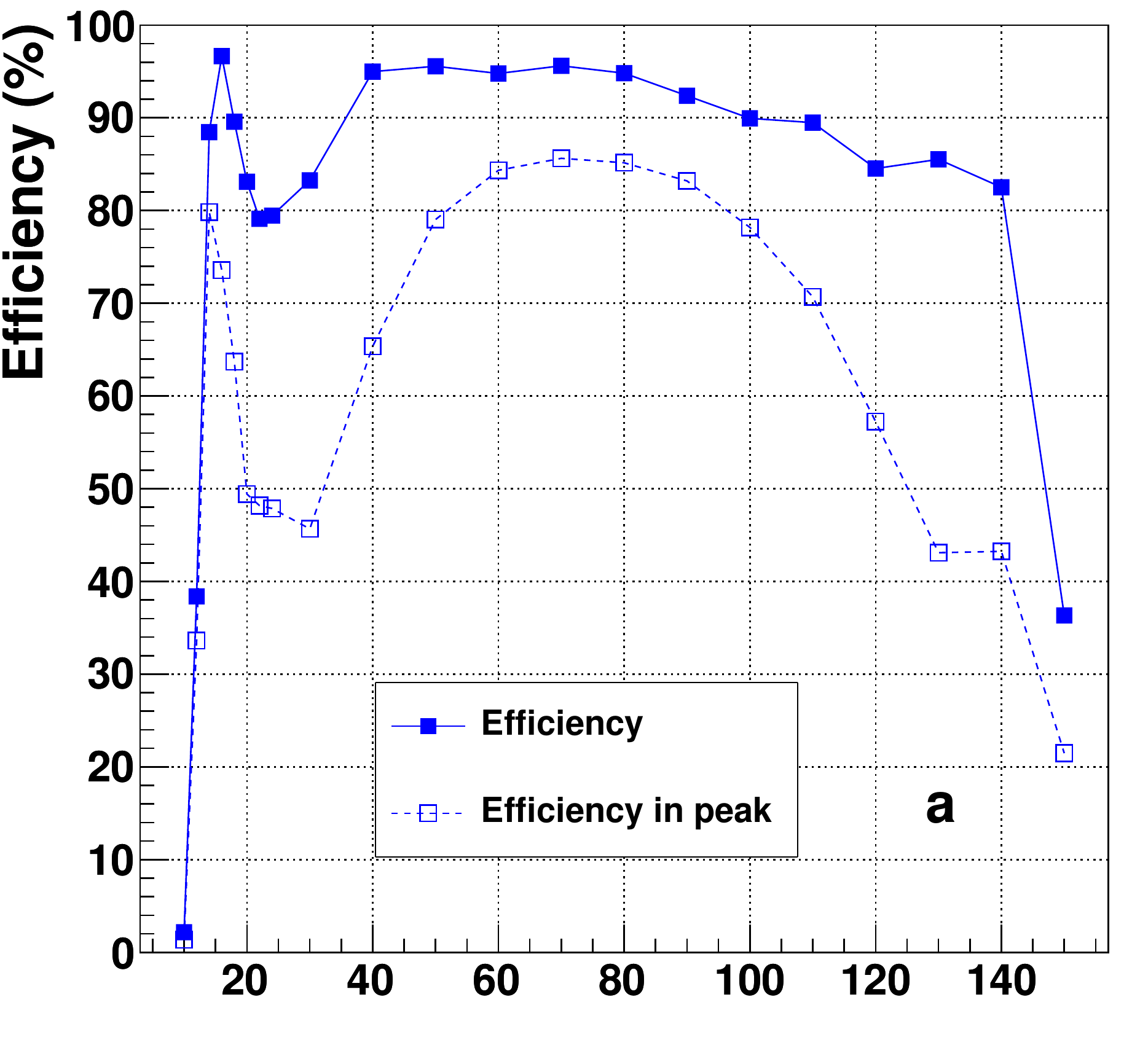}
\includegraphics[width=\swidth, height=\swidth]{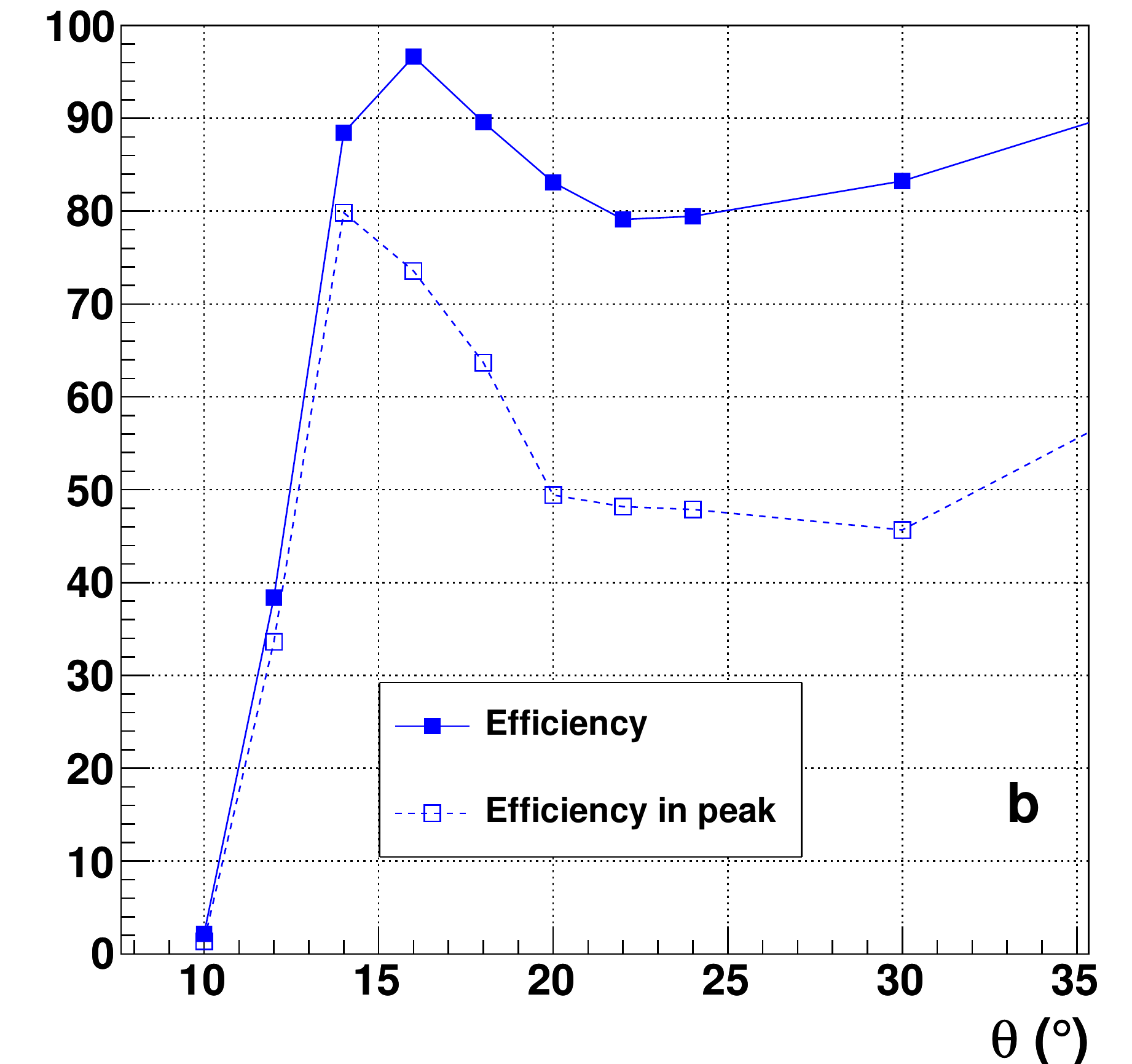}
\caption[Track reconstruction efficiency vs.~$\theta$ starting angle (0.3\,\gevc $mu^-$ single track events)]{Track reconstruction efficiency vs.~$\theta$ starting angle for {\bf 0.3
\gevc} {\boldmath $\mu^-$} single track events, in the full range {\boldmath $\theta \in [9^\circ, 
160^\circ]$} (a) and in the forward region {\boldmath $\theta \in [9^\circ, 35^\circ]$} (b)
(see \Reftbl{tab:stt:per:perf_03}).} \label{fig:stt:per:efficiency_03}
\end{figure*}
\begin{figure*}
\centering
\includegraphics[width=\swidth, height=\swidth]{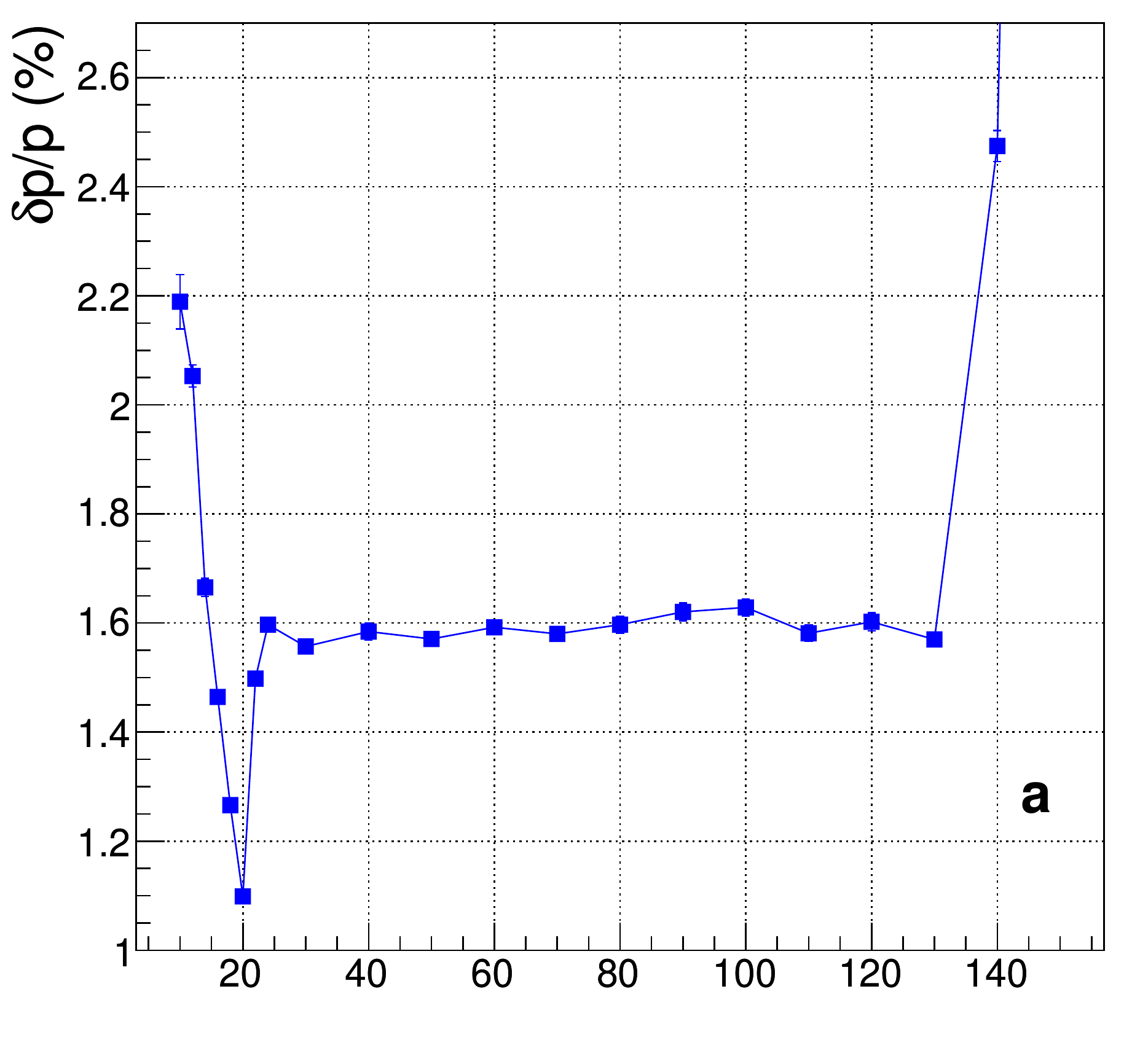}
\includegraphics[width=\swidth, height=\swidth]{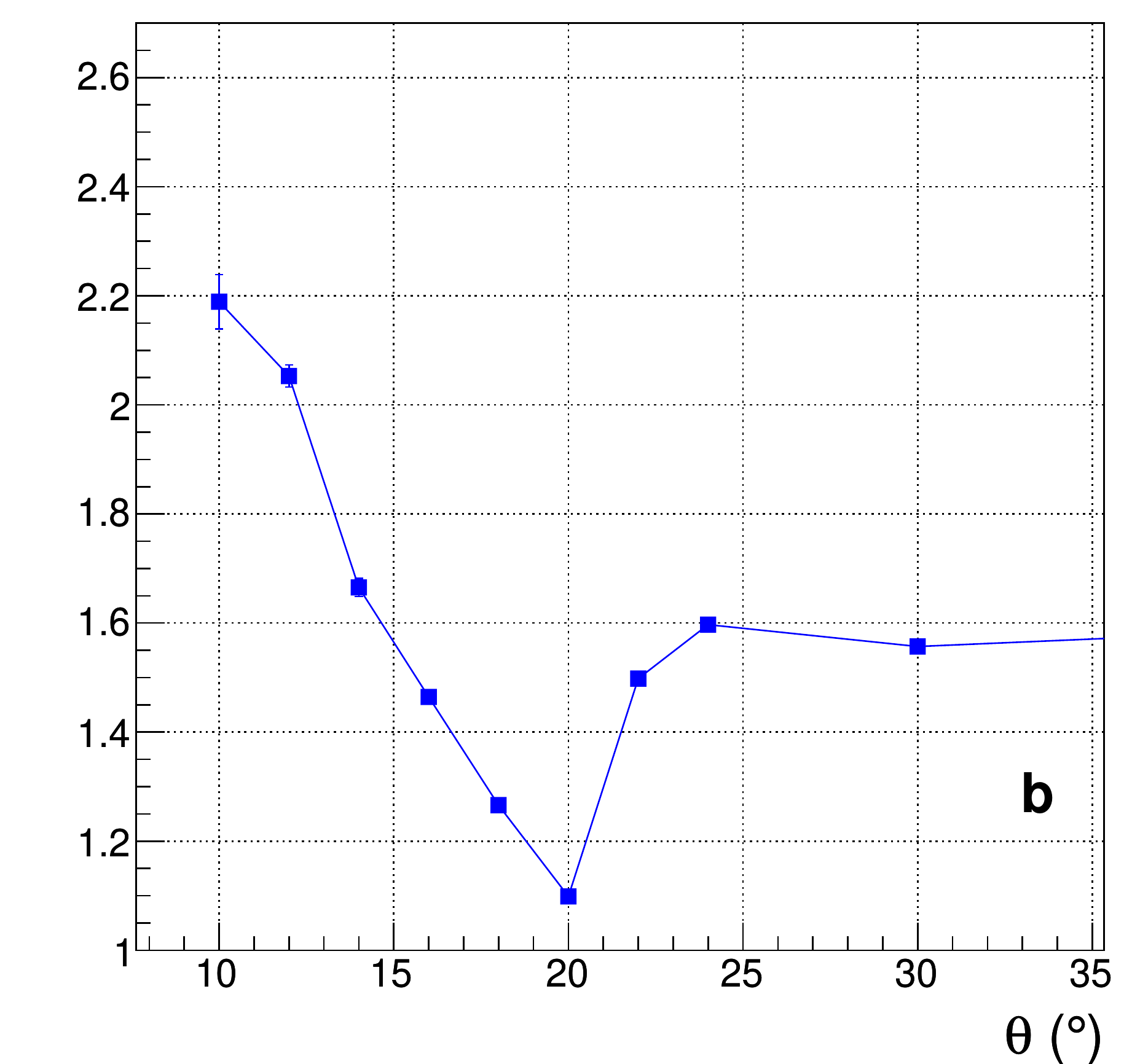}
\caption[Momentum resolution vs.~$\theta$ starting angle (1\,\gevc $mu^-$ single track events)]{Momentum resolution vs.~$\theta$ starting angle for {\bf 1 \gevc}
{\boldmath $\mu^-$} single track events, in the full angular range {\boldmath $\theta \in [9^\circ, 160^\circ]$} (a) and in the forward region {\boldmath $\theta \in [9^\circ, 35^\circ]$ }
(b) (see \Reftbl{tab:stt:per:perf_1}).} \label{fig:stt:per:resolution_1}
\end{figure*}
\begin{figure*}
\centering
\includegraphics[width=\swidth, height=\swidth]{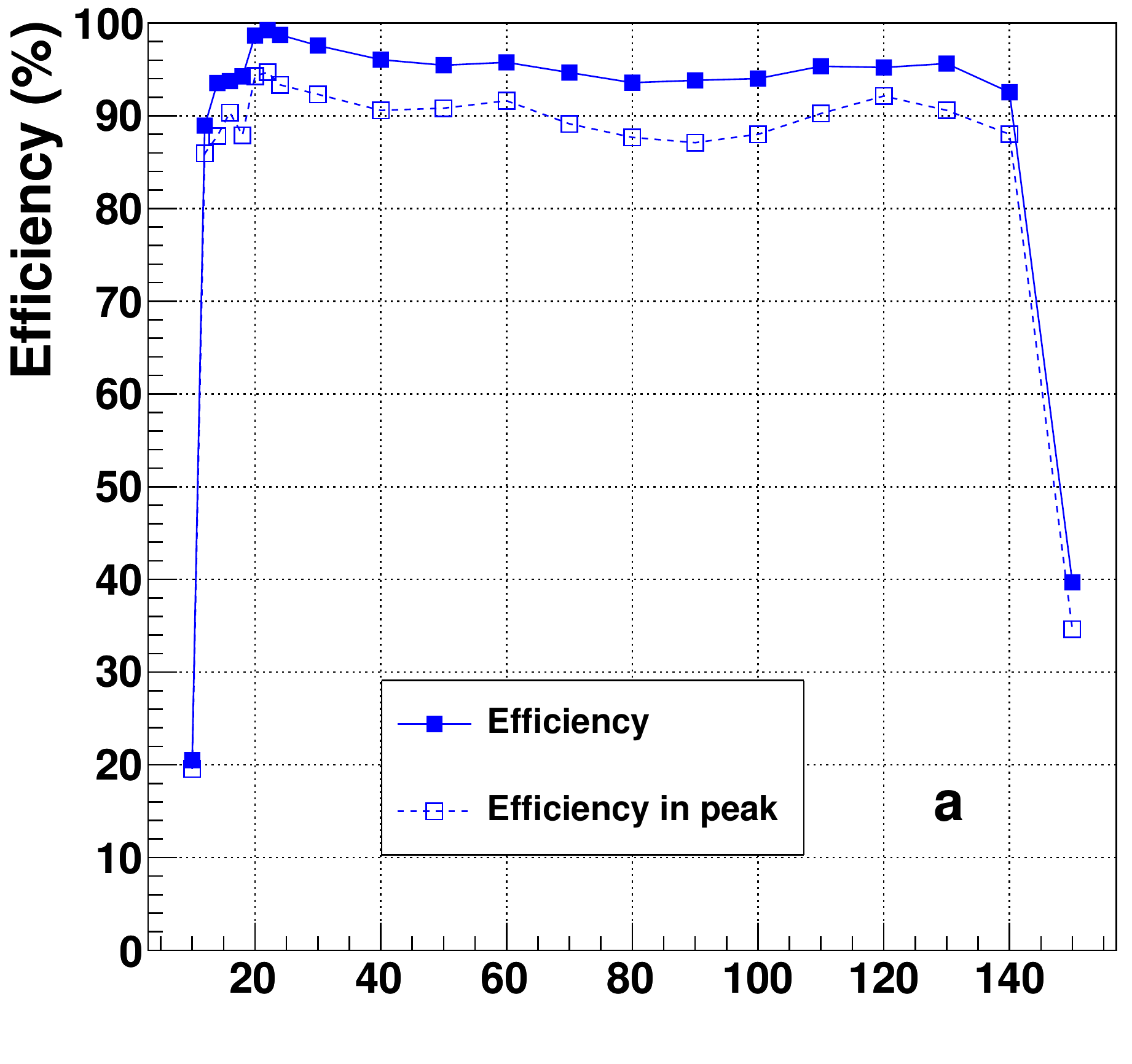}
\includegraphics[width=\swidth, height=\swidth]{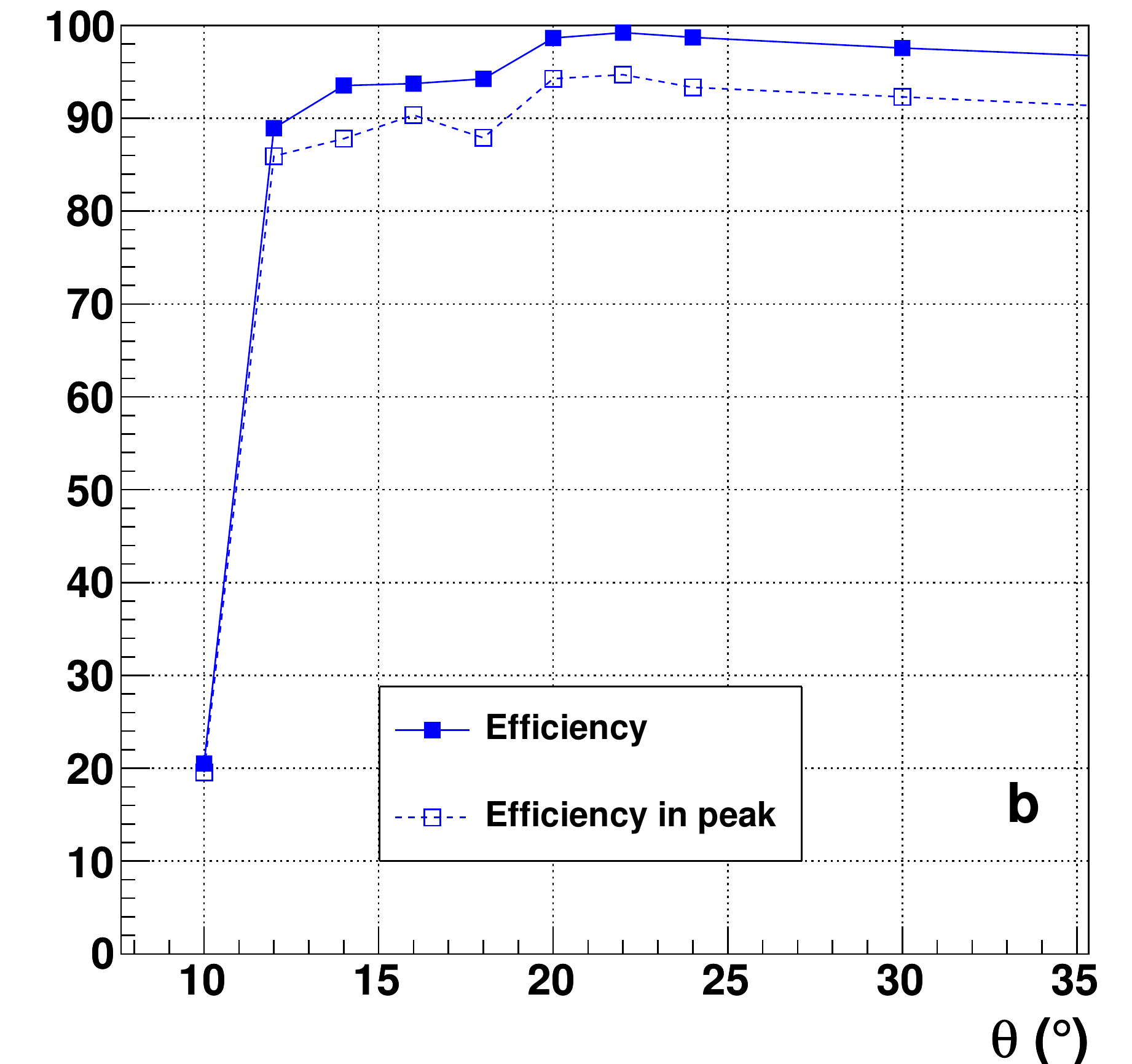}
\caption[Track reconstruction efficiency vs.~$\theta$ starting angle (1\,\gevc $mu^-$ single track events)]{Track reconstruction efficiency vs.~$\theta$ starting angle for {\bf 1
\gevc} {\boldmath $\mu^-$} single track events, in the full range {\boldmath $\theta \in [9^\circ, 
160^\circ]$} (a) and in the forward region {\boldmath $\theta \in [9^\circ, 35^\circ]$} (b)
(see \Reftbl{tab:stt:per:perf_1}).} \label{fig:stt:per:efficiency_1}
\end{figure*}
\begin{figure*}
\centering
\includegraphics[width=\swidth, height=\swidth]{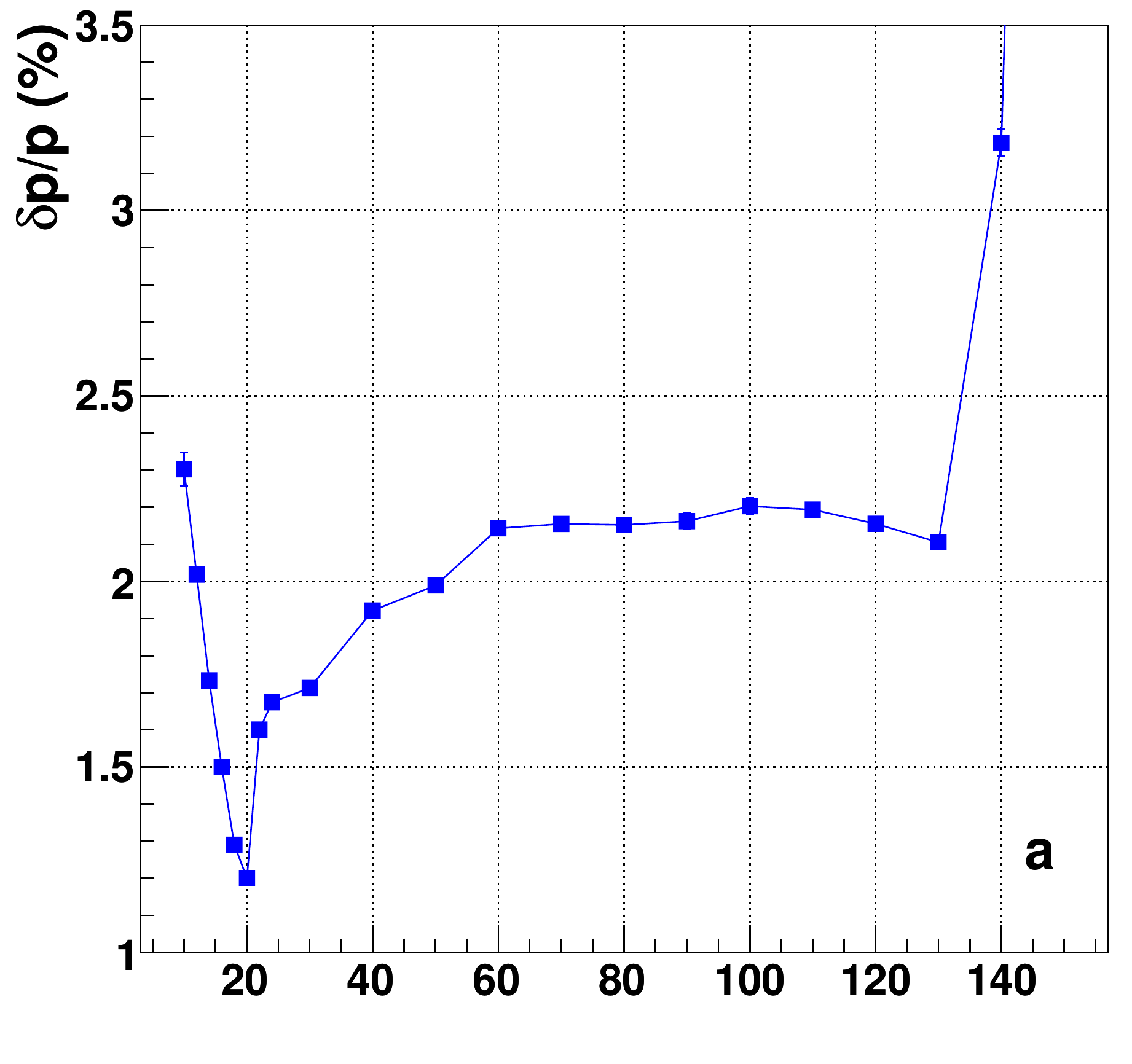}
\includegraphics[width=\swidth, height=\swidth]{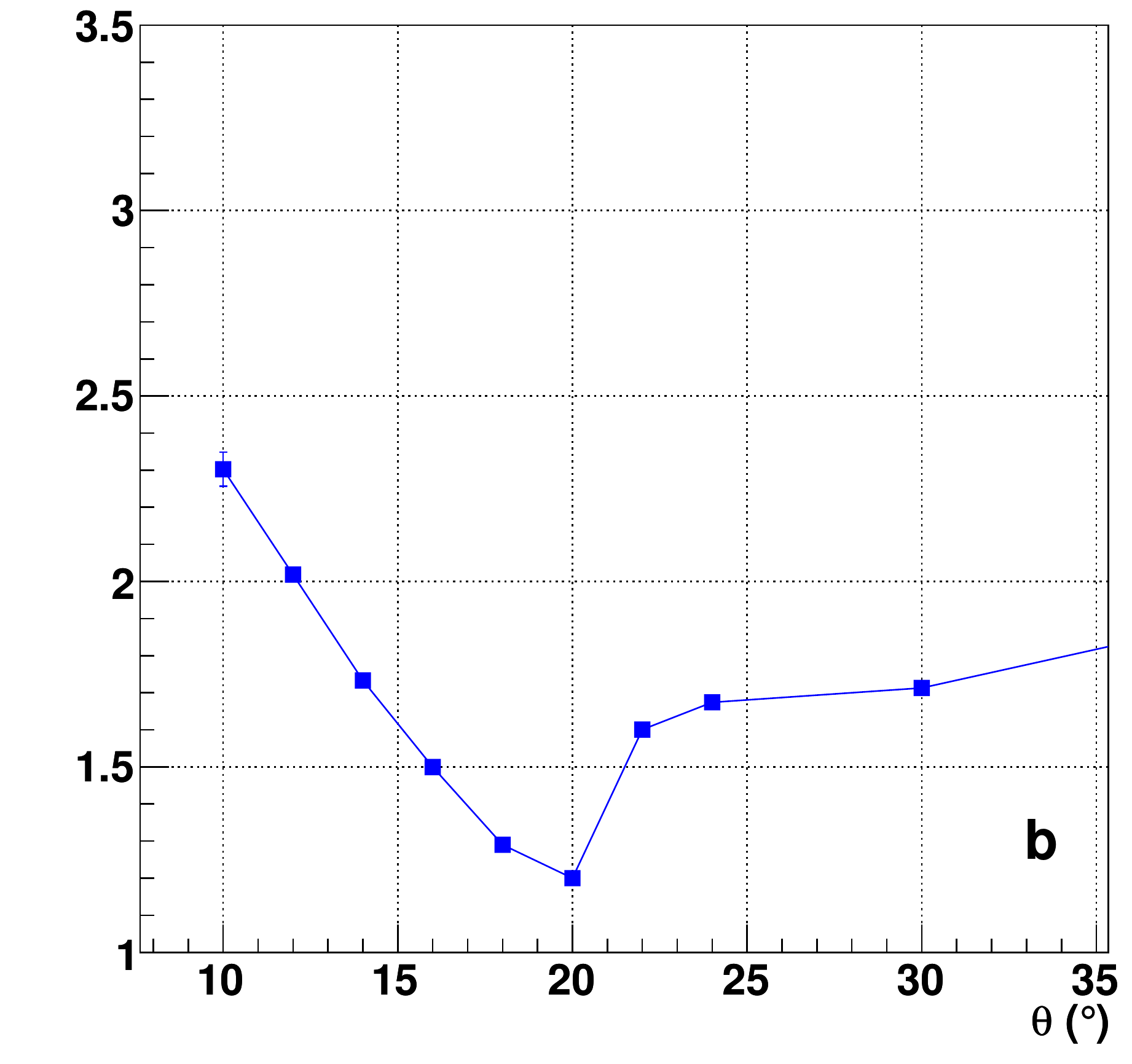}
\caption[Momentum resolution vs.~$\theta$ starting angle (2\,\gevc $mu^-$ single track events)]{Momentum resolution vs.~$\theta$ starting angle for {\bf 2 \gevc}
{\boldmath $\mu^-$} single track events, in the full angular range {\boldmath $\theta \in [9^\circ, 160^\circ]$} (a) and in the forward region {\boldmath $\theta \in [9^\circ, 35^\circ]$} 
(b) (see \Reftbl{tab:stt:per:perf_2}).} \label{fig:stt:per:resolution_2}
\end{figure*}
\begin{figure*}
\centering
\includegraphics[width=\swidth, height=\swidth]{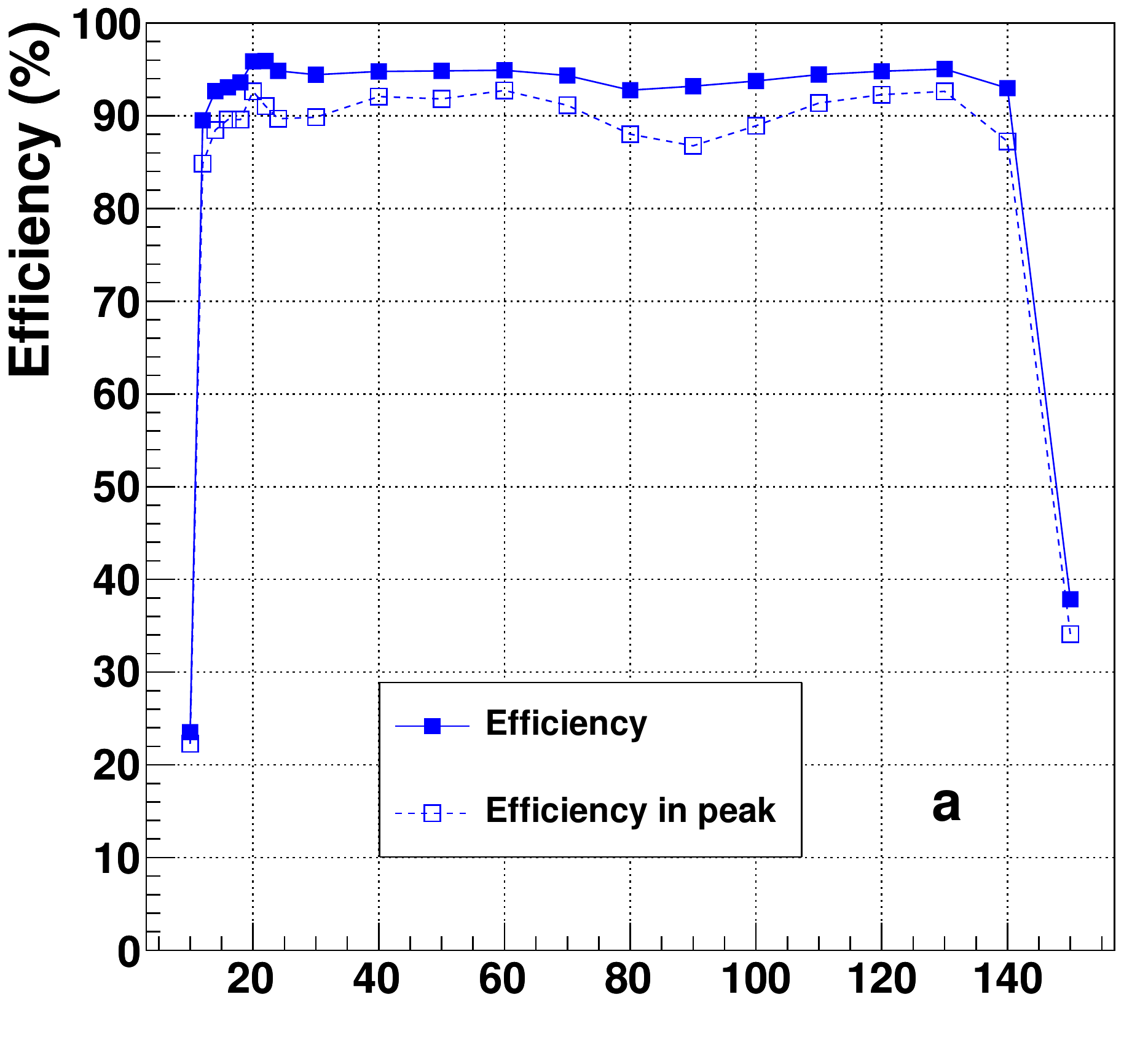}
\includegraphics[width=\swidth, height=\swidth]{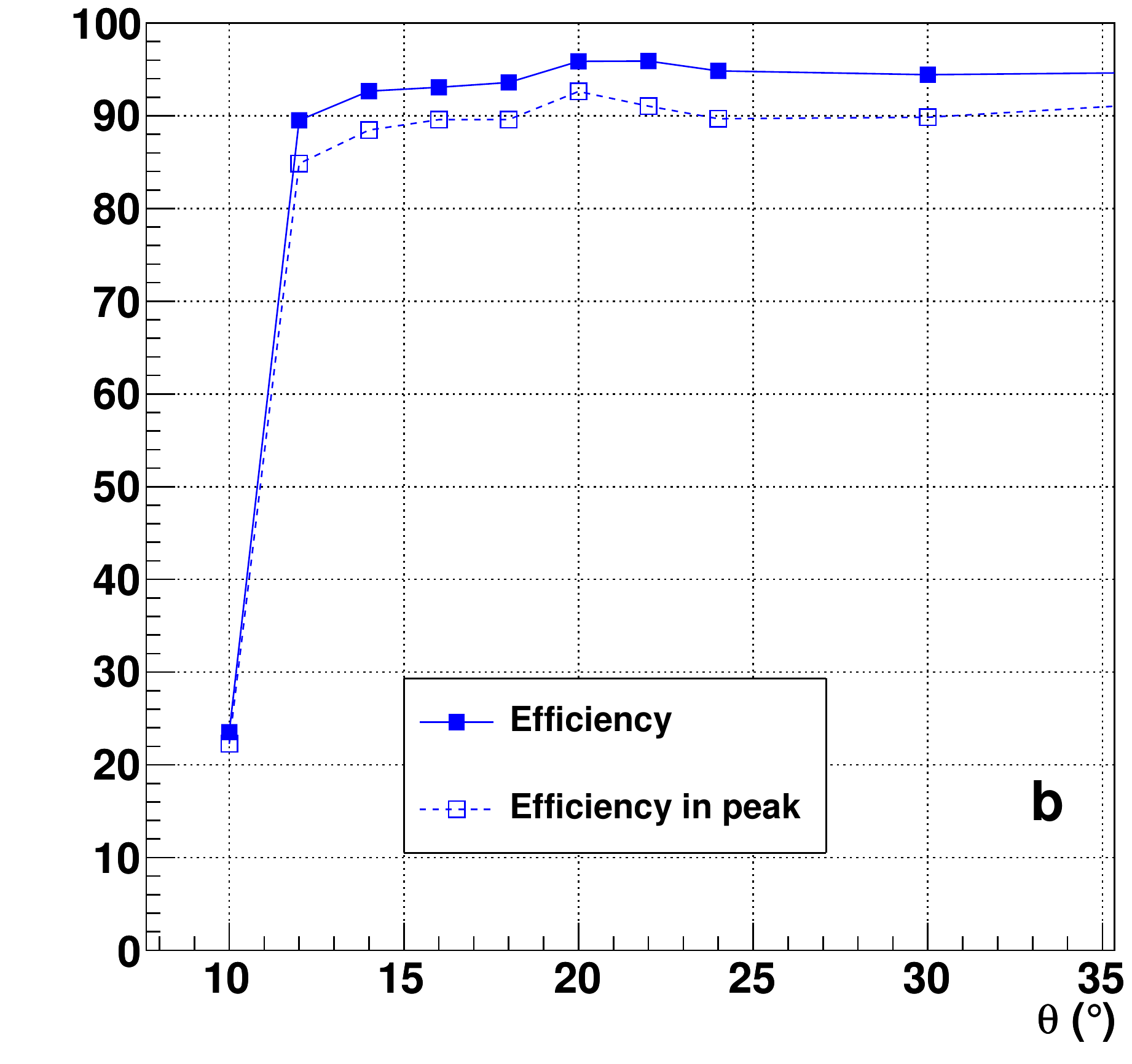}
\caption[Track reconstruction efficiency vs.~$\theta$ starting angle (2\,\gevc $mu^-$ single track events)]{Track reconstruction efficiency vs.~$\theta$ starting angle for {\bf 2
\gevc} {\boldmath $\mu^-$} single track events, in the full range {\boldmath $\theta \in [9^\circ, 
160^\circ]$} (a) and in the forward region {\boldmath $\theta \in [9^\circ, 35^\circ]$} (b)
(see \Reftbl{tab:stt:per:perf_2}).} \label{fig:stt:per:efficiency_2}
\end{figure*}
\begin{figure*}
\centering
\includegraphics[width=\swidth, height=\swidth]{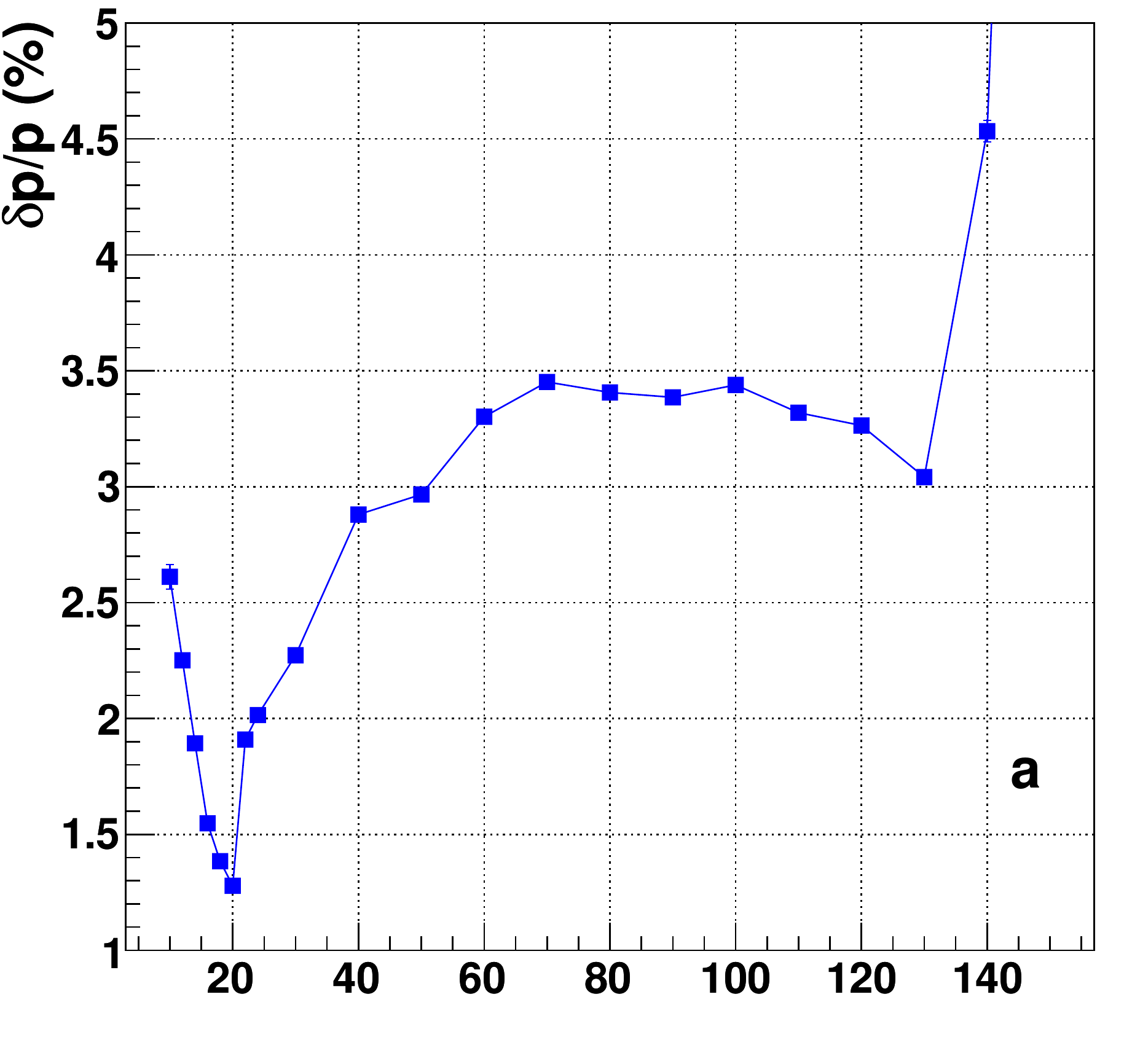}
\includegraphics[width=\swidth, height=\swidth]{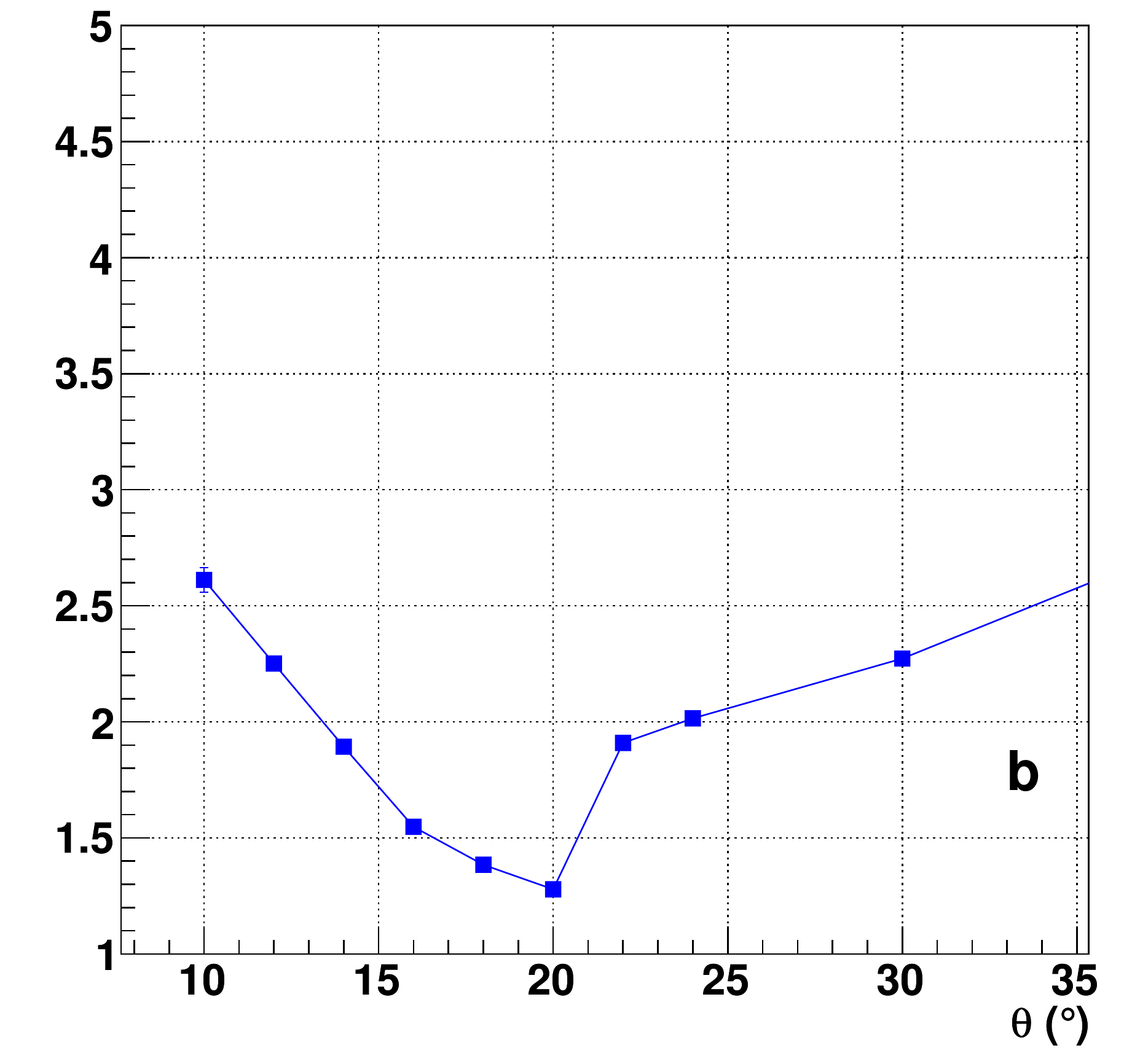}
\caption[Momentum resolution vs.~$\theta$ starting angle (5\,\gevc $mu^-$ single track events)]{Momentum resolution vs.~$\theta$ starting angle for {\bf 5 \gevc}
{\boldmath $\mu^-$} single track events, in the full angular range {\boldmath $\theta \in [9^\circ, 160^\circ]$} (a) and in the forward region {\boldmath $\theta \in [9^\circ, 35^\circ]$} 
(b) (see \Reftbl{tab:stt:per:perf_5}).} \label{fig:stt:per:resolution_5}
\end{figure*}
\begin{figure*}
\centering
\includegraphics[width=\swidth]{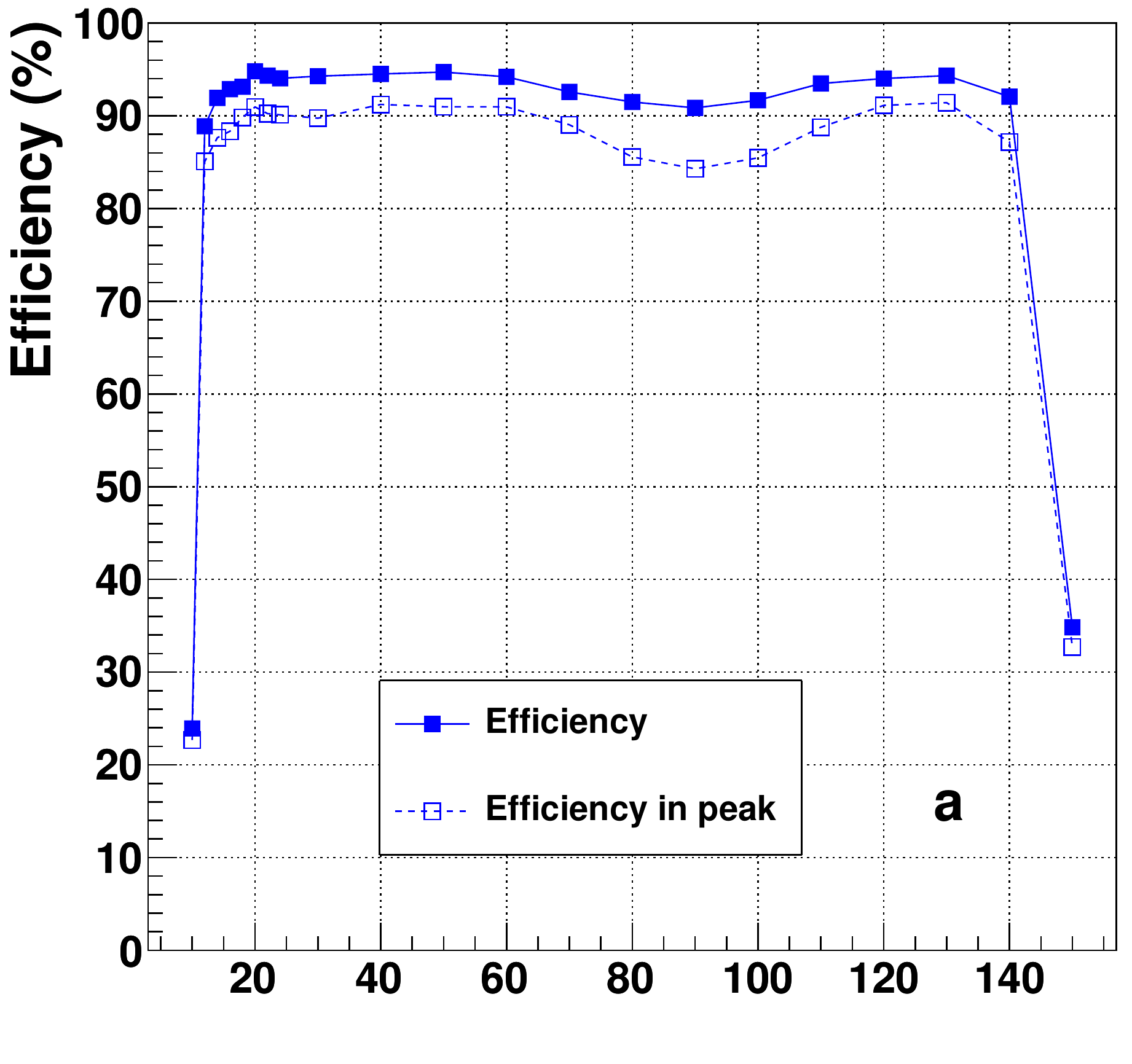}
\includegraphics[width=\swidth]{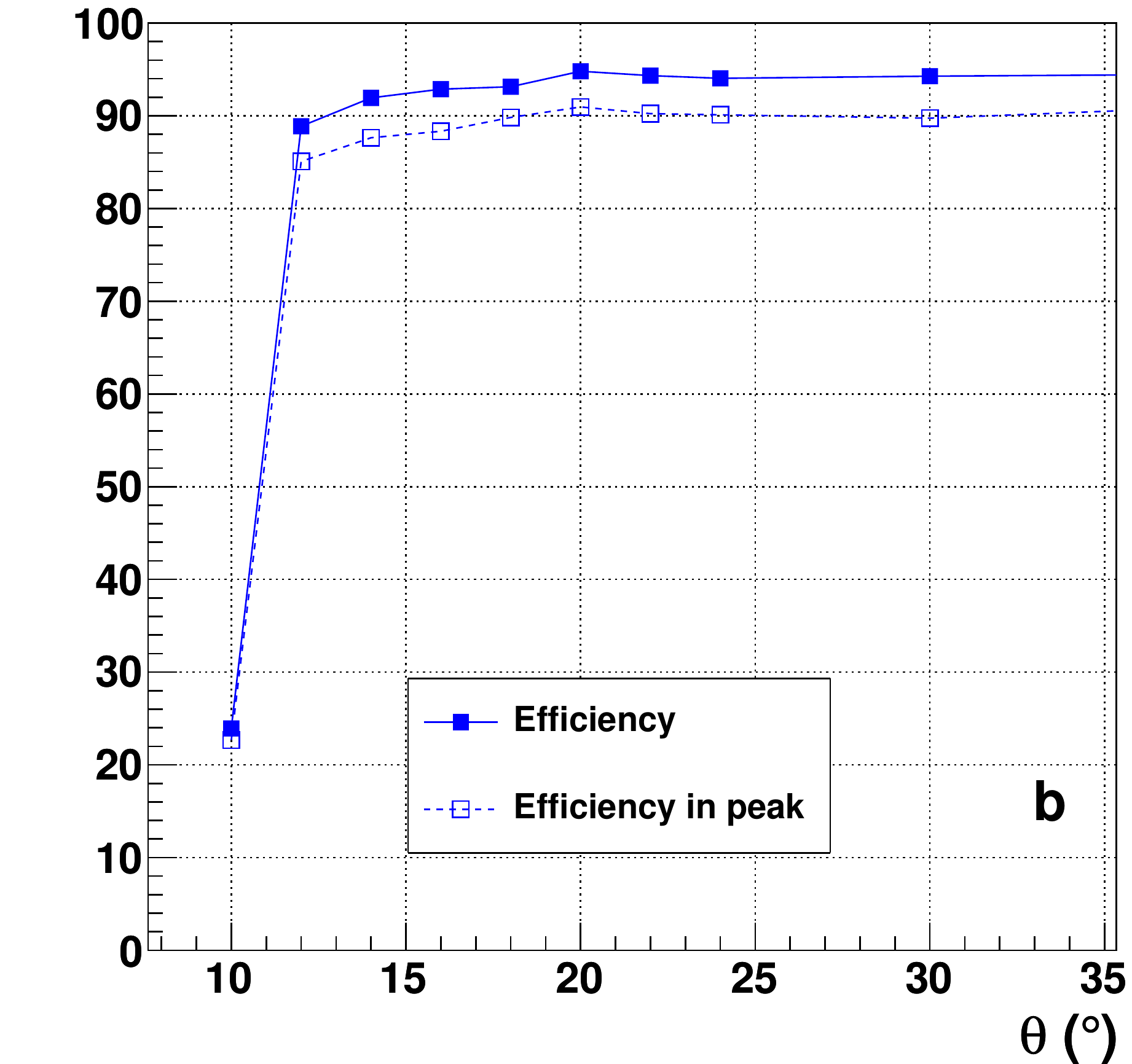}
\caption[Track reconstruction efficiency vs.~$\theta$ starting angle (5\,\gevc $mu^-$ single track events)]{Track reconstruction efficiency vs.~$\theta$ starting angle for {\bf 5
\gevc} {\boldmath $\mu^-$} single track events, in the full range {\boldmath $\theta \in [9^\circ, 
160^\circ]$} (a) and in the forward region {\boldmath $\theta \in [9^\circ, 35^\circ]$} (b)
(see \Reftbl{tab:stt:per:perf_5}).} \label{fig:stt:per:efficiency_5}
\end{figure*}
\begin{figure}
\centering
\includegraphics[width=\swidth]{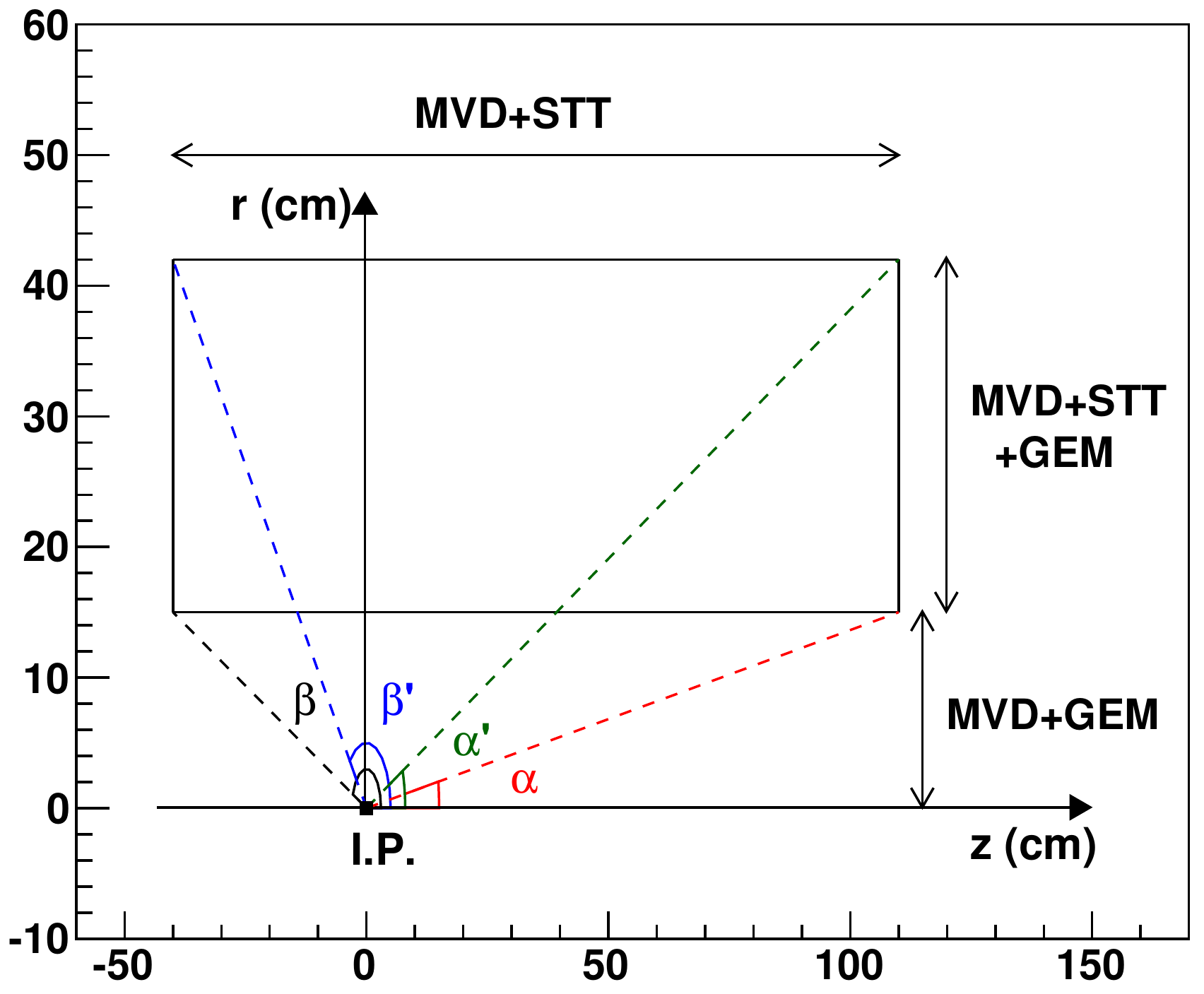}
\caption[Sketch of a section of the STT in the $(z,r)$ plane]{Sketch of a section of the STT in the $(z,r)$ plane. The marker 
corresponds to the interaction point (I.P.); the angle values are: $\alpha = 7.8^\circ$, $\alpha'=20.9^\circ$, $\beta=133.6^\circ$ and $\beta'=159.5^\circ$.}
\label{fig:stt:per:apparato}
\end{figure}

The values of momentum resolution and efficiency in peak (see \Refsec{sec:stt:resolstudies-uni}
 for the meaning) are summarized in \Reftbls{tab:stt:per:perf_03}--\ref{tab:stt:per:perf_5}. 
The momentum resolution and efficiency plots as a function of the $\theta$ angle 
are shown in \Reffigs{fig:stt:per:resolution_03}--\ref{fig:stt:per:efficiency_5}. 
% tables and figures
\begin{table*}[h!]
\begin{center}
\caption[Momentum resolution and reconstruction efficiency (0.3\,\gevc $mu^-$ single track events)]{Momentum resolution and reconstruction
efficiency for $10^4$ {\boldmath $\mu^-$} single track events generated at {\bf 0.3 \gevc} and {\bf fixed}
{\boldmath $\theta$} angle.} 
\label{tab:stt:per:perf_03}
\smallskip
\begin{tabular}{c|c|c|c}
\hline\hline
$\theta \, (^\circ)$ & Resolution (\%)   & Efficiency (\%)	& Efficiency in peak (\%) \\ 
\hline\hline
10	&$1.96 \pm 0.21$	&$2.17	\pm $0.15	&$1.36	\pm 0.16$\\
12	&$2.26 \pm 0.04$	&$38.39	\pm $0.49	&$33.61	\pm 0.47$\\
14	&$2.09 \pm 0.02$	&$88.45	\pm $0.32	&$79.84	\pm 0.40$\\
16	&$1.98 \pm 0.03$	&$96.65	\pm $0.18	&$73.56	\pm 0.44$\\
18	&$1.94 \pm 0.03$	&$89.58	\pm $0.31	&$63.69	\pm 0.48$\\
20	&$1.53 \pm 0.03$	&$83.10	\pm $0.37	&$49.42	\pm 0.49$\\
22	&$1.29 \pm 0.02$	&$79.11	\pm $0.41	&$48.18	\pm 0.49$\\
24 	&$1.34 \pm 0.02$	&$79.45	\pm $0.40	&$47.87	\pm 0.49$\\
30	&$1.64 \pm 0.03$	&$83.25	\pm $0.37	&$45.67	\pm 0.49$\\
40	&$1.47 \pm 0.02$	&$94.98	\pm $0.22	&$65.35	\pm 0.48$\\
50	&$1.35 \pm 0.01$	&$95.56	\pm $0.20	&$79.03	\pm 0.41$\\
60	&$1.29 \pm 0.01$	&$94.78	\pm $0.22	&$84.34	\pm 0.36$\\
70	&$1.24 \pm 0.01$	&$95.61	\pm $0.20	&$85.63	\pm 0.35$\\
80	&$1.23 \pm 0.01$	&$94.81	\pm $0.22	&$85.15	\pm 0.35$\\
90	&$1.24 \pm 0.01$	&$92.38	\pm $0.26	&$83.19	\pm 0.37$\\
100	&$1.25 \pm 0.01$	&$89.94	\pm $0.30	&$78.18	\pm 0.41$\\
110	&$1.22 \pm 0.01$	&$89.48	\pm $0.31	&$70.69	\pm 0.45$\\
120	&$1.27 \pm 0.01$	&$84.53	\pm $0.36	&$57.24	\pm 0.49$\\
130	&$1.33 \pm 0.02$	&$85.52	\pm $0.35	&$43.10	\pm 0.49$\\
140	&$2.16 \pm 0.05$	&$82.51	\pm $0.38	&$43.24	\pm 0.49$\\
150	&$6.47 \pm 0.17$	&$36.34	\pm $0.48	&$21.48	\pm 0.41$\\
\hline \hline
\end{tabular}
\end{center}
\end{table*}\noindent
%resolution @ 0.3 GeV
%efficiency @ 0.3 GeV
%
% table for 1 GeV/c tracks
\begin{table*}[h!]
\begin{center}
\caption[Momentum resolution and reconstruction efficiency (1\,\gevc $mu^-$ single track events)]{Momentum resolution and reconstruction
efficiency for $10^4$ {\boldmath $\mu^-$} single track events generated at {\bf 1 \gevc} and {\bf fixed}
 {\boldmath $\theta$} angle.} 
\smallskip
\label{tab:stt:per:perf_1}
\begin{tabular}{c|c|c|c}
\hline\hline
$\theta \, (^\circ)$ & Resolution (\%)   & Efficiency (\%)	& Efficiency in peak (\%) \\ 
\hline\hline
10	&$2.19	\pm 0.05$	&$20.51	\pm 0.40$	&$19.55	\pm 0.39$\\
12	&$2.05	\pm 0.02$	&$88.94	\pm 0.31$	&$85.93	\pm 0.35$\\
14	&$1.67	\pm 0.02$	&$93.53	\pm 0.25$	&$87.81	\pm 0.33$\\
16	&$1.46	\pm 0.01$	&$93.74	\pm 0.24$	&$90.35	\pm 0.30$\\
18	&$1.27	\pm 0.01$	&$94.25	\pm 0.23$	&$87.90	\pm 0.33$\\
20	&$1.09	\pm 0.01$	&$98.65	\pm 0.12$	&$94.27	\pm 0.23$\\
22	&$1.50	\pm 0.01$	&$99.23	\pm 0.09$	&$94.70	\pm 0.22$\\
24	&$1.60	\pm 0.01$	&$98.72	\pm 0.11$	&$93.33	\pm 0.25$\\
30	&$1.56	\pm 0.01$	&$97.57	\pm 0.15$	&$92.31	\pm 0.27$\\
40	&$1.58	\pm 0.01$	&$96.05	\pm 0.19$	&$90.58	\pm 0.29$\\
50	&$1.57	\pm 0.01$	&$95.45	\pm 0.21$	&$90.82	\pm 0.29$\\
60	&$1.59	\pm 0.01$	&$95.76	\pm 0.20$	&$91.62	\pm 0.28$\\
70	&$1.58	\pm 0.01$	&$94.66	\pm 0.22$	&$89.13	\pm 0.31$\\
80	&$1.60	\pm 0.01$	&$93.57	\pm 0.24$	&$87.67	\pm 0.33$\\
90	&$1.62	\pm 0.02$	&$93.82	\pm 0.24$	&$87.10	\pm 0.33$\\
100	&$1.63	\pm 0.02$	&$94.01	\pm 0.24$	&$88.00	\pm 0.32$\\
110	&$1.58	\pm 0.01$	&$95.34	\pm 0.21$	&$90.24	\pm 0.30$\\
120	&$1.60	\pm 0.02$	&$95.21	\pm 0.21$	&$92.13	\pm 0.27$\\
130	&$1.57	\pm 0.01$	&$95.63	\pm 0.20$	&$90.59	\pm 0.29$\\
140	&$2.47	\pm 0.03$	&$92.54	\pm 0.26$	&$88.01	\pm 0.32$\\
150	&$7.81	\pm 0.15$	&$39.69	\pm 0.49$	&$34.62	\pm 0.48$\\
\hline \hline
\end{tabular}
\end{center}
\end{table*}\noindent
%resolution @ 1.0 GeV
%efficiency @ 1.0 GeV
%
%table for 2 GeV/c tracks
\begin{table*}[h!]
\begin{center}
\caption[Momentum resolution and reconstruction efficiency (2\,\gevc $mu^-$ single track events)]{Momentum resolution and reconstruction
efficiency for $10^4$ {\boldmath $\mu^-$} single track events generated at {\bf 2 \gevc} and {\bf fixed}
 {\boldmath $\theta$} angle.} 
\label{tab:stt:per:perf_2}
\smallskip
\begin{tabular}{c|c|c|c}
\hline\hline
$\theta \, (^\circ)$ & Resolution (\%)   & Efficiency (\%)	& Efficiency in peak (\%) \\ 
\hline\hline
10	&$2.30	\pm 0.05$	&$23.53	\pm 0.42$	&$22.26	\pm 0.42$\\
12	&$2.02	\pm 0.02$	&$89.51	\pm 0.31$	&$84.84	\pm 0.36$\\
14	&$1.73	\pm 0.02$	&$92.66	\pm 0.26$	&$88.46	\pm 0.32$\\
16	&$1.50	\pm 0.01$	&$93.07	\pm 0.25$	&$89.58	\pm 0.31$\\
18	&$1.29	\pm 0.01$	&$93.59	\pm 0.24$	&$89.59	\pm 0.31$\\
20	&$1.20	\pm 0.01$	&$95.87	\pm 0.20$	&$92.62	\pm 0.26$\\
22	&$1.60	\pm 0.02$	&$95.90	\pm 0.20$	&$91.04	\pm 0.29$\\
24	&$1.67	\pm 0.02$	&$94.84	\pm 0.22$	&$89.68	\pm 0.30$\\
30	&$1.71	\pm 0.02$	&$94.43	\pm 0.23$	&$89.84	\pm 0.30$\\
40	&$1.92	\pm 0.02$	&$94.78	\pm 0.22$	&$92.07	\pm 0.27$\\
50	&$1.99	\pm 0.02$	&$94.84	\pm 0.22$	&$91.82	\pm 0.27$\\
60	&$2.14	\pm 0.02$	&$94.91	\pm 0.22$	&$92.73	\pm 0.26$\\
70	&$2.15	\pm 0.02$	&$94.34	\pm 0.23$	&$91.13	\pm 0.28$\\
80	&$2.15	\pm 0.02$	&$92.76	\pm 0.26$	&$88.01	\pm 0.32$\\
90	&$2.16	\pm 0.02$	&$93.19	\pm 0.25$	&$86.76	\pm 0.34$\\
100	&$2.20	\pm 0.02$	&$93.75	\pm 0.24$	&$88.92	\pm 0.31$\\
110	&$2.19	\pm 0.02$	&$94.44	\pm 0.23$	&$91.37	\pm 0.28$\\
120	&$2.16	\pm 0.02$	&$94.80	\pm 0.22$	&$92.28	\pm 0.27$\\
130	&$2.11	\pm 0.02$	&$95.04	\pm 0.22$	&$92.63	\pm 0.26$\\
140	&$3.18	\pm 0.04$	&$92.98	\pm 0.26$	&$87.23	\pm 0.33$\\
150	&$8.90	\pm 0.16$	&$37.86	\pm 0.49$	&$34.11	\pm 0.47$\\
\hline \hline
\end{tabular}
\end{center}
\end{table*}\noindent
%resolution @ 2.0 GeV
%efficiency @ 2.0 GeV
%
%table for 5 GeV/c tracks
\begin{table*}[h!]
\begin{center}
\caption[Momentum resolution and reconstruction efficiency (5\,\gevc $mu^-$ single track events)]{Momentum resolution and reconstruction
efficiency for $10^4$ $\mu^-$ single track events generated at {\bf 5 \gevc} and {\bf fixed}
 {\boldmath $\theta$} angle.} 
\label{tab:stt:per:perf_5}
\smallskip
\begin{tabular}{c|c|c|c}
\hline\hline
$\theta \, (^\circ)$ & Resolution (\%)   & Efficiency (\%)	& Efficiency in peak (\%) \\ 
\hline\hline
10	&$2.61	\pm 0.05$	&$23.93	\pm 0.43$	&$22.68	\pm 0.42$\\
12	&$2.25	\pm 0.02$	&$88.87	\pm 0.31$	&$85.07	\pm 0.36$\\
14	&$1.89	\pm 0.02$	&$91.93	\pm 0.27$	&$87.63	\pm 0.33$\\
16	&$1.55	\pm 0.02$	&$92.87	\pm 0.26$	&$88.33	\pm 0.32$\\
18	&$1.38	\pm 0.01$	&$93.13	\pm 0.25$	&$89.82	\pm 0.30$\\
20	&$1.28	\pm 0.01$	&$94.80	\pm 0.22$	&$90.95	\pm 0.29$\\
22	&$1.91	\pm 0.02$	&$94.33	\pm 0.23$	&$90.23	\pm 0.30$\\
24	&$2.01	\pm 0.02$	&$94.04	\pm 0.24$	&$90.10	\pm 0.30$\\
30	&$2.27	\pm 0.02$	&$94.27	\pm 0.23$	&$89.74	\pm 0.30$\\
40	&$2.88	\pm 0.03$	&$94.51	\pm 0.23$	&$91.22	\pm 0.28$\\
50	&$2.97	\pm 0.03$	&$94.71	\pm 0.22$	&$90.97	\pm 0.29$\\
60	&$3.30	\pm 0.03$	&9$4.20	\pm 0.23$	&$90.96	\pm 0.29$\\
70	&$3.45	\pm 0.03$	&$92.57	\pm 0.26$	&$89.03	\pm 0.31$\\
80	&$3.41	\pm 0.03$	&$91.49	\pm 0.28$	&$85.56	\pm 0.35$\\
90	&$3.38	\pm 0.03$	&$90.85	\pm 0.29$	&$84.28	\pm 0.36$\\
100	&$3.44	\pm 0.03$	&$91.67	\pm 0.28$	&$85.46	\pm 0.35$\\
110	&$3.32	\pm 0.03$	&$93.48	\pm 0.25$	&$88.76	\pm 0.32$\\
120	&$3.26	\pm 0.03$	&$94.02	\pm 0.24$	&$91.12	\pm 0.28$\\
130	&$3.04	\pm 0.03$	&$94.33	\pm 0.23$	&$91.41	\pm 0.28$\\
140	&$4.53	\pm 0.05$	&$92.06	\pm 0.27$	&$87.17	\pm 0.33$\\
150	&$11.36	\pm 0.23$	&$34.84	\pm 0.48$	&$32.70	\pm 0.47$\\
\hline \hline
\end{tabular}
\end{center}
\end{table*}\noindent
%resolution @ 5.0 GeV
%efficiency @ 5.0 GeV
Apart from the 0.3 \gevc set of simulated events, for which a dedicated comment
is needed, common conclusions can be drawn for the other event sets generated
at 1, 2 and 5 \gevc.
Concerning the momentum resolution, a common behavior can be identified by
looking at \Reffigs{fig:stt:per:resolution_1}, \ref{fig:stt:per:resolution_2} and \ref{fig:stt:per:resolution_5}: 
the resolution improves for $\theta$ values up to $\sim 21^\circ$, then starts 
to worsen again. The results can be interpreted on the basis of geometrical 
considerations, by looking at the sketch of the STT in the $(z,r)$ plane shown
in \Reffig{fig:stt:per:apparato}.
Tracks travelling with small $\theta$ values (but bigger than 7.8$^\circ$) 
hit just few straw layers, in particular only the axial ones if $\theta < 11.6^\circ$, preventing the reconstruction of the $z$
coordinate of the straw tube hits (\Refsec{sec:stt:sim:pr}); this results in 
a bad spatial (and hence momentum) resolution of the STT hits. On the other 
hand, the tracking in this forward angular region is performed mainly with the
hits produced in the MVD and in the GEM chambers. The very high precision of these two 
detectors improves the resolution, which becomes much better when including
also the spatial information coming from their hits.
As the $\theta$ value increases, tracks hit more and more straw layers, 
allowing a better track reconstruction in the tracker. This, combined with the
good resolution of the MVD and GEM hits, results in a better global momentum
resolution.
Then, for $21^\circ < \theta< 133^\circ$, tracks traverse the MVD and all the straw layers;
so the resolution obtained by the STT alone is improved with respect to that 
at lower $\theta$ values, but it suffers from the fact that there are no more 
hits in the GEM chambers. So the resolution is globally a bit worse.

Finally, for $\theta > 133^\circ$ tracks are going in the backward direction
and traverse a lower number of straw layers as the angle increases: 
Consequently, since the decreased number of hits is not compensated by any 
other outer tracking detector (like the GEMs in the forward direction),
the resolution becomes worse.

The reconstruction efficiency, shown in \Reffigs{fig:stt:per:efficiency_1}, 
\ref{fig:stt:per:efficiency_2} and \ref{fig:stt:per:efficiency_5}, is 
quite low around $\theta=10^\circ$ because the tracking procedure fails when
the number of reconstructed hits is too low. Then, it increases up to
more than $90\%$ in the central angular region. The efficiency presents a dip 
around $\theta=90^\circ$ due to the tracks that are lost because they go into 
the target pipe.
Finally, for tracks travelling in the backward direction, the efficiency starts
to decrease because of the reduced number of hits per track, that may cause
problems in the reconstruction.

Concerning the events at $0.3$ \gevc, the results are not so reliable as at
the others described above, in particular for small values of $\theta$. 
The reason is that the Kalman fit produces long tails in the momentum 
distributions (see \Reffig{fig:stt:per:totalmom}.a), even if the outcome of 
the prefit does not present these tails. 

This Kalman behaviour affects both the momentum resolution and the reconstruction efficiency, shown in 
\Reffigs{fig:stt:per:resolution_03} and \ref{fig:stt:per:efficiency_03}; it is probably due to a code bug, 
which has still to be deeply investigated and corrected.

% ================= comments on the old results =======================
%Concerning the region where the fine scan has been performed (\Reffig{fig:stt:per:resangles1530}), tracks with small $\theta$ values do not hit at all or hit 
%just few axial layers of straw tubes; this would prevent from reconstructing 
%the $z$ coordinate of the track in the STT. So the tracking is performed 
%mainly with the 
%hits produced in the MVD and the GEMs; hence the global resolution is dominated
%by the one of these two detectors. \\
%\Reffig{fig:stt:per:sttmvdresangles} shows that the momentum resolution in the central angular region, where almost all the straw layers are hit, is around 1.7\%.

\subsubsection{Studies at Fixed Transverse Momentum}
\begin{figure}
\centering
\includegraphics[width=\swidth, height=\swidth]{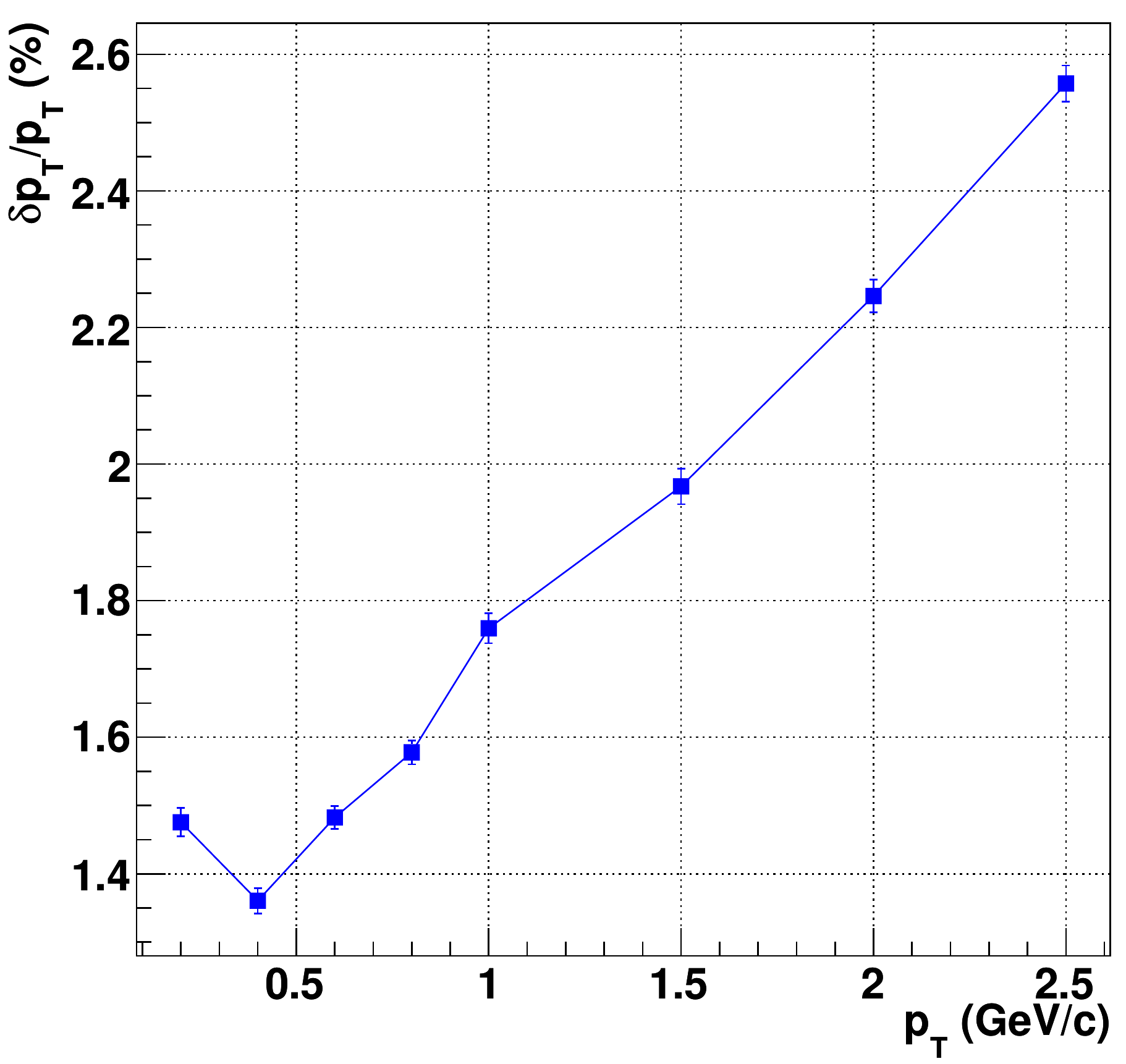}
\caption[Momentum resolution vs.~$p_T$ ($mu^-$ single track events)]{Momentum resolution vs.~$p_T$ for {\boldmath $\mu^-$} single track events, in the angular ranges {\boldmath $\phi \in [0^\circ, 360^\circ$} and {\boldmath $\theta \in [7^\circ, 160^\circ]$} (see \Reftbl{tab:stt:per:perf_pt}).} \label{fig:stt:per:resol_pt}
\end{figure}

\begin{figure}
\centering
\includegraphics[width=\swidth, height=\swidth]{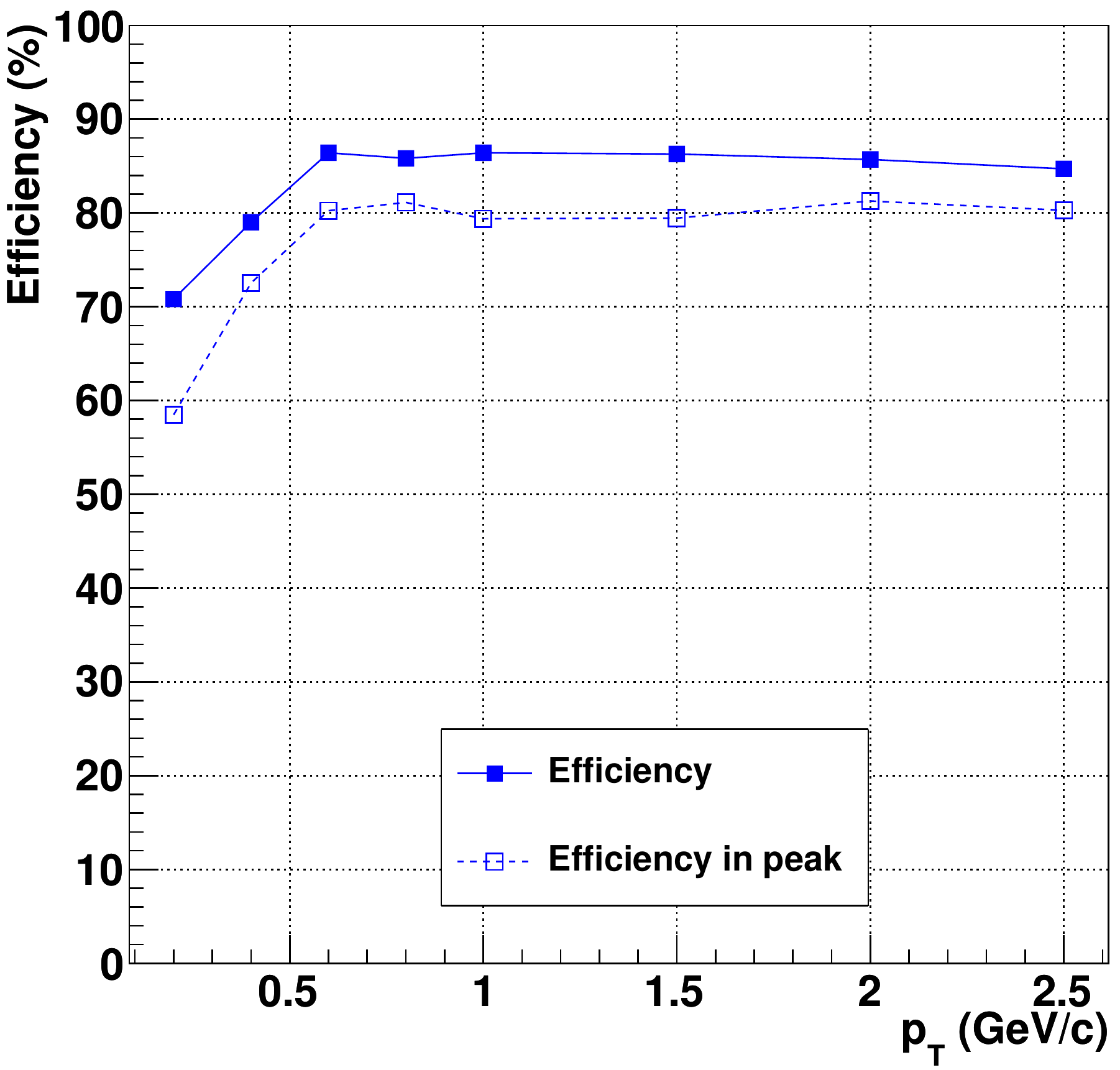}
\caption[Reconstruction efficiency vs.~$p_T$ ($mu^-$ single track events)]{Reconstruction efficiency vs.~$p_T$ for {\boldmath $\mu^-$} single track events, in the angular ranges {\boldmath $\phi \in [0^\circ, 360^\circ$} and {\boldmath $\theta \in [7^\circ, 160^\circ]$} (see \Reftbl{tab:stt:per:perf_pt}).}\label{fig:stt:per:efficiency_pt}
\end{figure}

The performances of the Straw Tube Tracker in terms of momentum resolution and 
reconstruction efficiency have been studied also through simulations of $10^4$ $\mu^-$ 
single track events generated at the interaction point I.~P., with $\phi \in [0^\circ, 360^\circ]$ 
and $\theta \in [7^\circ, 160^\circ]$.
The tracks have been generated at the following values of fixed $p_T$: 0.2, 0.4, 0.6, 0.8, 
1.0, 1.5, 2.0 and 2.5 \gevc. The momentum resolution and efficiency plots as function of the $p_t$ values
are shown in \Reffigs{fig:stt:per:resol_pt} and \ref{fig:stt:per:efficiency_pt}; the obtained
values are reported in \Reftbl{tab:stt:per:perf_pt}.
The momentum resolution is almost linear with $p_T$, as expected.
\begin{table*}[h!]
\begin{center}
\caption[Momentum resolution and reconstruction efficiency ($mu^-$ single track events)]{Momentum resolution and reconstruction
efficiency for $10^4$ {\boldmath $\mu^-$} single track events generated at {\bf fixed 
transverse momentum}.}
\label{tab:stt:per:perf_pt}
\smallskip
\begin{tabular}{c|c|c|c}
\hline\hline
$p_t$ (\gevc) & Resolution (\%)   & Efficiency (\%)	& Efficiency in peak (\%) \\ 
\hline\hline
0.2	&$1.48	\pm 0.02$	&$70.82	\pm 0.45$	&$58.48	\pm 0.49$\\
0.4	&$1.36	\pm 0.02$	&$79.01	\pm 0.41$	&$72.53	\pm 0.45$\\
0.6	&$1.48	\pm 0.02$	&$86.41	\pm 0.34$	&$80.24	\pm 0.40$\\
0.8	&$1.58	\pm 0.02$	&$85.82	\pm 0.35$	&$81.12	\pm 0.39$\\
1.0	&$1.76	\pm 0.02$	&$86.41	\pm 0.34$	&$79.38	\pm 0.40$\\
1.5	&$1.97	\pm 0.03$	&$86.28	\pm 0.34$	&$79.45	\pm 0.40$\\
2.0	&$2.25	\pm 0.02$	&$85.70	\pm 0.35$	&$81.27	\pm 0.39$\\
2.5	&$2.56	\pm 0.03$	&$84.70	\pm 0.36$	&$80.28	\pm 0.40$\\

\hline \hline
\end{tabular}
\end{center}
\end{table*}\noindent
%resolution @ various pt
%efficiency @ various pt

\subsubsection{Summary of the Results}
The performance of the STT has been investigated through the 
simulation of different sets of single track (muon) events, generated at the 
interaction point at different momentum values, polar angle $\theta$ and uniform
 azimuthal angle $\phi$.
The tracks have been fitted by applying the procedure summarised in 
\Refsec{sec:stt:sim}. The attention has then been focused on the 
momentum resolution of the generated particles and on the tracking efficiency.
In all the sets of simulations, the improvements due to the Kalman filter is 
evident, in particular in terms of momentum resolution: The mean values of the 
momentum distributions after the Kalman fit are more centered around the correct
 value than the ones obtained after the global helix fit. In addition, the 
Kalman distributions are narrower than the helix ones, resulting in better 
resolution values.
Tests with tracks generated with random $\theta$ and $\phi$ show that the 
momentum resolution ranges from $\sim 1.32$\% in case of 0.3 \gevc tracks, 
to $\sim 3.61$\% for 5 \gevc tracks (\Reffig{fig:stt:per:totalmom}, \Reftbl{tab:stt:per:totalmom}). 
\par
\begin{figure}
\centering
\includegraphics[width=\swidth, height=\swidth]{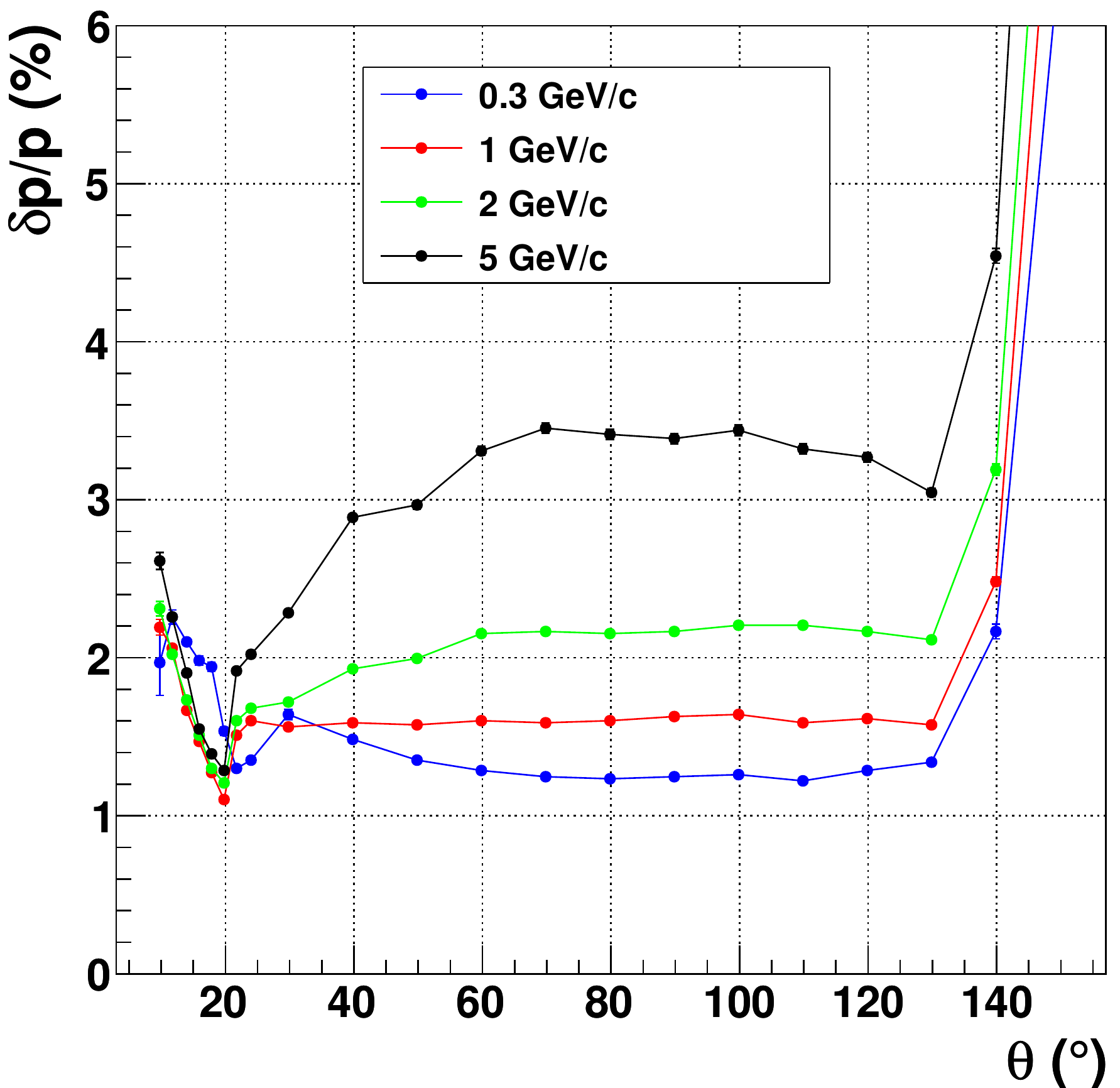}
\caption[Momentum resolution vs.~$\theta$ starting angle]{Momentum resolution vs.~$\theta$ starting angle for {\bf 0.3}, {\bf 1}, {\bf 2} and {\bf 5 \gevc}
{\boldmath $\mu^-$} single track events, in the full angular range {\boldmath $\theta \in [9^\circ, 160^\circ]$} (\Reffigs{fig:stt:per:resolution_03}, \ref{fig:stt:per:resolution_1}, \ref{fig:stt:per:resolution_2} and \ref{fig:stt:per:resolution_5}).}  \label{fig:stt:per:resol_sum}
\end{figure}

A more detailed investigation has been performed through the simulation of 
tracks scanning the whole CT angular region in fine steps.
The results are shown in \Reffigs{fig:stt:per:resolution_03}--\ref{fig:stt:per:efficiency_5} and reported in detail in \Reftbls{tab:stt:per:perf_03}--\ref{tab:stt:per:perf_5}. 
As shown also in the summary \Reffig{fig:stt:per:resol_sum}, the resolution improves up to $\sim 21^\circ$, due to the increasing number
of straw layers traversed by the tracks and to the high precision of the MVD 
and GEM hits. 
In the central angular region, the resolution is almost constant, ranging from $1.3\%$ at 0.3 \gevc, to $1.6\%$ at 1 \gevc, $2.2\%$ at 2 \gevc and $3.3\%$ at 5 \gevc.

In addition, a set of simulations at fixed values of transverse momentum has 
been performed. The obtained resolution is reported in \Reftbl{tab:stt:per:perf_pt} 
and in \Reffigs{fig:stt:per:resol_pt} and \ref{fig:stt:per:efficiency_pt}.
As shown in the plots, the resolution presents an almost linear behaviour as 
a function of the $p_T$ values, as expected.

\clearpage
% FILE: panda_tdr_stt_ben.tex
%
\section{Physics Channels Analysis}
%\COM{Author(s):P. Gianotti}
\label{sec:stt:ben}
In order to test that the proposed central straw tube tracker
fulfills the requirements of the \PANDA experiment, appropriate
physics channels have been identified to test the detector performance.
The set of channels proposed (see \Reftbl{tab:stt:ben:benchmark}) aims to test
the detector's capability to measure tracks and 
momenta of charged particles in an energy region from 100~MeV up to 15~GeV with high precision.
A special emphasis is also given to the capability to detect secondary vertices for
hadrons with $c-$ and $s-$ quark content.
\begin{table*}
  \centering
  \caption[Benchmark channels used to evaluate the performance of the central straw tube tracker]{Benchmark channels used to evaluate the performance of the central straw tube tracker.}
\smallskip
  \begin{tabular}{ll}\hline
    Channel & Final state  \\ \hline
    $\bar pp \rightarrow (n)\pi^+\pi^-$ & $(n)\pi^+\pi^-$  \\
    $\bar pp \rightarrow \psi(3770) \rightarrow D^+D^-$ & $2K\,4\pi$ \\
    $\bar pp \rightarrow \bar\Lambda\Lambda$ & $p\pi^-\bar p\pi^+$ \\
    $\bar pp \rightarrow \eta_c \rightarrow \phi\phi$ & 4K  \\
  \hline
  \end{tabular}

  \label{tab:stt:ben:benchmark}
\end{table*}
In the following sections the results of the performed data analyses are reported.
For the $\bar\Lambda\Lambda$ channel preliminary results are shown in \Refsec{sec:stt:sim:pr2} using the STT stand-alone pattern 
recognition.
A complete analysis of this channel will be possible only when the information of the forward tracking system will be included in the 
tracking code. 

\subsection{Simulation Environment}
\label{sec:stt:ben:env}
The analysis is performed within the \pandaroot framework using the EvtGen event generator for the event production,
Virtual Monte Carlo with Geant3 for the simulation, dedicated digitization and reconstruction code, and the rho analysis tool
for high level analysis. Event generation and analysis are performed on the \pandagrid. 
In the Monte Carlo simulations the primary vertex was generated according to the expected target beam interaction region, with 0.1\,cm 
size in transverse direction and distributed by a Gaussian function with FWHM\,=\,0.5\,cm along the $z$ axis. The full \PANDA geometry 
has been included in the simulation, and for the tracking, the MVD, STT and GEM detectors have been used.
An important remark here is that the analysis is performed with ideal particle identification,
i.e. for each reconstructed track its particle type is associated using the Monte Carlo information, in order to avoid possible 
bias from the detectors used for PID. At first only the reconstruction of the signal itself is considered without study of background suppression.
Final results of this study take mixing of the signal with generic background produced by the DPM event generator into account. The number 
of pile-up events is defined by the Poisson statistics. In this case complete tracks from the background events can be reconstructed as well as 
single hits from the background events can contribute to the tracks from the event of interest.

\subsection{$\bar pp \rightarrow (n)\pi^+\pi^-$ }
%\COM{Author(s):E. Fioravanti}
\label{sec:stt:ben:multipi}
In the $\bar p p$ annihilation process charged pions are the most abundant particles
produced.
Therefore, $\bar pp \rightarrow (n)\pi^+\pi^-$, with $n =$ 2, 4, are the basic channels to test the STT performance.
At an energy of 3.07~GeV in the center of mass system (CMS), the cross section of the channel
$\bar pp \rightarrow \pi^+\pi^-$ is $\sigma$~=~0.007~mb while
at a CMS energy of 2.954~GeV the cross section of the $\bar pp \rightarrow \pi^+\pi^-\pi^+\pi^-$ final state is $\sigma$~=~0.43~mb
\cite{bib:stt:ben:cs}.
The interesting figures of merit for these benchmark channels are:
\begin{itemize}
\item single pion track resolution,
\item momentum and invariant mass resolutions,
\item vertex resolution,
\item reconstruction efficiency.
\end{itemize}
The benchmark channel is simulated at a CMS energy of 3.07 GeV corresponding
to an antiproton beam momentum along the z direction of 4.0~\gevc.

\subsubsection{$\bar pp \rightarrow \pi^+\pi^-$}
\begin{figure*}%[ht!]
\begin{center}
\includegraphics[width=\swidth]{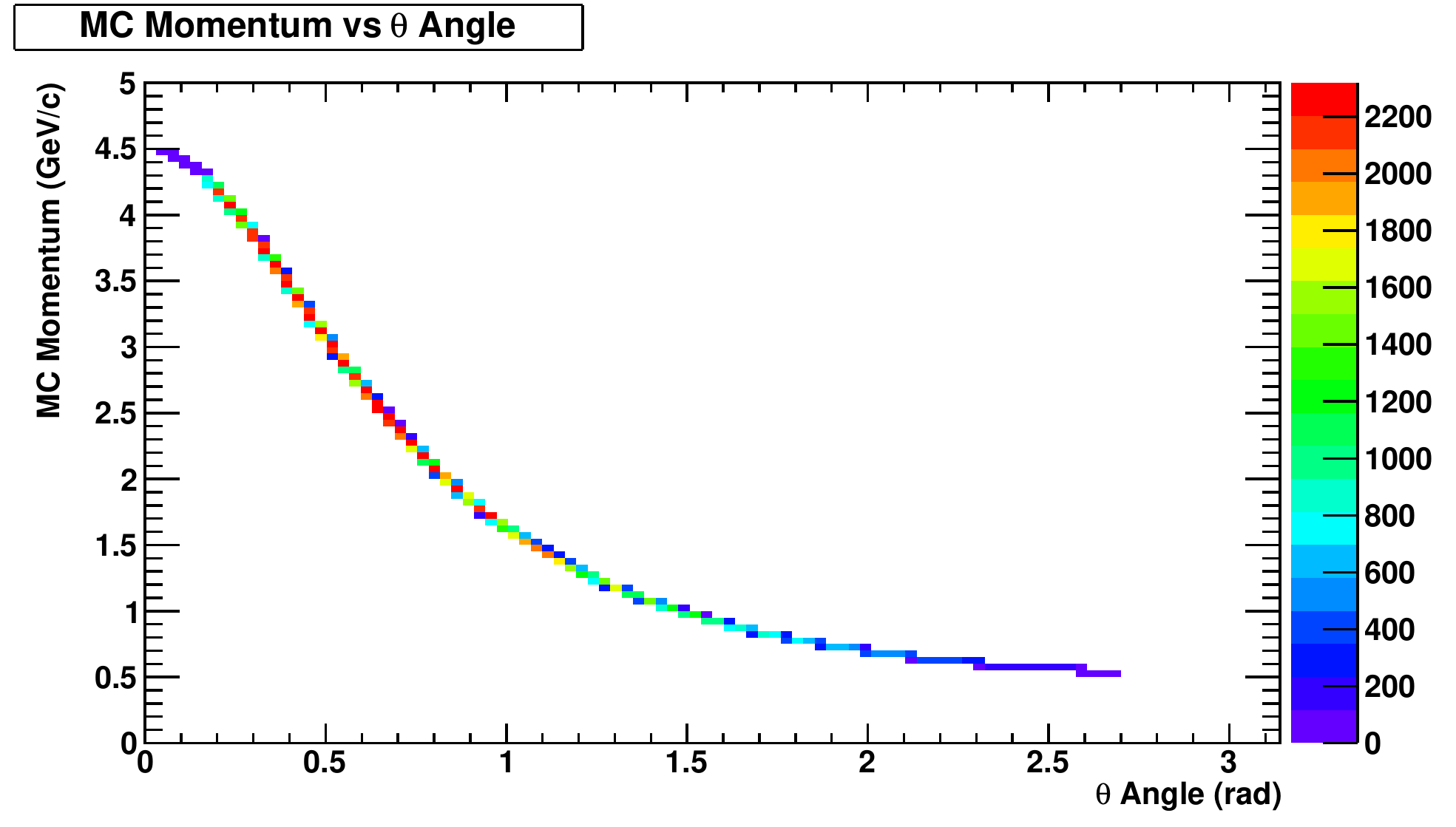}
\includegraphics[width=\swidth]{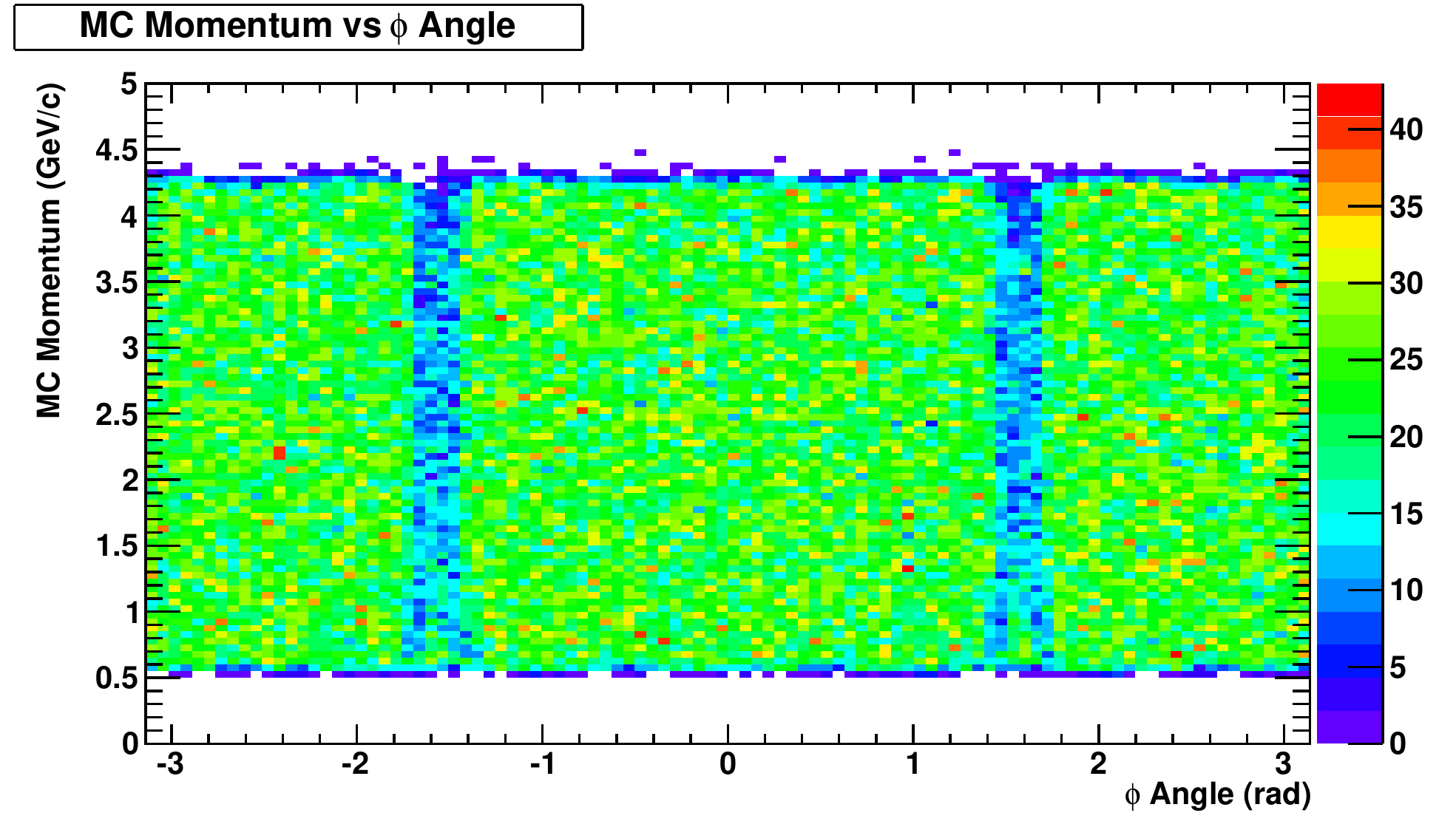}
\caption[$\bar{p}p\rightarrow\pi^+\pi^-$: Pion momentum distributions vs $\theta$ and $\phi$ angles]
{$\bar{p}p\rightarrow\pi^+\pi^-$: Pion momentum distributions vs $\theta$ (left) and $\phi$ (right) angles. }
\label{fig:stt:ben:momdisangle}
\end{center}
\end{figure*}

\Reffig{fig:stt:ben:momdisangle} shows the distribution
of the pion's momentum as a function of $\theta$ and $\phi$ angles.
The majority of the pions has a momentum between 1 \gevc and 4 \gevc and they are 
found within a polar angular range between 0.4 rad and 1.1 rad. 
\par
In the first step of the analysis we require that all reconstructed track candidates have at least one STT hit.
Events with 2.07 \gevcc $< m(\pi^+\pi^-)<$4.07 \gevcc are selected, then a vertex fit is performed and the best candidate in 
each event is selected using the minimal $\chi^2$ criterion.
\par
\begin{figure}%[h!]
\begin{center}
\includegraphics[width=\swidth]{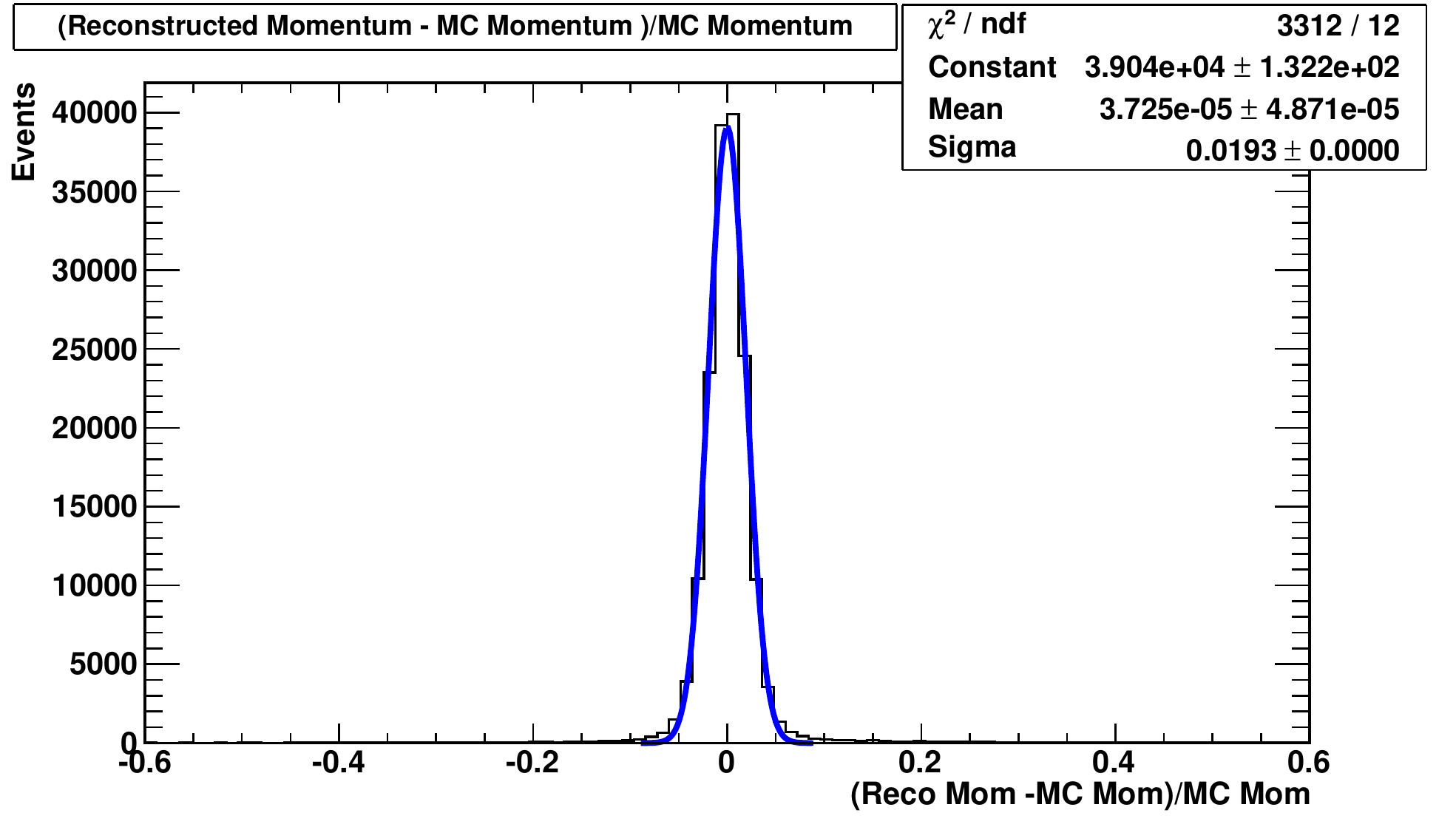}
\caption[$\bar{p}p\rightarrow\pi^+\pi^-$: Momentum resolution of the pion tracks]
{$\bar{p}p\rightarrow\pi^+\pi^-$: Momentum resolution of the pion tracks, without event mixing. The fit is done with a Gaussian function (see text for more details).}
\label{fig:stt:ben:momres2pi}
\end{center}
\end{figure}

\Reffig{fig:stt:ben:momres2pi} shows the difference between the reconstructed and the Monte Carlo generated momentum divided 
by the Monte Carlo one for the pion tracks. The distribution is fitted with a Gaussian function in order to extract the 
single pion track resolution which is 1.9\,\%.

\begin{figure*}%[ht!]
\begin{center}
\includegraphics[width=\dwidth]{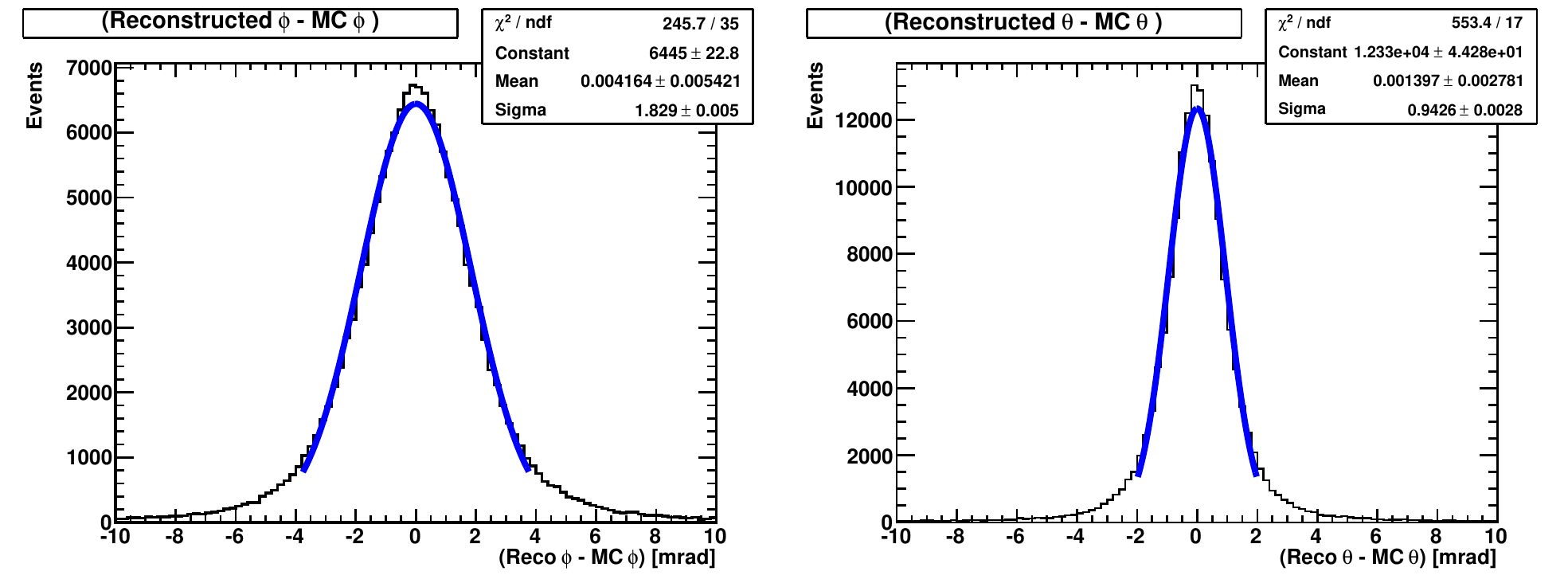}
\caption[$\bar{p}p\rightarrow\pi^+\pi^-$: Difference between the reconstructed and the Monte Carlo azimuthal and polar angle]
{$\bar{p}p\rightarrow\pi^+\pi^-$: Difference between the reconstructed and the Monte Carlo azimuthal angle (left) and polar angle 
(right), without event mixing. The fit is done with a Gaussian function (see text for more details).}
\label{fig:stt:ben:angleresolution2pi}
\end{center}
\end{figure*}

\Reffig{fig:stt:ben:angleresolution2pi} shows the distribution of the difference between the reconstructed azimuthal (polar) 
angle and Monte Carlo azimuthal (polar) angle of the single pion track. The distributions are fitted with a Gaussian function 
in order to extract the resolution of the two angles, which are 1.829~mrad for the azimuthal angle and 0.943~mrad for the polar angle.

\begin{figure}%[h!]
\begin{center}
\includegraphics[width=\swidth]{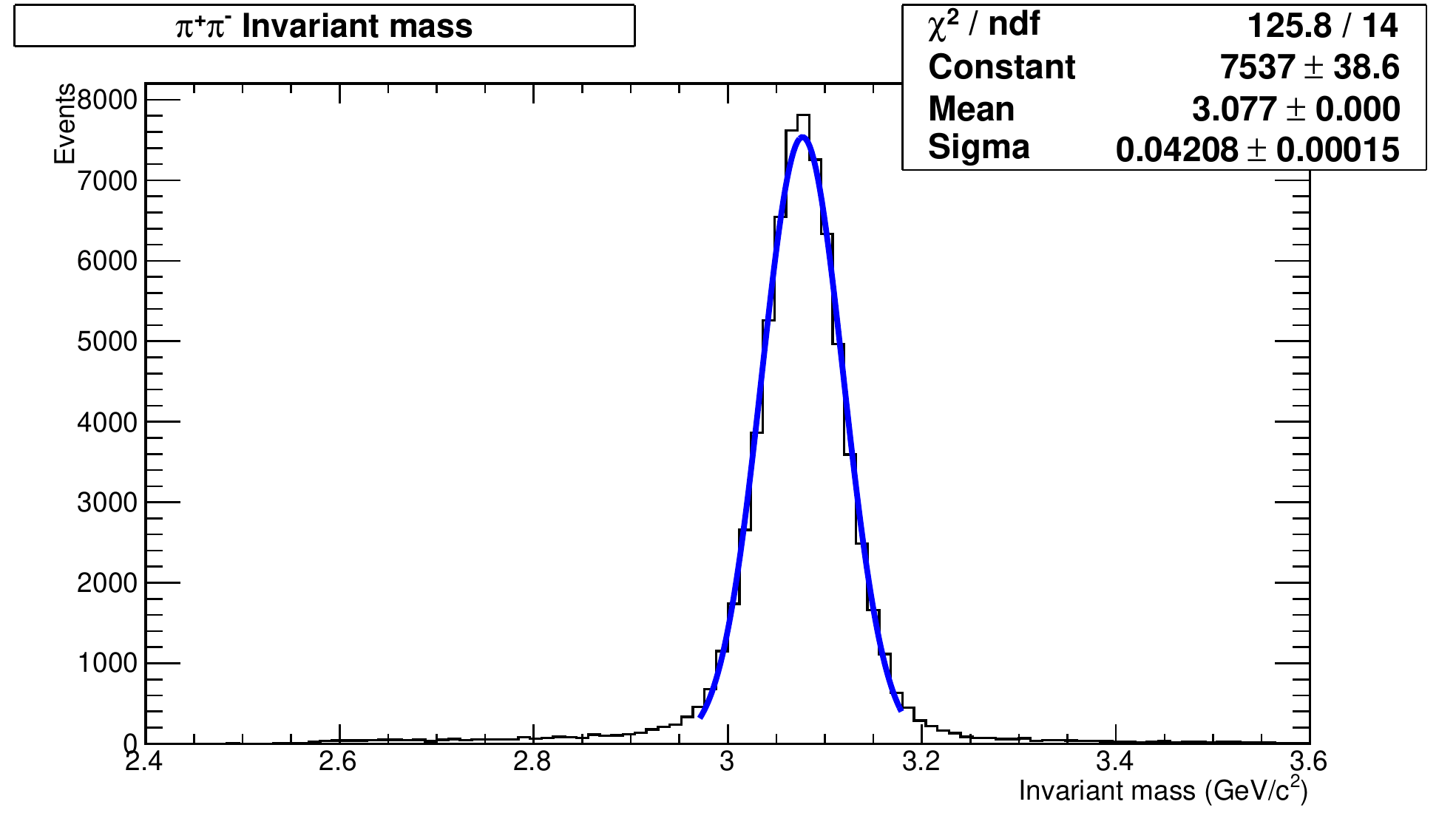}
\caption[$\bar{p}p\rightarrow\pi^+\pi^-$: Pions invariant mass distribution]
{$\bar{p}p\rightarrow\pi^+\pi^-$: Pions invariant mass distribution, without event mixing. The fit is done with a Gaussian function 
(see text for more details).}
\label{fig:stt:ben:invariantmass2pi}
\end{center}
\end{figure}

The two reconstructed pions are combined in order to reconstruct their invariant mass. The result is shown in 
\Reffig{fig:stt:ben:invariantmass2pi}; the distribution is fitted with a Gaussian function from which the estimation of the resolution is 
42~\mevcc and the global reconstruction efficiency, calculated as the ratio between the number of reconstructed events divided by the number 
of generated ones, is (70.9$\pm$0.3)\% and it comprises both reconstruction efficiency and geometrical acceptance.

\begin{figure*}[ht!]
\begin{center}
\includegraphics[width=0.32\dwidth, height=0.22\dwidth]{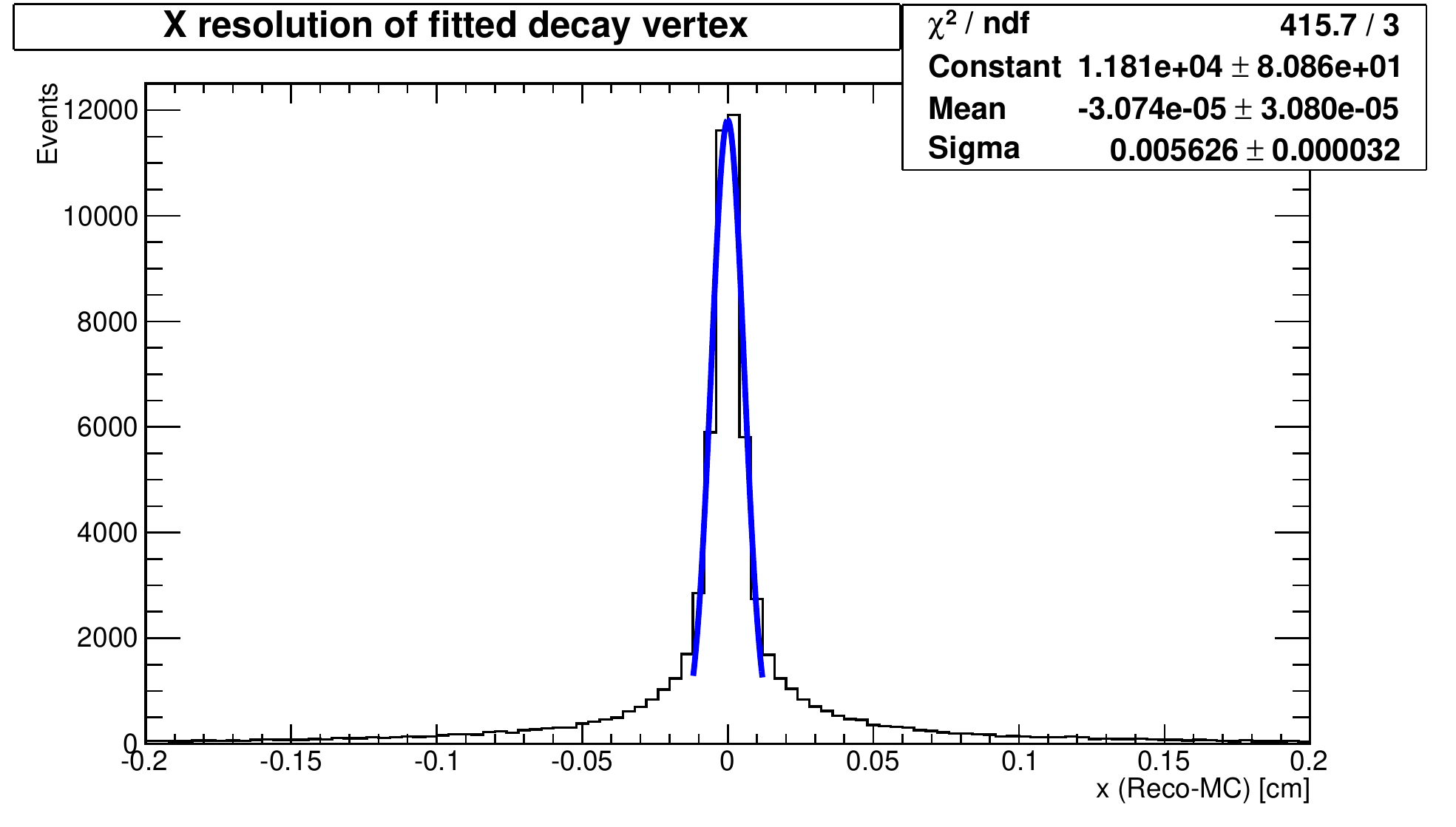}
\includegraphics[width=0.32\dwidth, height=0.22\dwidth]{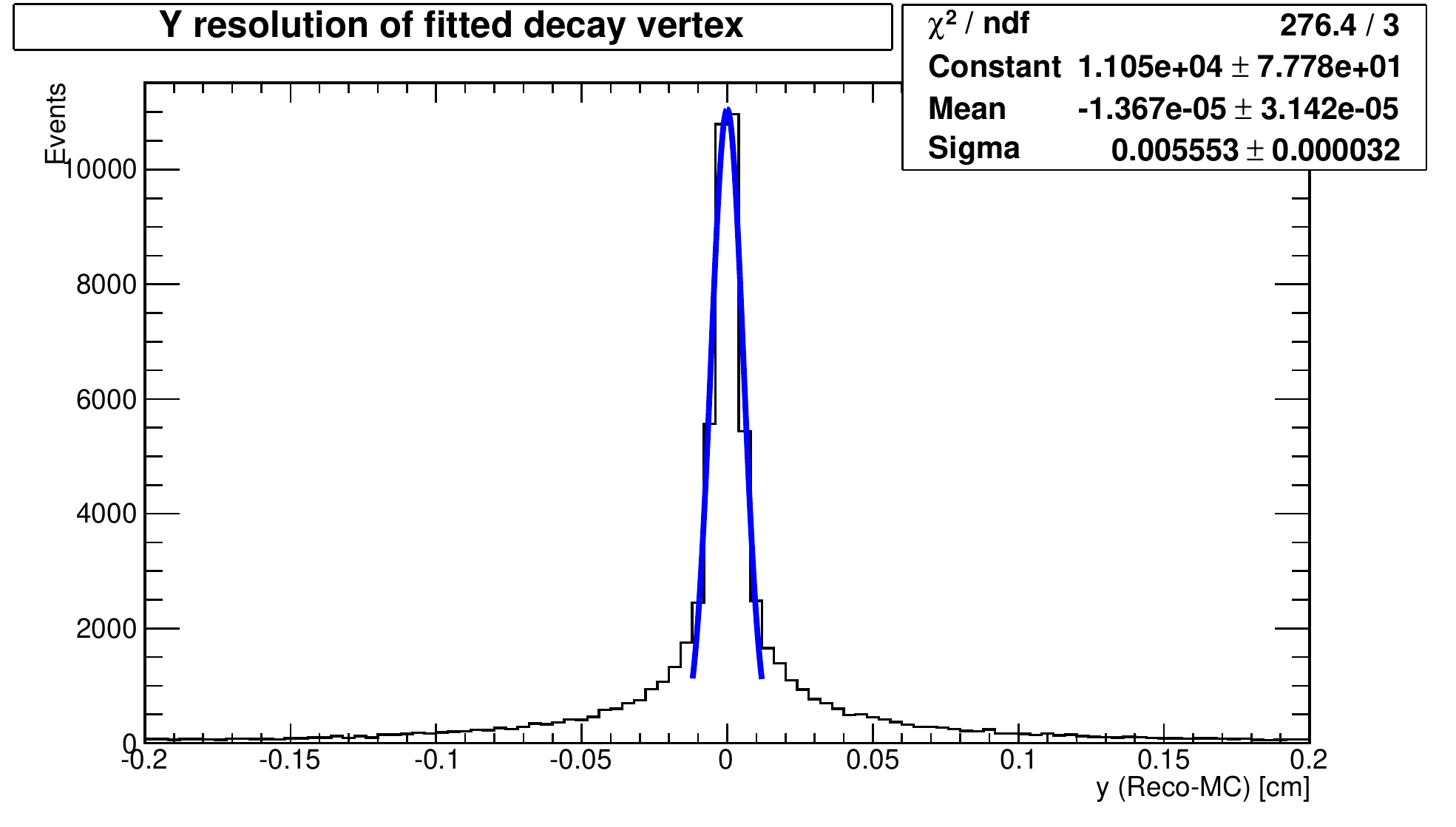}
\includegraphics[width=0.32\dwidth, height=0.22\dwidth]{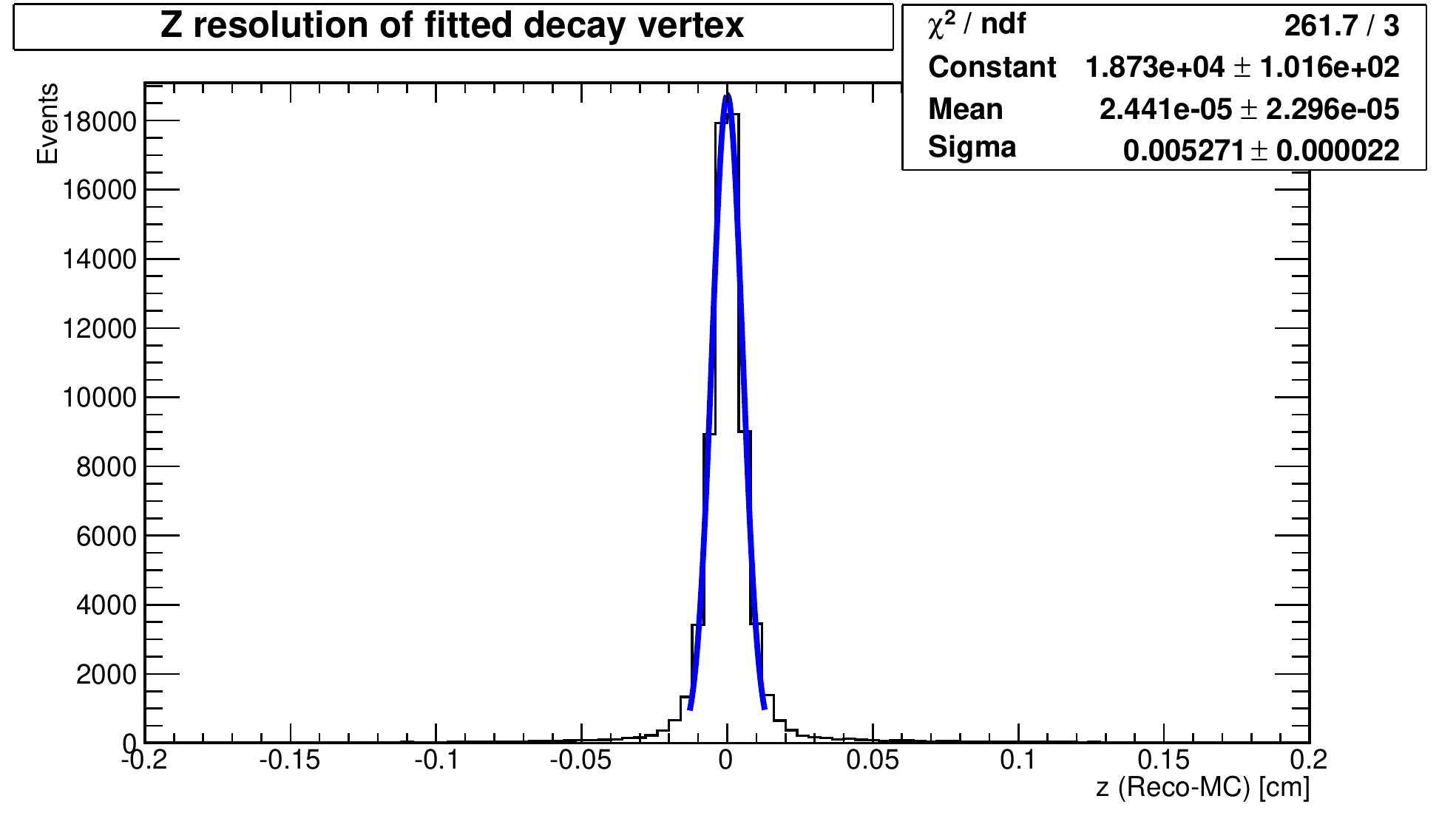}
\caption[$\bar{p}p\rightarrow\pi^+\pi^-$: Vertex resolution]{$\bar{p}p\rightarrow\pi^+\pi^-$: Vertex resolution, without event mixing (see text for more details). }
\label{fig:stt:ben:vertexFit2pi}
\end{center}
\end{figure*}

A vertex fit has been performed during the reconstruction of the final state, and the best candidate in each event has been selected by a minimal 
$\chi^2$ criterion. \Reffig{fig:stt:ben:vertexFit2pi} shows the resolution in x, y and z coordinates of the fitted decay vertex  
(e.g. difference between reconstructed vertex and Monte Carlo truth vertex position). The distributions are fitted with the Gaussian 
function in order to extract the resolutions which are: $\sigma_x$=56 $\mu$m, $\sigma_y$=56 $\mu$m and $\sigma_z$=53 $\mu$m.

\begin{figure}%[h!]
\begin{center}
\includegraphics[width=\swidth]{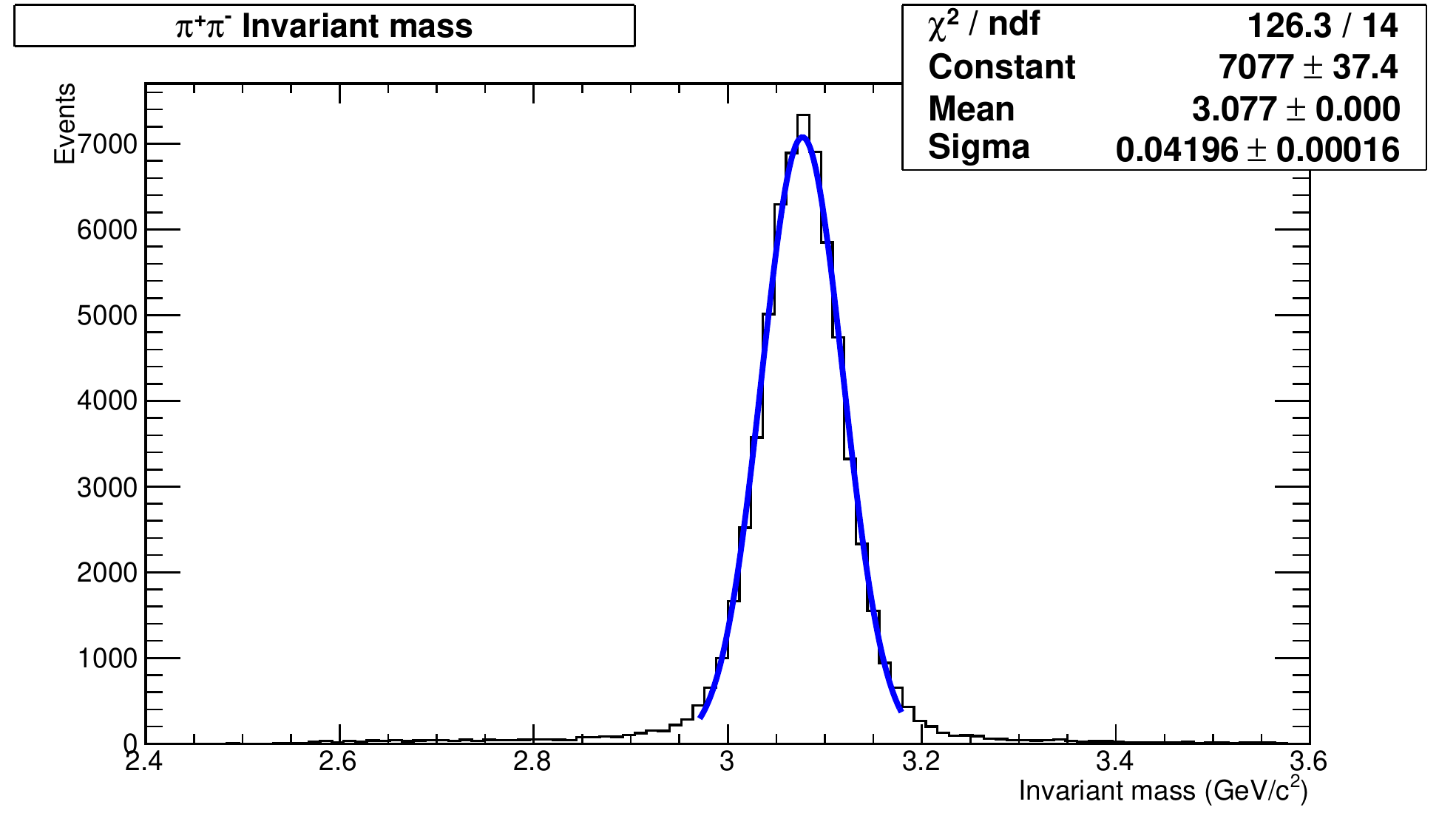}
\caption[$\bar{p}p\rightarrow\pi^+\pi^-$: Pions invariant mass distribution (after clean-up)]
{$\bar{p}p\rightarrow\pi^+\pi^-$: Pions invariant mass distribution after clean-up procedure. The fit is done with a Gaussian 
function (see text for more details).}
\label{fig:stt:ben:invmass2PiCU}
\end{center}
\end{figure}

For the pattern recognition in the presence of pile-up from the mixed background events, the clean-up procedure is applied to remove spurious hits.  
\Reffig{fig:stt:ben:invmass2PiCU} shows the two pions invariant mass distribution after the clean-up procedure; the global reconstruction 
efficiency is (65.9$\pm$0.3)\% and the resolution is 42 \mevcc. The single pion track resolution after the clean-up procedure is again 1.9\,\%.

\begin{figure}%[h!]
\begin{center}
\includegraphics[width=\swidth]{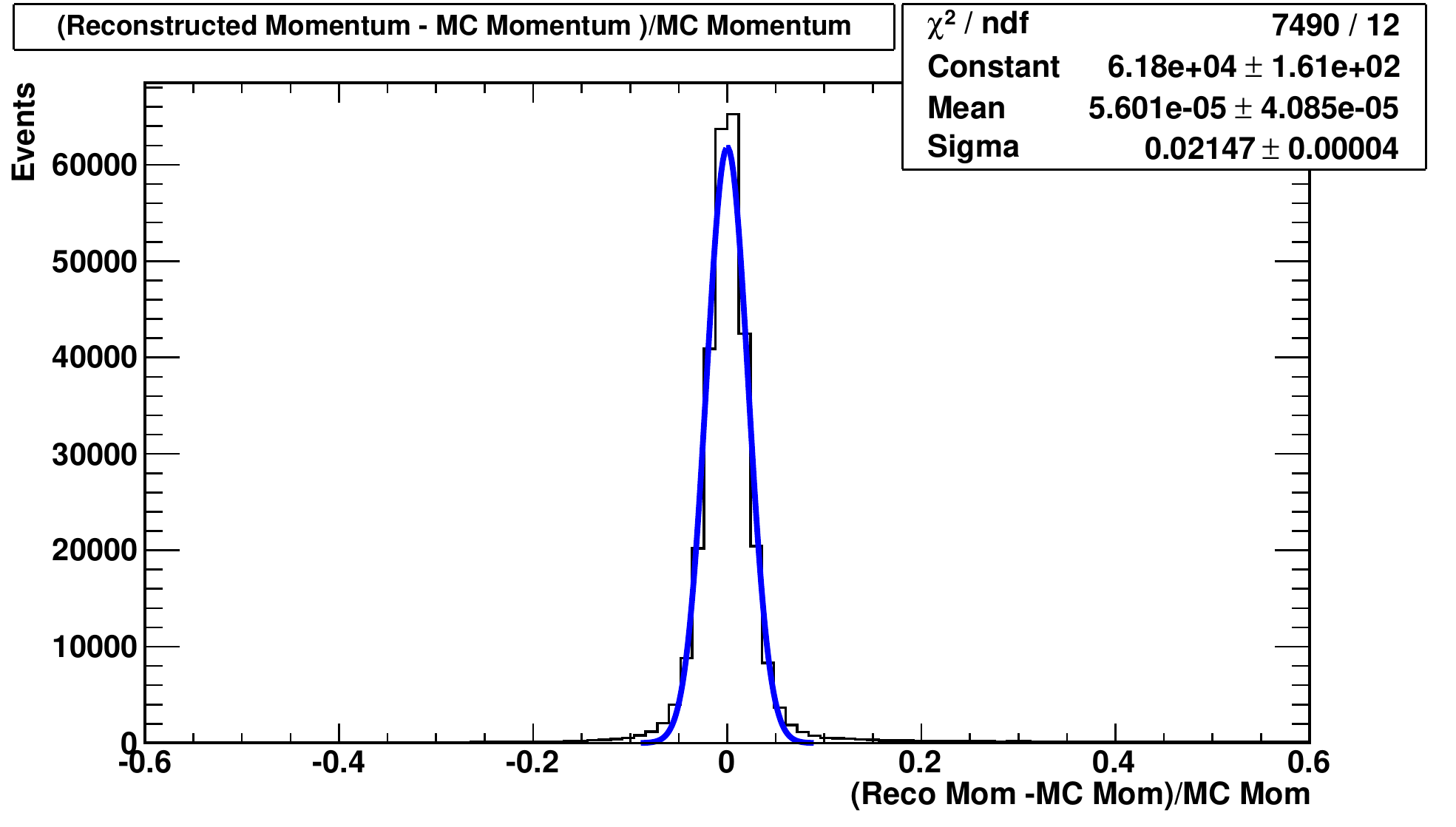}
\caption[$\bar{p}p\rightarrow\pi^+\pi^-$: Momentum resolution of the pion tracks (with event mixing)]{$\bar{p}p\rightarrow\pi^+\pi^-$: Momentum resolution of the pion tracks with event mixing. The fit is done with a Gaussian function 
(see text for more details).}
\label{fig:stt:ben:singlepion2piEMX}
\end{center}
\end{figure}

\begin{figure}%[h!]
\begin{center}
\includegraphics[width=\swidth]{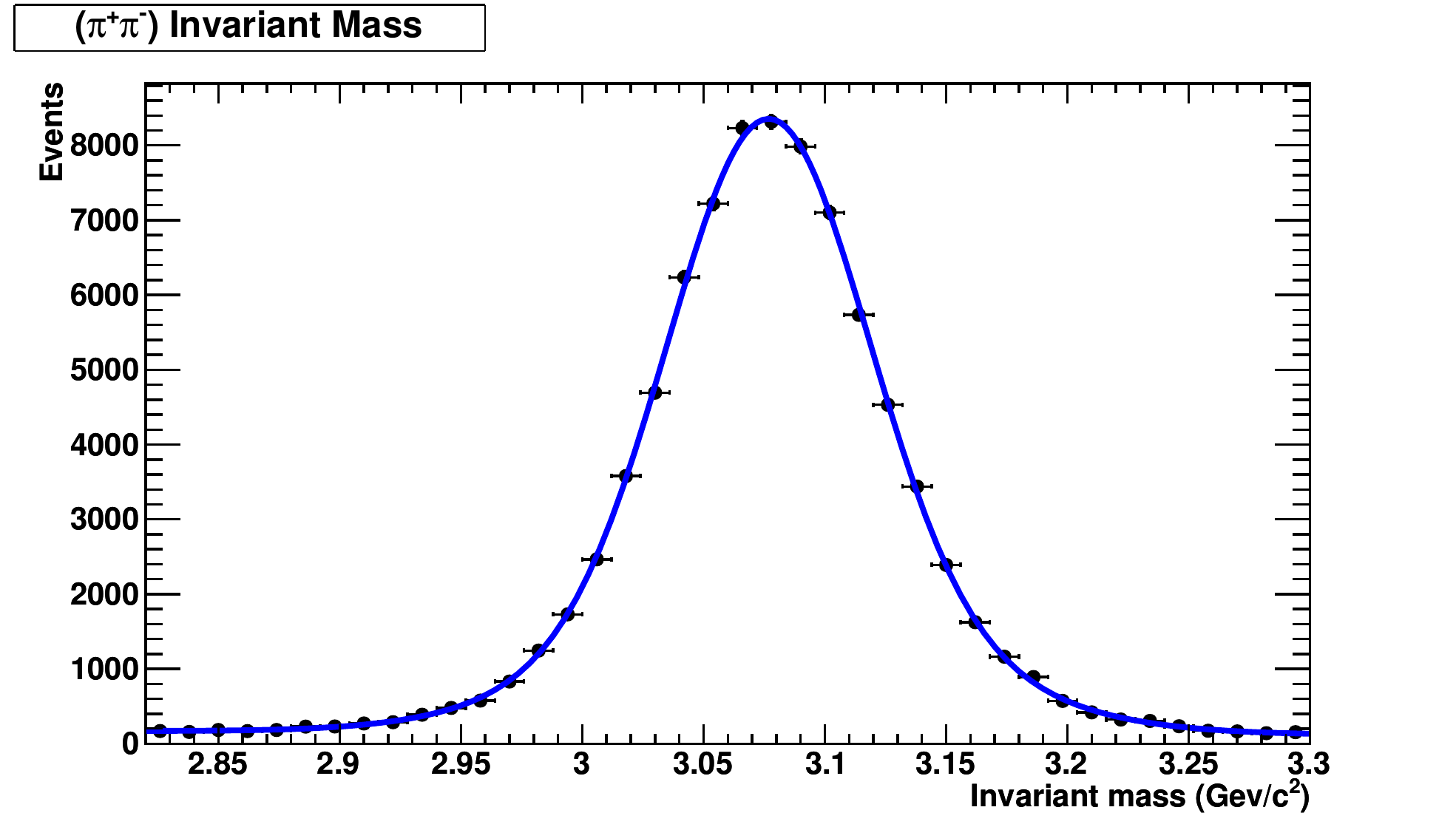}
\caption[$\bar{p}p\rightarrow\pi^+\pi^-$: Pions invariant mass distribution (with event mixing)]{$\bar{p}p\rightarrow\pi^+\pi^-$: Pions invariant mass distribution with event mixing. The fit is done with a double Gaussian function 
plus a polynomial. (see text for more details).}
\label{fig:stt:ben:invariantmass2piEMX}
\end{center}
\end{figure}

Taking mixing of the signal with generic background into account, \Reffig{fig:stt:ben:singlepion2piEMX} shows the difference between the 
reconstructed and the Monte Carlo generated momentum divided by the Monte Carlo one for the pion tracks. The single pion track resolution 
obtained from the Gaussian fit is 2.1\,\%. \Reffig{fig:stt:ben:invariantmass2piEMX} shows the two pions invariant mass distribution which is 
fitted with a double Gaussian function plus a polynomial to take the background into account. From the fit, the global reconstruction 
efficiency is (50.6$\pm$0.2)\% and the resolution is 47 \mevcc.

\begin{figure}%[h!]
\begin{center}
\includegraphics[width=\swidth]{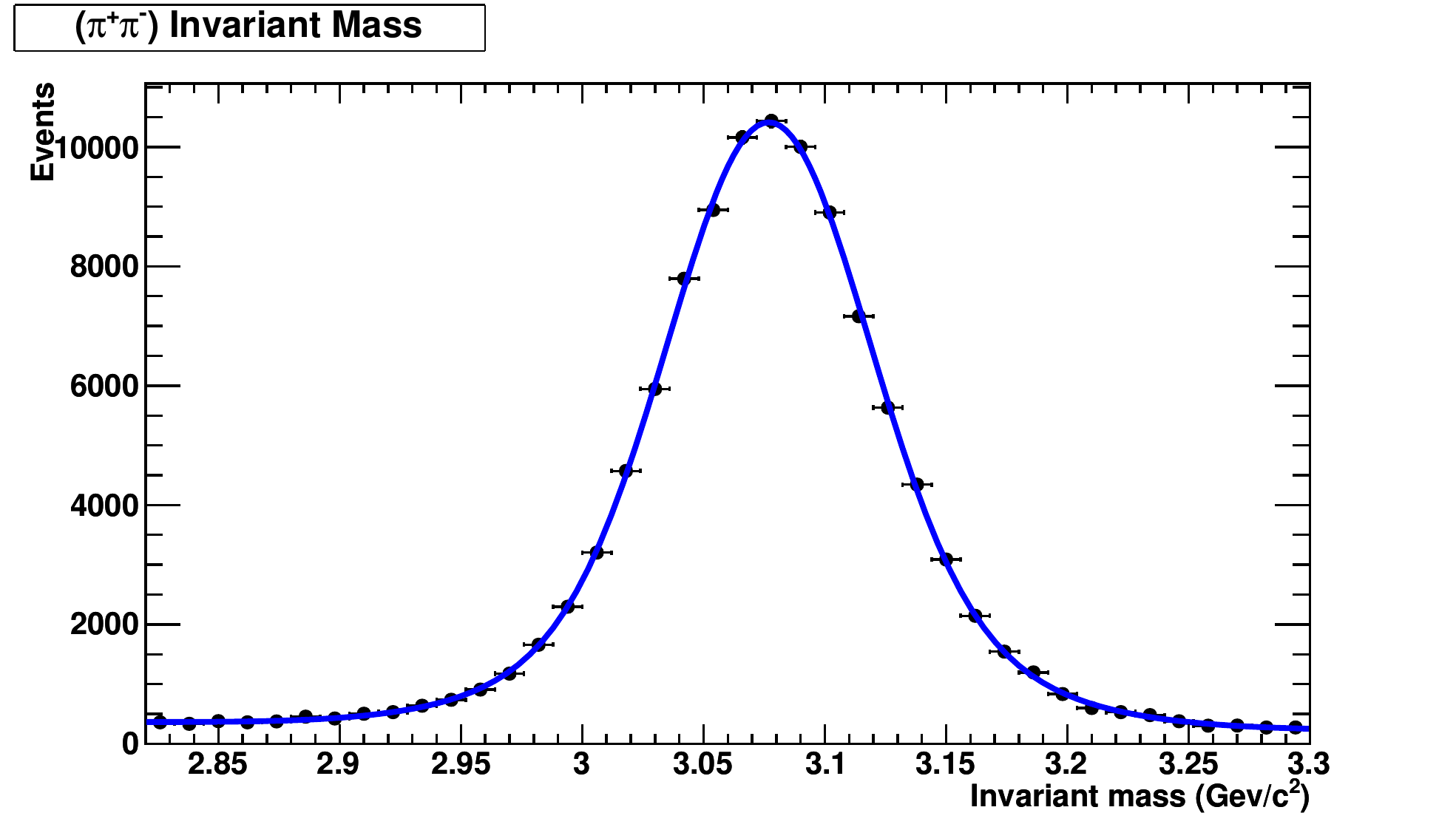}
\caption[$\bar{p}p\rightarrow\pi^+\pi^-$: Pions invariant mass distribution (with event mixing, no Monte Carlo truth PID)]{$\bar{p}p\rightarrow\pi^+\pi^-$: Pions invariant mass distribution with event mixing and without Monte Carlo truth PID. 
The fit is done with a double Gaussian function plus a polynomial. (see text for more details).}
\label{fig:stt:ben:invariantmass2piEMXNOPID}
\end{center}
\end{figure}
An additional study is to check to which extent the Monte Carlo based PID is relevant for this benchmark channel. So the two pions are reconstructed 
without any PID. The invariant mass is shown in \Reffig{fig:stt:ben:invariantmass2piEMXNOPID} and it looks unaffected. The 
two pions reconstruction efficiency is (49.0$\pm$0.2)\%; the resolution is 48 \mevcc.

\subsubsection{$\bar pp \rightarrow \pi^+\pi^-\pi^+\pi^-$}
\begin{figure*}%[ht!]
\begin{center}
\includegraphics[width=\swidth, height=5cm]{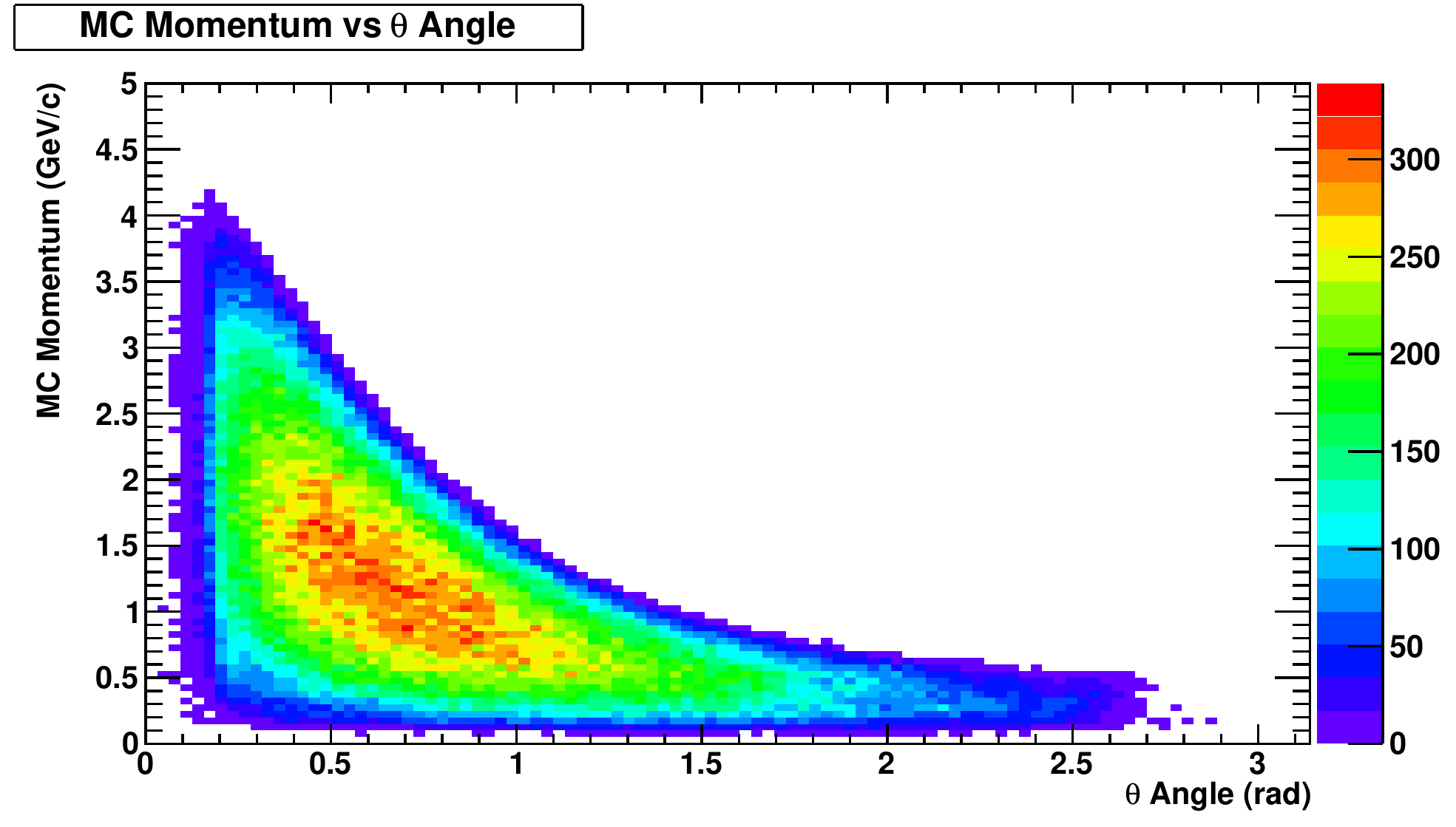}
\includegraphics[width=\swidth, height=5cm]{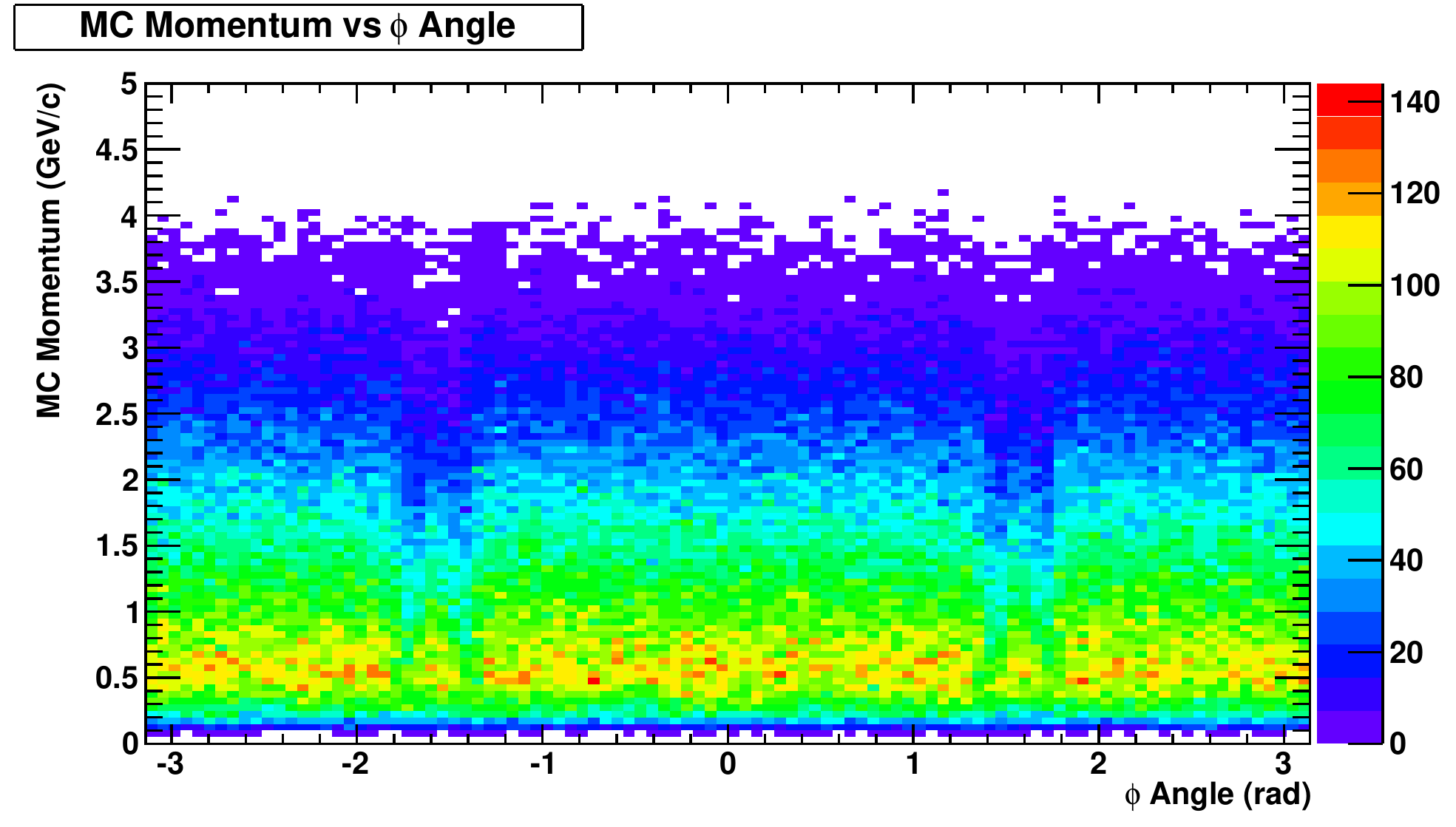}
\caption[$\bar{p}p\rightarrow 2(\pi^+\pi^-)$: Pion momentum distributions vs $\theta$ and $\phi$ angles]
{$\bar{p}p\rightarrow 2(\pi^+\pi^-)$: Pion momentum distributions vs $\theta$ (left) and $\phi$ (right) angles. }
\label{fig:stt:ben:momdisangle4Pi}
\end{center}
\end{figure*}
\Reffig{fig:stt:ben:momdisangle4Pi} shows the distribution
of the pion's momentum as a function of $\theta$ and $\phi$ angles.

The majority of the pions has a momentum between 0.5~\gevc and 2.5~\gevc and they are 
found within a polar angular range between 0.4~rad and 1.1~rad. 
In the first step of the analysis we require that all reconstructed track 
candidates have at least one STT hit.
Events with 2.57~\gevcc $< m(\pi^+\pi^-)<$3.57 \gevcc are selected, then
a vertex fit is performed and the best candidate in each event is selected
using the minimal $\chi^2$ criterion.

\begin{figure}[!h]
\begin{center}
\includegraphics[width=\swidth, height=7.5cm]{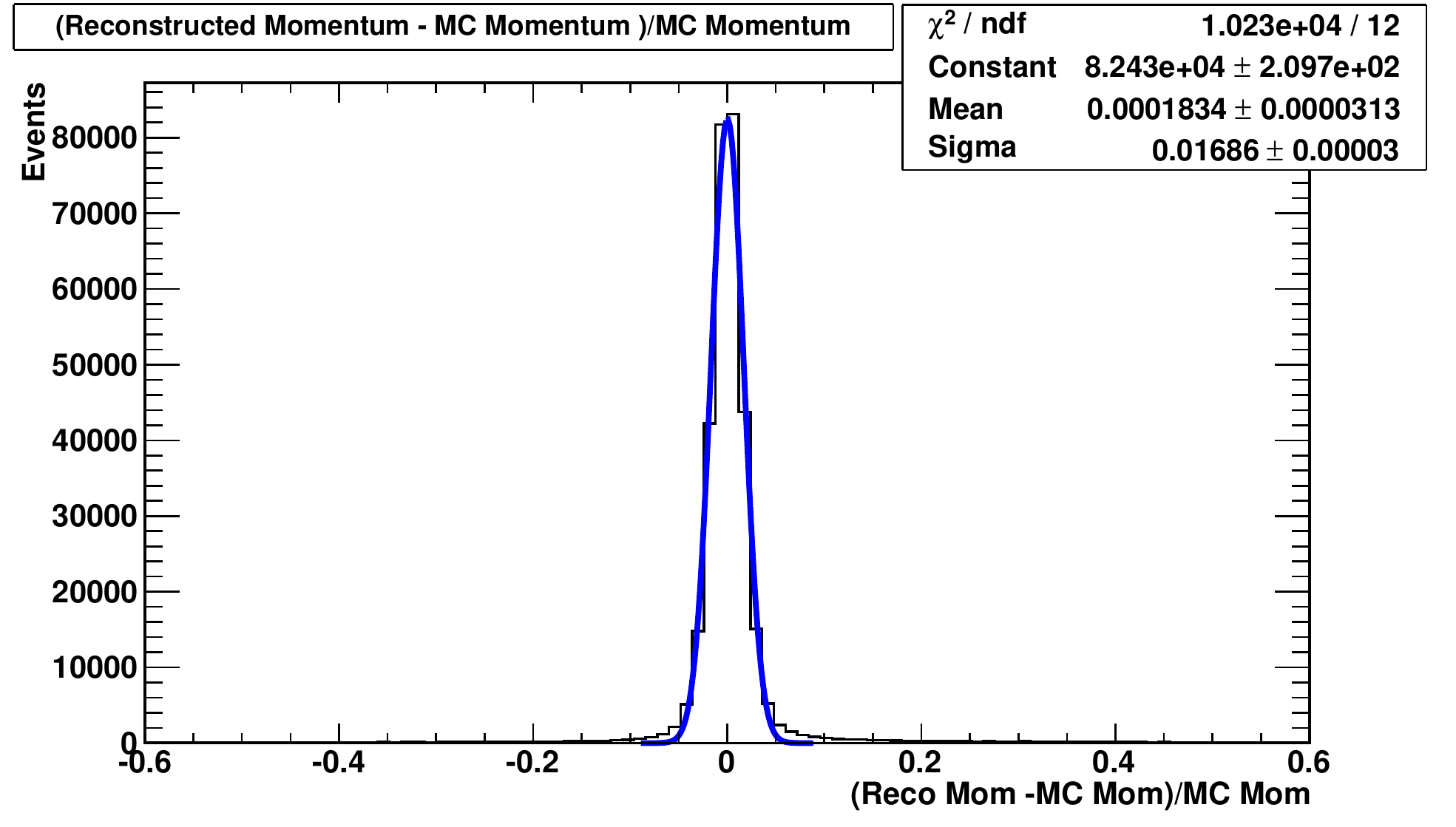}
\caption[$\bar{p}p\rightarrow 2(\pi^+\pi^-)$: Momentum resolution of the pion tracks]
{$\bar{p}p\rightarrow 2(\pi^+\pi^-)$:  Momentum resolution of the pion tracks, 
without event mixing. The fit is done with a Gaussian function (see text for more details).}
\label{fig:stt:ben:momres4pi}
\end{center}
\end{figure}

\Reffig{fig:stt:ben:momres4pi} shows the difference between the reconstructed and the Monte Carlo
generated momentum divided by the Monte Carlo one for the pion tracks. The distribution
is fitted with a Gaussian function in order to extract the single pion track resolution which
is 1.7\,\%.

\begin{figure*}%[ht!]
\begin{center}
\includegraphics[width=\dwidth, height=5cm]{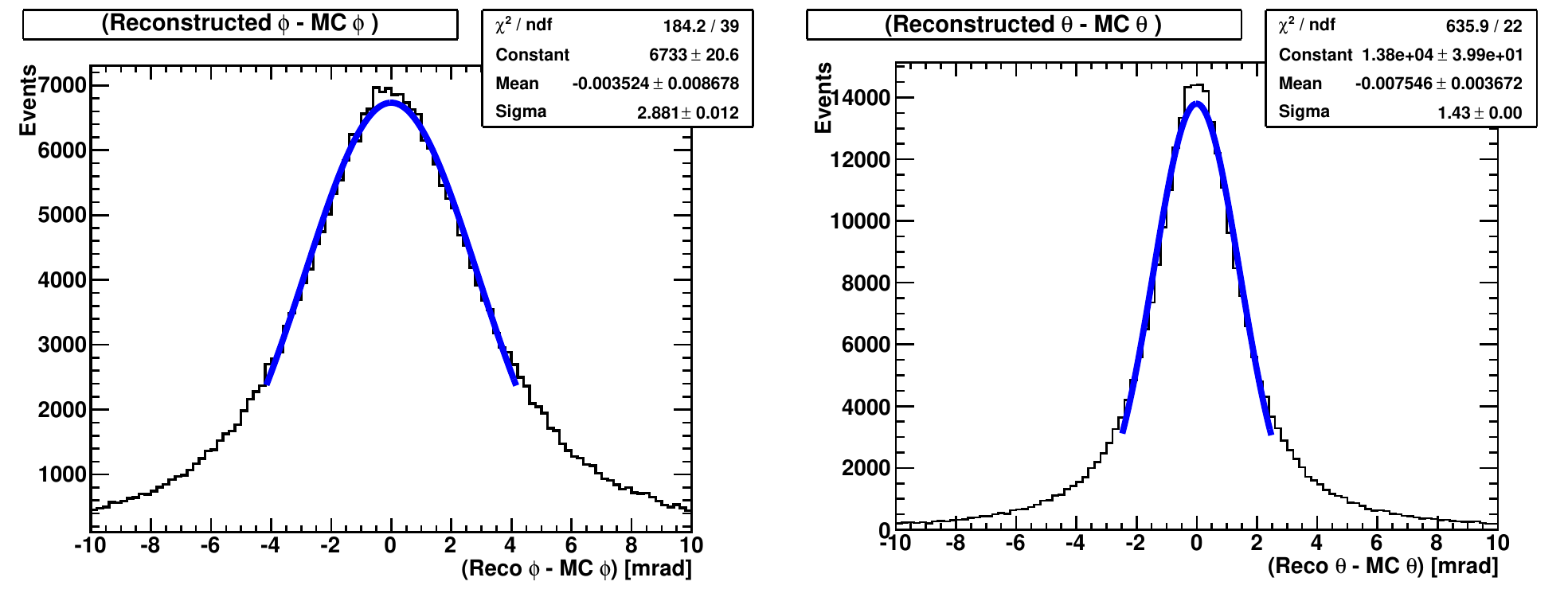}
\caption[$\bar{p}p\rightarrow 2(\pi^+\pi^-)$: Difference between the reconstructed and the Monte Carlo azimuthal and polar angles]{$\bar{p}p\rightarrow 2(\pi^+\pi^-)$: Difference between the reconstructed and the Monte Carlo azimuthal angle (left) and polar 
angle (right), without event mixing. The fit is done with a Gaussian function (see text for more details).}
\label{fig:stt:ben:angleresolution4pi}
\end{center}
\end{figure*}

\Reffig{fig:stt:ben:angleresolution4pi} shows the distribution of the difference between the reconstructed azimuthal (polar) angle 
and Monte Carlo azimuthal (polar) angle of the single pion track. The distributions are fitted with a Gaussian function in order 
to extract the resolution of the two angles, which are 2.881~mrad for the azimuthal angle and 1.430~mrad for the polar angle.

\begin{figure}
\begin{center}
\includegraphics[width=\swidth, height=7.5cm]{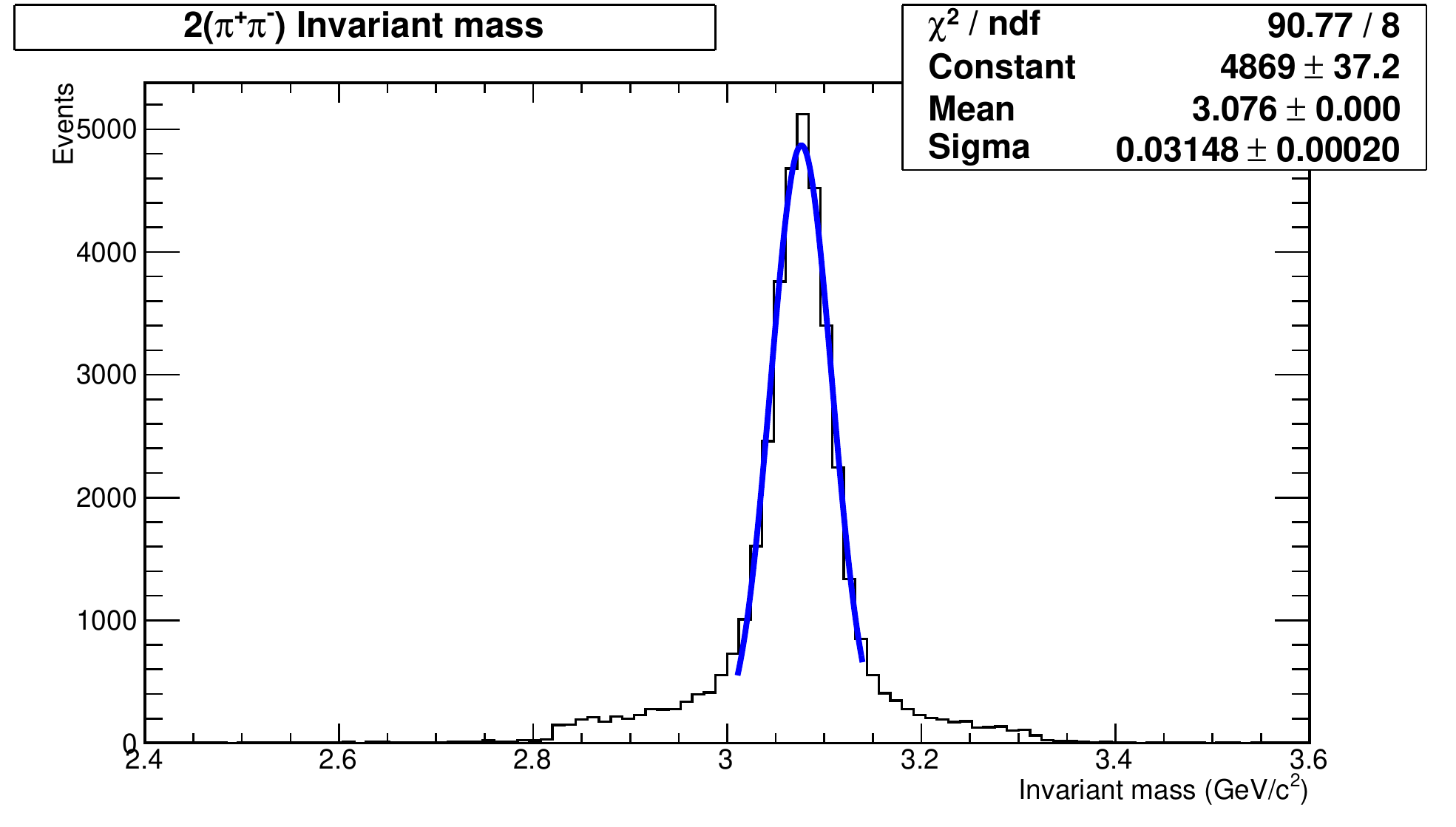}
\caption[$\bar{p}p\rightarrow 2(\pi^+\pi^-)$: Pions invariant mass distribution]
{$\bar{p}p\rightarrow 2(\pi^+\pi^-)$: Pions invariant mass distribution, without event mixing. The fit is done with a Gaussian 
function (see text for more details).}
\label{fig:stt:ben:invariantmass4pi}
\end{center}
\end{figure}

The four reconstructed pions are combined in order to reconstruct their invariant mass. The result is shown in 
\Reffig{fig:stt:ben:invariantmass4pi}; the distribution is fitted with a Gaussian function from which the estimation of the resolution is 
31 \mevcc and the global reconstruction efficiency, calculated as the ratio between the number of reconstructed events divided by the number 
of generated ones, is (43.1$\pm$0.2)\%, and it comprises both reconstruction efficiency and geometrical acceptance.

\begin{figure*}[ht!]
\begin{center}
\includegraphics[width=0.32\dwidth, height=0.22\dwidth]{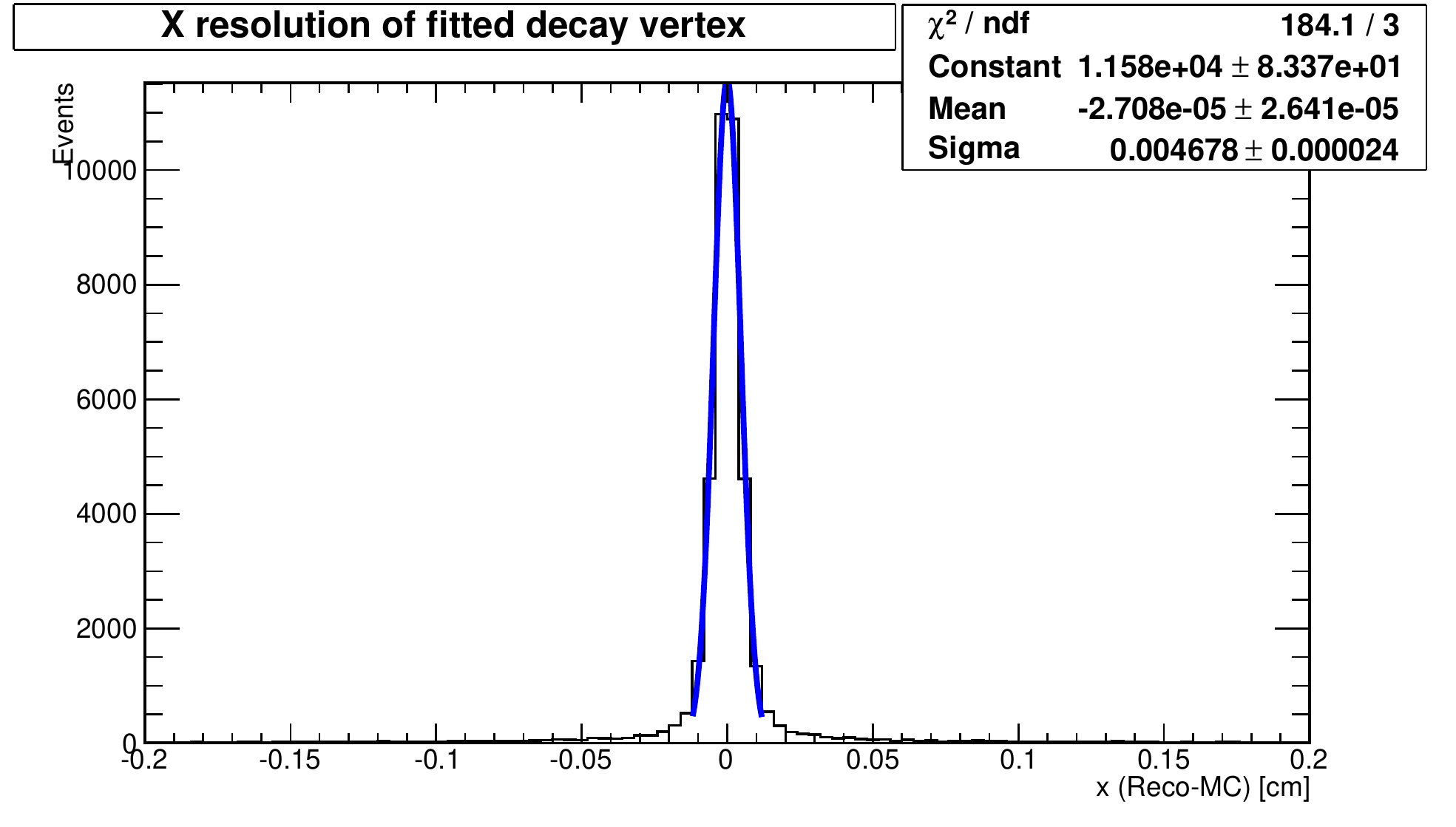}
\includegraphics[width=0.32\dwidth, height=0.22\dwidth]{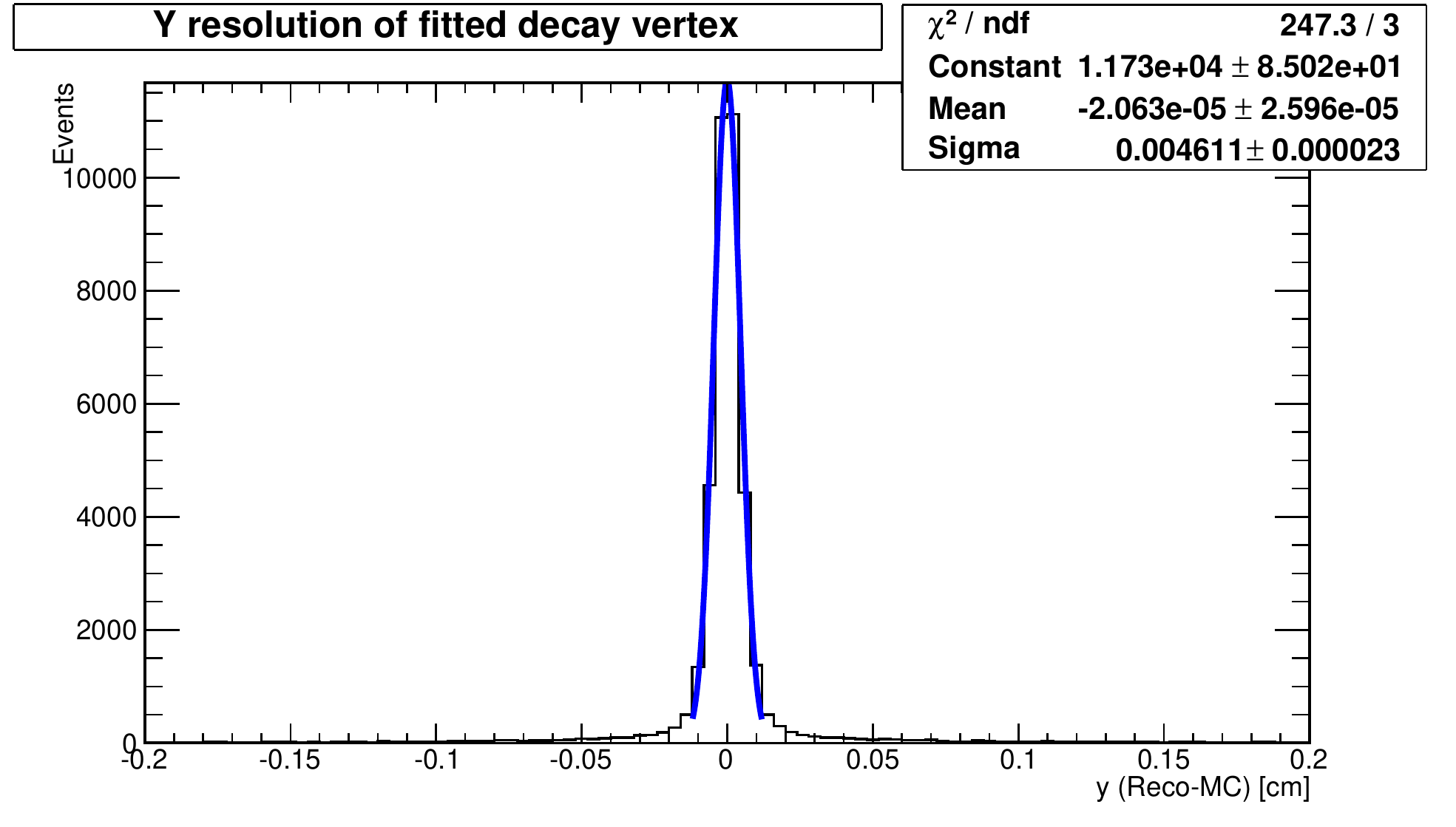}
\includegraphics[width=0.32\dwidth, height=0.22\dwidth]{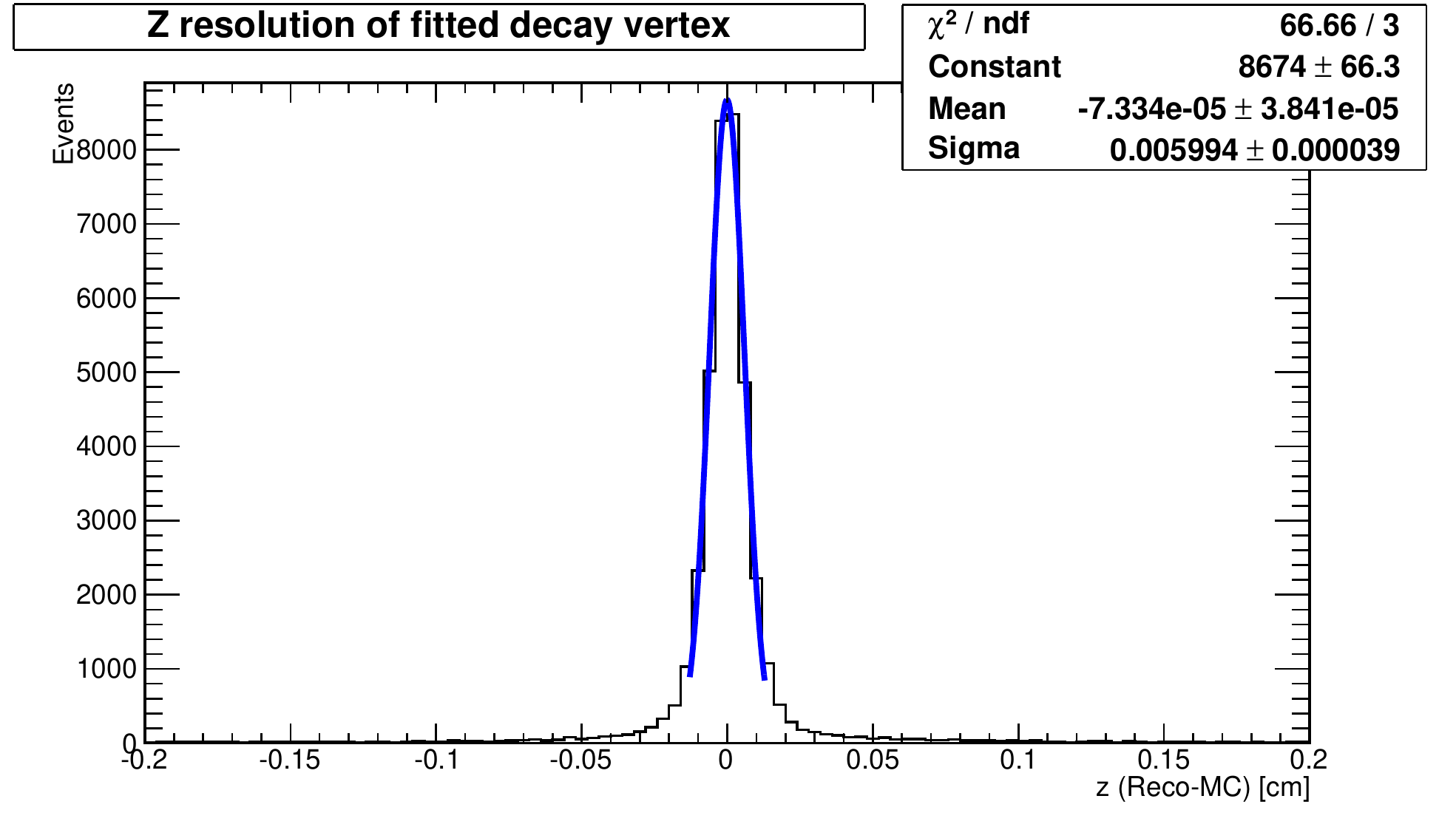}
\caption[$\bar{p}p\rightarrow 2(\pi^+\pi^-)$: Vertex resolution]
{$\bar{p}p\rightarrow 2(\pi^+\pi^-)$: Vertex resolution, without event mixing (see text for more details). }
\label{fig:stt:ben:vertexFit4pi}
\end{center}
\end{figure*}

A vertex fit has been performed during the reconstruction of the final state, and the best candidate in each event has been selected by the
minimal $\chi^2$ criterion. \Reffig{fig:stt:ben:vertexFit4pi} shows the resolution in x, y and z coordinates of the fitted decay vertex 
(i.e. difference between reconstructed vertex and Monte Carlo truth vertex position). The distributions are fitted with the Gaussian function in order to extract the resolutions which are: $\sigma_x$=47 $\mu$m, $\sigma_y$=46 $\mu$m and $\sigma_z$=60 $\mu$m.

\begin{figure}%[h!]
\begin{center}
\includegraphics[width=\swidth]{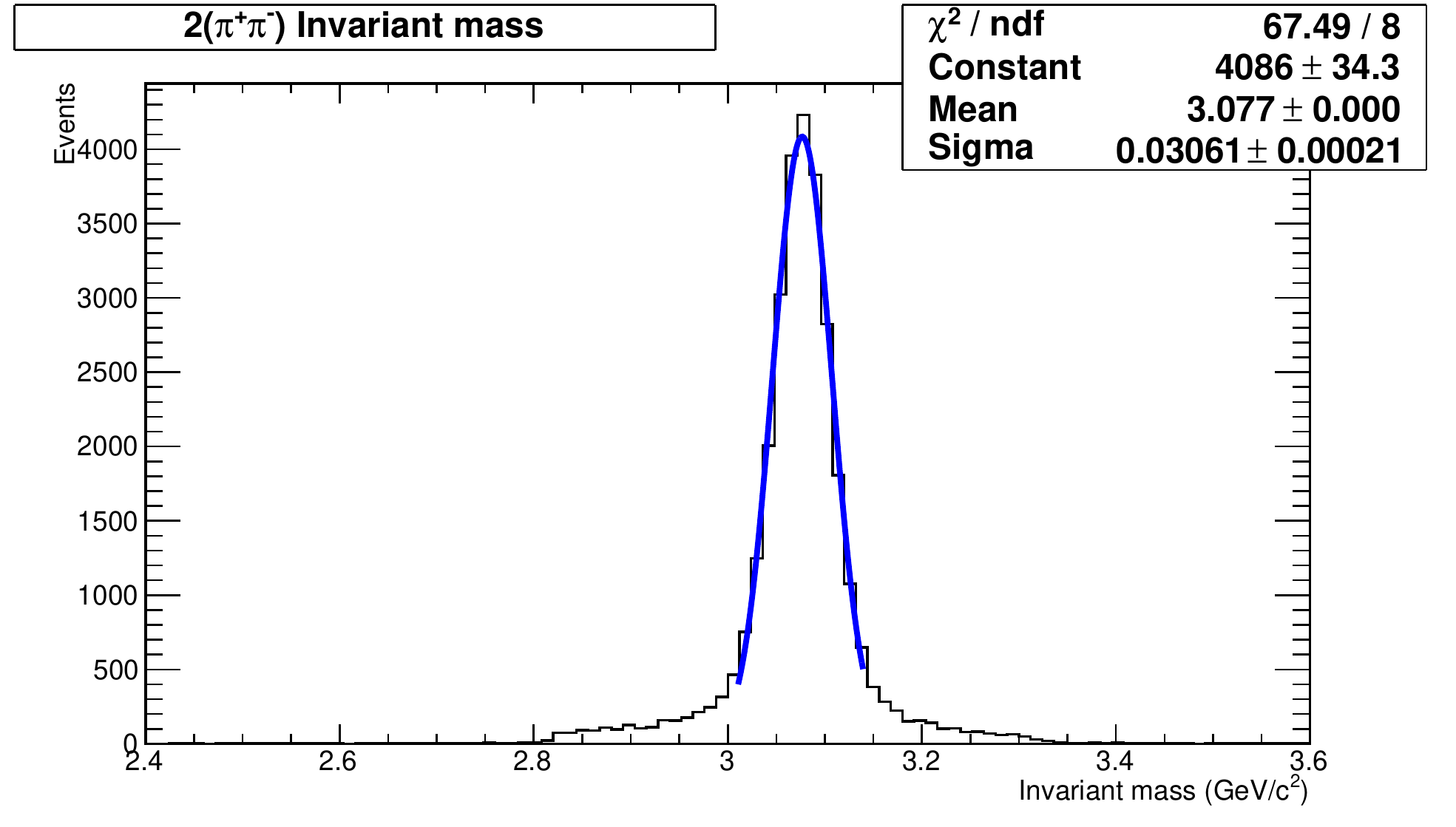}
\caption[$\bar{p}p\rightarrow 2(\pi^+\pi^-)$: Pions invariant mass distribution (after clean-up)]
{$\bar{p}p\rightarrow 2(\pi^+\pi^-)$: Pions invariant mass distribution after clean-up procedure. The fit is done with a Gaussian 
function (see text for more details).}
\label{fig:stt:ben:invmass4PiCU}
\end{center}
\end{figure}

For the pattern recognition in the presence of pile-up from the mixed background events, the clean-up procedure is applied to remove spurious hits.  
\Reffig{fig:stt:ben:invmass4PiCU} shows the four pions invariant mass distribution after the clean-up procedure; the global reconstruction 
efficiency is (31.7$\pm$0.2)\% and the resolution is 31~\mevcc. The single pion track resolution after the clean-up procedure is again 1.7\,\%.

\begin{figure}%[h!]
\begin{center}
\includegraphics[width=\swidth]{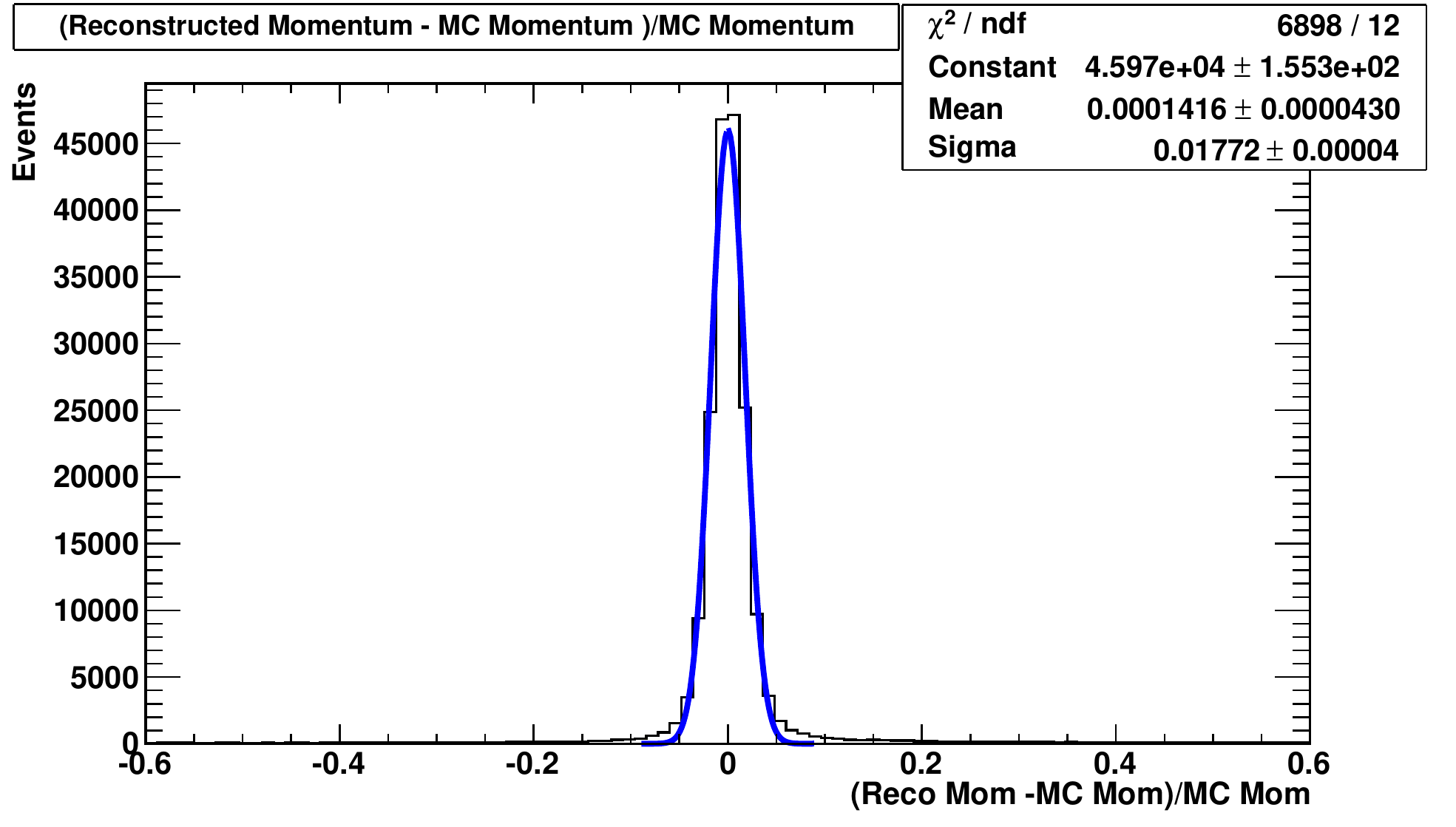}
\caption[$\bar{p}p\rightarrow 2(\pi^+\pi^-)$: Momentum resolution of the pion tracks (with event mixing)]
{$\bar{p}p\rightarrow 2(\pi^+\pi^-)$: Momentum resolution of the pion tracks with event mixing. The fit is done with a Gaussian 
function (see text for more details).}
\label{fig:stt:ben:singlepion4piEMX}
\end{center}
\end{figure}

\begin{figure}%[h!]
\begin{center}
\includegraphics[width=\swidth]{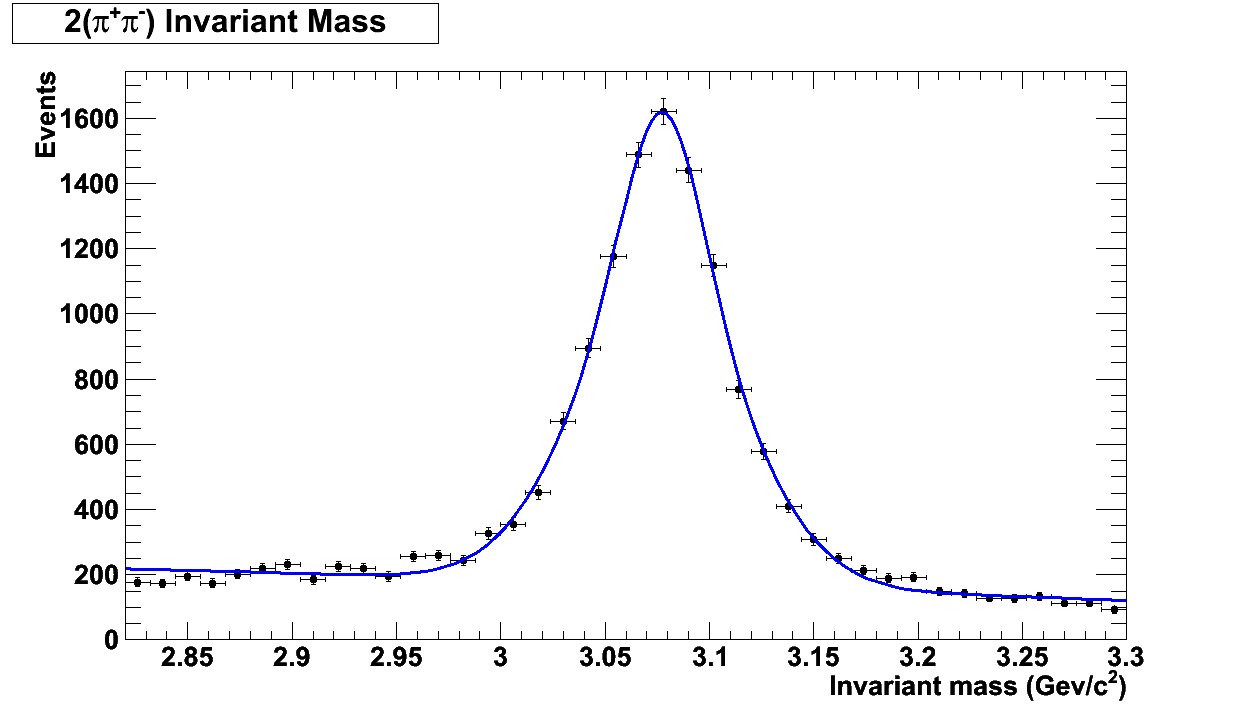}
\caption[$\bar{p}p\rightarrow 2(\pi^+\pi^-)$: Pions invariant mass distribution (with event mixing)]{$\bar{p}p\rightarrow 2(\pi^+\pi^-)$: Pions invariant mass distribution with event mixing. The fit is done with a double Gaussian 
function plus a polynomial. (see text for more details).}
\label{fig:stt:ben:invariantmass4piEMX}
\end{center}
\end{figure}

After mixing with background events, \Reffig{fig:stt:ben:singlepion4piEMX} shows the difference between the reconstructed and the Monte Carlo 
generated momentum divided by the Monte Carlo one for the pion tracks. The single pion track resolution obtained from the Gaussian fit is 1.8\,\%. \Reffig{fig:stt:ben:invariantmass4piEMX} shows the four pions invariant mass distribution which is fitted with a double Gaussian function plus a 
polynomial to take the background into account. From the fit, the global reconstruction efficiency is (17.2$\pm$0.2)\% and the resolution is 
39~\mevcc.

\begin{figure}%[h!]
\begin{center}
\includegraphics[width=\swidth]{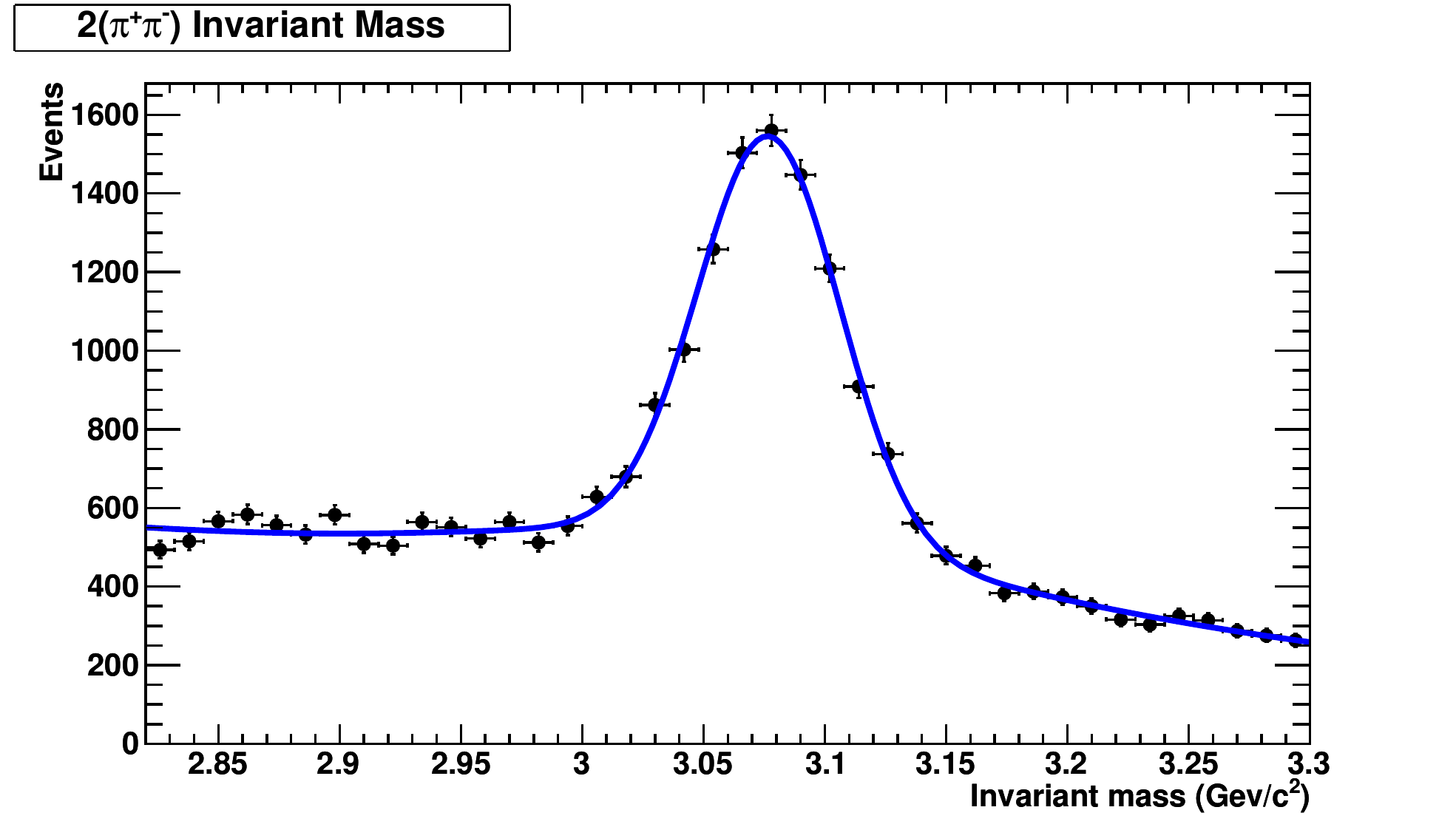}
\caption[$\bar{p}p\rightarrow 2(\pi^+\pi^-)$: Pions invariant mass distribution (with event mixing, no Monte Carlo truth PID)]
{$\bar{p}p\rightarrow 2(\pi^+\pi^-)$: Pions invariant mass distribution with event mixing and without Monte Carlo truth 
PID. The fit is done with a double Gaussian function plus a polynomial. (see text for more details).}
\label{fig:stt:ben:invariantmass4piEMXNOPID}
\end{center}
\end{figure}

An additional study is to check to which extent the Monte Carlo based PID is relevant for this benchmark channel. So the four pions are reconstructed without 
any PID. The invariant mass is shown in \Reffig{fig:stt:ben:invariantmass4piEMXNOPID} and it looks affected, infact the four pions 
reconstruction efficiency is (37.4$\pm$0.2)\%; the resolution is 39 \mevcc.

%\clearpage
\subsection{$\bar pp \rightarrow \eta_c \rightarrow \phi\phi$}
\begin{figure}[!h]
\centering
\includegraphics[width=0.95\swidth]{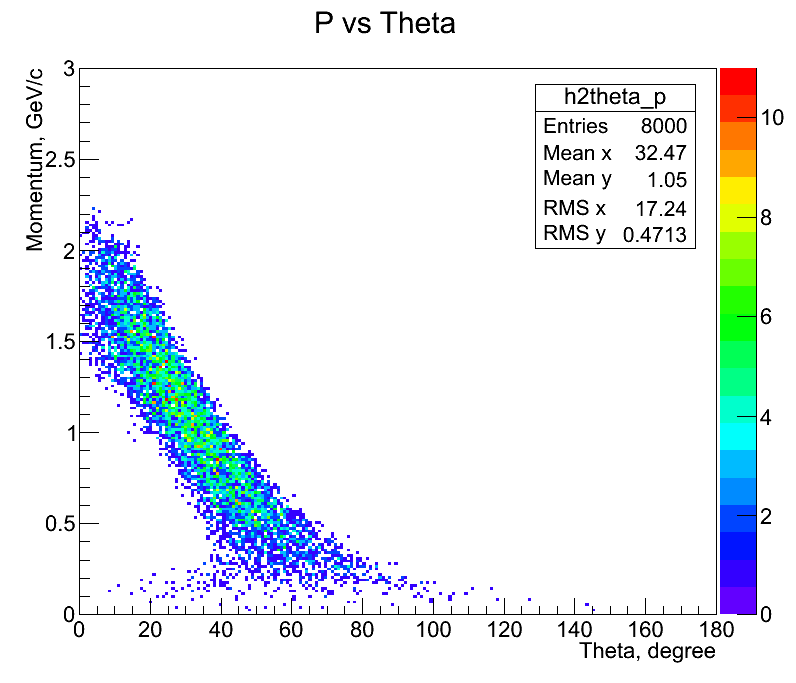}
\caption[Momentum vs polar angle for the kaons from the reaction
$\eta_c\rightarrow\phi \phi\rightarrow K^{+}K^{-}K^{+}K^{-}$]{Momentum vs polar angle for the kaons from the reaction
$\eta_c\rightarrow\phi \phi\rightarrow K^{+}K^{-}K^{+}K^{-}$.}
\label{fig:stt:per:etac:kinematics}
\end{figure}

%\section*{$\eta_c\rightarrow\phi \phi$ decay for central tracker studies}
Physics of charmonium is one of the main parts of the \PANDA experimental programme.
To study the performance of the central tracker with respect to charmonium physics
the $\eta_c$ state has been selected. The $\eta_c(1^1S_0)$ state of charmonium
with the mass $2980.4\pm 1.2$ \mevcc (according to the Particle Data Group (PDG))
was discovered more than thirty years ago. Being the ground state of charmonium it
represents an interest as a final state in decays of other charmonium states but
the resonance scan for precise determination of mass and width of $\eta_c$ is a
separate important task for the \PANDA experiment. The $\eta_c$ can be detected through
many exclusive decay channels, neutral or hadronic. For the study of the central
tracker performance the following decay mode has been selected: $\eta_c\rightarrow\phi \phi$
with the branching ratio $2.7\cdot10^{-3}$ with the subsequent decay $\phi\rightarrow K^+K^-$.
This decay mode has a very particular kinematics which simplifies its separation from the
general hadronic background. The small $Q$ value of $31$ \mev of the decay $\phi\rightarrow K^+K^-$
results in directions of the two kaons close to the direction of the $\phi$ meson. On the other hand
$\eta_c\rightarrow\phi \phi$ is a two-body decay and as a consequence the
directions of the two $\phi$ and therefore of $K^{\pm}$ are correlated. 
The kinematics of the final state kaons is shown in \Reffig{fig:stt:per:etac:kinematics}. 
The distribution of kaons covers a wide range of the central tracker
acceptance peaking between $20^{\circ}$ and $40^{\circ}$ and the covered momentum range is 
mainly from 200 \mevc to 2 \gevc.
The figures of merit of this analysis, which have to check the performance of the central tracker, are
the efficiency of the $\eta_c$ reconstruction and the resolution of the invariant mass for the $\eta_c$ and
the intermediate $\phi$ states. In addition, the vertex resolution is quoted, however this is not of primary
interest for the given channel.
The analysis is performed in the following steps:
\begin{itemize}
\item Charged candidates with opposite charge are combined to $\phi$ candidates
with $\phi$ mass preselection $1.02\pm0.1$ \gevcc.
\item A vertex fit is performed and the best $\eta_c$ candidate in each event is selected
by minimal $\chi^2$.
\item Events with $\phi$ candidates within a mass window $1.00$ \gevcc
  $< m(K^+ K^-) < 1.04$ \gevcc are selected.
\item $\eta_c$ is considered as reconstructed if it falls into the mass window [2.90;3.06] \gevcc.
\end{itemize}
It is important to note here that the parameters used in this analysis, such as the cut ranges for the invariant
mass, in the real experiment will be optimized for a best signal to background ratio while here they are based
on educated guess.
\begin{figure}[!h]
\centering
\includegraphics[width=\swidth]{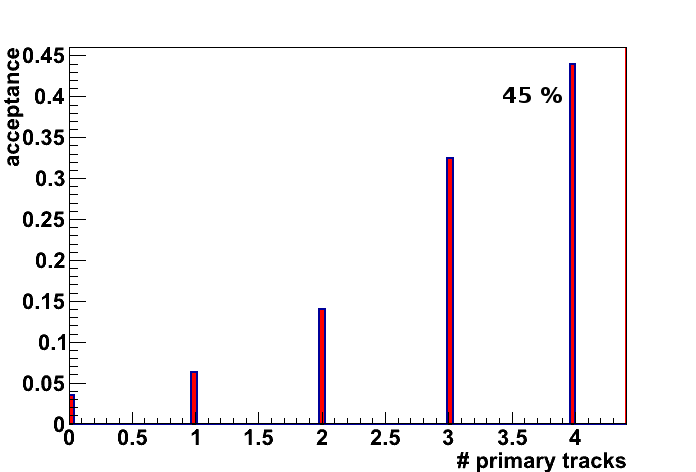}
\caption[Multiplicity of the kaons from the reaction $\eta_c\rightarrow\phi \phi\rightarrow K^{+}K^{-}K^{+}K^{-}$
within central tracker acceptance]{Multiplicity of the kaons from the reaction $\eta_c\rightarrow\phi \phi\rightarrow K^{+}K^{-}K^{+}K^{-}$
within central tracker acceptance.}
\label{fig:stt:per:etac:acceptance}
\end{figure}

The following results are presented without mixing the signal with background and results including
mixing come later. Before estimating the reconstruction efficiency it was studied how many $\eta_c$ events
have final state kaons within the central tracker acceptance. The study is based on Monte Carlo information and
a kaon is considered within detector acceptance if it creates there at least one Monte Carlo hit. Results
are summarized in \Reffig{fig:stt:per:etac:acceptance}, where the multiplicity of kaons within the
detector acceptance is presented. According to this plot 45\% of events have all 4 kaons within the acceptance
which defines an upper limit for the detector efficiency for $\eta_c$ reconstruction.
\begin{figure}[!h]
\centering
\includegraphics[width=0.95\swidth]{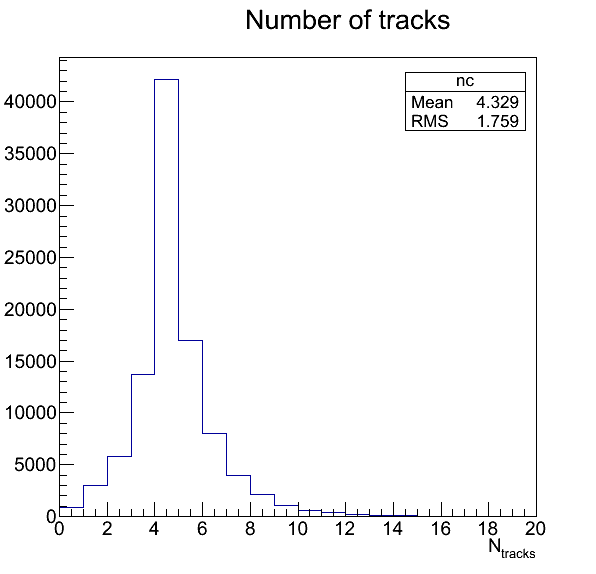}
\caption[Multiplicity of the reconstructed charged tracks from the reaction $\eta_c\rightarrow\phi \phi\rightarrow K^{+}K^{-}K^{+}K^{-}$]{Multiplicity of the reconstructed charged tracks from the reaction $\eta_c\rightarrow\phi \phi\rightarrow K^{+}K^{-}K^{+}K^{-}$.}
\label{fig:stt:per:etac:n_charged}
\end{figure}

At the beginning of the analysis the number of reconstructed charged
tracks was studied (\Reffig{fig:stt:per:etac:n_charged}).  From this
plot a tail in distribution is observed with a high number of
reconstructed tracks which arises due to secondaries and to ghost
tracks from the STT pattern recognition. In addition ghost tracks result in 72\%
of events having 4 or more reconstructed tracks which is higher than
the 45\% of estimated detector acceptance.

\begin{figure}[!h]
\centering
\includegraphics[width=0.95\swidth]{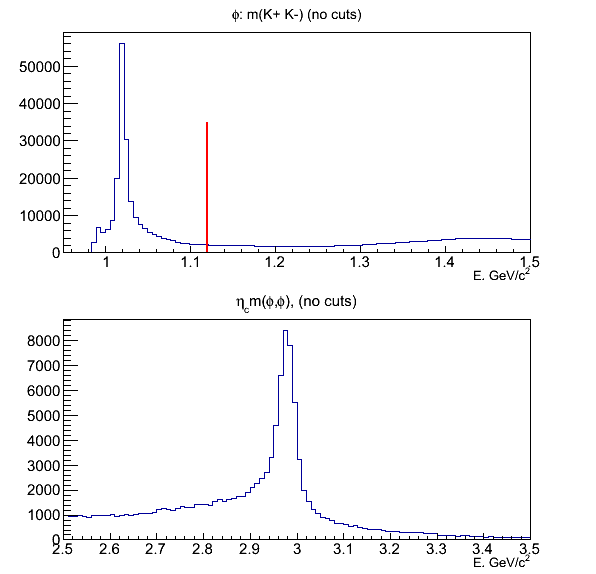}
\caption[Reconstructed invariant mass of $K^+K^-$ pairs and $\phi\phi$ pairs with preselection on $\phi$ invariant mass]{Reconstructed invariant mass of $K^+K^-$ pairs and $\phi\phi$ pairs with preselection on $\phi$ invariant mass.}
\label{fig:stt:per:etac:m_nocuts}
\end{figure}

\begin{figure}[!h]
\centering
\includegraphics[width=0.95\swidth]{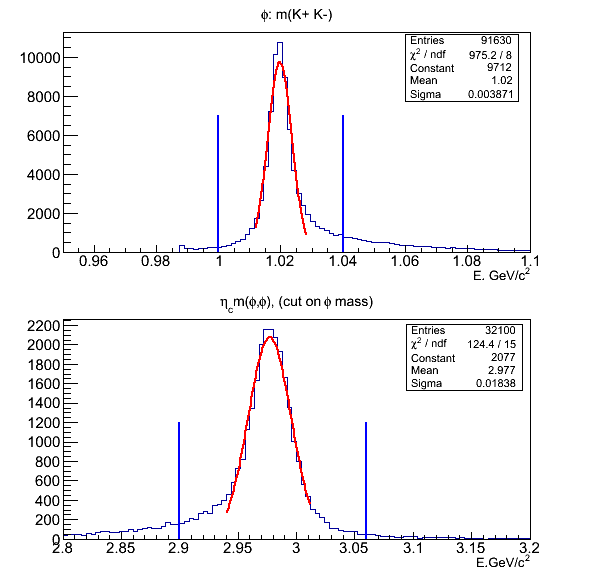}
\caption[Reconstructed invariant mass of $K^+K^-$ pairs and $\phi\phi$ pairs after full selection chain]{Reconstructed invariant mass of $K^+K^-$ pairs and $\phi\phi$ pairs after full selection chain.}
\label{fig:stt:per:etac:m_final_vtx}
\end{figure}

\begin{figure*}[!h]
\centering
\includegraphics[width=\dwidth]{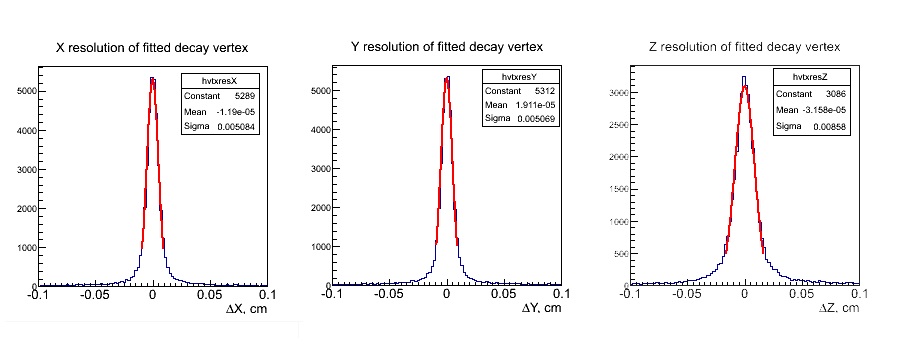}
\caption[Resolution of the reconstructed primary vertex of $\eta_c$ from the reaction $\eta_c\rightarrow\phi \phi\rightarrow K^{+}K^{-}K^{+}K^{-}$]{Resolution of the reconstructed primary vertex of $\eta_c$ from the reaction $\eta_c\rightarrow\phi \phi\rightarrow K^{+}K^{-}K^{+}K^{-}$.}
\label{fig:stt:per:etac:vertex_res}
\end{figure*}

Invariant mass distributions of $K^+K^-$ pairs of two $\phi$
candidates are presented in \Reffig{fig:stt:per:etac:m_nocuts}.  In
the upper plot a cut is indicated for the $\phi$ candidate's invariant mass
which is used for $\eta_c$ construction. The plot of the $\phi\phi$ invariant
mass has a significant tail on the left, which however is reduced after
requiring ideal PID.
Applying a vertex fit to four kaons combined to an $\eta_c$ candidate
the best $\eta_c$ in each event is selected from the minimum $\chi^2$
and those results are presented in
\Reffig{fig:stt:per:etac:m_final_vtx}. To extract the invariant mass
resolution of $\eta_c$ and $\phi$ the following two step approach was
applied. At the beginning each plot was fitted with a Gaussian
function and the extracted parameters $\mu_1$ and $\sigma_1$ were used
at the second step where a fit with a Gaussian function was performed
in the range $[\mu_1-1.6\sigma_1;\mu_1+1.6\sigma_1]$. The used range
satisfies that the full width of half maximum of the fitted peak is a
well defined quantity and the extracted width parameter of the
Gaussian $\sigma_2$ is quoted. The given approach allows to avoid
interference of the tails of the distribution with the extracted width
parameter. The obtained resolution for $\phi$ and $\eta_c$ are $3.9$
\mevcc and $18$ \mevcc correspondingly. Also, the given plot allows to
extract the $\eta_c$ reconstruction efficiency as a number of $\eta_c$
candidates within the mass range [2.90;3.06] \gevcc and it is
$27.3\pm0.2\%$.
In addition the space resolution of the primary $\eta_c$ vertex in $x,y,z$
coordinates is presented in \Reffig{fig:stt:per:etac:vertex_res}.  The
given plot represents the difference between reconstructed vertex and
Monte Carlo truth vertex position. The obtained resolutions in all
coordinates are $\sigma_x=51 \mum$, $\sigma_y=51 \mum$, $\sigma_z=86
\mum$.

\begin{figure}[!h]
\centering
\includegraphics[width=0.95\swidth]{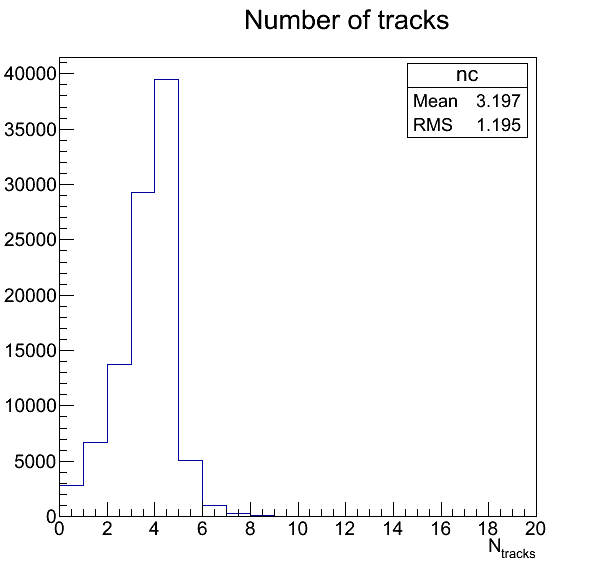}
\caption[Multiplicity of the reconstructed charged tracks from the reaction $\eta_c\rightarrow\phi \phi\rightarrow K^{+}K^{-}K^{+}K^{-}$ after
applying clean-up procedure]{Multiplicity of the reconstructed charged tracks from the reaction $\eta_c\rightarrow\phi \phi\rightarrow K^{+}K^{-}K^{+}K^{-}$ after
applying clean-up procedure.} \label{fig:stt:per:etac:n_charged_cu}
\end{figure}

For pattern recognition in the presence of pile-up from background
events the clean-up procedure is applied to remove spurious hits.  For
the case of signal without mixing with background this leads to a
deterioration of the $\eta_c$ reconstruction efficiency because some real
hits are removed by this procedure. The following change in the number of
reconstructed tracks is observed
(\Reffig{fig:stt:per:etac:n_charged_cu}), i.e. the number of events with
more than 4 reconstructed tracks is reduced significantly. The $\eta_c$
reconstruction efficiency after the clean-up procedure is $19.1\pm0.2\%$,
the resolution for $\phi$ and $\eta_c$ are slightly changed to 3.9 \mevcc
and 17 \mevcc correspondingly.

\begin{figure}[!h]
\centering
\includegraphics[width=0.95\swidth]{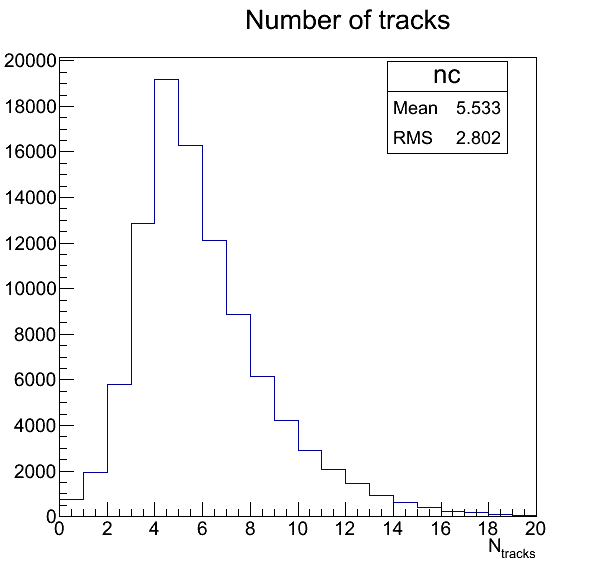}
\caption[Multiplicity of the reconstructed charged tracks from the reaction $\eta_c\rightarrow\phi \phi\rightarrow K^{+}K^{-}K^{+}K^{-}$ with 
background mixing]{Multiplicity of the reconstructed charged tracks from the reaction $\eta_c\rightarrow\phi \phi\rightarrow K^{+}K^{-}K^{+}K^{-}$ with 
background mixing.}
\label{fig:stt:per:etac:n_charged_mix}
\end{figure}

Final results of this study take into account the mixing of signal with
generic background produced by the DPM event generator, where the number of
pile-up events is defined by the Poisson statistics. In this case
all tracks from background can be reconstructed as well as
single hits from background can contribute to the tracks from events
of interest.  The number of reconstructed tracks in one event becomes
higher than for signal only (\Reffig{fig:stt:per:etac:n_charged_mix}).

\begin{figure}[!h]
\centering
\includegraphics[width=0.95\swidth]{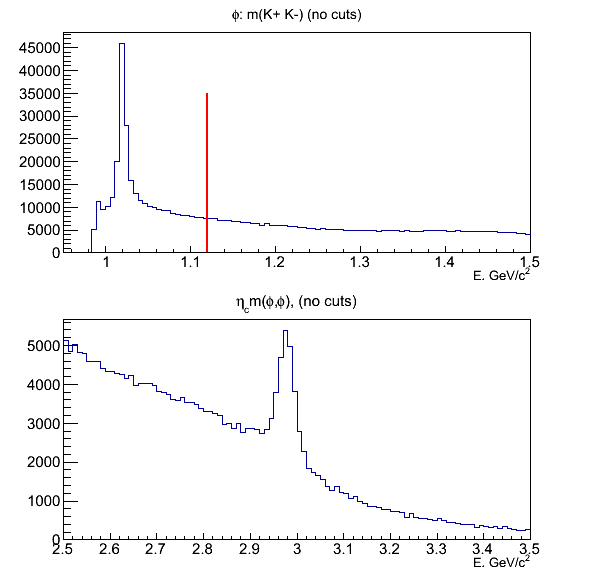}
\caption[Reconstructed invariant mass of $K^+K^-$ pairs and $\phi\phi$ pairs with preselection on $\phi$ invariant mass with background mixing]{Reconstructed invariant mass of $K^+K^-$ pairs and $\phi\phi$ pairs with preselection on $\phi$ invariant mass with background mixing.}
\label{fig:stt:per:etac:m_nocuts_mix}
\end{figure}

\begin{figure}[!h]
\centering
\includegraphics[width=0.95\swidth]{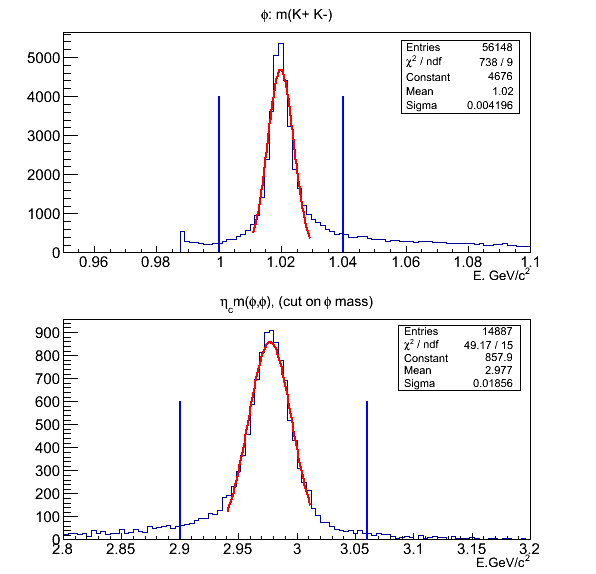}
\caption[Reconstructed invariant mass of $K^+K^-$ pairs and $\phi\phi$ pairs after full selection chain with background mixing]{Reconstructed invariant mass of $K^+K^-$ pairs and $\phi\phi$ pairs after full selection chain with background mixing.}
\label{fig:stt:per:etac:m_final_vtx_mix}
\end{figure}

\begin{figure}[!b]
\centering
\includegraphics[width=0.95\swidth]{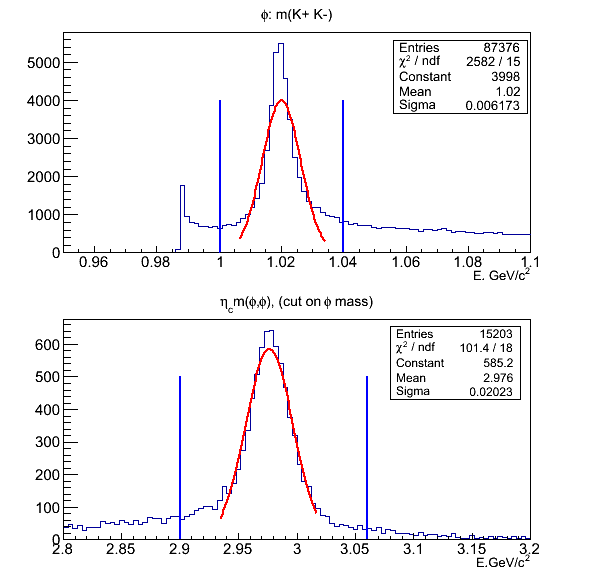}
\caption[Reconstructed invariant mass of $K^+K^-$ pairs and $\phi\phi$ pairs after full selection chain with background mixing and without 
Monte Carlo truth PID]{Reconstructed invariant mass of $K^+K^-$ pairs and $\phi\phi$ pairs after full selection chain with background mixing and without 
Monte Carlo truth PID.}
\label{fig:stt:per:etac:m_final_vtx_mix_nopid}
\end{figure}

Invariant mass distributions for $\phi$ and $\eta_c$ are presented in
\Reffig{fig:stt:per:etac:m_nocuts_mix}. Here the $\eta_c$ peak appears on
the top of a large combinatorial background. However after the whole
selection procedure invariant mass plots look very similar to the case
of signal only
(\Reffig{fig:stt:per:etac:m_final_vtx_mix}). The reconstruction efficiency
for $\eta_c$ however is lower in this case (11.6\%), but the presence of
``mixed'' background does not affect much the resolution of the
reconstructed invariant mass of $\eta_c$ and $\phi$, which are 19
\mevcc and 4.2 \mevcc respectively.
An additional important point of this study was to check how Monte
Carlo based PID was relevant here. For this case the $\eta_c$
reconstruction was performed without any PID as worst case
scenario. The corresponding invariant mass distributions for $\phi$
and $\eta_c$ are presented in
\Reffig{fig:stt:per:etac:m_final_vtx_mix_nopid}. Here, more pronounced
tails from combinatorics arise in the $\phi$ mass distribution but the
$\eta_c$ mass distribution looks not much affected ($\phi$ mass
resolution is 6.2 \mevcc and $\eta_c$ mass resolution is 20 \mevcc),
however $\eta_c$ reconstruction efficiency drops down to 9.6\%.
%\clearpage

%\clearpage
\subsection{$\bar pp \rightarrow \psi(3770) \rightarrow D^+D^-$}
With the $\psi(3770)$ benchmark channel, the STT's performance in the reconstruction of particles with short decay lengths is evaluated.
The figures of merit are the spatial resolution of the secondary vertices as well as the invariant mass resolution of the reconstructed $D$ mesons.

\subsubsection{Channel Description}

The reaction
\begin{equation*}
\bar{p}p \rightarrow \psi(3770) \rightarrow D^+ D^- \rightarrow K^-\pi^+\pi^+ K^+\pi^-\pi^-
\end{equation*}
at a beam momentum of 6.5788~\gevc
has a typical signature which is common for several of the channels within the scope of the \Panda physics program.

Within the scope of this benchmark, its following key features are of particular interest:

\begin{itemize}
	\item Secondary vertices with a short decay length (312~$\mum$ for the charged $D$ mesons),
	\item A relatively large number of ejectiles (6 for this channel) to be reconstructed in an exclusive analysis.
\end{itemize}

%\begin{figure}
%\centering
%\subfloat[$K^+$]{\includegraphics[angle=0,width=0.98\swidth]{stt/fig/run936cufix/psirun936cufix_hpthetakp.pdf}}
%\subfloat[$\pi^+$]{\includegraphics[angle=0,width=0.98\swidth]{stt/fig/run936cufix/psirun936cufix_hpthetapip.pdf}}
%\caption[Momentum and polar angle distribution of the generated kaons and pions]{Momentum and polar angle distribution of the generated kaons (a) 
%and pions (b).}
%\label{fig_hpthetap}
%\end{figure}

\begin{figure*}
\centering
\includegraphics[angle=0,width=0.98\swidth]{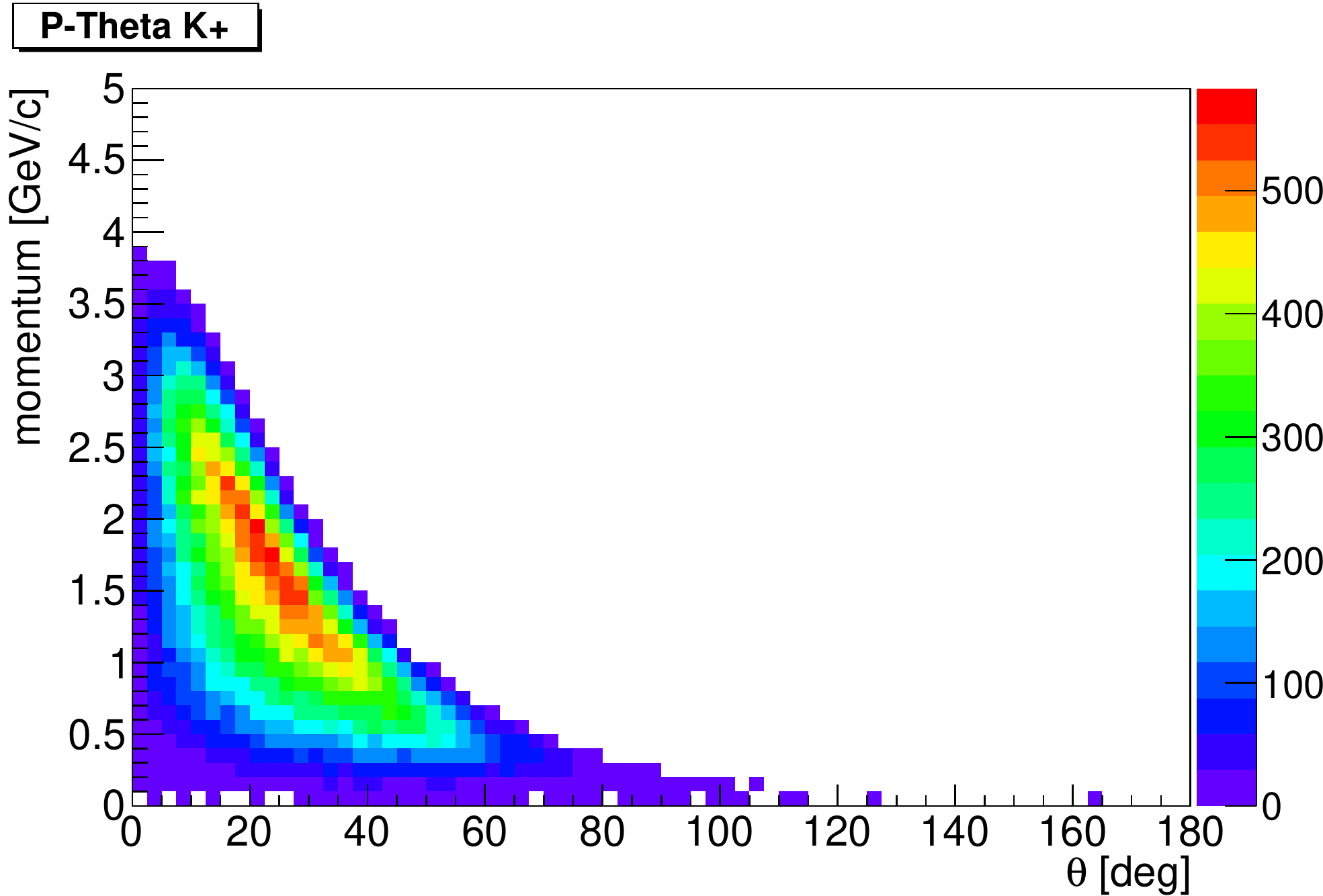}
\hspace{2mm}
\includegraphics[angle=0,width=0.98\swidth]{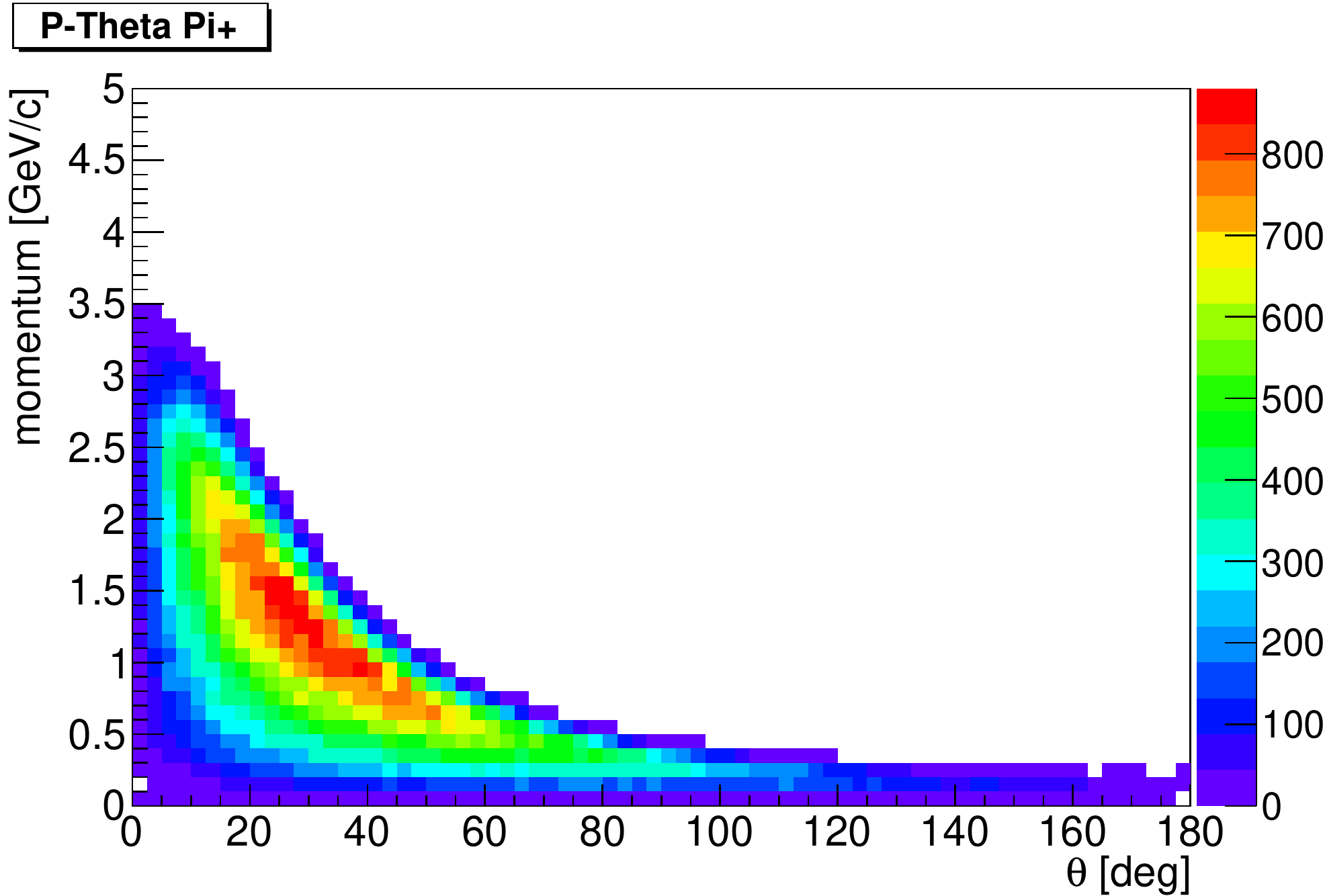}
\caption[Momentum and polar angle distribution of the generated kaons and pions]{Momentum and polar angle distribution of the generated kaons (left) 
and pions (right).}
\label{fig_hpthetap}
\end{figure*}
The distribution of momentum and polar angle is shown for the positive kaons and pions (\Reffig{fig_hpthetap}) The distribution 
of the negatively charged particles is identical and thus not shown here.
For both types the majority of particles have a momentum between 0.5~\gevc and 3~\gevc. While the kaons are only found in the forward 
hemisphere, the pions are also ejected at backwards angles due to their lower mass. However, the majority is found within a polar 
angle range between $5^{\circ}$ and $60^{\circ}$ in both cases.
There is a high probability that at least one of the decay particles is strongly forward peaked.

Reconstructing this class of events requires a good interplay of the central tracking detectors MVD and STT as well as additional 
information from the forward tracking to detect also those particles which are ejected at shallow angles below $10^{\circ}$.

%\begin{figure}
%\centering
%\includegraphics[angle=0,width=0.5\textwidth]{stt/fig/run936cufix/psirun936cufix_hpthetakm.pdf}
%\caption{Momentum and azimuthal angle distribution of the generated $K^-$.}
%\label{fig_hpthetakm}
%\end{figure}
%
%\begin{figure}
%\centering
%\includegraphics[angle=0,width=0.5\textwidth]{stt/fig/run936cufix/psirun936cufix_hpthetapip.pdf}
%\caption{Momentum and azimuthal angle distribution of the generated $\pi^+$.}
%\label{fig_hpthetapip}
%\end{figure}
%
%\begin{figure}
%\centering
%\includegraphics[angle=0,width=0.5\textwidth]{stt/fig/run936cufix/psirun936cufix_hpthetapim.pdf}
%\caption{Momentum and azimuthal angle distribution of the generated $\pi^-$.}
%\label{fig_hpthetapim}
%\end{figure}

\subsubsection{Resolution Study}

For the study of the achievable invariant mass and vertex resolutions, we simulate a large sample of signal events which have to 
pass an analysis chain similar to the procedure which would be used in a real experiment. 

%\subsubsubsection{Simulation Overview}

%A sample of $10^5$ signal events has been generated on the PandaGrid using PandaRoot
%revision 12725, in the following referred to as the july11 package.
%The simulation comprised the four typical steps of simulation, digitization, reconstruction and particle identification (compare Sec.~\ref{sec:stt:sim}). Note that for the particle identifaction the simulated Monte Carlo true information has been used in order to exclude any potential external influence on the tracking study.
%
%The relevant subdetectors which contribute hits to the reconstructed tracks are MVD, STT and forward GEM.

%The relevant numbers to be obtained are the mass, momentum and vertex resolutions for the D mesons.

%\subsubsubsection{Analysis Overview}

%\begin{figure}
%\centering
%\includegraphics[angle=0,width=0.5\textwidth]{stt/fig/run935rc5/run935oldnocu-rc5ana_combined_psi3770_hniceevents.pdf}
%\caption{}
%\label{fig_}
%\end{figure}
\begin{figure*}
\centering
\includegraphics[angle=0,width=0.98\swidth]{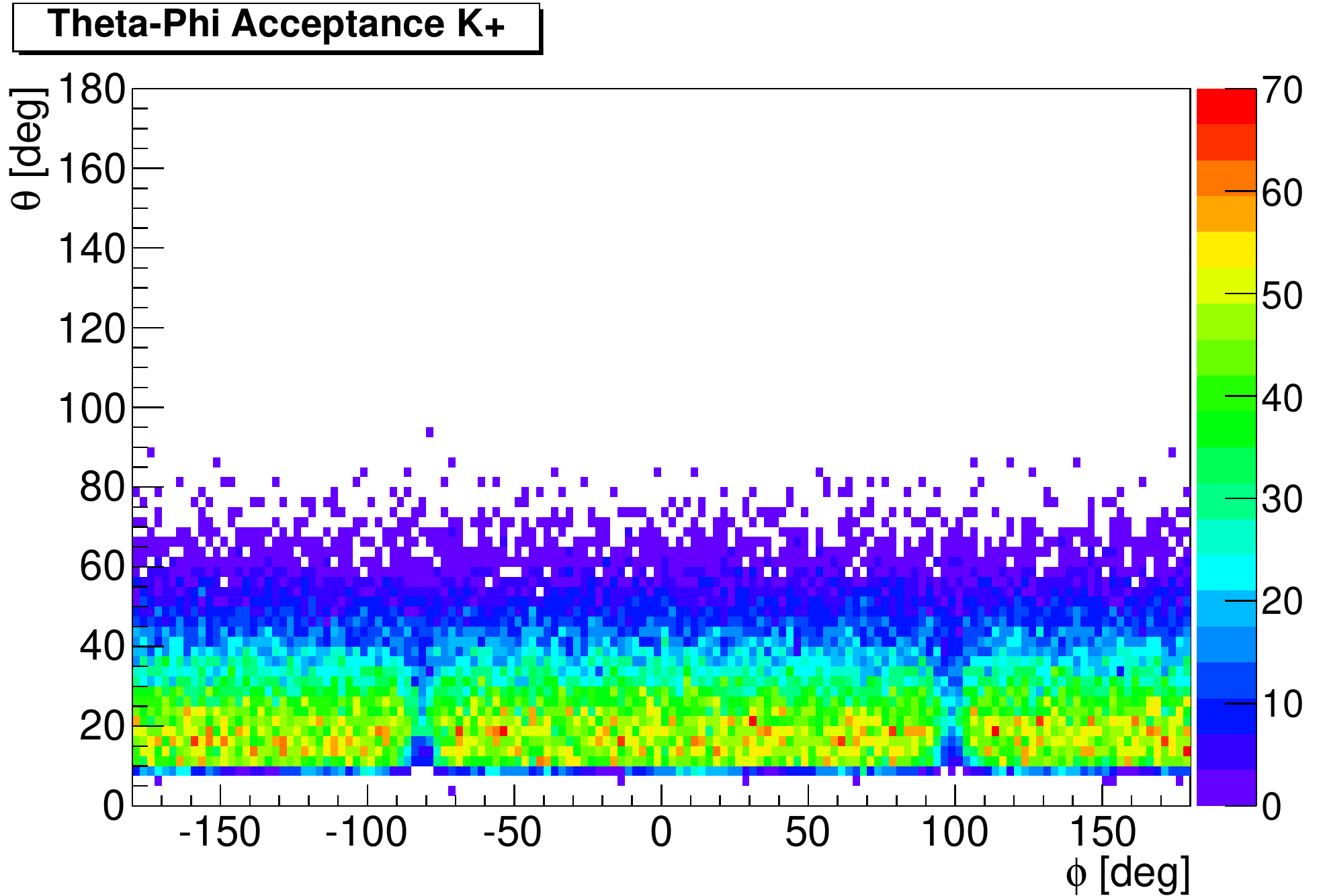}
\includegraphics[angle=0,width=0.98\swidth]{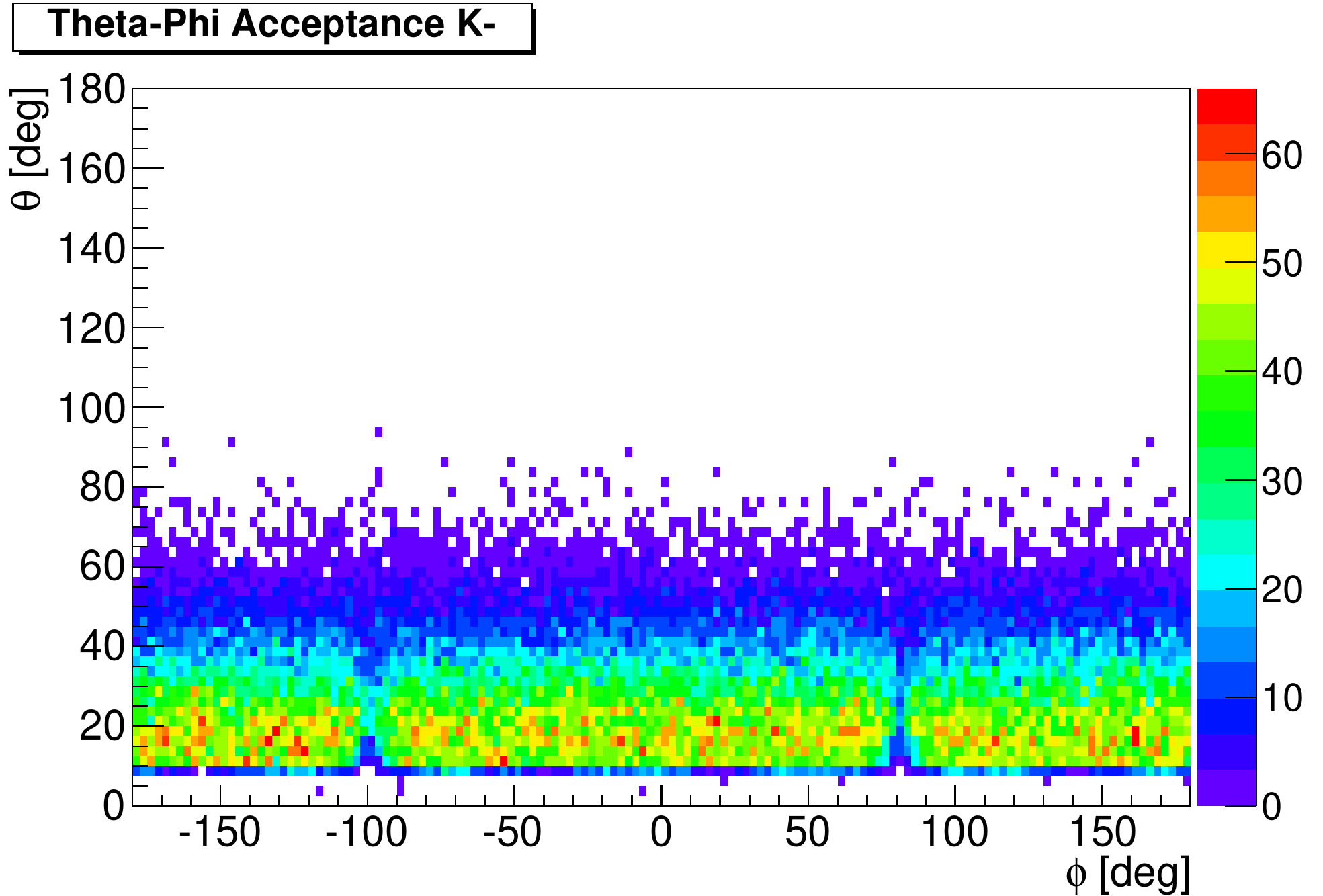}
\caption[Polar and azimuthal angle distribution of the reconstructed kaons]{Polar and azimuthal angle distribution of the reconstructed kaons (left) $K^+$ , (right) $K^-$.}
\label{fig_hacceptancephithetak}
\end{figure*}

\begin{figure*}
\centering
\includegraphics[angle=0,width=0.98\swidth]{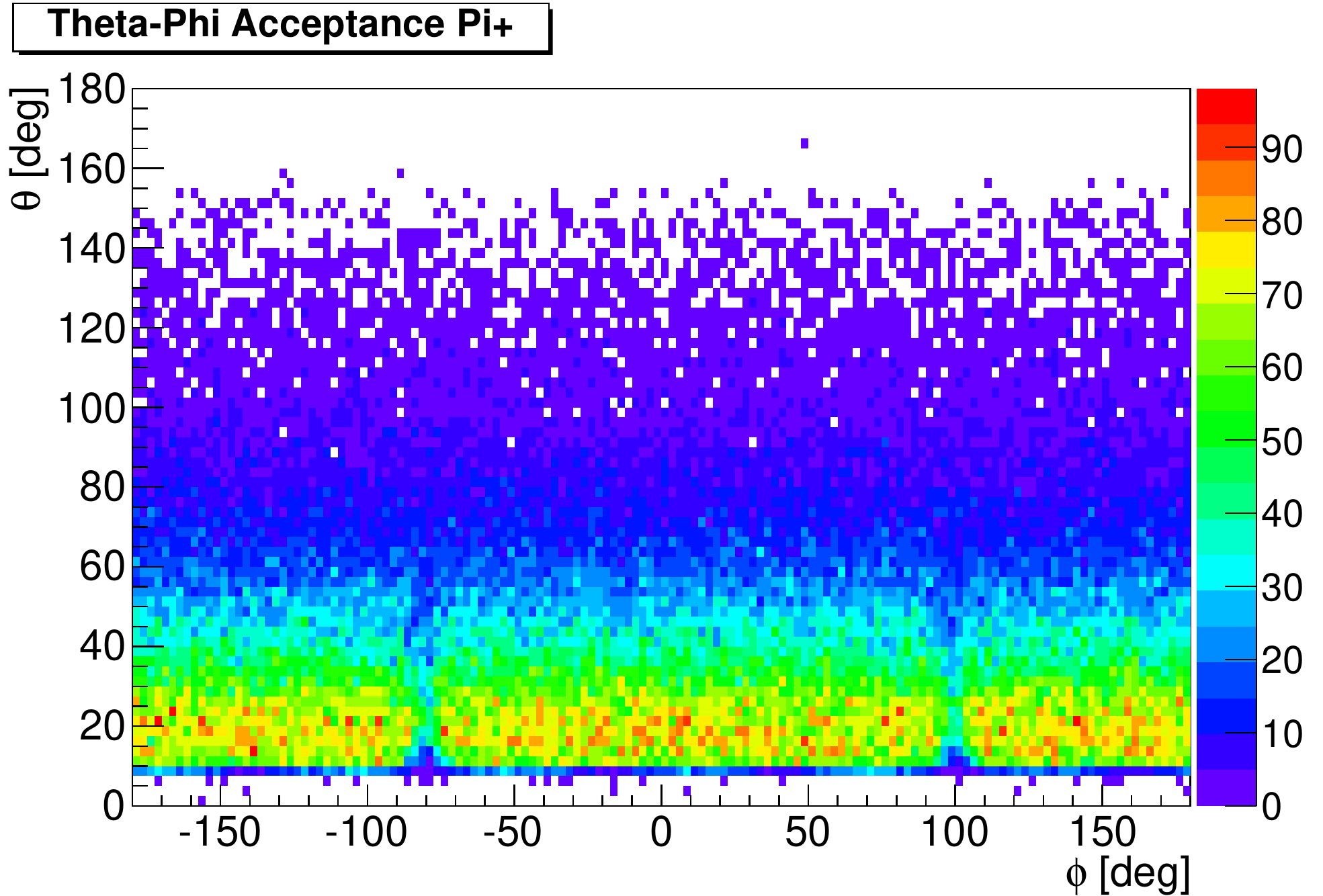}
\includegraphics[angle=0,width=0.98\swidth]{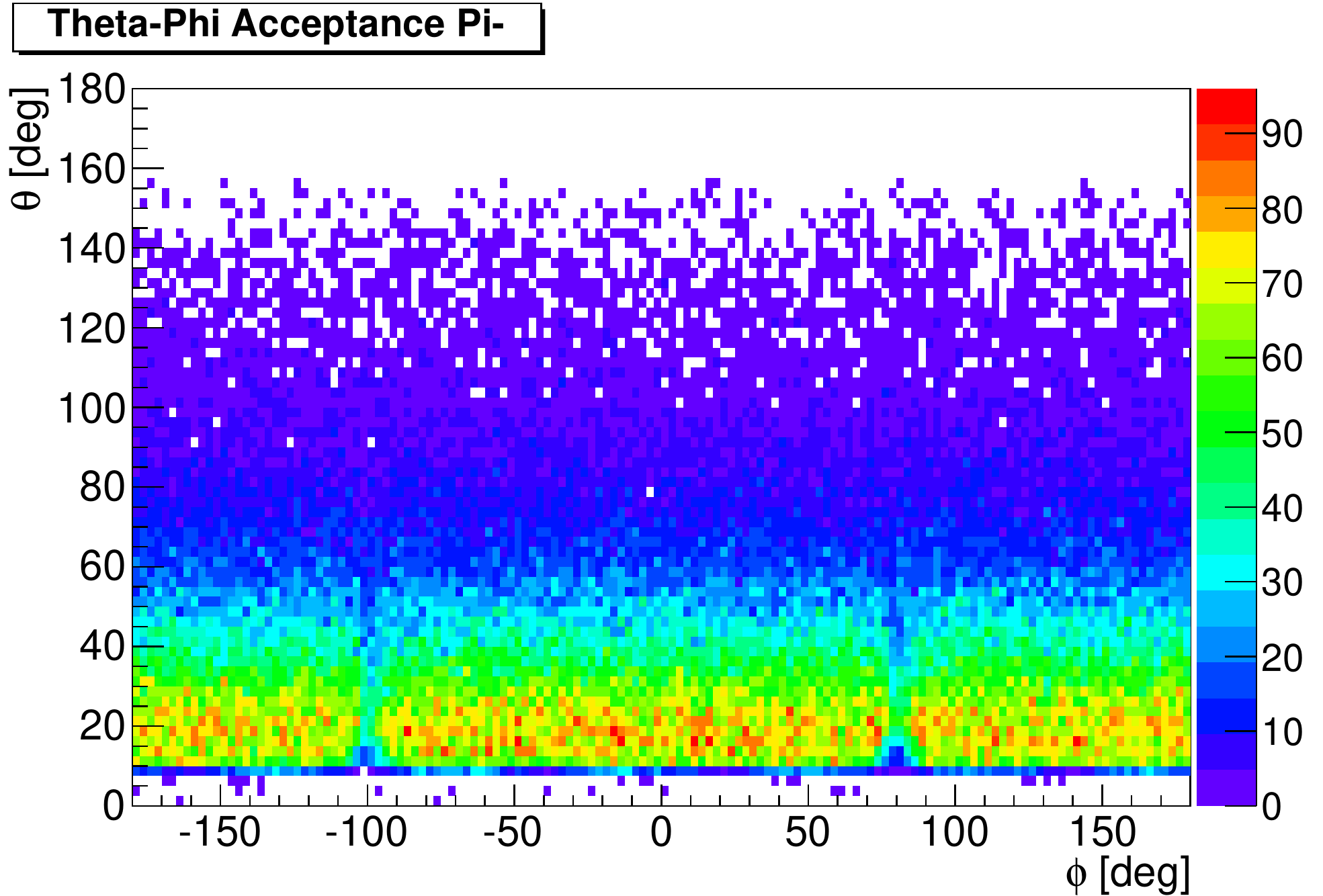}
\caption[Polar and azimuthal angle distribution of the reconstructed pions]{Polar and azimuthal angle distribution of the reconstructed pions, (left) $\pi^+$, (right) $\pi^-$.}
\label{fig_hacceptancephithetapi}
\end{figure*}

%\begin{figure}
%\centering
%\includegraphics[angle=0,width=0.5\textwidth]{stt/fig/run936cufix/psirun936cufix_hacceptancephithetakm.pdf}
%\caption{Polar and azimuthal angle distribution of the reconstructed $K^-$ tracks.}
%\label{fig_hacceptancephithetakm}
%\end{figure}
%
%\begin{figure}
%\centering
%\includegraphics[angle=0,width=0.5\textwidth]{stt/fig/run936cufix/psirun936cufix_hacceptancephithetapip.pdf}
%\caption{Polar and azimuthal angle distribution of the reconstructed $\pi^-$ tracks.}
%\label{fig_hacceptancephithetapip}
%\end{figure}
%
%\begin{figure}
%\centering
%\includegraphics[angle=0,width=0.5\textwidth]{stt/fig/run936cufix/psirun936cufix_hacceptancephithetapim.pdf}
%\caption{Polar and azimuthal angle distribution of the reconstructed $\pi^-$ tracks.}
%\label{fig_hacceptancephithetapim}
%\end{figure}

In the first step of the analysis all reconstructed track candidates which have at least one STT hit are considered for further 
processing. The distribution of polar and azimuthal angle of the reconstructed particles which pass the STT volume is shown in 
\Reffig{fig_hacceptancephithetak} and \Reffig{fig_hacceptancephithetapi}. The presence of the target pipe is clearly visible in 
these plots as well as the $\pm 10^{\circ}$ shift in the observed $\phi$ position due to the bending of the oppositely charged 
particles' tracks in the magnetic field.

The decay particles are then combined to $D^+$ and $D^-$ meson candidates. A first selection is done by requiring that the 
reconstructed $D$ meson mass values differ by no more than 750~\mevcc from the nominal D mass. After that, a vertex fit to the 
$D$-meson decay vertices is carried out.
%Note that this particular decay tree can be uniquely reconstructed without any ambiguities regarding the assignment of the 
%decay particles. In the signal channels any ambiguities 

\begin{figure*}
\centering
\includegraphics[angle=0,width=0.98\swidth]{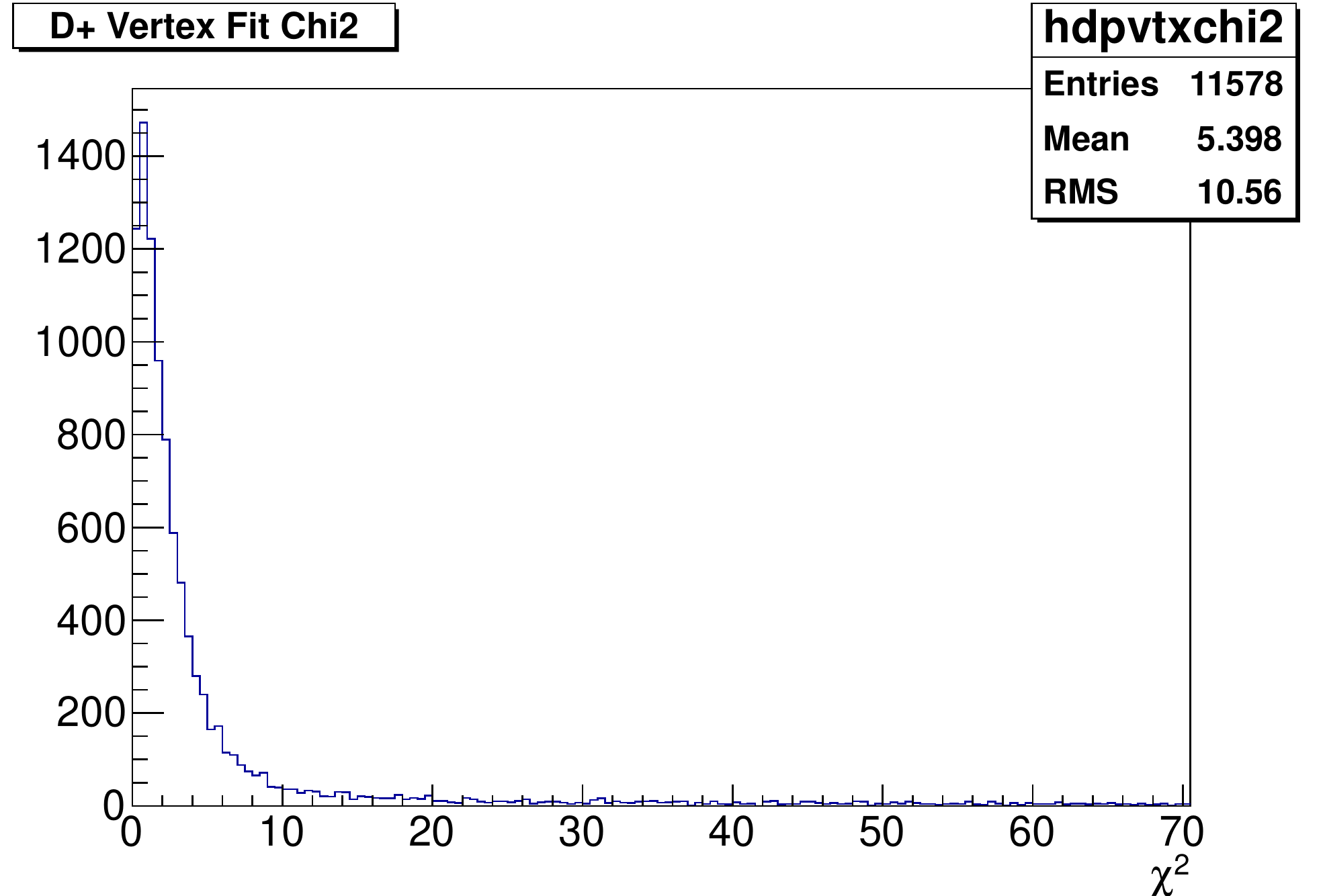}
\includegraphics[angle=0,width=0.98\swidth]{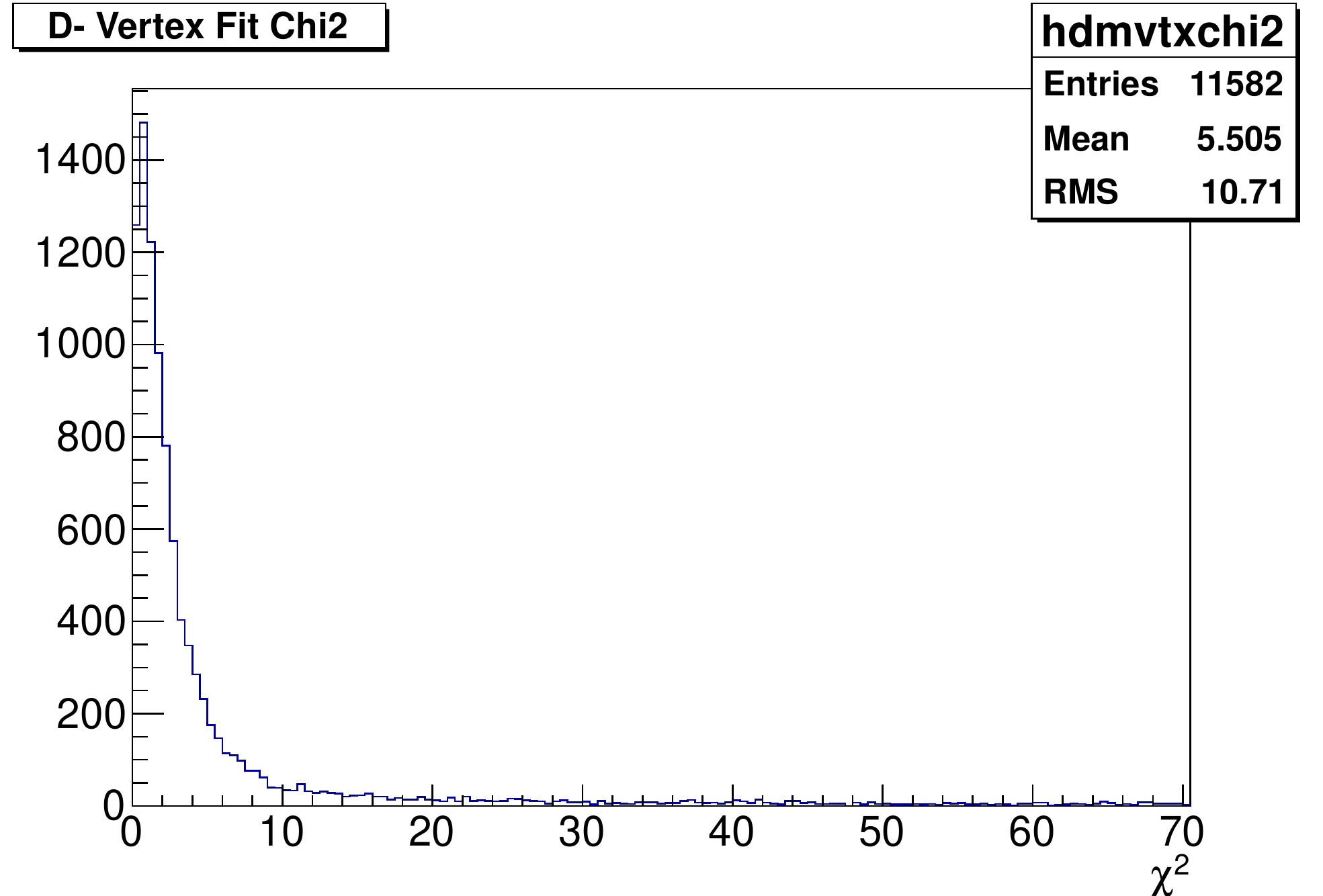}
\caption[$\chi^2$ distribution of the $D$ meson decay vertex fit]{$\chi^2$ distribution of the $D$ meson decay vertex fit, (left) $D^+$, 9right) $D^-$.}
\label{fig_hvtxchi2}
\end{figure*}

In the case of multiple $D$ meson candidates of the same charge within one event, the best one is selected based on the $\chi^2$ result from the
vertex fit. Additionally, an absolute maximum limit of $\chi^2 < 18$ has been applied as quality cut (compare \Reffig{fig_hvtxchi2}).
%As a final selection a mass difference of less than 250~\mevcc from the nominal D mass is required.

\subsubsection{Results}
%\begin{figure*}
%\centering
%\subfloat[x coordinate]{\includegraphics[angle=0,width=0.33\dwidth]{stt/fig/run936cufix/psirun936cufix_hdpvertexxfit.pdf}}
%\subfloat[y coordinate]{\includegraphics[angle=0,width=0.33\dwidth]{stt/fig/run936cufix/psirun936cufix_hdpvertexyfit.pdf}}
%\subfloat[z coordinate]{\includegraphics[angle=0,width=0.33\dwidth]{stt/fig/run936cufix/psirun936cufix_hdpvertexzfit.pdf}}
%\caption[Spatial resolution of the reconstructed $D^+$ decay vertex]{Spatial resolution of the reconstructed $D^+$ decay vertex.}
%\label{fig_hdpvertexfit}
%\end{figure*}
\begin{figure*}
\centering
\includegraphics[angle=0,width=0.33\dwidth]{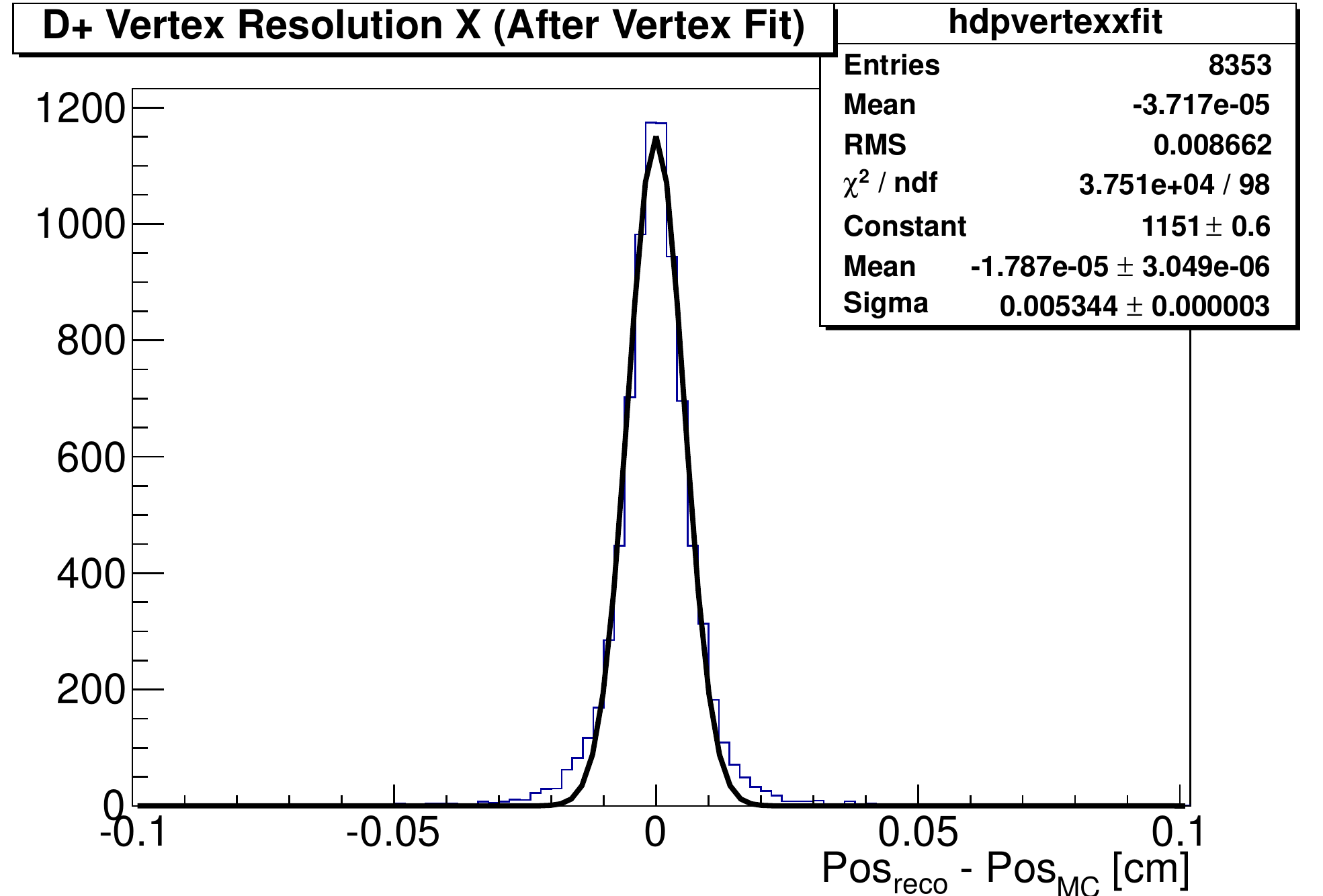}
\includegraphics[angle=0,width=0.33\dwidth]{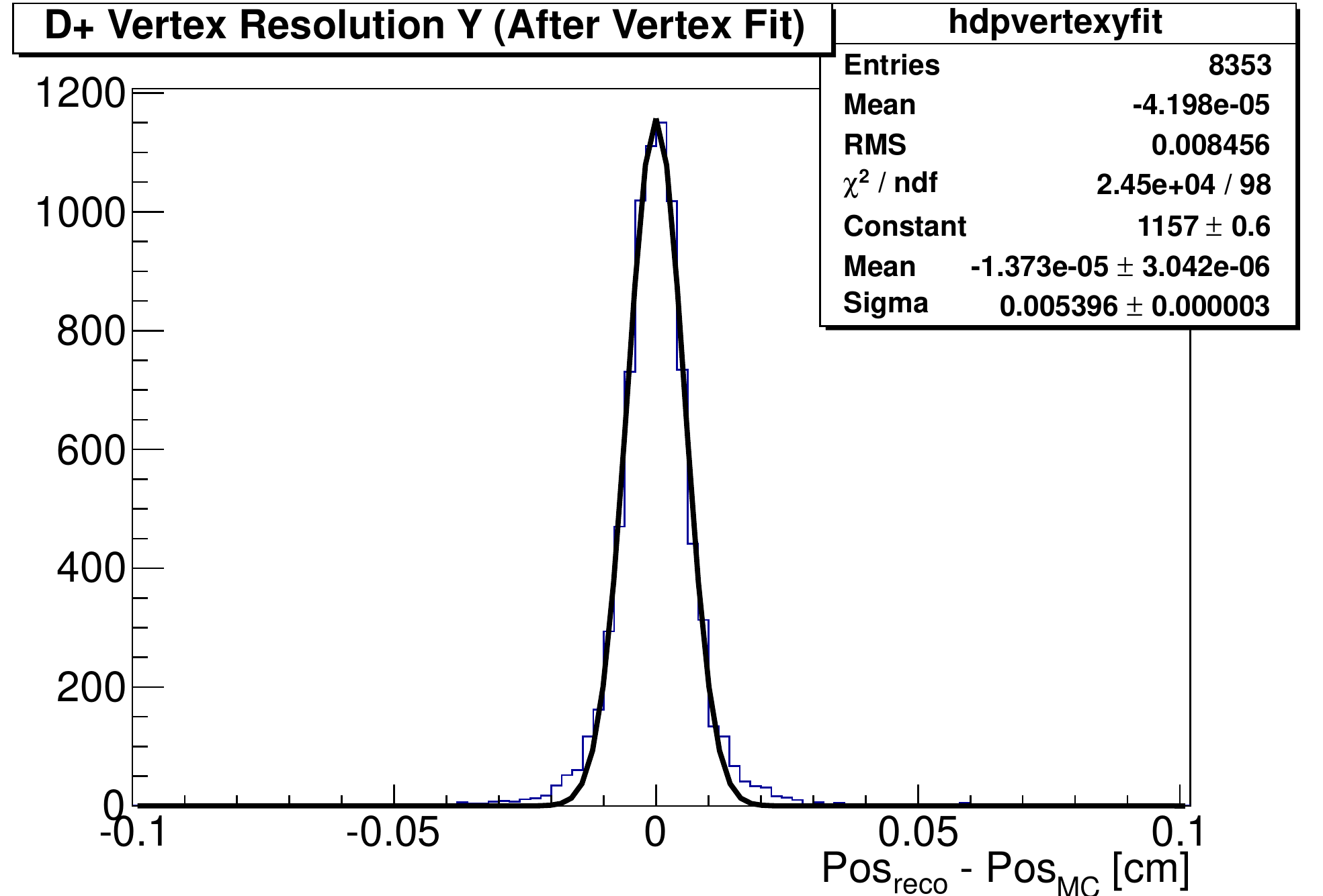}
\includegraphics[angle=0,width=0.33\dwidth]{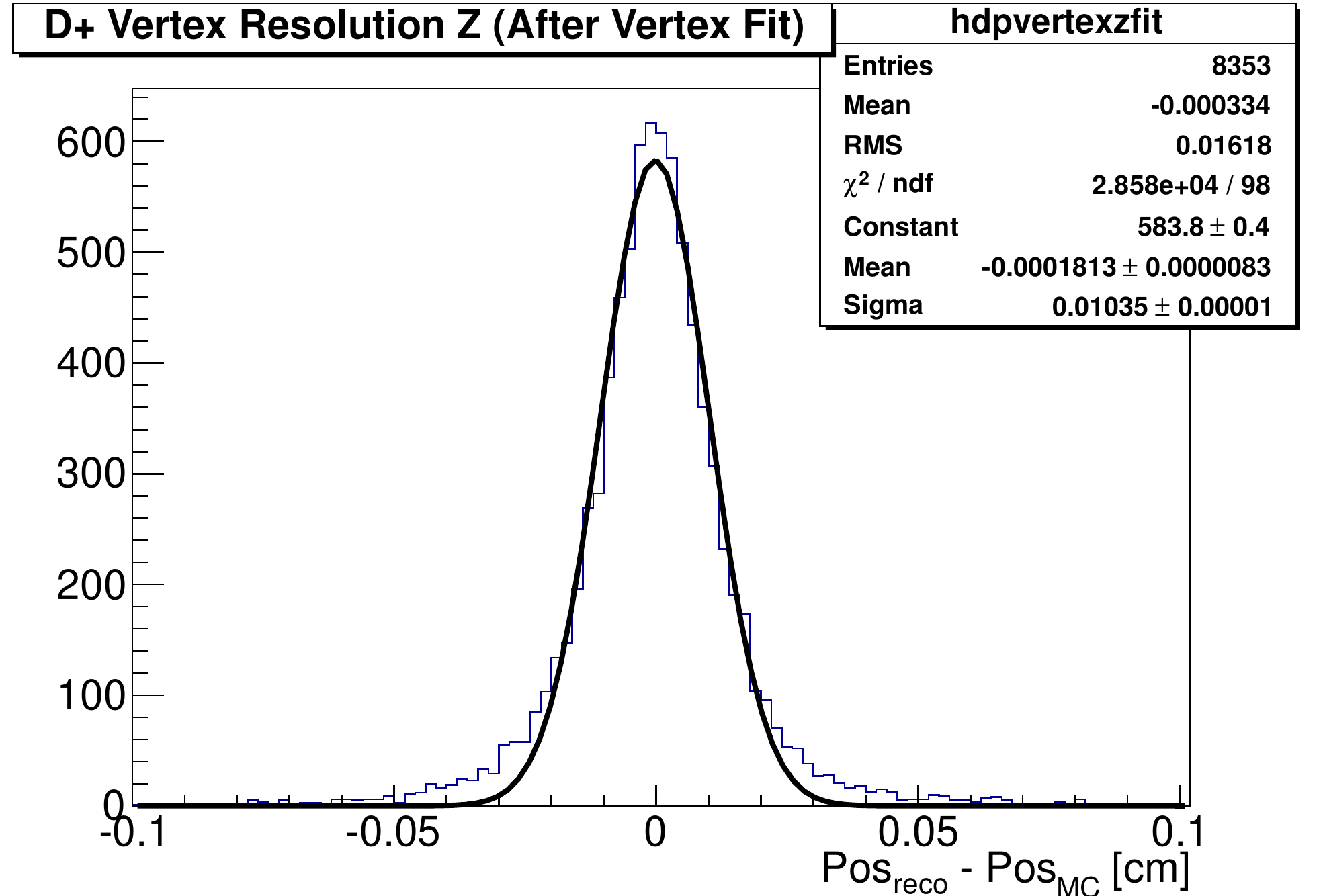}
\caption[Spatial resolution of the reconstructed $D^+$ decay vertex]{Spatial resolution of the reconstructed $D^+$ decay vertex.}
\label{fig_hdpvertexfit}
\end{figure*}
%\begin{figure}
%\centering
%\includegraphics[angle=0,width=0.5\textwidth]{stt/fig/run936cufix/psirun936cufix_hdpvertexzfit.pdf}
%\caption{Spatial resolution of the reconstructed $D^+$ decay vertex (z-coordinate).}
%\label{fig_hdpvertexzfit}
%\end{figure}
The obtained vertex resolution is in the order of 55~$\mum$ in xy-direction and 104~$\mum$ in z-direction (compare 
\Reffig{fig_hdpvertexfit}).
\begin{figure*}
\centering
\includegraphics[angle=0,width=0.98\swidth]{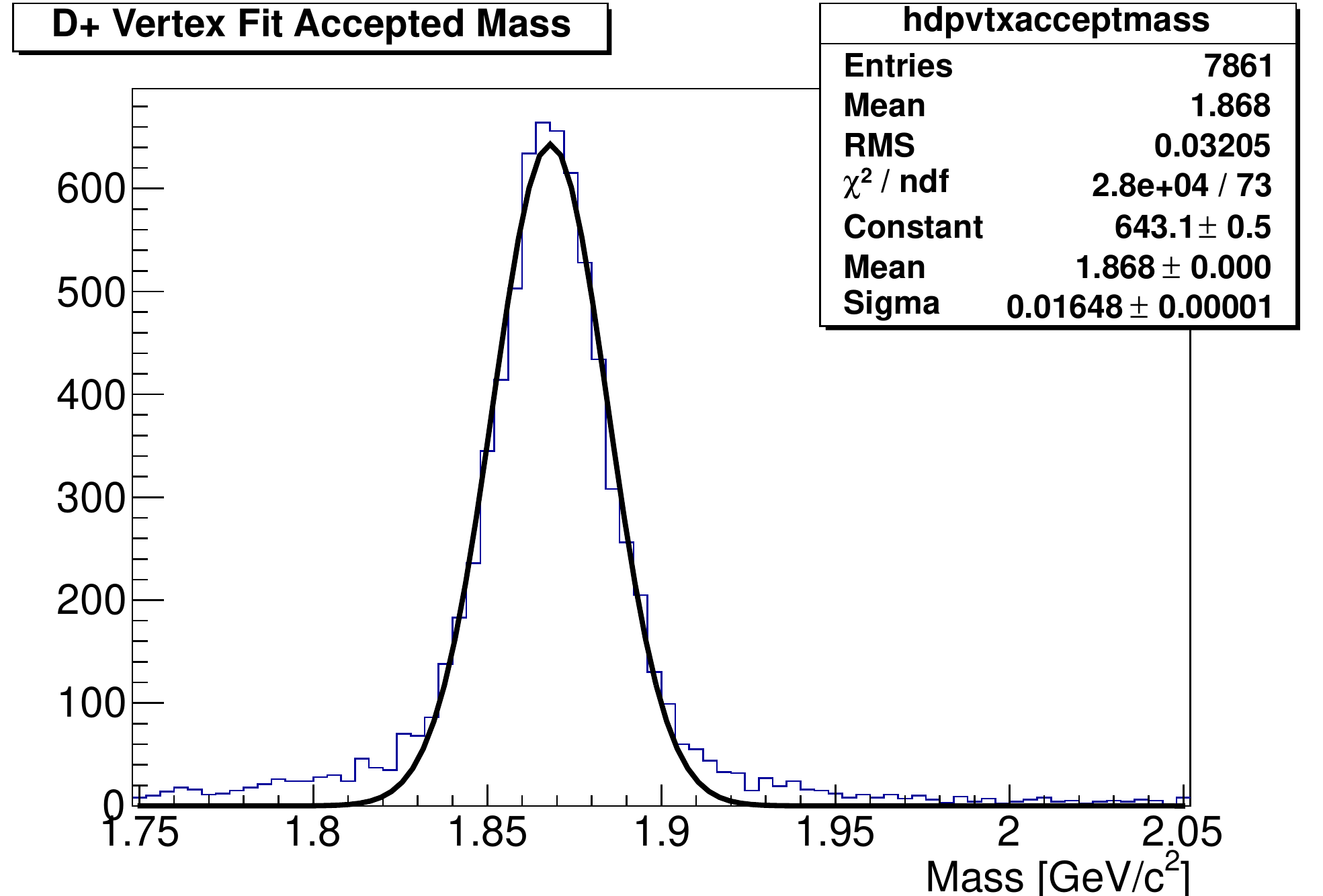}
\includegraphics[angle=0,width=0.98\swidth]{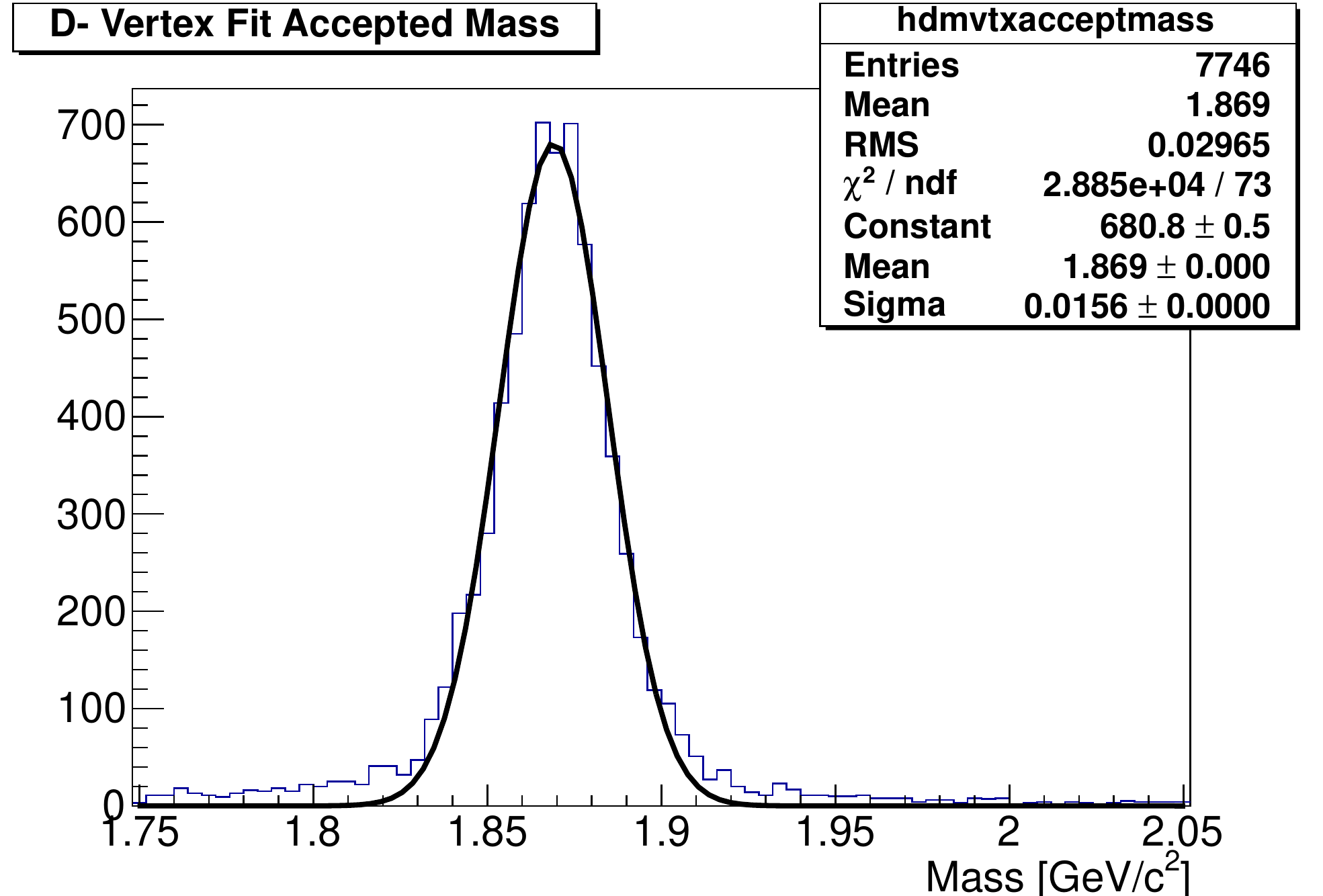}
\caption[Mass resolution of the reconstructed $D$ mesons after the vertex fit]{Mass resolution of the reconstructed $D$ mesons after the vertex fit.}
\label{fig_hdpvtxacceptmass}
\end{figure*}
%\begin{figure*}
%\centering
%\subfloat[$D^+$ mass resolution]{\includegraphics[angle=0,width=0.98\swidth]{stt/fig/run936cufix/psirun936cufix_hdpvtxacceptmass.pdf}}
%\subfloat[$D^-$ mass resolution]{\includegraphics[angle=0,width=0.98\swidth]{stt/fig/run936cufix/psirun936cufix_hdmvtxacceptmass.pdf}}
%\caption[Mass resolution of the reconstructed $D$ mesons after the vertex fit]{Mass resolution of the reconstructed $D$ mesons after the vertex fit.}
%\label{fig_hdpvtxacceptmass}
%\end{figure*}
The mass resolution is in the order of 16~\mevcc after the vertex fit (compare \Reffig{fig_hdpvtxacceptmass}).
The final reconstructed event sample consists of 5.9\,\% of the initially simulated signal events, i.e. this fraction of the events has all tracks within the STT's geometrical acceptance and also passes all quality cuts of the analysis resulting in both $D$ mesons of the event being successfully reconstructed. Thus, this number is a convolution of geometrical acceptance, reconstruction software performance and the settings of the quality cuts.

\begin{figure*}
\centering
\includegraphics[angle=0,width=0.98\swidth]{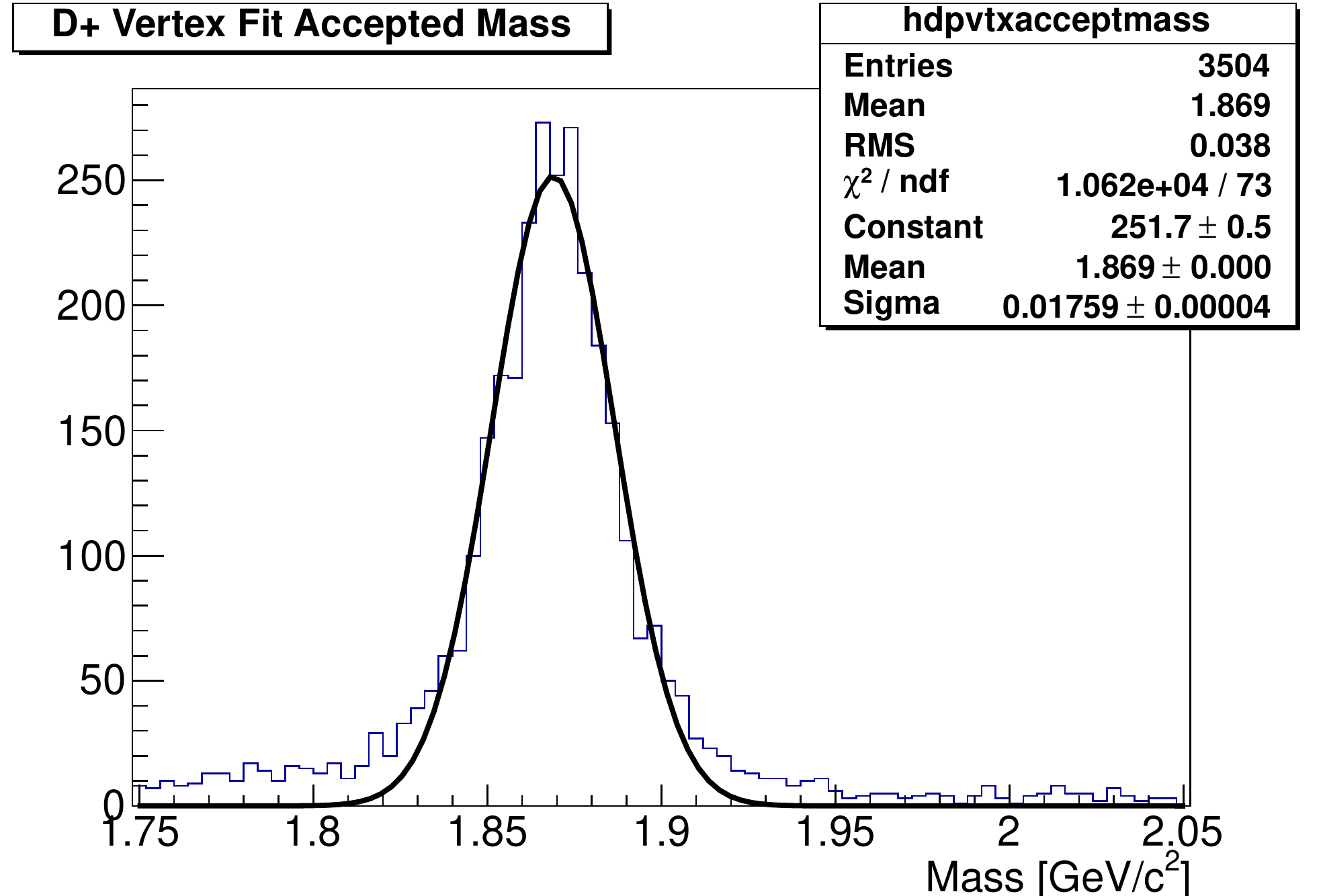}
\includegraphics[angle=0,width=0.98\swidth]{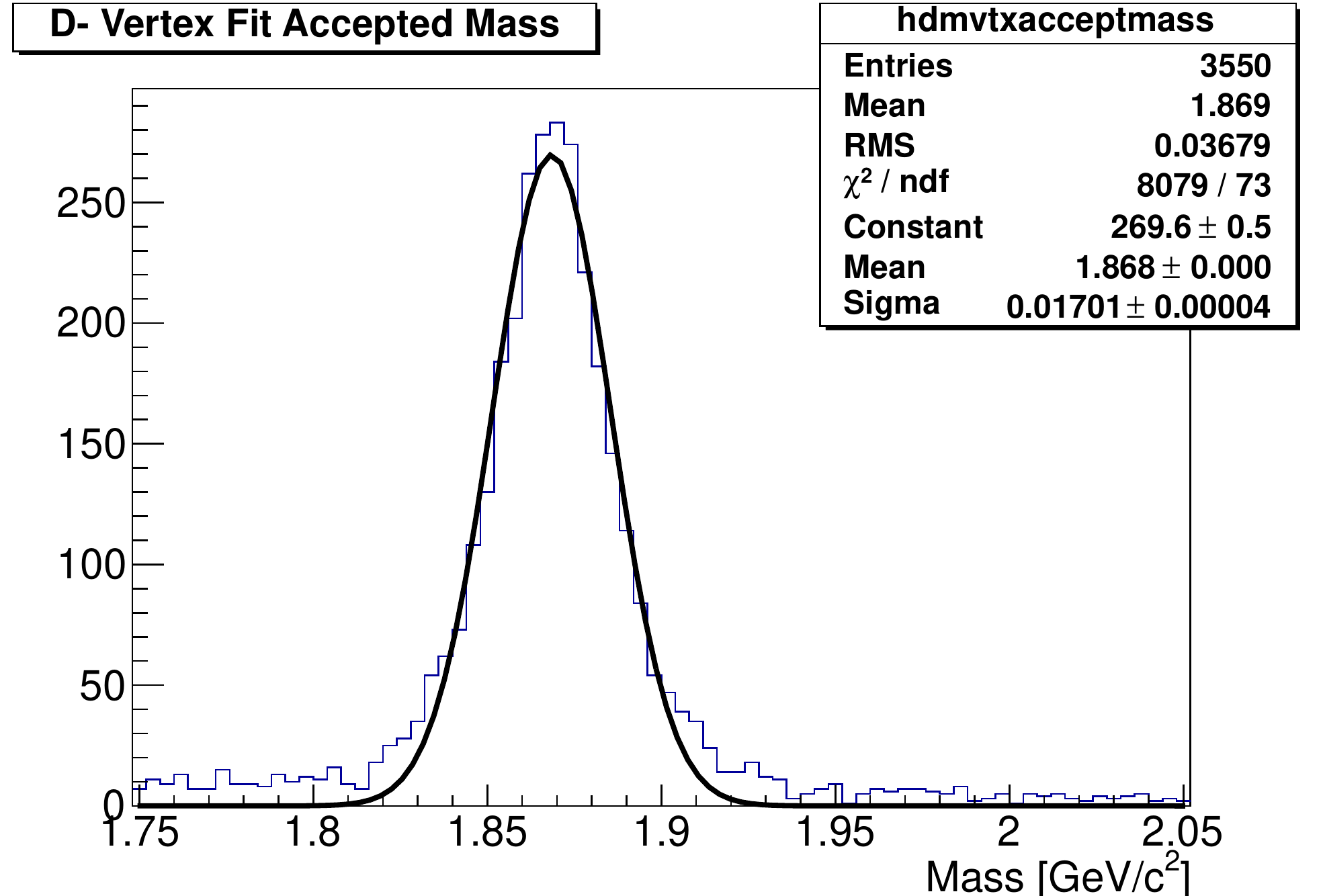}
\caption[Mass resolution of the reconstructed $D$ mesons after the vertex fit for mixed events]
{Mass resolution of the reconstructed $D$ mesons after the vertex fit for mixed events.}
\label{fig_mixhdpvtxacceptmass}
\end{figure*}

In addition to the pure signal, also mixed events where each signal event is overlayed with additional background events 
(compare \Refsec{sec:stt:ben:env}), have been analyzed to study the behavior of the reconstruction software under the influence of 
the additional particle tracks. The main influence of the mixed events is on the reconstruction software's efficiency which reduces the final reconstructed event sample to 3.3\,\% of the initially simulated signal events.
The resulting vertex resolutions are in the order of 61~$\mum$ (xy-direction) and 109~$\mum$ 
(z-direction). The mass resolution is in the order of 17~\mevcc. These resolutions are virtually the same as those obtained during the 
analysis of the pure signal and also the shapes of the distributions do not show any significant differences 
(compare \Reffig{fig_mixhdpvtxacceptmass}).
%\begin{figure*}
%\centering
%\subfloat[$D^+$ mass resolution]{\includegraphics[angle=0,width=0.98\swidth]{stt/fig/runmix403/psirunmix403_hdpvtxacceptmass.pdf}}
%\subfloat[$D^-$ mass resolution]{\includegraphics[angle=0,width=0.98\swidth]{stt/fig/runmix403/psirunmix403_hdmvtxacceptmass.pdf}}
%\caption[Mass resolution of the reconstructed $D$ mesons after the vertex fit for mixed events]{Mass resolution of the reconstructed $D$ mesons after the vertex fit for mixed events.}
%\label{fig_mixhdpvtxacceptmass}
%\end{figure*}

% EOF 

%EOF: panda_tdr_stt_ben.tex

%
% Bibliography for this chapter (remove %)
%
\bibliographystyle{panda_tdr_lit}
\bibliography{./stt/lit_stt}
% EOF

%
% STT TDR
% File for chapter 7
\chapter{Online Tracking}
% FILE: panda_tdr_stt_onl.tex
%
%\COM{Author(s): A. Tomaradze}
\label{sec:stt:onl}
Effective online track finding and reconstruction is crucial for fulfilling the 
goals of the \PANDA experiment.
Charged particles are used in most trigger objects, and robust, accurate charged 
particle reconstruction is essential for triggering on states, like $J/\psi$ and 
$D^0$, and on interesting topologies such as displaced vertices.

\section{Comparison with Existing Experiments}
The challenge for \PANDA is to determine track parameters from the charged particles 
produced in  $p\bar{p}$ collisions with an effective interaction rate of 20~MHz.  
The parameters for \PANDA are compared to other comparable recent experiments in 
\Reftbl{tbl:stt:onl:onlinetrk}.  The $e^+e^-$ experiments have higher bunch crossing 
rates, but a lower rate of physics events due to the lower cross-sections of those 
collisions.  However, even the older CLEO~III experiment performs pattern 
recognition over its entire drift chamber at a $\sim20$~MHz rate.  
The high--energy $p\bar{p}/pp$ experiments at the Tevatron and LHC have rates 
comparable to those expected for \PANDA, and so are a better comparison. CDF and 
D\O~both perform track finding and fitting in their trigger systems, and upgrades 
for CMS and ATLAS are being planned that would allow for full online track reconstruction 
and fitting, in much larger detectors and at twice the rate of \PANDA. 
A simple order-of-magnitude comparison is illustrative of the challenges \PANDA 
faces.  The \PANDA STT has roughly the same number of channels as D\O's fiber tracker, 
and an order of magnitude fewer channels than CDF's main drift chamber.  Although 
\PANDA's expected event rate is 2--3 times that seen at the Tevatron, the expected 
track multiplicities are an order of magnitude smaller.  Since the hardware used for 
online tracking in \PANDA is of similar or faster speed than CDF and D\O's systems, 
we fully expect that accurate online track reconstruction is possible at \PANDA.
\begin{table*}[!tb]
\begin{center}
\caption[Summary of event rate and tracking chamber parameters for various experiments comparable to PANDA]{Summary of event rate and tracking chamber parameters for various experiments comparable to \PANDA. For the \PANDA-STT the average number of layers is given.}
\smallskip
\begin{tabular}{lccccc}
\hline
      & event rate & trigger rate & avg. track & layers & cell size  \\  % & $p$ resolution \\
 &  &  (L1/(L2)/L3)     & multi.     &        &  (mm)              \\ %   & (online/final)\\ 
%% \multicolumn{2}{r}{ \quad\qquad ($\mathcal{L}$, $10^{33}~\mathrm{cm}^{-2}s^{-1}$)}  &  (L1/(L2)/L3)     & multi.     &        &  (mm)  &                  \\ %   & (online/final)\\ 
\hline
\multicolumn{6}{l}{$e^+e^-$ Experiments} \\
\hline
\\[-5pt]
CLEO~III \cite{bib:stt:onl:onlinecleo1,bib:stt:onl:onlinecleo2} &  250kHz    &   $<1$kHz/130Hz            &  \multirow{3}{*}{\parbox{1.8cm}{$\sim8~(B\overline{B})$\linebreak $2~(e^+e^-)$}}   
         &  47       &   7    \\ % &  /0.35\%    \\
BaBar \cite{bib:stt:onl:onlinebabar} &    2kHz  & 970Hz/120Hz   &    &   40   & $6-8$  \\  % &  1.9\%/0.47\% \\ 
Belle \cite{bib:stt:onl:onlinebelle} &    5kHz  &  500Hz/500Hz   &    &    50 & $8-10$  \\  % & /0.4\% \\
BES--III \cite{bib:stt:onl:onlinebesiii} &  $\sim3$kHz   &   $>4$kHz/3kHz            & $\sim4$  &    43    &   $6-8$    \\
%CLEO~III &  $10^{33}$       &   $>150$Hz/130Hz            &  \multirow{3}{*}{\parbox{1.4cm}{$\sim8~(B\overline{B})$\linebreak $2~(e^+e^-)$}}   
%         &  47       &   7    & $\sim99\%$ \\ % &  /0.35\%    \\
%BaBar &    $3\times10^{33}$        & 970Hz/120Hz   &    &   40   & $6-8$ &  $\sim94\%$ \\  % &  1.9\%/0.47\% \\ 
%Belle &  $>10^{34}$  &  500Hz/500Hz   &    &    50 & $8-10$ & $>90\%$ \\  % & /0.4\% \\
%BES--III &  $\sim10^{33}$       &   $>4$kHz/1kHz            & $\sim4$  &    43    &   $6-8$    & $\sim99\%$ \\ % &     /0.5\% \\
\\[-5pt]
\hline
\multicolumn{6}{l}{$ep$ Experiments} \\
\hline
\\[-5pt]
ZEUS  \cite{bib:stt:onl:onlinezeus1,bib:stt:onl:onlinezeus2} & \multirow{2}{*}{$\sim\!1$MHz}  &  600Hz/100Hz/20Hz  &  \multirow{2}{*}{$\sim10$} & 72  & $\sim25$ \\ 
H1  \cite{bib:stt:onl:onlineh11,bib:stt:onl:onlineh12,bib:stt:onl:onlineh13} &   &  1kHz/200Hz/50Hz/$\sim\!10$Hz  &  & 56  & $23\!-\!43$   \\ 
%ZEUS  & $\sim5-8\times10^{31}$  &  600Hz/100Hz/20Hz  &  & 72  & $\sim25$ & $\sim70\!-\!90\%$ \\ 
%H1  & $\sim5-8\times10^{31}$  &  1kHz/200Hz/50Hz/$\sim\!5$Hz  &  & 56  & $23\!-\!43$ &  \\ 
\\[-5pt]
\hline
\multicolumn{6}{l}{$pp+p\bar{p}$ Experiments} \\
\hline
\\[-5pt]
%CDF  & $\sim2\times10^{32}$  &  25kHz/350Hz/75Hz  &  & 96 & 8.8~mm &  96\% & $\sim1\%$/ \\
CDF  \cite{bib:stt:onl:onlinecdf1,bib:stt:onl:onlinecdf2,bib:stt:onl:onlinecdf3} & \multirow{2}{*}{7.5MHz}  &  30kHz/750Hz/75Hz  & \multirow{2}{*}{$\sim35$} & 96 & 8.8  \\ %& $\sim1\%$/ \\
D\O~\cite{bib:stt:onl:onlined01,bib:stt:onl:onlined02} &   &  10kHz/1.5kHz/50Hz &  & 32 & 0.4  \\[4pt]  %&   \\
CMS  \cite{bib:stt:onl:onlinecms} & \multirow{2}{*}{$\le\!40$MHz}  &  100kHz/$\sim$100Hz  &  \multirow{2}{*}{$>100$} & $\sim12$  & --- \\ % & $\sim\%$/    \\
ATLAS  \cite{bib:stt:onl:onlineatlas1,bib:stt:onl:onlineatlas2} &    &  100kHz/2kHz/200Hz  &  & $36$  & 2  \\ % & $\sim\%$/    \\
\\[-5pt]
\PANDA & $\sim$20MHz    & & $\sim4\!-\!6$ & 24 & 10  \\

%CDF  & $\sim2\times10^{32}$  &  30kHz/750Hz/75Hz  & $\sim35$ & 96 & 8.8 &  96\% \\ %& $\sim1\%$/ \\
%D\O  & $\sim2\times10^{32}$  &  10kHz/1.5kHz/50Hz &  & 32 & 0.4 &  $\sim95\%$ \\  %&   \\
%CMS  & $10^{33}-10^{34}$  &  100kHz/100Hz  &  $>100$ & $\sim12$  & --- &  85--98\% \\ % & $\sim\%$/    \\
%ATLAS &   $10^{32}-10^{34}$  &  100kHz/2kHz/200Hz  &  & $\sim36$  & 2 &  $>90\%$ \\ % & $\sim\%$/    \\
%%%PHENIX &  \\
\\[-5pt]
\hline
\end{tabular}
\end{center}

%\end{scriptsize}
%\end{footnotesize}
%\end{small}

\label{tbl:stt:onl:onlinetrk}
\end{table*}

\section{Online Strategy for \Panda}
The online track reconstruction will take advantage of the ``Compute Nodes'', which 
each contain several modern FPGAs with large associative memories that run at high 
clockspeeds.  
For example, track finding is easily parallelizable by segmenting the detector so 
multiple nodes can simultaneously search for tracks over the entire geometry.  The 
low track multiplicity further simplifies the problem, and should allow for a 
simplified fitting algorithm.
\par
Traditional trigger systems have a well--defined heirarchy of levels.  However, with 
\PANDA's ``triggerless'' design, it makes more sense to describe the series of tasks 
that are performed.  An example of such a flow is: Track Segment Finding $\to$ Track 
Linking $\to$ Track Fitting.  The algorithms used by the experiments given in 
\Reftbl{tbl:stt:onl:onlinetrk} are all generally similar, with tracks being found by 
algorithms such as a Hough transformation or road following algorithm, and then fitted 
by a simple $\chi^2$ fit.  The determining factor, then, is how these algorithms perform 
for the topologies that PANDA will eventually trigger on.  Several algorithms are being 
implemented, and will be tested for their efficiency and robustness against pathologies 
such as displaced vertices and low--momentum tracks that curl in the magnetic field 
of the detector, and will be benchmarked using the physics channels given in 
\Refsec{sec:stt:ben}.

%EOF: panda_tdr_stt_onl.tex

%
% Bibliography for this chapter (remove %)
%
\bibliographystyle{panda_tdr_lit}
\bibliography{./stt/lit_stt}
% EOF

%
% STT TDR
% File for chapter 8
\chapter{Organization}
% FILE: panda_tdr_stt_org.tex
%
%\COM{Author(s): P. Gianotti, J. Ritman}
\section{Production Logistics}
The \Panda-STT is a modular detector. It consists of a set of individual detector components which can be
produced and tested independently before the final installation. 
\par
The construction procedure (see \Refsec{sec:stt:lay:ass}) is such that the realization of the whole 
detector can be easily split on different sites. First of all, single tubes are assembled 
and tested individually; then multi-layer modules are realized; finally, the modules are mounted in the 
mechanical support.
This will be extremely useful to reduce the time needed for the detector construction.
\par
Even if the straw tubes are operated with an overpressure of about 1 bar of the gas mixture, they still have
sufficient strength to maintain the 50 g wire tension without overpressure. Therefore, single tubes can be 
constructed and stored in appropriate places to be ready for the multi-layer assembling later on.
\par
\begin{figure}
\centering
\includegraphics[width=0.95\swidth]{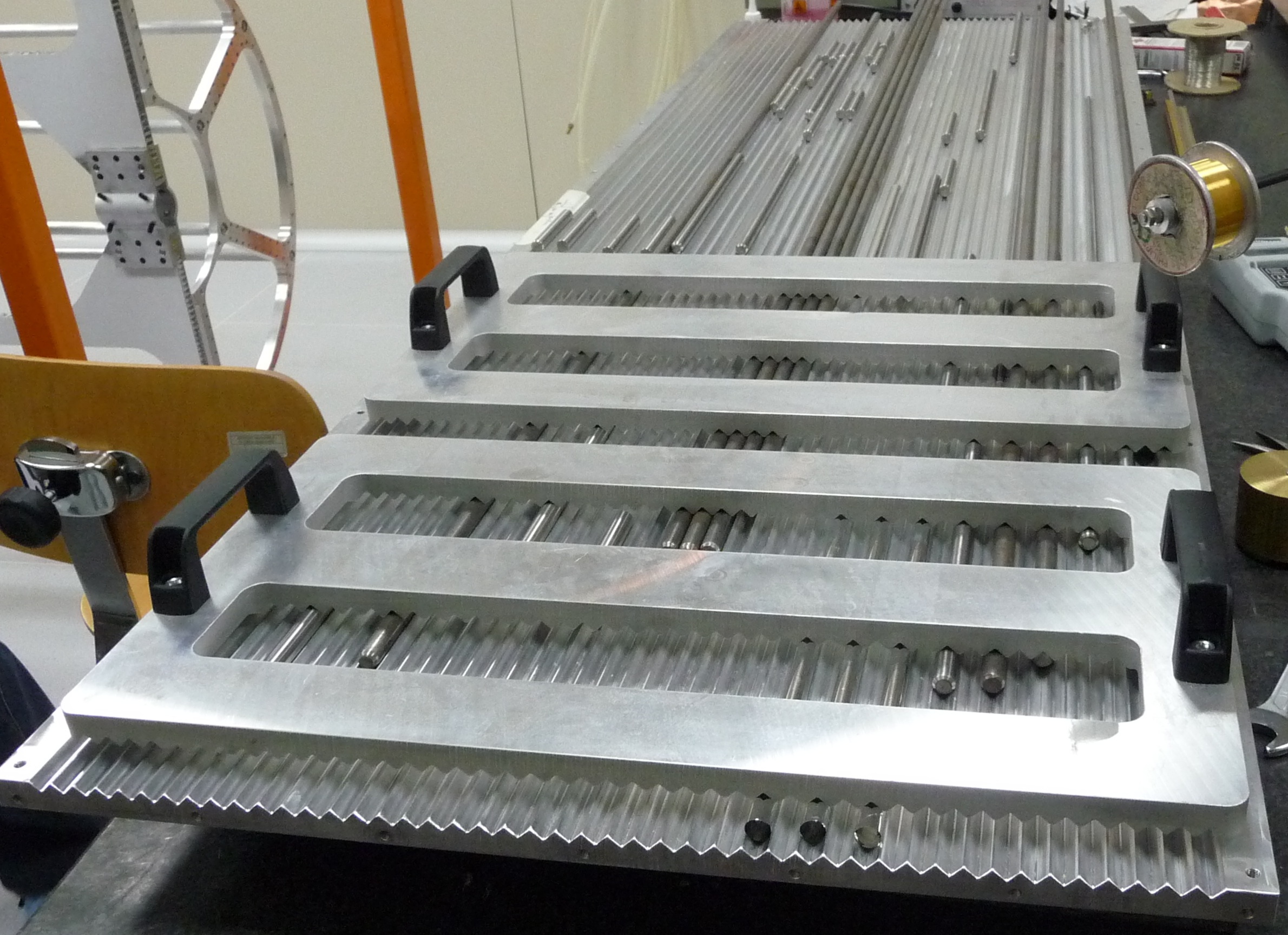}
\caption[v-shaped reference plate for multi-layer assembly]{v-shaped reference plate for multi-layer assembly.}
\label{fig:stt:org:pl}
\end{figure}

At present, both the Forschungszentrum J\"ulich and the LNF laboratories are equipped with the necessary
tools to start the production of straw tubes and modules. They both have a clean room (10000 class), a 
v-shaped reference plane (\Reffig{fig:stt:org:pl}) and the same expertise for starting a mass production.
However, a clean room is not absolutely necessary. The straw tube components (cathodes, pins, 
end-plugs, etc.) have to be cleaned before starting the assembly procedure. Thus, if this work is 
realized in an environment where the dust concentration is under control, this could be done in one step 
before starting the single tubes production. If a clean room is not available, greater care has to be placed 
in this cleaning phase, which has to precede the straw tube assembly.
%Therefore, if the mass production should be speed up, new sites could be added provided they acquire a 
%reference plate.
The only required tool to produce straw tube modules is the v-shaped reference plane. This object is a precisely 
manufactured mechanical plate that guarantees that the spacing of the straw wires within a multi-layer meets the
specifications ($\pm$ 50 $\mu$m).
%In any case, it could be decided that single straw production is performed on several sites, while the 
%assembling of multi-layer modules will be done only at LNF and FZ J\"ulich.

\section{Safety}
The design details and construction of the STT including the infrastructure for 
operation will be done according to the safety requirements of FAIR and the 
European and German safety regulations. All electrical equipment and gas systems 
will comply with the legally required safety code and concur to standards for 
large scientific installations to ensure the protection of all personnel working 
at or close to the components of the \PANDA experimental facility. Hazardous voltage 
supplies and lines will be marked visibly and protected from damage by any equipment 
which may cause forces to act on them. All supplies will be protected against 
over-current and over-voltage and have appropriate safety circuits and fuses. 
All cabling and connections will use non-flammable halogen-free materials according 
to up-to-date standards and will be dimensioned with proper safety margins to prevent 
overheating. A safe ground scheme will be employed throughout all electrical installations 
of the experiment. Smoke detectors will be mounted in all appropriate locations. 
The gas system is based upon non-flammable gases and thus does not pose a fire hazard. 
The maximum pressure of the gas will be regulated, and the system is designed such 
that a sudden failure of one tube (operating at maximally 2 bar) can not damage the 
adjacent tubes (that have equal or higher pressure than the escaping gas), and thus a
chain reaction is ruled out. Appropriate measures will be taken during installation 
and maintenance to avoid damage to or by the STT. The outer foil will protect the 
device against potential condensation risks from other components of \PANDA. More specific 
safety considerations are discussed in the respective sections throughout this document.

\begin{figure*}[!h]
\centering
%\vskip -1.5cm
\includegraphics[width=\dwidth]{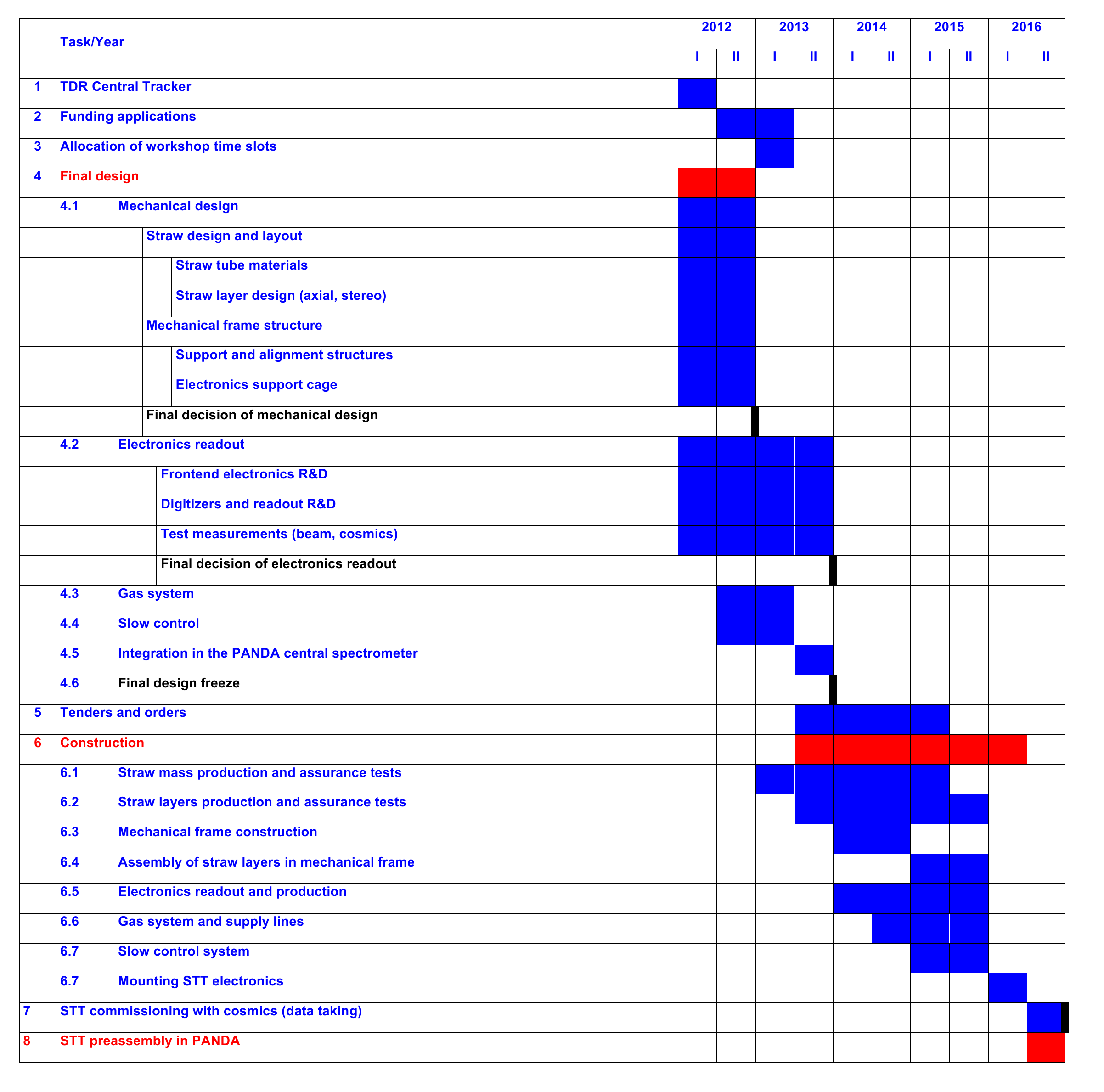}
%\vskip -4.5cm
\caption[Timeline for the STT realization]{Timeline for the STT realization. Milestones are marked in black.}
\label{fig:stt:org:tl}
\end{figure*}

\section{Timeline}
The projected timeline of the STT construction is based on the 
experience gained during the R$\&$D and prototyping phase of the STT project and
similar, former detector projects of the different group institutions. In particular
the construction experience from the Straw Tube Tracker at COSY, consisting of 
about 3200 straws in total and similar assembly techniques, is a major input for the 
definition of the construction packages
and time scales for the \Panda-STT realization. 
\par
The timeline of the STT construction consists of four
main periods: the completion of the final design; the main construction phase; 
the final assurance test of the assembled STT with cosmics; and the pre-assembly of the STT 
in the \PANDA target spectrometer. The detailed timeline shown in \Reffig{fig:stt:org:tl} lists the different 
construction items and their projected time slots. After the TDR is approved, the 
funding applications will start and the workshop time slots for the project will be booked.  
\par
The first phase of the STT construction with the definition of the complete final design has already started 
and will be finished by the middle of 2013. Here, all designs, the mechanical parts with straws 
and frame structure as well as the electronic readout parts, cables, and systems, have to be designed in detail. 
The integration of the detector in the \Panda central spectrometer will be made and the main dimensions, 
environment conditions, and requirements will be defined and assigned. Few aspects are still under discussion 
concerning the final mechanical layout, including cable and services routing, and the final electronic readout 
design. The latter decision between the two current electronic readout options will be based on further test 
measurements with existing prototype setups, consisting of up to 400 straws. Within the next 12 months this 
decision will be made, latest by the middle of 2013. 
\par
At the end of the design phase all technical specifications are fixed and technical drawings are prepared to 
start the tendering and order processes at the external production companies. Since most of the involved 
production companies have been already contacted during the R$\&$D phase of the project or during former 
similar projects, we estimate not more than about 18 months for sending out tenders, placing orders, and 
delivery times for the external components.    
\par
The main construction phase of the STT assembly will take about 30 months for the mass production of 6000 
straw tubes (about 30\,\% spare), gluing of the straw layers, construction of the mechanical frame, and 
insertion of the straw layers in the frame structure. The assembly will be split between the two main 
involved laboratories, Forschungszentrum J\"ulich and 
INFN Frascati. At both sites clean rooms (class 10000) are available for the construction. The straw mass 
production already includes assurance tests of gas leakage and wire tension measurements of each assembled 
straw. Individual straws showing gas leakage, deviation from the nominal wire tension, or broken wires are 
rejected. For the previous prototype constructions with about 1000 straws, the fraction of straws showing such 
failures was about one percent.
\par
The straws are glued to the axial and stereo layer modules with integrated 
gas manifolds and electronic coupling boards. For the final assurance test of all straws in a layer module, the 
module is flushed with an Ar/CO$_2$ gas mixture, straws are set on high voltage, a test board containing a 
preamplifier circuit is connected to the coupling boards and the signals from cosmic tracks are checked to 
identify dead or improper straws. Bad identified straws are removed from a layer module and replaced by single 
new straws. The modular layout of the STT allows to carry out the most time consuming construction steps of the 
straw mass production and layer module assembly highly in parallel. As soon as the first couple of hundred straws are 
produced and tested, we can start with the construction of the first layer modules. After the completion of the 
mechanical frame structure the layer modules are inserted and fixed to the frame. 
\par
In parallel to the mechanical STT assembly the electronic parts, cables, and readout boards will be produced and the 
complete readout system will be set up. After a test of all electronic channels with test pulses, the readout will be 
mounted in the STT mechanical frame structure and connected to the straws. By the first half of 2016 the construction phase 
will be finished including the setup of the gas system and slow control system.
\par
In the second half of 2016 the final commissioning of the full STT detector will be done with data-takings of cosmic ray 
tracks to set up the whole electronic readout and to calibrate the STT geometry with reconstructed tracks. 
After finishing these tests the detector will be ready for installation and pre-assembly in the PANDA central spectrometer.

\section{Work Packages and Contributing Institutes}
\begin{table}[t]
\begin{center}
\caption[Work package list with involved institutions]{Work package list with involved institutions.}
\begin{tabular}{c|c}
Work package & Involved institutes \\
\hline
                     &   {\bf FZ-J}  \\
Straw tube materials &   LNF   \\
\hline
                  &    {\bf LNF} \\
Mechanical frame  &    FZ-J \\ 
\hline
                     & {\bf FZ-J} \\
Front End Electronics, DAQ,    & AGH \\
Low and High voltages& {\bf JU} \\
                     & IFJ \\
                     & GSI \\
\hline
                & LNF \\ 
Crates, cables  & FZ-J\\
                & IFIN-HH\\
\hline
Slow control  & {\bf IFIN-HH} \\
\hline
Gas system    & {\bf LNF} \\
\hline
Online tracking & {\bf NU} \\
\hline
Monitoring calibration & {\bf PV} \\
                       & Fe \\
\end{tabular}
\label{tbl:stt:org:wp}
\end{center}
\end{table}
The design, construction and installation of the STT will be performed by a number of 
institutions which have gained specific expertise in past and ongoing large scale experiments 
at several accelerator facilities. The responsibilities for the various work packages are listed in \Reftbl{tbl:stt:org:wp}, 
in which the coordinating group of the task is denoted by boldface. A summary of the participating groups and of 
their members is given below:
\begin {itemize}
\item IFIN-HH Bukarest-Magurele, Romania (M.~Bragadireanu, M.~Caprini, D.~Pantea, D.~Pantelica, D.~Pietreanu, L.~Serbina, P.D.~Tarta) ({\bf IFIN-HH}); 
\item IFJ PAN, Cracow, Poland (B.~Czech, M.~Kistryn, S.~Kliczewski, A.~Kozela, P.~Kulessa, P.~Lebiedowicz, K.~Pysz, W.~Sch\"afer, R.~Siudak, A.~Szczurek) ({\bf IFJ});
\item Jagiellonian University of Cracow, Poland (S.~Jowzaee, M.~Kajetanowicz, B.~Kamys, S.~Kistryn, G.~Korcyl, K.~Korcyl, W.~Krzemien, A.~Magiera, P.~Moskal, Z.~Rudy, P.~Salabura, J.~Smyrski, A.~Wro\~nska) ({\bf JU});
\item AGH Cracow, Poland (T.~Fiutowski, M.~Idzik, B.~Mindur, D.~Przyborowski, K.~Swientek) ({\bf AGH});
\item Gesellschaft f\"ur Schwerionenforschung GmbH, Darmstadt, Germany (M. Traxler) ({\bf GSI});
\item INFN Frascati, Italy (N. Bianchi, D. Orecchini, P. Gianotti, C.~Guaraldo, V. Lucherini, E. Pace) ({\bf LNF}); 
\item FZ J\"ulich, Germany (A.~Erven, G. Kemmerling, H. Kleines, V.~Kozlov,  N.~Paul, M.~Mertens, R.~Nellen, H.~Ohm, S.~Orfanitski,  J.~Ritman,  T.~Sefzick, V.~Serdyuk, P.~Wintz, P.~W\"ustner) ({\bf FZ-J});
\item INFN and Univ. of Pavia, Italy  (G.~Boca, A.~Braghieri, S.~Costanza, P.~Genova, L.~Lavezzi, P.~Montagna, A.~Rotondi) ({\bf PV});
\item INFN and Univ. of Ferrara, Italy (D.~Bettoni, V.~Carassiti, A.~Cotta~Ramusino, P.~Dalpiaz, A.~Drago, E.~Fioravanti, I.~Garzia, M.~Savri\`e, G.~Stancari) ({\bf Fe});
\item Northwestern Univ., Evanston U.S.A. (S.~Dobbs, K.~Seth, A.~Tomaradze, T.~Xiao) ({\bf NU}).
\end{itemize}

% this section will rather be an appendix and not in the publically available document
%\section{Financing}

%
%EOF: panda_tdr_stt_org.tex

%
% Bibliography for this chapter (remove %)
%
%\bibliographystyle{panda_tdr_lit}
%\bibliography{./stt/lit_stt}
% EOF

%
\cleardoublepage
\onecolumn
% thanks.tex
% acknowledgements
%
\begin{center}
\vspace*{2cm}
{\Large\bf Acknowledgments}\addcontentsline{toc}{chapter}{Acknowledgements}
\vskip 2cm
\begin{minipage}[t]{8cm}
\sloppy\large
We acknowledge financial support from
the Bundesministerium f\"ur Bildung und Forschung (bmbf),
the Deutsche Forschungsgemeinschaft (DFG),
the Forschungszentrum J\"ulich GmbH,
the University of Groningen, Netherlands,
the Gesellschaft f\"ur Schwerionenforschung mbH (GSI), Darmstadt,
the Helmholtz-Gemeinschaft Deutscher Forschungszentren (HGF),
the Schweizerischer Nationalfonds zur F\"orderung der wissenschaftlichen
Forschung (SNF),
the Russian funding agency ``State Corporation for Atomic Energy Rosatom'',
the CNRS/IN2P3 and the Universit\'e Paris-sud,
the British funding agency ``Science and Technology Facilities
Council'' (STFC),
the Instituto Nazionale di Fisica Nucleare (INFN),
the Swedish Research Council,
the Polish Ministry of Science and Higher Education,
the European Community FP6 FAIR Design Study: DIRAC secondary-Beams,
contract number 515873,
the European Community FP7 Integrated Infrastructure Initiative:
HadronPhysics2, contract number 227431,
the INTAS, 
and the Deutscher Akademischer Austauschdienst (DAAD).\par
\end{minipage}
\end{center}
\vfill
%
% EOF
%

%
\cleardoublepage
\twocolumn
% acronyms.tex
%
\addcontentsline{toc}{chapter}{List of Acronyms}
\begin{acronym}
\acro{ADC}{Analog to Digital Converter}
\acro{ABS}{Acrylonitrile Butadiene Styrene}
\acro{ASIC}{Application Specific Integrated Circuit}
\acro{ASCII}{American Standard Code for Information Interchange}
\acro{BES}{BEijing Spectrometer}
\acro{BLH}{Base-line Holder}
\acro{CAD}{Computer Aided Design}
\acro{CBM}{Compressed Barionic Matter}
\acro{CDR}{Conceptual Design Report}
\acro{CERN}{Conseil European pour la Recherche Nucleaire}
\acro{CF}{Carbon Fiber}
\acro{CF}{Central Frame}
\acro{CMOS}{Complementary Metal-Oxide Semiconductor}
\acro{CN}{Computing Node}
\acro{COSY}{Cooler Synchrotron}
\acro{CPU}{Central Processing Unit}
\acro{CT}{Central Tracker}
\acro{CVS}{Code Versioning System}
\acro{DAC}{Digital-to-Analog Converter }
\acro{DAQ}{Data Acquisition}
\acro{DB}{Digital Board}
\acro{DCB}{Detector Concentrator Board}
\acro{DCS}{Detector Control System}
\acro{DESY}{Deutsches ElektronenSYnchrotron}
\acro{DIRC}{Detector for Internally Reflected Cherenkov Light}
\acro{DPM}{Dual Parton Model}
\acro{DVCS}{Deeply Virtual Compton Scattering}
\acro{ECS}{Experiment Control System}
\acro{EMC}{Electromagnetic Calorimeter}
\acro{ENC}{Equivalent Noise Charge}
\acro{EPICS}{Experimental Physics and Industrial Control System}
\acro{FADC}{Flash ADC}
\acro{FAIR}{Facility for Antiproton and Ion Research}
\acro{FE}{Front-End}
\acro{FEA}{Finite Element Analysis}
\acro{FEE}{Front-End Electronics}
\acro{FPGA}{Field Programmable Gate Array}
\acro{FS}{Forward Spectrometer}
\acro{FZ}{Forschungszentrum}
\acro{FZJ}{Forschungszentrum J\"ulich}
\acro{GEANT}{GEometry ANd Tracking}
\acro{GEM}{Gas Electron Multiplier}
\acro{GSI}{Gesellschaft f\"ur Schwerionenforschnung}
\acro{HADES}{High Acceptance DiElectron Spectrometer}
\acro{HESR}{High Energy Storage Ring}
\acro{HL}{High Luminosity}
\acro{HPTDC}{High Performance TDC }
\acro{HR}{High Resolution}
\acro{HV}{High Voltage}
\acro{IP}{Interaction point}
\acro{INFN}{Istituto Nazionale di Fisica Nucleare}
\acro{LHC}{Large Hadron Collider}
\acro{LVDS}{Low Voltage Differential Signaling}
\acro{MCP}{Micro-Channel Plate}
\acro{MDC}{Mini Drift Chamber}
\acro{MDT}{Mini Drift Tubes}
\acro{MIP}{Minimum Ionizing Particle}
\acro{MVD}{Micro Vertex Detector}
\acro{OPC}{Open Process Control}
\acro{OS}{Operating System}
\acro{PANDA}{AntiProton ANnihilation at DArmstadt}
\acro{PCB}{Printed Circuit Board}
\acro{PID}{Particle Identification}
\acro{PMT}{Photomultiplier}
\acro{PVC}{PolyVinyl Chlorid}
\acro{PWO}{Lead Tungstate}
\acro{PZC}{Pole-zero Cancellation}
\acro{QCD}{Quantum Chromo Dynamics}
\acro{RICH}{Ring Imaging Cherenkov Counter}
\acro{RMS}{Root Mean Square}
\acro{SODA}{Syncronization Of Data Acquisition}
\acro{STT}{Straw Tube Tracker}
\acro{TDC}{Time to Digital Converter}
\acro{TOF}{Time-of-Flight Detector}
\acro{TOT}{Time-over-Threshold}
\acro{TRB}{Time Readout Board}
\acro{TS}{Target Spectrometer}
\acro{VMC}{Virtual Monte Carlo}
\acro{WASA}{Wide Angle Shower Apparatus}
\end{acronym}
\vfill
%
% EOF
%

\cleardoublepage
\addcontentsline{toc}{chapter}{List of Figures}
\listoffigures
\cleardoublepage
\addcontentsline{toc}{chapter}{List of Tables}
\listoftables
%
% EOF

%
\end{document}